\documentclass[review]{elsarticle}
\usepackage[T1]{fontenc}
\usepackage{amsmath,amsfonts,amssymb}
\usepackage{booktabs}
\usepackage{dcolumn}
\usepackage[margin=2.5cm]{geometry}
\usepackage{siunitx}
\usepackage{float}
\usepackage{ulem}
\usepackage{bm}

\usepackage[table,usenames]{xcolor}
\usepackage[colorlinks=true, urlcolor=navyblue, linkcolor=navyblue, citecolor=navyblue]{hyperref}

\usepackage[utf8]{inputenc}
\usepackage{newunicodechar}
\usepackage{subcaption}
\usepackage{accents}

\usepackage{changes}
\usepackage[normalem]{ulem}

\providecommand{\+}{\cellcolor{green!70!black}}

\definecolor{blah}{RGB}{0,0,0}

\usepackage{array}

\newenvironment{EndorserList}{%
  \clearpage                       
  \thispagestyle{plain}
  \section*{Endorsements}
  \setlength{\parindent}{0pt}%
  \begin{tabular}{@{}p{0.35\textwidth}p{0.6\textwidth}@{}}%
}{%
  \end{tabular}%
}

\newcommand{\voorstander}[2]{#1 & #2\\[0.5ex]}

\newunicodechar{ }{~}

\usepackage{longtable}
\usepackage{booktabs}   
\usepackage{rotating}
\usepackage{hyperref}

\providecommand{\+}{\cellcolor{green!70!black}}
\journal{Progress in Particle and Nuclear Physics}
\biboptions{sort&compress}
\usepackage{titlesec}
\usepackage{sectsty}
\titleformat{\section}{\normalfont\Large\bfseries}{\thesection}{1em}{}
\titleformat{\subsection}{\normalfont\large\bfseries}{\thesubsection}{1em}{}
\titleformat{\subsubsection}{\normalfont\normalsize\bfseries}{\thesubsubsection}{1em}{}
\definecolor{navyblue}{rgb}{0,0.08,0.45}
\definecolor{purple}{rgb}{0.54, 0.0, 1.0}

\usepackage[mathscr,scaled=1.15]{urwchancal}
\DeclareFontFamily{OT1}{pzc}{}
\DeclareFontShape{OT1}{pzc}{m}{it}%
{<-> s * [1.15] pzcmi7t}{}
\DeclareMathAlphabet{\mathpzc}{OT1}{pzc}{m}{it}

\usepackage{graphicx} 
\usepackage{comment}

\usepackage[inline]{enumitem}
\newcommand{\diff}{\textrm{d}}

\begin{document}


\begin{frontmatter}

        \title{Hadron Physics Opportunities at FAIR}


\address[inst1]{GSI Helmholtzzentrum f{\"u}r Schwerionenforschung, Darmstadt, Germany} 
\address[inst9]{Ruhr University Bochum, Bochum, Germany}
\address[inst97]{Institut f\"{u}r Kernphysik, Goethe-Universit\"{a}t Frankfurt, Frankfurt, Germany}
\address[inst68]{Helmholtz Research Academy Hesse for FAIR (HFHF), GSI Helmholtz Centre for Heavy-Ion Research, Darmstadt, Germany}
\address[inst39]{Thomas Jefferson National Accelerator Facility (JLab), Newport News VA, USA} 
\address[inst37]{SUBATECH–Université de Nantes–IMT Atlantique–CNRS/IN2P3, Nantes, France}
\address[inst83]{Frankfurt Institute for Advanced Studies (FIAS), Frankfurt, Germany}
\address[inst14]{Instituto de Física Corpuscular (IFIC), CSIC–Universitat de València, Valencia, Spain}
\address[inst98]{School of Physics State Key Laboratory of Nuclear Physics and Technology, Peking University, Beijing, China}
\address[inst34]{High Energy Accelerator Research Organization (KEK), Tsukuba, Ibaraki, Japan} 
\address[inst52]{Stockholm University, Stockholm, Sweden}
\address[inst75]{School of Physics Engineering and Technology, University of York, Heslington, York, UK}
\address[inst40]{CERN, Geneva, Switzerland}
\address[inst33]{Variable Energy Cyclotron Centre (VECC), Kolkata, India}
\address[inst7]{AGH University of Krakow, Kraków, Poland} 
\address[inst98b]{Institute of Theoretical Physics,University of Wroclaw, Wroclaw, Poland}
\address[inst99]{Helmholtz-Zentrum Dresden-Rossendorf, Dresden, Germany}
\address[inst100]{Center for Advanced Systems Understanding (CASUS), Görlitz, Germany}
\address[inst16]{Institut f\"{u}r Theoretische Physik, Goethe-Universit\"{a}t Frankfurt, Frankfurt, Germany} 
\address[instSB]{IRFU, CEA, Universit\'{e} Paris-Saclay, Gif-sur-Yvette, France}
\address[inst58]{TUM School of Natural Sciences, Physics Department, Technical University of Munich, Garching, Germany}
\address[inst59]{Institute for Advanced Study, Technical University of Munich, Garching, Germany}
\address[inst60]{M\"unich Data Science Institute, Technical University of Munich,  Garching, Germany}
\address[inst82]{Institute of Nuclear Physics Polish Academy of Sciences, Krak\'ow, Poland}
\address[inst95]{Institut für Theoretische Physik, Universität Regensburg, Regensburg, Germany}
\address[inst3]{Florida State University, Tallahassee, Florida, USA} 
\address[inst61]{Bose Institute, Kolkata, India}
\address[inst41]{Johannes Gutenberg University Mainz, Germany} 
\address[inst69]{Justus-Liebig-Universit\"{a}t Gie\ss en, Gie\ss{}en, Germany}
\address[inst70]{University of Connecticut, Storrs, CT, USA}
\address[inst17]{George Washington University, Washington DC, USA} 
\address[inst18]{Thomas Jefferson National Accelerator Facility (JLab), Newport News VA, USA} 
\address[inst12]{Institute of Physics, University of Graz, NAWI Graz, Graz, Austria}
\address[inst11]{Departamento de F\'isica Interdisciplinar, Universidad Nacional de Educaci\'on a Distancia (UNED), Madrid, Spain} 
\address[inst667]{University of Notre Dame, Notre Dame, IN, USA}
\address[inst101]{Institut für Kernphysik, Technische Universität Darmstadt, Darmstadt, Germany}
\address[inst92]{Facility for Antiproton and Ion Research in Europe GmbH (FAIR), Darmstadt, Germany}
\address[inst53]{Institute of Physics, Jan Kochanowski University, Kielce, Poland}
\address[inst65]{Institute of Theoretical Physics, Chinese Academy of Sciences, Beijing, China}
\address[inst66]{School of Physical Sciences, University of Chinese Academy of Sciences, Beijing, China}
\address[inst76]{Joint Institute for Nuclear Research (JINR), Dubna, Russia}
\address[inst78]{Institute for Advanced Simulation (IAS-4), Forschungszentrum J\"ulich, J\"ulich, Germany}
\address[inst102]{Department of Physics, Technische Universität Darmstadt, Darmstadt, Germany}
\address[inst56]{School of Physics and Astronomy, University of Glasgow, Glasgow, United Kingdom}
\address[inst74]{Helmholtz-Institut f\"{u}r Strahlen- und Kernphysik, Universit\"{a}t Bonn, Bonn, Germany}
\address[inst93]{Department of Physics, The University of Osaka, Toyonaka, Osaka, Japan}
\address[inst24]{Nishina Center for Accelerator–Based Science, RIKEN, Wako, Saitama, Japan}
\address[inst104]{Department of Physics, Bergische Universität Wuppertal, Wuppertal, Germany}
\address[inst67]{Cyclotron Institute and Department of Physics and Astronomy, Texas
A\&M University, College Station, TX,  USA}
\address[inst73]{Helmholtz-Institut f\"{u}r Strahlen- und Kernphysik (Theorie) and Bethe Center for Theoretical Physics, Universit\"{a}t Bonn, Bonn, Germany}
\address[inst106]{Department of Physics and Astronomy, Uppsala University, Uppsala, Sweden}
\address[inst109]{National Centre for Nuclear Research, Warsaw, Poland}
\address[inst108]{Marian Smoluchowski Institute of Physics, Jagiellonian University, Kraków, Poland}
\address[inst666]{Helmholtz Institute Mainz (HIM), Mainz, Germany}
\address[inst668]{Michigan State University, East Lansing, MI, USA}
\address[inst79]{Albert Einstein Center for Fundamental Physics, Institute for Theoretical Physics, University of Bern, Bern, Switzerland}
\address[inst47]{Institut de Ci\`encies del Cosmos, Universitat de Barcelona, Barcelona, Spain} 
\address[inst84]{Departament de F\'\i sica Qu\`antica i Astrof\'isica, Universitat de Barcelona, Barcelona, Spain}
\address[inst111]{Faculty of Physics, Sofia University, Sofia, Bulgaria}
\address[inst13]{Technische Universität München (TUM), Munich, Germany}
\address[inst42]{Carnegie Mellon University, Pittsburgh, USA}
\address[inst81]{Research Center for Nuclear Physics, University of Osaka, Ibaraki, Osaka, Japan}
\address[inst90]{Departamento de F\'{\i}sica Te\'orica and IPARCOS, Universidad Complutense de Madrid, Madrid, Spain}
\address[inst85]{Instituto Superior Técnico (DF and DECN) , ULisboa, and  LIP, Laboratório de Instrumentação e Física Experimental de Partículas, Portugal}
\address[inst5]{ Dipartimento di Scienze Matematiche e Informatiche, Scienze Fisiche e Scienze della Terra, Universit\`a degli Studi di Messina, Messina, Italy}
\address[inst72]{Istituto Nazionale di Fisica Nucleare (INFN), Sezione di Catania, Catania, Italy}
\address[inst77]{Laboratoire de Physique des 2 infinis Irène Joliot-Curie, Université Paris-Saclay, CNRS-IN2P3, Orsay, France}
\address[inst62]{Instituto de Estructura de la Materia - (IEM-CSIC), Madrid, Spain}
\address[inst63]{School of Physics, Nanjing University, Nanjing, Jiangsu, China}
\address[inst64]{Institute for Nonperturbative Physics, Nanjing University, Nanjing, Jiangsu, China}
\address[inst110]{Cluster for Pioneering Research (CPR), RIKEN, Wako, Saitama, Japan}
\address[inst115]{Département de Physique Nucléaire et Corpusculaire, Université de Genève, Geneva, Switzerland} 
\address[inst86]{Department of Physics, Indiana  University, Bloomington,  IN,  USA}
\address[inst87]{Center for  Exploration  of  Energy  and  Matter, Indiana  University, Bloomington,  IN,  USA}
\address[inst88]{Theory Center, Thomas  Jefferson  National  Accelerator  Facility, Newport  News,  VA,  USA}
\address[inst89]{Department of Physics, University of Washington, Seattle, WA, USA}
\address[inst54]{Institute of Space Sciences (ICE, CSIC), Campus UAB, Carrer de Can Magrans, Barcelona, Spain}
\address[inst55]{Institut d'Estudis Espacials de Catalunya (IEEC), Barcelona, Spain}
\address[inst91]{Warsaw University of Technology, Faculty of Physics, Warsaw, Poland}
\address[inst30]{HUN-REN Wigner Research Centre for Physics, Budapest, Hungary} 

%
%

\author[inst1,inst9]{\underline{J.~G.~Messchendorp}\corref{cauthorjm}} 
\cortext[cauthorjm]{j.messchendorp@gsi.de}

\author[inst1,inst97,inst68]{\underline{F.~Nerling}\corref{cauthorfn}}
\cortext[cauthorfn]{f.nerling@gsi.de}

\author[inst39]{P.~Achenbach}
\author[inst37,inst83]{J.~Aichelin} 
\author[inst14]{M.~Albaladejo} 
\author[inst98]{L.~An} 
\author[inst34]{K.~Aoki} 
\author[inst52]{G.~Appagere} 
\author[inst9]{V.~Baru} 
\author[inst75]{M.~Bashkanov} 
\author[inst1]{A.~Bauswein} 
\author[inst1]{A.~Belias}
% ...
      

		
		



		
		
 
		\begin{abstract}

This White Paper outlines a coordinated, decade-spanning programme of hadron and QCD studies anchored at the GSI/FAIR accelerator complex. Profiting from intense deuteron, proton and pion beams coupled with high-rate capable detectors and an international theory effort, the initiative addresses 
fundamental questions related to the strong interaction featuring confinement and dynamical mass generation. This includes our understanding of hadron-hadron interactions and the composition of hadrons through mapping the baryon and meson spectra, including exotic states, and quantifying hadron structure. This interdisciplinary research connects topics in the fields of nuclear, heavy-ion, and (nuclear) astro (particle) physics, linking, for example, terrestrial data to constraints on neutron star structure.
A phased roadmap with SIS100 accelerator start-up and envisaged detector upgrades will yield precision cross sections, transition form factors, in-medium spectral functions, and validated theory inputs. Synergies with external programmes at international accelerator facilities worldwide are anticipated.
The programme is expected to deliver decisive advances in our understanding of non-perturbative (strong) QCD
and astrophysics,
and high-rate detector and data-science technology.
            
		\end{abstract}

			
        
	\end{frontmatter}
	
	\thispagestyle{empty}
    \newpage
    \begin{EndorserList}

\voorstander{Constantia Alexandrou}
{
University of Cyprus (Nicosia), Department of Physics, Nicosia, Cyprus
}

\voorstander{Gunnar Bali}{
Universität Regensburg, Regensburg, Germany
}

\voorstander{Geoffrey T. Bodwin}{
High Energy Physics Division,
Argonne National Laboratory, Lemont, Illinois, USA
}

\voorstander{Chen Chen}{
     Interdisciplinary Center for Theoretical Study, University of Science and Technology of China, Hefei, China; Peng Huanwu Center for Fundamental Theory, Hefei, China
}

\voorstander{Peng Cheng}{
    Department of Physics, Anhui Normal University, Wuhu, China
}
\voorstander{Zhu-Fang Cui}{
    School of Physics, Nanjing University, Nanjing, China
  }

\voorstander{Umberto d’Alesio}
{
Dipartimento di Fisica, Università degli Studi di Cagliari (University of Cagliari), Italy
}

\voorstander{Oleg Yu. Denisov}{
    Istituto Nazionale di Fisica Nucleare (INFN), Sezione di Torino, Torino, Italy
  }

\voorstander{Minghui Ding}{
    School of Physics and Institute for Nonperturbative Physics, Nanjing University, Nanjing, China
  }

\voorstander{Hannah Elfner}{
    Frankfurt Institute for Advanced Studies (FIAS), Institut für Theoretische Physik, Goethe Universität, Germany
  }

\voorstander{Tobias Frederico}{
    Instituto Tecnológico de Aeronáutica, São José dos Campos, Brazil
  }

%
%
\voorstander{Wolfgang Gradl}
{
Institute for Nuclear Physics, University of Mainz, Mainz, Germany
}

\voorstander{Jeremy Green}
{
Deutsches Elektronen-Synchrotron (DESY), Zeuthen, Germany
}

\voorstander{Laura Xiomara Gutiérrez-Guerrero}{
    Mesoamerican Centre for Theoretical Physics, Universidad Autónoma de Chiapas (MCTP–UNACH), Mexico
  }

\voorstander{Nasser Kalantar-Nayestanaki}{
    University of Groningen, ESRIG, Groningen, The Netherlands
  }

\voorstander{Michael Klasen}{
University of Münster, Institute for Theoretical Physics, Münster, Germany
}

\voorstander{Udo Kurilla}{
    GSI Helmholtzzentrum für Schwerionenforschung GmbH, Darmstadt, Germany
}

\voorstander{Inti Lehmann}{
    Facility for Antiproton and Ion Research in Europe GmbH (FAIR GmbH); GSI Helmholtzzentrum für Schwerionenforschung GmbH, Darmstadt, Germany
  }

\voorstander{Horst Lenske}
{
Justus Liebig University Giessen (JLU), Institute for Theoretical Physics, Gießen, Germany
}

\voorstander{Hai-Bo Li}{
    Institute of High Energy Physics, Beijing, China; University of Chinese Academy of Sciences, Beijing, China
  }

\voorstander{Zhun Lu}{
    School of Physics, Southeast University, Nanjing, China
}

\voorstander{Xiao-Rui Lyu}{
    University of Chinese Academy of Sciences, Beijing, China
  }

\voorstander{Yan-Qing Ma}
{
School of Physics, Peking University, Beijing, China
}

\end{EndorserList}

\begin{EndorserList}

\voorstander{Nilmani Mathur}
{
Department of Theoretical Physics, Tata Institute of Fundamental Research (TIFR), Mumbai, India
}

\voorstander{Volker Metag}{
    II. Physikalisches Institut, Justus-Liebig-Universität Giessen, Giessen, Germany
  }
  
\voorstander{Curtis A. Meyer}{
    Carnegie Mellon University, Pittsburgh, PA, USA
  }

\voorstander{Eulogio Oset}
{
Institute of Corpuscular Physics (UV-CSIC), University of Valencia (PCUV), Valencia, Spain
}

\voorstander{Juan Miguel Nieves Pamplona}
{
Instituto de Física Corpuscular, Parque Científico, Catedrático José Beltrán, Paterna, España
}

\voorstander{Simone Pacetti}
{
Department of Physics and Geology, University of Perugia, Perugia, Italy
}

\voorstander{Vaia Papadimitriou}{
Fermi National Accelerator Laboratory, Batavia, Illinois, USA
}
  
\voorstander{Joannis Papavassiliou}{
Department of Theoretical Physics and IFIC, University of Valencia and CSIC, Valencia, Spain
}

\voorstander{Marco Pappagallo}{
 Università di Bari \& INFN Sezione di Bari, Bari, Italy
}

\voorstander{Saša Prelovšek Komelj}
{
Faculty of Mathematics and Physics, University of Ljubljana, Slovenia
}

\voorstander{Khépani Raya Monta\~{n}o}{
    Department of Integrated Sciences and Center for Advanced Studies in Physics, Mathematics and Computation, University of Huelva, Huelva, Spain
  }

\voorstander{Anton Rebhan}
{
Institute of Theoretical Physics, Vienna University of Technology (TU Wien), Vienna, Austria
}

\voorstander{José Rodríguez-Quintero}{
    CEAFMC, Fac. Ciencias Experimentales, Universidad de Huelva, Huelva, Spain
}

\voorstander{Bernd-Jochen Schaefer}
{
Institut für Theoretische Physik, Justus-Liebig-Universität Gießen \& Helmholtz Forschungsakademie Hessen für FAIR (HFHF), GSI Helmholtzzentrum für Schwerionenforschung, Campus Gießen, Gießen, Germany 
}

\voorstander{Jochen Schwiening}{
    GSI Helmholtzzentrum für Schwerionenforschung GmbH, Darmstadt, Germany
  }

\voorstander{Jorge Segovia}{
    Área de Física Aplicada, Dpto. de Sistemas Físicos, Químicos y Naturales Facultad de Ciencias Experimentales Universidad Pablo de Olavide, de Sevilla, Spain
  }

\voorstander{Michael Unger}
{
Institute for Astroparticle Physics (IAP), Karlsruhe Institute for Technology (KIT), Karlsruhe, Germany
}

\voorstander{Marc Wagner}{
    Institut für Theoretische Physik, Goethe Universität Frankfurt am Main, Frankfurt am Main, Germany; Helmholtz Research Academy Hesse for FAIR, Frankfurt am Main, Germany
  }

\end{EndorserList}

\begin{EndorserList}

\voorstander{Xialong Wang}{
    Institute of Modern Physics, Fudan University, Shanghai, China
}

\voorstander{Hartmut Wittig}{
    Johannes Gutenberg-Universität Mainz, Mainz, Germany
}

\voorstander{Jia-Jun Wu}{
    School of Physical Sciences, University of Chinese Academy of Sciences (UCAS), Beijing, China
  }

\voorstander{Chang-Zheng Yuan}
    {
    Institute of High Energy Physics (IHEP), Chinese Academy of Sciences (CAS), Beijing, China
    }

\end{EndorserList}


    \newpage
	\tableofcontents
	



%
%
\newpage
\section*{Executive summary}
\addcontentsline{toc}{section}{Executive summary}

 \subsection*{A cross-community hadron-physics programme}
This document presents a scientific vision and strategic framework for a comprehensive, decade-long research programme centred at the GSI/FAIR accelerator complex in Germany. The initiative is the result of extensive discussions among researchers from a wide range of experimental and theoretical backgrounds. These discussions were facilitated via a series of international workshops \cite{SIS100Workshop2023, SIS100Workshop2024, QCDFair2024, QCDFair2025}, the establishment of a dedicated physics working group to evaluate (part of) the experimental feasibility of the programme \cite{PWGCBM}, and engagement with the broader nuclear and hadron physics community ({\it e.g.} Ref.\,\cite{Messchendorp02012025}).

\begin{figure}[ht]
\begin{center}
\vspace*{-0.3cm}
\includegraphics[width=0.82\textwidth]{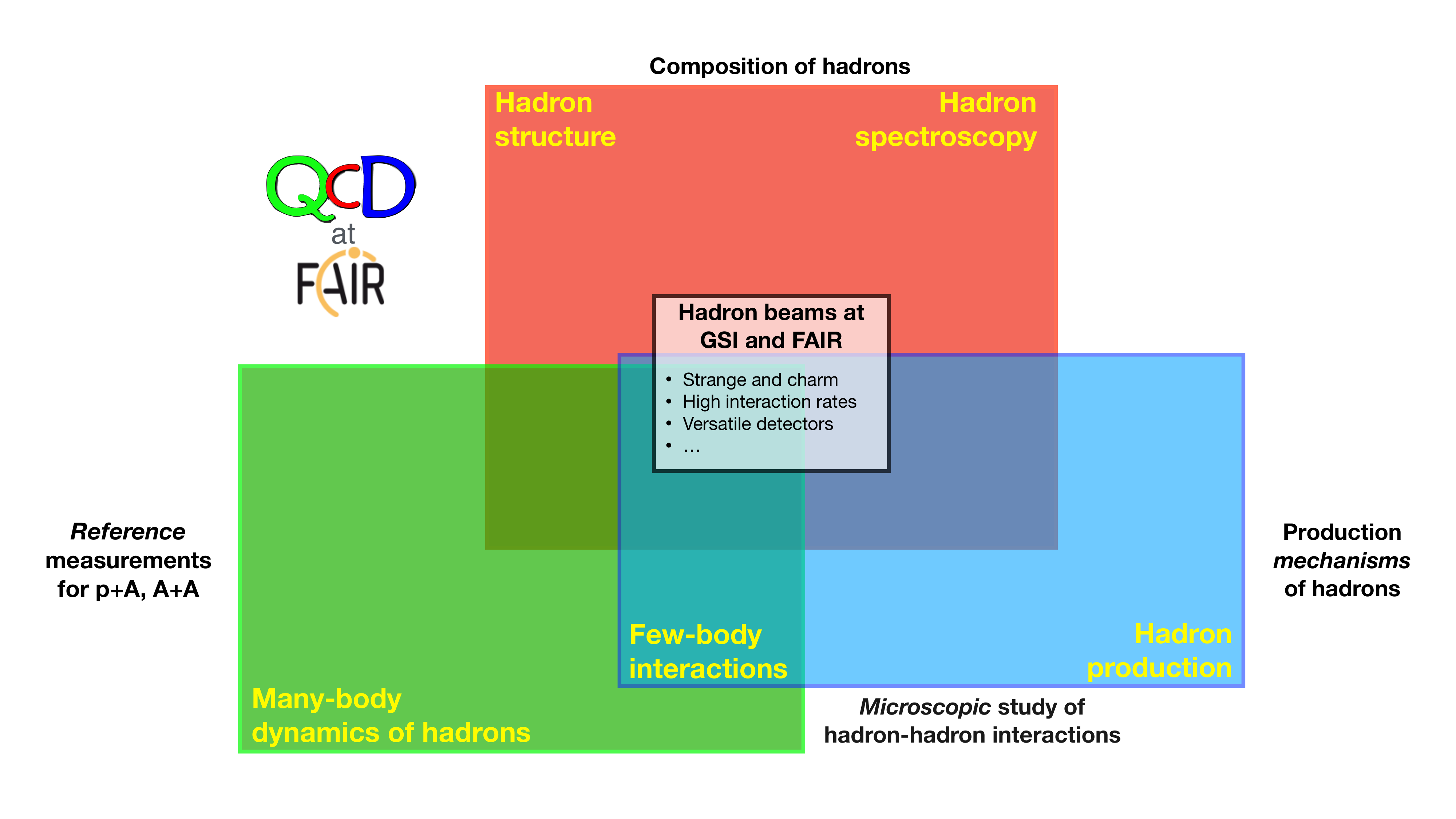}
\vspace*{-0.8cm}
\caption{\small 
Conceptual motivation of the proposed hadron physics programme at GSI/FAIR using hadronic beams. The programme aligns complementary experimental and theoretical efforts from various disciplines in a drive towards a unified microscopic understanding of strongly interacting matter.}
\label{fig:mindmap}
\end{center}
\end{figure}

The objective of the programme is to exploit high-intensity hadron beams, primarily pions, protons and deuterons, to unravel fundamental aspects of quantum chromodynamics (QCD), particularly in the non-perturbative regime. The initiative combines diverse experimental communities, with their cutting-edge methods, and a coordinated theoretical effort to explore the strong interaction and its implications across nuclear, hadron, heavy-ion, and astrophysical domains. Figure~\ref{fig:mindmap} sketches the conceptual motivation of the foreseen programme, highlighting its overarching objectives and their interconnections among the various disciplines addressing our understanding of strongly-interacting hadronic matter.

\noindent The hadron physics programme addresses aspects of some of the most profound questions in modern physics (Sec.\,\ref{sec.intro}): 
\begin{itemize}[itemsep=0.2em, parsep=0pt, topsep=0.3em,label=\(\circ\),left=0pt]
    \item How does the strong force give rise to massive hadrons like the proton?    \item Which symmetries govern the nature of matter in the non-perturbative QCD regime?
    \item How can the full hadron spectrum, including exotic and hybrid states be mapped and understood?
    \item At the most fundamental level, what are the essential degrees of freedom underlying hadronic interactions?
    \item What are the implications of these microscopic properties on phenomena that occur on a cosmic scale? 
\end{itemize}
These investigations address both the fundamental composition and interactions of hadrons as well as their properties and dynamics in extreme environments. Using hadron beams, the programme aims to map the spectrum of 
(conventional and exotic)
baryons and mesons, determine low-energy interaction parameters, extract their electromagnetic and weak transition form factors, and study their in-medium modifications under baryon-rich conditions. In particular, hadrons with multi-strange quark content will be explored with high precision and opportunities for investigating baryons and mesons with charm-quark content will be exploited. The resulting data will yield crucial insights into nucleon structure and advance our understanding of compact stellar objects (neutron stars, supernovae, binary neutron star mergers), the dense-matter equation of state (EoS), and phenomena relevant to dark matter and early-universe cosmology.

The experimental effort is complemented by a broad theoretical framework based on the quantum field theory of strong interactions, QCD, and ranging from effective field theories to  Dyson-Schwinger and Bethe-Salpeter equations techniques  and lattice QCD numerical simulations (Sec.~\ref{sec.Tools}).
Together these tools enable a detailed interpretation of hadron observables, from scattering amplitudes and spectral functions to binding energies and decay widths. In parallel, advanced data analysis techniques, including partial-wave analysis, dispersive methods and femtoscopy, as well as machine learning and artificial intelligence approaches are further being developed or deployed to maximise the scientific output from the experimental data.

\subsection*{Realisation of the hadron physics programme at GSI/FAIR}
The programme is implemented through a three-phase roadmap aligned with the staged realisation of FAIR (Sec.~\ref{sec:roadmap}). 

Phase-0 exploits the existing SIS18 capabilities (Sec.\,\ref{sec.FairGsi}) with HADES (Sec.\,\ref{sec.hades}) and hadron-physics-motivated activities with WASA at the Fragment Separator (FRS) (Sec.~\ref{sec.frs-wasa}). These studies are already providing valuable data on, {\it e.g.}, light-baryon structure, strangeness production and medium-induced effects in hadronic matter. Forthcoming experiments employing secondary pion beams at SIS18, in conjunction with HADES, will offer a unique opportunity for microscopic investigations of baryon structure and dynamics within the framework of SU(3)-flavour symmetry.

The next phase is anchored by the upcoming SIS100 synchrotron advancing the experimental capabilities (Sec.~\ref{sec.FairGsi}). With proton-beam momenta spanning a broad range -- from 5 to 30\,GeV/$c$ -- and a major increase in intensities, this stage will grant access to the high-mass resonance region, including the production of a wide spectrum of hadrons containing multiple strange and charmed quarks, and allows for high selectivity in studying specific final states and reaction channels. In addition, SIS100 can accelerate and deliver a wide range of beams, including deuterons and heavy-ion projectiles.

The Compressed Baryonic Matter (CBM) experimental setup (Sec.~\ref{sec.cbm}) will serve as the flagship detector system, providing high-rate data acquisition and precise tracking to resolve complex hadronic topologies and final states. Moreover, this phase will enable the use of the Super-Fragment Separator (Super-FRS), a high-acceptance in-flight rare-isotope separator with the capability to deliver secondary pion and antiproton beams as well (Sec.\,\ref{sec.frs-wasa}).

The final phase identified in this report envisions operation of the High-Energy Storage Ring (HESR) with high-intensity, momentum-cooled antiproton beams. In combination with a versatile detector system, HESR promises unprecedented resolution for the study of baryon–antibaryon annihilation, the spectroscopy of exotic states -- such as glueballs and hybrids, the production of numerous hyperon–antihyperon pairs, the formation of multi-strange hypernuclei, and much more (Sec.\,\ref{sec:hesr-physics}).

This paper primarily outlines hadron-physics opportunities enabled by pion-beam experiments at SIS18 during Phase-0, and by the subsequent use of SIS100 proton and deuteron beams.

\subsection*{From elementary interactions to properties of baryon-rich environments}
The aim is to explore the fundamental nature of hadronic matter and its interactions. This encompasses studies of the (low-energy) degrees of freedom of hadron–hadron interactions, to reveal the composition of hadrons, including exotic configurations, and hadronic reaction processes that serve as a reference for our understanding of baryon-rich matter. The following flagship topics have been identified, which are well suited to FAIR’s beam and detector capabilities:

\begin{itemize}[itemsep=0.2em, parsep=0pt, topsep=0.3em, left=0pt]
    \item Studies of {\bf hadron–hadron interactions in the SU(3)-flavour sector} (Sec.\,\ref{sec.HadrHadrInter}). This includes meson–meson, meson–baryon, and baryon–baryon systems. 
    Final-state interactions in light meson–meson systems ($\pi\pi$, $K\bar{K}$, $\pi K$) can rigorously be investigated via precision measurements, and hence combined with partial-wave analysis and deploying dispersive approaches to extract scattering parameters. 
    Pion beams will enable elastic and inelastic pion–nucleon scattering, providing access to the baryon spectrum and essential input for understanding many-body dynamics in heavy-ion reactions. 
    Another central goal is to obtain data on hyperon–nucleon and hyperon–hyperon interactions, crucial for understanding the properties of matter at high densities. 
    For near-threshold production with large momentum transfer, scattering lengths can be extracted from final-state interactions in a model-independent manner using dispersion techniques. 
    These methods are particularly valuable for hyperon–nucleon and hyperon-meson interactions where low-momentum scattering data are scarce. 
    For example, the $\Lambda p$ scattering length has been determined using this method from the $\vec{p}p\rightarrow pK\Lambda$ reaction. This programme aims to extend the technique to hyperon-hyperon systems. Complementary to this, the programme proposes to measure short-range particle correlations (femtoscopy) to extract information on hadron–hadron scattering parameters. 
    It is especially powerful for studying interactions of exotic hadrons, particularly those with open-charm, and for probing the microscopic dynamics of light nuclei formation, such as deuterons. 
    Femtoscopy will allow a direct comparison with results from hypernuclei and dispersive analysis for systems like $\Lambda\Lambda$.
    Precise measurements of hyperon–meson interactions will constrain theoretical models, clarify how coupled-channel dynamics shape the spectrum of excited states, and provide essential input for CP-violation tests in hyperon weak decays.
    
    \item {\bf Hadron spectroscopy}, through line-shape measurements and measurements of production and decay properties, offers a unique opportunity for the systematic study of hadron composition.
    Intense pion beams from SIS18 are ideally suited for precision baryon spectroscopy, enabling studies well within the third-resonance region and with various final states, including baryons and mesons in the strangeness $|S|=1$ domain ({\it e.g.} Sec.\,\ref{sec4:subsec:Doring}).     
    Proton beams from SIS100 will allow systematic studies of a spectrum of hadrons containing {\bf multi-strange} and {\bf charm} quarks and in particular the role of light-heavy quark correlations ({\it e.g.} Sec.\,\ref{sec:baryonspectra_exp}). 
    
    \item The discovery and systematic characterisation of {\bf exotic hadrons} will be addressed by measurements of the decays and line shapes with excellent resolution.
    For instance, the reaction $pp\rightarrow ppJ/\psi$ can be used to complete the search for hidden-charm pentaquarks ({\it e.g.} Sec.\,\ref{ch6:exotic_prospects}). FAIR is well positioned to discover not only the missing non-strange pentaquarks and thereby help complete the expected spin multiplets, but especially exotic states in the strange sector. 
    
    \item {\bf Form factors} provide essential information on the internal structure of hadron states, defining their size and shape based on the probe used. 
    FAIR experiments will enable measurements of electromagnetic transition form factors (ETFF) for a wide range of baryons, including excited hyperon states, providing insights into isospin and flavour symmetry breaking, as well as the role of meson clouds. 
    Dalitz decays and semi-leptonic weak decays offer complementary information for unstable hadrons (Sec.\,\ref{sec:el-transFF}). 
    HADES has already pioneered measurements of ETFFs for $N^*/\Delta(1232)$ in the time-like region from Dalitz decays, confirming the dominance of magnetic transitions and pion cloud effects. 
    Higher production cross sections expected at SIS100 energies and a better vertex resolution compared to that of HADES will extend these studies to hyperons, like $\Lambda(1405,1520)$ and $\Sigma(1385)$. 
    Furthermore, semi-leptonic weak decays of octet and decuplet hyperons will allow for the determination of vector and axial transition form factors, which are crucial for understanding SU(3)-flavour symmetry violations, improving the determination of the $V_{us}$ element of the CKM matrix and benchmarking future lattice QCD calculations. 
    
    \item A key puzzle is whether the proton contains non-negligible charm–anticharm ($c$+$\bar{c}$) content beyond that which is expected from QCD evolution. 
    Recent global fits provide some evidence for {\bf intrinsic charm of the proton}, but further experimental tests are needed. 
    FAIR, with its ability to explore $pp$ scattering into final states involving mesons with both open and hidden charm, can deliver critical new information, especially at low energies where intrinsic charm contributions might be relatively large ({\it e.g.} Sec.\,\ref{sec:intrinsic_charm}).
    
    \item {\bf Generalised Parton Distributions} (GPDs) provide access to hadron `gravitational' form factors, which characterise the in-hadron expectation value of the QCD energy–momentum tensor, yielding information on mass, spin, and pressure distributions. 
    With proton beams at SIS100, measurements of hidden- and open-charm final states will offer crucial input to this field, thereby advancing our understanding of the origin of the proton mass and further clarifying the role of gluons 
    ({\it e.g.} Sec.\,\ref{sec:intrinsic_charm}).

    \item Dilepton invariant-mass spectra provide a sensitive probe of {\bf in-medium modifications of vector-meson properties}, such as mass shifts and spectral broadening (Sec.\,\ref{subsub:inmediumpropvect}). HADES data from proton-induced reactions have already delivered the first measurements of low-momentum dielectron radiation off cold nuclear matter. At FAIR, the CBM experiment will extend these studies with high-precision measurements of dilepton spectra and angular distributions in $pA$ collisions, enabling the disentanglement of the underlying production mechanisms, including potential contributions from a quark–gluon plasma (QGP) in $AA$ reactions.
    Chiral symmetry restoration can be probed by testing the degeneracy of chiral-partner spectral functions ({\it e.g.}\ $\rho$/$a_1$). Key observables are thermal-dilepton invariant-mass spectra: the low-mass region $M_{e^+e^-}<1$~GeV/$c^2$ (in-medium $\rho$) and $1.1<M_{e^+e^-}<1.5$~GeV/$c^2$ ($\rho$–$a_1$ mixing). At FAIR energies, pre-equilibrium Drell Yan (DY)-like annihilation obscures the spectra; because a perturbative description is lacking, isolating thermal radiation requires precise constraints on DY-like contributions from $pp$ and $pA$ data.
    
    \item The programme will study the {\bf in-medium properties of strange pseudoscalar mesons ({\it e.g.}, $\bar{K}$, $K$) and hyperons ({\it e.g.}, $\Lambda$, $\Sigma$, $\Sigma(1385)$)}. 
    Knowledge of these properties is vital for understanding the early universe and neutron stars. 
    The experiments will explore how in-medium effects alter hadron spectral functions and cross sections, as exemplified by theoretical predictions for $K^{-}p$ cross sections (Sec.~\ref{subsub:inmediumpropstrbar}).
    
    \item Investigations into the {\bf in-medium properties of open and hidden charmed hadrons} are also foreseen. 
    These include studies of charmed meson ($D$, $D^*$) and charmonium ($J/\psi$) in nuclear matter, which is essential for understanding $J/\psi$ survival probability in the QGP and cold nuclear matter effects (Sec.\,\ref{subsub:inmediumpropopphidd}).
    
    \item The study of nuclei where one or more nucleons are replaced by hyperons or charmed baryons offers unique insights into strong interactions. 
    In particular, studies of {\bf hypernuclei} provide important information on hyperon–nucleon ($YN$) and hyperon–hyperon ($YY$) interactions, especially at low energies. 
    Precise determination of binding energies and lifetimes can constrain three-body forces, which are crucial for resolving the ``hyperon puzzle'' in neutron stars. 
    FAIR will enable high statistics analysis of light hypernuclei -- like hypertriton, hyperhydrogen-4, and hyperhelium-4 -- allowing for unprecedented accuracy in this sector (Sec.\,\ref{subsec.hypernuclei_in_piA}). 
\end{itemize}

\subsection*{Linking hadron physics to the cosmos}

The GSI/FAIR programme establishes strong synergies with (nuclear) astrophysics and astroparticle physics, providing vital constraints on the nuclear EoS and insights into extreme cosmic environments and dark matter (Sec.\,\ref{sec.AstroConnection}). 
High-intensity proton and pion beams at GSI/FAIR provide essential input for transport models, enabling precise extraction of nuclear matter properties from heavy-ion collisions, which are critical for interpreting gravitational wave signals from neutron star mergers.


Hyperon emergence in neutron-star cores --- which tends to soften the equation of state (EoS) and is potentially in conflict with the existence of $\sim 2 M_\odot$ pulsars (the ``hyperon puzzle'') --- will be constrained at FAIR by precision measurements of hyperon--nucleon and hyperon--hyperon interactions in production reactions and hypernuclear studies (Sec.\,\ref{subsubsect.astro}).


Improved measurements of, {\it e.g.}, the production cross sections of $^9{\rm Be}$ and $^{10}{\rm Be}$ in $pA$ reactions in the GeV domain are urgently needed to refine models of galactic cosmic ray (GCR) propagation. 
Although space-based data from AMS-02 have reached high precision, uncertainties in nuclear cross sections remain a limiting factor. 
Super-FRS at FAIR, with the high rigidity and acceptance, and capabilities of CBM at 10\,GeV/nucleon, will deliver crucial data, including those for key isotopes like $^{10}$Be, enhancing hadronic interaction models and extensive air shower simulations (Sec.\,\ref{subsec.hypernuclei_in_piA}). 

The programme also probes physics beyond the Standard Model. 
At GSI/FAIR, dark photon ($A^\prime$) searches in $e^+e^-$ pairs can be performed, constraining the existence in the 0.02–0.55\,GeV$/c^2$ mass range. 
Studies of $\eta/\eta'$ decays at SIS100 offer complementary paths to explore axion-like particles (ALPs), benefitting from higher luminosities and production cross sections (Sec.\,\ref{sec.astro_dm}).


\subsection*{A worldwide, complementary facility for hadron physics}

The proposed programme lies at the centre of a coordinated global effort to decode the strong interaction, extending, rather than duplicating, the capabilities of other flagship facilities. 
High‑precision electron and photon beams at Jefferson Lab and at ELSA illuminate nucleon spectroscopy, structure and exotic‑meson spectra; FAIR supplies the indispensable hadronic counterpart. 
J‑PARC’s beams map strangeness and flavour, while FAIR broadens the terrain in isospin and baryon density. 
The results align with CERN’s SPS and LHC campaigns on hadron structure and dense baryonic matter, and with BESIII’s detailed studies of, {\it e.g.}, charmonium-like and open‑charm states. 
Collectively, this network mounts a full‑spectrum assault on enduring puzzles of QCD ({\it e.g.} Sec.\,\ref{sec.IntContext}).

At FAIR’s core are the HADES (Sec.\,\ref{sec.hades}) and CBM (Sec.~\ref{sec.cbm}) spectrometers, versatile, high‑rate instruments purpose-built for the terra‑incognita energy window. 
With large acceptance, sub‑percent momentum resolution, powerful particle identification and, in the case of CBM, triggerless multi‑MHz readout, they provide kinematically exclusive studies of elementary $\pi p$ and $pp$ reactions and high‑statistics semi-inclusive measurements on complex nuclear targets. 
Their complementary energy ranges, lepton/hadron specialisations, and flexible trigger strategies give FAIR a uniquely powerful facility for precision exclusive measurements, delivering exactly the data needed to drive hadron physics into the next, QCD‑centric era. 
Strong synergy with theory and technological innovation underpins this effort. Cutting‑edge continuum and lattice Schwinger function calculations, chiral effective‑field theory, and dispersive analysis techniques now reach the precision where FAIR data can provide decisive tests of `strong' QCD.

\subsection*{What to expect from this White Paper}

This document presents the physics opportunities at GSI/FAIR over the coming decade, focusing on a programme driven by intense primary proton and secondary pion beams. 
It contextualises the proposed research within the current experimental and theoretical state of the art and, in doing so, provides both a review of the field and an outlook to the potential of this programme. 
As a White Paper, it offers indicative evidence of the feasibility for selected benchmark cases~\cite{cbm_feasibility_note}. 
More comprehensive feasibility studies will follow in dedicated publications.

\newpage
\section{Key questions in strong interaction physics}
\label{sec.intro}

{\small {\bf Convenors:} \it J.~G.\ Messchendorp, F.\ Nerling, C.~D.\ Roberts} 


%
\noindent Strong interactions in the Standard Model of particle physics (SM) are described by QCD, a Poincar\'e invariant quantum gauge field theory built upon the non-Abelian group SU$(3)$.  
In many ways analogous to quantum electrodynamics (QED), one might argue that the door to QCD was opened by the quark model \cite{Gell-Mann:1964ewy, Zweig:1964ruk, Zweig:1964jf} and the path thereafter paved by the proof of asymptotic freedom \cite{Gross:1973id, Politzer:1973fx}.  
The first comprehensive QCD review was presented almost fifty years ago \cite{Marciano:1977su}; and whilst much progress has been made since then, a great deal remains to be understood.

\begin{figure}[t]
\centering
\includegraphics[width=0.56\linewidth]{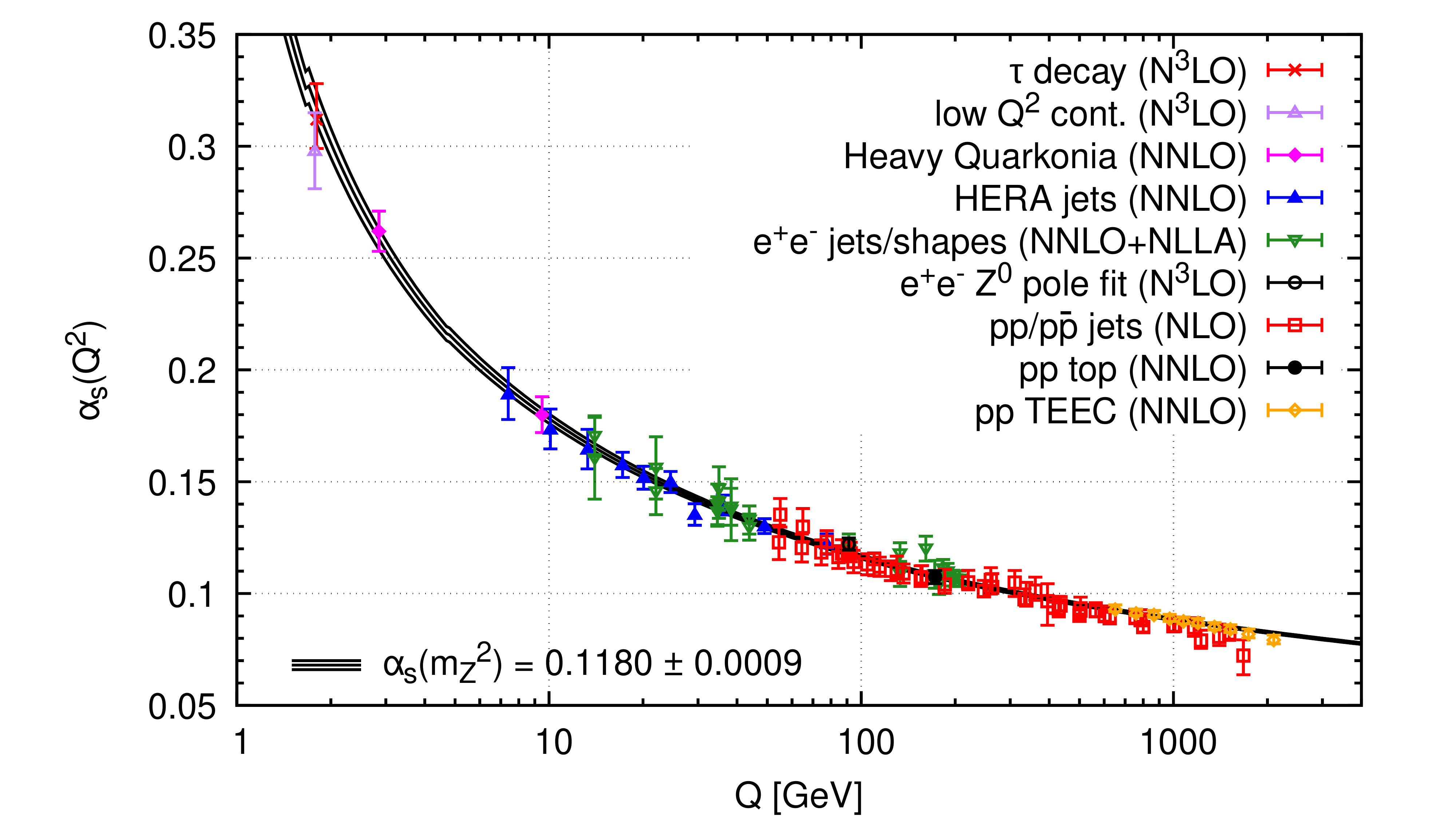}
\includegraphics[width=0.43\linewidth]{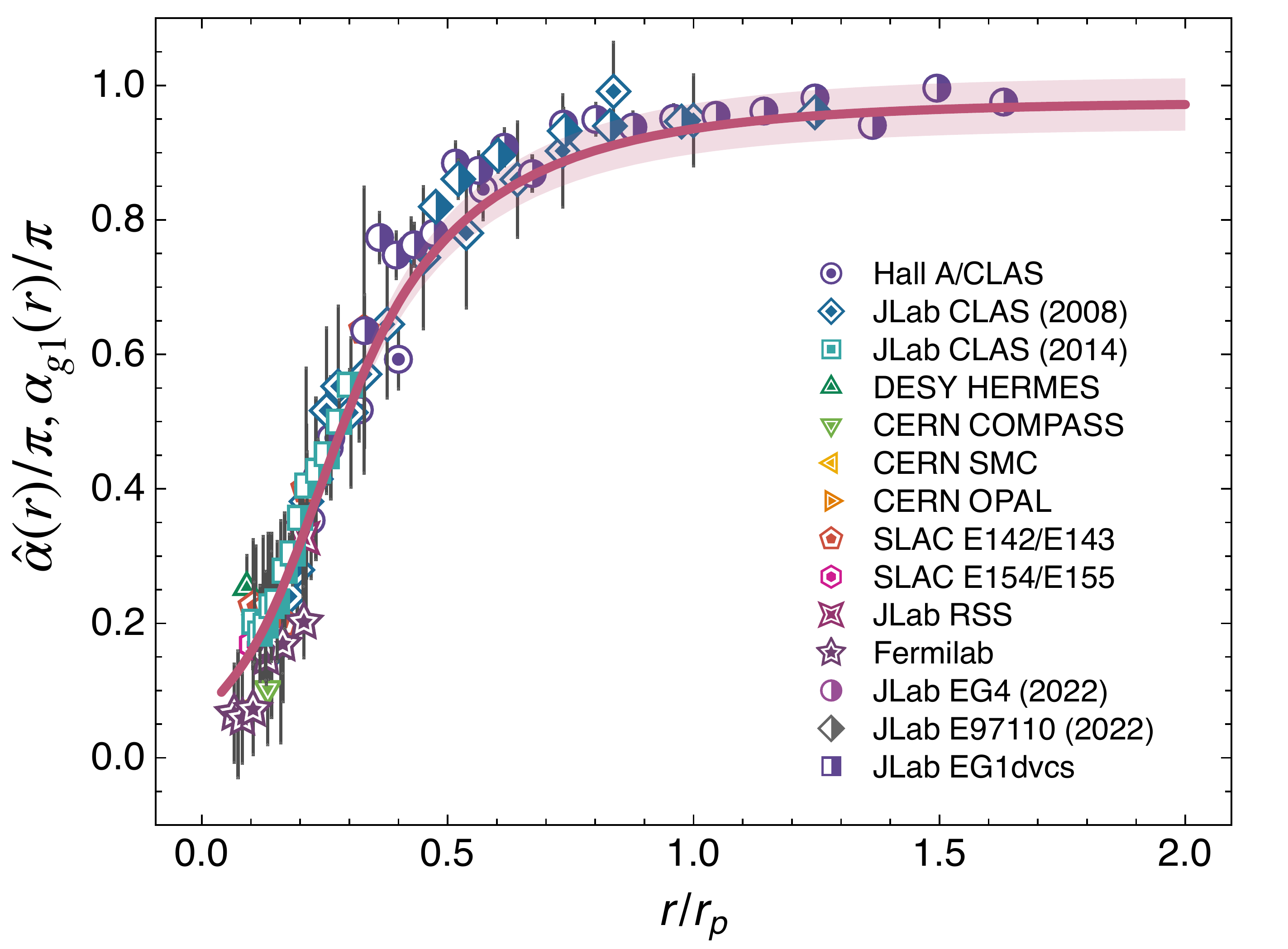}
\vspace*{-1cm}
\caption{\label{FigEffectiveCharge}
Left panel: Perturbative QCD prediction for the running coupling constant $\alpha_s(Q)$ compared to experimental determinations. 
(Figure by Particle Data Group \cite{ParticleDataGroup:2024cfk}).
Right panel: Process-independent effective charge, $\hat{\alpha}(r)/\pi$, calculated at all length-scales using results from non-perturbative continuum and lattice analyses of QCD's gauge sector~\cite{Cui:2019dwv, Deur:2023dzc} and plotted in units of $r_p$, the proton charge radius \cite{ParticleDataGroup:2024cfk}.
Also plotted are existing data on the process-dependent charge $\alpha_{g_1}$ -- see Refs.\,\cite{%
Deur:2005cf, Deur:2008rf, Deur:2014vea, Deur:2022msf,
Ackerstaff:1997ws, Ackerstaff:1998ja, Airapetian:1998wi, Airapetian:2002rw, Airapetian:2006vy,
Kim:1998kia,
Alexakhin:2006oza, Alekseev:2010hc, Adolph:2015saz,
Anthony:1993uf, Abe:1994cp, Abe:1995mt, Abe:1995dc, Abe:1995rn, Anthony:1996mw, Abe:1997cx, Abe:1997qk, Abe:1997dp, Abe:1998wq, Anthony:1999py, Anthony:1999rm, Anthony:2000fn, Anthony:2002hy}.
The QCD charge is weak at short distances (asymptotic freedom) and tends to saturate to a finite value at long range, opening a conformal window and providing for a stable infrared completion of the theory.
(Figure courtesy of D.\ Binosi.)
}
\end{figure}

In working to understand SM strong interactions today, one therefore begins with the 
Lagrangian density:
\begin{subequations}
\label{QCDdefine}
\begin{align}
{\mathpzc L}_{\rm QCD} & = \sum_{{\mathpzc f}=u,d,s,\ldots}
\bar{q}_{\mathpzc f} [\gamma\cdot\partial
 + i g \tfrac{1}{2} \lambda^a\gamma\cdot A^a+ m_{\mathpzc f}] q_{\mathpzc f}
 + \tfrac{1}{4} G^a_{\mu\nu} G^a_{\mu\nu},\\
%
%
\label{gluonSI}
G^a_{\mu\nu} & = \partial_\mu A^a_\nu - \partial_\nu A^a_\mu -
g f^{abc}A^b_\mu A^c_\nu,
\end{align}
\end{subequations}
written here with a Euclidean metric convention, as common in nonperturbative treatments \cite[Ch.\,1]{Ding:2022ows}, where $\{q_{\mathpzc f}\,|\,{\mathpzc f}=u,d,s,c,b,t\}$ are fields associated with the six known flavours of quarks; 
$\{m_{\mathpzc f}\}$ are their current masses, generated by Yukawa couplings of the quarks to the Higgs field
$\{A_\mu^a\,|\,a=1,\ldots,8\}$ represent the gauge (gluon) fields, generalisations of the photon, but with a non-Abelian character encoded in $\{\tfrac{1}{2}\lambda^a\}$, the generators of SU$(3)$ in the fundamental representation; 
and $g$ is the QCD coupling.
Following quantisation, $g$ increases with decreasing momentum transfer $Q$. 
The scale dependence, or ``running", of the strong coupling, conventionally defined as $\alpha_s=g^2/4\pi$, is depicted in Fig.\,\ref{FigEffectiveCharge}\,--\,left panel. 
In Eq.\,\eqref{QCDdefine}, $G^a_{\mu\nu}$ is the gluon field strength tensor, with $f^{abc}$ being the antisymmetric structure constants of the SU$(3)$ colour gauge symmetry group.  
The associated $A^b_\mu A^c_\nu$ term is the principal feature which distinguishes QCD from QED.  
Whilst appearing to be a small change, this modification completely alters the character of the theory because it introduces self-interactions among the gauge bosons, \textit{i.e}., $3$-gluon and $4$-gluon vertices.

It is such gluon self-interactions that deliver asymptotic freedom in QCD \cite{Gross:1973id, Politzer:1973fx}, \textit{i.e}., 
logarithmically weaker than $1/r$ forces between color charge carriers at short distances, $r$; 
and, according to predictions developed using a combination of continuum methods and numerical simulation of lattice-regularized QCD, ensure a stable long-wavelength completion of the theory~\cite{Deur:2023dzc, Brodsky:2024zev}.
These features are expressed in the QCD analogue of the Gell-Mann--Low QED running coupling, $\hat \alpha$, drawn in Fig.\,\ref{FigEffectiveCharge}\,--\,right panel.
The image compares the theory prediction for a process-independent running coupling \cite{Cui:2019dwv} with data on the directly measurable process-dependent effective charge defined via the Bjorken sum rule, $\alpha_{g_1}$ \cite[Sec.\,4.3]{Deur:2023dzc}.
On any length-scale domain for which QCD perturbation theory is valid, $r/r_p\lesssim 1/4$, the ratio of these two couplings is unity up to corrections with strength $\alpha_{\rm \overline{MS}}(r^2)/20$, where $\alpha_{\rm \overline{MS}}$ is a standard perturbative QCD coupling.
{\color{blah} On the complementary domain, since both charges are defined via operators that do not couple to isospin singlet channels, many dynamical contributions that might distinguish between them are eliminated.}
This explains why they continue to follow similar trajectories with increasing $r$.  
By definition, the Bjorken effective charge saturates: $r\gg r_p$, $\alpha_{g_1}/\pi = 1$.
In this same limit, $\hat\alpha/\pi = 0.97(4)$.

If one removes the Yukawa couplings of the quarks to the Higgs field, so that $\{m_{\mathpzc f}\equiv 0 \}$ in Eq.\,\eqref{QCDdefine}, then the classical action associated with ${\mathpzc L}_{\rm QCD} $ is scale-invariant. A scale-invariant theory cannot produce compact bound states.
Indeed, scale-invariant theories do not support dynamics, only kinematics \cite{Roberts:2016vyn}. So, if Eq.\,\eqref{QCDdefine} is really capable of explaining, among other things, the proton's mass, size, and stability, then remarkable features must emerge via the process of defining and completing quantum chromodynamics.

When attempting to explain observables using QCD, one basic fact is para\-mount, \textit{viz}.\ the gluon and quark degrees of freedom used to write ${\mathpzc L}_{\rm QCD}$ are not the objects that reach detectors.  
Confinement, an intricate concept \cite[Sec.\,5]{Greensite:2011zz,Ding:2022ows}
entails that only QCD-charge (colour) neutral bound states (hadrons such as protons or pions) can be detected. This means that all connections between QCD and experiment demand knowledge of non-perturbative QCD physics. At the same time, on the experimental side, 
processes involving hadron beams serve as a useful and sensitive probe to investigate this regime of strong interactions.

Arguably, the most fundamental QCD bound state is the proton. Uniquely, it appears to be absolutely stable; at least, the lower limit on its half-life is $10^{24}$-times longer than the time since Big Bang \cite{Ohlsson:2023ddi}. 
%
The proton mass, $m_p = 0.938\,$GeV$/c^2$ \cite{ParticleDataGroup:2024cfk}, is a defining scale in Nature. In fact, since all visible matter in the Universe is composed primarily of atomic nuclei, then one understands the origin of almost all perceptible mass if one grasps the source of $m_p$.
As a bound state seeded by three light valence quarks ($u+u+d$), one might expect $m_p \approx 2 m_u + m_d$.  Naively, 
this sum is just a few percent of the proton mass; for recent reviews see Refs.~\cite{Hoferichter:2025ubp,Horn:2016rip}. 
Looking back now at Eq.\,\eqref{QCDdefine}, the proton mass scale is conspicuously absent. These remarks highlight three of the most fundamental 
aspects in science: what is confinement; how does the mass of the proton emerge; and are these two phenomena -- confinement and emergence of the proton mass -- connected?

Related to the proton, there is what might seem to be a simpler system; namely, the $J^{PC}=0^{-+}$ (pseudoscalar) $\pi$-meson (pion), seeded by one light valence quark and one light valence antiquark partner, \textit{e.g}., $\pi^+ = u + \bar d$.  
Without the pion, nuclei could not form \cite{RevModPhys.21.474}.  Moreover, to effect binding in all known nuclei, the pion must be much lighter than the proton.  
Empirically, $m_\pi =0.14\,$GeV$/c^2 \approx \tfrac{1}{7} \, m_p$ -- the $\pi$-meson has a $\mu$-lepton-like mass \cite{ParticleDataGroup:2024cfk}.  
A conundrum is immediately apparent; namely, how is it that, in the same theory, a two valence-quark and a three valence-quark system can have such widely differing masses; 
especially when the mass of the $\rho$-meson, a pion-related $J^{PC}=1^{--}$ two valence body system, satisfies $m_\rho / m_p \approx \tfrac{4}{5}$ and
the first excited state of the pion ($\pi(1300)$) is nearly as heavy as the analogous excitation of the nucleon $N(1440)$? 
Indeed, current state-of-the-art numerical simulations of lattice-regularised QCD confirm this pattern \cite{Yan:2025mdm}.

Evidently, any attempt to describe how the large proton mass emerges in Nature must also explain why the pion remains unnaturally light, \textit{i.e}., 
how the same nonperturbative dynamics that generate the large proton mass simultaneously deliver a nearly massless pion.
The key to understand this is the dichotomous character of the pion, \textit{viz}.\ its dual status as both QCD bound state and Nature's most fundamental Nambu-Goldstone boson \cite{Nambu:1960tm, Goldstone:1961eq, Horn:2016rip}.  
Looking at it from the perspective of QCD Lagrangian symmetries, one realises that dynamical breaking of the chiral symmetry necessitates the presence of nearly massless degrees of freedom, the Goldstone bosons, \textit{i.e}. pions 
and other pseudoscalar mesons. 
Those in turn can serve as effective degrees of freedom for strong interaction analyses at low energies~\cite{Weinberg:1978kz, Gasser:1983yg, Gasser:1987rb, Bernard:1992qa}.
It is exactly those questions which define the field of hadron spectroscopy, which aims to provide a coherent understanding of the pattern of and creation mechanisms for the basic building blocks of nature; see Refs.\,\cite{Eichmann:2016yit, Mai:2022eur} and much of the discussion that follows.


Numerous questions emerge from these fundamentals.  
For instance, Yukawa couplings of the quarks to the Higgs field produce large differences between current-quark masses.
With a current mass on the order of $100\,$MeV$/c^2$, the $s$ (strange) quark may be described as the heaviest light quark; and the $c$ (charm) quark, whose current mass is roughly ten-times larger, is the lightest heavy quark. 
Unravelling the impact of these masses on hadron spectroscopy, stability, structure, and interactions is essential to delivering a comprehensive understanding of strong-interaction physics. SU$(3)$-flavour symmetry in the light-quark sector is broken: how is that breaking expressed and can it be predicted and understood? 
Indeed, much of the excited spectrum of hadrons with strangeness is not well explored, making new experimental facilities essential for advancing our understanding of strong interaction. 
Turning to the $c$ quark, whose current mass exceeds that of the proton, one is immediately inclined to ask: how does this change the emergent properties, character, and interactions of systems that contain one or more $c$ quarks? 
Then, suppose one forms hadrons into a cold, dense matter: how do the properties of the individual hadrons change and what does this reveal about the theory of strong interactions? 

The basic questions and huge array of contingent implications that are presented by Nature for explanation by a theory of strong interactions are now plain.  
What is the origin of the nuclear-physics mass scale, $m_p$, that characterises all visible matter; 
%
%
how is the underlying mechanism or phenomenon expressed in measurable quantities;
%
%
how is the impact of Higgs-induced quark current-masses expressed in observable quantities;
and what features do dense collections of hadrons possess that isolated hadrons do not?
Given the number and diverse character of hadrons, one can be certain that the expressions are manifold; indeed, often system specific.  
This is fortunate because it means that model predictions and putative explanations drawn more directly from QCD can be tested against a huge array of experimentally accessible observables. 
Nothing need therefore remain speculation when facilities exist that can test such predictions. 

Principal goals for theory today are to elucidate all observable consequences of putative answers to the basic questions posed by Nature and their numerous corollaries, and highlight paths to measuring them; and a primary challenge to experiment is to test those predictions so that science can close the book on QCD or move on to an improved theory of strong interactions and use the experience gained by solving QCD to develop a new approach to physics beyond the SM. 
Indeed, deciding the status of QCD is necessary before science can properly address the question of whether or not there is a theory of everything.



{\color{blah} The ensuing chapters identify a diverse programme of physics that exploits hadron beams produced at GSI and FAIR and demonstrate how this programme can contribute to answering, in particular, the following research questions:
\begin{enumerate}
    \item[A.] How are the strength and nature of hadron--hadron interactions connected to the underlying strong force, and how can scattering, final-state-interaction and femtoscopic studies, together with hypernuclear measurements, constrain the underlying interactions and many-body forces? 
    
    \item[B.] How does QCD generate the observed spectrum and internal composition of hadrons, and what can the discovery and characterisation of exotic hadrons reveal about the possible forms of hadronic matter? 
    
    \item[C.] How are the internal structure and dynamics of hadrons encoded in their form factors, intrinsic heavy-quark content, and generalized parton distributions, and what do these observables reveal about the origin of hadron mass and the role of quarks and gluons?
    
    \item[D.] How do the properties and interactions of hadrons change in the nuclear medium, and what can these modifications reveal about the emergence of hadronic properties from QCD and the behaviour of strongly interacting matter at finite density?
\end{enumerate}
This document discusses themes ranging from hadron--hadron interactions to hadron spectroscopy and structure, including exotic systems, and extending to leptonic and hadronic probes of strongly interacting matter. Connections and synergies with astroparticle physics are also highlighted.
The document closes with a constructive comparison of the QCD at FAIR programme with complementary programmes at other facilities worldwide, outlining how it contributes to the international effort to decipher the mysteries of QCD and highlighting its unique discovery potential.}

\newpage
\section{Exploiting hadronic beams at GSI/FAIR} 
\label{sec.HadronBeams}
{\small  {\bf Convenors:} \it T. Galatyuk, J.~G. Messchendorp, F. Nerling} 

\noindent The overarching goal in the realm of hadron physics is to gain a fundamental understanding of the nature of hadronic matter and interactions. This involves exploring how QCD manifests itself on the length and time scales relevant to the formation of hadrons. Identifying the underlying degrees of freedom and symmetries that dictate the observed properties of hadrons is crucial to this pursuit. One of the primary objectives is to provide comprehensive insights into the mechanisms responsible for generating the mass of hadrons, such as protons, neutrons, and pions, which constitute the fundamental entities for building and binding visible matter in the universe.

While the Higgs mechanism explains merely the individual quark current/Lagrangian masses, and thus accounts for only a few per cent of the mass of the compound object, \textit{i.e}., the hadron, the overwhelming contribution arises from internal dynamics, such as gluon self-interactions and their wide-ranging expressions. 

In striving to address these fundamental research objectives, manifold experimental and theoretical approaches are being employed, utilising a large array of methods. Experimentally, the focus is on exploiting the diverse spectrum of beams ranging from heavy ions to elementary hadronic and electromagnetic probes and the associated research communities. The facilities at GSI/FAIR are unique and complementary to electromagnetic accelerator complexes worldwide, offering a distinctive perspective.

\begin{figure}[t]
\centerline{\includegraphics[width=1.0\columnwidth]{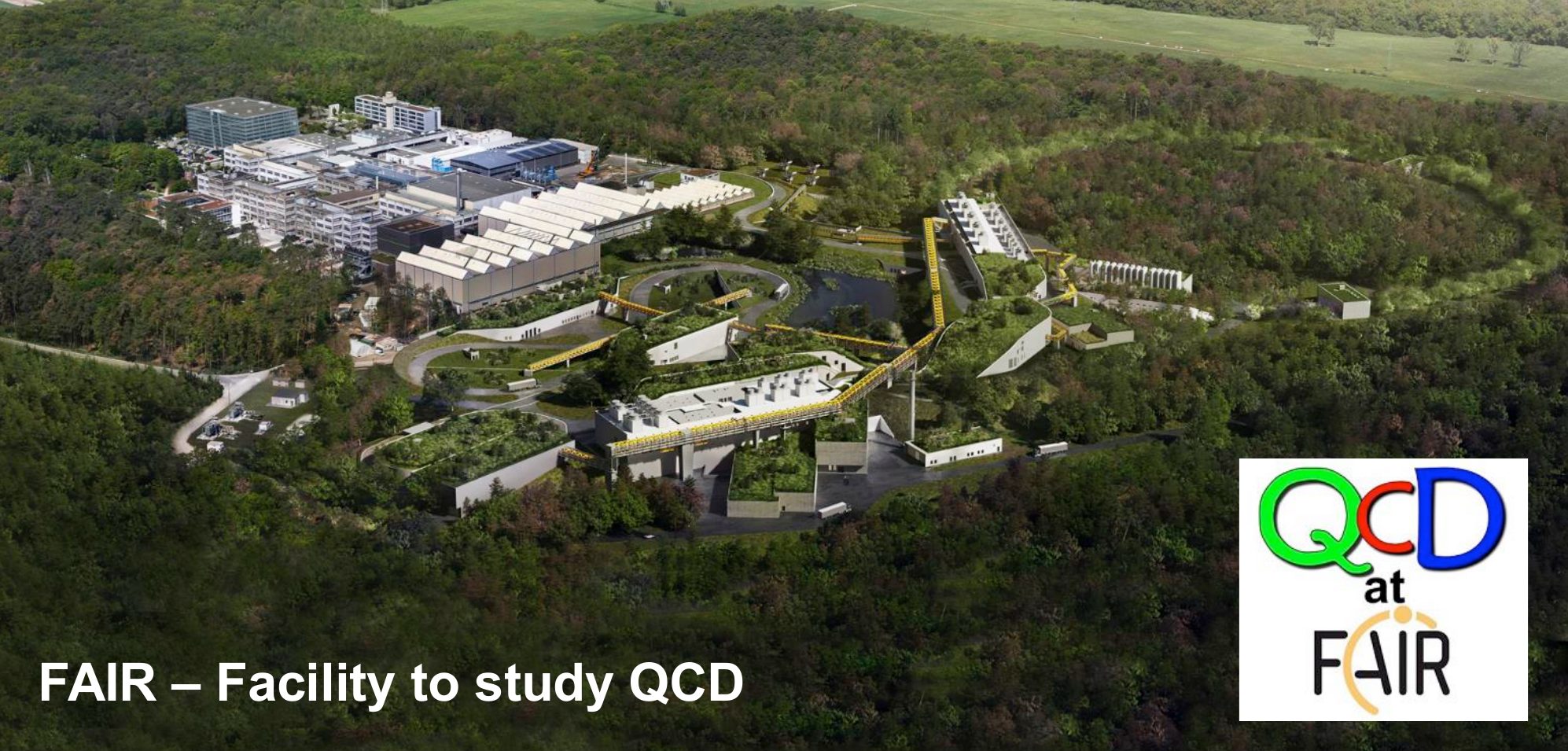}}
\caption{\label{FigFAIR}
A futuristic depiction of FAIR near Darmstadt, Germany. The infrastructure on the left-hand side represents the existing GSI site, while the right-hand side illustrates the various components that comprise FAIR. The details of GSI and FAIR are described in Sec.~\ref{sec.OverviewFacilities}. Picture provided D.~Fehrenz.}
\end{figure}

\subsection{Hadron physics at GSI/FAIR}

Hadron physics serves as a crucial link that connects seamlessly with both the heavy-ion and nuclear physics communities at GSI/FAIR. This connection establishes a comprehensive research facility -- see Fig.\,\ref{FigFAIR} -- capable of addressing various aspects of the strong interaction. The primary aim of hadron physics activities at GSI/FAIR is to unveil unequivocal insights into the properties and dynamics of baryons and mesons involving light, strange, and charm quarks. This is achieved through precision data obtained from diverse hadron beams, employing state-of-the-art experimental instruments and techniques, and a combination of \textit{ab initio} theoretical approaches.

The objective is to offer a microscopic framework, going beyond existing, successful phenomenology, to scrutinise the structure -- especially of baryons -- and their interactions. This, in turn, provides valuable input for describing the dynamics of baryon-rich matter at normal nuclear densities, extending to highly dense matter that is accessible, \textit{e.g.}, via the CBM experiment at SIS100 at FAIR.

Hadron physics activities focus on exploiting inclusive and exclusive reactions using elementary probes available at GSI/FAIR, including pion beams with momenta up to a few~GeV/$c$, proton beams up to 30\,GeV/$c$, and further developing the antiproton programme once the HESR becomes operational at FAIR. The goal is to cultivate an extensive physics programme and to foster a community driven by the science need for understanding the dynamics of QCD in the non-perturbative domain. These efforts aim to strengthen the impact in the closely interconnected realms of nuclear (astro)physics, hadron, and heavy-ion physics that are at the heart of GSI/FAIR.

\subsection{Roadmap -- Hadron physics from GSI, FAIR Phase-0 towards completing FAIR}
\label{sec:roadmap}

\begin{figure}[ht]
\begin{center}
\vspace*{-0cm}
\includegraphics[width=0.95\textwidth]{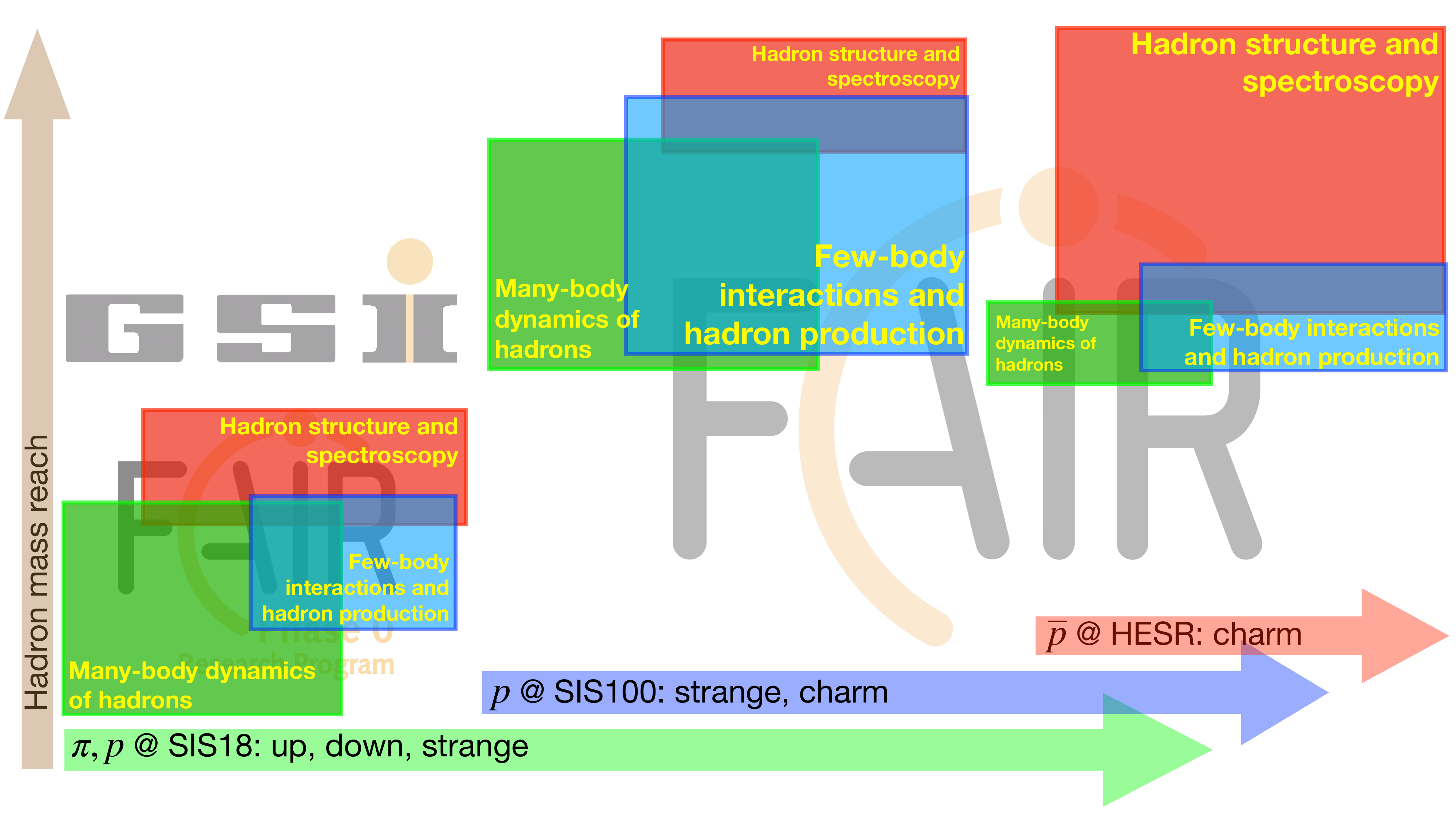}
\vspace*{-0.4cm}
\caption{\small Sketch of the roadmap for hadron physics activities at GSI and FAIR. The initial programme phase, indicated by GSI/FAIR Phase-0, will exploit pion and proton beams delivered by SIS18. {\color{blah} The next two programme phases at FAIR will make use of the SIS100. The first will exploit high-intensity proton beams, significantly extending the accessible hadron-mass reach compared with Phase-0. The subsequent phase will additionally exploit the antiproton storage ring HESR.} The three categories of boxes, distinguished (again) by green, blue, and red, symbolise the three research domains to be tackled {\color{blah} corresponding to the themes shown in Fig.~\ref{fig:mindmap}}. The size of each box reflects the respective contribution to the overarching programme.}
\label{fig:roadmap}
\end{center}
\end{figure}

The roadmap for the hadron physics programme at GSI and FAIR can be structured according to three strongly overlapping phases, as illustrated in Fig.~\ref{fig:roadmap}. These phases are primarily driven by the available primary and secondary beams and the corresponding energy ranges. Moreover, the availability of detector setups is a key factor determining the physics reach in each of the phases. We shall return to this aspect in a discussion at the end of this strategic document.

The hadron physics activities initiated within FAIR Phase-0 were largely driven by the use of proton beams delivered by SIS18 in combination with the high acceptance dielectron spectrometer (HADES), as well as by exploiting the WASA (wide angle shower apparatus) detector in conjunction with the fragment separator FRS. The role of strangeness in hadronic (hyperons) and nuclear matter (hypernuclei), respectively, has been the central theme in both programmes. Moreover, the production mechanisms and final-state interactions of light mesons with hadronic and nuclear matter are of great interest. First data were obtained in 2022.

\subsubsection{Pion and proton beams with SIS18}
\label{sec:pionbeams}
With recent advances in the operation of the SIS18 accelerator, providing high-quality beams with excellent micro- and macro-spill structures, it is now advantageous for the overall hadron-physics activities to exploit the use of secondary pion beams, with HADES as the backbone. 
Also, in light of ongoing and planned photo- and kaon-production experiments at various facilities worldwide, complementary precision pion-production data --particularly above the strangeness-production threshold -- will become very valuable. Globally, the combination of a pion-beam facility with a few-GeV energy beam and a high-precision dilepton spectrometer (HADES) is unique. 
The forward-tracking detector developments and analysis expertise acquired through the proton-driven hyperon-physics programme during FAIR Phase-0 provide an ideal basis for expanding this unique coverage across various physics topics in the strangeness domain.

With the current SIS18 accelerator and pion-beam facilities, $\pi N$ centre-of-mass energies up to 2\,GeV will be accessible with good pion-beam intensities ($>$10$^6$/spill). The initial state will be characterised by a ``simple'' system involving spinless and baryonless pions (near-Goldstone bosons) that directly excite the target nucleons. This will enable precision baryon spectroscopy studies well within the third-resonance region, with access to various final states including baryons and mesons in the strangeness $|S|=1$ domain. The reactions will predominantly involve two- and three-body final states -- a small enough number to permit rigorous partial-wave analyses. 
The excellent dilepton and baryon spectroscopy capabilities enable a physics programme that features \textit{precision} medium studies, thereby providing valuable input towards understanding many-body dynamics in heavy-ion reactions, and facilitating access to electromagnetic (transition) form factors of, \textit{e.g.}, light baryons and $|S|=1$ hyperons. An overview of its comprehensive physics programme can be found in Ref.~\cite{HADES:2026piQCD}.

Hence, during this phase, as illustrated by the lower-left panel in Fig.\,\ref{fig:roadmap}, the hadron physics programme will focus on hadron composition studies and on providing reference dynamics input for many-body systems such as those present in heavy-ion reactions.

\subsubsection{Hadron beams with SIS100}

SIS100 is expected to be commissioned by the end of 2028. Thereafter, high-energy proton or deuteron beam reactions will be used to significantly extend explorations of hadron physics. The maximum beam energy for protons will be 29\,GeV, with intensities of up to $2.5\times10^{13}$ protons/cycle. For a fixed proton-target experiment, the maximum centre-of-mass energy in the proton--proton system will reach around 7.5\,GeV, which is sufficient to produce baryonic matter with open strangeness $|S|=1,2,3$, open charm $|C|=1$, and meson-like matter with hidden strangeness and charm.

The foreseen high-end CBM detector will provide an excellent basis to reconstruct full reaction topologies via exclusive coincidence measurements of the momenta and particle types of final-state reaction products. This will open the possibility for precision measurements across a broad palette of reactions and over a wide energy range. These data will not only play a crucial role in contextualising the results from heavy-ion reactions at comparable energies --serving as an essential reference for the heavy-ion community -- but will also intersect significantly with the broader interests of the hadron and nuclear physics communities.

In particular, the nature of the underlying production mechanisms and interaction dynamics in elementary nucleon-nucleon scattering is of general interest. The choice of beam energy can be tailored to provide optimal kinematics for investigating (low-energy) final-state interactions of various hadrons, such as hyperons and charmonium, as well as for probing new, potentially exotic resonances that may not be accessible at facilities employing electromagnetic probes. These studies can be conducted via femtoscopy or, complementarily, by employing partial-wave analysis techniques of exclusively reconstructed reaction channels.

Finally, studies of reactions using the high-intensity proton beams from SIS100 offer opportunities to elucidate the structure of baryons through measurements of electromagnetic and weak transition form factors, as well as through resonance line-shape scans of excited hyperons. Moreover, the structure of the proton itself can be explored through the extraction of generalised parton distribution functions and investigations of the final-state interaction of charmonium with the nucleon.

\subsubsection{Towards opportunities with antiprotons at HESR}
\label{sec:hesr-physics}

The completion of FAIR remains the ultimate goal. Towards this realisation, a pivotal component is the implementation of the proton linac and high energy storage ring (HESR), specially designed to store beams of antiprotons within an energy range from 0.83\,GeV to 14\,GeV. These beams will undergo stochastic cooling to achieve a momentum spread of  $\Delta p/p \approx 5\times 10^{-5}$. Experiments using internal cluster-jet or pellet targets, in conjunction with a versatile detector system covering nearly the entire phase space -- as envisaged by the PANDA (antiProton annihilation at Darmstadt) experiment \cite{PANDA_GSI_Website, PANDA_FAIR_Website} -- offer a unique opportunity to investigate antiproton-proton and antiproton-nucleus annihilations with unprecedented precision, both in terms of resolution and statistical significance.

Compared to proton-proton collisions, reactions driven by antiproton--proton annihilations provide a clean final state with baryon number zero, well suited for precision partial-wave analyses. These reactions produce abundant ground-state and excited hyperon-antihyperon pairs with low background, establishing a truly remarkable strangeness factory conducive to a comprehensive exploration of their internal structures and production mechanisms, including the formation of hypernuclei with double strangeness.

Moreover, in the realm of charm physics, antiproton-proton collisions at HESR offer unique opportunities. A significant fraction of the hidden-charm sector can be systematically explored, particularly in the enigmatic XYZ mass regime. Data amassed over decades from experiments conducted at electron-positron collider facilities and $B$ factories have revealed an impressive array of hidden-charm ``exotic'' states, the nature of which remains elusive. Antiproton experiments at HESR will enable precise measurements of the line shapes of these resonances, shedding light on their internal structures; see, e.g., Ref.\,\cite{PANDA2019_resonanceX3872}.

Furthermore, annihilation processes involving antiprotons will open the door to Drell-Yan studies with the potential to deliver data of unprecedented precision that are relevant, \textit{e.g}., to the extraction of proton sea-quark densities.
They will also generate gluon-rich hadronic matter, providing an ideal environment for the search and identification of glueballs and hybrids, addressing another of the principal goals of research in the domain of strong QCD. 
An overview of the physics programme during the first phase of data taking can be found in Ref.\,\cite{Barucca2021PANDA}.

\newpage
\newcommand{\ols}[1]{\mskip.5\thinmuskip\overline{\mskip-.5\thinmuskip {#1} \mskip-.5\thinmuskip}\mskip.5\thinmuskip} 
\newcommand{\olsi}[1]{\,\overline{\!{#1}}} 

\section{Hadron-hadron interactions} 
\label{sec.HadrHadrInter}
{\small {\bf Convenors:} \it C.~Blume, C.~Hanhart}
%
%


\noindent

\textcolor{blah}{While QCD is formulated in terms of quarks and gluons, experiments observe only hadrons, which are composite particles formed from the fundamental building blocks. 
Which compositions are realised in nature and how they interact is still not fully understood.
Two key, widely used tools for gaining insights into those inner workings of QCD in the low-energy regime are spectroscopy and hadron–hadron scattering. }
In this chapter, we focus on the latter, although overlaps with the former, which are the subject of ensuing sections, are unavoidable.
As will subsequently become clear, hadron-hadron scattering offers additional complementary access to the following key issues.

\begin{itemize}
 \item The patterns of SU(3)-flavour symmetry breaking, which contain valuable information not only on the substructure of hadrons, but also on the QCD equation of state and even the structure of neutron stars. If the strangeness degree of freedom plays a role in their core -- as one would expect, given the option to circumvent the Pauli principle via the inclusion of hyperons -- it becomes difficult to explain the empirical fact that neutron stars with masses exceeding two solar masses exist ({\it hyperon puzzle}; see discussion in \ref{subsubsect.astro}).

 \item The subtle interplay between the chiral structure of QCD and the hadron spectrum: through non-perturbative hadron–hadron interactions, (quasi)bound states can be generated dynamically. While this mechanism is known to work for kaons scattering off heavy sources such as protons -- leading to the widely discussed $\Lambda(1405)$ -- the same does not occur for pions due to chiral suppression. Nevertheless, resonances may still exist in those systems.

 \item While the strange quark is the heaviest of the light quarks, the charm quark is the lightest of the heavy quarks; so, a systematic comparison of the dynamics of systems with strangeness or charm will provide deep insights into how QCD forms individual hadrons and more complex structures.

 \item Direct studies of non-leptonic weak decays of the $\Omega$ baryon, where SIS100 can act as an $\Omega$ factory via the  $p p \to \Omega^- \, N 3K$ process, which is expected to occur at a high rate.
 The $\Omega$ baryon is the only observed spin-\(3/2\) baryon that has no strong or electromagnetic decays. It is rendered unstable solely by flavour-changing weak interactions. Owing to its spin-\(3/2\), the information it provides about the electroweak interaction and its interplay with the strong force is complementary to that obtained from all other long-lived hyperons.

 \item Charge-Parity (CP) violation -- a fundamental feature of the Standard Model (and beyond) -- arises from the interference between CP-conserving strong phases, driven by hadron–hadron interactions, and CP-violating phases in the Standard Model, which emerge from the extremely short-ranged weak interactions. 
 This necessitates highly accurate information on the former. 
 In particular, both the sensitivity to and interpretation of CP-violating asymmetries in baryon decays require knowledge of strong interactions in specific baryon–meson systems. 
 For example, as detailed in Sec.~\ref{subsec.MesonBaryonInt}, the $\pi \Lambda$ scattering phase shifts are crucial for deducing CP violation in $\Xi$ decays.
\end{itemize}
 
 \noindent In this chapter, we present a perspective on the current status of our understanding of various hadron–hadron systems and identify the additional information that must be obtained from experiment to enable further progress. Moreover, we discuss what will already be possible during the early stages of the FAIR facility.

Direct scattering experiments are feasible for only a very limited class of systems: while protons can straightforwardly be scattered off each other, access to neutron observables requires the use of deuterons.

Charged pion beams allow for studies of elastic and inelastic pion–nucleon scattering. This provides access to the baryon spectrum. Although a significant amount of data exists from experiments conducted over 40 years ago, there remains a strong demand for data in additional channels and for more observables, as will become clear in the next subsection. FAIR will be able to provide such data.

Hyperon–nucleon scattering experiments can only be performed via secondary scatterings, which are technically demanding and do not allow the study of such systems at small relative momenta. Thus, while they yield valuable information on higher partial waves, alternative approaches are necessary to access low partial waves and small relative momenta.

It is therefore essential to employ production reactions to gather information on such systems. 
{\color{blah}  In this context, the $pp\to KKBB$ reaction discussed in Section~\ref{subsec.scatfromprod} is a particularly important opportunity at FAIR with CBM, exploiting its unique capabilities to access the interaction of baryons in the strangeness $S=-2$ system.}
A key issue here is the control of systematic uncertainties in the analysis and the reliable quantification of extraction uncertainties. The methods available in the literature will be reviewed in section~\ref{subsec.femto}.

One promising approach involves detailed studies of cusp structures, \textit{viz}.\ specific non-analyticities that emerge whenever a new channel opens in an $S$-wave coupled to the studied channel. The strength and shape of these cusps can directly be related to the underlying interaction strengths.

To date, the most extensively exploited experimental approach for accessing final-state interactions of hadrons over a broad kinematic range has been the analysis of two- and three-particle intensity correlation functions (``femtoscopy'').
In addition, by selecting different kinematics, it will also be possible at FAIR to extract information on hadron–hadron scattering through the application of dispersion theory in data analysis. The CBM experiment at FAIR thus provides a unique opportunity to investigate the same systems using a variety of complementary methods -- an important aspect for consolidating the results.

In the special case of hyperon–nucleon and hyperon–hyperon interactions, studies of hypernuclei offer further valuable information. A precise determination of the corresponding binding energies and lifetimes can impose stringent constraints on low-energy hyperon–nucleon and hyperon–hyperon interactions (in the case of double-$\Lambda$ hypernuclei), as well as on three-body forces that may be central to resolving the hyperon puzzle.

As discussed above, detailed studies of hadron–hadron interactions will provide new, much sought insights into the inner workings of QCD. In addition, knowledge of hadron–hadron interactions is also vital for fully exploiting CP violation studies involving more than one hadron in the final state, since CP violation manifests via interference effects. These interferences can either be enhanced or suppressed by hadronic final-state interactions; see, \textit{e.g}., \cite{Salone:2022lpt} and references therein.

A better understanding of the strong final-state interactions that occur in hyperon decays can serve two important purposes: first, to identify reactions and observables in which the CP violation signal is enhanced; and second, to enable a quantitative interpretation of weak decay results in terms of fundamental parameters. By providing information on hadron–hadron interactions, the hadron physics programme at FAIR can deliver crucial input towards improving our understanding of CP violation in the baryon sector, both within the Standard Model and beyond.


To summarise, the high interaction rates achievable at FAIR will enable a significant expansion of our knowledge of hadron interactions -- and, by extension, of QCD -- through the systematic investigation of the various tools and reactions described in detail below. These include $\pi p$ and $pp$ scattering, the measurement of hypernuclei, dispersive analyses of production reactions, and femtoscopy. The availability of high-statistics data will enable these studies to be carried out with high precision. Moreover, direct comparisons between different methods, \textit{e.g}., in the case of $\Lambda\Lambda$ interactions, will provide stringent control over systematic effects.

\subsection{Why experiments with pion beams are important} 
\label{sec4:subsec:Doring}

{\color{blah} Meson beams provide a versatile probe for hadron spectroscopy. The physics case for meson beams is discussed in Refs.,\cite{Briscoe:2015qia,Briscoe:2021cay}. Pion-beam facilities have been operated at several laboratories, including the CERN PS and SPS (see, \textit{e.g.}, \cite{NA61SHINE:2022tiz}), J-PARC (see, \textit{e.g.}, \cite{Ohnishi:2019cif}), and GSI with the HADES experiment (see, \textit{e.g.}, \cite{PionBeamHADES:2002,PionBeamHADES:2017}). To illustrate the physics potential of such facilities, we highlight below a few key topics discussed in these and related references, together with some new aspects. An overview of the facilities is provided in Chapters~\ref{sec.OverviewFacilities} and \ref{sec.IntContext}. The cited literature contains many additional opportunities, concerning not only baryon spectroscopy but also excited mesons.}


Pion beams were the first probe used in the discovery of excited baryons, leading to the establishment of many states by the early 2000s~\cite{Cutkosky:1979fy, Hohler:1984ux}, and continuing to this day via the GWU-SAID analysis group \cite{Arndt:1985vj, Arndt:1995bj, Arndt:1998nm, Arndt:2006bf, Workman:2008iv, Arndt:2008zz}. In recent years, photoproduction reactions as measured at ELSA, JLab, and MAMI have provided much more detailed information—on both the electromagnetic properties of resonances and the spectrum itself \cite{CB-ELSA:2003rxy, Anisovich:2004zz, CLAS:2006pde, CBELSATAPS:2007oqn, GRAAL:2008jrm, CLAS:2009rdi, CrystalBallatMAMI:2010slt, Anisovich:2011fc, CLAS:2015pjm, CLAS:2015ykk, CLAS:2017rxe}. Based on those measurements, many new states were added to the Review of Particle Physics by the Particle Data Group (PDG) \cite{ParticleDataGroup:2024cfk}; states found before the advent of photoproduction data have been upgraded in PDG star ratings, while a few others have been downgraded or removed; see Refs.\,\cite{Klempt:2009pi, Crede:2013kia, Ireland:2019uwn, JPAC:2021rxu} for reviews.

A major argument for using photons to discover excited baryons is the sensitivity to states that couple only weakly to $\pi N$; an analogous argument applies to searches for resonances in final states of $pp$-induced reactions, although the analyses are more complicated owing to possible crossed-channel effects. A resonance that couples substantially to the photon but only weakly to the $\pi N$ channel could be more easily detected in photoproduction reactions. In fact, reactions like $\gamma N \to \eta N$ are ideal for this purpose. The downside of photoproduction reactions is the larger number of amplitudes (multipoles) involved. In general, a resonance with given spin-parity $J^P$ couples to both an electric and a magnetic multipole, \textit{i.e}., significantly more degrees of freedom are present in photoproduction reactions than in pion-induced ones. This considerably increases the number of measurements needed to disentangle the amplitudes. Questions related to such ``complete experiments” are discussed in Refs.\,\cite{Wunderlich:2013iga, Workman:2016irf}. There are two variants of this problem: first, how many measurements are required to determine an amplitude at a given energy and scattering angle \cite{Kroenert:2020ahf}; and second, the ``truncated partial-wave” complete experiment \cite{Wunderlich:2013iga, Workman:2016irf}, which is somewhat more relevant for baryon spectroscopy involving resonances with given $J^P$.

Modern baryon spectroscopy analysis frameworks incorporate the real or virtual photon into an existing hadronic rescattering structure in one way or another \cite{Surya:1995ur, Drechsel:1998hk, Anisovich:2004zz, Matsuyama:2006rp, Drechsel:2007if, Julia-Diaz:2007mae, Durand:2008es, Anisovich:2011fc, Ronchen:2014cna, Ronchen:2015vfa, Ronchen:2018ury, Kamano:2019gtm, Mai:2021vsw, Mai:2021aui, Ronchen:2022hqk, Mai:2023cbp, Wang:2024byt, Doring:2025sgb}, as discussed in Sec.\,\ref{sec:PWAs}. The strong rescattering component, which also contains the resonance spectrum, is constrained by pion-induced reactions. Therefore, any data problems in pion-induced reactions will affect the analysis of photon-induced reactions. This is particularly relevant for inelastic pion-induced reactions, such as $\pi N \to \eta N$, $K\Lambda$, $K\Sigma$, $\omega N$, or three-body channels. On one hand, data in these reactions are known to be problematic, as discussed below; on the other hand, new resonances coupling weakly to $\pi N$ are still more visible in these reactions than in elastic $\pi N$ scattering.

For these reasons (\textit{a}) fewer amplitudes than in $\gamma^{(*)}$ production, (\textit{b}) bias in photoproduction analyses, (\textit{c}) sensitivity to channels other than $\pi N$, and (\textit{d}) the poor data situation,  remeasuring inelastic pion-induced reactions will decisively contribute to an improved and more quantitative understanding of the excited baryon spectrum and the discovery of new states.

Among the possible final states, the reactions $\pi N \to \eta N$, $K\Lambda$, and $\pi^+ p \to K^+\Sigma^+$ should be highlighted, as they isospin-filter $I=1/2$ $N^*$'s~\cite{Ronchen:2012eg} and $I=3/2$ $\Delta$'s~\cite{Doring:2010ap}, respectively. The reactions $\pi N \to K\Lambda$ and $\pi N \to K\Sigma$ provide highly nontrivial additional information, since the outgoing baryons are self-analysing, allowing access to spin observables in the final states. 
Take, for example, the reaction $\pi^+ p \to K^+\Sigma^+$, measured in Ref.\,\cite{Candlin:1982yv}. These measurements cover a substantial angular and energy range and include data on differential cross sections and recoil polarisation $P$. Together with less precise data on the spin-rotation parameter $\beta$ from Ref.\,\cite{Candlin:1988pn}, this set of observables is nearly complete; see, \textit{e.g}., the discussion in Ref.~\cite{Klempt:2009pi}.

An overview of the available data on inelastic pion-induced reactions is provided in Table~\ref{tab:datainelastic}.
\begin{table*}
\caption{Inelastic pion-induced reaction data. A full list of references to the different experimental publications can be found online~\cite{Juelichmodel:online}.}
\begin{center}
\renewcommand{\arraystretch}{1.9}
\begin {tabular}{l|l|r} 
\hline\hline
Reaction & Observables ($\#$ data points) & $\#$ data  \\ \hline
$\pi^-p\to\eta n$ &{$d\sigma/d\Omega$ (676), $P$ (79) }
& 755\\
$\pi^-p\to K^0 \Lambda$ &{$d\sigma/d\Omega$ (814), $P$ (472), $\beta$ (72) }
& 1,358\\
$\pi^-p\to K^0 \Sigma^0$ &{$d\sigma/d\Omega$ (470), $P$ (120) }
& 590\\
$\pi^-p\to K^+ \Sigma^-$ &{$d\sigma/d\Omega$ (150)}
& 150\\
$\pi^+p\to K^+ \Sigma^+$ &{$d\sigma/d\Omega$ (1124), $P$ (551) , $\beta$ (7)}
& 1,682\\ 
$\pi N\to \omega N$ & $\sigma(54)$, $d\sigma/d\Omega$ (124) & 178\\
\hline\hline
\end {tabular}
\end{center}
\label{tab:datainelastic}
\end{table*} 
The table is taken from Ref.\,\cite{Ronchen:2022hqk} and amended with the available $\pi N \to \omega N$ data collected and analysed in Ref.\,\cite{Wang:2022osj}. The number of data points should be compared to the tens of thousands of data available from photoproduction reactions \cite{Ronchen:2022hqk}, or the hundreds of thousands from electroproduction \cite{Mai:2023cbp}.

In addition to the much smaller data base, pion-induced reaction data typically carry large uncertainties and, in many cases, are contradictory owing to underestimated systematic effects. This leads to notoriously large $\chi^2$ values in global fits to all data, making the application of traditional statistical criteria, \textit{e.g}., to assess the significance of a signal for a new resonance, very difficult. The aim of a new pion-induced experiment is not only to provide more data, but, for the first time, to provide {\it consistent} data. This is indeed possible: for instance, the EPECUR experiment has drastically improved the data situation in elastic pion–nucleon scattering \cite{EPECUR:2014wrs}; see, \textit{e.g}., Fig.\,4 in Ref.\,\cite{Briscoe:2021cay}). See also Ref.\,\cite{Shirotori:2012ka} for a demonstration of how drastically measurements of the $K^+\Sigma^+$ final state can be improved with modern experimental techniques. For studies of pion-induced reactions at HADES, see Refs.\,\cite{Salabura:2015yot, HADES:2024}.

Planned reactions with meson beams at J-PARC, which is the only alternative facility for pion beams, are discussed in Ref.\,\cite{Ohnishi:2019cif}.

\begin{figure}[h!]
\begin{center}
\hspace*{-1em}\includegraphics[width=0.215\textwidth]{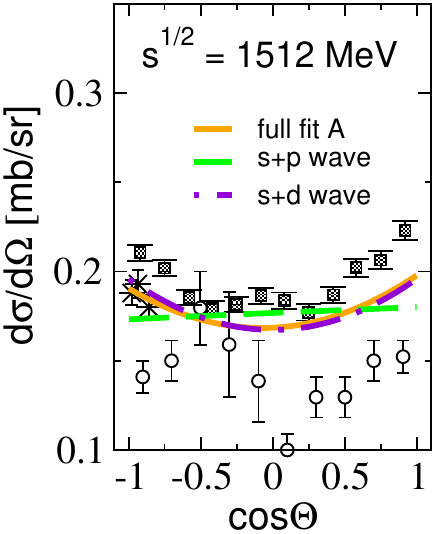}
\hspace*{-0.cm}
\includegraphics[width=0.160\textwidth]{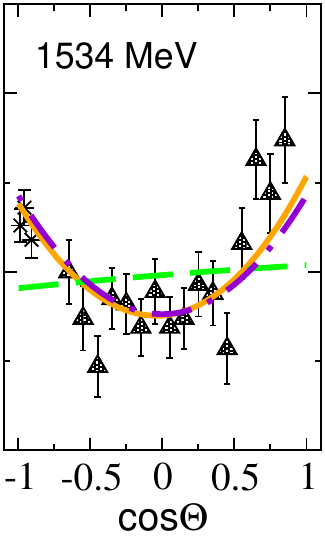}
\hspace*{1em}
\includegraphics[width=0.33\textwidth]{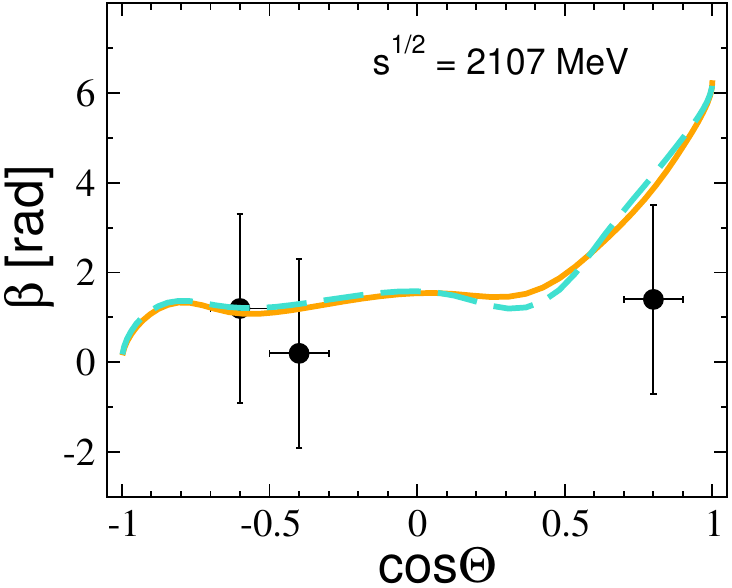}
\hspace*{1em}
\includegraphics[width=0.20\textwidth]{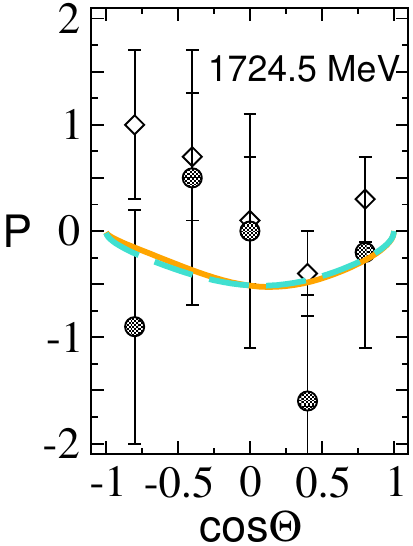}
\end{center}
\vspace*{-0.4cm}
\caption{Left two panels: Differential cross section of the reaction $\pi^- p \to \eta n$ at scattering energies $\sqrt{s} = 1512$ and $\sqrt{s} = 1534\,$MeV. Central panel: Spin-rotation parameter $\beta$ for $\pi^+ p \to K^+ \Sigma^+$.
Right panel: Polarisation $P$ for the reaction $\pi^- p \to K^0 \Sigma^0$.
For the data references and the analysis solutions shown (A in orange, B in turquoise), see Ref.~\cite{Ronchen:2012eg}. The different symbols correspond to different experiments.}
\label{fig:etaN_s+dwave}     
\end{figure}

\begin{figure}
    \centering
\includegraphics[width=0.48\linewidth]{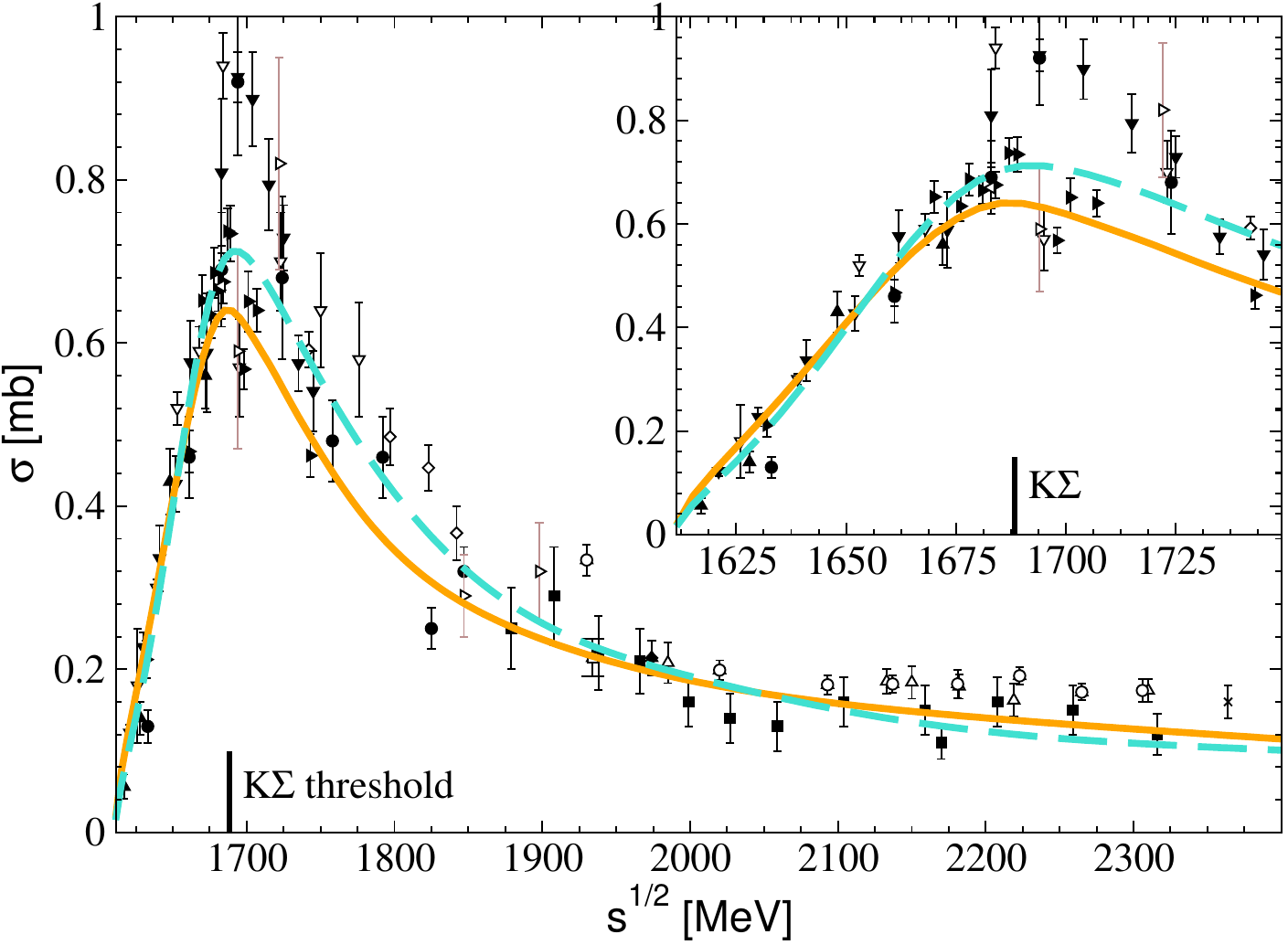}
    \includegraphics[width=0.49\linewidth]{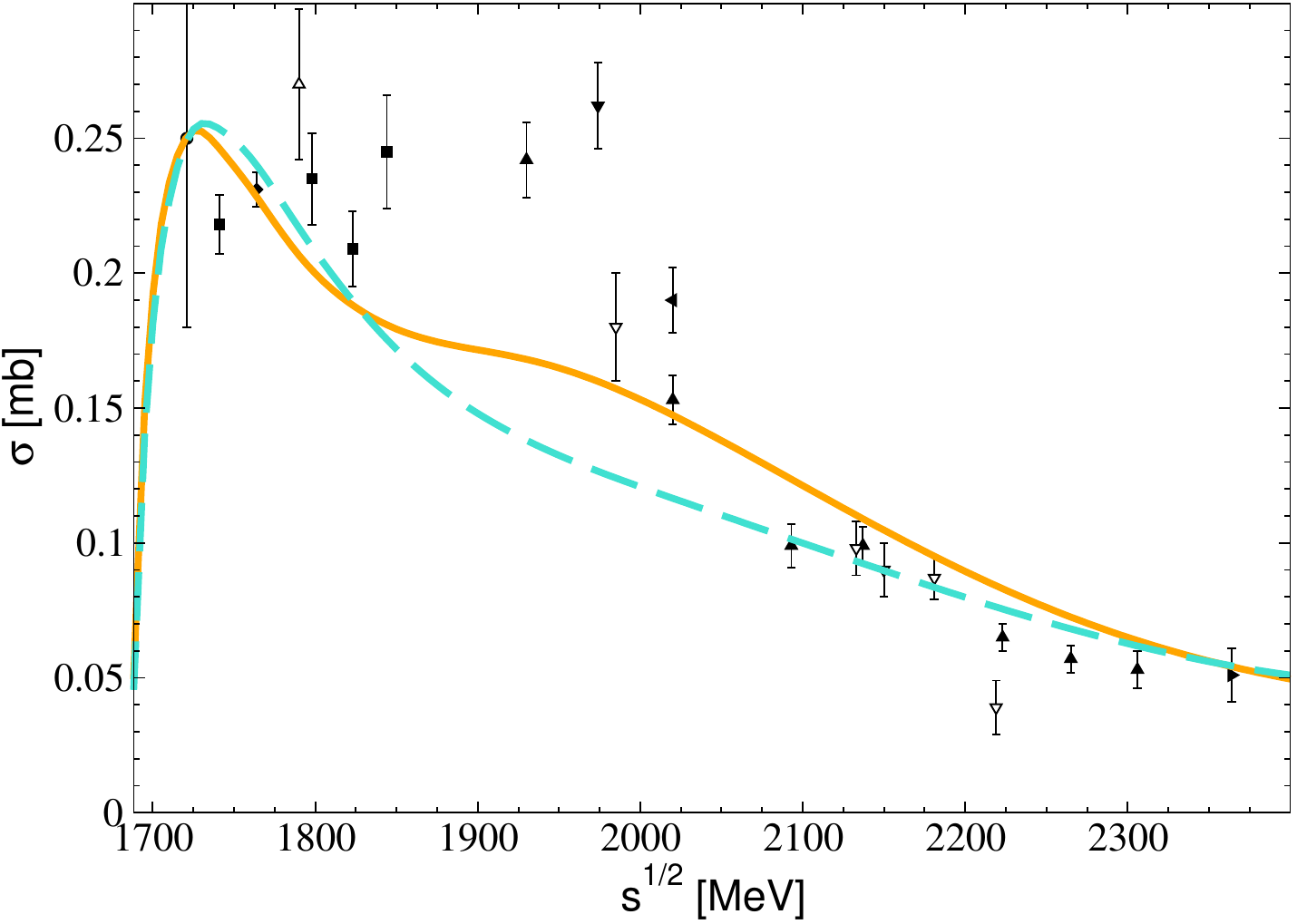}
    \vspace*{-0.2cm}
    \caption{Total cross sections for the reactions $\pi^-p\to K^0\Lambda$ (left) and $\pi^-p\to K^+\Sigma^-$ (right) as a function of scattering energy $\sqrt{s}$. For the shown fits A (orange) and B (turquoise) and data references, see Ref.~\cite{Ronchen:2012eg}.}
    \label{CH4:figMM-3}
\end{figure}

The situation for inelastic reactions is illustrated in Fig.~\ref{fig:etaN_s+dwave}.
The differential cross section data shown on the left demonstrate that the systematic problems do not simply owe to multiplicative factors that might be adjusted \cite{Ball:2009qv}; rather, the overall shapes of the cross sections do not match. The plot of the spin-rotation parameter $\beta$ illustrates how few data are available for this particularly valuable experimental observable; and the right-hand plot shows the typical quality of polarisation data, which are expected to be bounded by $|P| \leq 1$.

The total cross section data shown in Fig.~\ref{CH4:figMM-3} exhibit similar issues, \textit{i.e}., clearly contradictory measurements. The reaction $\pi^- p \to K^+ \Sigma^-$ is almost impossible to describe owing to an apparent structure around a scattering energy of $\sqrt{s} \approx 2\,$GeV, which, however, is based on only one or two data points. More conclusive data with improved energy and angular coverage from a single, consistent experiment would be especially valuable, as this reaction forbids $t$-channel meson exchange and is thus of particular interest for studying hadron dynamics. Moreover, no polarisation data are available for $\pi^- p \to K^+ \Sigma^-$.

In summary, inconsistencies dominate the data uncertainties for pion-induced reactions. This has serious consequences for baryon spectroscopy, affecting the reliability of spectrum determination, the significance of newly identified states, and the extraction of their properties. Indirectly, these issues also affect the analysis of photoproduction reactions, in which significant experimental and theoretical effort has been invested over recent decades. Acquiring a large, consistent dataset of pion-induced reactions -- particularly inelastic reactions -- would allow the field to better exploit the intrinsic advantages of meson beams: fewer amplitudes than in photoproduction, simpler conditions for a ``complete experiment”, larger cross sections, and, in some cases, enhanced sensitivity to resonances that couple only weakly to $\pi N$.

The impact of such new data on different theoretical analyses, once available, could be quantified through joint efforts across analysis groups, as was done in Ref.\,\cite{Anisovich:2016vzt} for photoproduction.

\subsection{Remarks on hadronic systems}

In this subsection, we present the current status of our understanding of various hadronic systems relevant to the physics programme at FAIR. Particular emphasis is placed on those areas where additional data -- ideally to be collected at FAIR -- are essential for making further progress in the field. These include questions ranging from the existence of new forms of hadron bound states to the hyperon puzzle in neutron stars.

\subsubsection{Meson-meson systems}
\label{MesMes}


{\bf $\pi\pi$, $K\bar K$ and $\pi K$ interactions below 1.5\,GeV}
\label{subsec.MesonMesonInt}




 \noindent
Two-body interactions of pions and kaons are important in the context of FAIR for several reasons. The most relevant is that they provide all, or a significant part, of the final-state interactions (FSI) in many hadronic processes. Being well known from data and well understood theoretically, they can be used to test or calibrate the formalisms and approximations employed to extract information on other hadrons, \textit{e.g}., from production, femtoscopy, or effective theories. In addition, they can be described using Chiral Perturbation Theory (ChPT, \cite{Gasser:1983yg,Gasser:1984gg}), which is the low-energy effective theory of QCD, thus providing a rigorous link to the fundamental interactions that FAIR aims to explore.

Scattering data on $\pi\pi \rightarrow \pi\pi$\cite{Hyams:1973zf,Durusoy:1973aj,Losty:1973et,Cohen:1973yx,Protopopescu:1973sh,Grayer:1974cr,Hyams:1975mc,Hoogland:1977kt,Kaminski:1996da,Batley:2010zza}, $\pi\pi \rightarrow K\bar{K}$\cite{Cohen:1980cq,Etkin:1981sg}, and $\pi K \rightarrow \pi K$~\cite{Estabrooks:1977xe,Aston:1987ir} were mostly obtained in the 1970s and 1980s, indirectly from reactions of the form $M_1 N \rightarrow M_2 M_3 N^\prime$, where $M_i = \pi, K$ and $N, N^\prime$ are nucleons, assuming the dominance of one-pion exchange. This generic approach led to large systematic uncertainties and often inconsistent data sets, even within the same experiment.

Fortunately, two-body scattering involving pions and/or kaons is subject to strong dispersive constraints derived from first principles: analyticity (from causality), unitarity (incorporating probability conservation), and crossing symmetry. These constraints can be used to discard inconsistent data sets, restrict the allowed data descriptions, or be solved for certain partial waves in specific energy regions using data from other channels and regions as input.

\begin{figure}
    \centering
    \includegraphics[width=1\linewidth]{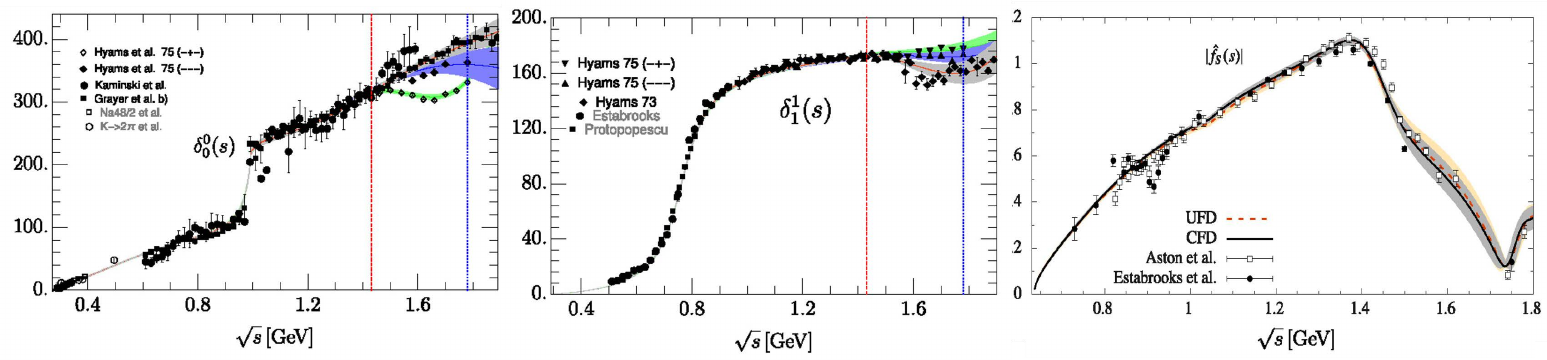}
    \vspace*{-1cm}
    \caption{Illustration of the precision achieved in dispersively constrained descriptions of $\pi\pi$ and $\pi K$ scattering data \cite{Hyams:1973zf, Protopopescu:1973sh, Grayer:1974cr, Hyams:1975mc, Kaminski:1996da, Batley:2010zza, Estabrooks:1977xe, Aston:1987ir}. Shown are the $\pi\pi \to \pi\pi$ phase shifts for the S0 (left) and P-waves (centre), as well as the modulus of the $K\pi \to K\pi$ $S$-wave amplitude for the isospin combination $f_S = f_0^{1/2} + f_0^{3/2}/2$ (right). Figures are taken from Refs.\cite{Pelaez:2019eqa, Pelaez:2020gnd}, where constrained parametrisations for these and other waves can be found.
\label{CH4:fig_CFDplots}}
\end{figure}

In particular, Roy-like dispersion relations for the lowest partial waves below 1.1\,GeV have either been solved with ChPT input \cite{Ananthanarayan:2000ht, Colangelo:2001df, DescotesGenon:2001tn, Buettiker:2003pp, Moussallam:2011zg}, or imposed as constraints in fits to data within that region \cite{Garcia-Martin:2011iqs, Pelaez:2020gnd}. In the latter case, forward dispersion relations have also been used to constrain the full amplitudes up to around 1.5\,GeV, although the data fits extend even further \cite{Pelaez:2019eqa,Pelaez:2020gnd}. These complementary approaches lead to fairly consistent, constrained parametrisations of the existing data for many partial waves. Some examples of these are shown in Fig.~\ref{CH4:fig_CFDplots}.

Relatively simple constrained parametrisations are available for $\pi\pi \rightarrow \pi\pi$ \cite{Garcia-Martin:2011iqs, Pelaez:2019eqa}, $\pi\pi \rightarrow K\bar{K}$ \cite{Pelaez:2020gnd}, and $K\pi \to K\pi$ \cite{Pelaez:2020gnd}. These can be used as input for other processes or to test the viability and reliability of different techniques relevant to FAIR. The pion-beam facility at FAIR will enable meson-meson FSI studies through reactions of the type $\pi^- p\rightarrow \pi\pi N/K\bar{K}N/K\pi\Lambda$, complemented by partial-wave analysis.

Roy and Roy-like dispersion relations have also been employed to gain direct spectroscopic insights, particularly for determining the mass and width -- defined via pole positions in the complex energy plane -- of broad, light scalar resonances. This includes the $f_0(500)$ \cite{Caprini:2005zr, Garcia-Martin:2011nna} and the $K_0^*(700)$ \cite{Descotes-Genon:2006sdr, Pelaez:2020uiw}, both determined with high precision. Even for comparatively simple resonances, such as the $\rho(770)$, the accuracy of modern production data can only be fully exploited using theoretical tools of sufficient rigour \cite{Colangelo:2018mtw, Hoferichter:2023mgy}.

Heavier $f_0$ resonances have been studied using forward dispersion relations (as in the case of the $f_0(1370)$ \cite{Pelaez:2022qby}), and through coupled-channel analyses of heavy-meson decays for even heavier states \cite{Ropertz:2018stk}. Pole parameters of strange resonances up to 1.8\,GeV$/c^2$ were extracted in Ref.\,\cite{Pelaez:2016klv}. Establishing model-independent relations between pole parameters and observable line shapes in this context remains of great importance \cite{Heuser:2024biq}.

Concerning the question of how to incorporate these meson–meson partial waves into the analysis of other processes, such as three-body decays of baryons with pion pairs in the final state, there are varying levels of rigour in the literature. For decades, particularly in spectroscopy studies, simple models such as isobars (sometimes neglecting non-resonant waves, coupled channels, or relying only on na\"ive Breit–Wigner resonant shapes) were sufficient to describe existing data semi-quantitatively.

However, given FAIR’s focus on precision and its connection to QCD, the goal should now be to achieve the highest level of rigour, which can be achieved through dispersive analyses and effective theories. Examples of such studies exist for $\gamma\gamma \to \pi\pi$ \cite{Garcia-Martin:2010kyn, Hoferichter:2011wk}, $\eta' \to \pi\pi\gamma$ \cite{Kubis:2015sga, Holz:2022hwz}, and for various $\Upsilon$ decays \cite{Chen:2015jgl, Chen:2016mjn, Baru:2020ywb}, where the level of theoretical control has improved significantly compared to earlier phenomenological studies, such as Ref.\,\cite{Chen:2011jp}. This method can be transferred to other systems.

As noted above, well-understood interactions can also serve as benchmarks for testing experimental methods aimed at extracting, \textit{e.g}., scattering parameters from processes other than direct scattering. A clear illustration of this idea is provided in Ref.\,\cite{Albaladejo:2025lhn}, where the femtoscopy data of Ref.\,\cite{ALICE:2023eyl} were reanalysed, yielding valuable insights.

\vspace{0.3cm}

\noindent
{\bf Reactions involving singly charmed hadrons}

 
\noindent
As a consequence of the strong scale dependence of the strong coupling constant at energy scales above $\Lambda_{\rm QCD} \sim 200{-}400\,$MeV -- see Fig.\,\ref{FigEffectiveCharge} on $r/r_p \lesssim 1$ -- the physics of systems containing heavy charm and/or bottom quarks (the lifetime of the top quark is too short for it to hadronise and thus top is not relevant here) differs significantly from that of purely light-quark systems.

In recent years, there have been significant experimental developments and a large number of discoveries of exotic hadronic structures in the charm sector, such as the $D^*_0(2300)$, $D^*_{s0}(2317)$, $D_{s1}(2460)$, $X(3872)$ (also known as $\chi_{c1}(3872)$), the family of $Z^+_c$ states, the pentaquarks $P_c$, and, most recently, the $T_{cc}(3875)^+$; for reviews emphasising different aspects, see Refs.\,\cite{Hosaka:2016pey, Esposito:2016noz, Chen:2016spr, Guo:2017jvc, Ali:2017jda, Olsen:2017bmm, Karliner:2017qhf, Liu:2019zoy, Brambilla:2019esw, Yang:2020atz, Chen:2022asf, Meng:2022ozq}).

These discoveries challenge conventional quark model pictures and have led to a renaissance in hadron spectroscopy, renewing interest in multi-hadron systems. The understanding of exotic hadrons is among the most prominent topics in contemporary hadron physics. A common feature shared by many of these exotic states is their proximity to hadron–hadron thresholds, such as $DK$, $D^*K$, $D\bar{D}^*$, and $DD^*$, which correspond to the $D^*_{s0}(2317)$, $D_{s1}(2460)$, $X(3872)$, and $T_{cc}(3875)$, respectively. 
Some of these discoveries must necessarily be multiquark states, so do not fit within standard formulations of the quark model, although multiquark configurations were already considered sixty years ago \cite{Gell-Mann:1964ewy, Zweig:1964ruk}.
The proximity of the masses with the thresholds suggests that residual interactions between the two hadrons may bind them into molecular states, where colour-singlet hadrons act as the relevant building blocks \cite{Hosaka:2016pey, Swanson:2006st, Guo:2017jvc}. 
When located very close to an $S$-wave threshold, such states typically have a spatial extension that is significantly larger than that of a compact hadron, \textit{i.e}., $r_{\rm compact} \approx  1\,$fm, a property which leaves striking imprints on observables.

Besides hadronic molecules, alternative quark configurations have also been discussed in the literature, including compact tetraquarks \cite{Ali:2019roi, Maiani:2024quj} built from (anti-)diquarks \cite{Barabanov:2020jvn}, and hadrocharmonia \cite{Dubynskiy:2008mq}. 

On the other hand, for hidden-charm systems below open-charm thresholds, the agreement between quark model predictions and experimental observations is remarkably good \cite{Godfrey:1985xj}. The presence of heavy-quark degrees of freedom thus offers a valuable tool for exploring exotic hadrons.
For instance, in 2003, the discovery of the $D^{\ast}_{s0}(2317)$ \cite{BaBar:2003oey} and $D_{s1}(2460)$ \cite{CLEO:2003ggt} in the charm–strange meson sector, and that of the $D_0^{\ast}(2300)$ (then referred to as $D^{\ast}_0(2400)$) \cite{Belle:2003nsh}, challenged the notion that mesons could be accurately described within a constituent quark–antiquark model, such as outlined in Ref.\,\cite{Godfrey:1985xj}. Specifically, three main issues were identified:
\begin{enumerate*}[label=(\textit{\roman*})]
\item the $D_{sJ}$ masses are significantly lower than quark model predictions\textcolor{blah}{---for example, the quark model of Ref.~\cite{DiPierro:2001dwf} puts the lightest
charm-strange states with $J^P=0^+$ ($J^P=1^+$)
170~MeV (145~MeV) too high with respect to experiment};
\item the mass splitting $M_{D_{s1}(2460)} - M_{D^\ast_{s0}(2317)} \simeq M_{D^\ast} - M_D$ (within uncertainties) appears fine-tuned;
\item $M_{D^{\ast}_0(2300)} \approx M_{D^{\ast}_{s0}(2317)}$, in contrast to the expectation that a strange meson should be heavier than its non-strange counterpart.
\end{enumerate*}

Unsurprisingly, these discoveries and the problems they pose have sparked a surge of theoretical investigations seeking to explain the states, interpreting them in various ways as $c\bar{q}$ mesons, compact tetraquarks, combinations thereof, or heavy–light meson–meson molecules; see Refs.\,\cite{Chen:2016spr,Guo:2017jvc} for reviews. These resonances are often regarded as charm counterparts to the light scalar mesons, with the $D_0^*(2300)$ and $D_{s0}^*(2317)$ corresponding to the $f_0(500)$ and $f_0(980)$, respectively \cite{vanBeveren:2003kd}.

Some lattice QCD computations \cite{Mohler:2013rwa, Lang:2014yfa, Bali:2017pdv} have found charmed–strange meson masses consistent with experimental values when meson–meson-type interpolators are employed. 
The first detailed lattice QCD study of the $D^\ast_0(2300)$ \cite{Moir:2016srx}, performed at $M_\pi \simeq 391\,$MeV$/c^2$ and incorporating the necessary $D\pi$, $D\eta$, and $D_s \bar{K}$ ($I=1/2$) coupled channels, identified a pole below the $D\pi$ threshold. 
Later studies include Ref.\,\cite{Cheung:2020mql} on the $D^{\ast}_{s0}(2317)$; Refs.\,\cite{Gayer:2021xzv, Yan:2024yuq} on the $D^{\ast}_0$; Refs.\,\cite{Gregory:2021rgy,Yeo:2024chk} in the SU(3)-symmetric quark mass limit, following the proposal in Ref.\,\cite{Du:2017zvv}); and Refs.\,\cite{Lang:2022elg, Lang:2025pjq, Gregory:2025ium} on the $D_1$ states.

A major advance in understanding the low-lying positive-parity charmed mesons came from unitarised heavy-meson chiral perturbation theory (UChPT); see Ref.\,\cite{Guo:2023wkv} for a recent review. The relevant amplitudes up to next-to-leading order (NLO) were computed in Refs.\,\cite{Kolomeitsev:2003ac, Hofmann:2003je, Guo:2006fu, Guo:2009ct}, with the associated low-energy constants (LECs) fixed in Ref.\,\cite{Liu:2012zya} by fitting to scattering lengths obtained via lattice QCD at various pion masses \cite{Liu:2008rza, Liu:2012zya}.

At NLO, UChPT successfully describes the finite-volume energy levels from the aforementioned lattice QCD studies in the non-strange $I = 1/2$ sector (with coupled channels $D\pi$, $D\eta$, and $D_s \bar{K}$) \cite{Albaladejo:2016lbb,Guo:2018tjx}, and in the $(I, S) = (0, 1)$ sector (with $DK$ and $D_s \eta$ coupled channels), where the $D_{s0}^*(2317)$ resides \cite{MartinezTorres:2014kpc,Albaladejo:2018mhb}.

\begin{figure}[t]
  \centering
  \includegraphics[width=0.6\textwidth]{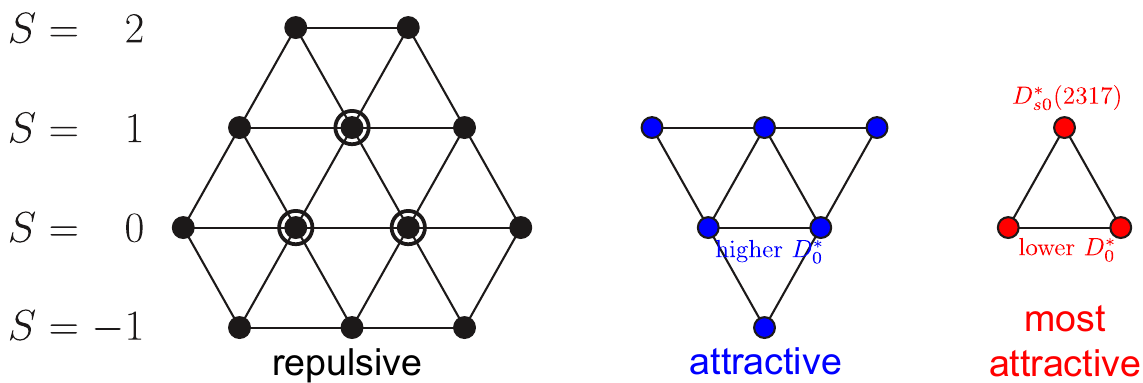}
  \caption{Decomposition of charmed-meson--light-meson pairs into different SU(3) flavour multiplets \cite{Albaladejo:2016lbb}.}
  \label{fig:irreps}
\end{figure}

The unique feature of the framework is that, among the irreducible representations of the SU(3) flavour decomposition $\mathbf{\overline{3}} \otimes \mathbf{8}=\mathbf{\overline{15}} \oplus \mathbf{6} \oplus \mathbf{\overline{3}}$ (see Fig.\ref{fig:irreps}), chiral symmetry, considered using leading-order ChPT, dictates that both flavour anti-triplet ($\mathbf{\overline{3}}$) and sextet ($\mathbf{6}$) charm–light meson scatterings exhibit attractive interactions, with the anti-triplet being the most attractive \cite{Kolomeitsev:2003ac, Hyodo:2006kg, Albaladejo:2016lbb}. 
The scalar $D_{s0}^*(2317)$ is in the anti-triplet, while there are two $I=1/2$ isospin doublets, one in the anti-triplet and the other in the sextet. Their mixing produces two $I=1/2$ $D_0^*$ resonances with poles at $\sqrt{s_L} = 2105^{+6}_{-8} - i 102^{+10}_{-12},\text{MeV}/c^2$ and $\sqrt{s_H} = 2451^{+36}_{-26} - i 134^{+7}_{-8},\text{MeV}/c^2$ \cite{Albaladejo:2016lbb, Du:2017zvv}, in contrast to the single $D_0^*(2300)$ listed in the Review of Particle Physics \cite{ParticleDataGroup:2024cfk}.
(See also Refs.\,\cite{Kolomeitsev:2003ac, Guo:2006fu, Guo:2009ct} and Ref.\,\cite{Meissner:2020khl} for a discussion of two-pole structures.)

The flavour sextet is exotic because it cannot be formed by a charm quark and a light antiquark. In addition to the higher $D_0^*$ with $(I,S)=(1/2,0)$, it also has a $D\bar K$ virtual state with $(I,S)=(0,-1)$ \cite{Albaladejo:2016lbb}, which has been supported by lattice QCD calculations \cite{Cheung:2020mql}, and an exotic resonance with $(I,S)=(1,1)$ \cite{Guo:2009ct}, deep in the complex plane owing to mixing with the repulsive $\mathbf{\overline{15}}$. A recent LHCb measurement of $B \rightarrow \bar{D}^{(*)} D_s^{+} \pi^{+} \pi^{-}$ may have observed a signal for the latter exotic resonance \cite{LHCb:2024iuo}.

Evidence for the two-$D_0^*$ picture also comes from comparison with experimental data. For instance, data for the processes $B^{-} \rightarrow D^{+} \pi^{-} \pi^-$ \cite{LHCb:2016lxy}, $B_s^0 \to \bar{D}^0 K^{-} \pi^{+}$ \cite{LHCb:2014ioa}, $B^0 \rightarrow \bar{D}^0 \pi^{-} \pi^{+}$ \cite{LHCb:2015klp}, $B^{-} \rightarrow D^{+} \pi^- K^-$ \cite{LHCb:2015eqv}, and $B^0 \rightarrow \bar{D}^0 \pi^{-} K^{+}$ \cite{LHCb:2015tsv} can all be well described using UChPT amplitudes \cite{Du:2017zvv, Du:2019oki, Du:2020pui}.
In addition, Ref.\,\cite{Du:2020pui} showed that the LHCb data for $B^{-} \rightarrow D^{+} \pi^{-} \pi^-$ \cite{LHCb:2016lxy} disfavour a Breit–Wigner description of the $D_0^*(2300)$ \cite{ParticleDataGroup:2024cfk}.

The same flavour structure also applies to axial-vector charmed mesons, featuring an antitriplet and a sextet in the molecular picture \cite{Albaladejo:2016lbb,Du:2017zvv}. 
A recent model, which considers spin-0 and spin-1 $cq$ diquarks and a spin-0 $\bar{q} \bar{q}$ anti-diquark, also predicts a similar flavour structure \cite{Maiani:2024quj}. 
However, when including a spin-1 light $\bar{q} \bar{q}$ anti-diquark, the $\mathbf{\overline{15}}$ would also appear in the same framework \cite{Dmitrasinovic:2004cu, Dmitrasinovic:2005gc, GuoHanhart2025}. Therefore, the recent lattice results in Ref.\,\cite{Gregory:2025ium} appear to favour the molecular picture over the diquark–antidiquark interpretation for the lowest-lying positive-parity open-charm states.

A crucial test to distinguish the molecular picture from the chiral doublet model of the $D_{s0}^*(2317)$ and $D_{s1}(2460)$ \cite{Bardeen:2003kt, Nowak:2003ra} lies in the prediction of the total width of the $D_{s0}^*(2317)$: it is around $\sim 100\,$keV$/c^2$ in the former \cite{Faessler:2007gv, Liu:2012zya, Guo:2018kno, Fu:2021wde}, but only a few tens of keV$/c^2$ in the latter \cite{Bardeen:2003kt}. 
Another useful observable is the shape of the dipion invariant mass distribution in the decay $D_{s1}(2460) \to D_s \pi^{+} \pi^{-}$ \cite{Tang:2023yls}, which exhibits a double-bump structure if the $D_{s1}(2460)$ has a molecular character, and a single broad bump if it is a compact state. A double-bump feature was recently observed by LHCb \cite{LHCb:2024iuo}.

As outlined above, for the positive-parity open-charm states, a coherent picture has emerged from lattice QCD and UChPT, which finds some support from experiment. 
However, among the most direct observables that can confirm or disprove the scenario advocated above are the scattering lengths in the relevant channels, as these are known to be sensitive to the subtle interplay between chiral dynamics and nearby states.

Information on these scattering lengths can be obtained via femtoscopic studies, with theoretical predictions provided in Refs.\,\cite{Liu:2012zya, Albaladejo:2023pzq, Torres-Rincon:2023qll}. 
On the experimental side, recent results from the ALICE Collaboration \cite{ALICE:2024bhk} for the correlation functions agree with neither the chiral amplitude predictions nor the lattice QCD results. 
In contrast, preliminary results from the STAR Collaboration \cite{Gwizdziel:2024xpl} show better agreement. Further measurements of this system are highly desirable.

Alternatively, $\pi D$ scattering information can be extracted from semileptonic $B$ decays \cite{Gustafson:2023lrz}, in analogy with semileptonic $K$ decays used to determine the $\pi\pi$ phase shifts \cite{Rosselet:1976pu}. 
A comparison of the values obtained through these independent methods would not only provide insights into the inner workings of strong interactions, but also play an important role in refining experimental tools.

{\color{blah} As the $\pi D$ system consists of non-identical particles, the correlation function is mostly sensitive to the final state interaction, attributed to the Coulomb and the strong force, and thus should to first be insensitive to the chaoticity of the source.  FAIR will therefore complement these studies through femtoscopy of the $\pi D$ system, providing additional access to the scattering lengths in a lower-multiplicity environment than ALICE and STAR, with tunable beam energies to suppress feed-down contributions.}

\subsubsection{Meson-baryon interactions}
\label{subsec.MesonBaryonInt}


\begin{figure}[t]
    \centering
    \includegraphics[width=0.99\linewidth]{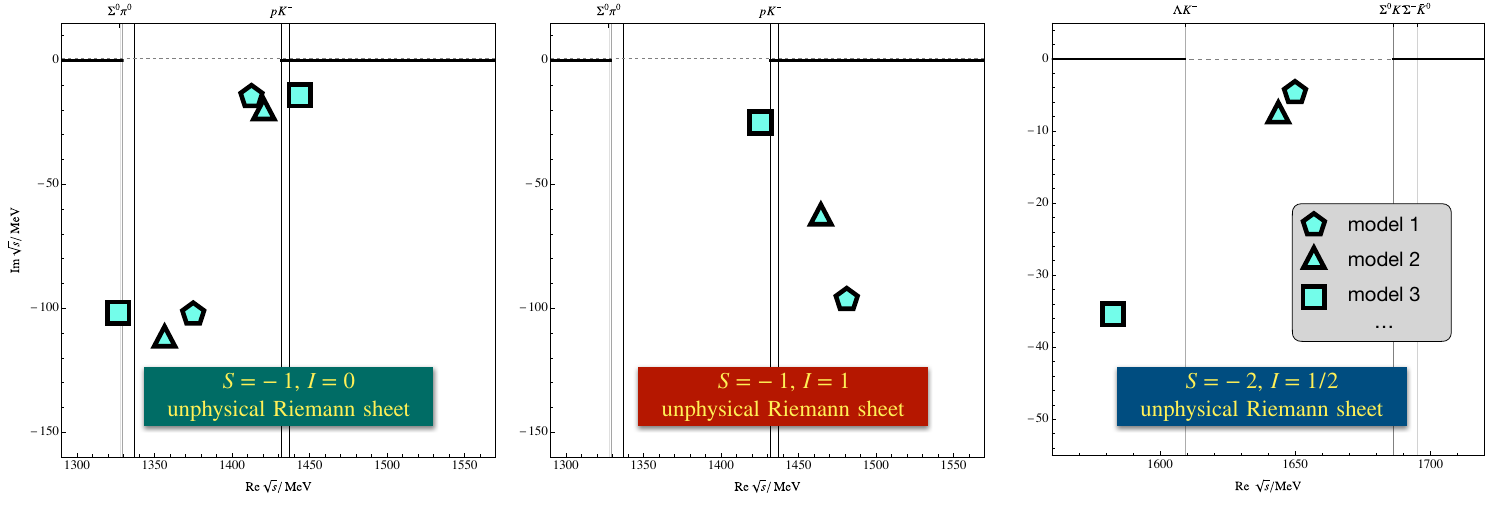}
    \caption{\textcolor{blah}{
    Sample of UChPT (NLO) model predictions for various quantum numbers, based on currently available meson–baryon scattering data. [Preliminary study based on Refs.~\cite{Mai:2014xna,Sadasivan:2018jig}.] A two-pole structure in the isoscalar, single-strangeness sector emerges consistently across all models, whereas the pole structures in other channels exhibit significantly greater variation, which owes primarily to the limited available data in the non-zero strangeness sector.}
    \label{CH4:figMM-2}}
\end{figure}

The dispersion-theoretical tools discussed in Sec.\,\ref{MesMes}, specifically Roy (or Roy–Steiner) equations, have so far only been applied to the simplest meson–baryon system, \textit{viz}.\ pion–nucleon scattering \cite{Hoferichter:2015hva}. 
Their rigorously justified range of validity in the $s$-channel is limited, allowing only for the precise extraction of pole parameters for the $\Delta(1232)$ and the Roper resonance $N(1440)$ \cite{Hoferichter:2023mgy}.
Beyond spectroscopic considerations, the strongly constrained pion–nucleon phase shifts are also important for other areas, such as investigations of CP violation in hyperon decays \cite{Zheng:2025tnz}.

Unitarised coupled-channel models in the light meson–baryon sector have been highly successful in describing several experimentally observed states as dynamically generated baryon molecules \cite{Doring:2025sgb}. For instance, in Ref.\,\cite{Krehl:1999km}, the well-known Roper resonance is identified as a candidate for a dynamically generated $f_0(500)N$ state. (For an alternative perspective, see Ref.\,\cite{Burkert:2019bhp}.) Another prominent case is the two-pole structure of the $\Lambda(1405)$ in the strangeness sector \cite{Meissner:2020khl, ParticleDataGroup:2024cfk,Mai:2020ltx,Hyodo:2020czb}.
Although there is a growing qualitative consensus on the latter, various open questions remain concerning meson–baryon systems involving strangeness. As an example, predictions of $S=-2$ resonance poles in the baryon sector based solely on $S=-1$ data cannot be made reliably, as demonstrated in Fig.\,\ref{CH4:figMM-2}. That figure shows three (among many possible) solutions that describe the $MB(S=-1) \to MB(S=-1)$ data with comparable quality. Future data expected from the upcoming SIS100 facility, {\it e.g.} by studying $pp\rightarrow pKK+\Xi\pi/\Lambda \bar{K}$ with CBM, will help resolve this issue by eliminating current ambiguities. In turn, this will also improve the precision with which known states, like the long-debated $\Lambda(1380)$, represented by the poles in the first row of Fig.\,\ref{CH4:figMM-2}, can be determined.
Addressing these challenges, which are crucial for hadron spectroscopy, will simultaneously deepen our understanding of meson–baryon interactions. This improved knowledge will be a key input for CP violation studies in $\Xi$ and $\Omega$ baryon decays. Other aspects of the light quark sector were discussed in Sec.\,\ref{sec4:subsec:Doring}.

Switching from systems with strange quarks, the heaviest of the light quarks, to those with charm quarks, the lightest of the heavy quarks, useful illustrations can be provided via open-charm states treated within unitarised coupled-channel approaches \cite{Tolos:2004yg, Tolos:2005ft, Lutz:2003jw, Lutz:2005ip, Hofmann:2005sw, Hofmann:2006qx, Lutz:2005vx, Mizutani:2006vq, Tolos:2007vh, Jimenez-Tejero:2009cyn, Haidenbauer:2007jq, Haidenbauer:2008ff, Haidenbauer:2010ch, Wu:2010jy, Wu:2010vk, Wu:2012md, Oset:2012ap, Garcia-Recio:2008rjt, Gamermann:2010zz, Romanets:2012hm, Garcia-Recio:2013gaa, Tolos:2013gta, Xiao:2013yca, Liang:2014kra, Nieves:2019kdh, Nieves:2024dcz}. 
This framework allows one to address some experimentally observed charmed baryon states, such as the $\Lambda_c(2595)$, which was proposed in Ref.\,\cite{Tolos:2004yg} as a candidate for a meson–baryon molecular state.
Subsequent models were developed based on a bare meson–baryon interaction, saturated via $t$-channel exchange of vector mesons in the zero-range approximation \cite{Lutz:2003jw, Lutz:2005ip, Hofmann:2005sw, Hofmann:2006qx, Mizutani:2006vq}. This approximation was revisited in later work by incorporating the full energy dependence of the $t$-channel vector-exchange term \cite{Jimenez-Tejero:2009cyn}.
Further developments can be found in Refs.\,\cite{Haidenbauer:2007jq, Haidenbauer:2008ff, Haidenbauer:2010ch, Wu:2010jy, Wu:2010vk, Wu:2012md, Oset:2012ap, Garcia-Recio:2008rjt, Gamermann:2010zz, Romanets:2012hm, Garcia-Recio:2013gaa}, with particular emphasis in the last three on incorporating heavy-quark spin symmetry (HQSS) \cite{Isgur:1989vq, Neubert:1993mb, Manohar:2000dt}. 
HQSS constraints at leading order were also respected in later works that extended the SU(3) local hidden gauge formalism to include the charm heavy flavour sector \cite{Xiao:2013yca, Liang:2014kra}.
An insightful discussion of the extended local hidden gauge approach and the $\mathrm{SU}(6){\mathrm{lsf}} \times \mathrm{SU}(2){\mathrm{HQSS}}$ framework is given in Ref.\,\cite{Nieves:2019kdh}. 
Most recently, Ref.\,\cite{Nieves:2024dcz} emphasised the importance of considering the interplay between constituent-quark model degrees of freedom and chiral baryon–meson interactions. These are coupled via a unitarised scheme consistent with lowest-order heavy-quark symmetries to determine the structure of the lowest-lying $1/2^-$ and $3/2^-$ $\Lambda_Q$ states in the strange, charm, and bottom sectors, including the notable case of the $\Lambda_c(2595)$.



The $D^-p$ interaction is characterised by a combination of Coulomb and strong interactions. The $D^-p$ system is coupled to the $\bar{D}^{0}n$ channel, whose threshold lies approximately 4\,MeV lower.
Recent femtoscopic measurements performed by the ALICE Collaboration \cite{ALICE:2022enj} have been compared with various theoretical models, including descriptions of the strong interaction via meson exchange \cite{Haidenbauer:2010ch, Yamaguchi:2011xb}, an SU(4) contact interaction \cite{Guichon:2004xg}, and a chiral quark model \cite{Fontoura:2012mz}. The data are compatible with the Coulomb-only hypothesis within 1.5 standard deviations, although the level of agreement improves slightly when an attractive $D^-p$ strong interaction is included.
Interestingly, the first lattice QCD calculations at the physical point by the HAL QCD Collaboration also suggest that the $D^-p$ interaction is moderately attractive \cite{Yamada:2025}.
The study in Ref.\,\cite{Yamaguchi:2011xb}, which is the model more compatible with experimental measurement, predicts a
rather attractive $\bar DN$ interaction and it also suggests the formation of a $\bar D N$ bound state in the isospin $I=0$ channel with a binding energy of the order of $2\,$MeV. 
According to this study, this attraction originates from the tensor-force-induced coupling to the $D^*N$ channel.
All other cited model calculations predict a repulsive interaction and agree less well with the experimental measurement. 
However, owing to the large experimental uncertainties, no model can yet be ruled out.

This first exploratory study paves the way for precision investigations of final-state strong interactions involving charm hadrons. Future measurements of these interactions will offer unique and valuable guidance for theoretical efforts aimed at advancing our understanding of the strong force in the charm sector. Complementary information can also be obtained by exploiting, \textit{e.g}., the cusp effect induced in the $D^0p$ excitation function by the $D^+n$ threshold \cite{Sakai:2020psu}, which points to the presence of a nearby quasi-bound state; see Sec.\,\ref{subsubsec.cusp}. The future physics programme at FAIR thus holds significant potential for deepening our understanding of interactions between charm hadrons and nucleons.


There has also been sustained interest in the low-energy $J/\psi N$ interaction \cite{Peskin:1979va, Bhanot:1979vb} for several reasons: it can offer vital insights into the role of gluons in nucleon structure \cite{Kharzeev:1995ij}; the hidden-charm $P_c$ pentaquarks were discovered in the $J/\psi p$ channel \cite{LHCb:2015yax,LHCb:2019kea}; it is relevant to the properties of $J/\psi$ in nuclear matter and the possibility of $J/\psi$–nucleus bound states \cite{Luke:1992tm, Sibirtsev:2005ex}; and the $J/\psi N$ scattering lengths provide a measure of how strongly two hadrons without any common valence quarks can interact.  Such interactions are Okubo–Zweig–Iizuka (OZI) suppressed. 

Different theoretical studies have reported significant variations in the $S$-wave $J/\psi N$ scattering lengths. For instance, a deep $J/\psi N$ bound state was obtained in lattice QCD with $m_\pi \approx 805\,$MeV$/c^2$ \cite{Beane:2014sda}, whereas a very weak $J/\psi N$ interaction was reported in Ref.\,\cite{Skerbis:2018lew} with $m_\pi \approx 266\,$Me$/c^2$V. Ref.\,\cite{Wu:2024xwy} argues that $J/\psi N$ scattering at low energies should be dominated by soft-gluon exchanges, rather than by open-charm coupled-channel effects, in contrast to $J/\psi$ near-threshold photoproduction \cite{Du:2020bqj, JointPhysicsAnalysisCenter:2023qgg}. 
The $J/\psi N$ scattering length has been estimated as $a_{J / \psi N} \lesssim -0.16\,$fm, consistent with the determination by the HAL QCD lattice collaboration, which places it in the range between $-0.42\,$ and $-0.28\,$fm for $m_\pi \approx 146\,$MeV$/c^2$ \cite{Lyu:2024ttm}. These scattering lengths could be investigated  experimentally at FAIR from $pp \to pp  J/\psi$ using the method described in Sec.\,\ref{subsubsec.dispersive}.
(Possible interaction mechanisms contributing to $J/\psi N$ scattering are also discussed in Sec.\,\ref{JpsiNCSM}.)


\vspace{0.3cm}

\noindent {\bf Role of baryon-meson interactions in non-leptonic weak decays of hyperons}


\noindent The various strengths of the non-leptonic decay branches are sensitive to the bound-state properties of quarks in hyperons. 
For $\Lambda$ and $\Sigma$ hyperon decays, the final states consist of a pion and a nucleon. 
The initial weak process can mix the hyperon with the nucleon and its excitations. 
Such a virtual state may then decay into the final hadron pair via the strong interaction. 
The mass of the $\Sigma$ hyperon lies approximately equidistant between the nucleon and the Roper resonance $N(1440)$. 
Consequently, the physics of the Roper, and possibly also of the $N(1535)$, which can be studied in detail through the pion beam programme at FAIR, influences these decays. 
A better understanding of the relevant scattering amplitudes is therefore essential.

In weak $\Xi$ decays, the final state contains a $\Lambda$ and a pion. This suggests a connection to the rather poorly known spectrum of excited $\Sigma$ states. 
The decay $\Xi^- \to \Lambda \pi^-$ is the golden channel for observing CP violation in hyperon decays \cite{Salone:2022lpt}; thus, a solid understanding of the associated strong phase is a prerequisite for such studies. 
Measurements of $\Lambda$-$\pi^-$ strong interactions near the $\Xi(1321)$ mass region are crucial for determining the relevant phase shifts needed in CP tests of cascade decays. 
This interaction can be studied at HADES using pion beams in the reaction $\pi^- p \rightarrow \Lambda \pi^- K^+$.

For the $\Omega$ baryon, literally every decay is interesting.
As remarked in Sec.\,\ref{subsubsec:hnu}, the semi-leptonic decays are flavour-related to the electroweak direct and transition form factors of the $\Delta(1232)$ baryon.  

In the main decay channel, $\Omega^- \to \Lambda K^-$, with a branching fraction of $67.7(7)$\%, the final state can carry orbital angular momentum $L = 1$ (p-wave) or $L = 2$ (d-wave). 
The general expectation is that the d-wave component is very small \cite{Jenkins:1991bt}, but this has recently been challenged by BESIII measurements \cite{BESIII:2020lkm}. 
The decay asymmetry parameter $\alpha_{\Omega \to \Lambda K}$ is known to be small \cite{ParticleDataGroup:2024cfk}. 
A significantly large d-wave would therefore imply a strong difference between the scattering phase shifts of the two partial waves. 
To obtain unambiguous information on the d-wave contribution, polarised $\Omega$ baryons are needed.
Currently, it is unknown whether $\Omega^-$ production in $pp$ interactions leads to significant polarisation.

In addition to the dominant decay channel, one can study the less well-known decays into $\Xi \pi$ final states, where even the $\alpha$ decay parameters are poorly known. 
The decays $\Omega^- \to \Xi^0 \pi^-$ and $\Omega^- \to \Xi^- \pi^0$ are the second and third most prominent decay modes, respectively. 
In weak hyperon decays, isospin can change via $\Delta I = 1/2$ or $\Delta I = 3/2$. 
For spin-$1/2$ hyperons and kaons, the $\Delta I = 1/2$ rule is observed to dominate \cite{Jenkins:1991bt, Cirigliano:2011ny}. 
Applied to $\Omega$ decays, this would predict the ratio
$\mathrm{Br}(\Omega^- \to \Xi^0 \pi^-) / \mathrm{Br}(\Omega^- \to \Xi^- \pi^0) = 2$,
which deviates from the experimental result. 
A recent BESIII analysis \cite{BESIII:2023ldd} reported $\mathrm{Br}(\Omega^- \to \Xi^0 \pi^-) / \mathrm{Br}(\Omega^- \to \Xi^- \pi^0) = 2.97 \pm 0.19 \pm 0.11$.

The $\Delta I = 1/2$ rule can further be tested via the three-body decays $\Omega \to \Xi \pi \pi$. 
While the branching ratios for the $\Xi \pi$ final states are relatively well known, the decay asymmetry parameters remain poorly measured \cite{ParticleDataGroup:2024cfk}, with the most recent data dating from 1984. 
Improved measurements are essential to disentangle the p- and d-wave contributions.

More generally, the $\Delta I = 1/2$ amplitude for $\Omega \to \Xi \pi$ forms a coupled-channel system with $\Omega \to \Lambda \bar{K}$. Exploring the associated strong scattering amplitudes is therefore of interest. 
Resonances with matching quantum numbers appear nearby, such as the $\Xi(1530)$, which lies above the $\Xi \pi$ but below the $\Lambda \bar{K}$ threshold and contributes to the p-wave, and the $\Xi(1820)$, which contributes to the d-wave \cite{Duplancic:2004dy}. 

Further insight can be gained by studying the decays $\Xi(1820) \to \Lambda \bar{K}$, $\Xi \pi$, $\Xi(1530) \pi$. 
The latter can proceed via both s- and d-waves, so determining separate branching ratios would be valuable. 
GlueX at JLab has initiated studies on the $\Xi(1820)$ \cite{Pauli:2022ehd} (see also Sec.\,\ref{sec:baryonspectra_exp}), and FAIR can provide complementary information.

The above discussion regarding final-state interactions in weak hyperon decays applies here, too. 
A better understanding of hadron-hadron scattering and the dynamics of resonance formation, \textit{e.g}., of the $\Xi(1820)$), as discussed in Ref.\,\cite{Kolomeitsev:2003kt}, will benefit the interpretation of two-body $\Omega$ decays.


\subsubsection{Baryon-baryon interactions}
\label{subsec.BaryonBaryonInt}

Thanks to a vast database and significant theoretical advances, our understanding of the nucleon–nucleon interaction is now highly developed; see Refs.\,\cite{Epelbaum:2008ga, Machleidt:2011zz, Epelbaum:2019kcf} for reviews. 
In contrast, much less is known about systems involving strangeness. This makes FAIR a particularly promising facility for systematically exploring the implications of QCD’s SU(3) flavour symmetry and its breaking.

\subsubsection{Baryon-baryon bound and virtual states (dibaryons)}
\label{subsubsec.Dibaryons}

In analogy with the isospin-zero proton–neutron bound
state, the deuteron, and the isospin-one proton–proton, proton–neutron and neutron–neutron virtual states that are signalled by the large respective scattering lengths, many two-baryon systems are predicted to develop poles dynamically that might appear in data as pronounced cusps or resonant structures. 
These are sometimes called dibaryons. 
Searches for such states have a long history, dating back almost 90 years \cite{Chadwick:1934}; however, a first positive signal was only found 14 years ago, when Ref.\,\cite{WASA-at-COSY:2011bjg} reported a clear peak structure in the reaction $pn\to d\pi^0\pi^0$, the $d^*(2380)$ dibaryon \cite{WASA-at-COSY:2014dmv, WASA-at-COSY:2012seb, A2:2019arr, Bashkanov:2018ftd, A2:2022kkx}. 
The pattern of the emergent states with baryon number
two promises new insights into the SU(3) structure of QCD and 
its inner workings. There are even claims that dibaryons modify the behaviour of matter at the high densities achievable in neutron stars \cite{Vidana:2017qey, Mantziris:2020xwi, Celi:2023gtj, Celi:2025wnc}.

With baryon number $B=2$, dibaryons have a high quark content, $N_{q}-N_{\bar{q}} =6$, which makes them difficult to produce and reconstruct. There are several approaches to accessing dibaryon degrees of freedom: searches for dibaryons in heavy quarkonia decays, photoproduction reactions on the deuteron, and nucleon–nucleon reactions.

The need to produce dibaryon–antidibaryon pairs with large particle multiplicities in the final state limits the possibility of studying baryon-baryon dynamics in heavy quarkonia decays. 
Small cross sections for photo-induced reactions and limited availability of mesonic beams make dibaryon production on deuteron targets with photo-, pion- and kaon-induced reactions challenging. 
These constraints leave nucleon–nucleon, or more generally baryon–baryon, collisions as the most promising source of information on dibaryon physics, making FAIR a unique facility for investigating various aspects of dibaryon dynamics. 
A large-acceptance, hermetic detector, such as CBM, can ensure that all decay products are reconstructed, reducing unwanted combinatorial contributions from background channels.

Various pairs of baryons from the baryon octet can produce (quasi-)bound states or virtual states: for instance, the presence of a quasi-bound state in the $\Sigma N$ system reveals itself as a pronounced cusp in the $\Lambda N$ cross section \cite{Haidenbauer:2021smk}. 
Besides the bound/virtual states between the (strong interaction) stable
baryons, one can consider the formation of deuteron-like bound/virtual states between higher-lying $N^\ast$ and $\Delta^\ast$ states. 
Such structures remain largely unexplored. 
There is a six-decades-old discussion of whether the spin $S=2$ $N\Delta$ system forms a quasi-bound state \cite{Gal:2013dca, Kukulin:2022gze, Niskanen:2023dsm}. 
Recently, the ELPH experiment reported evidence for additional structures in the dibaryon sector through an exclusive analysis of $\gamma d\to\pi^0\pi^0 d$ data~\cite{Ishikawa:2018wkv}. In particular, structures were identified in the relevant invariant-mass distributions and interpreted as possible additional $NN$-like dibaryon states, complementing the established $d^*(2380)$ signal. These observations have been partially supported by independent measurements from the BGOOD Collaboration~\cite{Jude:2022atd}. Together, these results point to a potentially richer spectrum of $B=2$ states than is currently established from $NN$ scattering alone. However, the interpretation of the observed structures remains unsettled, as a consistent theoretical description of both datasets is still lacking. This highlights the need for complementary measurements with controlled hadronic probes to establish the nature, quantum numbers, and dynamics of such candidate $B=2$ states.
 

In Ref.\,\cite{Gal:2013dca}, the $\pi N\Delta$ system is treated as a dynamical three-body problem, with the $\pi N$ interaction driven by the $\Delta(1232)$ and the $\Delta N$ interaction consistent with the resonance structure seen in the $^1 \! D_2$ partial wave of $NN$ scattering. 
It was found that the mass of the $d^*(2380)$ could be reproduced in this way. 
More differential cross section data on the $d^*(2380)$, \textit{e.g}., $d^*(2380)\to NN\pi$, $d^*(2380)\to d\gamma$ and $d^*(2380)\to de^+e^-$,
will allow this model to be tested further.

Analogously to the cusp mentioned above in the $\Lambda N$ cross section, driven by a $\Sigma N$ quasi-bound state, theory also predicts pronounced cusps in the $\Lambda\Lambda$ system at the thresholds of the $\Xi N$ channels owing to a nearby virtual state \cite{Haidenbauer:2015zqb, Fujiwara:2006yh}. (See also Sec.\,\ref{subsubsec.cusp}.)
In fact, the latest lattice calculations performed by the HAL QCD 
Collaboration close to the physical point~\cite{HALQCD:2019wsz}, 
predict the existence of a state, commonly referred to as the  
H-dibaryon, at exactly the same location. 
A recent calculation of the H-dibaryon at unphysical masses~\cite{Green:2021qol} revealed that systematic uncertainties due to the lattice discretisation can be large and  for the first time obtained a continuum limit result, with a weakly bound dibaryon at a SU(3) symmetric point with $m_\pi=m_K=420$~MeV~\footnote{A review of lattice QCD calculations of baryon-baryon scattering with a focus on prospects and challenges can be found in Ref.~\cite{Green:2025rel}.}.
To clarify the situation, high-statistics experimental studies of the $\Lambda\Lambda/\Xi N$ systems are highly desirable. 
Such studies can be performed at FAIR with reactions like $pp\to K^+K^+\Lambda\Lambda$ and $pp\to \Xi NK^+K^+$. 
We return to these systems below.
In addition, measurements of the type $pp\to \Lambda NK$ and $pp\to \Lambda N\pi K$ would enable searches for dibaryon structures in the $\Delta\Lambda/\Delta\Sigma/N\Sigma^*/\Delta\Sigma^*$ and $\Delta\Xi/\Sigma^*\Lambda/\Sigma^*\Sigma/\Sigma^*\Sigma^*/\Delta\Sigma^*$ systems.

There is an ongoing theory discussion of the possible existence of deuteron-like dibaryon states with charmed baryons, such as $\Lambda_c N$,
$\Lambda_c \Lambda_c$, and $\Xi_{cc}^{(*)} \Sigma_{c}^{(*)}$ \cite{Meguro:2011nr, Liu:2011xc,Pan:2020xek}. 
Experimental data on open-charm production are essentially non-existent. Any data on $D^{(*)}$ production in $pp$ collisions would be very valuable.



\noindent
\subsubsection{Secondary hyperon-nucleon interactions}

The most direct access to hyperon--nucleon scattering is via secondary interactions, where hyperon beams are generated in the collisions of some probe with a suitable target. {\color{blah} Experimentally, however, such measurements are particularly challenging. Since hyperons are unstable and decay on short timescales, they cannot be transported over macroscopic distances like stable hadron beams, and the interaction vertex has to be reconstructed from their decay products. At the same time, the probability for a hyperon to undergo a secondary interaction before decaying is small, requiring high-intensity secondary beams, sufficiently thick targets, and excellent tracking and vertexing capabilities. The resulting event rates are therefore typically low, while the identification of the primary hyperon, its decay, and the subsequent scattering process requires careful control of backgrounds and detector acceptance.}
Over the past few years, new measurements of $\Lambda N$ scattering by CLAS \cite{CLAS:2021gur} and $\Sigma N$ scattering by the E40 experiment at J-PARC \cite{J-PARCE40:2021qxa, J-PARCE40:2021bgw, J-PARCE40:2022nvq} have been reported, including the first more extensive data on 
$\Sigma^+p$ and $\Sigma^-p$ differential cross sections away from threshold.

The CLAS data were obtained from the reaction 
$\gamma p \to  [K^+]\Lambda,\; \Lambda p \to \Lambda' p' \to \pi^- pp'$,
covering the $\Lambda$ momentum range $0.9-2.0\,$GeV$/c$. 
In the J-PARC experiment, the $\Sigma^\prime$ baryons were produced in the
reactions $\pi^{\pm} p \to K^+\Sigma^{\pm}$, and the 
data were obtained for $\Sigma$ momenta between $0.44\,$GeV$/c$ and approximately $0.85\,$GeV$/c$. 
In addition, a $\Lambda p$ differential cross section at 
$\approx1\,$GeV has been reported by the BESIII Collaboration \cite{BESIII:2024geh}, where the $\Lambda$ is produced in the reaction $J/\psi \to \bar \Lambda\Lambda$. 
BESIII has also reported cross sections for $\Xi^0 n \to \Xi^-p$ \cite{BESIII:2023clq}, where the $\Xi^0$ was produced in $J/\psi \to \bar \Xi^0\Xi^0$, and for $\Sigma^+n$ scattering \cite{BESIII:2025bft} from 
$J/\psi \to \bar \Sigma^-\Sigma^+$. 
In the future, $\Lambda p$ angular distributions and even polarisations
will become available from the J-PARC E86 experiment \cite{Miwa:2022coz},
for momenta $0.4-0.8\,$GeV$/c$, exploiting $\Lambda$ production via $\pi^-p \to K^0\Lambda$. 
There is also a new project at SPring-8 to measure the $\Lambda p$ 
scattering cross section using $\gamma p\to K^+\Lambda$,
targeting the momentum range $0.3-0.6\,$GeV$/c$.

As evident from the above, none of the recent measurements covers momenta very close to threshold. 
Thus, characteristic quantities, such as the scattering lengths, cannot be determined from these data. 
For that, one must use production reactions, as discussed in 
Sec.\,\ref{subsec.scatfromprod} below. 
Still, data at moderate momenta remain very valuable, particularly when they include differential cross sections or other differential observables. 
Such information provides insight into more detailed features of baryon-baryon dynamics, especially regarding the relative importance of various partial waves. 
Indeed, this point was already exploited in the recent
extension of the hyperon-nucleon potential to next-to-next-to-leading order in ChEFT \cite{Haidenbauer:2023qhf}.
However, additional data at significantly lower energies are needed to complete the overall picture. 
In particular, angular-dependent observables would be especially useful for clarifying the actual onset of higher partial waves.







\subsubsection{Hypernuclei}
\label{subsec.hypernuclei}


While secondary scatterings provide valuable data on hyperon-nucleon interactions at moderate energies, and therefore for higher partial waves, there is little information at very low energies. 
Moreover, this approach offers no insights into hyperon-hyperon interactions. 
For those systems, valuable information can be extracted either from production reactions, as detailed in Secs.\,\ref{subsec.scatfromprod} and \ref{subsec.femto}, or from hypernuclei. 
These are nuclei in which at least one nucleon is replaced by a hyperon; see Ref.\,\cite{HypernuclearDataBase} for an overview. 
Here, details are provided on how the theoretical understanding of these systems can be improved, as well as the experimental status and the opportunities at FAIR. A brief perspective on systems with charm is also provided.

\vspace{0.3cm}

\noindent{\bf Theory aspects}
\label{yn_theory}


\noindent In the past decade, theoretical and computational tools for performing microscopic calculations of hypernuclei have been significantly improved~\cite{Haidenbauer:2025zrr}. 
Specifically, besides the conventional treatment within the Faddeev and Faddeev–Yakubovsky approaches \cite{Miyagawa:1993rd, Nogga:2001ef} and the variational approach \cite{Hiyama:2001zt, Nemura:2002fu}, new {\it ab initio} methods, like the No-Core Shell Model (NCSM) have been developed \cite{Wirth:2014apa, Le:2020zdu}, allowing one to compute binding energies for hypernuclei well beyond the $S$-shell. 
So far, studies of hypernuclei up to $^{13}_{\ \Lambda}$C have been reported \cite{Wirth:2014apa}.

Moreover, YN interactions based on ChEFT have been employed. 
Originally suggested for the nucleon–nucleon interaction, ChEFT can be extended to baryons with strangeness by exploiting the approximate SU(3)- flavour symmetry of QCD \cite{Haidenbauer:2013oca, Haidenbauer:2019boi, Haidenbauer:2023qhf}. 
One particular merit of this approach is that two-body and three-body forces (3BFs) can be derived and treated consistently \cite{Petschauer:2015elq}.

Hypernuclei have also been studied in the framework of nuclear lattice effective field theory (NLEFT) \cite{Hildenbrand:2024ypw}. 
This framework is a powerful quantum many-body method that combines aspects of effective field theories with lattice methods. 
Using this scheme, results for hypernuclei up to $^{16}_{\ \Lambda}$O have been obtained. 
Other approaches, like the auxiliary field diffusion Monte Carlo algorithm, have allowed investigations to extend to much heavier hypernuclei \cite{Lonardoni:2013gta}, though only for simplified representations of the YN interaction.

\begin{figure}
  \centering
  \includegraphics[width=0.5\linewidth]{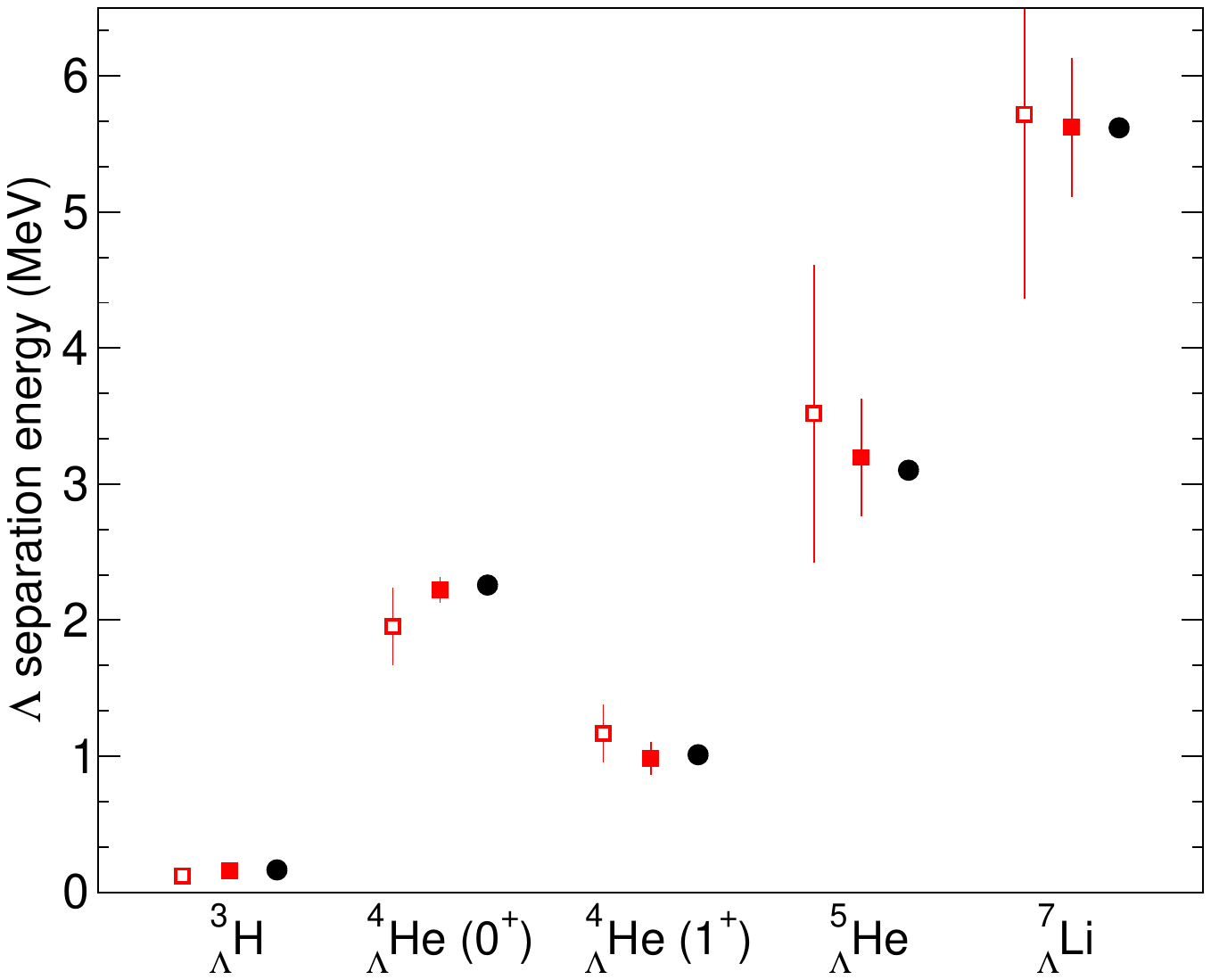}
  \vspace*{-.5cm}
  \caption{Separation energies for light $\Lambda$ hypernuclei \cite{Le:2024rkd}. Results without (open squares) and with (filled squares) three-body forces are shown, including theoretical uncertainties at NLO and N$^2$LO in the chiral expansion. Experimental values are indicated by circles.
  \label{CH4:JH_sepE}}
\end{figure}

Recent calculations of light hypernuclei within {\it ab initio} approaches and with realistic $\Lambda N$–$\Sigma N$ potentials as input show that the binding energies and mass spectra can be qualitatively reproduced \cite{Haidenbauer:2019boi, Le:2023bfj}. 
However, they also revealed noticeable deviations from the experimental values. 
Whether these deviations point to shortcomings in the employed YN interactions or indicate a need to include 3BFs remains unclear at present. 
A partial answer has been given by investigations employing NN and YN interactions derived using ChEFT. 
In this case, the inherent power counting not only determines the order at which 3BFs start to contribute, thereby enabling conclusions on their relative importance, but also allows an estimate of their magnitude \cite{Le:2023bfj}. 
Calculations of $\Lambda$ hypernuclei up to $A = 7$, guided by such principles, suggest that 3BFs alone could be sufficient to achieve agreement with the experimental binding energies \cite{Le:2024rkd}; see Fig.\,\ref{CH4:JH_sepE}. 
Similar conclusions can also be drawn from the NLEFT work, where even hypernuclei up to $A = 16$ were considered \cite{Hildenbrand:2024ypw}.

The binding energies of different hypernuclei also provide valuable input to understand the amount of charge-symmetry breaking (CSB) in the hyperon-nucleon interaction.
To date, there are no data on $\Lambda n$ scattering that could be compared with those for $\Lambda p$, or for $\Sigma^- n$ compared with $\Sigma^+p$, to establish possible CSB in the $\Lambda N$ and/or $\Sigma N$ systems. 
However, clear experimental evidence for CSB has long been available from the separation energies of the hypernuclei $^4_\Lambda$H and $^4_\Lambda$He, in both the ground state ($J^P=0^+$) and the excited state ($1^+$) ~\cite{HypernuclearDataBase}. 
The differences in the separation energies are on the order of $100-200\,$keV, \textit{i.e}., much larger than those known from ordinary nuclei, such as $^3$H and $^3$He, when excluding the trivial effect of the Coulomb interaction.

There are also noticeable CSB effects in heavier systems, such as the isospin triplet $^7_\Lambda$He, $^7_\Lambda$Li$^*$, $^7_\Lambda$Be, and the $T=1/2$ doublet $^8_\Lambda$Li, $^8_\Lambda$Be \cite{Botta:2016kqd}. CSB effects in light $\Lambda$ hypernuclei have recently been studied using phenomenological CSB potentials \cite{Gazda:2015qyt}, as well as a CSB interaction derived within ChEFT \cite{Haidenbauer:2021wld}. 
In the latter case, CSB in the energy levels can be well described, leading to the conclusion that the $\Lambda n$ interaction in the $^1S_0$ state must be noticeably more attractive than that in $\Lambda p$, whereas the difference in the $^3\! S_1$ state is much smaller and of opposite sign.

An investigation has also been conducted to assess whether the CSB effects established from the $^4_\Lambda$H/$^4_\Lambda$He systems are consistent with those in the $A=7,8$ hypernuclei \cite{Le:2022ikc}. 
However, firm conclusions remain difficult to draw owing to the large experimental uncertainties in the latter cases. 
New, higher-precision measurements are desirable for these systems, and also for other light hypernuclei where calculations within {\it ab initio} approaches remain feasible.
 
Finally, it should be mentioned that {\it ab initio} approaches have also been successfully applied in calculations of $\Lambda\Lambda$ and $\Xi$ hypernuclei \cite{Hiyama:2019kpw, Contessi:2019csf, Le:2021wwz, Le:2021gxa}. 
Candidates for the possibly lightest bound systems have been identified, for example $^{\ \ 5}_{\Lambda\Lambda}$He \cite{Contessi:2019csf,Le:2021wwz} or $^4_\Xi$H \cite{Hiyama:2019kpw,Le:2021gxa}. 
These await experimental validation, although it is clear that the corresponding experiments are challenging. 
However, the high beam intensities and variety of beams provided by FAIR, including the long-term antiproton programme, offer a very promising experimental perspective.

The general theoretical picture revealed by phenomenological models and lattice QCD (HAL QCD Collaboration) computations shows that the $\Lambda_c$-nucleon ($\Lambda_c$N) interaction is attractive \cite{Miyamoto:2017tjs, Haidenbauer:2017dua, Garcilazo:2019ryw, Vidana:2019amb, Maeda:2015hxa}. 
However, the $\Lambda_c$N interaction 
determined from initial lattice QCD calculations~\cite{Miyamoto:2017tjs} is significantly weaker than that suggested by most phenomenological studies and very recent lattice results~\cite{Zhang:2024} indicate even a repulsive $\Lambda_c N$ interaction. Contrary to the latter, preliminary femtoscopic results from the ALICE Collaboration suggest that the $\Lambda_c N$ interaction could be indeed weakly attractive~\cite{Zhang:2025szg}.
Thus, it is generally believed
that bound $\Lambda_c N$ states should exist, even for a moderately attractive $\Lambda_c N$ interaction \cite{Haidenbauer:2020uci}.

\vspace{0.3cm}

\noindent
{\bf Experimental status: nuclei with strangeness}


\noindent
Hypernuclei can provide valuable information on the hyperon–hyperon and hyperon–nucleon interaction at low energies. 
They can, for instance, be formed by coalescing a hyperon into a nucleus or transforming a nucleon inside the nucleus into a hyperon, \textit{e.g}., by a strangeness-exchange reaction $K^{-}+{}^{A}Z\rightarrow \pi^{-}+{}^{A}_{\Lambda}Z$ or an associated production reaction $\pi^{+}+{}^{A}Z\rightarrow K^{+}+{}^{A}_{\Lambda}Z$ \cite{Feliciello:2015dua,Gal:2016boi}. 
The production in heavy-ion collisions is normally modelled either by coalescence or within a statistical-thermal approach \cite{Braun-Munzinger:2018hat}. 
A rather large number of single-$\Lambda$ hypernuclei are known experimentally and, for some, even excited states are well studied by, \textit{e.g}., $\gamma$-ray spectroscopy \cite{Hashimoto:2006aw, HypernuclearDataBase}. 
Interestingly, a stable hypernucleus of mass number $A=5$ is known, \textit{viz}.\ ${}^{5}_{\Lambda}$He, where the normal nuclei of the same mass number are all unstable, only being visible as quickly decaying resonances. 
Here, the $\Lambda$ provides additional attraction. 
There is currently a debate over whether the $\Lambda nn$ bound state has been observed experimentally \cite{HypHI:2013sxa, WASA-FRS:2023yro}. 
A few double-$\Lambda$ hypernuclei have been observed experimentally, one candidate for a single-$\Sigma$ hypernucleus is known, and there is even a suggestion of a $\Xi$ hypernucleus~\cite{HypernuclearDataBase}. 
Heavy-ion collisions at RHIC and LHC have also revealed a few anti-hypernuclei, namely anti-hypertriton, ${}^{3}_{\bar{\Lambda}}\overline{\rm H}$~\cite{STAR:2010gyg}, anti-hyperhydrogen-4, ${}^{4}_{\bar{\Lambda}}\overline{\rm H}$~\cite{STAR:2023fbc}, and anti-hyperhelium-4, ${}^{4}_{\bar{\Lambda}}\overline{\rm He}$ \cite{ALICE:2024ilx}.

Since there is no known hypernucleus of mass number $A=2$, the hypertriton ${}^{3}_{\Lambda}$H is the lightest known hypernucleus. 
Its binding energy is only about 2.3\,MeV, and usually the value of the so-called $\Lambda$-separation energy $B_{\Lambda}$ is reported. 
This is the energy needed to separate the $\Lambda$ from the remaining nucleus: in the case of the hypertriton, from a deuteron core, with $B_{\Lambda} = 174 \pm ^{110} _{34}\,$keV~\cite{HypernuclearDataBase}.

Higher-mass hypernuclei are all more strongly bound, \textit{i.e}., $B_{\Lambda} \approx 2\,$MeV for $A=4$ hypernuclei and $B_{\Lambda} \approx 3\,$MeV for ${}^{5}_{\Lambda}$He \cite{HypernuclearDataBase}. 
The heaviest known hypernucleus is ${}^{208}_{\;\;\;\Lambda}$Pb, with a $\Lambda$-separation energy of $B_{\Lambda} = 26.5 \pm 0.5\,$MeV \cite{HypernuclearDataBase}.

The ground states of all known hypernuclei decay weakly with lifetimes on the order of $10^2\,$ps. 
The hypertriton is a special case.
Since its $\Lambda$-separation energy is only a few hundred keV, it is believed to decay as practically a free $\Lambda$ hyperon \cite{Gal:2016boi,Hildenbrand:2020kzu}. 
The most recent and precise experimental results from heavy-ion collisions \cite{STAR:2021orx, ALICE:2022sco} yield values very close to the free $\Lambda$ lifetime, \textit{i.e}., around 263\,ps \cite{ParticleDataGroup:2024cfk}. 
Heavier hypernuclei have lifetimes generally around $140 - 280\,$ps.

The ground-state mass or, more precisely, the $\Lambda$-separation energy of hypernuclei can directly be connected to an effective $\Lambda$-nucleon interaction. 
This has been studied within phenomenological approaches and shell-model calculations, resulting in a typical single-$\Lambda$ potential depth of $U_{\Lambda} \approx 30\,$MeV in infinite nuclear matter \cite{Gal:2016boi}. 
ChEFT calculations use the limited YN scattering data to constrain their low-energy constants and are then employed to describe $\Lambda$-separation energies via so-called {\it ab initio} methods.
The production of light hypernuclei in reactions with $\pi$ beams on various targets \cite{Kittiratpattana:2023atz,Ergun:2024jwi} will be discussed in Sec.\,\ref{subsec.hypernuclei_in_piA}.

For heavier hypernuclei, the measurement of weak mesonic decays is feasible up to masses around that of carbon hypernuclei, where the mesonic weak decay branch is estimated to be around 10\% of the total decay rate \cite{Motoba:1988sk}. 
Beyond this, non-mesonic weak decay $\Lambda N \to N N$ becomes the only experimentally accessible mode. 
These decays are not Pauli-blocked and have a $Q$-value of approximately 176\,MeV, enabling the two nucleons to escape the residual unstable nucleus, and thus permitting investigation of the interaction at short distances within the hyper-bound-state. 
Furthermore, non-mesonic weak decays resemble the $\Delta S = 0$ nucleon--nucleon interaction, which can also be studied via $NN$ scattering. 
Unlike $NN$ scattering, where the parity-conserving component of the weak interaction is masked by the strong force, non-mesonic weak decays offer a clearer picture: both the parity-violating and parity-conserving components are accessible \cite{Parreno:1996if}. 
Hence, measurements of non-mesonic decays of heavy hypernuclei provide access to hitherto unexplored aspects of the weak $NN$ interaction.

With meson (pion or kaon) or electron beams, only elementary processes enable the production of hypernuclei near the stability line of the target nuclei. 
The use of proton-rich and neutron-rich exotic beams on fixed targets is one of the primary experimental approaches for producing proton- or neutron-rich hypernuclei \cite{Rappold:2016pdf}. 
In this context, the study of $^9_\Lambda$B is of particular interest, as it has the proton-halo nucleus $^8$B as its core, which will significantly influence the binding energy of $^9_\Lambda$B. 
Thus, the opportunity to study this hypernucleus at FAIR will offer unique, direct insights into the $\Lambda N$ interaction in a proton-rich environment.

\vspace{0.3cm}

\noindent{\bf Status of charmed nuclei}


\noindent
The poorly constrained interaction of nucleons with charmed hadrons, particularly baryons, allows for the hypothetical existence of charmed nuclei. 
Indeed, there have been some tentative claims of experimental observation. Three ambiguous candidates were reported by an emulsion experiment conducted in Dubna around forty-five years ago \cite{Batusov1976, Batusov:1981pu, Batusov1981je}.
{\color{blah}However, the experimental production and identification of charmed nuclei remain highly challenging due to their low production rates and the short lifetime of charmed hadrons.}

The experimental discovery of charmed nuclei would be striking and could provide crucial insights into the interactions between charmed baryons and nucleons, through comparison of the $\Lambda_c$ separation energies with theoretical models. 
As discussed herein, the CBM experiment is well suited for the reconstruction of charmed hadron decays as well as hypernuclei, making the search for charmed nuclei feasible. 
The most promising candidates are the charmed deuteron (${\rm c_d \rightarrow dK^-\pi^+}$) and charmed triton (${\rm c_t \rightarrow {^{3}H}K^-\pi^+}$) \cite{Andronic:2021erx, ALICE-PUBLIC-2023-002}, assuming the decay $\Lambda_c^+ \rightarrow pK^-\pi^+$ as the basis, with the subsequent conversion behaving similarly to that in hypernuclei.

\subsection{Scattering parameters from short ranged production reactions}
\label{subsec.scatfromprod}

Analyticity and unitarity provide a strong connection between scattering and production reactions. 
This section details how this connection can be exploited to extract scattering parameters from production reactions with short-ranged production mechanisms.

Weak decays of heavy mesons are typical examples of short-ranged production mechanisms, but the production of heavy final states in $pp$ collisions, which can be studied at FAIR, also qualifies. 
To see this, note that the initial energy of the proton pair must be sufficiently high to put the final-state particles on their mass shell. This requires large initial momenta, $p_i$, and, in addition, for near-threshold production, momentum transfers that are much larger than the final momenta. 
To make this statement more quantitative, observe that to access a final state of total mass $M_f$, the initial proton momentum in the centre-of-mass frame must be as large as $p_i = \frac{1}{2} \sqrt{M_f^2 - 4M_p^2}$.
For near-threshold production, this value serves simultaneously as an estimate for the typical momentum transfer. 
On the other hand, one may estimate the final momentum range necessary to study the effect of a scattering length $a$ of the order of 1\,fm as $p_f \approx 1/a \approx 200\,\mbox{MeV}/c^2$.
For a reaction such as $pp \to pK\Lambda$ with $M_f = M_\Lambda + m_K + M_p \approx 2550$\,MeV$/c^2$, one finds $p_i \approx 860\,$MeV$/c$. 
The impact of the production operator on the final-state distributions may thus be estimated as $(p_f/p_i)^2$, \textit{i.e}., of the order of 5\%. For heavier final states, the effect is even smaller. 
With an accuracy of 5\% or better, one may therefore treat the production of hidden-strangeness final states from $pp$ collisions as production from a pointlike source.

\begin{figure}[t]
    \centering
    \includegraphics[width=0.58\textwidth]{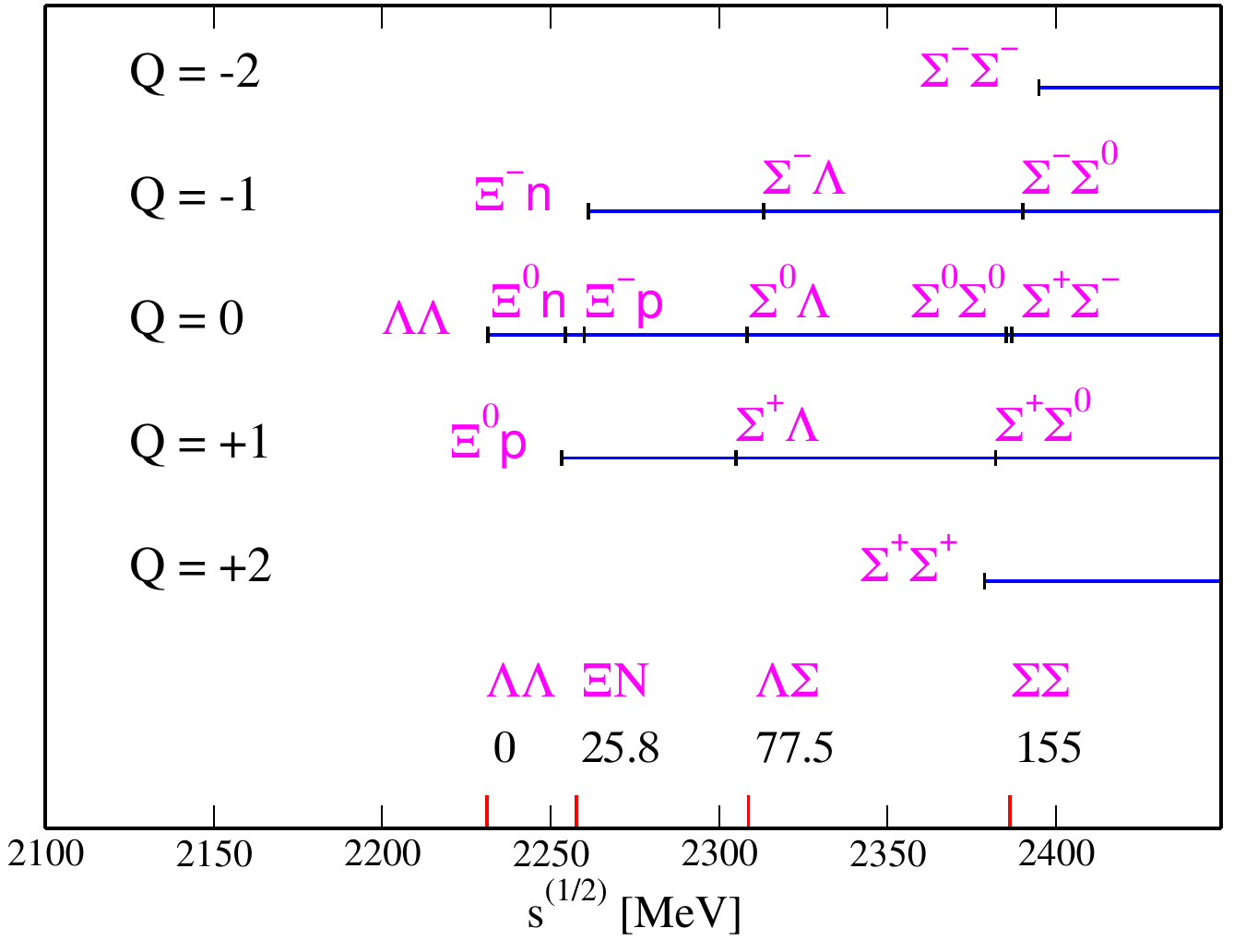}
    \caption{Thresholds for the various strangeness minus-two two-baryon systems. Those with charges $Q = 0$, $1$, and $2$ are, in principle, accessible at FAIR.} 
    \label{fig:S2}
\end{figure}

As argued below, particularly for experimental studies of two-baryon systems with strangeness $S = -2$, \textit{i.e}., those containing two strange quarks, FAIR offers excellent opportunities. 
Not only can a variety of systems be studied, but a range of tools is also available, allowing for systematic uncertainties to be investigated simultaneously. 
For illustration, Fig.\,\ref{fig:S2} shows the location of the thresholds for different $S = -2$ channels, sorted by their electric charge. In $pp$ collisions, through reactions of the type $pp \to K^+K^+BB'$, with $B^{(\prime)} = N$ or $Y$ and under the condition that the total strangeness of the $BB^\prime$ system is $-2$, one gains access to neutral two-baryon systems. Meanwhile, $pp \to K^+K^0BB^\prime$ and $pp \to K^0K^0BB^\prime$ allow access to systems with charges $Q = 1$ and $Q = 2$, respectively.

\subsubsection{Scattering lengths from threshold cusps} 
\label{subsubsec.cusp}


Unitarity of the $S$-matrix dictates that amplitudes must exhibit a square-root branch point at each two-particle threshold. 
Consequently, reaction rates must show a specific non-analytic behaviour -- a cusp -- at the thresholds of coupled two-particle channels, when the systems that
open up are in an $S$-wave. There is no corresponding cusp for higher partial waves owing to the centrifugal barrier, nor for the opening of multi-particle thresholds owing to their different phase-space behaviour. 
Since the cusp occurs precisely at the threshold energy, its shape directly encodes the transition strength between the two coupled channels, \textit{i.e}., the one at threshold and the observed channel \cite{Budini:1961bac, Cabibbo:2004gq, Meissner:1997fa}; for a review, see Ref.\,\cite{Guo:2019twa}. 
This holds provided that their relative production strengths are known; see the discussion of Eq.\,\eqref{FKGEq2} below.

Pronounced cusps appear particularly when there is either one pole or even several nearby.
A good example is the double cusp near the charged and neutral $K\bar{K}$ thresholds, enhanced by isospin-violating $a_0(980)$-$f_0(980)$ mixing, which was predicted theoretically \cite{Achasov:1979xc, Achasov:2003se, Hanhart:2007bd} and confirmed experimentally by BESIII in $J/\psi \to \phi \pi \eta$ \cite{BESIII:2018ozj,BESIII:2023zwx}.

The connection between cusp strength and scattering parameters in the threshold region can be quantified with high precision in a model-independent manner using non-relativistic effective field theory (NREFT). Indeed, the cusp effect has enabled the most precise determination of the $\pi\pi$ scattering lengths to date \cite{Batley:2009ubw}, based on $6.031 \times 10^7$ $K^{\pm} \rightarrow \pi^{\pm} \pi^0 \pi^0$ decays collected by the NA48/2 experiment, using the Bern-Bonn NREFT framework \cite{Colangelo:2006va, Bissegger:2008ff, Gasser:2011ju}. A scattering length as small as 0.3\,fm was extracted with a few percent accuracy.

Consider two channels, channel~1 and channel~2, with thresholds at $E_{\rm th1}$ and $E_{\rm th2}>E_{\rm th1}$, respectively. At leading order (LO) in NREFT, the $T$-matrix in the non-relativistic normalisation is given by \cite{Cohen:2004kf, Dong:2020hxe, Sakai:2020psu}
\begin{align}
    T^{\rm NR} & = - \frac{2 \pi}{\operatorname{det}}\left(\begin{array}{cc}
    \frac{1}{\mu_1}\left(\frac{1}{a_{22}}+i p_2\right) & \frac{1}{a_{12} \sqrt{\mu_1 \mu_2}} \\
    \frac{1}{a_{12} \sqrt{\mu_1 \mu_2}} & \frac{1}{\mu_2}\left(\frac{1}{a_{11}}+i p_1\right)
    \end{array}\right), \quad 
    \operatorname{det} = \frac{1}{a_{12}^2}-\left(\frac{1}{a_{11}}+i p_1\right)\left(\frac{1}{a_{22}}+i p_2\right),
    \label{FKGEq2}
\end{align}
where $p_i = \sqrt{\lambda(s,m_{i1}^2, m_{i2}^2)}/(2\sqrt{s})$ is the CM momentum of a particle in channel~$i$, with $m_{i1}$ and $m_{i2}$ being the masses of the two particles; $\mu_i = m_{i1}m_{i2}/(m_{i1}+m_{i2})$ is the reduced mass. The parameters $a_{11}$ and $a_{22}$ describe the interaction strength within each individual channel, while $a_{12}$ characterises the inter-channel coupling.

The production amplitudes for channels~1 and~2 are then given by
\begin{align}
   \mathcal{A}_1  = P_1 \left(T_{11}^{\rm NR} + \frac{P_2}{P_1} T_{21}^{\rm NR} \right) \quad \text{and} \quad \mathcal{A}_2 = P_1 \left(T_{12}^{\rm NR} + \frac{P_2}{P_1} T_{22}^{\rm NR} \right), \label{eq:Aprod}
\end{align}
respectively, where $P_1$ and $P_2$ are constants describing the production strengths of the two channels. The parameters $a_{11}$, $a_{22}$, $a_{12}$, and $P_2/P_1$ can be determined by fitting the invariant mass distributions of both channels ($P_1$ acts as a normalisation factor).

Once the parameters are fixed, the scattering length $a_i$ can be obtained as: $a_i = \left.-\frac{\mu_i}{2 \pi} T_{ii}^{\rm NR}\right|_{\sqrt{s} = m_{i1} + m_{i2}}$. The sign convention used here implies $p_i \cot \delta_i = -1/a_i + \mathcal{O}(p_i^2)$, which differs from that adopted in Ref.\,\cite{Sakai:2020psu}.

\begin{figure}[t]
    \centering
    \includegraphics[width=0.58\textwidth]{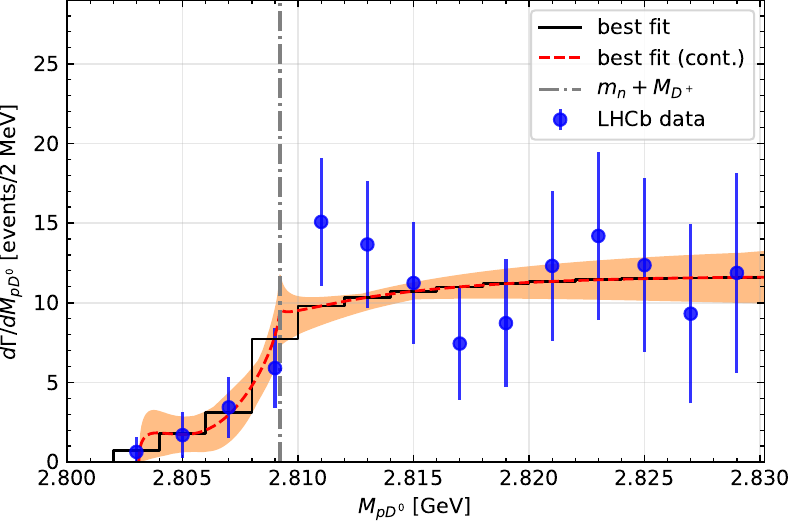}
    \caption{Fit to the $pD^0$ invariant mass distribution of the $\Lambda_b \rightarrow \pi^{-} p D^0$ decay using NREFT at leading order (LO). The data are from the LHCb experiment~\cite{LHCb:2017jym}. Figure adapted from Ref.\,\cite{Sakai:2020psu}.} 
    \label{fig:ND}
\end{figure}

This formalism has been employed to extract the isoscalar and isovector $ND$ scattering lengths as $-0.79_{-0.61}^{+0.66}$~fm for $I=0$ and $-3.8_{-2.0}^{+1.4}+i\,2.7_{-2.7}^{+1.6}$~fm for $I=1$, respectively~\cite{Sakai:2020psu}, by fitting to the $pD^0$ invariant mass distribution of the $\Lambda_b \rightarrow \pi^{-} p D^0$ decay; see Fig.\,\ref{fig:ND}.

The method can be used to obtain new information on strangeness $S = -2$ systems, namely on the $\Lambda\Lambda$ to $N\Xi$ transition strength from the $pp \to K^+K^+ \Lambda\Lambda$ and $pp \to K^+K^+ p\Xi^-$ reactions. The thresholds of the $\Lambda\Lambda$ and $N\Xi$ channels are at about 2231\,MeV and 2260\,MeV, respectively, with the $\Xi N$ threshold splitting into $n\Xi^0$ and $p\Xi^-$, separated by around 5\,MeV. Thus, the formalism mentioned above needs to be extended to three channels and should allow for the presence of the Coulomb interaction in the $p\Xi^-$ channel. As already pointed out, theoretical studies of the $\Lambda\Lambda$ -- $\Xi N$ systems definitely suggest the presence of a pronounced cusp structure \cite{Fujiwara:2006yh, Haidenbauer:2015zqb}.


A measurement of the $\Lambda\Lambda$ and $p\Xi^-$ invariant mass distributions with sufficiently good energy resolution around the $p\Xi^-$ threshold can be used to extract the $N\Xi \to \Lambda\Lambda$ transition strengths. 
The next-to-leading-order corrections are expected to be at the percent level (estimated as $\mathcal{O}(v^2)$, where $v$ is the magnitude of the relative two-particle velocity). 
It should be emphasised that a full analysis requires high-quality data for both the $\Lambda\Lambda$ and the $\Xi N$ final states, especially since the latter contain both spin-triplet and spin-singlet contributions, which could be disentangled by exploiting the self-analysing polarisation of the hyperon decays and/or, in the long term, by employing a polarised target.

\subsubsection{Constraining final state interactions via the Jost function method}

\begin{figure}[t]
  \centering
  \begin{minipage}[t]{0.5\linewidth}
    \centering
    
    \raisebox{0.8cm}{\includegraphics[width=\linewidth]{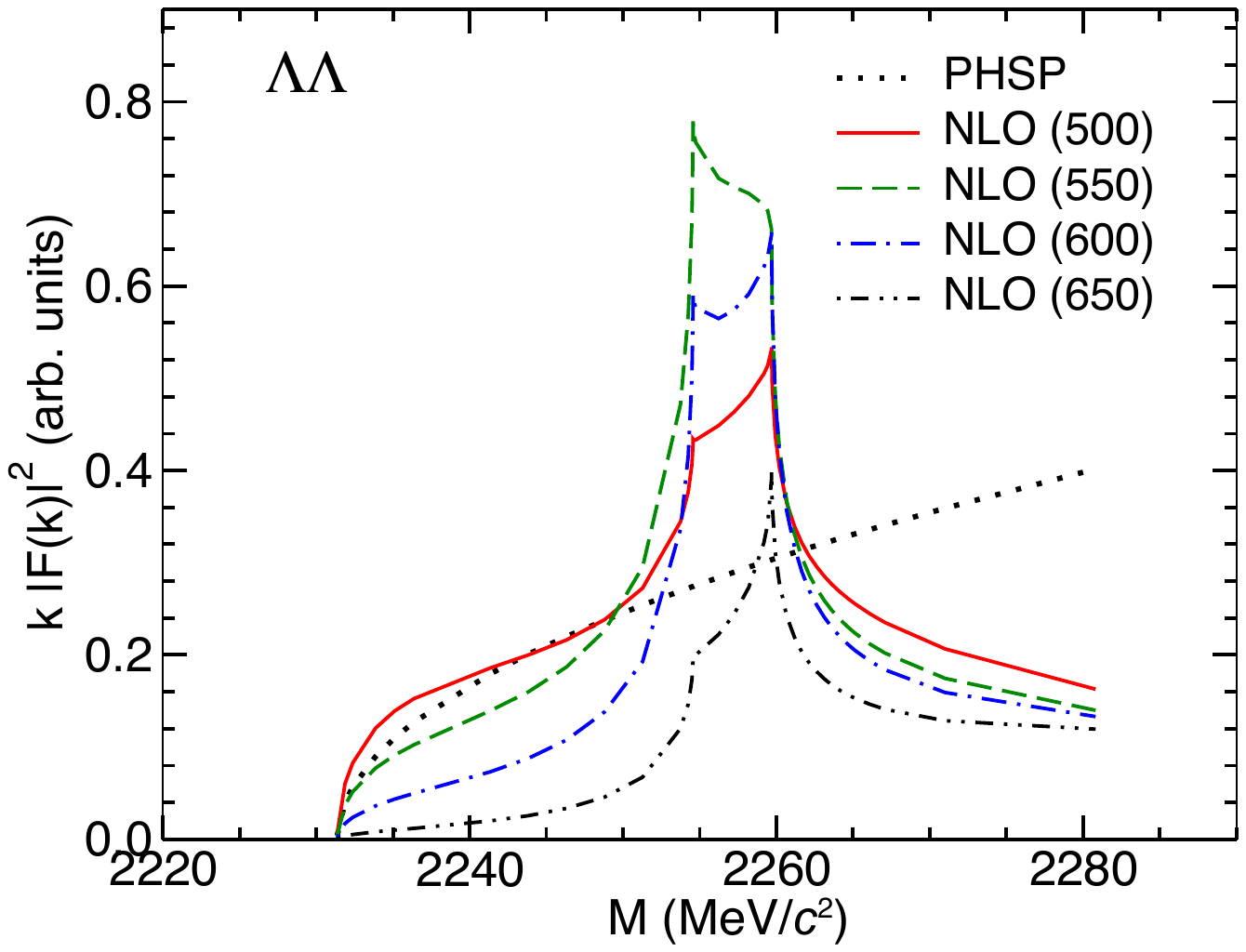}}
  \end{minipage}%
  \begin{minipage}[t]{0.53\linewidth}
    \centering
    \includegraphics[width=\linewidth]
    {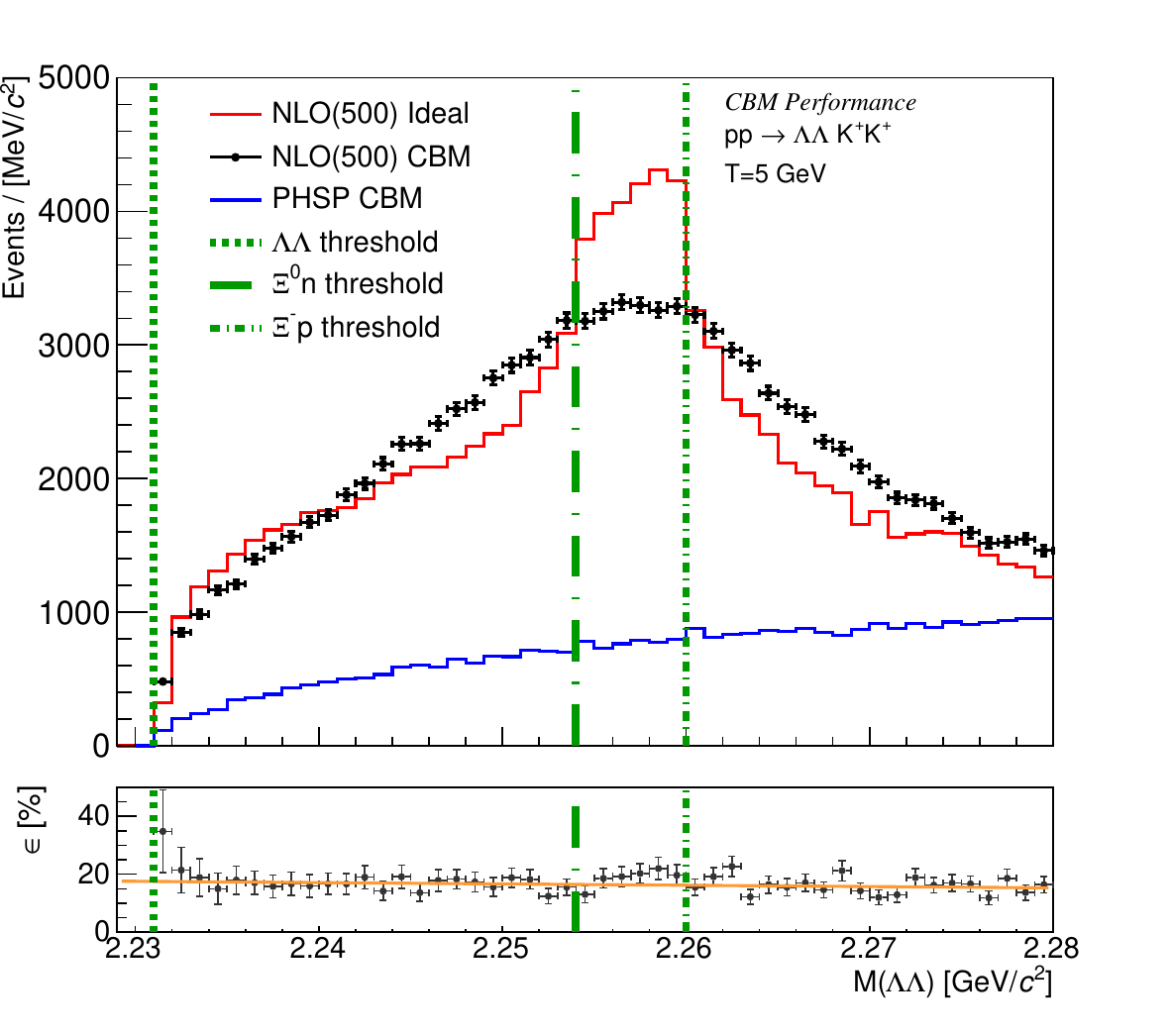}
  \end{minipage}
   {\setlength{\abovecaptionskip}{-0.8cm} 
  \caption{
Left: Excitation function of the $\Lambda\Lambda$ system in $pp \to K^+K^+\Lambda\Lambda$ from ChEFT up to NLO \cite{Haidenbauer:2018gvg}, illustrating theoretical uncertainties via cutoff values of 500 (red), 550 (green), 600 (blue), and 650\,MeV (black). The dotted line indicates pure phase space.
Right: Top panel shows the expected $\Lambda\Lambda$ mass distribution (black points) from CBM after 10 days of data taking at a 1\,MHz interaction rate with a 5\,GeV proton beam and a 5\,cm thick LH$_2$ target. Simulations are based on the NLO ChEFT (500\,MeV cutoff) amplitude, assuming an estimated cross section of 0.35\,$\mu$b for the $pp \to \Lambda\Lambda K^+K^+$ reaction. The blue (red) histogram represents the CBM response for a phase-space (PHSP) model (ideal detector resolution). The bottom panel shows the mass-dependent total efficiency (efficiencies $\times$ acceptances). The resolution is uniform over the presented mass range and found to be $\approx$6~MeV ($\sigma$). Take note of the following disclaimer~\cite{cbm_feasibility_note}.
\label{fig:lamlam}
  }}
\end{figure}

Here, it is worth recalling that ChEFT provides a systematic framework for analysing hadron–hadron as well as multi-hadron interactions; but with increasing order in the framework's expansion scheme, new low-energy constants appear, which must be determined from experiment. 
To constrain these parameters with sufficient accuracy across various partial waves, high-quality data at different energies are required. 
In this context, production reactions with short-ranged interaction mechanisms offer a suitable alternative to scattering data, particularly in situations where data at low relative momenta are unavailable, as is the case for some hyperon–nucleon systems and virtually all hyperon–hyperon interactions.

As argued in opening Sec.\,\ref{subsec.scatfromprod}, for the production of hidden strangeness or heavier final states in $pp$ collisions, the production operator can, to a good approximation, be regarded as pointlike. In such a scenario, the final-state interactions are not captured by the scattering amplitude itself but rather by the Jost function, which corresponds to the final-state wave function at the origin. 
Therefore, one may propose constraining the low-energy constants of hyperon–nucleon and hyperon–hyperon systems not only by fitting to scattering data from secondary interactions and hypernuclei, but also by including production data.

To illustrate that, in the strangeness $S=-2$ system, even data of moderate quality would significantly enhance understanding of hyperon–nucleon and hyperon–hyperon interactions, the left panel of Fig.\,\ref{fig:lamlam} displays predictions for the $\Lambda\Lambda$ excitation function obtained from ChEFT up to NLO \cite{Haidenbauer:2015zqb, Haidenbauer:2018gvg}. 
Evidently, a clear signal for cusp effects appears at the opening of the $\Xi N$ thresholds. 
At the same time, the results show a pronounced variation with the cutoff parameter used in the regularising function (varied here between 500 and 650\,MeV). 
This variation directly reflects today's limited knowledge of the interaction in the $\Lambda\Lambda$ and $\Xi N$ systems. 
In a well-constrained application of ChEFT, the regulator dependence would be largely absorbed into the renormalisation of the low-energy constants, thereby ensuring that low-energy observables remain practically unchanged. However, in the present case, where relevant data are lacking, the low-energy constants had to be inferred from a study of the $\Lambda N$ and $\Sigma N$ systems \cite{Haidenbauer:2013oca}, exploiting the approximate SU(3) flavour symmetry of the strong interaction.

%
%
It is worth emphasising that the variation in the results for the considered cutoffs is particularly pronounced because there exists a virtual state whose pole lies very close to the $\Xi N$ thresholds and whose exact position varies slightly depending on the potential used. 
The actual location of this pole is not yet constrained by any empirical information; however, as can be seen, it has a significant impact on the magnitude of the excitation function.
In this context, one may also mention that lattice QCD studies by the HAL~QCD Collaboration find a $\Xi N$ interaction of comparable strength \cite{HALQCD:2019wsz}, with scattering lengths of the order of one fermi or larger \cite{Kamiya:2021hdb}. 
Notably, these studies also reveal a pronounced cusp in the $\Lambda\Lambda$ system at the opening of the $\Xi N$ channel. 
One can expect that understanding of the hyperon–hyperon interaction, as well as the hyperon–nucleon interaction with strangeness $S=-2$, will improve significantly once data from CBM for this channel become available.

The right panel of Fig.\,\ref{fig:lamlam} shows the expected detector response of CBM for the fully reconstructed $pp \rightarrow K^+K^+ \Lambda\Lambda$ reaction, with $\Lambda \rightarrow \pi^-p$, comparing a ChEFT prediction with a phase-space model (arbitrarily normalised), while accounting for detector resolution, acceptance, and efficiency. 
A clear distinction is observable, revealing the cusp signature. 
Moreover, the resolution and efficiencies are relatively uniform across the mass range of interest. 
A kinematic fitting procedure was applied in the analysis, including total four-momentum and $\Lambda$-mass constraints, and taking into account the expected beam momentum resolution ($\Delta p/p \approx 10^{-3}$) and detector resolution. {\color{blah}A more detailed feasibility study is foreseen to establish the experimental sensitivity and to assess the potential physics impact of this measurement.}

An alternate method that allows direct access to scattering parameters from data, without relying on fits to Jost functions, is introduced in Sec.\,\ref{subsubsec.dispersive}. 
While this approach is formally appealing, it is limited by the requirement that the interactions must remain elastic over a sizeable energy range. 
Consequently, although the method is easier to apply, it is suitable for a smaller number of systems compared to the approach discussed in this subsection.

\subsubsection{Dispersive analysis of short ranged production reactions} 
\label{subsubsec.dispersive}

 
For large momentum transfer reactions, there is a method based on dispersion theory that allows one to extract the scattering lengths of the produced two-body system from the pertinent FSIs, at least as long as the
interactions among the final-state particles are elastic \cite{Gasparyan:2003cc}.  
Moreover, the scheme allows for an estimation of the theoretical uncertainty. 
The basic idea of the method is to exploit the scale separation between 
the short-ranged production process and a long-ranged FSI. 
In such cases, the production mechanism can be regarded as pointlike, and the FSI can be factored out. 
This prerequisite is fulfilled in reactions with large momentum transfer, $q_t$, \textit{e.g}., in the reaction $pp \to pK^+\Lambda$, the momentum transfer in the relevant kinematics is of the order of 900\,MeV$/c$ and thus significantly larger than the momentum range that needs to be probed in the final state, which is of the order of $100 - 200\,$MeV$/c$,
estimated from the inverse scattering length. 
Sufficiently large scattering lengths are expected not only in the baryon–baryon sector, but also for some meson–baryon systems, specifically for channels involving $D$ ($D^*$) mesons, where near-threshold molecules and/or pentaquark states have been reported; for discussions of such states, see, \textit{e.g}., Refs. \cite{Guo:2017jvc, Ali:2017jda, Olsen:2017bmm, Karliner:2017qhf, Liu:2019zoy, Brambilla:2019esw, Yang:2020atz, Chen:2022asf, Meng:2022ozq}.

With the above conditions fulfilled, and by imposing analyticity and unitarity constraints on the amplitude and assuming that there are no bound states, one arrives at the following expression
for the reaction amplitude \cite{Omnes:1958hv, Gasparyan:2003cc}:
\begin{eqnarray}
A_S(s,t,m^2)=\exp\left[{\frac{1}{\pi}\int_{m_{0}^2}^{m_{\text{max}}^2}\frac{\delta_S(m'^2)}{m'^2 - m^2 - i0}\, dm'^2}\right]\Phi(s,t,m^2),
\label{OM}
\end{eqnarray}
where $m$ is the invariant mass of the produced two-particle system with threshold value $m_0$,
$s$ is the total centre-of-mass (CM) energy squared, and $t$ represents all the remaining kinematic variables on which the amplitude depends.
The function $\Phi(s,t,m^2)$ varies slowly with $m^2$, which is a consequence of the assumed large momentum transfer. 
The cutoff $m_{\text{max}}$ must be chosen such that the integral extends over the entire region where FSI effects are expected to be significant.
Based on scale arguments, a condition for $m_{\text{max}}$ was derived in Ref.\,\cite{Gasparyan:2003cc}, which reads, when reformulated in terms of the maximum kinetic energy in the two-body system,
$\epsilon_{\text{max}} = m_{\text{max}} - m_0 \gtrsim 1/(2 a_S^2 \mu)$.
Here, $a_S$ is the scattering length in question and $\mu$ is the reduced mass of the two-particle system.
As exemplified in Ref.\,\cite{Gasparyan:2003cc} for the hyperon–nucleon interaction, the cutoff should be on the order of $\epsilon_{\text{max}} \approx 40\,$MeV.
Note that the two-body scattering process must be essentially elastic in this region, \textit{i.e}., there should be no other open channels coupled via the strong interaction, and it should be dominated by the $S$-wave amplitude, parametrised by the phase shift $\delta_S(m^2)$. 
The index ``$S$'' on the quantities above (and below) serves as a reminder that Eq.\,\eqref{OM} can only be applied to amplitudes corresponding to a specific spin state $S$.
Thus, it can be applied directly to channels such as $pp$, $\Lambda\Lambda$, or $\Sigma^+\Sigma^+$,
where there is only a single $S$-wave channel, or to meson–baryon systems involving pseudoscalar mesons ($\pi$, $K$, $D$, \ldots). 
In other cases, one must find ways to separate the possible states experimentally, \textit{e.g}., by measuring spin-dependent 
quantities \cite{COSY-TOF:2016qxd}.

As shown in~\cite{Gasparyan:2003cc}, one can employ the formalism of Ref.~\cite{Geshkenbein:1998gu} to invert Eq.~(\ref{OM}). This allows one to express the scattering
length in terms of the reaction amplitude squared (or the differential cross section
${d^2\sigma_S}/{(dm'^2dt)}$)
\begin{eqnarray}
a_S
=
\lim_{{m}^2\to m^2_{0}}\frac{1}{2\pi}\left(\frac{m_a+m_b}
{\sqrt{m_a m_b}}\right){\bf P}
\int_{m^2_{0}}^{m^2_{max}}dm'^2\sqrt{\frac{m^2_{max}-{m}^2}{m^2_{max}-m'^2}}
\frac{1}{\sqrt{m'^2-m^2_{0}}\,(m'^2-{m}^2)}
\log{\left\{\frac{1}{p'}\left(\frac{d^2\sigma_S}{dm'^2dt}\right)\right\}} \, ,
\label{a}
\end{eqnarray}
where $m_a$ and $m_b$ are the masses of the two particles, $m_0 = m_a + m_b$,
$p'$ is the CM momentum in the two-body system, and {\bf P} indicates that
the principal value of the integral must be taken.
An analogous equation can be derived for the effective range, although the 
uncertainty in its extraction will be larger.
A generalisation to the situation where the Coulomb interaction is present
is described in Ref.~\cite{Gasparyan:2005fk}.

Possible theoretical uncertainties of the method originate from the following sources:  
(i) energy dependence of the production operator (already discussed at the beginning of this section),  
(ii) influence of scattering at higher energies ($m > m_{\mathrm{max}}$),  
(iii) final-state interaction among other pairs of particles, and (iv) contributions from inelastic channels (e.g.\ in the case of $\Lambda p$, from 
the $\Sigma N \leftrightarrow \Lambda N$ transition).
For the $\Lambda p$ FSI, the theoretical uncertainty in the determination of the scattering
length, including the various sources mentioned above, was estimated to be $0.3$~fm at most~\cite{Gasparyan:2003cc}. This estimate was
confirmed by exploratory calculations of production amplitudes using hyperon--nucleon 
interactions with different strengths, i.e.\ with $\Lambda p$ triplet and singlet 
scattering lengths varying from $-0.7$ to $-2.5\,$fm.
Finally, it should be stressed that a fairly good experimental mass resolution is required for the application of the method. 
The data considered in the exemplary study \cite{Gasparyan:2003cc} had a resolution of $2\,$MeV$/c^2$, which is ideal. 
However, even from data with a resolution of the order of
$\approx 5\,$MeV$/c^2$, meaningful information can be extracted.
The method was used in Refs.\,\cite{COSY-TOF:2013uqx, COSY-TOF:2016qxd} to extract information on the $\Lambda p$ scattering lengths from the reaction $\vec pp\to pK^+\Lambda$. While the unpolarised data cannot be used to
extract a meaningful scattering length without additional assumptions, it still demonstrates that in high-statistics experiments
the method described allows for a determination of
the scattering length with a statistical and systematic accuracy 
of 0.05 and 0.2~fm, respectively. The lower statistics of the polarised measurement gave, for the spin-triplet 
scattering length, a value of 
$-2.6^{+0.7}_{-1.4 \ \rm stat.}\pm 0.6_{\rm syst.}\pm 0.3_{\rm theo.}$~fm~\cite{COSY-TOF:2016qxd}. This value
provides the first model-independent direct determination
of this quantity, which may be combined with the
somewhat smaller value inferred from femtoscopic data (discussed
in the next section) to give
$1.4\pm 0.2$~fm, with a theoretical uncertainty that is difficult to assess~\cite{Mihaylov:2023ahn}. {\color{blah} We note that polarised proton beams at SIS100 are technically conceivable, but would require a dedicated polarised source, spin-preserving acceleration, and additional spin-manipulation infrastructure; such operation is not part of the present SIS100 baseline. The possibilities for employing a polarised target in the long term are considered.}

\subsubsection{Summary for studies of production reactions with large momentum transfer}

It is here worth including a brief summary of the most promising systems to be studied at FAIR, which could significantly improve our understanding of hadron–hadron interactions beyond the current level.

The high initial energy at the SIS100 ($\sqrt{s}_{\mathrm{max}} = 7.5\,$GeV) provides access to FSIs in several interesting channels, \textit{e.g}.,
\begin{itemize}
\item
$\Sigma^+\Sigma^+$. Here, only the $^1S_0$ partial
wave contributes and the scattering length is expected to be fairly large 
\cite{Haidenbauer:2015zqb,Haidenbauer:2018gvg}. 
A possible reaction is $pp \to \Sigma^+\Sigma^+K^0 K^0$ ($\sqrt{s}_{\rm min} > 3.4\,$GeV).
The presence of the Coulomb interaction distorts the signal, especially 
at low momenta. However, 
methods to address this have already been developed \cite{Gasparyan:2005fk}. 
It remains to be seen whether the mass resolution can be pushed to values better than 5\,MeV$/c^2$.

\item
$\Lambda\Lambda$. Again, only the $^1S_0$ partial wave contributes. 
A possible reaction is $pp \to \Lambda\Lambda K^+ K^+$.
This system has already been investigated using the dispersive method \cite{Gasparyan:2011kg}, 
based on data from the $\Lambda\Lambda$ invariant mass
in the reaction $^{12}C(K^-,K^+\Lambda\Lambda X)$.
Current empirical information from 
measurements of $\Lambda\Lambda$ correlation functions, as well as
lattice simulations, points to scattering lengths of $a\approx -0.8\,$fm. 
Given the uncertainty inherent in the dispersive method, which is further exacerbated by the proximity of the first inelastic channels, $\Xi N$, 
this lies at the lower limit of where the method remains useful.
However, the direct fits using the Jost function approach offer a way to extract important information, not only on the $\Lambda\Lambda$ system, but also on the $\Xi N$ system.

\item
$\Lambda_c p$. In this case, the $S$-wave can be in both spin-0 and spin-1
states. Lattice QCD calculations by the HAL QCD collaboration 
\cite{Miyamoto:2017tjs}
(for large pion masses but extrapolated to the physical point within ChEFT \cite{Haidenbauer:2017dua}) suggest that the interaction could be nearly spin-independent, \textit{i.e}., $a_0\approx a_1$. 
While hadronic calculations predict scattering lengths around $a\approx -1\,$fm \cite{Haidenbauer:2017dua}, recent lattice results point to a repulsive interaction \cite{Zhang:2024}.
This makes further study of this system especially important.
A measurement of the invariant mass can be performed in the reaction
$pp\to \Lambda_c p \bar D^0$ ($\sqrt{s}_{\rm min} > 5\,$GeV). 
Coulomb effects are also present here.

\item $\bar D^0p$. The same reaction gives access to the $\bar D^0p$ interaction, which is also of considerable interest. The expected scattering length is sufficiently large to allow for application of the dispersive method. 

\item $\Sigma_c^{(*)} D^{(*)}$. Such systems can be studied via $pp\to p \Sigma_c^{(*)}D^{(*)}$ ($\sqrt{s}_{\rm min} > 5.6\,$GeV).
These two-body systems are particularly interesting owing to the 
formation of $P_c$ pentaquarks. 
Although strictly speaking, they do not meet the requirements for a dispersion-theoretic treatment because inelastic channels
are open, they can still be explored using the Jost function method \cite{Du:2019pij}, which has already been applied to LHCb data \cite{LHCb:2019kea}, where the pentaquarks were discovered. 
Moreover, various exotic mesons are known to couple to inelastic channels only rather weakly, such that an analysis of these channels assuming purely elastic interactions might be justified phenomenologically.

\item 
$pp\to pp J/\psi$ and the $p J/\psi$ interaction ($\sqrt{s}_{\rm min} > 5\,$GeV).
The interest in this channel stems from its relevance as a discovery channel for $P_c$ pentaquarks and for probing the role of $\Lambda_c D^{(*)}$ intermediate states; see Refs.\,\cite{Du:2020bqj, JointPhysicsAnalysisCenter:2023qgg} for a discussion in the context of photoinduced reactions.

\item $\pi^-p\to K^+\Lambda\pi^-$.
In this process, the $\Lambda\pi^-$ scattering phase shift will be studied at $\sqrt{s}$ equal to the $\Xi^-$ mass, which is a prerequisite for CP violation studies in hyperon decays.
\end{itemize}

\subsection{Femtoscopy}
\label{subsec.femto}

The key difference between the method described above and femtoscopy, as a means of extracting information on the interaction between hadrons, lies in the spatial and temporal extent of the region from which the particles are emitted, usually referred to as the source size. While in the former case the source is effectively point-like, femtoscopy is characterised by sources comparable to or much larger than the interaction range. 
Especially in the case of heavy-ion collisions, one can safely assume that the interacting particles originate from spatially separated points and essentially undergo secondary scattering, thereby providing direct access to scattering parameters. However, the large source sizes observed in heavy-ion collisions offer only very limited sensitivity to short-range interactions \cite{Fabbietti:2020bfg}. 
Consequently, much effort has recently been devoted to femtoscopic studies in high-energy proton–proton ($pp$) collisions, which significantly enhance the correlation signal \cite{ALICE:2020mfd}. Nevertheless, the smaller source sizes characteristic of $pp$ collisions introduce specific challenges, which become particularly pronounced at lower centre-of-mass energies, as will be discussed below.

\subsubsection{Opportunities in $pp$ collisions at FAIR}

FAIR will offer several opportunities to use femtoscopy to complement measurements from previous experiments, both in small and large collision systems. The potential for performing correlation studies in heavy-ion collisions is well established and usually relies on simultaneously extracting the source size and the scattering parameters from fits to the correlation function, using established models, such as the Lednick\'y–Lyuboshitz (LL) approach \cite{Lednicky:1981su, Lisa:2005dd}. 
Small collision systems, $pp$ in particular, have also proven to be a very useful tool for studying final-state interactions, as the associated sources have widths of $1-2\,$fm, \textit{i.e}., several times smaller than typical sources in heavy-ion collisions and comparable to the range of the strong force \cite{Fabbietti:2020bfg}. 
On one hand, this results in a very strong correlation signal, which compensates for the reduced number of measured pairs. On the other hand, there are several challenges associated with correlation studies in small collision systems, which are discussed in detail in Sec.\,\ref{subsubsec.femtosmall}. 

Once such issues are understood and controlled, $pp$ collisions at FAIR will offer a wide range of opportunities to study the strong interaction. In particular, owing to the low collision energies, the correlation signal will not be significantly contaminated by feed-down contributions from higher-mass states to the particles of interest (\textit{e.g}.\ $\Xi^{-} \rightarrow \Lambda + \pi^{-}$), which complicate femtoscopic studies at high energies, like those available at RHIC and LHC. These contributions can weaken and distort the measured correlation functions and require careful correction, thereby introducing additional sources of systematic uncertainty. 

Performing such studies at centre-of-mass energies closer to the production thresholds, as foreseen at FAIR, will suppress such feed-down contributions and thus facilitate the analysis of the correlation functions. It will therefore be possible to test and constrain interactions derived within, \textit{e.g}., ChEFT or deduced from lattice QCD calculations, with high precision. 
Moreover, measurements of particle pairs that involve coupled channels across small and large systems could offer opportunities to disentangle the effects of different initial states. This may enable the study of dynamically generated states, such as the $\Lambda(1405)$, through $p K^-$ femtoscopy \cite{ALICE:2019gcn}, and the $\Xi(1620)$ through $\Lambda K^-$ correlations \cite{ALICE:2023wjz}. 
The $K^-p$ and $\Lambda K^-$ correlation functions have been studied theoretically in Refs.~\cite{Kamiya:2019uiy,MantovaniSarti:2023lbl}.
The measured correlations can be related to interaction parameters extracted from dispersive analyses, as demonstrated for $\pi K$ interactions in Ref.\,\cite{Albaladejo:2025lhn}.


Experiments at FAIR will also provide access to more exotic particles, such as open-charm mesons. These particles are of particular interest owing to their early production in high-energy collisions, making them ideal probes of QCD properties, such as some QCD factorisation theorems \cite{Collins:1989gx}, studied at RHIC and the LHC \cite{ALICE:2020wfu}. However, the measured spectra of charmed hadrons are influenced by FSIs with the surrounding medium, making it essential to properly constrain the interaction between charmed and non-charmed hadrons \cite{He:2011yi}. Femtoscopy is one of the most suitable tools to achieve this, and the clean measurements that can be obtained at FAIR will contribute significantly to this effort.

Another interesting topic that will be accessible for study at FAIR is the production process of light nuclei, such as the microscopic properties of deuteron formation. 
So-called coalescence models describe the formation of a deuteron from an initial state consisting of two nucleons that subsequently bind into a deuteron \cite{Butler:1963pp,Kapusta:1980zz}. 
A key observable in coalescence studies is the ``$B_A$'' parameter, with $A = 2$ for the deuteron. This parameter can be obtained experimentally from the deuteron-to-nucleon yield ratio, typically measured as a function of the transverse momentum $p_\mathrm{T}$. 

Substantial effort has been made to exploit small collision systems for coalescence studies, demonstrating a direct link between the nucleon emission source, measured via femtoscopy, and the $p_\mathrm{T}$ spectra and $B_2$ parameter of the produced deuterons. 
Advanced coalescence models make use of the Wigner formalism, which describes the coalescence probability based on the deuteron wave function and the initial spatial separation of the nucleons \cite{Blum:2019suo,Mrowczynski:2019yrr}. 
Recent studies show that using a realistic deuteron wave function and a source determined from femtoscopic measurements, one can successfully reproduce the measured $B_2$ values and the deuteron spectra in $pp$ collisions at the LHC \cite{Mahlein:2023fmx}. 

If the source realised in $pp$ collisions at FAIR is well understood, it will enable an extension of the physics programme by testing these advanced coalescence models, thereby enhancing the understanding of light-nuclei production.
 
Finally, an interesting hypothesis for deuteron formation has been proposed by the ALICE Collaboration, based on $\pi d$ correlations \cite{ALICE:2025byl}. 
This study reveals a peak structure in the $\pi d$ correlation function measured in $pp$ collisions at $\sqrt{s} = 13\,$TeV, which is consistent with a residual signal from a $\Delta \rightarrow N \pi$ decay. 
One suggested explanation is the fusion of two nucleons into a deuteron via a reaction of the type $X + N \rightarrow (NN) + \pi \rightarrow d + \pi$, where $X$ denotes any resonance that can decay into $N + \pi$. 
Such a process naturally addresses a key limitation of most coalescence models, namely, the lack of energy conservation, by effectively involving a $3 \rightarrow 2$ particle transition.
Moreover, the results in~\cite{ALICE:2025byl} indicate that this production mechanism may be the dominant one, with at least $78.4 \pm 5.5\%$ of deuterons reportedly originating from such processes. 

The lower energies accessible at FAIR will provide a much cleaner environment in which to study this reaction and will allow for a decisive test of the role of resonances in deuteron formation. Furthermore, if deuterons are indeed formed with the assistance of resonances, residual peaks from higher-mass states, such as the $N(1520)$, could be reconstructed at FAIR, benefitting from the low combinatorial background and the constrained kinematics of the collision energy.

The application of femtoscopy will also provide valuable complementary information to some of the topics already outlined in this chapter:
\begin{itemize}
    \item $dd$, $dp$, and $d\pi$: these systems will offer essential insights relevant for dibaryon studies; see Sect.\,\ref{subsubsec.Dibaryons}.
    \item $\Lambda_c p$: as discussed in Sect.\,\ref{subsec.hypernuclei}, determining whether the $\Lambda_c p$ interaction is repulsive or attractive is crucial for assessing the existence of charmed hypernuclei.
    \item $\Lambda \Lambda$: this channel enables a direct comparison of the different approaches outlined in this paper (hypernuclei studies, dispersive analyses of production reactions, and femtoscopy), thus offering a unique opportunity to evaluate the strengths and limitations of each method.
\end{itemize}

\subsubsection{Formalism and assumptions}

Femtoscopy \cite{Lisa:2005dd}, as a technique for studying particle interactions, is based on measuring correlations in momentum space. The corresponding correlation function in the two-particle case is, therefore, defined as:
\begin{equation}
  CF(\vec{q}) = \frac{P(\vec{p}_1, \vec{p}_2)}{P(\vec{p}_1)P(\vec{p}_2)} \, ,
  \label{eq:cf}
\end{equation}
where $\vec{q} = \vec{p}_1 - \vec{p}_2$ is the relative momentum of the two particles 
involved {\color{blah}, defined in the pair rest frame}.

To derive a theoretical relation between the observable correlation function $CF(\vec{q})$, the underlying source function $S(\vec{r})$, and the relative wave function $\Psi(\vec{q},\vec{r})$, which encodes the two-particle FSI, several assumptions must be made (for a detailed discussion, see Refs.\,\cite{Pratt:1997pw, Lisa:2005dd}):\\[-5ex]
\begin{itemize}
    \item \textit{Higher-order (anti)symmetrisation can be neglected.} Ignoring possible multi-particle quantum statistical effects is considered a reasonable assumption when phase-space saturation is low, as expected in $pp$ collisions at FAIR.
    
    \item \textit{The emission process is initially uncorrelated.} This allows factorisation of the two-particle matrix elements, enabling the two-particle probability to be expressed in terms of one-particle source functions. Whether this assumption holds at FAIR $pp$ collision energies may require further scrutiny.
    
    \item \textit{Smoothness approximation.} Since source functions must, in principle, be evaluated off-shell, it is commonly assumed that $S(\vec{p},\vec{x})$ varies slowly with $\vec{p}$, allowing for on-shell treatment. This assumption is more stringent in the presence of two-particle FSIs \cite{Lisa:2005dd}. While it appears valid for large source sizes, such as those realised in heavy-ion collisions, it may not apply to small sources \cite{Pratt:1997pw}. Its validity should therefore be carefully evaluated before interpreting femtoscopic measurements in $pp$ collisions at FAIR.
\end{itemize}

Under the listed assumptions, the correlation function can theoretically be approximated by the Koonin--Pratt formula \cite{Koonin:1977fh,Pratt:1986cc}:
\begin{equation}
  CF(\vec{q}) = \int S(\vec{r})\,|\Psi(\vec{q},\vec{r})|^2 \, d^3r \, ,
  \label{eq:koonin}
\end{equation}
where $S(\vec{r})$ represents the source function, depending on the spatial coordinates $\vec{r}$.


Experimentally, the correlation function is usually constructed as
\begin{equation}
  CF(\vec{q}) = \frac{{\it Same}(\vec{q})}{{\it Mixed}(\vec{q})} \, ,
\end{equation}
where {\it Same}$(\vec{q})$ corresponds to the two-particle distribution measured in the same event, and \textit{Mixed}$(\vec{q})$ to that obtained from mixed-event pairs.

{\color{blah}
The traditional approach to utilizing the femtoscopy formalism is to study the spatio-temporal properties of particle emission using a system with a known wave function. A prominent example is identical pion femtoscopy and the detected Bose-Einstein correlations, which strongly depend on the properties of the emitting source. The outcome of this approach is a single unknown source function, often modeled by a Gaussian distribution with only one free parameter: the source size. However, Eq.~\ref{eq:koonin} can also be used to constrain the properties of final-state interactions (FSI). The Lednicky--Luboshitz (LL) formalism provides an analytical expression to compute the correlation function from the source size, scattering length, and effective range of the interaction~\cite{Lednicky:1981su}. The LL model uses the asymptotic form of the wave function and the effective range expansion, but is only applicable to large source sizes. More accurate numerical femtoscopy tools do exist~\cite{Mihaylov:2018rva}, typically utilizing effective potentials, but suffer from heavy computational costs. Nevertheless, the main systematic uncertainty in femtoscopic studies of the FSI is related to the source function, as the correlation signal is a convolution of source-induced and FSI effects. For this reason, the source function requires dedicated treatment, either by using the common hadron hypothesis in small collision systems~\cite{ALICE:2023sjd} or relying on transport models to provide reasonable constraints on the allowed profile and size of the source~\cite{HADES:2016dyd}. }

Femtoscopic correlations in hadron and lepton collisions have been widely investigated within the high-energy physics community. However, the theoretical interpretation of the results remains less established and underdeveloped. A meaningful quantitative comparison between femtoscopic observables from hadron–hadron and heavy-ion collisions is hindered by the use of inconsistent and often undocumented analysis techniques, varying detector acceptances, and diverse fitting functions historically adopted in the field.
Many of these systematic uncertainties can be significantly mitigated by employing a unified approach, using the same detector, reconstruction software, collision energies, and consistent analysis methods (\textit{e.g}., event mixing). 
Such an approach enables more robust comparisons and may provide valuable insight into the origin of space–momentum correlations in small systems.

The current physics programme provides a rare opportunity to perform a consistent and direct comparison of femtoscopic observables across $pp$, $pA$, and $AA$ collisions. 
Given the critical role of femtoscopic systematics in interpreting the results of $AA$ collisions, it is essential to explore whether similar scaling behaviours in $pp$ collisions reflect a shared underlying mechanism. 
Any similarities observed may point to a deeper connection between the space–time structure of the particle-emitting source in systems comparable to the confinement scale and in systems of considerably larger size.

\subsubsection{Femtoscopy in small reaction systems}
\label{subsubsec.femtosmall}

Using femtoscopy in $pp$ collisions at low centre-of-mass energies offers many unique physics opportunities, as outlined above. However, these analyses also present significant challenges that must be addressed to determine whether the method can be developed into a precision tool for hadron physics.

To extract the values of strong interaction parameters, the LL~formalism \cite{Lednicky:1981su, Lednicky:2005tb} is commonly applied. In the case of two spin-$1/2$ particles, this formalism involves {\color{blah} five} free parameters: $R_0$ denotes the femtoscopic source radius, $a^s_0$, $a^t_0$ and $r^s_0$, $r^t_0$ the scattering lengths and effective ranges for the singlet and triplet states, respectively. In many studies, a spin-averaged approach is adopted, reducing the number of parameters to {\color{blah} three}. 
{\color{blah} For spin-$1/2$ particles, it is also necessary to distinguish between identical and non-identical particles. Quantum-statistical effects arise only for identical particles, such as $pp$ and $\Lambda\Lambda$. In these cases, the scattering length and effective range are associated exclusively with the spin-singlet channel. Separate scattering parameters for the singlet and triplet channels arise only for non-identical particles, such as $p\Lambda$, which is discussed below.}


{\color{blah}
The LL formula derived from Eq.~\ref{eq:koonin} relies on three main assumptions: (i) the modification of the two-particle wave function is dominated by the $S$-wave interaction; (ii) the exact wave function can be approximated by its asymptotic form in the relevant region; and (iii) the source function is a Gaussian with a radius of $\sqrt{2} \, R_0$, where the factor $\sqrt{2}$ arises when transforming from the single-particle to the relative source. In such cases, the interaction can be described by the asymptotic form of the scattering wave function, governed by the phase shifts, which in turn are typically well parametrised using the effective range expansion at low momenta. However, he validity of the second assumption requires the interaction region, within which the exact wave function differs from its asymptotic form, to be significantly smaller than the source size. Consequently, the relevant length scale, to be compared with the source size, is the interaction range, rather than the effective range. In particular, the effective range does not provide a direct measure of the spatial extent of the interaction and may even take negative values, while remaining constrained by the interaction range. However, in $pp$ collisions, the source size is often smaller than the interaction range of the investigated system.  Although the smoothness (on-shell) approximation is therefore difficult to justify for small sources, the assumption of uncorrelated particle emission is still expected to remain valid for sufficiently high-multiplicity $pp$ collision events.
Generally,} the analysis must account for the full scattering wave function of the interacting particles, which requires specification of a potential to generate it. This has been implemented in the non-relativistic approximation within the CATS framework \cite{Mihaylov:2018rva}, which also considers the crucial effect of resonance decays on the effective source distributions. Clearly, this introduces a model dependence into the analysis, which must be carefully quantified. 

Potentials constructed with similar dynamical content, \textit{e.g}., incorporating one-pion exchange in the nucleon–nucleon interaction, typically lead to nearly identical correlation functions. Nevertheless, as highlighted in a recent study \cite{Epelbaum:2025aan}, the source description necessarily depends, at least to some extent, on the chosen basis for the relative wave function. Consequently, extracting distorted interaction parameters is possible if the same source is used across different potentials and/or final states. The arguments presented in Ref.~\cite{Epelbaum:2025aan} clearly underlines that any purported discrimination between on-shell-equivalent potentials \cite{Gmitro:1986ay} lacks physical meaning.

\begin{figure}[t]
  \centering
  \includegraphics[width=0.85\linewidth]{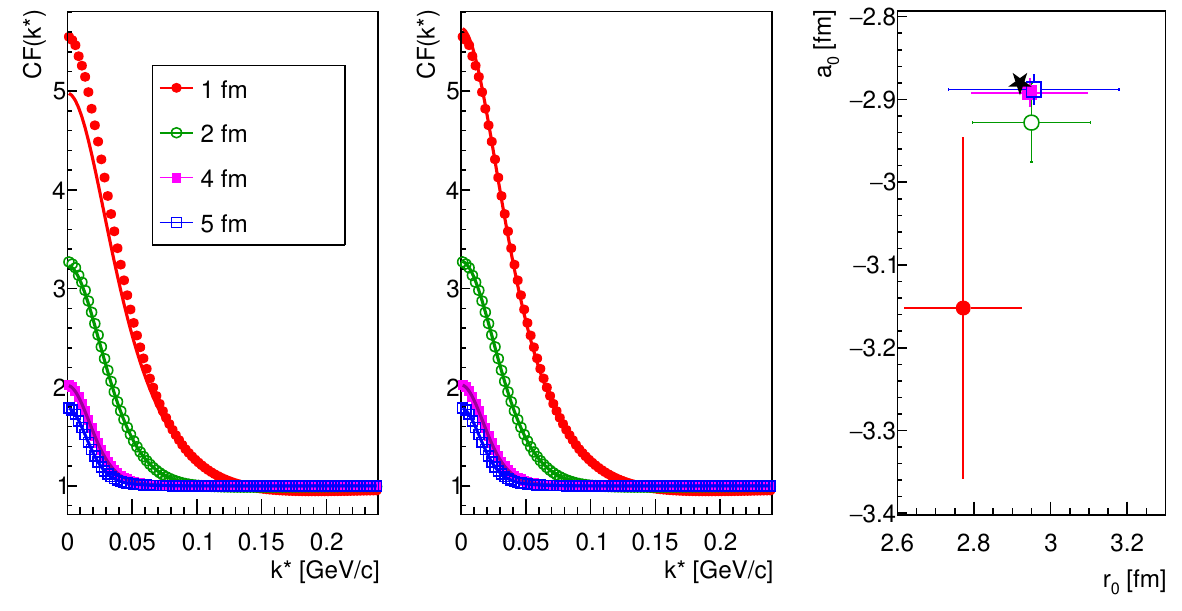}
    \caption{
Left panel: Correlation functions for $p\Lambda$ pairs calculated with CATS \cite{Mihaylov:2018rva} for four different source sizes using the Usmani potential \cite{usmani}, shown as symbols. 
The lines correspond to calculations within the LL formalism \cite{Lednicky:1981su,Lednicky:2005tb}, using the parameters $a_0$ and $r_0$ optimised for a source size of 4\.fm.
Middle panel: The same CATS calculations, but with $a_0$ and $r_0$ fitted to reproduce the calculated correlation functions.
Right panel: The values of $a_0$ and $r_0$ obtained from the fits shown in the middle panel. The black star indicates the interaction parameters extracted directly from the underlying potential.
}
  \label{fig:hbt2}
\end{figure}

As an example illustrating the difference between an LL-type analysis and one employing potentials in small systems, Fig.\,\ref{fig:hbt2} displays the $p\Lambda$ correlation function generated with CATS using the Usmani potential \cite{usmani}. 
The strong interaction parameters, $a_0$ and $r_0$, parametrised within the LL~approach, were obtained through an iterative fitting procedure. This involved generating sets of correlation functions for various source sizes and interaction parameters, followed by a $\chi^2$ minimisation to determine the best-fit values. 

The stability of the LL formalism was tested by recalculating the correlation functions using CATS for different source sizes to assess the robustness of the extracted scattering parameters. The results indicate that, for smaller source sizes, the extracted strong interaction parameters vary significantly. This suggests that, in such cases, femtoscopic correlations should be analysed using potential-based approaches rather than relying solely on the LL~formalism.

For small source sizes, even the basic Koonin-Pratt formula, Eq.\,(\ref{eq:koonin}), may no longer be universally valid, as some of the underlying assumptions discussed above -- such as an initially uncorrelated emission process or the smoothness approximation \cite{Pratt:1997pw,Lisa:2005dd} -- are potentially no longer strictly fulfilled. 
Nevertheless, recent studies performed by the ALICE collaboration at the LHC indicate that, for $pp$ collisions at several TeV, not only are the measured correlation functions well described by the Koonin-Pratt formula, but also the corresponding hadron emission source is practically independent of the particle species \cite{ALICE:2020ibs, ALICE:2023sjd}. This has been confirmed through modelling the source function using either the resonance source model (RSM) \cite{ALICE:2020ibs} or the CECA model \cite{Mihaylov:2023pyl}, both of which distinguish between the emission of primordial particles and production from strongly decaying resonances.

This observation suggests that one can constrain the emission source using systems with well-known final-state interactions and correlations, \textit{e.g}., $pp$ or $\pi\pi$, and then use this knowledge to analyse systems with less well-understood interaction properties, thereby reducing systematic uncertainties. 
However, such a ``common'' source remains, for now, a purely phenomenological observation. Its range of applicability and the corresponding uncertainties in the extracted scattering parameters are largely unexplored. 
There may exist as-yet undetected correlations in systems with small sources and low multiplicities, which could break the factorisation of source functions and relative wave functions beyond the considerations presented above. These effects will be particularly relevant for the physics programme described here.

Existing studies demonstrate a scaling of the source size with both the pair transverse mass, $m_T = \sqrt{p_T^2 + m_0^2}$, where $p_T$ is the transverse momentum and $m_0$ the particle rest mass, and the event multiplicity, with hints of an eventual saturation of the minimum source size. 
The $pp$ collision energies of several GeV at FAIR will result in lower average event multiplicities compared to the LHC, posing a significant challenge to the Koonin-Pratt formalism owing to extremely small emission sources.

To address this, benchmark correlation functions -- such as those from $pp$, $\pi\pi$, and $\pi K$ pairs, for which the hadronic interactions are well understood -- should be fitted to rigorously test the applicability of the Koonin-Pratt formalism, the validity of the common emission source hypothesis, and the saturation behaviour of $m_T$-multiplicity scaling. Furthermore, systematic and multi-differential measurements involving various combinations of particle species with differing masses (including nuclei and hyperons) should be conducted, accounting for different source profiles. 
This will help reduce uncertainties in the extracted scattering parameters. 
In particular, for sufficiently high-quality data, it is possible to simultaneously fit the parameters governing the FSI and those describing the emission source. This reduces model dependence and allows for systematic studies of the relationship between the initial and final states.

\begin{figure}[t]
  \centering
  \includegraphics[width=0.85\linewidth]{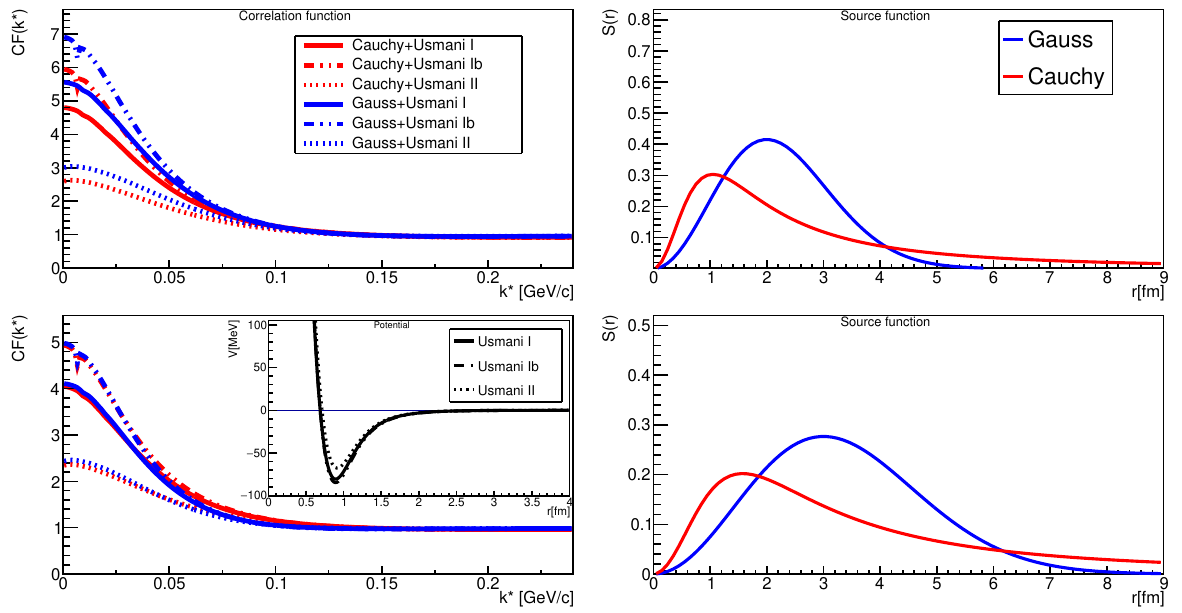}
  \caption{
    Correlation functions for source sizes of $1.0\,$fm (top panels) and $1.5\,$fm (bottom panels) for three different potentials describing the strong interaction of $p\Lambda$ pairs~\cite{usmani}. Two source parameterisations are compared: Gaussian and Cauchy.
  \label{fig:potentials}}
\end{figure}

Another aspect is the choice of source parameterisation. As already mentioned, a Gaussian shape is assumed in most cases, but for small systems, other parameterisations may be required. The sensitivity of the correlation function to the assumed shape of the source depends on the source size. 

To illustrate this, let us consider two commonly used source parameterisations: Gaussian and Cauchy distributions. The non-Gaussian shape of the latter can, for example, arise owing to feed-down from strong decays; see the discussion in Ref.\,\cite{ALICE:2023sjd}). Two representative source sizes, as typically observed in elementary collisions ($1.0\,$fm and $1.5\,$fm), are examined.
Figure~\ref{fig:potentials} presents examples of the resulting correlation functions calculated using three variants of the Usmani potential \cite{usmani}, relevant for $p\Lambda$ femtoscopy studies: Usmani~I (corresponding to the singlet state), Usmani~II (triplet state), and Usmani~Ib (a version of Usmani~I scaled by a factor of 1.05). The upper panels show results for a source size of $1.0\,$fm, while the lower panels correspond to $1.5\,$fm.
The comparison reveals that for a source size of $1.5\,$fm or larger, the correlation functions obtained using different source shapes remain consistent for a given interaction potential. In contrast, for smaller sources, the correlation functions exhibit noticeable differences depending on the chosen parameterisation. This suggests that particular care must be taken when modelling the shape of the source function in analyses involving small sources, as it can significantly affect the interpretation of the correlation function.

\begin{figure}[t]
  \centering
  \includegraphics[width=0.85\linewidth]{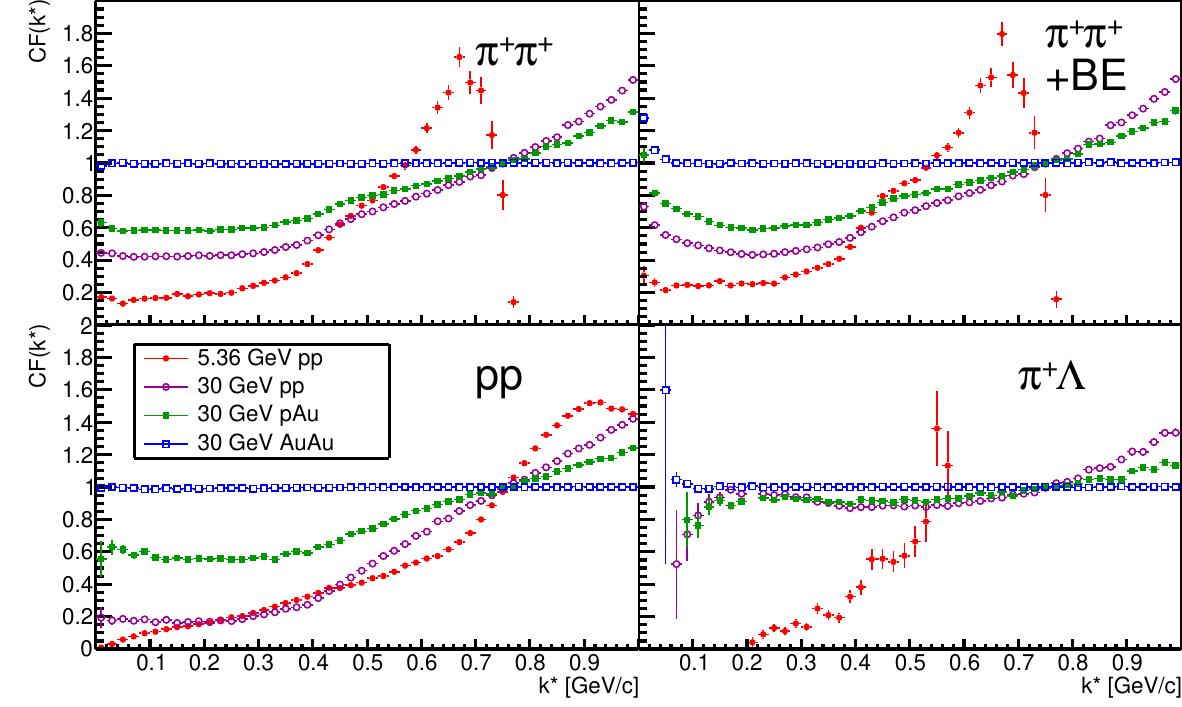}
  \caption{
    Correlation functions for $\pi^{+} \pi^{+}$, $pp$, and $\pi^{+} \Lambda$ pairs in $pp$ (at proton beam momenta of $5.36$ and $30\,$GeV$/c$), $p$Au, and Au+Au collisions (both at a beam momentum of 30\,GeV$/c$), simulated with the UrQMD model. 
    All correlation functions are calculated without any FSI between the pairs. 
    $\pi\pi$ correlations are shown both without (top left panel) and with Bose–Einstein (BE) correlations only (top right panel). For heavy-ion collisions (top right), the BE effects manifest as a deviation of the correlation function from unity for $k^{*} < 50\,$MeV$/c$. In lighter systems, this effect is less pronounced owing to the dominance of non-femtoscopic effects.
  \label{fig:hbt1}}
\end{figure}

Other challenges arise from contributions known as non-femtoscopic effects. These include, among others, the influence of global momentum conservation, resonance decays, and correlations between the space-time coordinates of emission points and particle momenta, which may be present in small systems. 
Generally, these effects depend on particle multiplicity and are thus more pronounced in smaller systems. 
In hadron–hadron collisions, fitting approaches that assume only femtoscopic effects, caused by FSI and/or quantum statistics, often fail to adequately describe the measured correlation functions, particularly at higher relative momenta $q$.

The behaviour of the correlation function at large $q$ is typically either neglected or accounted for using \textit{ad hoc} or semi-empirical terms in the fit functions. However, the parameters and functional forms introduced in this way are usually not directly linked to any physical mechanism. 

This approach presents two main issues. 
First, these functions are designed to fit the correlation structure at high $q$, where femtoscopic effects are negligible, and therefore their behaviour at low $q$ is unconstrained. As a result, they may introduce distortions in the extracted femtoscopic radii, especially for the small source sizes characteristic of $pp$ collisions. Even seemingly straightforward effects, such as momentum conservation, can produce non-trivial correlation structures at low $q$, complicating interpretation.
Second, \textit{ad hoc} fitting functions provide no means of assessing the physical plausibility of their parameters. If non-femtoscopic correlations are indeed dominated by conservation laws, as is often assumed, then it should be possible to derive physically motivated analytical expressions for their contributions.

Three-dimensional correlation functions carry more information about the geometry and dynamics of the emission source than their one-dimensional counterparts. However, they require significantly higher statistics, which has historically limited their application in most high-energy particle experiments.

Figure~\ref{fig:hbt1} presents examples of simulated correlation functions, calculated in the pair rest-frame system (where $k^*$ is the momentum of the first particle in the pair), for different collision systems ($pp$, $p$Au, and AuAu) and various particle combinations. The particle sources were generated using the UrQMD model \cite{Bass:1998ca, Bleicher:1999xi}.
The upper left panel shows the correlation functions for $\pi^{+} \pi^{+}$ pairs without the effects of FSIs and Bose–Einstein (BE) statistics. In this case, $CF(q)$ is expected to be unity, independent of $q$. 
While this is indeed observed for Au\,Au collisions, strong deviations from unity are evident for the smaller systems ($pp$ and $p$Au), signalling the presence of non-femtoscopic effects. The magnitude of these effects increases as the particle multiplicity of the system decreases.

To isolate genuine femtoscopic effects, such as those arising from BE quantum statistics, such non-femtoscopic contributions must be subtracted. This subtraction introduces dependence on modelling choices and parameterisations \cite{STAR:2011zch}. For comparison, the upper right panel of Fig.\,\ref{fig:hbt1} shows the same $\pi^{+} \pi^{+}$ correlation functions, this time including BE effects. These lead to an enhancement in $CF$ at very low $k^*$, which is clearly visible in the first data points for Au\,Au collisions but remains obscured in smaller systems owing to the dominating non-femtoscopic effects.
Similar observations can be made for the $pp$ correlations (lower left panel) and for $\pi^+ \Lambda$ pairs (lower right panel).

All the above aspects contribute to theoretical uncertainties, and a key issue that must be addressed in future studies is how to reliably estimate the magnitude of these uncertainties, which affect the strong interaction parameters extracted from measured correlation functions. However, the high event statistics accessible with CBM will enable multi-differential analyses of various correlation functions. This will help to quantify and eventually control systematic effects, thereby providing a deeper understanding of the relationship between correlation functions and hadron FSIs.
{\color{blah}Indeed, there are currently thriving efforts to quantify the possible limits of applicability of the formalism. For instance, a recent study~\cite{Smith:2026ocz} demonstrates the robustness of the smoothness and on-shell approximations, which play a central role in the derivation of Eq.~\ref{eq:koonin}. The necessity to perform analyses over large data sets requires improvements over the original LL model, and ongoing research~\cite{Murase:2025nlo} highlights the possibility of expanding the analysis beyond $S$-wave scattering. The analysis of existing femtoscopy data within theoretical studies, such as~\cite{Chizzali:2022pjd}, shows the possibility of extracting unambiguous information related to the scattering parameters from data. These developments motivate and justify a dedicated femtoscopy program at FAIR, combining advanced theoretical formalisms with precision measurements across uncharted kinematic regimes.}

\newpage
\section{Composition of hadrons}
\label{sec.Composition}

{\small {\bf Convenors:} \it C. S. Fischer, P. Salabura } 

\noindent Baryons are three-quark bound states, but in full QCD their structure is richer than a simple three-body picture. Their conserved quantum numbers are carried by three valence quarks, while the full baryon wavefunction includes gluons and sea quark–antiquark pairs that dress the valence core. Modern nonperturbative approaches—such as lattice QCD, Dyson–Schwinger/Faddeev frameworks (i.e. functional methods), and effective field theories—seek to connect this valence description with the effective quasi-particle degrees of freedom used at low energies. This connection is highly nontrivial and provides deeper insight into confinement and dynamical chiral symmetry breaking than conventional constituent-quark models alone.
In full detail, \textit{e.g}., it includes all the intricate scale-dependent internal structure components associated with form factors and (generalised) parton distributions. 

Understanding the spectrum, composition, and structure of baryons in terms of QCD is a formidable but crucial task that requires concerted efforts from both theory and experiment. 
Here, the SIS100 proton and the secondary pion beams at SIS18 offer many unique physics possibilities. 
The energy of the pion beams enables the study of non-strange baryons up to the fourth resonance region, low-energy hyperon-meson interactions, and excited hyperon states with masses up to $1.5-1.6\,$GeV$/c^2$. 
The available proton energy is sufficient to generate baryons with one or more strange- and/or charm-quark components, paving the way for systematic studies of the spectrum of baryons with strangeness and charm.
Furthermore, as sketched below, it offers possibilities to test important structural elements, such as vector- and axial form factors via measurements of leptonic and semi-leptonic decays in the strange baryon sector, the intriguing aspect of intrinsic charm quark content of the nucleon,
and also its structural core in terms of gravitational form factors and generalised parton distributions (GPDs). 
Furthermore, there are excellent opportunities to access the rich physics of (strange) baryon weak decays, and therewith provide a cross-relation
to neutrino physics. 
The aforementioned physics goals are reflected in experimental programmes at several of the world's premier facilities. 
They are discussed briefly, concentrating on those operating in an energy range similar to that of SIS100.



\subsection{Baryon spectroscopy: light sector, strangeness and charm}
\label{subsec.Spectroscopy}

One of the key elements in unraveling the structure of the strong interaction is understanding the baryon excitation spectrum of QCD. 
Significant experimental progress has been made in recent years by the analysis of data from photo- and electroproduction at ELSA, JLab, MAMI and {\color{blah} from pion induced reactions at FAIR/GSI (for recent review see \cite{Thiel:2022xtb})} resulting in a substantial number of additions to the review of particle properties \cite{ParticleDataGroup:2024cfk}.
Despite this progress, numerous long-standing issues are still not well understood. 

One of these questions concerns which effective degrees of freedom are appropriate for deriving the hadron reaction dynamics observed in the laboratory from QCD.
Are all baryon excitations efficiently and well described using a sophisticated three-body picture (or a diquark-quark approximation); or, instead, are some such states generated by coupled-channel effects in hadron-hadron interactions? 
The answer may depend on the state in question. 

There is a remarkable global one-to-one correspondence between the theoretical baryon spectrum calculated with functional methods \cite{Eichmann:2016yit} (three-body Faddeev equation; see below) and the experimental spectrum in the mass region below 2\,GeV$/c^2$. 
However, the characteristics of individual excited states, like the Roper resonance and the $\Lambda(1405)$, are known to be influenced to a lesser or greater extent by coupled-channel effects; see, \textit{e.g}., Refs.\,\cite{Suzuki:2009nj, Burkert:2019bhp, Mai:2020ltx}.

In the strange quark sector, in particular for baryons with two or three strange quarks, the available experimental data set is sparse  and does not support a systematic study of these issues. 
For the triple strange baryons there is not even a clean identification of any excitation in terms of well determined quantum numbers. 
This is clearly an important task for future experiments, since baryons with non-zero strangeness are especially interesting. 
They bridge the gap between the relativistic light quarks and the (more) non-relativistic heavy quarks and serve as probes for the flavour dependence of the underlying effective QCD interaction.
In this respect, it is also mandatory to extend the range of studies to the charm baryon spectrum, which pose their own challenges, both theoretically and experimentally, as discussed hereafter. 

\subsubsection{Theoretical approaches}\label{sec:baryonspectra}
In the following, three QCD-connected approaches to calculating the spectrum of baryons, including those with strangeness and charm, are discussed: lattice-regularised QCD (LQCD); functional methods using Dyson-Schwinger and/or Bethe-Salpeter equation analyses (DSE/BSE); and effective field theory (EFT).  

\medskip

\noindent{\bf LQCD:}
{\color{blah} For baryons that are stable under the strong interaction (i.e. the ground 
states in the $N$, $\Lambda$, $\Sigma$, $\Xi$, $\Omega$ and charm/bottom sector), 
masses have been obtained using the methods outlined in Sec.\,\ref{sec:toolslatqcd}. 
These calculations are performed at physical quark masses and extrapolated to the 
continuum limit and have errors below one percent \cite{Fodor:2012gf}. 
This situation, however, changes completely for resonances. 

LQCD computations of excited states in the light baryon sector have mostly
been performed at larger than physical up/down quark masses corresponding to pion masses (far) 
in excess of the measured value. Early studies treat all states as stable, which (taking into
account finite-volume effects) amounts to identifying finite volume energy levels with bound 
baryon states and resonances. In addition, only single-hadron operators were used, see e.g.
Refs.\,\cite{Edwards:2011jj, Edwards:2012fx, Engel:2013ig}. Qualitatively, these calculations 
show a quark-model like pattern of excitations. Further, a distinct pattern of hybrid baryons
emerges \cite{Dudek:2012ag, Edwards:2012fx}. It is important to bear in mind that these studies 
were typically also limited to a single lattice spacing, eliminating the possibility of
extrapolation to the continuum limit. Nevertheless, the techniques introduced in these studies 
have proved to be valuable in the refining of the LQCD approach.

Modern studies of resonances employing the L\"uscher finite-volume formalism  \cite{Luscher:1990ck,Luscher:1991cf} are beginning to emerge for low-lying baryon 
resonances, see e.g. \cite{Briceno:2017max,Mai:2022eur} for recent review articles.
In its original formulation the method serves to describe excited states as resonances 
in hadron-hadron scattering, and there is an ongoing effort to adapt it also to 
account for three-body scattering effects. For the lightest nucleon excitation, 
the $\Delta(1232)$, some early computations using other methods had already obtained 
estimates of the width and associated coupling \cite{Alexandrou:2013ata, Alexandrou:2015hxa}. 
More recent analyses with the L\"uscher method 
\cite{Andersen:2017una, Silvi:2021uya, Bulava:2022vpq, Alexandrou:2023elk} 
established its general usefulness for studying baryon resonances in a channel where
two-body effects are dominant (for the pion masses used in the study). Studying negative 
and positive parity $J=\frac{1}{2}$ nucleon excitations, however, turns out to be much 
more challenging due to the presence of three-body cuts. Exploratory results have been  
obtained a decade ago \cite{Lang:2012db, Lang:2016hnn} and await to be updated.}

Recently, the quantum numbers of the $\Lambda(1405)$ were studied in coupled-channel 
$\Sigma\pi$-$N\bar{K}$ scattering. Models based on unitarised chiral perturbation theory 
obtain two poles in the vicinity of the $\Lambda(1405)$. This picture is supported by 
LQCD computations at heavier-than-physical pion mass in 
Refs.\,\cite{BaryonScatteringBaSc:2023zvt, BaryonScatteringBaSc:2023ori}, 
where all successful parameterisations of the scattering amplitude contained two poles. 
Interestingly, this is also in at least qualitative agreement with a result obtained using
unitarised ChPT, see below and Sec.\,\ref{subsec.MesonBaryonInt}, when changing the hadron 
masses to those used in the lattice simulation \cite{Guo:2023wes}.

For charm baryons, many ground states have been calculated with reasonable control of 
systematic uncertainties associated with taking the continuum and physical quark current-mass 
limits 
\cite{Briceno:2012wt,PACS-CS:2013vie, Alexandrou:2014sha, Brown:2014ena, Perez-Rubio:2015zqb}. 
While the experimental spectrum of ground-states with given quantum numbers is a theory 
post-diction, these studies also predict many double- and even triple-charm hadrons not 
seen in experiment at the time. They can also serve by providing predictions for the 
quantum numbers of observed states, as exemplified for $\Omega_c^0$ baryons in 
Ref.\,\cite{Padmanath:2017lng}.

Future calculations for baryons with strangeness will need to map out the light and strange 
quark mass dependences, aim at quantifying the systematic uncertainties associated with 
discretisation and (residual) finite-volume effects, and extend the systems studied to 
other quantum numbers and higher energies. For charm baryons, there are opportunities to 
study resonances with the approaches now employed in the light-quark spectrum. The methods 
used in studies of the baryon spectrum will be useful for improving control of excited-state
contamination in baryon structure calculations, and for future explorations of the structure 
of exotic hadron candidates.

Furthermore, LQCD can provide inputs for experimentally inaccessible scenarios, 
\textit{e.g}., $\pi\Lambda$, $\pi\Xi$, etc. Using SU(3)-flavour and chiral symmetry 
this can then close gaps in the understanding of low-energy meson-baryon dynamics; 
see \textit{e.g}., Refs.\,\cite{Tiburzi:2008bk, Mai:2009ce, Torok:2009dg}, ultimately 
providing better constrained predictions for the hadron spectrum.

\medskip

\noindent{\bf DSE/BSE:}
The spectra of strange and charm baryons have also been explored with functional methods \cite{Roberts:2015lja, Eichmann:2016yit, Ding:2022ows, Binosi:2022djx, Ferreira:2023fva}. 
In this approach, one solves non-perturbative Faddeev equations, whose elements are determined self-consistently from their Dyson-Schwinger equations. 
One advantage of such methods is that they can map the entire baryon spectrum from light to heavy quarks \cite{Qin:2019hgk}. 
Current efforts focus on systematic improvements and applications to a wider range of quantum numbers. 
Furthermore, the baryon wave functions, which are obtained along with the spectrum, can be used for a variety of form factor and other structure calculations for light baryons, hyperons and heavy baryons.

Two strategies are typically employed: one is to solve the genuine three-body equation, where three dressed-quarks interact via the exchange of dressed-gluons \cite{Eichmann:2009qa};
and the other is to consider its reduction to a quark-diquark system \cite{Cahill:1988dx, Reinhardt:1989rw, Efimov:1990uz}, where the quarks and diquarks interact via a ``ping-pong'' or back-and-forth quark exchange. 
Originally, such diquark correlations were thought of as frozen two-body clusters without internal structure \cite{Anselmino:1992vg}. 
If this picture were true, then it would greatly restrict the number of reaction mechanisms and final states, something which a few contemporary analyses wish to exploit \cite{Barabanov:2020jvn, Aoki:2021cqa}. 
On the other hand, with dynamically active diquarks, much more diverse reaction patterns are accessible. 
The interesting observation is that the well-explored spectrum of light baryons is insensitive to whether the baryons are treated as three-quark or dynamical quark-diquark systems \cite{Eichmann:2016yit, Barabanov:2020jvn}. In the mass region below 2\,GeV$/c^2$, there is a one-to-one correspondence between the theoretical and experimental spectrum. 
Notably, the presence of both Dirac-scalar (``good'') and Dirac-axialvector (``bad'') diquarks is necessary to achieve this success.
\label{GoodBad}
The mass region above 2\,GeV$/c^2$ remains to be explored.

In the strange baryon sector, hyperons and their double and triple-strange counterparts are especially interesting because their experimentally known excitation spectrum is still very sparse.
With the unique possibility of tuning the energy of the proton beam to scan the relevant mass region from the production thresholds, FAIR provides excellent opportunities to explore their spectrum. 
Here, DSE/BSE predictions for the spectrum of baryons with strangeness exist \cite{Eichmann:2018adq, Chen:2019fzn, Yin:2019bxe} and those for excited states of $\Sigma$ and $\Lambda$ hyperons \cite{Eichmann:2018adq} have been confirmed by a Bonn-Gatchina analysis \cite{Sarantsev:2019xxm}. 
This demonstrates the capabilities of the approach, especially given that the 
resulting wave functions provide further information on the internal structure of those baryons. 

The existence of dynamical diquark clusters within baryons has many observable consequences \cite{Barabanov:2020jvn} that can be extracted, \textit{e.g.}, from systematic comparisons of calculations including full diquark dynamics with ones involving static approximations. 
This is especially clear in discussions of the spectrum and decay-modes of baryons containing one or two  $c$ quarks, and transitions between them. 
For example, in the case of heavy $\Sigma_c$ and $\Lambda_c$ baryons it is sometimes assumed that the light quarks cluster into a light diquark which then forms a two-body system with the remaining charm quark. 
DSE/BSE calculations with fully dynamical diquarks show that this is not the dominant configuration; instead, the clustering of light and charm quarks to a heavy-light diquark that interacts with a light spectator quark is favoured \cite{Yin:2019bxe, Yin:2021uom, Torcato:2023ijg}. 
In general, such outcomes speak against the treatment of singly-heavy baryons as effectively two-body light-diquark + heavy-quark systems and doubly heavy baryons as two-body heavy-heavy diquark + light-quark systems. 
Instead, it is a dynamical question which component of a baryon’s Faddeev amplitude is dominant, to be decided in explicit calculations on a case-by-case basis. 
In this respect, it is also interesting to compare spectra, form factors and decay channels between systems with one or two strange quarks and systems with one or two charm quarks. 
Exploiting its reach in energy and high luminosity, the SIS100 proton beam, coupled with the strengths of the CBM detector, would enable stringent tests of diquark-based explanations of baryon structure, particularly via high-statistics studies of strange (${\rm S}=-2$, ${\rm S}=-3$) baryon production and decays. 
Comparisons with charm baryons would also be possible.
They would, however, demand higher interaction rates. 

\medskip

\noindent{\bf Effective Field Theory:}
EFT and related coupled-channel dynamical and unitarisation approaches are valuable tools for studying the (excited) spectrum of baryons. 
Beyond many successful applications to mesons -- see Sec.\,\ref{subsec.MesonMesonInt} -- some excited baryons, like $N(1535)$, $N(1650)$, $\Lambda(1405)$ or $\Lambda(1670)$, emerge in these types of approaches through meson-baryon dynamics. 
Such schemes also possess considerable predictive power. 
This is showcased, \textit{e.g}., by the predicted emergence of the second broad pole with the same quantum numbers as the $\Lambda(1405)$, which is now referred to as the $\Lambda(1380)$; see Refs.\,\cite{Mai:2020ltx,Hyodo:2020czb}.


There are several particularly interesting ways in which the new SIS100 accelerator can provide novel constraints on these EFT approaches:\\[-6ex]
\begin{itemize}
    \item
    With an option to generate low-energy pion beams, an avenue opens to constrain $\pi N\to \{\pi N,K\Lambda,...\}$ scattering. 
    As discussed in Sec.\,\ref{sec4:subsec:Doring} this can improve partial-wave analyses that then feed into the UChPT programmes. 
    The key point in this regard is that SU(3)-flavour symmetry is manifestly preserved throughout the procedure depicted in the flowchart of Eq.\,\eqref{eq:ch5:maxim:UCHPT2}; note also the discussion in Ref.\,\cite{Lutz:2001yb}. 
    Thus, data in the strangeness $S=0$ sector also constrains $|S|=1,2,3$ results and vice versa. 
    As an example, see Fig.\,\ref{CH4:figMM-1} from a recent NNLO-kernel UChPT model \cite{Lu:2022hwm}.

\begin{figure}[t]
    \centering
    \includegraphics[width=1.\linewidth]{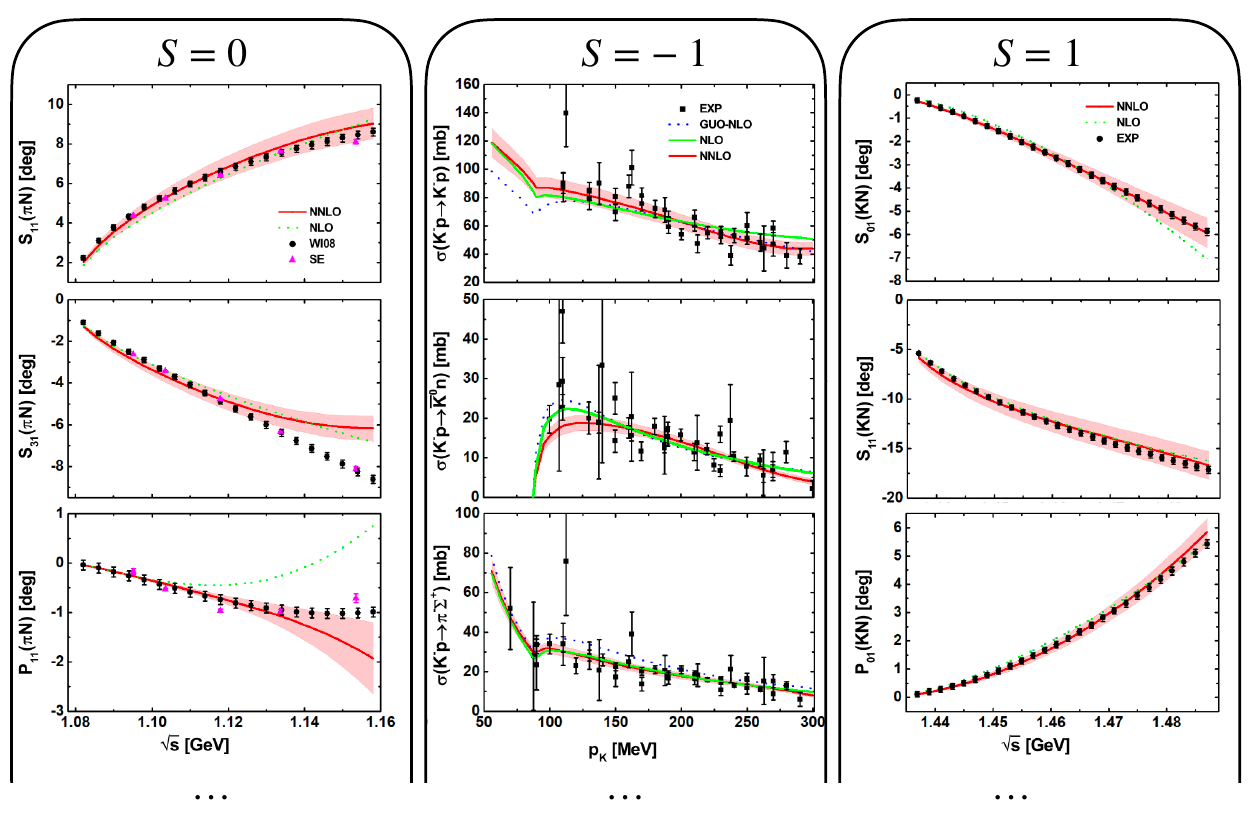}
    \caption{Compilation of fit results of the NNLO UCHPT approach from Ref.~\cite{Lu:2022hwm}. Parameters are adjusted to simultaneously fit cross sections for $\{K^-p\to X|X=K^-p,\bar{K}^0n, \pi^-\Sigma^+, \pi^0\Sigma^0,\pi^+\Sigma^-,\pi^0\Lambda,\eta\Lambda\}$ with $p_K$ denoting the $K^-$ laboratory momentum as well as phase-shifts of $\pi N$ and $K N$ elastic scattering compared to (black dots) SAID WI08~\cite{Arndt:2006bf} and SP92 solutions~\cite{Hyslop:1992cs}. Comparison is also made to the NLO-UCHPT results. For more details see Ref.~\cite{Lu:2022hwm}.
    \label{CH4:figMM-1}}
\end{figure}

    \item The possibility of producing baryonic matter with open strangeness or charm is another fascinating opportunity. 
    So far, very little is known about higher-strangeness baryons. 
    As mentioned above, UChPT models provide a connection between different strangeness sectors, which in principle allows one to accommodate $S=-2$ states; see, \textit{e.g}., Refs.\,\cite{Feijoo:2023wua, Ramos:2002xh}. However,   considerable uncertainty exists in this owing to limited data and model dependence of UChPT approaches. 
    An example of this was shown in Fig.\,\ref{CH4:figMM-2}.
    \item In the open charm sector, many states have been identified \cite{ParticleDataGroup:2024cfk}, but only a few have received a four-star status by the PDG; see, also, Ref.\,\cite{Guo:2017jvc}. 
    Still, on the theory side, unitary models exist; see, \textit{e.g}., Refs.\,\cite{Lutz:2003jw, Lin:2023iww}, akin with the strangeness sector discussed above.     
    Here, a controlled scan over the relevant energy ranges $\sqrt{s}\sim 3\,{\rm GeV}$ would be very valuable.
\end{itemize}

These things can have various implications for understanding the composition of baryons. 
For instance, by construction, UChPT approaches define the dominant degrees of freedom to be hadrons, through which interactions are expressed. 
Notably, however, certain states -- like $\Delta(1232)$ and $\Sigma(1385)$ with $J^P=3/2^+$ -- are not part of the UChPT-predicted {\it dynamically generated} spectrum. 
(In the DSE/BSE framework, they appear readily as solutions of the Faddeev equation.)
Instead, one might need to add such states to UChPT through a two-potential formalism as discussed, \textit{e.g}., in Ref.\,\cite{Sadasivan:2018jig}. 
Ultimately, a formalism accounting for chiral and SU(3)-flavour symmetries, but also allowing for the presence of explicit (compact) states may allow for the identification of a minimal spectrum of hadrons. 
Derived amplitudes could then also be put into a larger framework, including virtual photon probes, to study the compositeness of excited states, as discussed below.

Once all systematic uncertainties are pinned down through new data, there is a prospect for placing realistic constraints on $\Omega \to \bar{K} \Lambda$, $\Xi \pi$, related to CP-violation tests; see Sec.~\ref{Sec:hyp_nu}. 
The main idea is to estimate the initial production vertex of the meson–baryon pair from ChPT, with the FSI then determined using UChPT.

\begin{figure}
    \centering
    \includegraphics[height=5.4cm]{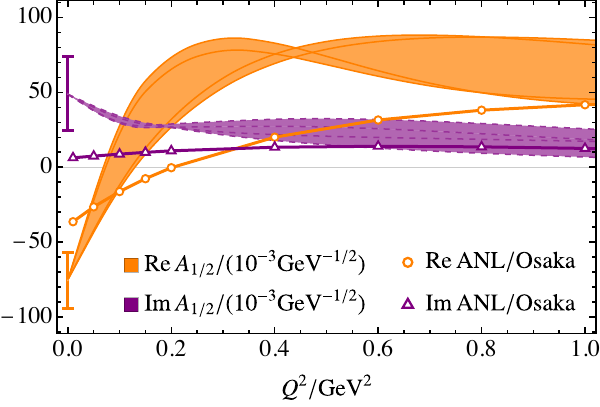}
    \includegraphics[height=5.4cm]{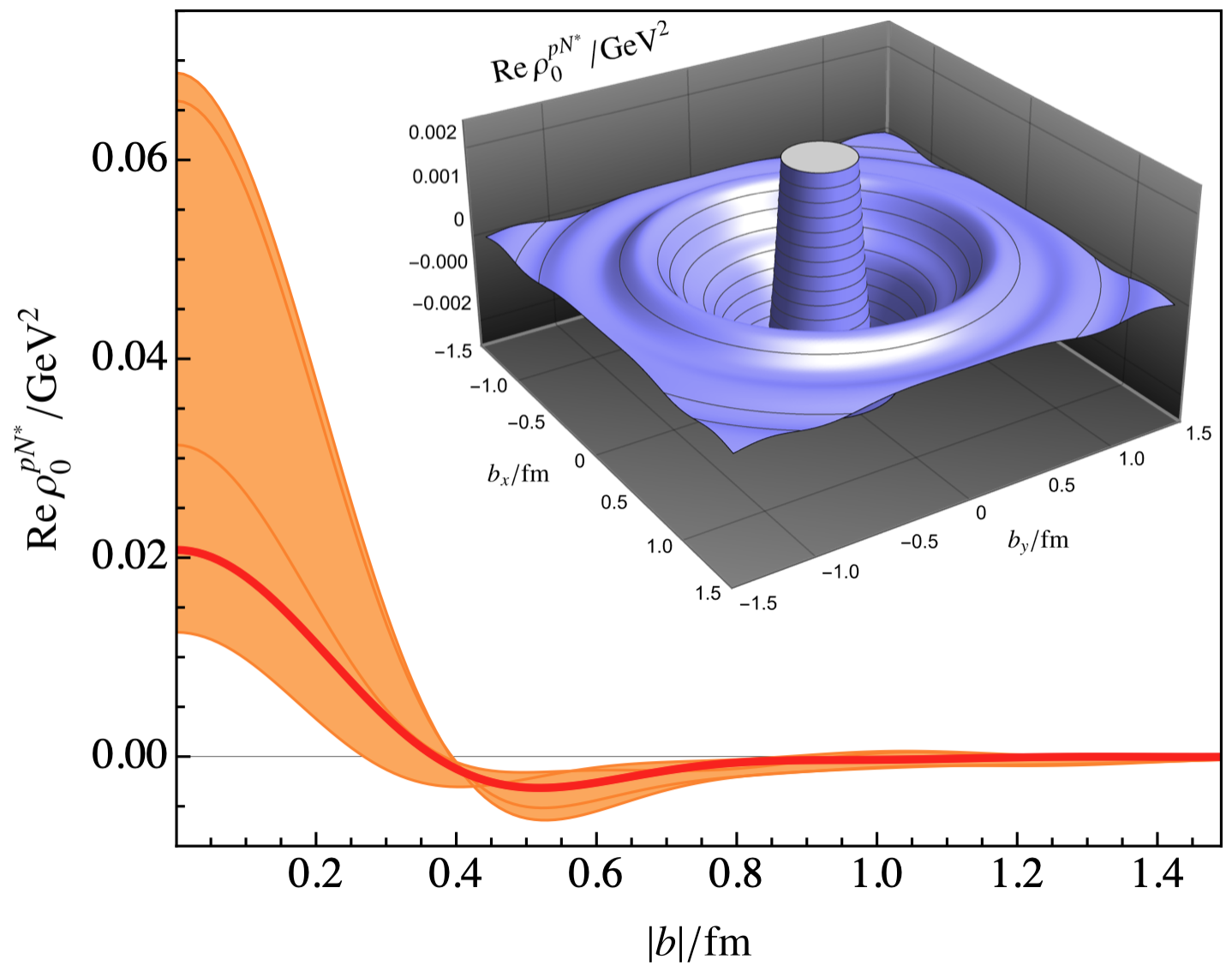}
    \vspace*{-0.2cm}
    \caption{Left: Transition form factor of $N(1440)$ from Ref.\,\cite{Wang:2024byt} (colored bands) in comparison with the ANL-Osaka solution \cite{Kamano:2018sfb}. 
    Right: Corresponding transverse charge density in the $p\to N(1440)$ transition.}
    \label{fig:enter-label1}
\end{figure}

Finally, one can also consider studying transition form factors of strangeness resonances.
Here, hadronic input is also relevant for defining the photon-induced transition of baryons to their excited states. 
Useful methods related to hadron resonance models are described in Refs.\,\cite{vanAntwerpen:1994vh, Phillips:1997tj, Ruic:2011wf, Mai:2012wy, Mai:2014xna, Haberzettl:2021wcz, Bruns:2022sio}. 
In some cases, no new parameters need to be introduced, so that a baryon ($B$) $\gamma B \to B^*$ excitation amplitude can be deduced. 
Such amplitudes are key inputs for calculating transition form factors of resonances, allowing, for instance, the study of the transverse charge density $B \to B^*$ in a specific reference frame \cite{Tiator:2009mt, Ramalho:2023hqd}. 
Examples of recent calculations of such a charge density for the Roper resonance are available in Refs.\,\cite{Roberts:2018hpf, Wang:2024byt}.  
Particular examples are shown in Fig.\,\ref{fig:enter-label1}.

\medskip

\subsubsection{Experiment}
\label{sec:baryonspectra_exp}
\noindent{\bf Collider experiments:}
In recent years, new opportunities for studying multi-strange hyperons have emerged from the high-energy collider experiments, \textit{e.g}., Belle, BESIII, and LHCb. 
At these facilities, light-flavor baryon resonances are copiously produced in the decays of heavy baryons containing a charm or bottom quark, or in the decay of heavy-flavor mesons, and the available high-statistics data samples provide an interesting alternative for studying these resonances. 

For instance, excited $\Xi$~baryons are produced and have been studied in the
decay of the charmed $\Lambda_c^+$ into $(\Sigma^+ K^-)_{\Xi(1690)}\,K^+$ by the Belle Collaboration~\cite{Belle:2001hyr} and into $(\Xi^- \pi^+)_{\Xi^\ast}\,K^+$ by the BaBar Collaboration~\cite{BaBar:2008myc}, and also by Belle in the decay $\Xi_c^+\to (\Xi^-\pi^+)_{\Xi^\ast}\,\pi^+$~\cite{Belle:2018lws} with high statistical
quality. 
Clear signatures are observed for the decuplet ground state, $\Xi(1530)$, and for the lowest-mass excitations, $\Xi(1620)$ and $\Xi(1690)$. 

The BESIII Collaboration studied $\Xi^\ast$~production in reactions such as $e^+e^- \to \psi(3686) \to (K^-\Lambda)_{\Xi^\ast} \bar{\Xi}^+$ and has observed structures associated with the $\Xi(1690)$ and $\Xi(1820)$ resonances \cite{BESIII:2023mlv}. 
A partial wave analysis (PWA) was even performed to study the properties of these intermediate state $\Xi^\ast$~hyperons. 
However, such studies from $c\bar{c}$~decays are still statistically limited and nothing of significance has been observed for masses beyond 1.9\,GeV$/c$. 

Finally, the LHCb Collaboration has recently reported the observation of $\Xi^\ast$~resonances in high-energy $pp$~collisions. 
The results are based on an analysis of $\Xi_b\to  J/\psi \,\Lambda K^-$ and the first observation of $\Xi(1690)$ and $\Xi(1820)$ in the decay of $\Xi^-_b$ was announced \cite{LHCb:2020jpq}. 
Masses and widths have been determined with improved precision in an amplitude analysis of the $K^- \Lambda$ mass distribution. 

The $\Omega(2012)^-\to \Xi^0\,K^-~(\Xi^-\,K_S^0)$ was discovered by Belle in 2018 in the decays of the heavy $b\bar{b}$~mesons $\Upsilon(1S), \Upsilon(2S)$, and $\Upsilon(3S)$ \cite{Belle:2018mqs}.
The ground-state~$\Omega^-$ is copiously produced in the decay $\Xi_c^0\to \Omega^-\,K^-$ and with lower statistics in $\Omega_c^0\to\Omega^-\,\pi^+$. Such data samples were used by the BaBar Collaboration for a spin measurement of the $\Omega^-$~hyperon \cite{BaBar:2006omx}, for instance.

\smallskip


\noindent{\bf ELSA, JLab, J-PARC:}
Recent years have seen a renaissance in hyperon spectroscopy. 
Several new facilities -- INSIGHT at ELSA, KLF at JLab, and J-PARC, see Sec.\,\ref{sec.OverviewFacilities} -- are being constructed to explore excited states of strange baryons. 
These projects will complement rather that compete with a hyperon programme at FAIR.

The future INSIGHT experiment, performing photoproduction experiments at ELSA \cite{Hillert:2017nzr}, will have an important focus on $\Lambda^*$- and $\Sigma^*$-hyperons, studying their spectrum and properties in detail and searching for multiquark-states in the strangeness sector. 
The experiment features a combination of almost complete angular coverage for high-resolution photon measurements, charged-particle detection, and the ability to perform measurements using a transversally or longitudinally polarised target. 
Therefore, it will offer the possibility to perform measurements not only with a polarised beam but also with a polarised target. 
In addition, in the case of strange baryons, the self-analysing weak decay of the $\Lambda$- and $\Sigma$ baryons provides access to the recoil polarisation. 
This will enable the measurement of different polarisation observables that are indispensable for an unambiguous and precise determination of the spectrum of excited states. 

The $K_L$ facility (KLF) will utilise a broad-momentum ($p_{K_L} \in [0,10]$ GeV/$c$) beam of neutral kaons to elucidate the hyperon spectrum. 
This neutral kaon beam will enable measurement of the line shapes of excited $\Lambda$ and $\Sigma$ baryons with unprecedented accuracy, as both these states are produced directly via the $s$-channel. 
The production of $\Sigma^+$ particles on a hydrogen target is particularly appealing from an experimental point of view, and the KLF can be regarded as a $\Sigma$ factory.

This high-purity kaon beam, generated via a two-step production process using electromagnetic probes, comes at the cost of beam intensity: only around $10^4$ kaons per second will be available at KLF across the full momentum range. 
While this intensity is sufficient for studying $\Lambda$ and $\Sigma$ production channels, it becomes a limiting factor for investigating excited cascade states ($K_L p \to K^+ \Xi^*$). 
Although such events can still be tagged, the available statistics would restrict studies to decay channels of only the most prominent states, rendering a comprehensive $\Xi^*$ scan virtually impossible. 
The situation is even more constrained for $\Omega^*$ states owing to the sharp decline in high-energy kaon beam flux, \textit{e.g}., only a few hundred events for the first excited $\Omega^*$ state are expected over a two-year measurement period.

This situation might improve if Jefferson Lab undergoes an energy upgrade to a 22\,GeV electron beam, currently under discussion \cite{Accardi:2023chb}, although this is not expected before the late 2030s. 
The hyperon spectroscopy programme at FAIR, which will focus on rare hyperon decays and the production of hyperons with high strangeness and weak coupling to a single kaon, will be a critical extension of the studies planned at KLF.


GlueX, despite being a photon-beam experiment and hence subject to low production cross-sections, has demonstrated its capability to measure a wide range of $S=-1$ and $S=-2$ states \cite{Pauli:2022ehd}. 
To investigate the production mechanisms of hyperons, GlueX has measured the beam asymmetry 
for the reaction $\gamma p \rightarrow K^+\Sigma^0$ \cite{GlueX:2020qat}, as well as spin-density matrix elements for $\gamma p \rightarrow K^+\Lambda(1520)$~\cite{GlueX:2021pcl}. 
GlueX is also examining the double-pole structure of the $\Lambda(1405)$ \cite{Wickramaarachchi:2022mhi}, which is supported by preliminary analyses.

The production of $S=-2$ states must proceed via the decay of a heavy $S=-1$ resonance, such as in the reaction $\gamma p \rightarrow K^+Y^* \rightarrow K^+K^+\Xi^-$. 
This makes the production process more complex and renders a full partial-wave analysis -- necessary to determine the nature of a state -- much more challenging. 
GlueX is currently searching for a range of excited $\Xi$ states by identifying peaks in the invariant mass spectra of their decay products. 
Both charged and neutral ground states, as well as three excited states, have so far been identified \cite{Pauli:2022ehd}. 
The measurement of their production cross-sections is ongoing.

Owing to their different production mechanisms, the INSIGHT and GlueX hyperon efforts are complementary to the programme planned at FAIR. 
Most notably, FAIR will have the capability to measure not only $S=-1$ and $S=-2$, but also $S=-3$ resonances. 
These highly strange resonances are not accessible at INSIGHT and are most likely also inaccessible to GlueX owing to their low cross-sections in photoproduction.

\smallskip

Introducing heavy flavours into a baryon adds an important degree of freedom, offering insights into quark correlations. 
According to some quark models \cite{PhysRevD.92.114029}, the collective motion of the light-quark pair relative to the heavy quark (the $\lambda$-mode), and the relative motion between the light quarks themselves (the $\rho$-mode), are separated in excitation energy. 
This separation resembles the so-called isotope effect. 
Since the colour-magnetic interaction between quarks is inversely proportional to the quark mass, the heavy quark acts as an inertial anchor. 
If true, this would allow the extraction of information about light-quark (diquark) correlations. 

Of particular interest is the so-called ``good diquark'' with spin-parity $0^+$, as it is expected to contribute to diquark condensation in highly dense matter -- several times the normal nuclear matter density -- where a colour-flavour locked superconducting phase may emerge. 
Although diquark correlations in hadrons in vacuum may differ significantly from those in dense matter, studying them in vacuum is a crucial first step towards understanding quark dynamics in extreme environments.

A physics programme at J-PARC has been proposed to investigate singly-charmed and multi-strange baryons, $\Lambda_c^\ast/\Sigma_c^\ast$ (represented as $Y_c^\ast$), and $\Xi^\ast$ and $\Omega^\ast$, respectively, by means of missing mass techniques in the reactions $\pi^-p \rightarrow D^{\ast-}Y_c^\ast$, $K^-p \rightarrow K^\ast\Xi^\ast$, and $K^-p \rightarrow K^\ast K^+\Omega^\ast$~\cite{J-PARC_E50,J-PARC_P85,J-PARC_E97}. 
Measurements of the mass spectra, production rates, and decay branching ratios of $Y_c^\ast$, $\Xi^\ast$, and $\Omega^\ast$ are planned. 
These observables carry information on the internal structure of the baryons. In order to conduct these experiments, the $\pi20$ and K10 beam lines will be used; see Sec.\,\ref{sec.OverviewFacilities}.

\smallskip

\noindent{\bf FAIR:}
{\color {blue} 
Spectroscopy of non-strange baryons and $S=-1$ hyperons in pion-induced reactions is conducted with the HADES spectrometer at SIS18. As discussed in Sec.~\ref{sec:pionbeams}, pion-induced reactions are particularly well suited for baryon spectroscopy, and in combination with high-precision dilepton detection, HADES offers unique opportunities. The energy range spanned by the GSI pion beam facilitates non-strange baryon spectroscopy up to the fourth resonance region and studies of hyperon excited states with masses up to 1.6--1.7~$\text{GeV}/c^2$. The first round of experiments was focused on studies of the production of baryon resonances in the second resonance region via decays into two-pion final states~\cite{HADES:2020kce}. A partial wave analysis performed with the Bonn-Gatchina framework on combined pion- and photon-induced reactions provided a new determination of decay branches into $\Delta\pi$, $\sigma N$, and $\rho N$. These results also appeared essential for the extraction of electromagnetic transition form factors from measured Dalitz decays (see next section). In the coming campaigns,  measurements will be extended to higher energies and will include also studies of hyperon production and polarization.} 

Extension of baryon spectroscopy to excited to $S=-2,-3$ sector and charm baryons will be achieved by means of proton--proton reactions at SIS100 and CBM detector. This programa will benefit from the detector’s excellent mass resolution, vertexing capabilities, large acceptance, and high count-rate performance and higher enrgies of proton beams.  At the maximum beam momentum of 30\,GeV$/c$, the energy available above the production threshold amounts to approximately 3.5\,GeV for $\Omega$ ($S=-3$) and 2.5\,GeV for $\Lambda_c$. 
This enables access to the excited spectrum of charm and multi-strange baryons. 
From experiment commencement, owing to the comparatively large production cross sections for hyperons, CBM at SIS100 will be particularly well suited for high-statistics spectroscopy of excited hyperons in the $S=-2$ and $S=-3$ sectors.

An energy scan with reconstruction of exclusive final states, such as $pp \rightarrow pK^+K^+\Xi^{(*,-)}$ and $pp \rightarrow pK^+K^+K^0\Omega^{*,-}$, will allow for a systematic study of charged excited states as a function of excitation energy. 
Various decay modes and their respective line shapes can be measured in exclusive final states, \textit{e.g}., $\Xi^{(*,-)} \rightarrow \Xi^- \pi^0$ or $\Xi^{(*,-)} \rightarrow (\Lambda/\Sigma) K^-$, with excellent mass resolution of around $\sigma_{M_{\Xi}} = 6$~MeV/$c^2$ and high reconstruction efficiencies, \textit{e.g}., $20\%$ for $\Lambda K^-$ at 30~GeV/$c$.

Line shape measurements in such decays are particularly interesting for excited states with masses close to meson-baryon thresholds, where coupled-channel effects are expected to play a significant role \cite{Feijoo:2023wua}. These studies can be complemented by measurements of electromagnetic Dalitz decays into ground states and dilepton pairs (see next section), offering further insights into the internal structure of the states, \textit{e.g}., helping to distinguish between baryon-meson molecular configurations and more compact quark-based systems.

Measurements of open-charm channels will be more challenging owing to much lower cross sections and the demanding requirements for background suppression. 
However, this can be achieved in CBM with the aid of the Micro Vertex Detector, which offers a resolution on the order of tens of $\mu$m, enabling discrimination of the secondary decay vertex of charm hadrons.
Three-body reactions of the type $pp \rightarrow p\, \bar{D}^0\, \Lambda_c^{*,+}$ would allow for studies of Dalitz distributions of excited baryons and investigations of the $\Lambda_c$-$p$ FSI, providing important constraints on the existence of hyper-charm nuclei. 
Similar to the programme at J-PARC, decay modes of excited charm baryons ($Y^*$) into charm meson-nucleon ($Dp$) and charm hyperon-pion ($Y\pi$) final states can be measured.  This will assist in shedding light on quark-diquark correlations in light-heavy quark systems and studying FSI effects.


 \subsection{Form factors of baryons}
\label{sec:el-transFF}

(Transition) form factors offer a wealth of information on the structure of hadrons. 
For a pointlike particle, the form factor is a constant; but all hadrons are composite and therefore exhibit non-trivial form factors, $F(q^2)$, when probed at momentum transfer $q$. 
The radius of the composite object is determined from the derivative of the form factor via 
\begin{eqnarray}
	\label{eq:def-radius-gen}
	\langle r^2 \rangle := \left. \frac{6}{F(0)} \frac{dF(q^2)}{dq^2}  \right\vert_{q^2=0} \,, 
    \label{radius}
\end{eqnarray}
and its size depends on the probe, \textit{i.e}., on the type of form factor: electromagnetic or gravitational, etc. 
(If $F(0) =0$, then the $1/F(0)$ factor is omitted from Eq.\,\eqref{radius}.)
Hence, different probes yield complementary structural information about the subject hadron. 
It is also of interest to study how the structure of a given state changes if one of its valence constituents is replaced, for instance, by changing the flavour content of a given type of hadron. 

On the theoretical side, several frameworks (\textit{e.g}., LQCD, DSE/BSE methods, chiral perturbation theory, and heavy-quark effective theory) allow the investigation of how a continuous variation of quark masses affects hadron structure. 
This discussion will highlight the impacts of flavour dependence and/or approximate flavour symmetry and show how strangeness and charm can serve as diagnostic tools for probing hadron structure. 
 
The primary tools for studying nucleon structure are electromagnetic probes, such as those employed in elastic $e^-p$ scattering. 
In addition, exclusive inelastic processes, like $e^-p \to e^-\Delta^+$, allow for the study of hadron excitation via transition form factors (TFFs). 
All systems with nonzero spin are characterised by $N>1$ form factors (and associated radii). The number of form factors, $N$, changes depending of the spins of the initial and final states.  

Elastic and transition processes on $q^2 \simeq 0$ are accessible using LQCD and EFT methods; so they are currently receiving much attention from these communities, \textit{e.g}., in connection with a precise determination of the proton charge radius \cite{Gao:2021sml, Cui:2022fyr}.
At larger momentum transfer, $|q^2|>m_p^2$, where $m_p$ is the proton mass, other tools are required.  Data on this domain can be used to validate solid QCD predictions for elastic hard scattering cross sections and deep spacelike\,$\leftrightarrow$\,timelike universality \cite{Lepage:1980fj}, and search for signals of diquark correlations.  In this context, DSE/BSE methods are proving valuable insights \cite{Yao:2024drm, Yao:2024uej, Cheng:2024cxk, Cheng:2025yij}.

Transition form factors at $q^2>0$, $q^2\simeq 0$ can also be accessed at hadron facilities via radiative and Dalitz decays, such as $\Delta^+ \to p\gamma$ and $\Delta^+ \to p e^+e^-$, respectively. These processes have been investigated, for example, by the HADES experiment, as discussed below.

Additional information on transition structure -- accessible to both experiment and theory -- can be obtained via neutrino scattering and semileptonic decay processes, \textit{i.e}., using weak probes. 
In these reactions, both vector and axial form factors are accessed \cite{pesschr}. 
Examples include the scattering process $\nu_e n \to e^- \Delta^+$ and the semileptonic decay $\Lambda \to p e^- \bar{\nu}_e$.

Each of these probes reveals different, yet complementary, information about the internal structure of baryons, with particular sensitivity to their valence composition. 
However, a non-relativistic language borrowed from atomic physics cannot do justice to the intricate dynamics of relativistic many-body systems. 
In the presence of meson clouds, one can argue that the meaningful question is not whether a given baryon is a three- or five-quark state, but rather how large the respective contributions are. 

Moreover, Poincar\'e covariance in quantum field theory entails that even a low-mass, low-spin state must contain relative orbital angular momentum between its constituents; see, e.g., Refs.\,\cite{Llewellyn-Smith:1969bcu, Oettel:1998bk}.
Therefore, a key question concerns the extent of overlap between a given state and configurations with or without intrinsic orbital angular momentum, which leads to observable differences in form factors. 
Sophisticated theoretical approaches are required to provide quantitative predictions. 
These will be reviewed in Sec.\,\ref{subsec:el-transFF-theory}. 

Here, it is worth presenting a few illustrative examples related to approximate symmetries and genuine relativistic effects.
To begin with a relatively straightforward case: for the electromagnetic transition form factors (ETFFs) of $\Delta \to N \gamma^{(*)}$, isospin symmetry dictates that the photon is purely isovector. 
Consequently, the form factors for $\Delta^+ \to p \gamma^*$ and $\Delta^0 \to n \gamma^*$ must be nearly identical, despite one system being charged and the other neutral. 
Extending this reasoning to three flavours, one can systematically compare 
electromagnetic transitions from decuplet to octet baryons and make semi-quantitative predictions based on SU(3)-flavour symmetry. 
For instance, a suppression of 
$\Sigma^{-}(1385) \to \Sigma^- \gamma^{*}$ relative to 
$\Sigma^{+}(1385) \to \Sigma^+ \gamma^{*}$ is predicted owing to approximate U-spin symmetry.  (This is an approximate SU$(2)$ symmetry of the QCD Lagrangian under a unitary rotation of $d$ and $s$ quarks.)

Corresponding reactions involving virtual photons would reveal the full transition form factors and provide further insight into the quantitative breaking of flavour symmetry. 
So far, electromagnetic transitions in the cascade sector have not been observed. 
Approximate U-spin symmetry predicts a suppression of 
$\Xi^-(1530) \to \Xi^- \gamma^{(*)}$ relative to 
$\Xi^0(1530) \to \Xi^0 \gamma^{(*)}$. 
Thus, the electrically neutral system is expected to couple more strongly 
to the electromagnetic field than the charged system \cite{Kaxiras:1985zv, Jenkins:2011dr, Holmberg:2018dtv}. 
This behaviour could, in turn, serve as a diagnostic tool for newly observed excited cascade states, helping to determine whether they  belong to a decuplet or an octet multiplet.

Turning to an example of states with a potentially large meson-baryon component, one can study the transition form factors of the excited $\Lambda^*$ states in the region of $1.4\,\mathrm{GeV}$ \cite{Dalitz:1967fp, Kaiser:1995eg, Garcia-Recio:2003ejq, Mai:2014xna}. 
In the following, we refer to these states by their spin-parity assignment as $\Lambda(1/2^-)$. 
One can consider isovector transitions, $\Lambda(1/2^-) \to \Sigma^0 \gamma^{(*)}$, and isoscalar transitions, 
$\Lambda(1/2^-) \to \Lambda \gamma^{(*)}$; see also Ref.\,\cite{Geng:2007hz}.
If a $\Lambda(1/2^-)$ state has a large $\Sigma\pi$ component, an isovector photon can couple to the lightest state, the pion, involving the pion vector form factor. 
Given the low mass of the pion, the peripheral structure should be 
dominated by this pion-cloud effect. 
In contrast, an isoscalar photon does not couple to a pion pair, making the two transitions quite distinct. 
If a $\Lambda(1/2^-)$ state has a large nucleon-antikaon component, the photon can couple to the antikaon as a semi-light state. 
This is allowed for both isovector and isoscalar transitions. 
It would be particularly interesting to measure the various transition radii and compare them with each other.

ETFFs contain information about not only the size but also the shape and dynamical properties of hadrons. For example, in the space-like kinematic regime ($q^2 \le 0$ or $Q^2 = -q^2 \ge 0$), the form factors reveal the charge and magnetic distributions. 
However, the properties of relativistic quantum field theory make this interpretation more nuanced, with a rigorous understanding requiring that one consider the distributions as corresponding to tomographic images of transverse densities in the impact-parameter plane \cite{Miller:2010nz}.

One of the successes of the non-relativistic quark model is the prediction of the dominance of the magnetic over the electric ETFF for the nucleon-to $\Delta(1232)$ transition \cite{Pascalutsa:2006up, Ramalho:2008aa}.
This result is consistent across all phenomenological approaches and confirmed by QCD-based calculations. 
This ETFF provides insight into deviations from spherical shape. 
Moving beyond the non-relativistic picture, QCD-based calculations attribute such deformation to intrinsic orbital angular momentum \cite{Eichmann:2011aa, Lu:2019bjs}.
Moreover, ETFFs also encode information on the dynamical properties of hadrons. In the time-like regime ($q^2 > 0$, $Q^2 < 0$), form factors reflect the signatures of particle creation and are closely tied to spectroscopy, exhibiting characteristic peaks near the photon point $Q^2 = 0$.
 

This discussion makes clear that the study of transition form factors should not be limited to scattering reactions, especially not for unstable hadrons. 
For transitions between hadrons $h_1$ and $h_2$, the decay region provides complementary information via Dalitz decays $h_1 \to h_2 \, e^+ e^-$ and semi-leptonic weak decays $h_1 \to h_2 \, \ell \nu_\ell$ (with a charged lepton $\ell$ and its associated neutrino $\nu_\ell$).
Producing hadrons in hadronic reactions yields sufficiently high production rates to enable the study of rare Dalitz and semi-leptonic decays, thereby offering valuable insights into the vector and axial-vector transition form factors of hadrons. For instance, information about the radius (see Eq.\,\eqref{eq:def-radius-gen}) can be extracted from either hadron–electron scattering, $h_1 e^- \to h_2 e^-$, in the spacelike region ($q^2 \to -0$), or from the Dalitz decay $h_1 \to h_2 \, e^+ e^-$ by analysing the slope of the $q^2 \to +0$ distribution in the timelike region ($(m_{e^+e^-})^2 < q^2 < (M_{h_1} - M_{h_2})^2$).

The above discussion highlights the strong connection between form factors and the study of dense and hot baryonic matter. 
ETFFs of baryons provide crucial insight into virtual massive photon-baryon couplings, which are essential for understanding the emissivity of baryonic matter in heavy-ion collisions, as measured via dilepton emission. 
In particular, the role of the $\rho$ meson and the modifications of its spectral function in nuclear matter, obtained from hadronic many-body calculations \cite{Rapp:1999ej}, play a dominant role in the low-mass region, influencing both the total thermal yield and the determination of the temperature. 
Therefore, future high-statistics experiments that combine hadron and dielectron measurements with pion and proton beams at FAIR are of special importance for the CBM heavy-ion programme \cite{Salabura:2020tou}.

 \subsubsection{Theoretical approaches}
\label{subsec:el-transFF-theory}
There are several theoretical approaches to the calculation of baryon elastic and transition form factors, some of which are now sketched.
\smallskip

\noindent\textbf{Quark Models}.
  Constituent quark models have long been used.  They have been developed to predict or analyse the wealth of experimental findings on baryon electromagnetic transition amplitudes in the spacelike kinematic regime and can broadly be divided into four classes:\\[-6ex]
\begin{itemize}
    \item[i)] The single quark transition model (SQTM) \cite{Burkert:2002zz}, which assumes SU(6) spin-flavour and $O(3)$ orbital symmetries, and an SU(6)$\otimes O(3)$ structure of the baryon wave functions. 
    It is based solely on valence constituent-quark degrees of freedom and avoids parametrisations potentially contaminated by meson-cloud effects. The model is supposed to be valid in the intermediate to high four-momentum transfer region, $Q^2 > m_p^2$.

    \item[ii)] The covariant spectator quark model, a field-theoretic framework in which baryons consist of three quarks, with two forming a non-pointlike diquark of average mass \cite{Ramalho:2008ra, Gross:2012si}. 
    The baryon vertices are derived from symmetry constraints rather than a confining potential, and the radial wave functions are informed by LQCD computations or experimental data. This model gives reliable predictions at large $Q^2$.

    \item[iii)] The hypercentral quark model family \cite{Bijker:2015gyk}, which employs a confining potential $V(x)$ within a hypercentral approach to describe baryon structure.

    \item[iv)] The light-front quark model family, defined in the infinite momentum frame and typically restricted to $qqq$ states. 
    Inspired and informed by DSE/BSE analyses, it incorporates the momentum dependence of quark masses phenomenologically and achieves good agreement with data for $Q^2 > 2 m_p^2$ \cite{Aznauryan:2015zta,Aznauryan:2017nkz}.
\end{itemize}


A stringent test for any model developed for electron scattering data analysis is its predictive power across different kinematic regimes of hadron scattering reactions, including Dalitz decays such as $N^*/\Delta \rightarrow N e^+ e^-$~\cite{Ramalho:2016zgc, Ramalho:2015qna}. 
The $\Delta(1232)\, \frac{3}{2}^+$ serves as a prime example where effects associated with $q\bar{q}$ excitations, manifesting through the pion cloud, extend the baryon quark core and, in some models, can be as significant as the pure valence constituent-quark contributions \cite{Ramalho:2008aa}. Similar meson-cloud effects are also prominent in other resonances, such as the $N(1440)\, \frac{1}{2}^+$, which may involve a strong interplay between meson-baryon and quark degrees of freedom \cite{Suzuki:2009nj, Burkert:2019bhp}.
Commonly, models that incorporate meson-cloud dressing of the three-quark core offer improved descriptions of transition form factor data, resonance masses, and decay modes. 

A critical issue in interpreting experimental data on helicity amplitudes and multipole form factors is the presence of correlations among them. These correlations stem from their common derivation from a set of independent, kinematic-singularity-free form factors. Such correlations are especially relevant at low $Q^2$, close to the pseudo-threshold, and must not be neglected when parametrising data.
The importance of these pseudo-threshold constraints becomes apparent near the photon point in the $\gamma^\ast N \to \Delta(1232)$ transition, as well as for other low-lying nucleon excitations. Consequently, the determination of nucleon-resonance electrocouplings demands that data analyses begin from a foundation built on the independent, kinematic-singularity-free form factors.

Looking ahead, measurements of baryon electromagnetic transition form factors via Dalitz decays using pion beams present a promising avenue for disentangling the initial resonance production and final decay mechanisms from the intermediate resonance propagation. Progress in this direction will be essential for reducing theoretical uncertainties in the understanding of resonance electrocouplings and their evolution with momentum transfer.

Moreover, the experimental and theoretical results obtained thus far primarily concern the electrocouplings of low-lying nucleon resonances. 
Timelike experiments at higher energies with pion beams, such as those planned at FAIR, and their extension to $pp$ reactions offer highly promising opportunities to explore analogous quantities for baryons containing $s$ and $c$ quarks.

\smallskip

\noindent{\bf Dispersive methods}.
In the past decade, significant efforts have been made in both experiment and theory to study nucleon electromagnetic form factors (EMFFs) \cite{Denig:2012by, Pacetti:2014jai, Punjabi:2015bba}. 
In this context, dispersion theory has played, and continues to play, a valuable role in connecting available data to parametrisation of nucleon EMFFs \cite{Lin:2021umz, Lin:2021xrc}. 
The model-independent dispersion theory framework incorporates all constraints from unitarity, analyticity, and crossing symmetry, as well as aspects of the asymptotic behaviour of the form factors dictated by perturbative QCD.

The successful dispersive treatment of nucleon EMFFs also holds significant promise for a model-independent description of the electromagnetic structure of hyperon states. 
These form factors are far less well known than those of the nucleon, and experimental data are only available in the timelike region \cite{Schonning:2023hnv}. 
Reference~\cite{Granados:2017cib} considered once-subtracted dispersion relations for the electromagnetic $\Sigma$-$\Lambda$ TFFs and expressed them in terms of the pion EMFF and the two-pion $\Sigma$-$\Lambda$ scattering amplitudes. 
The analysis predicted the ETFFs of $\Sigma$-$\Lambda$ and studied the role of pion rescattering and the explicit inclusion of decuplet baryons within three-flavour ChPT. 
In Ref.~\cite{Lin:2022dyu}, the dispersive determination of the electromagnetic $\Sigma$--$\Lambda$ transition form factors was extended to include the $K\bar{K}$ intermediate state in a 
$\pi\pi$-$K\bar{K}$ coupled-channel treatment in SU(3) ChPT. 
This led to a shift in the electric $\Sigma$-$\Lambda$ transition form factor $G_E$, while the magnetic form factor $G_M$ remained essentially unchanged. 

At present, the dispersion-theory determination of the ETFFs for $\Sigma$--$\Lambda$ suffers from sizeable uncertainties owing to limited knowledge of certain low-energy constants (LECs) in SU$(3)$ ChPT. 
A precise determination of these three-flavour LECs from future experiments will be essential to better constrain the transition form factors. 
ETFFs for the spin-$\tfrac{3}{2}$ $\Sigma$ to the $\Lambda$ hyperon were considered in Ref.\,\cite{Junker:2019vvy}. 
Furthermore, dispersion theory has been applied to analyse the full set of cross section data for the reaction $e^+e^- \to \Lambda\bar{\Lambda}$ \cite{Lin:2022baj}. 
As more experimental data on hyperon electromagnetic structure become available, dispersion-theoretical methods will provide a powerful tool for analysing these data and for developing predictions of spacelike form factors based on timelike data. 
They may also offer valuable information about the underlying physical mechanisms.

In combination with EFT, dispersive methods have also been used to study the quark-mass dependence of the nucleon isovector form factors \cite{Alvarado:2023loi}, and to investigate the ETFFs of the 
$\Delta(1232)$ \cite{Aung:2024qmf} and the $N(1520)$ \cite{An:2024pip}$\to N$ transitions. 
Predictions for the Dalitz decays $\Delta(1232), N(1520) \to N \, e^+ e^-$ have been provided and scrutinised by HADES. 
They will also be studied using SIS100 at FAIR. 
The same applies to the transition form factors that enter 
the Dalitz decays $\Sigma^0 \to \Lambda \, e^+ e^-$ \cite{Granados:2017cib, Husek:2019wmt,Lin:2022dyu} and 
$\Sigma^0(1385) \to \Lambda \, e^+ e^-$ \cite{Junker:2019vvy}.


\smallskip

\noindent{\bf DSE/BSE}.
There exists a substantial body of results on light baryon elastic and transition form factors within the DSE/BSE approach; 
see, \textit{e.g}., Refs.\,\cite{Eichmann:2016yit, Barabanov:2020jvn, Roberts:2020hiw, Chen:2021guo, Chen:2023zhh, Carman:2023zke, Yao:2024uej, Achenbach:2025kfx}. 
The most extensively studied systems are those involving 
nucleons and $\Delta$ baryons, including their respective elastic form factors and the $\Delta \to N$ transition. 
Results for axial form factors and hyperons are also available. 
As in spectroscopy calculations, both three-body and quark-diquark approaches yield consistent outcomes, showing good agreement with experimental data where such data exist. 
Discrepancies typically arise at low $Q^2$, where meson-baryon effects become significant. 
As discussed earlier, these effects have been estimated in quark models.
Corresponding treatments have been carried out for mesons; see, \textit{e.g}., Ref.\,\cite{Williams:2018adr, Miramontes:2021xgn}. 
However, for baryons, only exploratory studies have thus far been performed \cite{Hecht:2002ej, Flambaum:2005kc}. 
While most attention has so far been devoted to systems involving light quarks, in principle it is straightforward to extend these calculations to strange and charm baryons.
  
Most of the existing work concerns spacelike photon momentum transfer. 
However, the main region of interest in the context of the $pp$ programme at FAIR lies at $q^2>0$ (small $q$). 
This region is dominated by physical poles, such as those associated with the $\rho$, $\omega$, $\phi$, and $J/\psi$ mesons in the electromagnetic case. 
A key observation -- and one of the principal strengths of DSE/BSE approach -- is that these poles are dynamically generated ar the quark-gluon level through the underlying $n$-point Schwinger functions, establishing a deep connection between the spacelike and timelike regimes. 

Accessing the timelike region is technically challenging owing to the presence of singularities in the $n$-point functions that contribute to the matrix elements. 
A promising strategy for overcoming this difficulty involves contour deformations combined with analytic continuation. 
An illustrative example is the calculation of the timelike pion form factor presented \cite{Miramontes:2021xgn}, which not only dynamically generates the $\rho$-meson pole but also captures its width, \textit{i.e}., 
the resonance dynamics displace the pole into the complex plane. 
Implementing such techniques will be essential for future systematic studies of timelike baryon form factors.

\smallskip

\noindent{\bf Lattice QCD}.
In recent years, significant efforts have been devoted to determining the axial and electromagnetic form factors of the nucleon on the lattice in the spacelike region. 
In particular, $Q^2 \lesssim 1~{\rm GeV}^2$ calculations of the isovector flavour axial form factor have reached a relatively mature stage, with several studies available in which systematic uncertainties are under control; see, \textit{e.g}., Ref.\,\cite{Gupta:2024krt}. 
Good agreement is observed among the results for the range of $Q^2$ considered. 
The lattice results tend to lie somewhat above fits to neutrino-deuterium scattering data~\cite{Meyer:2016oeg}, but show better consistency with anti-neutrino-hydrogen scattering measurements \cite{MINERvA:2023avz}.

Progress has also been made in the calculation of electromagnetic form factors, where the primary focus lies in extracting the magnetic moment, as well as the electric and magnetic radii. Since only discrete momentum values are accessible on the lattice, probing the low-$Q^2$ region remains challenging and requires large spatial volumes. Nevertheless, techniques are being developed for the direct determination of the radii~\cite{Hasan:2017wwt,Alexandrou:2020aja,Ishikawa:2021eut}.

Furthermore, computing the form factors for the proton and neutron individually, rather than the isovector flavour combination, is especially difficult owing to the necessity of evaluating disconnected quark-line diagrams. 
These introduce additional stochastic noise, compounding the usual Monte Carlo gauge noise. 
Precision is gradually improving towards the level required to resolve differences between experimental results for the radii~\cite{Djukanovic:2023beb,Djukanovic:2023jag,Alexandrou:2025vto}.

Other isoscalar electromagnetic and axial form factors (for strange and even charm quarks) have also been calculated -- see, \textit{e.g}., Ref.\,\cite{Djukanovic:2021qxp} for a review -- and there are ongoing efforts to determine electromagnetic form factors at large $Q^2$ \cite{Syritsyn:2025fiu}. 
Beyond the nucleon, the form factors of hyperons have so far received less attention. 
However, the isovector axial charges have recently been computed with all systematics under control~\cite{Bali:2023sdi}, and the extension to $Q^2 > 0$ is straightforward.

In the near future, recent theoretical and technical developments will enable the study of $N \to \Delta(1232)$ and $N \to N^*$ transitions on the lattice. The relevant matrix elements are relatively straightforward to compute at unphysically heavy pion masses, where the initial and final states are stable~\cite{Alexandrou:2007dt,Alexandrou:2008bn,Lin:2011da}. However, a nucleon-to-resonance transition requires the application of the finite-volume formalism, both for $N\pi \to N\pi$ scattering and for the $N \to N\pi$ transition. 
Several scattering studies have already been performed -- see, \textit{e.g}., Ref.\,\cite{Romero-Lopez:2022usb}, while the first evaluations of the $N \to N\pi$ matrix elements are currently underway. 

In addition, the first lattice calculation of transition form factors for the semi-leptonic decay $\Lambda \to p$ was recently completed by the ETMC Collaboration; see Ref.\,\cite{bacchio2025studylambdatop}.

\subsubsection{Experimental programme at GSI/FAIR}

\begin{figure}[t]

\centerline
{
\includegraphics[width=0.45\columnwidth,height=0.44\columnwidth]{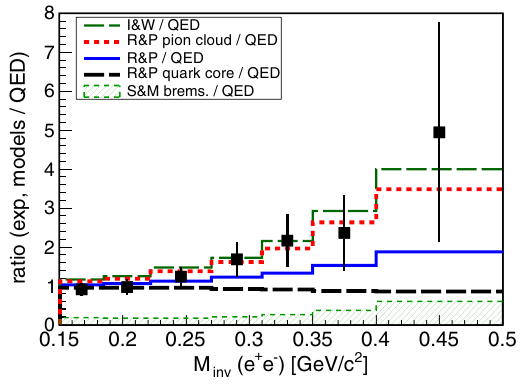}

\vspace*{1ex}

\includegraphics[width=0.45\columnwidth,height=0.44\columnwidth]{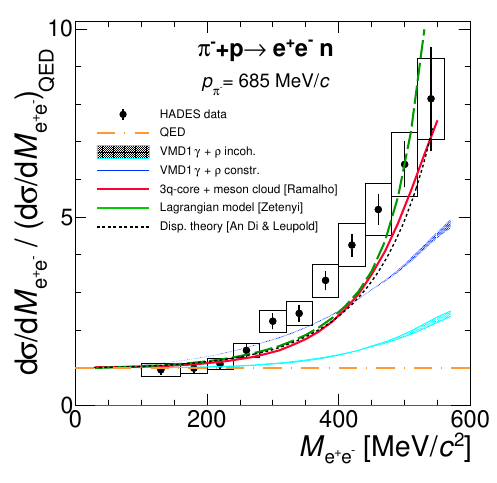}
}

\vspace*{-1ex}

\caption{\label{eTFF_HADES}
The ratio of the dielectron yield measured by HADES from Dalitz decays of the $\Delta(1232)$ \cite{HADES:2017jkt} (left) and $N^*(1520)/N(1535)$ \cite{HADES:2022vus} (right), shown as squares with error bars as a function of the invariant mass, is compared with the simulated contribution from resonances using a point-like form factor (QED model). 
Various lines represent results from different model calculations of the form factors: the Ramalho-Pe\~na covariant quark model (R\&P) \cite{Ramalho:2015qna}, the Iachello-Wan model for the $\Delta \to N\gamma$ transition \cite{PhysRevC.69.055204}, the effective Lagrangian model \cite{Zetenyi:2020kon}, a quark model \cite{Ramalho:2016zgc}, and a dispersive approach \cite{An:2024pip}.}
\end{figure}

Measurements of ETFFs of baryons in the timelike region have been pioneered at SIS18 for $N^*/\Delta(1232)$ resonances by the HADES experiment. 
The result for the $\Delta$ resonance was obtained from the Dalitz decay $\Delta^+ \rightarrow p e^+ e^-$ (see Fig.~\ref{eTFF_HADES}), measured exclusively in $p+p$ reactions at $\sqrt{s} = 2.43\,$GeV. 
The dielectron invariant mass ($M^2_{e^+e^-} = q^2$) covered a range up to 400\,MeV/$c^2$ \cite{HADES:2017jkt}. 
The results confirmed the expected dominance of the magnetic character of the transition and provided evidence for the significant role of pion cloud effects; see, e.g., Refs.\,\cite{Julia-Diaz:2006ios, Ramalho:2015qna}.
The $\Delta$ production amplitude was controlled via a partial-wave analysis of single-pion production, allowing for the determination of the Dalitz decay branching ratio. 

A similar strategy was employed in $\pi^- p$ reactions to measure $N^*$ Dalitz decays in the second resonance region \cite{HADES:2022vus}. 
Pion-induced reactions offer a significant advantage over proton-induced reactions, as resonances are produced directly and their mass can be selected by tuning the beam momentum (formation process).

The effective squared sum of the electric, magnetic, and Coulomb transition form factors for the dominant $N(1520)$ resonance was measured, demonstrating the crucial role of the intermediate $\rho$ vector meson—consistent with predictions from the extended Vector Meson Dominance Model. The results confirm the dual coupling mechanism of baryons to virtual photons, comprising both direct and vector meson couplings, where the latter vanishes at $q^2 = 0$.

The extracted effective ETFF was found to agree with calculations using dispersive theory \cite{An:2024pip}, effective Lagrangian approaches \cite{Zetenyi:2020kon}, and the covariant quark model~\cite{Ramalho:2016zgc}. Furthermore, the analysis of triple-differential polar and azimuthal angular distributions of the emitted lepton pairs as a function of the invariant mass enabled the extraction of spin-density matrix elements, which are predicted to depend on the spin and parity of the transition.

High-statistics measurements of these distributions are planned with pion beams at FAIR, targeting the third resonance region. Spin-density matrix elements provide access to helicity amplitudes and allow the separation of electric, magnetic, and Coulomb form factor contributions in non-strange baryons \cite{Zetenyi:2012hg, Krivoruchenko:2001jk}. 
They also serve as an important reference for measurements of photon polarisations in heavy-ion collisions.

These types of measurements can be extended to the hyperon sector, taking advantage of the higher production cross sections at SIS100 and the narrower widths of hyperon states, which make them easier to identify. 
For instance, Dalitz decays of $\Lambda(1520)$, $\Lambda(1405)$, $\Sigma(1385)$, etc., to both $\Lambda$ and $\Sigma^0$ accompanied by a dilepton pair can be studied. 
These decays provide information on the electromagnetic structure of the hyperon resonances and the role of the meson cloud \cite{HADES:2020pcx}.

A first step in this direction has already been made by the HADES collaboration, which successfully measured dilepton production tagged by $\Lambda$ in $pp$ interactions at a beam kinetic energy of 4.5\,GeV. 
Using this technique, the Dalitz decay $\Sigma^0 \rightarrow \Lambda e^+e^-$ could be reconstructed, albeit with limited statistics for studying higher-mass hyperon decays.

\begin{figure}[h]   
\centerline
{
\includegraphics[width=0.5\columnwidth,height=0.42\columnwidth]{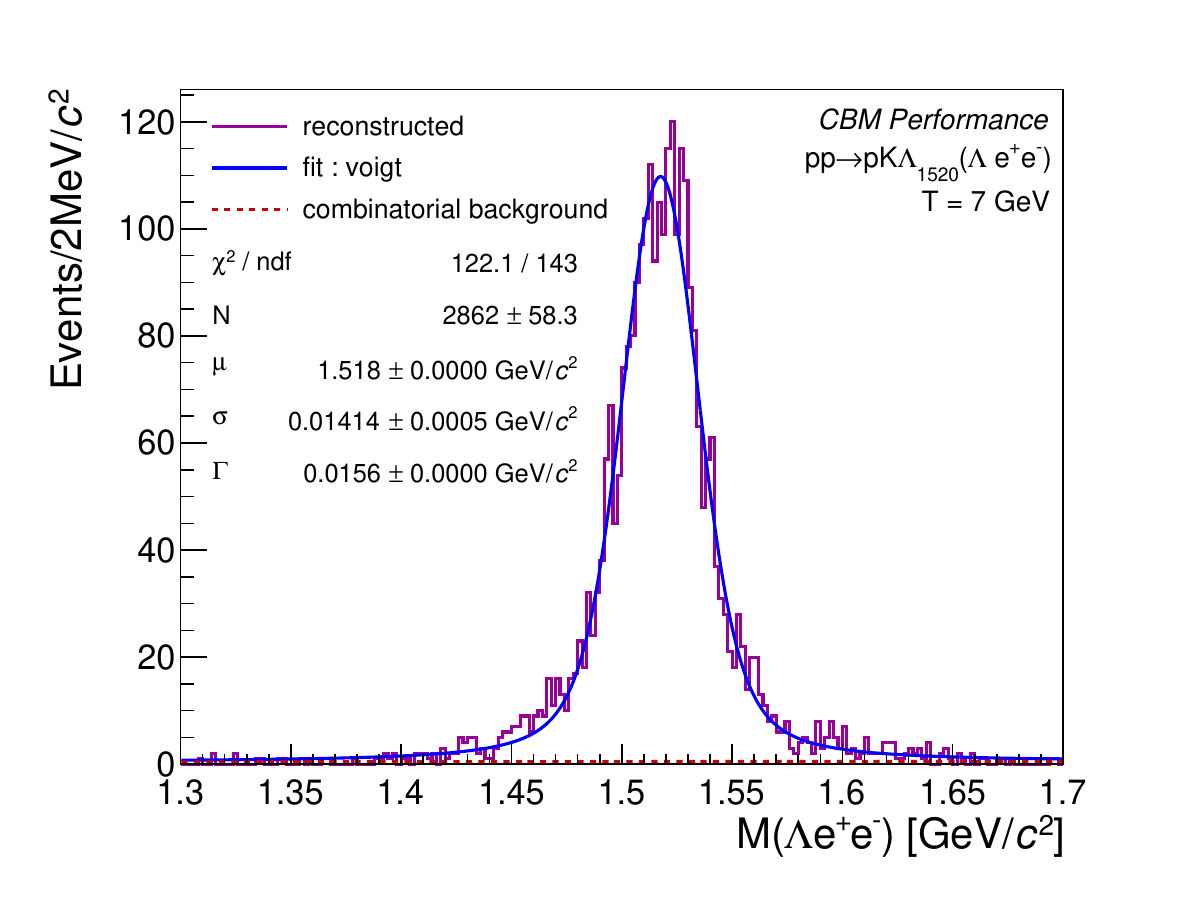}
\hspace*{-1cm}
\includegraphics[width=0.6\columnwidth,height=0.5\columnwidth]{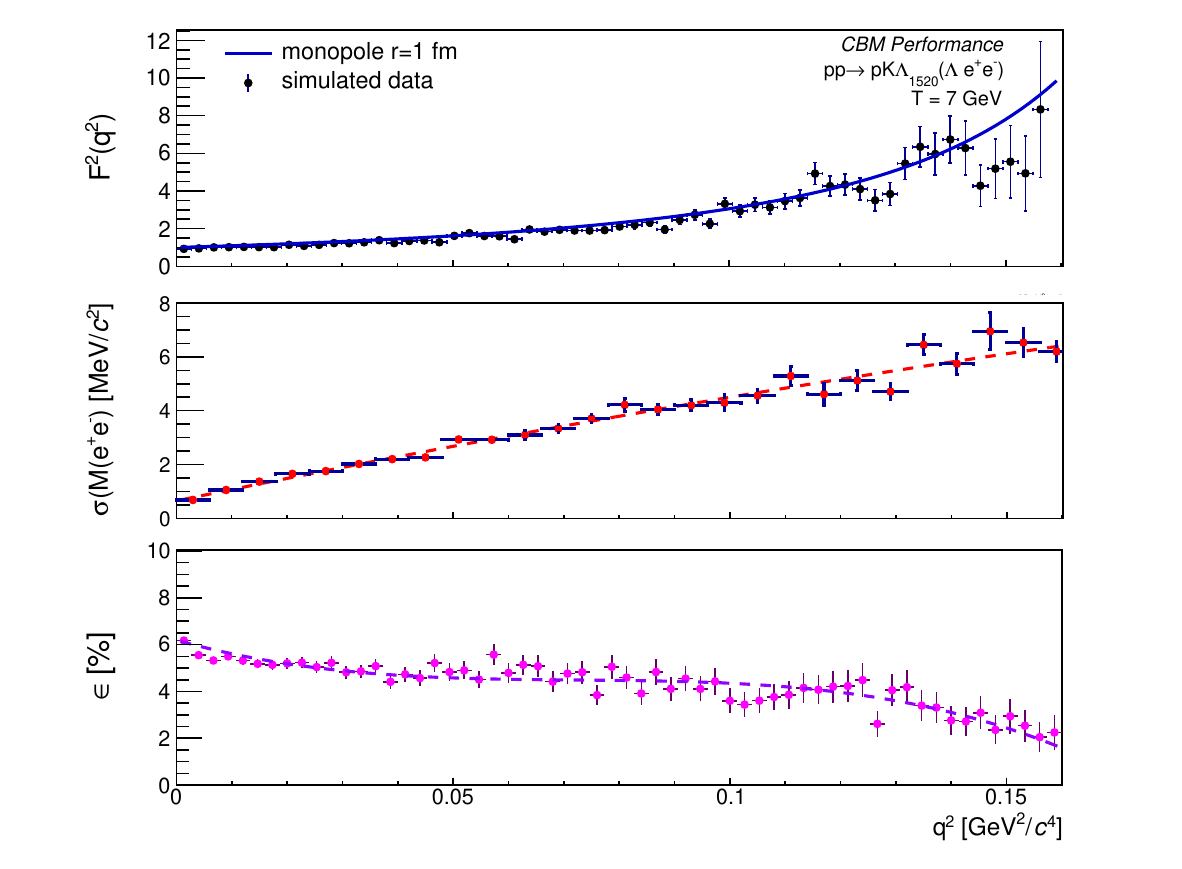}
}
\vspace*{-0.3cm}
\caption{\label{fig:L1520_CBM}
Simulation results for the $pp \rightarrow pK^+\Lambda(1520) \rightarrow pK^+\Lambda e^+e^-$ reconstruction in CBM at a kinetic energy of $T=7\,$GeV; assuming 10 days of beam time at 1\,MHz total interaction rate with $10^2\,\mu$b cross section for the reaction of interest and with a $\Lambda(1520)\rightarrow \Lambda e^+e^-$ branching fraction of $10^{-4}$. 
Left: invariant mass distribution of the $\Lambda e^+e^-$ system with the results of a fit based on a Voigt function. The natural width of the $\Lambda(1520)$ was fixed to $\Gamma =15.6\,$MeV in the simulation and in the fit. 
Right: expected measured square of the form factor, $F^2(q^2)$, based on a monopole form factor with $r=1\,$fm as input to the event generator (top panel), $e^+e^-$-mass resolution (middle panel), and total efficiency (bottom panel) as a function of $q^2$.  
The form factor $F^2(q^2)$ is obtained by dividing the simulated yield by the expected yield for a point-like scenario, taking into consideration the experimental resolution and efficiency obtained from the fits as presented in the middle and lower panels, respectively. 
The results are based on an exclusive analysis, \textit{i.e}., $pp\rightarrow \Lambda(1520)K^+p$ with $\Lambda(1520)\rightarrow\Lambda e^+e^-$ and $\Lambda\rightarrow p\pi^-$ whereby a 5C refit has been applied (4-momentum and $\Lambda$ mass constraints). Take note of the following disclaimer~\cite{cbm_feasibility_note}.} 
\end{figure}

At the higher energies available at SIS100 and with the CBM detector, which offers significantly improved vertex resolution (crucial for tagging $\Lambda$ from hyperon Dalitz decays), such measurements could be performed with the necessary precision. 
In particular, precise studies will be enabled by an order-of-magnitude larger acceptance and the expected excellent mass resolutions: (\textit{i}) for dileptons, about $\sigma_{m} = 1-6\,$MeV/$c^2$; and (\textit{ii}) for the combined $\Lambda e^+e^-$ system, about $\sigma = 14\,$MeV/$c^2$.
Of particular importance is good coverage and resolution at low dilepton invariant mass ($<500\,$MeV/$c^2$), as demonstrated in the simulation results of the reconstruction of the reaction $pp \rightarrow pK^+ \Lambda(1520) \rightarrow pK^+ \Lambda e^+e^-$, shown in Fig.\,\ref{fig:L1520_CBM}. This low-mass region is crucial for Dalitz decay measurements and for determining the slopes of the mass-dependent transition form factors.

\begin{figure}[t]
    \centering
    \includegraphics[width=0.9\linewidth]{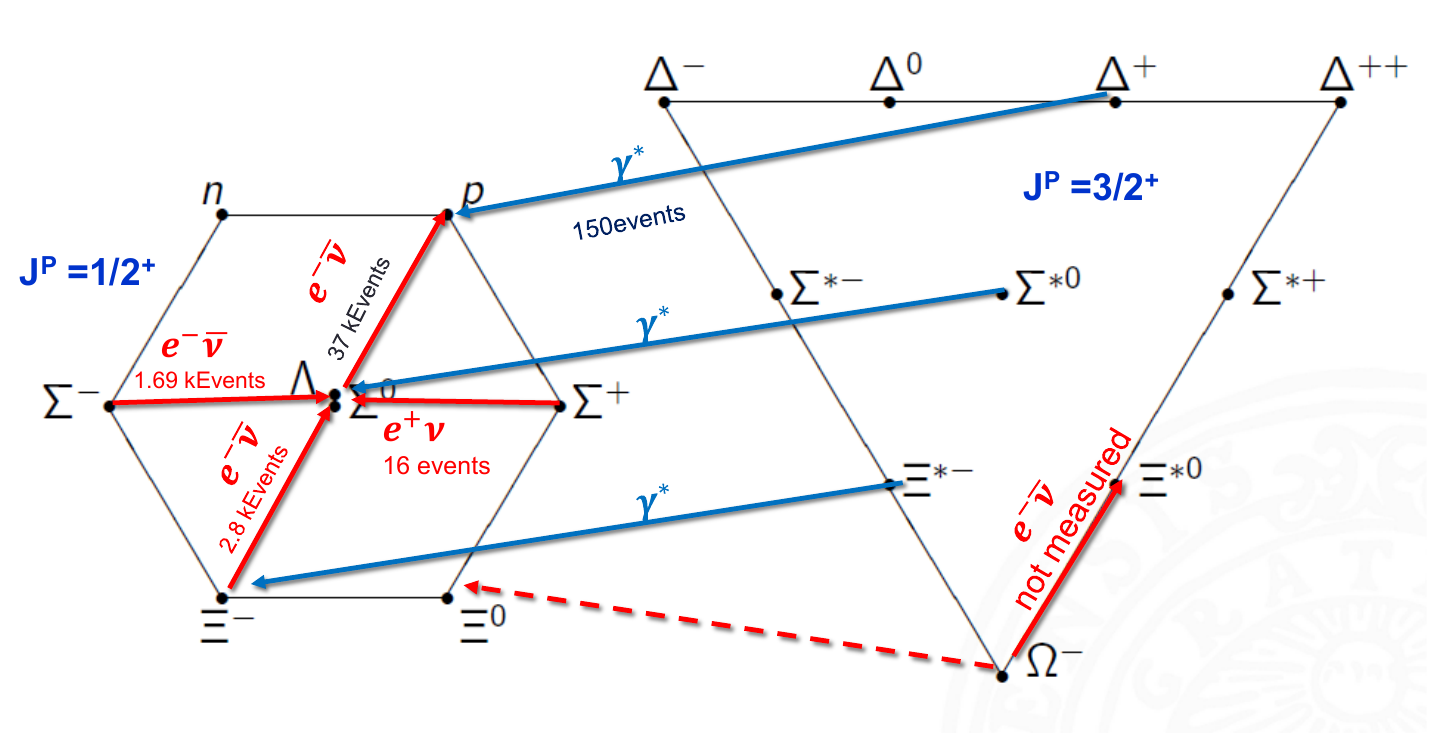}
    \caption{Semileptonic transitions between octet and decuplet baryons that can be reconstructed in exclusive production channels with only charged particles in the final state (excluding the neutrino) are indicated by red arrows. Available statistics from previous experiments at the SPS~\cite{Bristol-Geneva-Heidelberg-Orsay-Rutherford-Strasbourg:1983jzt} (labels in red) and Fermilab~\cite{Dworkin:1990dd} (labels in black) are shown. Electromagnetic transitions that can also be reconstructed at FAIR are indicated by blue arrows.
    \label{fig:SL_transitions}}
\end{figure}

As a next step, it will be of great interest to extend these measurements to the weak sector, with the aim of determining the vector and axial transition form factors in the semileptonic decays of octet and decuplet hyperons, as illustrated in Fig.\,\ref{fig:SL_transitions}. The branching ratios for Dalitz and semileptonic decays are of a similar order, $\mathcal{O}(10^{-4} - 10^{-5})$, and the production cross sections are on the order of tens of microbarns. Combined with the 10~MHz data-taking capability of CBM, this could yield approximately $10^5$ reconstructed weak decays per day. To suppress the substantial background, the decays should be reconstructed in exclusive channels where all particles except the neutrino are detected. Examples include $pp \rightarrow p\,\Xi^-\,K^+\,K^+$ with $\Xi^- \rightarrow \Lambda\,e^-\,\bar{\nu}$ or $pp \rightarrow p\,\Lambda\,K^+$ with $\Lambda \rightarrow p\,e^-\,\bar{\nu}$.
     
The vector and axial form factors constitute a valuable dataset that can be directly compared with the theory predictions described above. 
These measurements would form a cornerstone of the broader effort to understand baryon structure. 
Since the form factors are approximately related by SU(3) flavour symmetry, studying their deviations in the hyperon sector offers a unique opportunity to probe SU(3)-flavour symmetry breaking effects.

%

Moreover, transitions between octet hyperons are of particular interest for the precise determination of the Cabibbo--Kobayashi--Maskawa (CKM) matrix element $|V_{us}|$~\cite{Mateu:2005wi}. 
In total, six form factors (three vector and three axial) are involved in transitions between octet baryons, but only three -- the vector form factors $f_1$, $f_2$ and the axial form factor $g_1$ -- are significant. 
Their ratios can be extracted from lepton differential distributions and decay asymmetries. Combined with measured branching ratios and assuming SU(3)-flavour symmetry conservation they provide input to the current determination of $|V_{us}| = 0.2250 \pm 0.0027$~\cite{Cabibbo:2003ea,ParticleDataGroup:2024cfk} from hyperon decays. This value should be compared to the values from kaon decays $|V_{us}| = 0.22431 \pm 0.00085$~\cite{ParticleDataGroup:2024cfk} and $\tau$ decays $|V_{us}| = 0.2207 \pm 0.0014$~\cite{HFLAV:2022esi}.
The result from hyperon decays is based on just a few experiments conducted in the 1980s using hyperon beams at the SPS~\cite{Bristol-Geneva-Heidelberg-Orsay-Rutherford-Strasbourg:1983jzt} and at Fermilab~\cite{Dworkin:1990dd}. 
It is worth noting that these analyses relied on the $\Lambda$ polarisation using a decay asymmetry parameter of $\alpha_\Lambda = 0.64 \pm 0.014$, whereas the currently accepted value is $\alpha_\Lambda = 0.746 \pm 0.008$.  The most recent progress includes the first lattice QCD calculation of the form factors~\cite{Bacchio:2025auj} allowing to extract $|V_{us}|= 0.2300 \pm 0.0060$  using the ab-initio calculated form factors and experimental value for the $\Lambda\to p e^-\bar\nu_e$ branching fraction. On the experimental side BESIII and LHCb studied $\Lambda\to p\mu^-\bar{\nu}_{\mu}$~\cite{BESIII:2021ynj,LHCb:SLHD} and new BESIII results on the absolute branching fraction and weak coupling strengths for $\Lambda\rightarrow pe^-\bar{\nu}_e$ and $\bar{\Lambda}\rightarrow \bar{p}e^+\nu_e$ are presented in Ref.~\cite{BESIII:2025hgj}. These results are based on the exclusively reconstructed data sample of about 2000 events. Further precise measurements and lattice QCD calculations of the form factors are needed if the hyperon semileptonic decays are to significantly contribute to the test of the first row of the CKM unitarity.
The connection between these form factors and neutrino physics is discussed in Sec.\,\ref{Sec:hyp_nu}.

\subsection{Nucleon structure and intrinsic charm}
\label{sec:intrinsic_charm}
Several decades of experimental and theoretical studies and analyses have established that, like all hadrons, the proton is a complex system composed of quarks, antiquarks, and gluons (partons), whose interactions are governed by the principles of QCD. 
Nevertheless, despite intensive efforts, many aspects of the proton's internal structure remain incompletely understood.

One modern approach to describing proton structure employs Wigner quasi-probability distributions, which are five-dimensional functions that encode the light-front longitudinal and transverse momentum, as well as transverse spatial distributions, of partons inside the proton. 
Integration over transverse position yields transverse momentum-dependent parton distributions (TMDs), while integration over transverse momentum gives access to generalised parton distributions (GPDs) \cite{Diehl:2023nmm}.

Studies of GPDs and TMDs can aid in resolving the so-called ``spin puzzle'' \cite{EuropeanMuon:1987isl,Aidala2013SpinStructure}, which concerns the magnitude of the quark-parton contribution to the nucleon spin. 
In this context, beyond their intrinsic spin, the orbital motion of gluons and quarks also contributes \cite{Jaffe:1989jz} and must therefore be identified and included in the total spin budget. 
Indeed, as stressed above, orbital angular momentum is an essential feature of any Poincar\'e-invariant treatment of hadron bound states \cite{Eichmann:2016yit, Liu:2022ndb, Yu:2024ovn}.

To date, empirical access to GPDs has primarily been pursued via deeply virtual reactions in lepton scattering experiments; however, theory suggests that GPD-relevant data may also be obtained from meson- and proton-induced reactions~\cite{Berger:2001zn, Sawada:2016mao, Kumano:2009he}. For instance, with proton beams of approximately $30\,\mathrm{GeV}$, as will be available at SIS100, $2 \to 3$ reactions, $pp \to p\pi B$, with $B$ being either a proton or a baryon resonance, may offer access to GPDs and transition GPDs~\cite{Diehl:2024bmd}. Such reactions have sizeable cross sections, of the order of a few~$\mu$b, and can therefore be studied without stringent luminosity requirements. Nevertheless, further analyses are needed to determine the kinematic conditions necessary for reliable GPD extraction. 

A principal motivation for GPD studies is their capacity to provide access to hadron gravitational form factors, \textit{i.e.}, the form factors that characterise the in-hadron expectation value of the QCD energy--momentum tensor \cite{Polyakov:2018zvc, Burkert:2018bqq, Burkert:2023wzr}: mass, spin, and pressure. 
Since the Higgs-boson-generated quark current masses are two orders of magnitude smaller than the proton mass, much attention has been devoted to understanding the nature of QCD dynamics and the role of gluons in generating the proton mass \cite{Yang:2018nqn, Roberts:2021nhw, Ding:2022ows, Binosi:2022djx, Ferreira:2023fva}. 
Experimental data on the production of hidden- and open-charm final states obtained with proton beams at SIS100 could significantly enhance our understanding of the origin of the proton mass and the contribution of gluons to that mass.

Additionally, the energy coverage of SIS100 proton beams provides access to charm quark distributions on the proton's valence quark domain, about which little is currently known. 
Charm production data should be able to discriminate between sea-like and valence-like profiles, thereby offering excellent potential to resolve the longstanding puzzle of ``intrinsic charm'' \cite{Brodsky:1980pb}.

Related to these themes, some researchers have argued that vector-meson ($V$) production in proton-proton collisions ($pp \to pp\,V$), along with measurements of strangeness- or charmonium-nucleon final-state interactions, can provide unique insights into the gluon contribution \cite{Lappi:201467a,Cisek:2019bmu} to the proton mass \cite{Kharzeev:2021qkd,Bertulani:2021gqi}. 
In this context, studies of in-medium modification of charmonium via proton-nucleus ($pA$) collisions are also of significant interest. 
For clean signals of the effects sought, near-threshold production is expected to be optimal.
These and related issues are discussed below.

\subsubsection{Theoretical approaches}
\label{JpsiNCSM}

On the theory side, information about baryon structure can be accessed using a range of different methods. In the following, we outline recent advances made using LQCD, the DSE/BSE framework, and phenomenological models. We also discuss specific opportunities for testing and exploiting these predictions at SIS100.

\smallskip

\noindent{\bf Lattice QCD}.
Many nucleon structure observables can be studied on the lattice through the evaluation of matrix elements with local and nonlocal operators. 
The former are the most straightforward to compute and are connected, via the operator product expansion, to moments of the unpolarized, helicity, and transversity generalised parton distribution functions, \textit{viz}.\ the generalised form factors (GFFs). 
Lower moments have been well studied in the forward limit~\cite{FlavourLatticeAveragingGroupFLAG:2024oxs} and, \textit{e.g}., the quark (up, down, strange, and charm) and gluon contributions to the proton’s momentum have been determined at physical pion masses \cite{Alexandrou:2020sml, Wang:2021vqy}. 
Furthermore, both forward and off-forward helicity matrix elements have been calculated, enabling a complete decomposition of the proton’s spin.

In general, both statistical and systematic uncertainties, such as those owing to discretisation effects, increase with $Q^2=-t$. 
Nevertheless, the quark and gluon gravitational form factors, encoded in the second moment of the unpolarized GPDs, have been extracted at or near physical pion masses for $Q^2$ up to $1-2\,\mathrm{GeV}^2$; see, \textit{e.g}., Ref.\,\cite{Burkert:2023wzr,Hackett:2023rif}. 
Higher moments are more difficult to evaluate owing to signal deterioration and operator mixing under renormalisation with operators of lower dimension. New approaches, such as those employing the gradient flow method, are currently being investigated to overcome these limitations \cite{Shindler:2023xpd}.

Rapid progress has been made in providing more direct constraints on GPDs through the calculation of quasi- and pseudo-GPDs extracted from matrix elements of non-local operators. 
These distributions are matched to the standard GPDs either in the limit of large nucleon momentum or short distance. 
In the forward limit, the quark and gluon parton distribution functions (PDFs) have been studied intensively, with, \textit{e.g}., continuum-limit, physical pion mass results already available for the isovector unpolarised PDFs~\cite{Lin:2025hka}.

GPDs have been less thoroughly investigated, owing partly to the computational cost associated with using symmetric frames to access finite momentum transfer. 
To date, quark GPDs have mostly been determined at unphysical quark masses in the isovector flavour combination~\cite{Bhattacharya:2025cia}. 
However, an alternative approach employing asymmetric frames has been developed to alleviate this issue \cite{Bhattacharya:2022aob}. 

Significant challenges remain for these direct approaches, including the need to achieve sufficiently large nucleon momenta, suppress higher-twist contributions, and handle power divergences arising in the renormalisation of nonlocal currents for quasi-PDFs and GPDs. 
These are in addition to the standard systematic uncertainties inherent in the lattice approach, such as excited-state contamination, finite-volume effects, discretisation (cut-off) effects, and unphysical quark masses.

More advances are expected and several studies have explored the use of lattice results to constrain GPDs in phenomenological fits to experimental data; see, \textit{e.g}., Ref.\,\cite{Cichy:2024afd,Riberdy:2023awf,Guo:2023ahv}. 
In particular, flavour separation is straightforward on the lattice, and the moments can be extracted without contamination from higher-twist matrix elements.

\smallskip

\noindent{\bf DSE/BSE.}
Functional methods, in particular continuum Schwinger function methods (CSMs) provide another widely used approach to nucleon (baryon) structure \cite{Eichmann:2016yit, Burkert:2019bhp, Binosi:2022djx, Ding:2022ows, Ferreira:2023fva, Deur:2023dzc}. 
In this case, one begins with a well-defined approximation to all quantum field equations -- DSEs -- relating to the nucleon bound-state problem, including the Faddeev equation. 
The first three-body study of this type was reported in Ref.\,\cite{Eichmann:2009qa}. 
In the ensuing years, numerous refinements have been made; today, a unified set of predictions exists for pion, kaon, and nucleon electromagnetic form factors \cite{Yao:2024drm, Yao:2024uej}, which compare favourably with available data and offer predictions that can be tested at modern high-luminosity, high-energy facilities.

As noted in Sec.\,\ref{sec.intro}, the origin of the nucleon mass, $m_N$, is one of the most fundamental unanswered questions in science today. 
It can be addressed by studying the expectation value of the QCD 
energy--momentum tensor in the nucleon, which yields the following current:
\begin{align}
m_N \Lambda_{\mu\nu}^{Ng}(Q)  & = - \Lambda_+(p_f)
\left[
K_\mu K_\nu A(Q^2)
 + i K_{\left\{\mu\right.}\!\sigma_{\left.\nu\right\}}\,\!_\rho Q_\rho J(Q^2)
 + \tfrac{1}{4} (Q_\mu Q_\nu - \delta_{\mu\nu} Q^2) D(Q^2)
\right]
\Lambda_+(p_i)  \,,
\label{EMTproton}
\end{align}
where
$p_{i,f}$ are the momenta of the incoming/outgoing nucleon, $p_{i,f}^2=-m_N^2$,
$K=(p_i+p_f)/2$, $Q=p_f-p_i$;
all Dirac matrices are standard~\cite[Sec.\,2]{Roberts:2000aa}, with $\sigma_{\mu\nu}= (i/2)[\gamma_\mu,\gamma_\nu]$;
$\Lambda_+$ is the projection operator that delivers a positive-energy nucleon;
and $a_{\left\{\mu\right.}\!b_{\left.\nu\right\}}=(a_\mu b_\nu + a_\nu b_\mu)/2$.

Importantly, in addition to the nucleon mass distribution form factor, $A$, such studies also provide access to form factors associated with the distribution of total angular momentum, $J$, and in-nucleon pressure and shear forces, $D$. 
In the forward limit, $Q^2=0$, symmetries entail $A(0)=1$, $J(0)=1/2$.
$D(0)$ is also a conserved charge, but, like the nucleon axial charge, $g_A$, its value is a dynamical property.
The value of $D(0)$ has been described as the last unknown global property of the nucleon \cite{Polyakov:2018zvc}; hence, its determination is a focus of considerable attention.

\begin{figure}[t]

\centerline{\includegraphics[width=0.65\columnwidth]{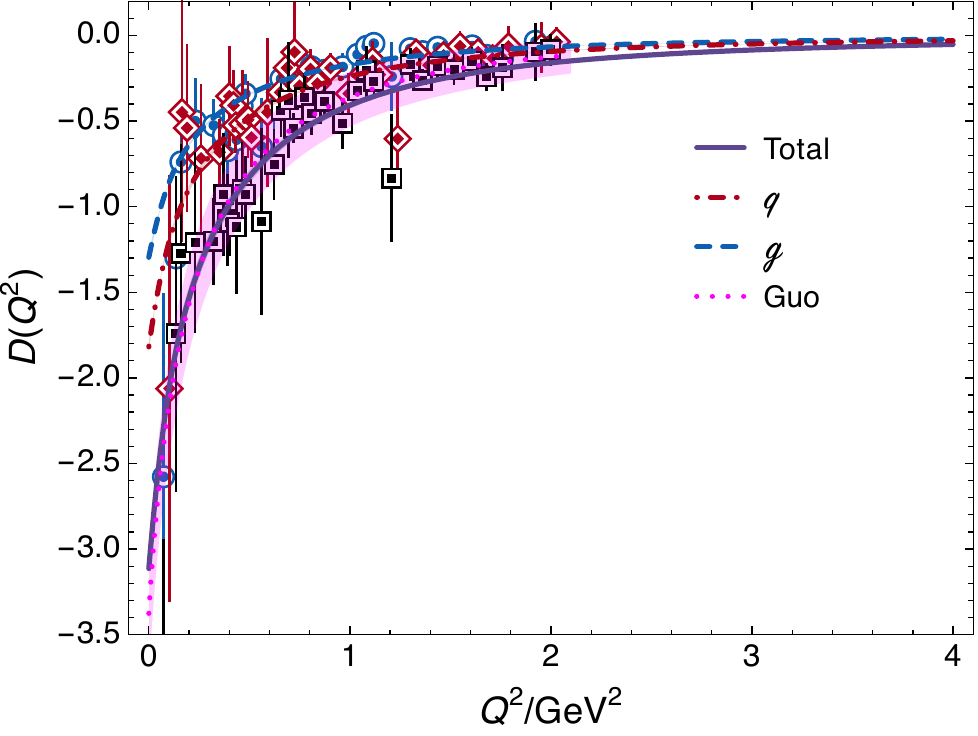}}

\caption{\label{FigGFFD}
Nucleon gravitational pressure form factor; see Eq.\,\eqref{EMTproton}.
Curves: CSM predictions \cite{Yao:2024ixu}; bracketing bands indicate the extent of the $1\sigma$ statistical uncertainty in the numerical analysis.
The value of the ``last unknown global property of the nucleon'' is predicted to be $D(0) = -3.11(1)$.
Displayed as the dotted magenta curve within a like-coloured uncertainty band, a dispersion relation analysis, informed by $\pi\pi$ and $K\bar{K}$ scattering data, yields a consistent result \cite{Cao:2024zlf}: both pointwise and in the value of $D(0) = -3.38^{+0.34}_{-0.35}$.
While the total form factor is independent of the resolving scale, $\zeta$, the species decompositions evolve with $\zeta$.
The LQCD points are reproduced from Ref.\,\cite{Hackett:2023rif}:
black squares --- total form factor;
red diamonds --- quark component; 
blue circles --- glue.
At $\zeta = 2\,$GeV, shown in this image, the quark contribution to both the form factor and the radius exceeds that of the glue.
}
\end{figure}

Following an exploratory study using lattice-regularised QCD \cite{Hackett:2023rif}, CSM predictions for all three 
gravitational form factors have recently become available \cite{Yao:2024ixu}. The pressure form factor is displayed 
in Fig.\,\ref{FigGFFD}. With systematic errors considered and controlled, statistical errors associated with numerical 
algorithms small, and in unifying the nucleon gravitational form factors with data-consistent predictions for pion, 
kaon, and nucleon electromagnetic form factors \cite{Yao:2024drm, Yao:2024uej}, the CSM results are fully benchmarked.  
A notable prediction is that the core pressure inside the nucleon is at least an order of magnitude greater than that 
in a neutron star; moreover, that in-pion and in-kaon core pressures are a factor of two larger than that in the 
nucleon \cite{Xu:2023izo}.

Empirical access to gravitational form factors may be provided by measurements of deeply virtual Compton scattering 
(DVCS) or analogous meson production and a reliable interpretation of the resulting data in terms of nucleon GPDs; see, \textit{e.g}., Ref.\,\cite[Sec.\,V]{Polyakov:2018zvc}.
Currently, such data are sparse and the bridge between data and theory involves significant model dependence \cite{Dutrieux:2024bgc}.
Both issues can be addressed by SIS100, which is capable of providing additional data in the ERBL kinematic regime, wherein 
skewness, $\xi$, exceeds the light-front momentum fraction, $x$.
The ERBL domain is inaccessible with lepton beams.
Meanwhile, the CSM prediction for the in-proton near-core pressure is consistent with an inference based on extant 
DVCS data \cite{Yao:2024ixu, Burkert:2018bqq}.

Returning to the nucleon mass problem, key insights may be obtained by considering the $p_f \to p_i=p$ limit of the 
trace of the energy-momentum tensor in the nucleon:
\begin{equation}
 m_N \Lambda_{\mu\mu}^{Ng}(0) = \Lambda_+(p) m_N^2 \,.
 \label{TraceAnomaly}
\end{equation}
Detailed discussions of this and related issues are provided elsewhere \cite{Roberts:2016vyn}.
The basic point, again highlighted in Sec.\,\ref{sec.intro}, is that, absent Higgs boson couplings into QCD, 
the right-hand side of Eq.\,\eqref{TraceAnomaly} must be zero unless the trace of the QCD energy momentum tensor 
exhibits an anomaly, \textit{i.e}., the scale invariance of classical chromodynamics is broken by quantisation.
This is the case, as can most directly be seen through the emergence of a mass scale in the gluon two-point 
Schwinger function (gluon propagator).
That possibility was first elucidated in Ref.\,\cite{Cornwall:1981zr}. Since then, continuum and lattice Schwinger 
function methods have been refined to a point from which this outcome is seen to be indisputable 
\cite{Binosi:2022djx, Ferreira:2023fva, Deur:2023dzc}.
There is no clearer expression of emergent hadron mass: owing primarily to their own self-interactions, the 
massless gluon partons in the QCD Lagrangian come to be described by a mass-scale that is approximately $m_N/2$.

For roughly thirty years, it was imagined that empirical access to the highly nontrivial gluodynamics that underly Eq.\,\eqref{TraceAnomaly} could be provided by data on the near-threshold photoproduction of $J/\psi$ 
mesons from the proton \cite{Kharzeev:1995ij}. 
However, that path assumes validity of vector meson dominance (VMD) in developing the $\gamma + p \to J/\psi + p$ reaction mechanism.
In this connection, the VMD assumption is now known to be untenable \cite{Du:2020bqj, Xu:2021mju, Sun:2021pyw}.
Nevertheless, discarding VMD, some analyses still attempt to draw a connection between $\gamma + p \to J/\psi + p$ and in-proton gluodynamics using GPDs \cite{Guo:2023pqw} or holographic models \cite{Mamo:2022eui}.
Yet, even here, there are counterarguments, which continue to be developed \cite{Du:2020bqj, Sakinah:2024cza, Tang:2024pky}.
Today, one can only conclude that the reaction mechanism which explains $J/\psi$ photoproduction is poorly understood and no objective connections can be drawn between that process and in-proton gluodynamics.
Here, too, SIS100 can make a crucial contribution with new, precise data on $pp$ scattering into final states involving vector mesons. 
With unique kinematics and a clean signal, the data can be critical in developing a sound understanding of reactions that may fairly be considered to proceed via strong gluodynamics. 

With the ability to explore $pp$ scattering into final states involving mesons with both open and hidden charm, SIS100 can also contribute significantly to resolving the puzzle surrounding intrinsic charm in the proton \cite{Brodsky:1980pb}; namely, whether the proton has a non-negligible $c+\bar c$ content in excess of that which may be expected based on QCD evolution equations \cite{Dokshitzer:1977sg, Gribov:1971zn, Lipatov:1974qm, Altarelli:1977zs} and whether the profile of the associated PDF has 
significant support on the valence quark domain, $x\gtrsim 0.2$.

A recent global fit to data \cite{Ball:2022qks} has provided evidence at the $2.5\sigma$ level of significance in support of the intrinsic charm hypothesis; see below.
However, the outcomes reported are influenced by practitioner-method bias and, by usual standards, the level of significance is well below that required to support a discovery claim. 
Furthermore, the momentum fraction found to be lodged with $c+\bar c$ is readily explained by evolution \cite{Lu:2022cjx, Yu:2024qsd}. 
Thus, the questions surrounding intrinsic charm remain open.  
Here, too, SIS100 and CBM can deliver critical new information owing to the excellent coverage of both $pp$ and $pA$ inclusive reactions at forward and mid-rapidity where, should it exist, intrinsic charm is predicted to have the greatest influence.

Existing theory predictions for the proton's $c+\bar c$ content \cite{Lu:2022cjx, Yu:2024qsd} were obtained using a quark + dynamical diquark simplification of the nucleon Faddeev equation.  
A modern perspective on diquark (quark + quark) correlations is provided elsewhere \cite{Barabanov:2020jvn}.
It is nevertheless worth mentioning that QCD-connected predictions for diquark properties show that such correlations must be nonpointlike, fully dynamical, \textit{viz}.\ subject to continual breakup and reformation and interacting with all quark-sensitive probes, and characterised by mass scales that are heavier than their natural mesonic analogues.
Moreover, owing to the character of SU$(3)$-colour, baryons are the most likely systems within which diquarksmight play a dominant role, see also the discussion in Sec.\,\ref{sec:baryonspectra}.   
In putative four- and five-body systems, \textit{i.e}., exotic or hybrid mesons and pentaquarks, (molecule-like) composites of colour singlet states can exist.
They are potentially more likely because diquark subcomponents would be heavier; thus, diquark subclusters would only be favoured in exceptional circumstances; see, \textit{e.g}., Ref.\,\cite{Hoffer:2024alv}.

\smallskip

\noindent{\bf Model approaches to intrinsic charm}.
As already noted, the existence of intrinsic charm in the nucleon was postulated more than forty years ago \cite{Brodsky:1980pb}; yet despite many attempts, it has not been unambiguously confirmed experimentally. 
It was argued recently \cite{Maciula:2021orz, Goncalves:2024elt} that including intrinsic charm PDFs can improve the description of the LHCb cross sections for $D^0 + {\bar D}^0$ production in fixed-target p + $^{4}$He 
(at $\sqrt{s} = 86.6$\,GeV), p + $^{20}$Ne (at $\sqrt{s} = 68.5$\,GeV), and p + $^{40}$Ar (at $\sqrt{s} = 110.4$\,GeV) collisions~\cite{LHCb:2018jry,LHCb:2022cul}. 
Nuclear effects for $^{20}$Ne and even $^{40}$Ar should not be too large, but they are not fully under control and this introduces material uncertainties.

High-energy neutrino data from the IceCube experiment \cite{IceCube:2017cyo} are consistent with a small component of $c \bar c$ in the nucleon wave function \cite{Goncalves:2021yvw}. 
Global fits to world data can also accommodate the existence of an intrinsic charm component at the level of 1\% \cite{Guzzi:2022rca,Ball:2022qks}. 
It has been shown that measurement of the $\nu_{\tau}$ neutrino in the forward direction in proton--proton collisions in collider mode may provide valuable new information on intrinsic charm in the nucleon \cite{Maciula:2022lzk}. 
In that case, the recombination contribution is expected to be small.

\begin{figure}[t]
    \centering
    \includegraphics[width=0.5\linewidth]{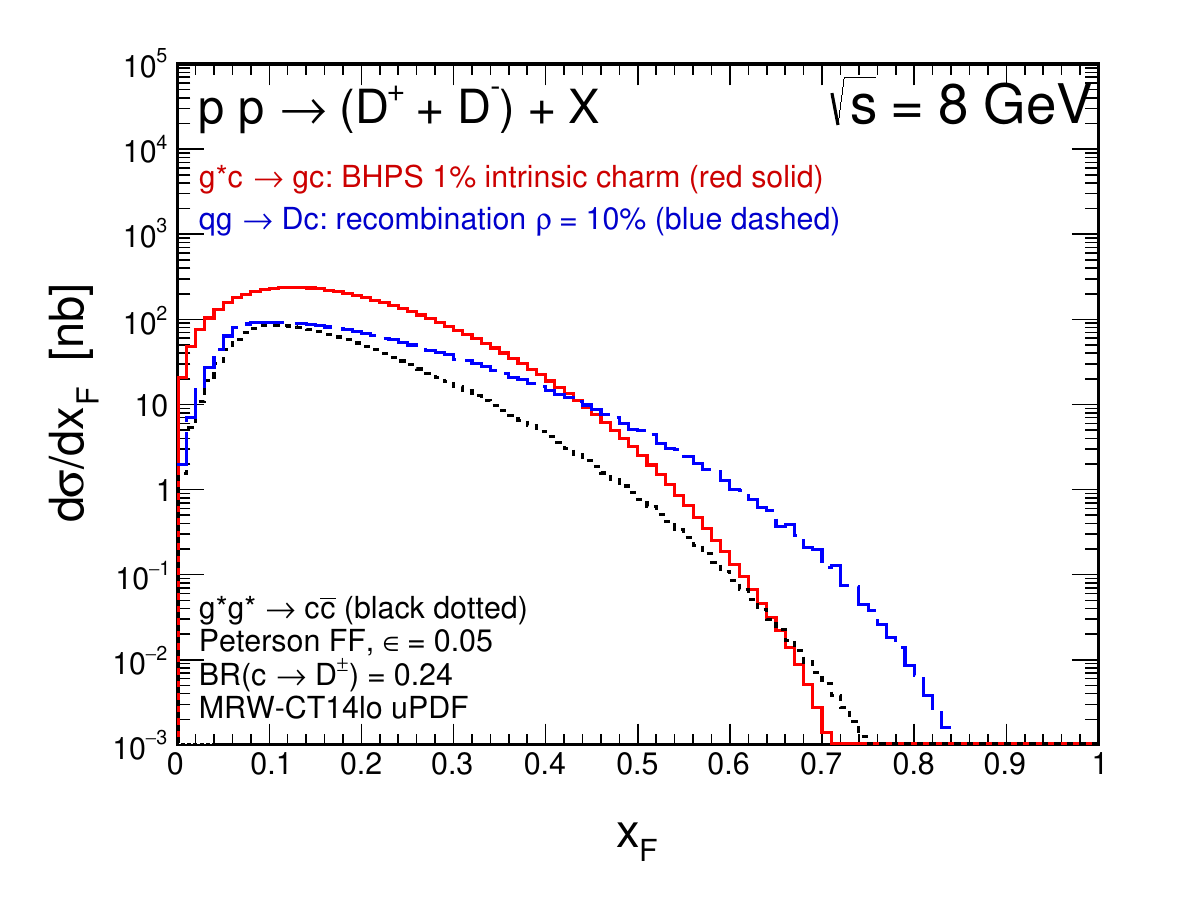}
    \vspace*{-0.5cm}
    \caption{Differential cross section for charged open charm meson production in $pp$ collisions at $\sqrt{s} = 8$\,GeV, shown as a function of the meson Feynman-$x$ ($x_F$). Different contributions from various charm production mechanisms are displayed separately. Details are specified in the figure.
    \label{Fig-Dmes-xF-8GeV}}
\end{figure}

Intrinsic charm, in the form of higher Fock components of the nucleon wave function, may lead to increased production of $D$ and $J/\psi$ mesons, detectable using the SMOG device at LHCb \cite{Vogt:2022glr, Vogt:2023plx}. 
Reactions in which intrinsic charm contributions may play a relatively more important role occur at low energies. 
At SIS100 energies, the situation is expected to be similar; see Fig.~\ref{Fig-Dmes-xF-8GeV}.

\begin{figure}[t]
    \centering
    \includegraphics[width=0.32\linewidth]{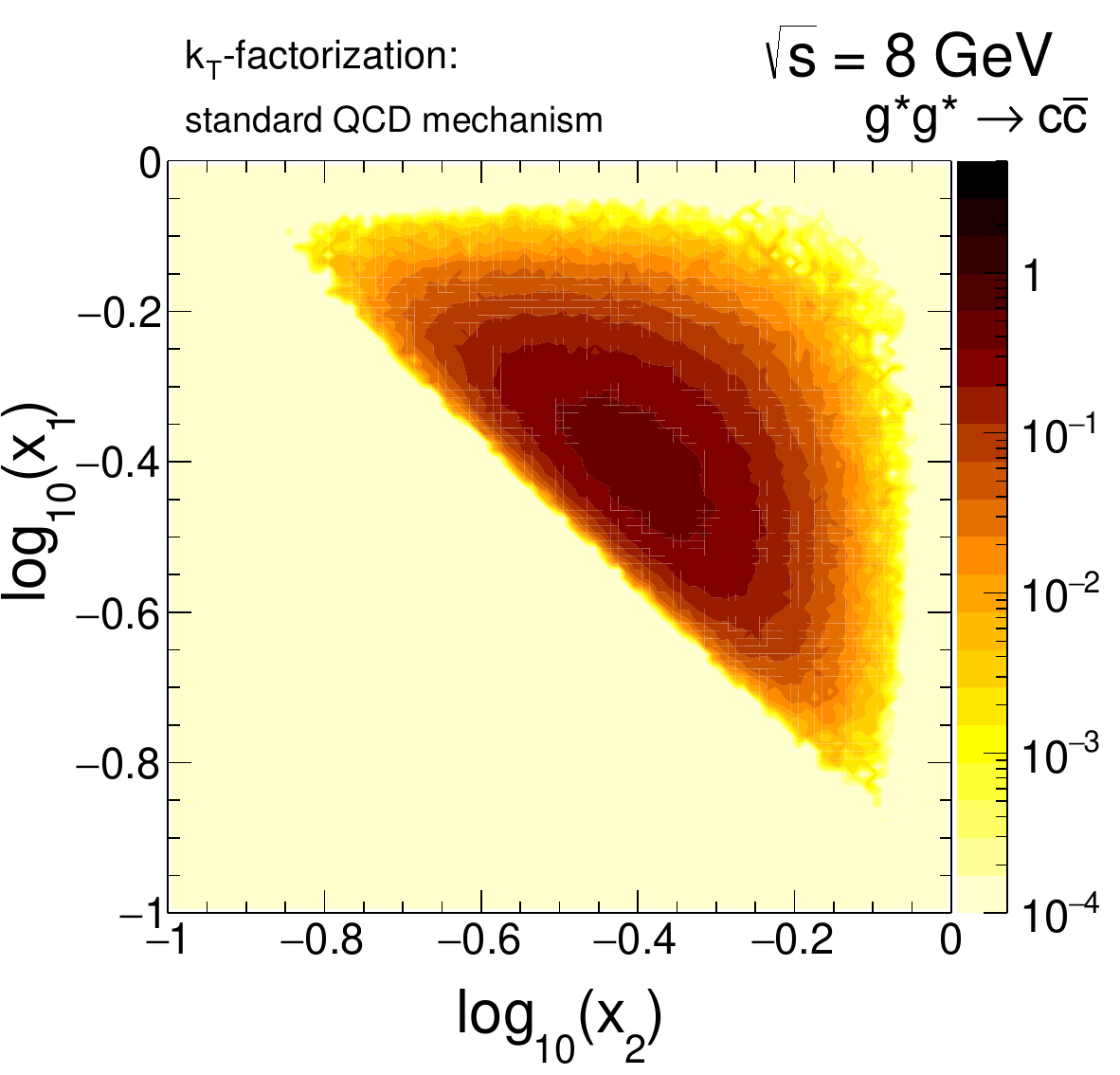}
    \includegraphics[width=0.32\linewidth]{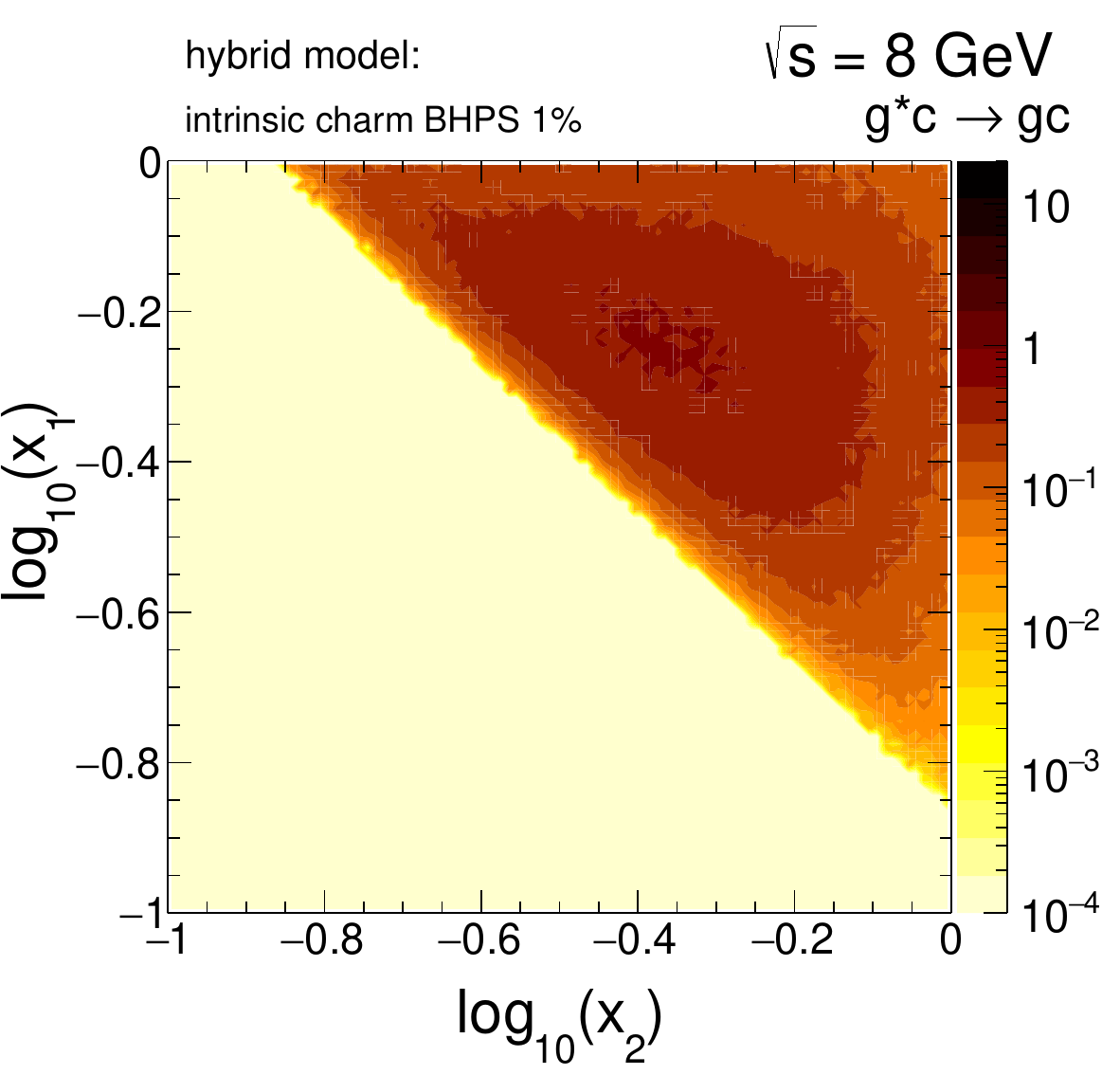}
    \includegraphics[width=0.32\linewidth]{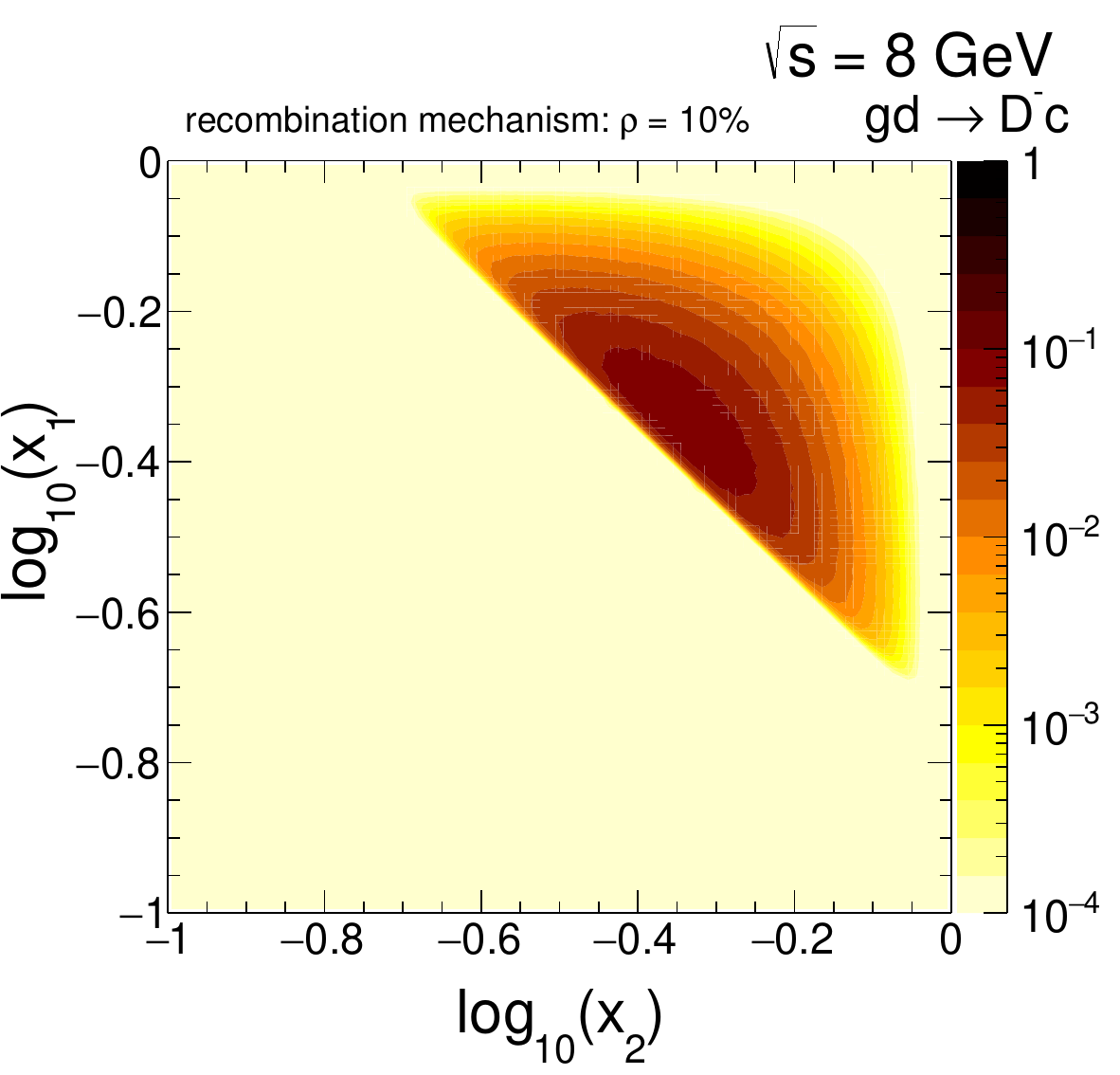}
    \caption{Two-dimensional distributions as a function of the longitudinal momentum fractions $\log_{10}(x_1)$ and $\log_{10}(x_2)$ probed via charm production in $pp$ scattering at $\sqrt{s} = 8$~GeV. The cross-sections for the $g^*g^* \to c \bar{c}$ (left panel), $g^*c \to gc$ with intrinsic charm (middle panel), and recombination $gd \to D^-c$ (right panel) mechanisms are shown separately. Further details are specified in the figure.
    \label{Fig-x1x2-8GeV}}
\end{figure}

This discussion indicates that measurements of $D^0$ and $\bar D^0$ (decaying into $K^{\mp} + \pi^{\pm}$) at SIS100 energies may deliver important new information, which could place tighter limits on any intrinsic charm component of the nucleon. 
As shown in Fig.~\ref{Fig-x1x2-8GeV}, such measurements should also help constrain gluon distributions at larger values of longitudinal momentum fractions ($x \gtrsim 0.3$). 
A detailed knowledge of the gluon PDF at large $x$ is crucial for evaluating the outcomes of searches for new massive particles at the LHC. 
The same applies to $J/\psi$ production \cite{Cisek:2025wnu}. 

It would be valuable to make predictions before the experiments are completed. 
At low energies, one may anticipate a $D^0$--$\bar D^0$ asymmetry, arising either from recombination~\cite{Maciula:2022otw} or from a complex structure of higher Fock components~\cite{Vogt:2022glr, Vogt:2023plx}. None of the existing approaches provides perfect agreement with the SMOG data; therefore, there is room for further improvement.

\subsubsection{Experiment}
\label{ch5:charmproduction}
The production of hadrons containing charm quarks near their production threshold has long been suggested as a process that can probe aspects of nucleon structure, including the distribution of gluons within the nucleon, elements of the nucleon energy-momentum tensor, and the charm-quark content of the nucleon wave function. 
Photoproduction of $J/\psi$ mesons near threshold was for a long time considered sensitive to these effects; however, as previously discussed, this interpretation is now regarded as dubious. 
The near-threshold production of $J/\psi$ in $pp$ collisions may offer an environment more directly connected to nucleon structure. 
Measuring open-charm production near threshold, particularly the production of $D^{(*)0}\Lambda_c^+$, has become a crucial topic owing to the relatively large expected cross section -- roughly an order of magnitude greater than that of $J/\psi$ -- and the potential for contributing to the $J/\psi$ yield via rescattering. 
Measuring both closed- and open-charm production near threshold is essential for uncovering the production mechanisms and identifying any associated observables linked to nucleon structure.



\medskip

\begin{figure}[t]
    \centering
    \includegraphics[width=0.32\linewidth]{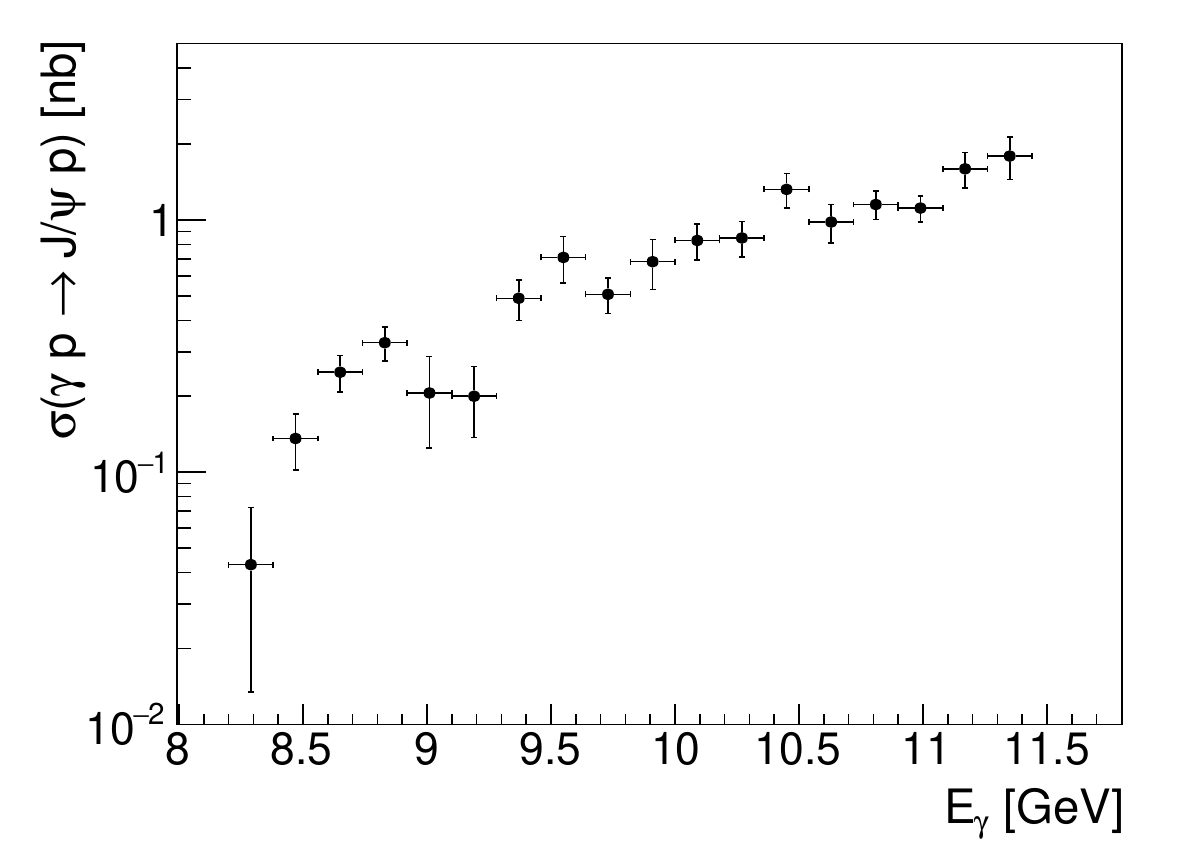}
    \includegraphics[width=0.3\linewidth]{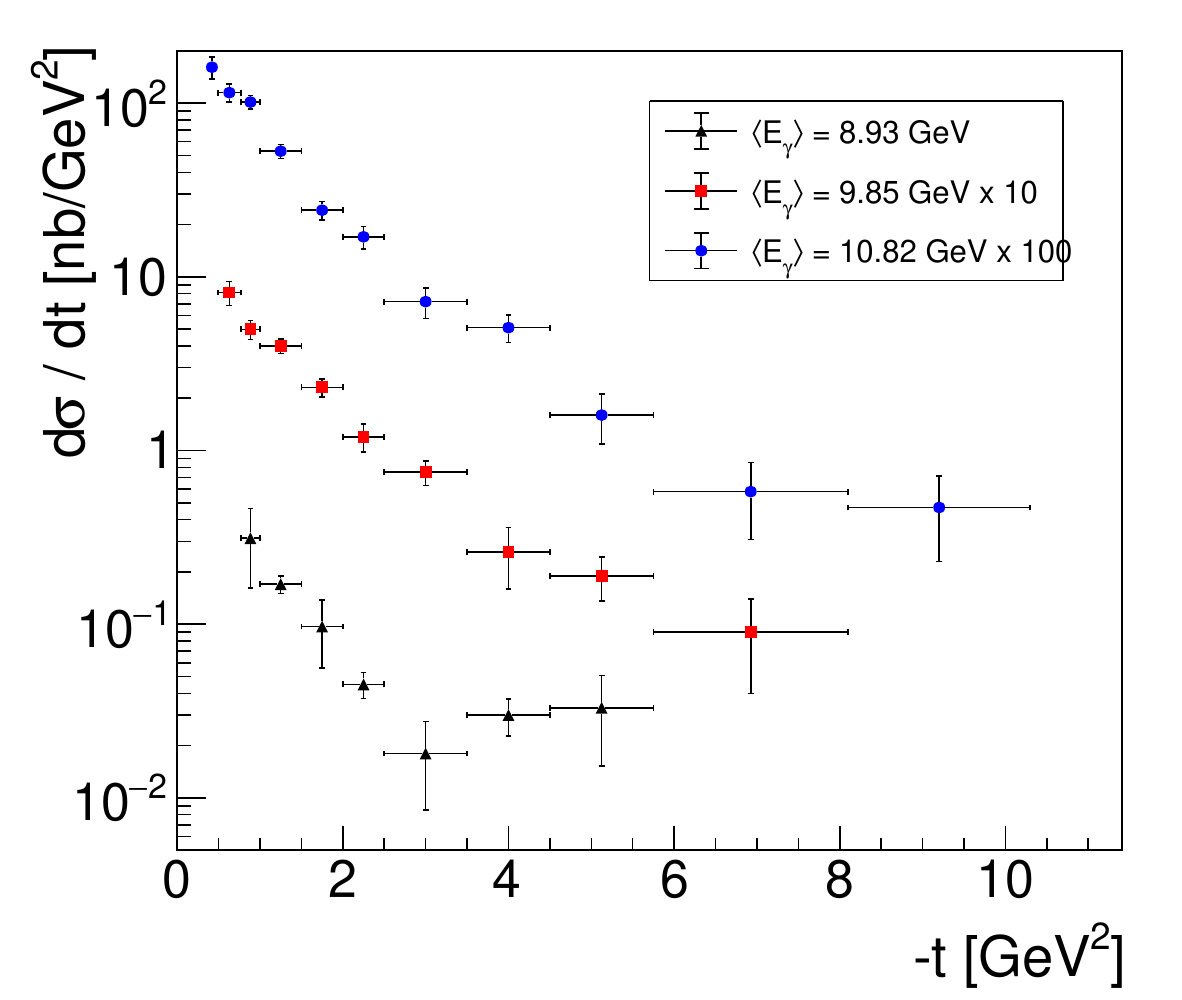}
    \includegraphics[width=0.32\linewidth]{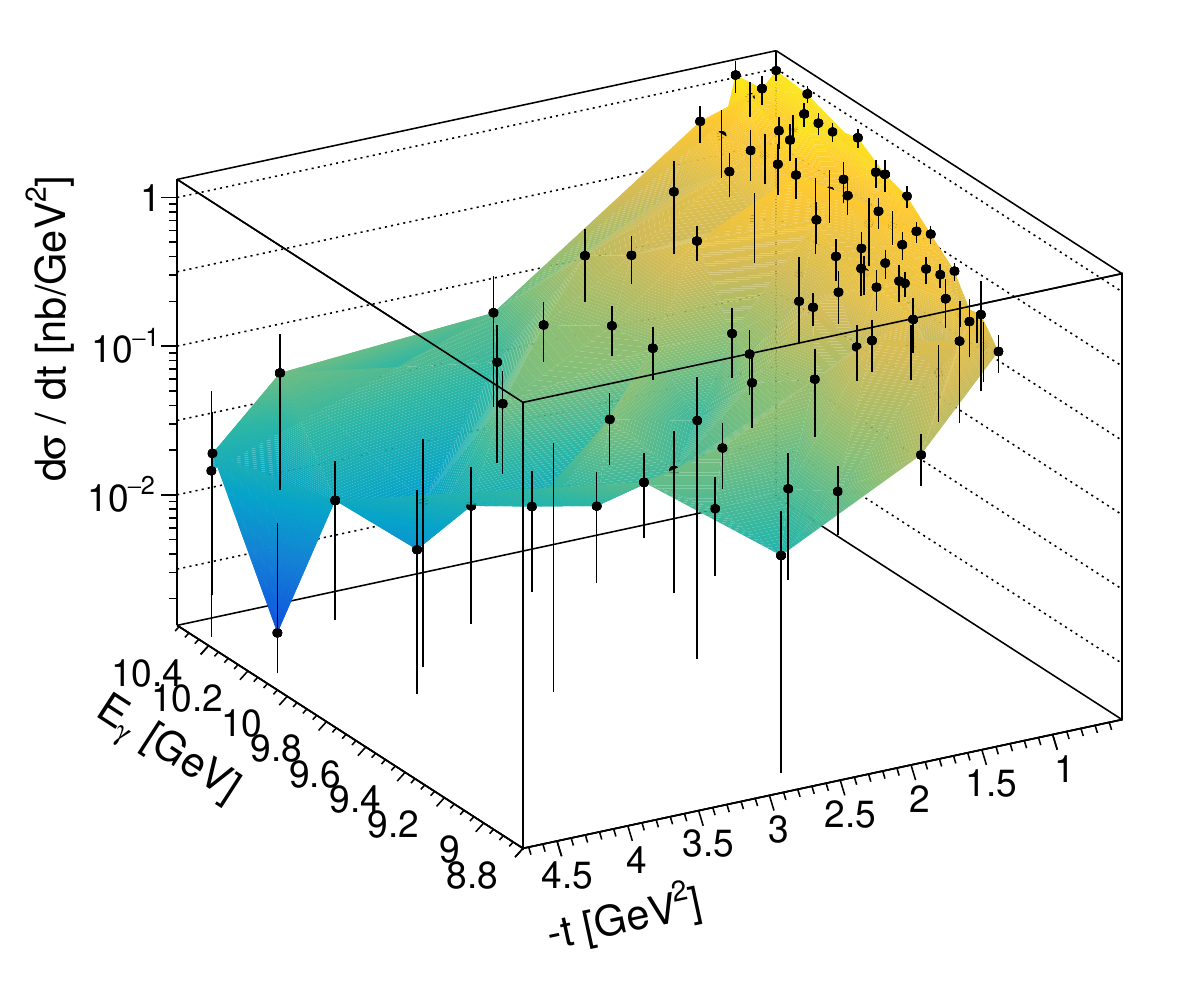}
    \caption{Measurements of $J/\psi$ photoproduction cross-sections at Jefferson Lab: (left) total cross-sections from GlueX~\cite{GlueX:2023pev}; (middle) differential cross-sections from GlueX~\cite{GlueX:2023pev}; (right) differential cross-sections from $J/\psi$-007~\cite{Duran:2022xag}.}
    \label{fig:jlabjpsixsec}
\end{figure}

\noindent{\bf JLab}.
The real or quasi-real photoproduction of the $J/\psi$ meson near its production threshold, $E_\gamma = 8.2\,$GeV, is an integral part of the physics programme of the 12~GeV era at JLab \cite{JLab12GeVProgram}. 
All four experimental halls have completed or planned experiments aimed at measuring $J/\psi$ photoproduction near threshold. 
The GlueX, CLAS12, and $J/\psi$-007 experiments have all collected data on this topic, and the published results are shown in Fig.\,\ref{fig:jlabjpsixsec}.


GlueX has produced two publications on $J/\psi$ photoproduction.  
The first total cross-section measurements were published in Ref.\,\cite{GlueX:2019mkq}; using four times more data, both total and differential cross-sections were reported in Ref.\,\cite{GlueX:2023pev}.
The results are shown in Fig.\,\ref{fig:jlabjpsixsec} (left and middle).
The main features observed include the dominance of diffractive production, with deviations from this behaviour evident at large Mandelstam-$t$, particularly close to the production threshold. 
The total cross section may exhibit some structure near the $D^{(*)}\Lambda_c^+$ thresholds, suggesting a possible contribution from open-charm rescattering to this reaction \cite{Du:2020bqj, JointPhysicsAnalysisCenter:2023qgg}, although only at the $\approx 2\sigma$ level.  
No evidence of the $P_{cc}^+$ states seen by LHCb was found in these measurements, although any upper limits on their production depend strongly on the models used to describe the reaction.

The $J/\psi$-007 experiment used a two-armed spectrometer to detect the $J/\psi$ decay products at specific kinematic points in photon energy and Mandelstam-$t$, enabling high-precision two-dimensional measurements of the $J/\psi$ differential cross-section as a function of $t$ in ten bins of photon energy \cite{Duran:2022xag}. 
These results are shown in Fig.\,\ref{fig:jlabjpsixsec}~(right). 
The GlueX and $J/\psi$-007 measurements agree well within experimental uncertainties. 
The total cross section as a function of beam energy was not released.

Measurements of $J/\psi$ photoproduction at CLAS12 cover the quasi-real photon energy range from threshold up to 10.6\,GeV on several different targets. Current analyses focus on production off deuterium and helium targets. Preliminary results for the total and differential cross sections show overall good agreement with other measurements of near-threshold $J/\psi$ photoproduction at JLab.
  
These cross section data have been analysed to extract quantities such as the effective mass radius of the nucleon and gluonic form factors, although the validity of these interpretations has been questioned; see the discussion following Eq.\,\eqref{TraceAnomaly}. 
One important question to be addressed is the size of the open-charm cross section in this near-threshold region, since the $J/\psi$ could also be produced through the rescattering of open-charm hadron pairs, in addition to normal diffractive production.
 
            

Beyond $J/\psi$ photoproduction, the excited charmonium states up to the $\psi(2S)$ are accessible at GlueX. 
Preliminary mass spectra for the reaction $\gamma p \to \chi_{c1,2} p$, $\chi_{c1,2} \to \gamma J/\psi$ have been presented \cite{Pentchev:2024ffv}. The $\chi_{cJ}$ states have positive charge conjugation and provide a complementary view of charmonium photoproduction compared to the $J/\psi$, which has negative charge conjugation.

The GlueX and CLAS12 experiments are still in the data collection phase. The current GlueX data-taking campaign is expected to conclude in 2026, resulting in at least a threefold increase in the available dataset. 
A high-intensity ``GlueX-III'' run was recently approved \cite{GlueXPhaseIII}. Although not yet scheduled, it promises to deliver an eightfold increase in statistics, along with improved particle identification through the addition of a transition radiation detector. 
The CLAS12 experiments have completed about half of their scheduled beam time. For both GlueX and $J/\psi$-007, in addition to the dielectron decay of the $J/\psi$, the dimuon channel is also under analysis.
        
Beyond the ongoing experimental programmes at CLAS12 and GlueX, the SoLID experiment will provide a next-generation measurement of $J/\psi$ production near threshold \cite{Chen:2014psa}. 
SoLID will employ a large-acceptance spectrometer and an electron beam with an intensity two orders of magnitude higher than that of CLAS12, enabling precision measurements of the photo- and electroproduction of $J/\psi$ for beam energies below 11\,GeV. 
This experiment is not expected to begin data-taking before 2029.

An energy upgrade to the CEBAF accelerator at JLab has also been proposed, raising the current maximum electron beam energy to 22\,GeV \cite{Accardi:2023chb}. 
These energies would allow for precision measurements of the known charmonium states and also permit access to charmonium-like states, such as the $X(3872)$ and $Z_c(3900)$. 
Planning for this upgrade remains in its early stages and is currently anticipated to proceed towards the end of the 2030s.

\smallskip
                

\noindent{\bf J-PARC}.
At this facility, $J/\psi$ production will be measured in $pA$ collisions at a maximal beam energy of 30~GeV, with coverage ranging from mid to backward rapidity. 
Therefore, by measuring the forward region in $pA$ collisions at NA60+ or CBM, it will be possible to access the entire rapidity range, from forward to backward, within this energy regime.

Full-range $J/\psi$ measurements in $pA$ collisions at these energies are considered both important and interesting for the following two reasons.

One important point is the effect of intrinsic charm (IC) on $J/\psi$ production. 
At $\sqrt{s} \sim 8\,$GeV, the contribution from hard gluon-gluon collisions in nucleon-nucleon interactions is rather small -- on the order of a few nb --whereas this process dominates at higher energies. 
On the other hand, $J/\psi$ production from the IC contribution is also expected to be of the same magnitude, even for $P_{\mathrm{IC}} = 0.1\%$ \cite{Vogt:2022glr}.
The impact of IC on $J/\psi$ production is anticipated to be relatively large, particularly in the very forward and backward rapidity regions. This effect is expected to manifest both in the absolute value of the production cross section and in its dependence on the target mass number $A$.

The second point concerns the $J/\psi$-nucleon interaction. 
In the case of $pA$ collisions, the production cross section naturally depends on the final-state interaction with the nucleons in the target nucleus.
When considering the formation time of the $J/\psi$ and its Lorentz boost, the dominant interaction mechanism differs between the forward and backward rapidity regions. 
In the case of forward production, the interaction between the pre-resonant charm quark pair and the nucleons is important, as the $c\bar{c}$ pair traverses the nucleus before forming the $J/\psi$. 
Conversely, in the case of backward production, the fully formed $J/\psi$ meson passes through the nucleus, and thus the $J/\psi$-nucleon interaction becomes dominant.
These effects should manifest as differences in the $A$-dependence of the production cross section between the forward and backward directions.



\smallskip


\noindent{\bf FAIR}.
The production of charm can be investigated at FAIR in both proton-proton and proton-nucleus collisions, from threshold up to $\sqrt{s} = 7.6\,$GeV. 
In particular, the exclusive final states $\bar{D}^0\Lambda_c^+p$ (open charm) with hadronic decays, and $ppJ/\psi$ with dilepton decays (both dielectron and dimuon), can be reconstructed with uniform acceptance over the Dalitz plot.  (This is discussed further in connection with Fig.\,\ref{fig:dalitz_Jpsi_sim} below.)

The main challenge lies in the low production cross sections: on the order of a few tens of nanobarns for open charm, and a small fraction of that for $J/\psi$. 
The CBM experiment is designed to handle high interaction rates of up to 10~MHz and is thus capable of performing such measurements. 
High-statistics measurements of the Dalitz distributions for both reactions, with a mass resolution of $4-6\,$MeV/$c^2$ achievable with CBM, would allow for extraction of the scattering amplitudes of meson-baryon interactions using methods described in Sec.\,\ref{sec:intrinsic_charm}.
This would be of great importance for searches of pentaquark states (see next section) and may also provide access to the trace anomaly, thereby contributing to our understanding of the origin of the proton mass \cite{Krein:2020yor}.

Furthermore, these measurements can be extended to the $pA$ system to study cold matter effects on $J/\psi$ production. 
They can be performed on various targets at forward and central rapidities (CBM), and at backward rapidities using the HADES detector at FAIR (and at J-PARC). 
Cold matter effects can be quantified via nuclear modification factors as functions of rapidity and transverse momentum.

In addition, measurements of the asymmetry in $D^0/\bar{D}^0$ production in $pA$ collisions can provide insights into nucleon PDFs at large $x$, as well as into the absorption mechanisms of $J/\psi$ and $D$ mesons in nuclear matter. 
In the latter case, nuclear effects on nucleon PDFs, such as anti-shadowing and the EMC effect at large $x$, play an important role.

These measurements are also relevant in the context of the IC content of the nucleon, as discussed in the previous section. 
According to model predictions (see, \textit{e.g}., Ref.\,\cite{Vogt:2022glr}), the low-energy domain is particularly well suited for such studies. 
While at LHC energies the effects of IC can only be observed at forward rapidities, at the lowest SPS and highest SIS100 energies the relative IC contributions -- compared with ``standard'' gluon-gluon and quark-gluon charm production processes -- are larger and located around mid-rapidity. 
This leads to a characteristic flattening of the $\bar{D}^0$ and $J/\psi$ rapidity distributions.

Measurements of $J/\psi$ production in $pp$ and $pA$ collisions at low energies at FAIR, J-PARC, and SPS (NA60+), covering a wide range of rapidities, will provide important information on the observable effects of a (possible intrinsic) charm content in the nucleon.

\subsection{Connections to neutrino physics}
\label{Sec:hyp_nu}

The high production rates of hyperons and the versatile detectors at SIS100 make it feasible to study rare decays of hyperons, in particular electromagnetic and semileptonic weak processes.
Both in experiment and in theory, a basic question concerns how the structure of hadrons depends on the current-masses of the quarks and how do these properties evolve when shifting from lighter to heavier quark masses. 
On the experimental side, this can be addressed by comparing hyperons with nucleons and also their respective excitations. 
Such studies should not be restricted to electromagnetic decays but should also include semileptonic weak decays, with the dual purpose of studying the structure of hadrons and providing input for precision electroweak physics. 

The Sakharov conditions \cite{Sakharov:1967dj}, explaining the dynamical origin of the baryon asymmetry of the Universe, indicate that nature contains a significant amount of CP violation. 
Quantitatively, it needs to be larger than the small amount accounted for by the Standard Model of particle physics via the phase in the CKM matrix \cite{Bernreuther:2002uj}. 
The additional CP violation might reside in the quark sector, in the lepton sector, or in both. 
Two search strategies related to the physics discussed herein are the study of strong phases in weak hyperon decays, particularly in $\Xi\rightarrow \Lambda\pi^-$ and $\Omega\rightarrow \Lambda K^-$, detailed in Sec.\,\ref{subsec.MesonBaryonInt}, 
and the comparison of the numbers of neutrinos and antineutrinos emerging from oscillations \cite{T2K:2019bcf}, which is discussed below.



\label{subsubsec:hnu}

To reveal CP violation in the neutrino sector requires counting neutrinos and antineutrinos with high accuracy. 
In turn, this demands a good understanding of neutrino–matter and antineutrino–matter interactions with comparable 
precision. 
In practice, the scattering involves atomic nuclei and is therefore influenced by the physics of nuclei, nucleons, and quarks \cite{NuSTEC:2017hzk}. 
As one concrete example, consider scattering on neutrons at relatively low energies \cite{T2K:2019bcf, Alekou:2022emd}. 
For neutrinos, the relevant scattering events are the quasi-elastic processes $\nu n \to \ell^- p$, where $\ell$ denotes a charged lepton. 
Antineutrinos, on the other hand, lead to the scattering events $\bar\nu n \to  \ell^+ \Sigma^-$, which are Cabibbo suppressed, and to the Cabibbo-allowed but energetically somewhat higher-lying events 
$\bar \nu n \to \ell^+ \Delta^-$ \cite{Singh:2006xp}. 
Both produced baryons, $\Sigma$ and $\Delta$, lead to pion production.
 
Thus, the task for hadron physics is to obtain an understanding of vector and axial-vector transition form factors for $n \to \Sigma$ and $n \to \Delta$ (and more generally of pion production) that is of comparable depth to the understanding of the quasi-elastic form factors $n \to p$. 
On the theory side, this is challenging, particularly for the broad $\Delta$ resonance. 
On the other hand, several theoretical approaches provide predictions of better quality for systems including strangeness. 
Such predictions can be tested by data on weak semileptonic decays of hyperons. 
Of course, by approximate flavour symmetry together with crossing symmetry, these decay processes are related to the neutrino–matter scattering process $\nu N \to e \Delta$.  
Branching ratios for semi-leptonic decays are on the order of $10^{-4} - 10^{-3}$, hence of similar size to those for electromagnetic Dalitz decays.

The $\Omega$ is the long-living flavour partner of the $\Delta$. 
The process $\Omega^- \to \Xi^0 \ell^- \bar\nu_\ell$ has been observed for $\ell = e$, based on a sample of only 14 signal events \cite{BRISTOL-GENEVA-HEIDELBERG-ORSAY-RUTHERFORD-STRASBOURG:1984jku}; therefore, it is almost unknown experimentally.
A kinematically complete large data sample could reveal the interplay of vector and axial-vector transition form factors \cite{Holmberg:2019ltw,Mommers:2022dgw}. 
Another weak decay, $\Omega^- \to \Xi^0(1530) \, \ell^- \bar\nu_\ell$, has not yet been observed. Here, too, differential data would address the interplay of vector and axial-vector transition form factors. 
Flavour and crossing symmetry relate these decays to the form factors of $\Delta$ baryons.  
One gains access to the $\Delta$--$\Delta$--$\pi$ coupling constant via a Goldberger-Treiman relation, \cite{Bertilsson:2023htb}.

In this spirit, the process $\Omega^- \to \Xi^0 \, \ell^- \bar\nu$ \cite{Holmberg:2019ltw,Mommers:2022dgw} is related to the process $\bar\nu n \to \ell^+ \Delta^-$, and the process 
$\Sigma^{*0}(1385) \to \Lambda \, e^+ e^-$ \cite{Sanchis-Alepuz:2017mir,Junker:2019vvy,Ramalho:2023epv} is related 
to the vector part of $\bar\nu n \to \ell^+ \Delta^-$. 
The process $\Sigma^0 \to \Lambda \, e^+ e^-$~\cite{Granados:2017cib,Sanchis-Alepuz:2017mir,Husek:2019wmt,Lin:2022dyu} is related to the vector part of $\nu n \to \ell^- p$. 
Of course, these reactions provide complementary information to the electron-scattering studies of $e^- N \to e^- \Delta$ and the Dalitz decays 
$\Delta \to N \, e^+ e^-$ \cite{Pascalutsa:2006up, Ramalho:2023hqd}. 
For all the three-body decays mentioned, fully differential data, \textit{i.e}., Dalitz distributions, are required to draw conclusions about the corresponding vector and/or axial-vector transition form factors. 
Experimental input (differential decay rates) can serve as guidance for all the theory methods discussed in Sec.\,\ref{subsec:el-transFF-theory}: quark models, DSE/BSE, lattice QCD, and effective field theory (EFT). 
As an example, a combination of lattice and EFT methods might be used for extrapolation to the $N \to \Delta(1232)$ transitions.

Another connection between neutrino-matter scattering and hadron interactions is related to the fact that the axial-vector currents can couple directly to Goldstone bosons when chiral symmetry is spontaneously broken. As a consequence, neutrino-matter scattering probes also induced pseudoscalar (transition) form factors. For the previously discussed example of the $N \to \Delta(1232)$ transitions (and subsequent pion production), this form factor is probed in the spacelike region by neutrino-nucleon scattering; see e.g. \cite{Unal:2021byi} and references therein. But if one turns to the timelike region of the transition form factors, one will run over the pion pole. This ties the single-pion production (via $\Delta(1232)$) in neutrino-nucleon scattering to elastic pion-nucleon scattering and correspondingly neutrino-nucleus to pion-nucleus scattering. For somewhat larger energy transfers, the production of an intermediate $\Delta(1232)$ resonance might be replaced by higher-lying baryon resonances until one reaches energies where the neutrinos scatter on single quarks instead of whole nucleons or nuclei \cite{NuSTEC:2017hzk}. Such high energy transfers can be relevant for DUNE, but they are only marginally tested at the Tokai-to-Kamioka (T2K) experiment \cite{T2K:2011qtm,T2K:2018rhz} and by the proposed European Spallation Source neutrino Super Beam (ESSnuSB) \cite{ESSnuSB:2023ogw}. Instead, the production of low-lying baryon resonances, particularly the $\Delta(1232)$, and the corresponding transition form factors are very significant. The relation between neutrino physics and pion-nucleon and pion-nucleus scattering will be further addressed in Section \ref{sec.lblref}.
\newpage
\section{Exotic hadrons}
\label{sec.Exotics}

{\small {\bf Convenors:} \it N. Brambilla, S. Dobbs} 

\noindent One of the outstanding questions in the physics of strong interactions is which quark and gluon degrees of freedom are manifested in the observed spectrum of hadrons. 
Beyond the well-known $q\bar{q}$ mesons and $qqq$ baryons, QCD allows for colour-singlet combinations of four, five, or more quarks, as well as ``hybrids'' in which gluon degrees of freedom contribute to defining hadron properties, and even bound states of gluons known as ``glueballs''. 
The search for such ``exotic'' hadrons has been underway for many decades, and evidence for several varieties of such states has been building, particularly since the advent of several high-luminosity experimental facilities in the early 2000s.

\begin{figure}[tbhp]
    \centering
    \includegraphics[width=\linewidth]{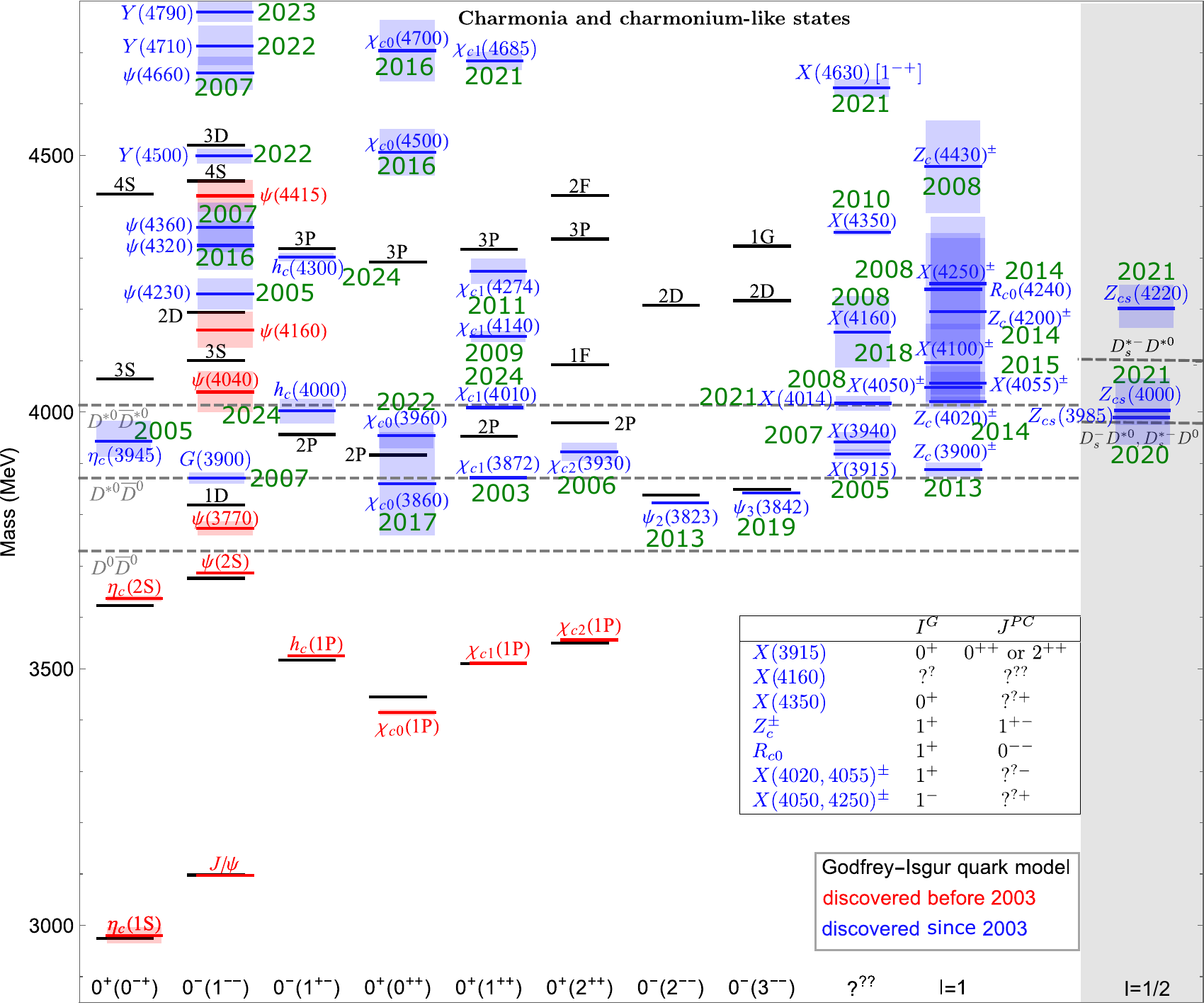}
    \caption{Mass spectrum of charmonia and charmonium-like states. Predictions from the Godfrey-Isgur quark model \cite{Godfrey:1985xj} are denoted by black lines, while structures discovered before 2003 and since 2003 are shown as red and blue lines, respectively. For those discovered since 2003, the year of discovery is also indicated. Figure adapted from Ref.\,\cite{Guo:2023wkv}. }
    \label{fig:charmomiumlike}
\end{figure}

In the light-quark sector, most efforts have focused on searching for hybrid mesons, glueballs, and several multiquark candidates. 
Many candidates for heavy-quark exotic states also exist, which can be classified as either open-flavour or closed-flavour exotics. 
The closed-flavour candidates include heavy-quarkonium-like states and hidden-charm $P_c$ pentaquarks. 
For open-flavour exotics, candidates with one or two charm quarks have been identified. 
Exotic mesons containing a heavy quark and a heavy antiquark are generally referred to as XYZ states.
Figure~\ref{fig:charmomiumlike}, shows the spectrum of charmonia and charmonium-like states observed in experiments, compared to the charmonium spectroscopy predicted by the Godfrey--Isgur quark model \cite{Godfrey:1985xj}.

Although numerous exotic-hadron candidates have been identified, a systematic understanding of their nature has been difficult to achieve, especially owing to the limited information available for many of the states: they are often observed in only one decay channel and by just one or two experiments. 
In order to more firmly establish the existence and properties of such exotic hadrons, it is imperative to observe them in multiple decay channels and via different production mechanisms. 
The experimental programme envisioned at FAIR has the energy reach, luminosity, and detector capabilities to observe many of these exotic states.

This chapter first reviews some of the recent experimental evidence for exotic hadrons, with a focus on states that can be studied at the FAIR experiments. 
It then examines several frameworks for understanding their structure, considering the light- and heavy-quark sectors separately. Finally, some potential highlights of an exotic hadron spectroscopy programme at FAIR are discussed. 

\subsection{Exotic hadrons in experiments}

\subsubsection{$e^+e^-$ experiments}
\label{ch6:subsec:electron_positron}

With the BEPCII, SuperKEKB and VEPP-2000 colliders, there are currently three operating machines delivering $e^+e^-$ collisions to experiments with a hadron physics programme: BESIII at BEPCII, Belle II at SuperKEKB, and SND and CMD-3 at VEPP-2000. 
The three machines and their experiments work in different regions of CM energy: BESIII is collecting data in the $\tau$-charm region with an energy reach from $2-5\,$GeV, Belle II collects data in the bottomonium region largely on the $\Upsilon(4S)$ resonance, and SND and CMD-3 collect data at low energies between 0.3\,GeV and 2\,GeV. 
Past experiments whose data remain relevant to the field include KLOE, operating on the $\phi$ resonance, CLEO, operating in both the charmonium and bottomonium regions, and the $B$-factories BABAR and Belle.

In all cases, the main mechanism to produce hadrons in $e^+e^-$ experiments is annihilation into a virtual photon, $e^+e^-\to \gamma^* \to q\bar q$, with $q = u, d, s, c, b$ depending on the available CM energy. Moreover, initial-state radiation (ISR) allows access to lower effective CM energies through the radiative-return process, in which one or more photons are emitted by the incoming electron or positron, reducing the invariant mass available to the hadronic system. As such, $e^+e^-$ machines are perfectly suited to study vector mesons, conventional and exotic alike, that can be resonantly produced in the collision. 
Hadrons with $J^{PC}$ other than $1^{--}$ are more difficult to produce, but can still be studied in $e^+e^-$ experiments by other means: two-photon collisions, $\gamma^{(*)} \gamma^{(*)} \to \textrm{hadrons}$, give access to $C=+$ resonances, with the two-photon coupling and form factor being indicators of the nature of a resonance. 
Large samples of $J/\psi$ and $\psi(2S)$ mesons at BESIII allow for detailed studies of hadrons composed of the light $u$, $d$, and $s$ quarks in radiative and hadronic decays. 
Similarly, transitions from vector charmonium-like states give access to, \textit{e.g}., the $\chi_{c1}(3872)$ and $Z_c(3900)$. 
In addition, hadronic weak decays of $D$ and $B$ mesons, copiously produced in $e^+e^-$ collisions above the respective open-flavour thresholds, allow the study of hadron resonances with well-defined quark flavours.

\smallskip

\noindent {\bf Light quark exotics}.
A good example of a potentially exotic light-quark state directly produced in $e^+e^-$ annihilation is the $\phi(2170)$, first discovered by BABAR in the process $e^+e^- \to \gamma_\textrm{ISR} \, \phi f_0(980)$ \cite{BaBar:2006gsq}. 
It has since been confirmed in multiple final states by BABAR, Belle, and BESIII in the direct annihilation process \cite{BaBar:2007ptr, BaBar:2007ceh, Belle:2008kuo, BaBar:2011btv, BESIII:2018ldc, BESIII:2017qkh, BESIII:2021yam, BESIII:2020vtu, BESIII:2020gnc, BESIII:2021bjn, BESIII:2020xmw, BESIII:2022wxz, BESIII:2021aet}, and also in the hadronic decay $J/\psi \to \eta \phi(2170)$~\cite{BES:2007sqy,BESIII:2014ybv}. 
Hypotheses proposed to explain the $\phi(2170)$ include a conventional $s\bar{s}$ state, but also a $\bar{s}sg$ hybrid meson, a tetraquark configuration, or a $\Lambda \bar{\Lambda}$ bound state. 
For a comprehensive list of interpretations and references, see Ref.\,\cite{BESIII:2021aet} and sources cited therein.

Radiative $J/\psi$ decays have long been considered particularly gluon rich and thus an ideal hunting ground for glueballs and hybrid mesons. 
A recent analysis of the decay $J/\psi \to \gamma \eta \eta^\prime$ \cite{BESIII:2022riz,BESIII:2022iwi}, based on the world-record dataset of 10 billion $J/\psi$ decays collected with the BESIII experiment, revealed a new spin-exotic resonance, $\eta_1(1855)$, with $J^{PC}=1^{-+}$, which could be the isospin-singlet partner of the hybrid meson candidate $\pi_1(1600)$. 
This state has yet to be confirmed by another experiment or observed in a different decay or production mode.

In addition, analyses of pseudoscalar mesons produced in radiative $J/\psi$ decays have yielded several intriguing results. 
The $X(1835)$ was first observed as a near-threshold enhancement in the radiative decay to a $p\bar{p}$ pair \cite{BES:2003aic}, with a corresponding peak in the reaction $J/\psi \to \gamma \eta^\prime \pi\pi$ exhibiting a lineshape with a sudden drop right at the $p\bar{p}$ threshold \cite{BES:2005ega, BESIII:2010gmv, BESIII:2016fbr}. 
At larger $\eta^\prime\pi\pi$ invariant masses, the same spectrum reveals three additional broad peaks, corresponding to the $X(2120)$, $X(2370)$, and $X(2600)$ \cite{BESIII:2016fbr}, with the $X(2370)$ recently proposed as a candidate for the pseudoscalar glueball. 
More information is needed to substantiate such a claim.

A final example of the significance of $e^+e^-$ experiments for light-quark exotic hadrons is the $a_0(1710)$. 
It was discovered in a partial-wave analysis of $\eta_c \to \eta \pi\pi$ decays using two-photon fusion data from BABAR \cite{BaBar:2021fkz}. 
An isovector resonance was later confirmed by BESIII using $D_s$ decays: the decays $D_s^+ \to K\bar{K} \pi^+$ show an intensity in the $K\bar{K}$ invariant mass region around 1.7\,GeV that is markedly inconsistent with expectations from isospin symmetry when comparing $K\bar{K} = K^+K^-$ and $K_S^0 K_S^0$ \cite{BESIII:2021anf}. 
This observation is explained by interference between the well-known $f_0(1710)$ and the isovector $a_0(1710)$, with the interference term having opposite sign for $K^+K^-$ and $K_S^0 K_S^0$. 
The observation of a charged partner in $D_s^+ \to K_S^0 K^+ \pi^0$ \cite{BESIII:2022npc,LHCb:2023evz} later confirmed these claims. 
If the $a_0(1710)$ is indeed an isospin-1 partner of the $f_0(1710)$ with a nearly degenerate mass, it would be difficult to retain the $f_0(1710)$ as a scalar glueball candidate.

\smallskip

\noindent {\bf Heavy quark exotics}.
These hadrons were introduced earlier as XYZ states. 
While the PDG now employs a more precise naming scheme, this generic terminology is still sometimes used. 
In both the charmonium and bottomonium regions, potentially exotic vector mesons have been observed directly in $e^+e^-$ annihilation: the $\psi(4230)$ and $\psi(4360)$ in the charmonium region, and the $\Upsilon(10753)$, $\Upsilon(10860)$, and $\Upsilon(11020)$ in the bottomonium region. 
These states are observed in cross sections of the type $e^+e^- \to (c\bar{c})\pi\pi$ or $e^+e^- \to (b\bar{b})\pi\pi$, respectively, where $(c\bar{c})$ or $(b\bar{b})$ denotes a compact charmonium or bottomonium state; see, \textit{e.g}., Refs.\,\cite{BaBar:2005hhc, BESIII:2016bnd, Belle:2019cbt, Belle:2015tbu}. 
The signals associated with these states are much larger than those observed for the conventional $\psi(3773)$ and $\Upsilon(4S)$ resonances. Moreover, they often feature charged charmonium-like ($Z_c$) or bottomonium-like ($Z_b$) intermediate states \cite{Belle:2013yex, BESIII:2013ris, Belle:2011aa, Belle:2014vzn}. 
These properties support their interpretation as exotic hadron candidates.

Outside of the vector states, production of further exotic heavy-quark states in $e^+e^-$ collisions is difficult. 
As discussed, two $Z_c$ states were observed in $e^+e^- \to Z_c \pi$, with decays $Z_c(3900) \to J/\psi\,\pi$, $D^*\bar{D}$ and $Z_c(4020) \to h_c\,\pi$, $D^*\bar{D}^*$ \cite{Belle:2013yex, BESIII:2013ris, BESIII:2013qmu, BESIII:2015ntl, BESIII:2013ouc, BESIII:2014gnk, BESIII:2013mhi, BESIII:2015tix}. 
Charged charmonium-like $Z_c$ states are also seen in weak $B$ decays at the $B$-factories \cite{Belle:2007hrb, Belle:2014nuw, Belle:2013shl} and at LHCb \cite{LHCb:2014zfx,LHCb:2015sqg}, but currently no consistency has been found between the two production mechanisms.

The $\chi_{c1}(3872)$ can be produced at $e^+e^-$ machines in the process $e^+e^- \to \gamma\,\chi_{c1}(3872)$, with a cross section that peaks near the $\psi(4230)$ mass \cite{BESIII:2019qvy}, and in $e^+e^- \to \omega\,\chi_{c1}(3872)$ \cite{BESIII:2024ext}. 
For the latter process, the CM energy dependence of the cross section is not yet precisely known. 
Recently, first evidence for production of the $\chi_{c1}(3872)$ via two-photon fusion was reported by Belle \cite{Belle:2020ndp}.

Many of the other heavy-quark exotic hadron candidates discovered by experiments, such as LHCb, are currently out of reach for $e^+e^-$ experiments, either owing to kinematical constraints or because of the lack of a sizeable transition from a vector meson to a non-vector exotic state. A future Super Tau-Charm Factory \cite{Achasov:2023gey}, with energies up to 7\,GeV and luminosity increased by up to two orders of magnitude compared to BESIII, could change that. In the meantime, the community must rely on other production processes to investigate these exotic hadron candidates.

\subsubsection{Hadron-beam experiments}

With the exception of LHC experiments, which are discussed in Sec.\,\ref{ch6:subsec:lhc} below, recent hadron beam experiments have contributed mostly to the advancement of light-meson spectroscopy by studying diffractive dissociation reactions of secondary high-energy pion or kaon beams on proton or nuclear targets. 
The most prominent experiments of this kind are E852 at BNL, COMPASS at CERN, and VES at IHEP, Protvino. 
They are all fixed-target magnetic forward spectrometers equipped with electromagnetic calorimeters, enabling them to measure a variety of decay channels with final states containing $\pi^\pm$, $\pi^0$, $K^\pm$, $K^0_\text{S}$, $\eta$, and $\eta^\prime$. 
The high beam momenta of \qty{18}{GeV} (E852), \qtyrange[range-units=single]{29}{37}{GeV} (VES), and \qty{190}{GeV} (COMPASS) ensure good kinematic separation of possible baryonic target excitations from the forward-going mesonic final state.

The above experiments extensively studied pion diffraction into various final states, primarily in the search for light-quark spin-exotic mesons, in particular the $\pi_1$ states with $J^{PC} = 1^{-+}$.
%
Up to about 5~years ago, the experimental situation regarding the $\pi_1$ states was plagued by puzzling and seemingly contradictory results. 
One of the biggest puzzles was the observation of two $\pi_1$ states, $\pi_1(1400)$ and $\pi_1(1600)$. 
In particular, the $\pi_1(1400)$ exhibited properties that were difficult to explain: it appeared to decay only into $\eta\pi$ and, compared to most predictions, was significantly lighter than expected. 
Moreover, the $\pi_1(1600)$ was too close in mass to be considered an excitation of the $\pi_1(1400)$. 

This longstanding issue was most likely resolved by a coupled-channel analysis performed by the JPAC collaboration \cite{JPAC:2018zyd}, using $P$- and $D$-wave amplitudes for the $\eta\pi$ and $\eta^\prime\pi$ channels measured by COMPASS \cite{COMPASS:2014vkj}. 
In contrast to most previous analyses, which were based on sum-of-Breit-Wigner-type models, this analysis employed a more advanced approach based on $S$-matrix principles. 
Using only a single resonance pole in the $\eta^{(}{'}\protect\vphantom{\eta}^{)}\pi$ $P$-waves, with pole parameters consistent with those measured for the $\pi_1(1600)$, the JPAC model successfully described the $\eta\pi$ and $\eta^\prime\pi$ partial-wave amplitudes. 
This result was confirmed by another coupled-channel analysis performed by Kopf et al., which also included data from $\bar{p}p \to \pi^0\pi^0\eta$, $\pi^0\eta\eta$, and $K^+K^-\pi^0$ measured by the Crystal Barrel experiment, as well as $\pi\pi$ scattering data \cite{Kopf:2020yoa}. 
Since the issue of the Review of Particle Physics \cite{ParticleDataGroup:2024cfk}, the entry for the $\pi_1(1400)$ was therefore merged into that of the $\pi_1(1600)$.

One of the deepest puzzles surrounding the $\pi_1(1600)$ state was the seemingly conflicting results obtained by two analyses of E852 data on diffractively produced $3\pi$ events. 
The first study performed a PWA of about 250\,000 events and claimed the first observation of the $\pi_1(1600)$ in the $\rho(770)\pi$ decay mode \cite{E852:1998mbq,Chung:2002pu}. 
However, a later analysis performed by Dzierba et~al.\ \cite{Dzierba:2005jg}, using a data sample roughly 20 times larger and an improved PWA model, concluded that the E852 data did not provide evidence for a $\pi_1(1600)$ signal in the $\rho(770)\pi$ channel. 

Only the analyses of the significantly larger COMPASS $\pi^-\pi^-\pi^+$ data sample \cite{COMPASS:2015gxz, COMPASS:2018uzl, COMPASS:2021ogp} could reconcile the E852 results. The sample of $4.6 \times 10^7$ events allowed the PWA to be performed in narrow bins of the squared four-momentum~$t$ transferred from the beam pion to the target proton, and with a model including a much larger set of partial waves.
COMPASS was able to demonstrate that the $\rho(770)\pi$ $P$-wave with $J^{PC} = 1^{-+}$ can only be described by including a resonance characterised by parameters consistent with the $\pi_1(1600)$. 
However, for $t \lesssim \qty{0.5}{GeV^2}$, the $\pi_1(1600)$ signal is masked by non-resonant contributions and only becomes visible at higher $t$. 
Applying the analysis models of the two E852 studies to the COMPASS data \cite{COMPASS:2021ogp} confirmed the finding of Dzierba et~al.\ \cite{Dzierba:2005jg} that the signal claimed by the first E852 result was largely an analysis artefact caused by an overly restrictive PWA model. 
It also showed that the non-observation of the $\pi_1(1600)$ by Dzierba et~al.\ was due to the chosen analysis range $t < \qty{0.53}{GeV^2}$, where the $\pi_1(1600)$ signal is overwhelmed by non-resonant background.

The $\pi_1(1600)$ is now considered an established state \cite{ParticleDataGroup:2024cfk}, and the main focus of current research is to study its couplings to various decay and production channels, and to search for its orbital and radial excitations as well as for its SU(3)$_\text{flavor}$ partner states. 
A recent breakthrough was the discovery of the $\eta(1855)$ with $J^{PC} = 1^{-+}$ by BESIII; see Sec.~\ref{ch6:subsec:electron_positron}.

The PWA of the highly precise COMPASS $\pi^-\pi^-\pi^+$ data also revealed a surprising resonance-like signal with conventional $J^{PC} = 1^{++}$ quantum numbers, the $a_1(1420)$ \cite{Adolph:2015pws}. 
This state is too close in mass and too narrow to be an excitation of the ground-state $a_1(1260)$. 
The signal is also about 20-times smaller than that of the $a_1(1260)$, disfavouring a $q\bar{q}$ interpretation. 
The $a_1(1420)$ is only observed in the $f_0(980)\, \pi$ decay mode and lies close to the $K \bar{K}^*(892)$ threshold. 
In a recent analysis \cite{COMPASS:2020yhb}, COMPASS showed that a triangle singularity \cite{Ketzer:2015tqa, Aceti:2016yeb} describes the $a_1(1420)$ signal slightly better than a resonance hypothesis. 
Triangle singularities could also play a role in the $X$, $Y$, $Z$ states in the hidden-charm sector and other hadronic reactions; see Ref.\,\cite{Guo:2019twa}.

COMPASS also studies kaon diffraction into the $K^-\pi^-\pi^+$ final state \cite{Wallner:2022prx, Wallner:2023rmn}. 
A comprehensive PWA yields the most complete measurement of the strange-meson spectrum from a single analysis so far, covering 11 strange-meson resonances in the mass region from \qtyrange[range-units=single]{1.2}{2.4}{GeV}. 
In addition to improving the resonance parameters for many known states, the results suggest that COMPASS has observed the $K_3$ and $K_4$ ground states for the first time. 
Most interestingly, COMPASS identifies a supernumerary state with $J^P = 0^-$, the $K(1690)$, that lies in mass between the two states predicted by the quark model in the \qty{1.7}{GeV} region, which are also seen in the COMPASS data. 
This makes the $K(1690)$ the first clear candidate for a crypto-exotic pseudoscalar strange meson, \textit{viz}.\ an exotic state with the same quantum numbers as an ordinary quark-model state.

\subsubsection{Photon-beam experiments}
\label{ch6:subsec:exoticphotoproduction}
High-intensity photon beams in the several-GeV energy range have been used for many years to improve knowledge of baryon spectroscopy. 
With the recent upgrade of the CEBAF accelerator at JLab, beams of real and quasi-real photons of up to 12\,GeV are available, enabling the study of diffractive meson photoproduction and the photoproduction of charm-quark hadrons. 
These measurements are complementary to the pion beam experiments described above. 
The spin-1 nature of the photon beam should lead to a different pattern of couplings to the produced hadrons than the spinless pion beam, and studying the transfer of polarisation from the beam to the hadrons can provide additional insight into their nature. 
On the other hand, the multiple photon polarisation states increase the complexity of the partial-wave analyses required to interpret these data.

The GlueX and CLAS12 experiments are the primary contributors to the spectroscopy programme at JLab and have been collecting large sets of photoproduction data, with approved experimental programmes that will extend into the next decade. 
GlueX has excellent acceptance for charged and neutral particles, allowing for the study of a wide range of final states. 
One initial focus is to identify the $\pi_1(1600)$ in its decay to $\eta\pi$ and $\eta^\prime\pi$, similarly to the COMPASS analysis described above \cite{Albrecht:2024qdh}. 
Thus far, upper limits on the $\pi_1(1600)$ have been established \cite{GlueX:2024erj}, and the $a_2(1320)\to\eta\pi^0$ differential cross section has been measured \cite{GlueX:2025kma}, providing the first use of partial-wave amplitudes for polarised meson photoproduction.

Other ongoing analyses include the search for vector hybrid mesons in decays to $\pi\pi$ and $K\overline{K}$, the search for the $\phi(2170)$ in $\phi\pi^+\pi^-$ \cite{Gotzen:2024fal,gluexcollaboration2025searchy2175photoproductioncross}, and the study of vector–pseudoscalar channels, such as $\omega\pi$, $\omega\eta$, and $\phi\eta$, which facilitate the search for several different hybrid mesons and other exotics \cite{Scheuer:2024sad}. 

CLAS12 has excellent particle identification and tracking resolution, and is currently in the midst of collecting data for exotic meson studies. Presently, the focus is on the analysis of final states with multiple charged particles, such as $\pi^+\pi^-\pi^-$ and $K^+K^-\pi^+$.

The threshold for the photoproduction of charm/anticharm-quark pairs is 8.2\,GeV, and is therefore accessible at JLab, as detailed in Sec.\,\ref{ch5:charmproduction}. 
The study of $J/\psi$ photoproduction enables searches for $s$-channel production of the $P_{c\bar{c}}$ states observed by LHCb, as discussed in the next section. 
Although no signal for these states has yet been seen, interpretation of the data has been limited by statistical precision and the need for improved reaction models to more accurately describe near-threshold photoproduction. 
An upgrade of the CEBAF accelerator to an electron beam energy of up to 22\,GeV would allow the study of most charmonium-like exotic states in photoproduction, although, as already noted, this upgrade is not currently expected before the late 2030s \cite{Accardi:2023chb}.

\subsubsection{LHC experiments} 
\label{ch6:subsec:lhc}
There are four large experiments at the LHC, among which the LHCb experiment is dedicated to heavy-flavour physics and has played a major role in heavy hadron spectroscopy studies, while the CMS and ATLAS experiments are also competitive owing to their large datasets. 
At the LHC, exotic hadrons can either be produced directly from proton-proton collisions -- in which, in principle, all species of exotic hadrons with all possible quantum numbers can be generated -- or from $b$-hadron decays, from which most of the exotic hadrons observed at the LHC to date have been identified.
So far, 24 new exotic hadron candidates have been observed by the LHC experiments, all belonging to the category of heavy hadrons. These candidates can be grouped into four categories: 
\begin{enumerate}[itemsep=0.2em, parsep=0pt, topsep=0.3em] 
    \item Charmonium-like tetraquark (or hybrid) states;
    \item Charmonium-like pentaquark states;
    \item Tetraquark candidates containing two charm quarks;
    \item Singly charmed tetraquark states,
\end{enumerate}
which will be discussed in detail below.

\smallskip

\noindent {\bf Charmonium-like tetraquark states}.
A rich set of charmonium-like tetraquark candidates has been discovered in the decay $B^+ \to J/\psi \phi K^+$. 
States of this type, being electrically neutral, may also be candidates for hybrid mesons.

The study of the $B^+ \to J/\psi \phi K^+$ decay mode dates back to 2009, when CDF found evidence for the $\chi_{c1}(4140)$ state \cite{CDF:2009jgo}. In 2014, CMS confirmed its existence with a direct observation \cite{CMS:2013jru}. 
In 2017, the LHCb experiment performed the first full amplitude analysis of this decay \cite{LHCb:2016axx}, identifying four exotic tetraquark candidates in the $J/\psi \phi$ system: the previously known $\chi_{c1}(4140)$, as well as $\chi_{c0}(4500)$, $\chi_{c0}(4700)$, and $\chi_{c1}(4274)$.

Using the full Run~2 dataset, LHCb updated this analysis with significantly increased statistics, revealing several additional resonant contributions \cite{LHCb:2021uow}, namely $\chi_{c1}(4685)$, $X(4630)$, $T_{c\bar{c}\bar{s}1}(4000)^+$, and $T_{c\bar{c}\bar{s}1}(4220)^+$. 
The last two states possess manifestly exotic quark content, $c\bar{c}\bar{s}u$.

A subsequent study of the isospin-symmetric decay mode $B^0 \to J/\psi \phi K_S^0$ established the existence of the isospin partner of $T_{c\bar{c}\bar{s}1}(4000)^+$ \cite{LHCb:2023hxg}. 
In addition, an amplitude analysis of the decay $B^+ \to D_s^+ D_s^- K^+$ revealed a significant threshold enhancement, interpreted as a new resonance, $X(3960)$, which is likely to be an exotic state \cite{LHCb:2022aki}.

In a recent study of $B^+ \to D^{*\pm} D^{\mp} K^+$ decays \cite{LHCb:2024vfz}, several new charmonium-like states were observed in the $D^{*\pm} D^{\mp}$ system. 
Among them, the $\chi_{c1}(4010)$ state is an exotic candidate.

\smallskip

\noindent {\bf Charmonium-like pentaquark states}.
The story of pentaquark states begins with the amplitude analysis of the $\Lambda_b^0 \to J/\psi p K^-$ decay \cite{LHCb:2015yax}, 
which marked a milestone discovery in exotic hadron spectroscopy. 
A narrow pentaquark state, $P_{c\bar{c}}(4450)^+$, and a broader structure, $P_{c\bar{c}}(4380)^+$, were initially observed. 

Four years later, with the addition of Run~2 data, the $P_{c\bar{c}}(4450)^+$ state was resolved into two distinct states, 
$P_{c\bar{c}}(4440)^+$ and $P_{c\bar{c}}(4457)^+$, and a new state, $P_{c\bar{c}}(4312)^+$, was discovered \cite{LHCb:2019kea}. 
Further evidence for a new pentaquark candidate, $P_{c\bar{c}}(4337)^+$, was observed in the analysis of $B_s^0 \to J/\psi p \bar{p}$ decays \cite{LHCb:2021chn}. 

A search for the strange partner of the $P_{c\bar{c}}^+$ states was performed in the $\Xi_b^- \to J/\psi \Lambda K^-$ decay \cite{LHCb:2020jpq}, analogous to the $\Lambda_b^0 \to J/\psi p K^-$ channel but with the $u$ quark replaced by an $s$ quark. 
This revealed evidence for a neutral state, $P_{c\bar{c}s}(4459)^0$, although further confirmation is required.

Two years later, a pentaquark candidate containing an $s$ quark, $P_{c\bar{c}s}(4338)^0$, was discovered in the decay 
$B^- \to J/\psi \Lambda \bar{p}$ \cite{LHCb:2022ogu}, marking the first confirmed observation of a pentaquark state of this type.

\smallskip

\noindent {\bf Tetraquark candidates containing two (or more) charm quarks}.
In 2020, LHCb performed a measurement of the promptly produced $J/\psi$-pair mass spectrum \cite{LHCb:2020bwg}, resulting in the first observation of a four-charmed tetraquark candidate, $T_{c\bar{c}c\bar{c}}(6900)$,
marking another breakthrough in exotic hadron studies.
This observation was later confirmed by both the ATLAS and CMS experiments \cite{ATLAS:2023bft,CMS:2023owd}.

In addition, the CMS experiment reported the observation of a new state, $T_{c\bar{c}c\bar{c}}(6600)$, and evidence for $T_{c\bar{c}c\bar{c}}(7100)$.
The latter was confirmed in an updated study using a significantly larger data sample \cite{CMS:2025xwt}, in which a spin-parity measurement of the $T_{c\bar{c}c\bar{c}}$ states was also performed.
Following this discovery, the mass spectrum of prompt $D^0 D^0 \pi^+$ candidates was studied by LHCb \cite{LHCb:2021vvq,LHCb:2021auc},
leading to the first observation of the doubly charmed tetraquark state $T_{cc}(3875)^{+}$ -- the longest-lived exotic state thus far discovered -- 
which has attracted significant attention from the community.

\smallskip

\noindent{\bf Singly charmed tetraquark states}.
The rich set of $B \to DDh$ decay modes provides plentiful opportunities for exotic hadron spectroscopy studies.
The first high-impact results came from the amplitude analysis of $B^+ \to D^+ D^- K^+$ \cite{LHCb:2020bls,LHCb:2020pxc}, where two states were observed in the $D^- K^+$ system, named $T_{cs1}^{*}(2900)^0$ and $T_{cs0}^{*}(2870)^0$.
These states have a quark content of $cs\bar{u}\bar{d}$, representing the first exotic hadron candidates with four different quark flavours, and are therefore unambiguously exotic.
In a subsequent analysis of the $B^{+} \to D^{*+} D^{-} K^+$ decay \cite{LHCb:2024vfz}, their existence was confirmed.

The decay $B^{-} \to D^{-}D^{0}K_{S}^{0}$ was recently studied \cite{LHCb:2024xyx} to search for $T_{cs1}^{*}(2900)^0$ and $T_{cs0}^{*}(2870)^0$ in the isospin-symmetric $D^{0}K_{S}^{0}$ decay mode.
The $T_{cs0}^{*}(2870)^0$ state was observed, while no evidence was found for $T_{cs1}^{*}(2900)^0$.
Motivated by this observation, the decays $B^0 \to \bar{D}^0 D_s^+ \pi^-$ and $B^+ \to \bar{D}^- D_s^+ \pi^+$ were investigated \cite{LHCb:2022sfr}.
These two modes are isospin symmetric, and two new states -- $T_{c\bar{s}0}^{*}(2900)^{++}$ and $T_{c\bar{s}0}^{*}(2900)^{0}$ -- were observed in the $D_s^+ \pi^+$ and $D_s^+ \pi^-$ systems, respectively.

\subsection{Light quark exotics}

\subsubsection{Lattice QCD studies}
\label{ch6:sec:lqcdlightexotics}

Lattice QCD (LQCD) studies of the spectrum of hadrons involving light quarks and gluons are far too numerous to list individually here. Therefore, only a few recent studies will be recorded. 
Further details on many other investigations can be found in the review articles cited below.

\begin{figure}
\begin{center}
\includegraphics[width=0.48\textwidth]{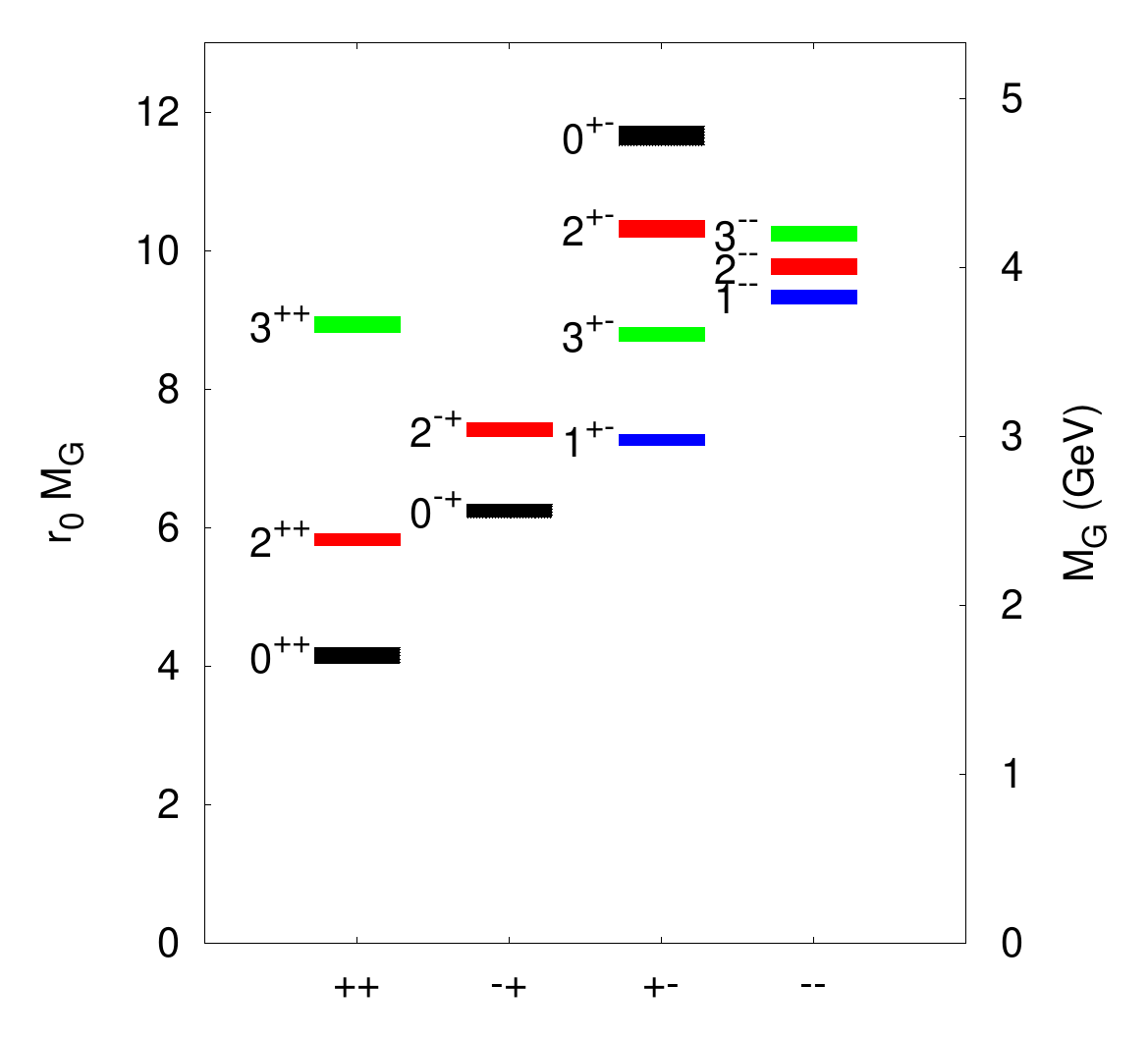}
\includegraphics[width=0.46\textwidth]{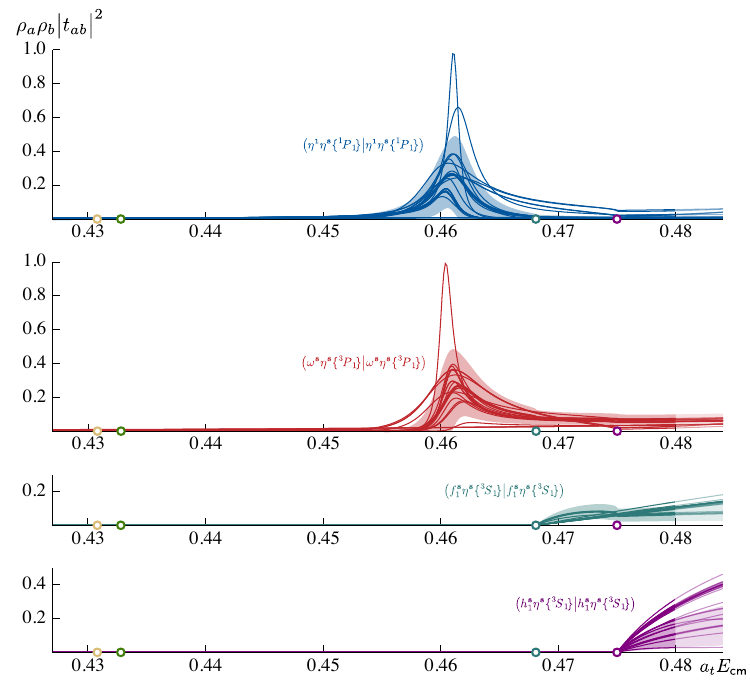}
\end{center}
\vspace*{-0.5cm}
\caption{(Left) The pure-gauge Yang--Mills theory glueball mass spectrum from Ref.\,\cite{Chen:2005mg}, shown in terms of the Sommer parameter $r_0$ and in GeV. 
(Right) Diagonal transition $t$-matrix elements against CM energy times the temporal lattice spacing, plotted as $\rho_a\rho_b|t_{ab}|^2$, where $\rho_{a,b}$ are threshold momentum factors, for each parametrisation successfully describing the finite-volume spectra for non vector-vector channels. 
From top to bottom: $\eta^1\eta^8\{{}^1P_1\},\ \omega^8\eta^8\{{}^3P_1\},\ f_1^8\eta^8\{{}^3S_1\},\ h_1^8\eta^8\{{}^3S_1\}$.
\label{fig:latticelight}}
\end{figure}

The spectrum of glueballs in the pure-gauge Yang--Mills theory has been determined \cite{Morningstar:1999rf, Chen:2005mg} and is shown in the left-hand plot of Fig.\,\ref{fig:latticelight}. 
Such Monte Carlo calculations require a large number of gauge configurations for adequate statistics, but the lack of quark loops in the pure gluon theory makes acquiring such large numbers straightforward with today's computers, and the fact that all the states shown in this plot are stable in the absence of quarks makes the determination of this spectrum fairly easy in LQCD. 
The current status of LQCD computations of glueballs was recently reviewed in Ref.\,\cite{Morningstar:2024vjk}, which shows some evidence that the lowest-lying scalar glueball may not be a well-defined resonance in a theory including quarks.

Studying the hadron spectrum in full LQCD including light quarks is much
more difficult. 
Incorporating the fermion determinant into the Monte Carlo updating and inverting the Dirac fermion matrix in evaluating the needed Wick contractions are computationally expensive, and only a handful of states are stable, with the vast majority being unstable resonances. 

To reliably study resonances in LQCD requires the use of the so-called
L\"uscher method to extract the resonance poles from the scattering $K$-matrix deduced from a large number of finite-volume multi-hadron energies. Because of these difficulties, only a handful of low-lying resonances have been reliably studied in LQCD to date; see, \textit{e.g}., Refs.\,\cite{Briceno:2017max, Prelovsek:2025gmd, Morningstar:2025xmi}). 
An example was recently presented in Refs.\,\cite{BaryonScatteringBaSc:2023zvt, BaryonScatteringBaSc:2023ori}, in which two poles in the energy region of the $\Lambda(1405)$ were found using a large lattice with near-physical pion mass $m_\pi \sim 200$\,MeV.   

Higher-lying resonances, such as those having exotic quantum numbers, have only been studied qualitatively in LQCD on small lattices and at very heavy pion masses. 
A recent example is Ref.\,\cite{Woss:2020ayi}, in which the decays of an exotic $1^{-+}$ hybrid meson resonance were investigated at the SU(3)-flavour-symmetric point with an extremely heavy pion mass 
$m_\pi \sim 700$\,MeV. 
Diagonal transition matrix elements squared for non vector-vector channels are shown in the right-hand plot of Fig.\,\ref{fig:latticelight}. 
Partial widths to 10 decay channels are determined and the $b_1\pi$ is found to be the largest by far.

\subsubsection{Light exotic mesons: hybrids, glueballs, and the $\phi(2170)$ }

\textbf{Lightest hybrid states.} 
As discussed previously, the isovector meson $\pi _{1}(1600)$ with exotic quantum numbers $J^{PC}=1^{-+}$ is well established \cite{ParticleDataGroup:2024cfk}. 
The resonance $\pi_{1}(1400)$ is not considered an independent state, since there is only one pole in the nearby energy region of the complex plane \cite{JPAC:2018zyd}. 
The mass of $\pi_{1}(1600)$ fits well with model and lattice predictions for a hybrid isotriplet state $u\bar{d}g,\, d\bar{u}g,\,$ $(u\bar{u}-d\bar{d})g/\sqrt{2}$ \cite{Meyer:2015eta}. 

The novel exotic ($J^{PC}=1^{-+}$, $I=0$) state $\eta_{1}(1855)$ found by BESIII can be interpreted as being predominantly $s\bar{
s}g$. 
It is then natural to expect a full nonet of $q\bar{q}g$ states, which
includes a light exotic $\eta _{1}(\approx 1660)$ as $(u\bar{u}+d\bar{d})g/\sqrt{2}$, as well as a kaonic member $K_{1}$ ($u\bar{s}g,\, \bar{u}s g,\, d\bar{s}g,\,
\bar{d}s g$) with a mass between $\pi _{1}(1600)$ and $\eta _{1}(1855)$ 
\cite{Shastry:2022mhk,Qiu:2022ktc}; see also Ref.\,\cite{Giacosa:2024epf}. Hybrids are usually broad, but $\eta_{1}(\approx 1660)$ could be narrower. The identification of hybrids, in particular the search for the latter two novel ones, is possible via their decays, in modes such as $\rho \pi$, $b_{1}\pi$, and $a_{1}\pi$.

\smallskip

\noindent\textbf{Lightest glueball states:} The search for glueballs has long been underway; see, \textit{e.g}., Refs.\,\cite{Llanes-Estrada:2021evz,Gross:2022hyw}. 
LQCD shows that the scalar, tensor, and pseudoscalar
glueballs are the lightest (below $\sim 2.5\,$GeV) \cite{Athenodorou:2020ani}.
In all three cases, there are certain resonances that could potentially
contain a substantial glueball component.
(\textit{i}) $f_{0}(1500)$ and $f_{0}(1710)$ in the scalar sector, the latter being favoured by different
approaches \cite{Cheng:2006hu, Janowski:2014ppa, Gui:2012gx}, though the scalar glueball could be smeared across multiple scalar-isoscalar resonances \cite{Klempt:2021wpg}.  Scalar glueballs can be investigated via their decays into pseudoscalar mesons, such as $\pi \pi$ and $\bar{K}K$. 
(\textit{ii}) Various tensor-isoscalar $f_{2}$ states with mass close to 2\,GeV \cite{Klempt:2022qjf}; besides $\pi \pi$ and $\bar{K}K$ decay modes, the tensor glueball may also couplestrongly to $\rho \rho \rightarrow 4\pi$, as suggested by recent models~\cite{Hechenberger:2023ljn, Vereijken:2023jor, Giacosa:2024epf}. 
(\textit{iii}) The previously mentioned resonances $X(2370)$ and $X(2600)$ in the pseudoscalar domain; the pseudoscalar glueball is expected to couple sizeably to the $\eta$ and $\eta^{\prime}$ mesons owing to the chiral anomaly, enhancing the decay channels $\eta^{(\prime)}\pi \pi$ and $\eta^{(\prime)}KK$ \cite{Eshraim:2012jv}.
(An instanton-based estimate of the coupling is presented in Ref.\,\cite{Giacosa:2023fdz}.)

\smallskip

\noindent\textbf{The case of }$\phi (2170)$. 
This resonance is an example of an exotic vector candidate (tetraquark, molecule, or hybrid; see \textit{e.g}., Ref.\,\cite{Zhao:2019syt}). 
The most recent ``quark model summary'' in Ref.\,\cite[RPP]{ParticleDataGroup:2024cfk} identifies it as a D-wave orbitally excited $\bar{s}s$ meson and partner of the well-established $\omega(1650)$, yet this interpretation may be disputed because the mass seems to be at least 200\,MeV too large: the mass difference between the
two isoscalar states amounts to about 500\,MeV, whereas it is usually about
200\,MeV for other nonets, and it is also significantly larger than the quark-model results \cite{Godfrey:1985xj}. 
No $\bar{K}K$ decay mode has been seen, contrary to model predictions \cite{Piotrowska:2017rgt} (the nonet partners 
$\omega(1650)$, $K^{\ast}(1680)$ show strong two-pseudoscalar
decays). 
The search for $\bar{K}K$ and $KK^{\ast}$ modes of the resonance $\phi(2170)$ can help to clarify its character.

\smallskip

\noindent\textbf{Feasibility of the search for light exotic mesons}. 
The study of meson resonances below $\sim 2.5\,$GeV, such as those listed above, requires the analysis of reactions of the type $pp \rightarrow ppK^{+}K^{-}$, $pp \rightarrow pp\pi^{+}\pi^{-}$, \ldots; thus a necessary step is the clear identification of both charged pions and kaons together with their separation. 
These processes and the related analyses are challenging for the $pp$ scattering discussed in Sec.\,\ref{sec.Tools} because of the rather involved final state, the difficulty of the related PWA, and the broadness of some intermediate meson resonances. 
The use of a pion beam instead of a proton beam would considerably improve the search for light exotics.
An antiproton beam represents the best option for accessing gluon-rich states, such as glueballs and hybrids. 
In particular, the direct formation of some exotic states, \textit{e.g}., glueballs, in $\bar{p}p$ annihilation processes would also facilitate such searches.

\smallskip

\noindent{\bf Other four-quark states}.
In the previous section, the $\phi(2170)$ was highlighted as a potential candidate for an exotic meson that can be studied with the SIS100 proton beam using the CBM detector. 
In fact, considering this meson as a potential four-quark state with quark content $s\bar{s}l\bar{l}$ (with $l \in \{u,d\}$), there are many more states in flavour $SU(3)$ multiplets with quantum numbers $J^{PC} = 0^{-+},\ 1^{--},\ 1^{-+},\ 1^{++}$ that could be produced. 
From a theoretical viewpoint, it is therefore important to determine the mass pattern of these multiplets, the internal content of such states, \textit{e.g}., in terms of potential two-body clusters, and their decay patterns.

Which types of internal structure can be expected, based on, for example, existing observations of heavy-quark exotics (which are discussed in more detail in the following section)?  This is particularly relevant for four-quark states containing strange quarks, which sit near the transition between heavy and light hadronic degrees of freedom.  Establishing the existence or non-existence of such states would shed light onto this aspect of the strong interaction.

For simplicity, one may first consider a four-quark state with two heavy quarks and two light quarks.  
For hidden-flavour states, the quarks and antiquarks could arrange themselves into tightly bound diquark–antidiquark pairs $(Qq)(\bar{Q}\bar{q})$ interacting via colour forces \cite{Esposito:2016noz}. Alternatively, the heavy quark and antiquark could cluster together into a compact core surrounded by a light $q\bar{q}$ pair, forming a $(Q\bar{Q})(q\bar{q})$ configuration, which is the hadro-quarkonium picture \cite{Voloshin:2007dx}. 
Or the state could cluster into two heavy–light meson components $(Q\bar{q})(q\bar{Q})$, which is particularly relevant for states close to meson–meson thresholds, then corresponding to a meson–molecule picture \cite{Guo:2017jvc}.  

For open-flavour states, the possible clusters are different owing to their quark content: the heavy quarks could form a diquark pair bound to an antidiquark, $(QQ)(\bar{q}\bar{q})$, or the state could consist of two degenerate meson–meson configurations $(Q\bar{q})(Q\bar{q})$.  
Which of these configurations is realised in Nature is a dynamical question. 
In general, these possibilities are also not mutually exclusive: in principle, every experimental state may be a superposition of components with different structure, where the leading component may differ on a case-by-case basis.  
If the quantum numbers allow it, substantial mixing effects with `ordinary’ quark–antiquark mesons may also occur.
In principle, the same types of configuration may occur in the light-quark sector with two strange quarks present in the four-quark 
state, \textit{i.e}., with quark content $s\bar{s}l\bar{l}$ (hidden flavour) or $ss\bar{l}\bar{l}$ (open flavour). 


An important question to ask is then what may be expected from theoretical approaches that can describe the expected spectrum of such states, such as LQCD or the DSE/BSE approach, before the start of SIS100.
LQCD has focused mainly on heavy-light four-quark states so far, with results summarised, \textit{e.g}., Ref.\,\cite{Francis:2024fwf}. 
The double-strange sector is less explored so far.

The situation is somewhat similar in the DSE/BSE approach. 
Here, one solves four-body Bethe-Salpeter equations with input generated by solutions of Dyson-Schwinger equations; see Sec.\,\ref{sec:functional-methods}. 
The approach has already been applied to the lightest four-quark 
systems, \textit{i.e}., light and strange four-quark states with scalar quantum numbers \cite{Eichmann:2015cra}, and has recently been generalised 
and advanced to a quantitative level in applications within the heavy-light quark sector \cite{Hoffer:2024fgm, Hoffer:2024alv}; see also Ref.\,\cite{Eichmann:2025tzm}. 
Since the approach is equally well applicable in the light and heavy quark sectors, then, in principle, there is no impediment to the exploration of the above-mentioned flavour $SU(3)$ multiplets with quantum numbers $J^{PC} = 0^{-+}, 1^{--}, 1^{-+}, 1^{++}$. Such studies are underway.

\subsection{Heavy quark exotics: theoretical description}
\label{ch6:Sec:Heavy_Ex}

Our search for exotic states started in the sixties. 
Presently, most of the firm evidence comes from the sector with two heavy quarks, where the XYZ states have been discovered. 
The search for QCD exotics in the light-quark sector is difficult, because in order for a multiquark state to be clearly identifiable, it needs to be narrow enough to stand out on top of the experimental background. 
Multiquark states containing only light quarks typically have many open
decay channels, with a large phase space, so they tend to be wide. Moreover, they mix significantly with conventional hadrons, making them difficult to pin down. 

Multiquark states with heavy quarks, charm or bottom, are different. 
The two large quark masses make the system somewhat more-or-less nonrelativistic (NR), reducing the mixing and creating special circumstances that may support narrow multiquark states. 
They were not seen in early searches because, to be produced, they need the huge luminosities provided by the recent experiments. 
It is worth emphasising that the XYZ states emerge at or above the strong decay thresholds, which is the energy level above which heavy quarkonium states can decay into a pair of heavy-light mesons. 
Still, many of them have a small decay width, highlighting the fact that they are not merely composite particles, but have unique properties and interactions that differentiate them from conventional mesons and baryons.
Consequently, the study of these exotics in the realm of QCD could give us 
novel information about the strong fundamental force.

As remarked in the experimental introduction, the spectroscopy of mesons and baryons containing hidden and open heavy-flavour quarks has made tremendous progress in recent years, providing evidence for multiquark exotic states, including tetraquarks (for mesons) and pentaquarks (for baryons); see, \textit{e.g}., Refs.\,\cite{Esposito:2016noz, Lebed:2016hpi, Guo:2017jvc, Brambilla:2019esw, Guo:2019twa, Chen:2022asf, Meng:2022ozq, Ali:2019roi, A2:2019arr, Brambilla:2010cs}.

For some of these states, their multiquark content is unambiguously determined by their quantum numbers, for instance the isotriplet $Z_c$ and $Z_b$ states, which are electrically charged and decay to a heavy quarkonium and a single pion. 
Some of these states lie almost precisely at one of the heavy-light decay thresholds.

Notably, {\bf the $X(3872)$}, whose mass is very close to the sum of the masses of the $D^0$ and $\bar D^{\ast 0}$ mesons, displays exceptional properties, including a very small binding energy and a large scattering length. 
It is also the only state observed in all of the abovementioned processes, \textit{viz}.\ in direct hadroproduction, $B$ decays, $e^+e^-$ collisions, and even in heavy-ion collisions, which opens new avenues for the investigation of exotics and their study in the Quark–Gluon Plasma.

Another example is the isospin-zero {\bf $T_{cc}^+$} state, with a mass just below the $D^{*+} D^{0}$ threshold and the longest lifetime thus far observed for an exotic state. 
Pentaquarks also appear in the proximity of baryon–meson thresholds.

A large theoretical community has set out to calculate the properties of
XYZ states by building dedicated models.
Given that they lie within the purview of nonperturbative QCD, it is difficult to make first-principles predictions of the spectrum, widths, and production cross sections of such multiquark states. 
\textit{A priori}, the simplest system consisting of only two quarks and two antiquarks (tetraquarks) is already a very complicated object, and it is unclear whether any kind of clustering occurs in such states.
To simplify the problem, a focus is placed on certain substructures and their implications. 
They include {\bf molecular and compact diquark descriptions}, which are nowadays among the most prominent approaches:
hadronic molecules \cite{Tornqvist:2004qy, Close:2003sg}, composed of colour-singlet mesons bound together by residual nuclear forces; and
tetraquark bound-states between a diquark and an
antidiquark \cite{Jaffe:1976ig, Maiani:2004vq, Jaffe:2003sg, Lebed:2015tna}.
Note that where states are particularly close to threshold, like the $X(3872)$, owing to the small binding energy or large scattering length, universal characteristics emerge and a {\bf molecular EFT} description can be employed to methodically explore their properties \cite{Guo:2017jvc, Braaten:2007dw, Braaten:2004rn, Fleming:2011xa, Fleming:2007rp}.
Moreover, employing chiral dynamics allows one to extend the range of applicability of the EFT \cite{Baru:2021ldu, Du:2019pij, Du:2021zzh}. 
On the other hand, various model approaches depend, \textit{e.g}., on a cut-off, because they cannot systematically constrain the short-distance physics.

Hadronic molecules are interpreted as loosely bound states of two or more hadrons, held together by the strong interaction, analogously to nuclei. These are spatially extended systems, with their size controlled by the inverse binding momentum $\gamma = \sqrt{2\mu E_B}$, where $E_B$ denotes the binding energy and $\mu$ the reduced mass of the hadrons. 
While the main focus is typically placed on bound states, shallow virtual states \cite{Matuschek:2020gqe} and even unstable states \cite{Baru:2003qq, Braaten:2007dw} can also belong to this category, as long as their constituents are not too broad \cite{Filin:2010se,Guo:2011dd}.
Although both the molecular and the compact diquark models predict mostly the same structure of the exotic tetraquark multiplets, the difference lies in the dynamics and, \textit{e.g}., the breaking pattern of the approximate QCD symmetries. 
For instance, considering shallow near-threshold states, the molecular size can reach several femtometres, far exceeding the typical size of compact configurations, which are constrained to $\lesssim 1\,$fm by confinement dynamics. 
This leaves imprints in hadronic line shapes. 
In addition, the amount of SU(3)-flavour and heavy-quark spin symmetry violations are, for molecular states, dominated by the violation already realised at the level of the constituents. This is not the case for compact structures.

One may argue that a rigorous description of heavy-quark exotic hadrons should be based only on QCD itself, 
should contain a description of all states in the sector with two heavy quarks (quarkonia, hybrids, tetraquarks, pentaquarks),
and should embrace all dynamical scales.
A description of this type, obtained from QCD on the basis of symmetries and scale separation, is the 
{\bf Born–Oppenheimer (BO) Effective Field Theory (BOEFT)} 
\cite{Berwein:2015vca, Brambilla:2017uyf, Soto:2020xpm, Berwein:2024ztx, Brambilla:2024thx}, which is implemented in a systematic field theory framework via the Born–Oppenheimer approximation \cite{Braaten:2014qka, Juge:1997nc} applied to the two heavy quarks (the slow degrees of freedom) with respect to the light degrees of freedom (light quarks and gluons) at the nonperturbative scale $\Lambda_{\rm QCD}$.
At leading order, it produces coupled Schr\"odinger equations whose potentials are the QCD BO static energies, which may be calculated using, \textit{e.g}., LQCD. 
While contemporary applications of BOEFT use some LQCD input, see Sec.\,\ref{ch6:sec:lqcdheavyexotics}, this is in the form of a few universal nonperturbative correlators that are, within this framework, the same for the charmonium and bottomonium sectors, thus providing improved predictive power.

In BOEFT, {\bf the $X(3872)$} emerges from the solution of three coupled Schr\"odinger equations. 
It is composed of two tetraquarks at 90\% and quarkonium at 10\%, has a radius of about 14\,fm, in agreement with the molecular description, and a compact core in colour octet in agreement with the compact diquark description \cite{Brambilla:2024thx}.
The quarkonium component allows a prediction for the radiative transition in agreement with experiment.  (Within the molecular EFT the radiative decays can also be described \cite{Guo:2014taa}, however, they are controlled by a contact term parameter at leading-order.)

The fact that the hadroproduction of the $X(3872)$ is sizeable has long been used by compact diquark model practitioners to claim that it cannot be described as a molecule. 
However, these descriptions can be complementary at different energy scales. 
Hadroproduction is a short-distance process that can be calculated in a factorised form using BOEFT \cite{Lai:2025tpw}, showing that the heavy quark–antiquark core is indeed in a colour octet but not excluding 
that the state has a large radius.
This makes a strong case for the study of the $X(3872)$ at SIS100.

Regarding {\bf Pentaquarks}, the proximity of the $\Sigma_c\bar{D}^{(*)}$ thresholds to these narrow pentaquark structures suggests that the corresponding two-hadron states play an important role in their dynamics, supporting an interpretation of their structure as hadronic molecules. 
In the most common molecular picture, the $P_c(4312)$ is an $S$-wave $\Sigma_c\bar{D}$ bound state, while the $P_c(4440)$ and $P_c(4457)$ are bound states of $\Sigma_c\bar{D}^*$ with different spin structures \cite{Liu:2019tjn, Xiao:2019aya, Sakai:2019qph, Du:2019pij, Xiao:2020frg, Du:2021fmf}. 
A virtual state interpretation of $\Sigma_c\bar{D}$ has also been proposed \cite{Fernandez-Ramirez:2019koa}. 
However, within the molecular picture, in addition to the three established pentaquarks mentioned above, there should be four more: as a consequence of 
heavy-quark spin symmetry, there should also exist $\bar D^{(*)}\Sigma_c^*$ bound systems. 
Although the analysis of Ref.\,\cite{Du:2019pij} found weak evidence for a narrow $D\Sigma_c^*$ state at 4380~MeV in the LHCb data, as yet there is no evidence for the three predicted $D^*\Sigma_c^*$ states.

The quantum numbers of the $P_c(4440)$ and $P_c(4457)$ are still under debate in the molecular picture, with $1/2^-$ and $3/2^-$ assignments (or vice versa) being largely consistent with the data. 
However, it has been argued that the inclusion of one-pion exchange could eliminate this degeneracy \cite{Du:2021fmf,Du:2019pij}. 
As a general feature of the molecular picture, all $P_c$ states are expected to have negative parity, with $P_c(4312)$ unambiguously identified as the $1/2^-$ $\Sigma_c\bar{D}$ molecule. 
This may differ from the compact diquark prediction but agrees with the BOEFT prediction in terms of the number of states, masses, quantum numbers and parity \cite{Brambilla:2025xma}; see Fig.\,\ref{fig:pp_Jpsi_sim} below. 
(In this context, see also Ref.~\cite{Alasiri:2025roh}.)
A strong case it thus made for a study of the pentaquarks at SIS100; especially the search for the missing states needed to complete the spin-symmetry multiplet.

If the $P_c$ states are molecular in character, they should manifest as strong threshold enhancements in the corresponding $\Sigma_c^{(*)}\bar{D}^{(*)}$ channels. 
Moreover, they should also appear as visible peaks near the $\Sigma_c^{(*)}\bar{D}^{(*)}$ thresholds in the invariant mass spectra of more distant channels, such as $\eta_c p$ and $\Lambda_c \bar{D}^{(*)}$, through coupled-channel effects \cite{Du:2021fmf}.


\textit{Ab initio} calculations of XYZ spectra using LQCD remain challenging because they require studying the coupled-channel scattering of hadrons on the lattice; see Sec.\,\ref{sec:toolslatqcd}). 
Direct LQCD calculations of exotic-state production and propagation in medium are not currently possible, whereas the BOEFT description can be used to study these processes.
As noted above, DSE/BSE methods also show promise in this area.

\subsubsection{LQCD studies}
\label{ch6:sec:lqcdheavyexotics}

\noindent{\bf Spectra}.
The spectrum of hadrons involving heavy quarks has been the subject of numerous LQCD studies. 
Many recent investigations have focused on tetraquark systems containing heavy quarks. 
It is not feasible to provide an exhaustive list of all such computations here, so only a few recent examples will be highlighted. 
Further details of many other works can be found, \textit{e.g}., in Refs.\,\cite{Prelovsek:2023sta, Francis:2024fwf, Hudspith:2024lsj, Lewis:2025xtu}.

\begin{figure}[t]
\parbox{0.55\linewidth}{
\includegraphics[width=0.54\textwidth]{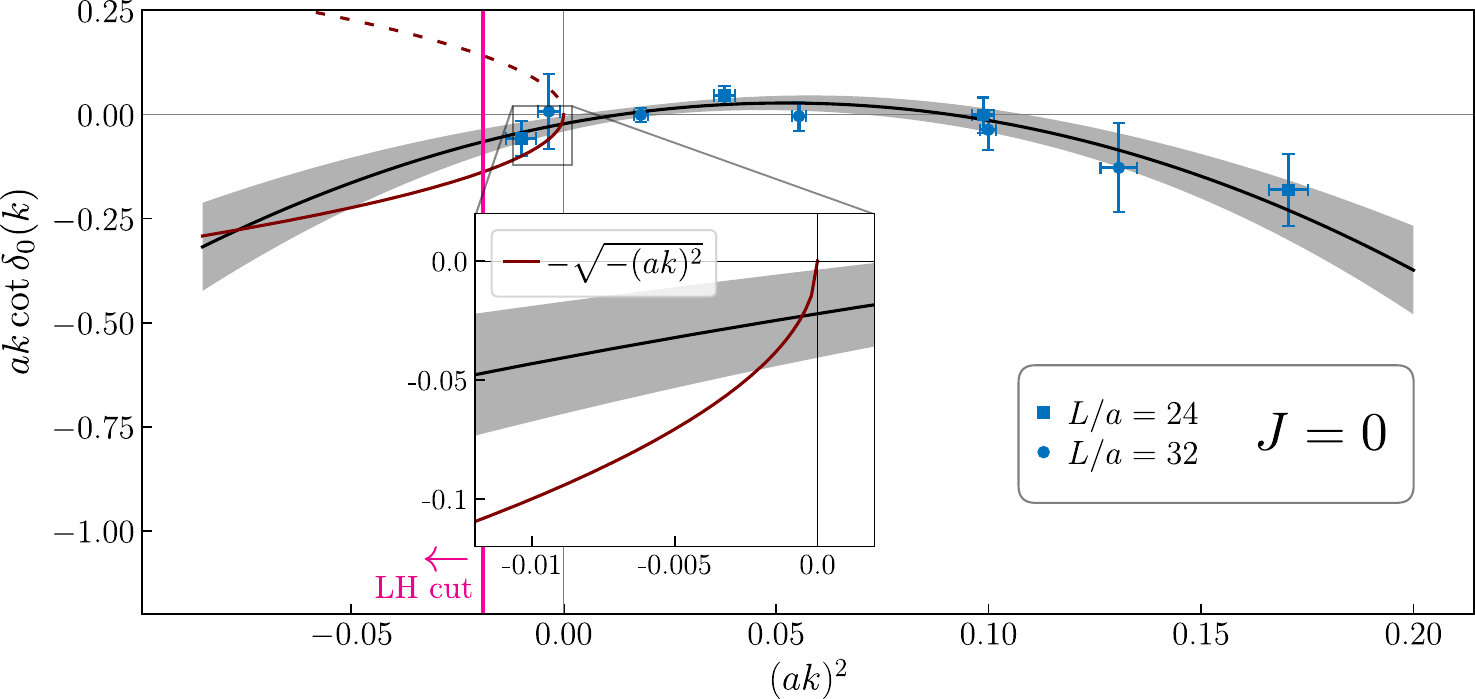} 
\vspace{3ex}
\includegraphics[width=0.54\textwidth]{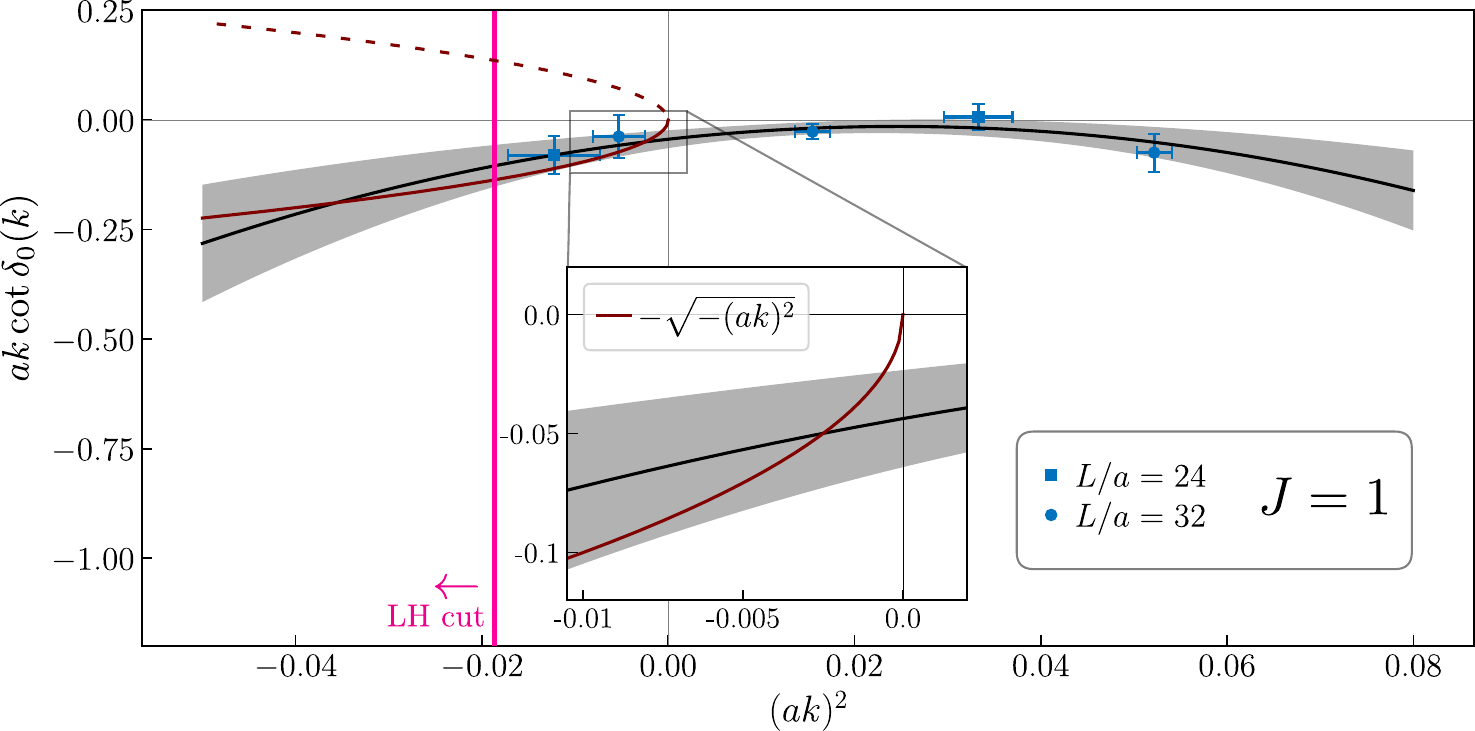}}
\qquad
\parbox{0.40\linewidth}{
\quad
\includegraphics[width=0.36\textwidth]{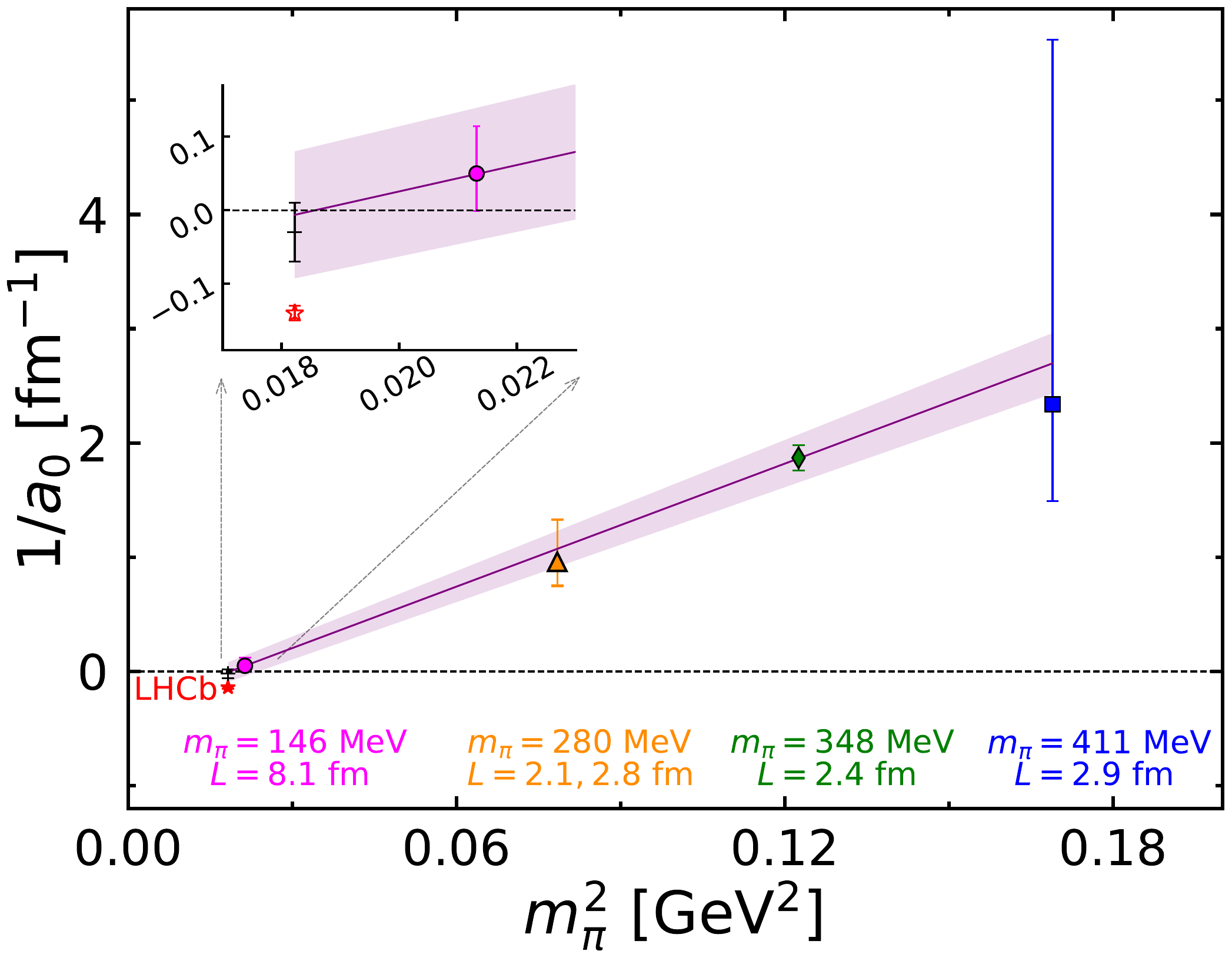} 
\vspace*{-0.2cm}
\includegraphics[width=0.39\textwidth]{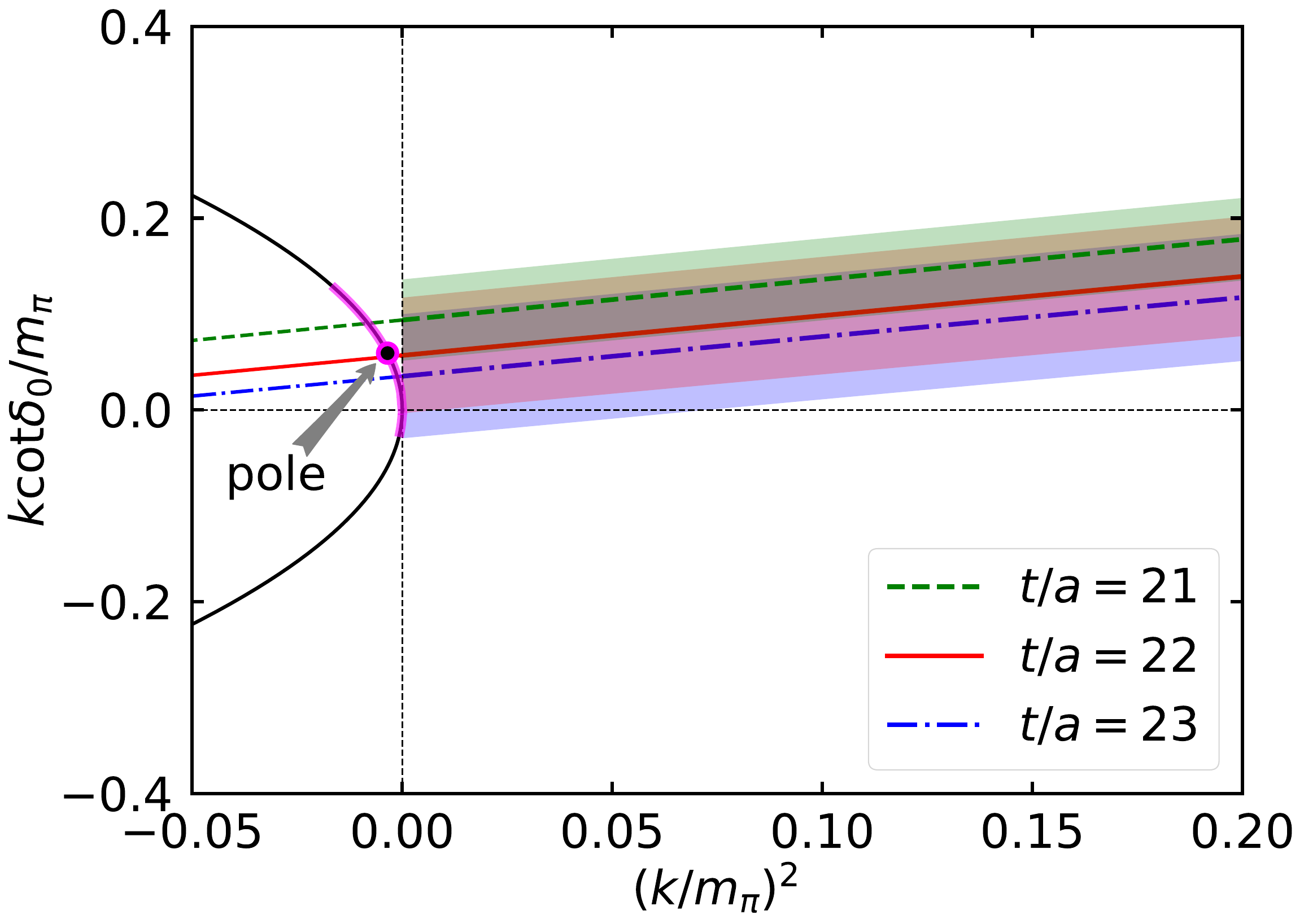}}
\caption{
(Left) Results from Ref.\,\cite{Alexandrou:2023cqg} for the functions 
$ak \cot \delta_0(k)$ for $S$-wave $B\overline{D}$ scattering (top left) and $S$-wave $B^\ast\overline{D}$ scattering (bottom left), where $k$ is the scattering momentum, $\delta_0(k)$ is the scattering phase shift, and the lattice spacing is $a = 0.11887(80)$\,fm.  
The points were obtained directly from the lattice energy levels, and the curves correspond to ERE fits through order $k^4$.  
The intersections of $ak \cot \delta_0(k)$ with the solid (dashed) red parabolas correspond to shallow (virtual) bound states.
(Top right) Inverse of the scattering length, $1/a_0$, for $D^\ast D$
scattering in the $I = 0$ and $S$-wave channel obtained from
LQCD simulations in Ref.\,\cite{Lyu:2023xro} (magenta circle) is compared with results from Refs.\,\cite{Ikeda:2013vwa} (blue square), \cite{Chen:2022vpo} (green diamond), and \cite{Padmanath:2022cvl} (yellow triangle),as well as the real part of the experimental value from LHCb (red star) \cite{LHCb:2021auc}.
(Bottom right) $k \cot\delta_0 /m_\pi$ from Ref.\,\cite{Lyu:2023xro} for $D^\ast D$ scattering in the $I = 0$ and $S$-wave channel, as a function of $(k/m_\pi)^2$, where $k$ is the scattering momentum. 
The intersection of $k \cot \delta_0 /m_\pi$ with the black solid line indicates the pole in the scattering amplitude. 
The pole position is shown by the black point, with the magenta line indicating statistical and systematic errors combined.
\label{fig:latticeheavy}}
\end{figure}

One recent study of $B\overline{D}$ and $B^\ast\overline{D}$ isospin-0 $S$-wave scattering in Ref.\,\cite{Alexandrou:2023cqg} addressed the issue of whether mixed bottom-charm $\overline{b}\overline{c}ud$ tetraquarks exist as bound states or resonances. 
Applying the L\"uscher method with energy spectra in two different volumes, sharp peaks in the $B^\ast \overline{D}$ scattering rates close to threshold were found, which are associated with shallow bound states -- either genuine or virtual -- lying a few~MeV or less below the $B^\ast\overline{D}$ thresholds. 
In addition, hints of resonances with masses around 100\,MeV above threshold and decay widths of order 200\,MeV were found. 
Results for the $S$-wave scattering phase shift are shown in the left-hand plots of Fig.\,\ref{fig:latticeheavy}.

Another recent study \cite{Lyu:2023xro} focused on the doubly-charmed 
tetraquark $T_{cc}^+$ system near the physical point, where the quark masses yield pion and kaon masses at their observed values. 
In this study, the HALQCD method was employed to obtain the $D^\ast D$ interaction potential from a hadronic space-time correlation determined in LQCD. 
From the interaction potential, the isoscalar $S$-wave scattering length and scattering phase shift were extracted. 
Results for the scattering length are shown in the top right image
of Fig.\,\ref{fig:latticeheavy}, and the bottom right image displays the result for the scattering phase shift. 
The interaction is found to be attractive at all distances, leading to a near-threshold virtual state and a large scattering length. 
The potential determined in this study provides a semi-quantitative 
description of the LHCb data on the $D^0 D^0 \pi^+$ mass spectrum.

\noindent{\bf Lattice QCD computations of the BO QCD static energies}.
\label{ch6:BO_static}
BOEFT \cite{Berwein:2015vca, Brambilla:2017uyf, Soto:2020xpm, Berwein:2024ztx, Brambilla:2024thx} is formulated as an expansion in terms of the inverse heavy-quark mass $1/m_Q$. 
QCD static energies, or adiabatic surfaces, appear at each order in $1/m_Q$. 
These can be computed using LQCD and serve as potentials in the BOEFT coupled Schr\"odinger equations. 
The leading order -- $(1/m_Q)^0$ -- corresponds to the static potential.

Static QCD energies are characterised by quantum numbers $\Lambda_\eta^\epsilon$, where $\Lambda$ is the projection of the angular momentum of the light degrees of freedom onto the quark-antiquark axis, $\epsilon=\pm 1$ is the eigenvalue under reflection through a plane containing the quark-antiquark axis, and $\eta = \pm 1$ is the symmetry quantum number under the combined operations of charge conjugation and spatial inversion about the midpoint
between the quark and antiquark. 
The heavy quark-quark case is studied in the same way, excluding charge conjugation.

States with $\Lambda = 0, 1, 2, \dots$ are denoted by 
the capital Greek letters $\Sigma$, $\Pi$, $\Delta$, and so on. 
States which are even (odd) under the abovementioned parity-charge-conjugation operation are denoted by the subscripts $g$ ($u$). 
There is an additional label for the $\Sigma$ states: $\Sigma$ states
which are even (odd) under a reflection in a plane containing the $QQ$ axis are denoted by a superscript $+$ ($-$). 
These static energies can be computed from the large-time behaviour of generalised Wilson loop correlation functions, where appropriate creation
operators -—superpositions of various spatial paths of smeared links from the quark to the (anti)quark location -- replace the straight spatial lines of ordinary Wilson loops.

The static energies have been studied extensively for quarkonium and for hybrids using LQCD at both small and large quark-antiquark 
separations; see Refs.\,\cite{Griffiths:1983ah, Campbell:1984fe,Campbell:1987nv, Perantonis:1990dy, Juge:1997nc, Morningstar:1998xh, Juge:1999ie, Juge:1999aw, Bali:2000vr,
Juge:2002br, Bali:2003jq, Bicudo:2018jbb, Capitani:2018rox, Schlosser:2021wnr, Brambilla:2022het, Sharifian:2023idc}), as well as 
for $Q\bar{Q}$ tetraquarks in Ref.\,\cite{Sadl:2021bme} and $QQ$ tetraquarks in Refs.\,\cite{Lyu:2023xro,Bicudo:2024vxq}.

Using these leading-order static energies and inserting them into the appropriate BOEFT single or coupled Schr\"odinger equations 
yields the spin-degenerate spectra of heavy exotics, as in
Refs.\,\cite{Perantonis:1990dy, Juge:1997nc, Juge:1999ie, Berwein:2015vca, Oncala:2017hop, Capitani:2018rox}
for quarkonium hybrids, 
in Refs.\,\cite{Bicudo:2015vta, Bicudo:2022ihz, Lyu:2023xro, Braaten:2024tbm, Brambilla:2024thx} 
for tetraquarks, and in Ref.\,\cite{Brambilla:2025xma} for pentaquarks.

Spin corrections to the $\Sigma_g^+$ static potential appear at order $(1/m_Q)^2$ for quarkonium within pNRQCD/BOEFT \cite{Brambilla:2004jw}, and have been studied using LQCD in 
Refs.\,\cite{Bali:1997am, Koma:2006si, Koma:2006fw}.
The first LQCD results for the hybrid spin-dependent and hybrid-quarkonium 
mixing potentials at order $(1/m_Q)^1$ appeared recently in Ref.\,\cite{Schlosser:2025tca}.
Spin-dependent contributions to heavy hybrid mesons, such as spin-dependent potentials and hybrid-quarkonium mixing potentials, also appear at order $(1/m_Q)^1$ \cite{Brambilla:2018pyn, Brambilla:2019jfi, Oncala:2017hop}. The same applies to tetraquarks and pentaquarks \cite{Soto:2020xpm}.

The avoided level crossing between static energies with the same BO quantum numbers was studied in Refs.\,\cite{Bulava:2019iut, Bulava:2024jpj}. 
This phenomenon is related to string breaking and induces mixing between quarkonium and tetraquark configurations with the appropriate BO quantum numbers.
This mixing naturally induces a quarkonium component inside the $X(3872)$ \cite{Brambilla:2024thx}, which has phenomenological implications, for example in the calculation of radiative transitions.

\subsection{Prospects at SIS100}
\label{ch6:exotic_prospects}

The high intensity of the proton beam from SIS100 and the maximum CM energy of 7.6\,GeV raise the possibility of searching for most of the exotic hadron states discussed in this section. 
However, little is known about the processes through which these hadrons may be exclusively produced, which limits the ability to make firm projections. 
Light-quark exotic hadrons tend to have large widths and overlap in mass with other hadrons, making PWA a necessity and introducing a reliance on well-understood reaction models. 
As discussed in Chapter~\ref{sec.Tools} below, central exclusive production has been effective at higher proton beam energies in isolating meson production, and could also be effective at SIS100 energies, with contributions from Pomeron and other Regge exchanges expected.

Heavy-quark exotic hadrons are generally narrower and easier to identify, but their cross sections are much smaller than those for light-quark hadrons, presenting a different challenge. 
Until recently, of the heavy-quark exotics, only the $X(3872)$ has been observed in hadroproduction, and only in high-energy $pp$ and $p\bar{p}$ collisions at the LHC and Tevatron \cite{CDF:2003cab, D0:2004zmu, LHCb:2011zzp, LHCb:2020sey}. 
The recent identification of the $T_{c\bar{c}c\bar{c}}$ and $T_{cc}$ states at the LHC, as described above, are the only other heavy-quark exotics observed in hadroproduction. 

Identifying any additional heavy-quark exotic states in $pp$ collisions at SIS100, which provides a very different production mechanism compared to the $e^+e^-$ collisions and hadron decays in which most of these states have been discovered, would offer very valuable insights into their internal structure and formation. 

Finally, exotics with baryon number greater than one are expected to be produced more abundantly in $pp$ interactions, and FAIR could offer a uniquely optimal environment for their discovery.

Hereafter, the discussion identifies several possible exotic hadron candidates that both can be addressed at FAIR and represent promising cases for early CBM measurements. 
These could serve as benchmarks for developing the exotic hadron search programme in greater detail. 
The focus is on final states with only charged particles, since the CBM experiment excels at their reconstruction.

\begin{figure}[t]
    \centering
    \includegraphics[width=0.45\linewidth]{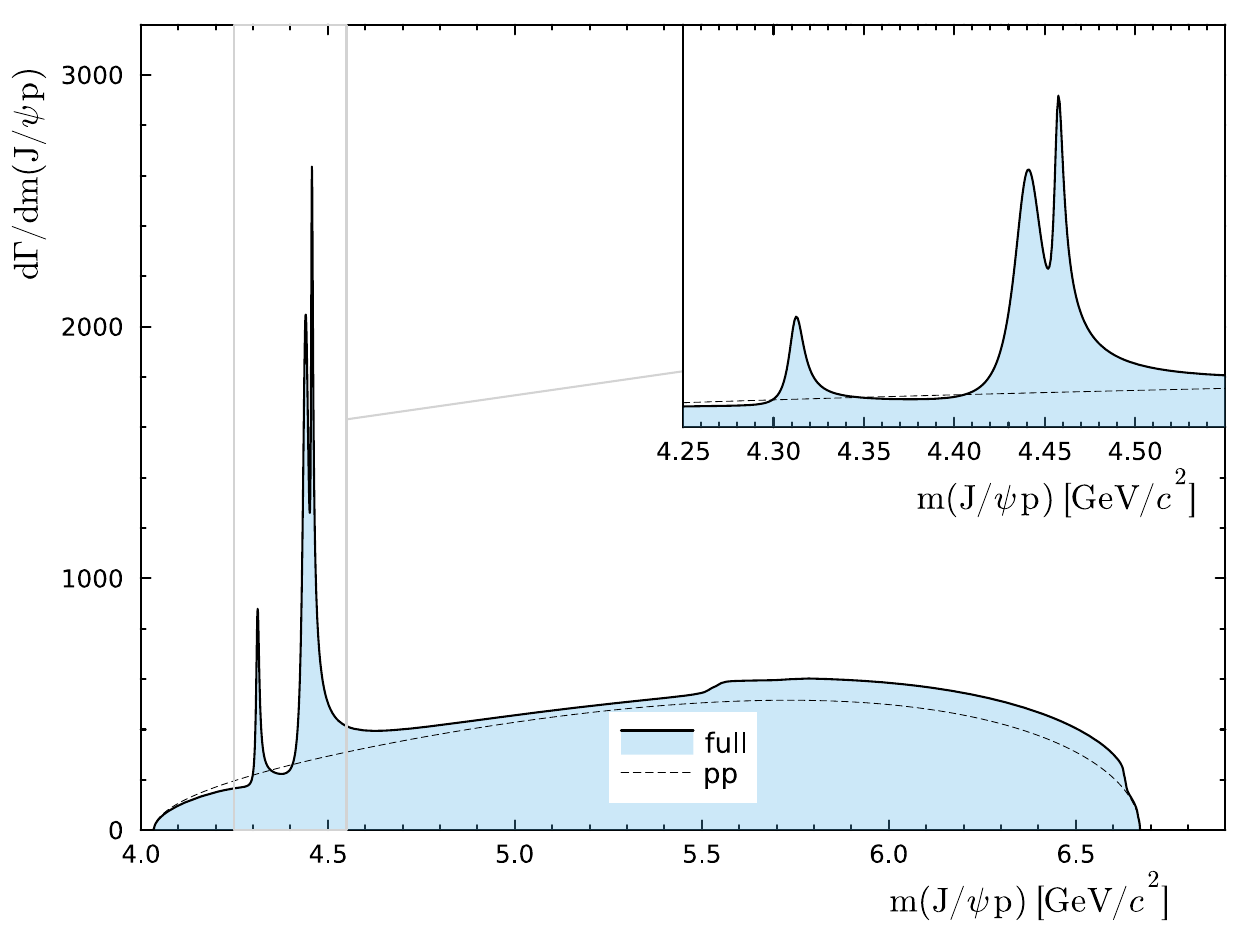}
    \includegraphics[width=0.54\linewidth]{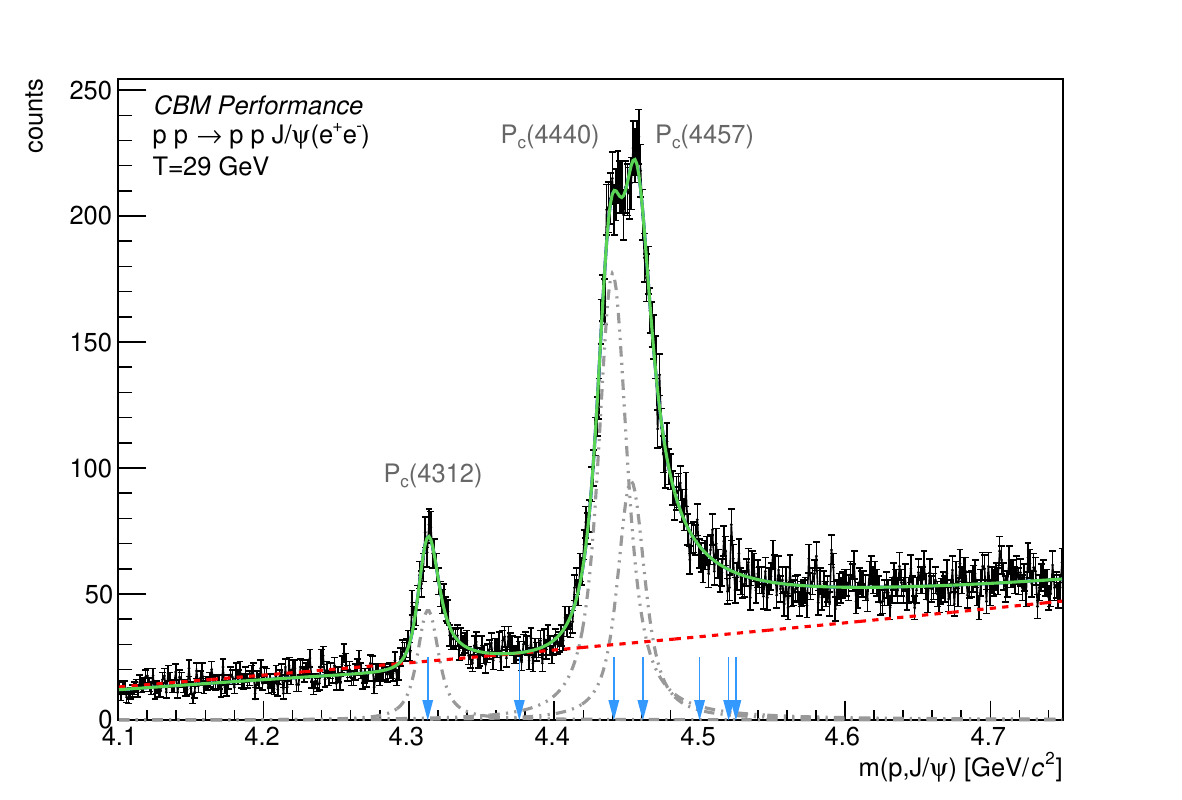}
    \caption{Simulated invariant mass distribution of $J/\psi\,p$ for $pp \to pp\, J/\psi$, $J/\psi \to e^+e^-$ events, using a model that includes the three narrow $P_{c\bar{c}}^+$ states, parametrised by Breit–Wigner amplitudes and a proton–proton scattering amplitude modelled using the scattering-length approximation. 
    Distributions are shown for (left) generated events and (right) events reconstructed in CBM and corresponding to about 100 days of data taking at a 100\,MHz total interaction rate. 
    The reconstructed momenta of the final-state particles have been refitted using kinematical constraints imposing energy and momentum conservation and constraining the $e^+e^-$ mass to the nominal $J/\psi$ mass.
    The reconstructed events exhibit a good mass resolution of $4-6\,$MeV, and an example fit is shown that resolves all three states. 
    The blue arrows indicate the positions of predicted states in molecular models as well as within the BOEFT framework (see main text). Please note, that in some places the more formal notation $P_c$ is used, whereas also notations further specifying the quark content, {\it i.e.} $P_{c\bar{c}}$ are used. Take note of the following disclaimer~\cite{cbm_feasibility_note}.
    \label{fig:pp_Jpsi_sim}}
\end{figure}

\begin{figure}[ht]
    \centering
    \includegraphics[width=0.45\linewidth]{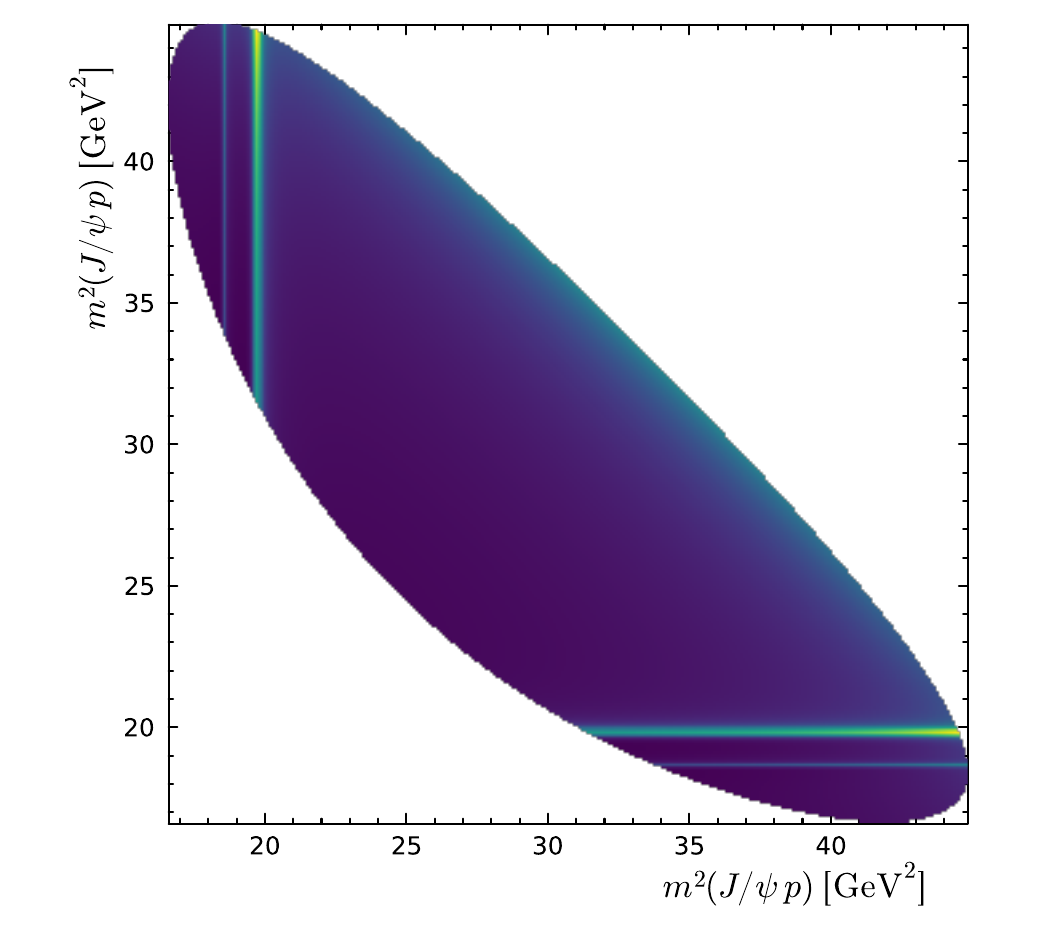}
    \includegraphics[width=0.44\linewidth]{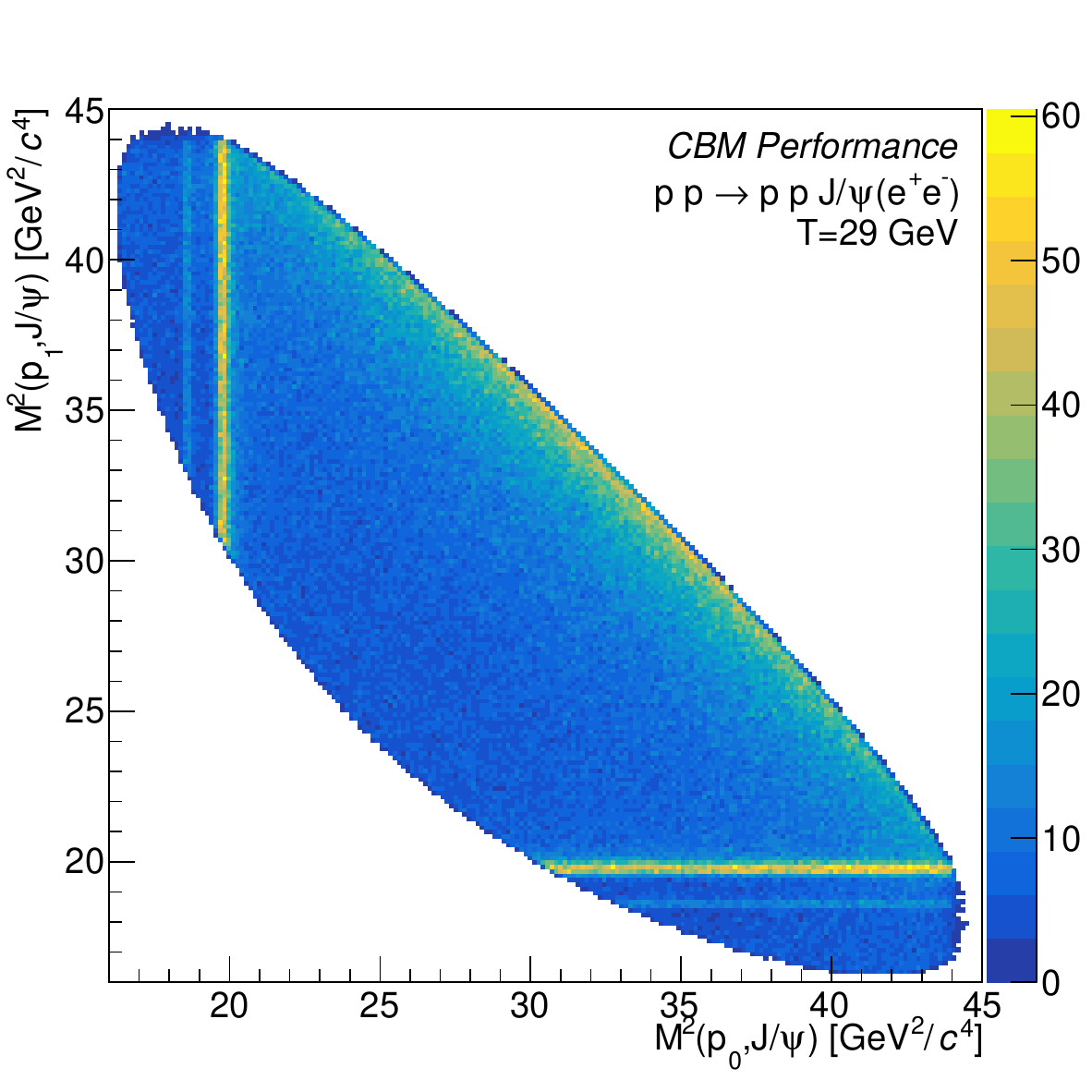}
    \vspace*{-0.3cm}
    \caption{Dalitz plot of simulated $pp \to pp\, J/\psi$, $J/\psi \to e^+e^-$ events, obtained using a model that includes the three narrow $P_{c\bar{c}}^+$ states, parametrised by Breit–Wigner amplitudes, and a proton–proton scattering amplitude modelled using the scattering-length approximation. 
    Distributions are shown for (left) generated events, and (right) events reconstructed in CBM, illustrating the good acceptance across the available phase space. Take note of the following disclaimer~\cite{cbm_feasibility_note}.
    \label{fig:dalitz_Jpsi_sim}}
\end{figure}

\begin{figure}[t]
    \centering
    \includegraphics[width=0.45\textwidth]{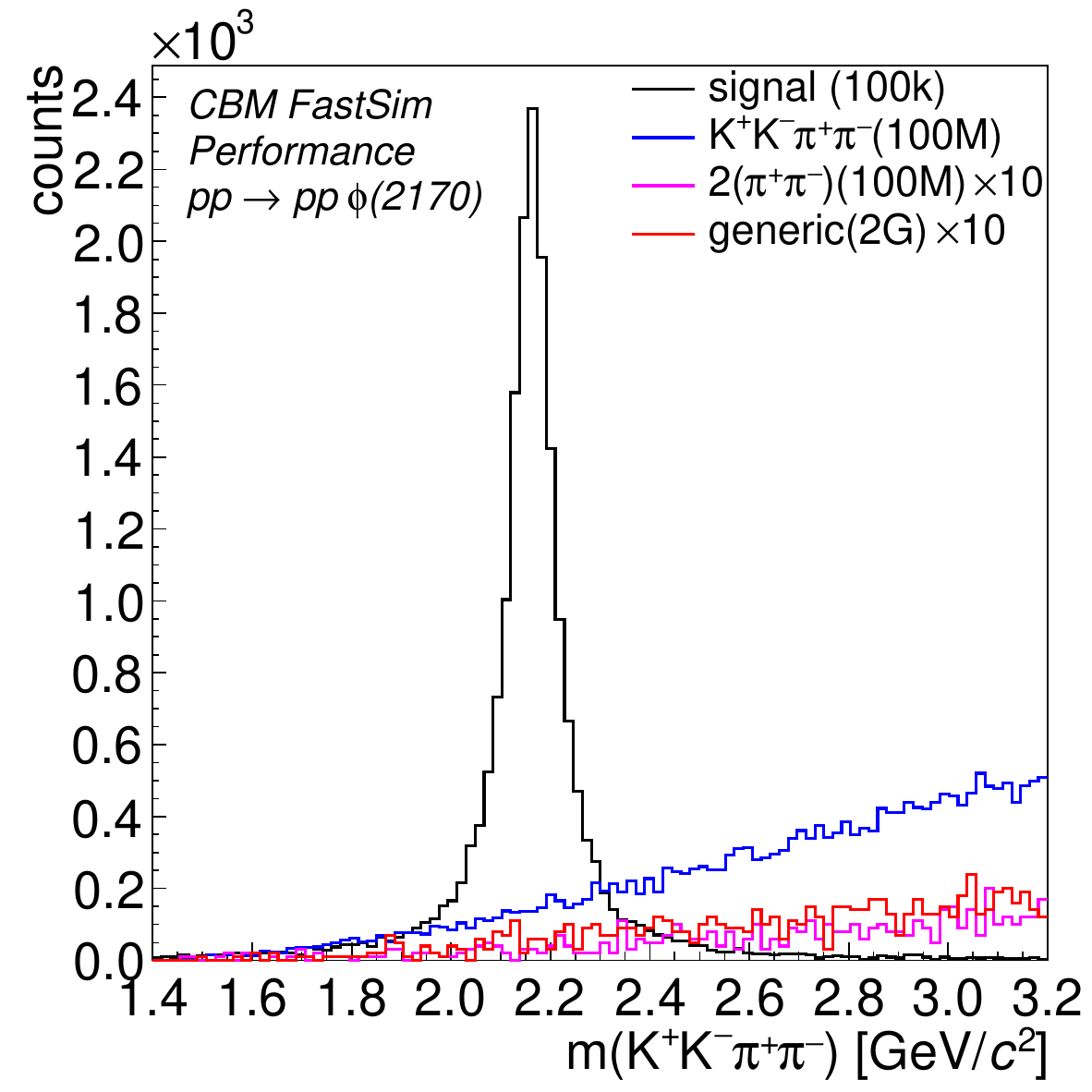}
\includegraphics[width=0.45\textwidth]{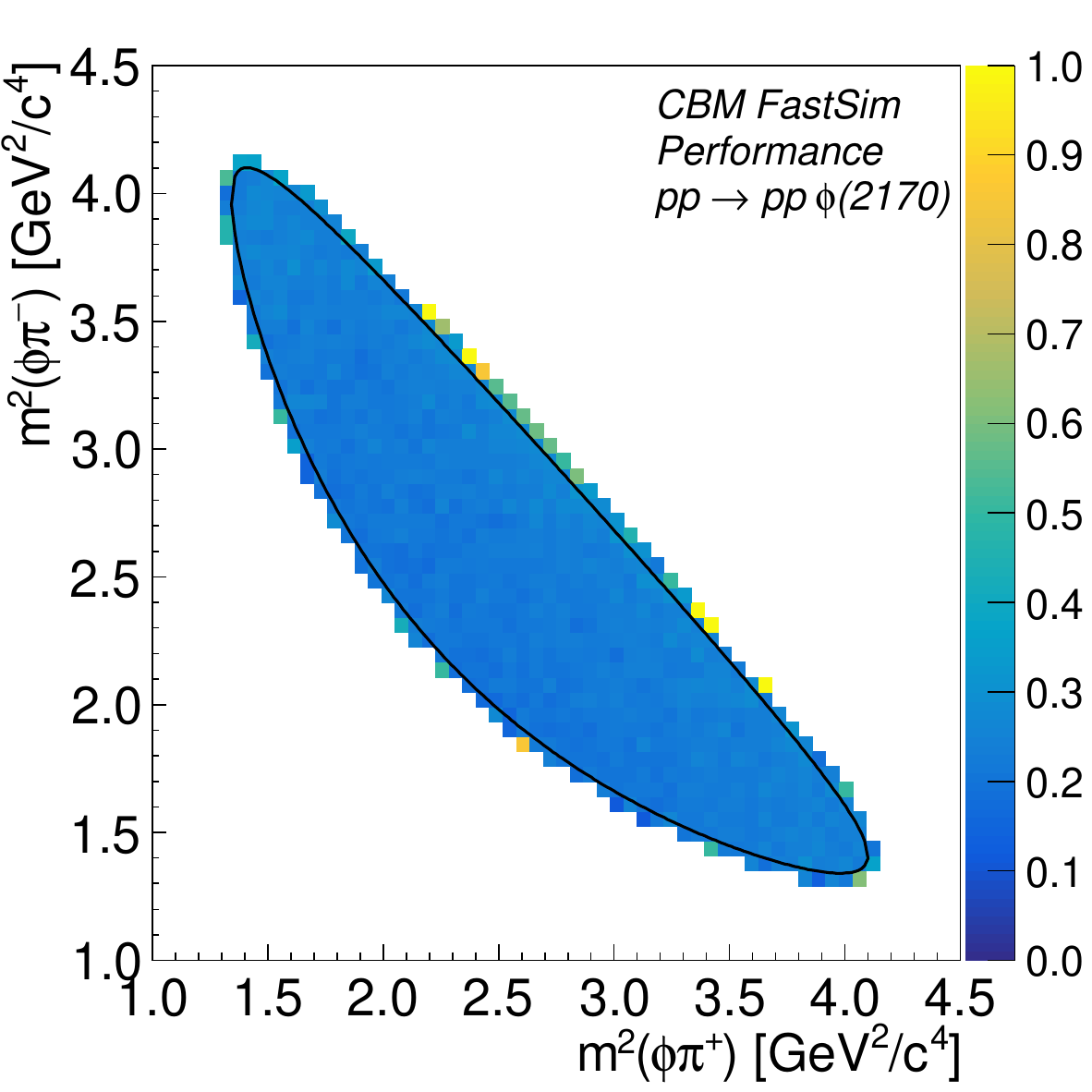}
\vspace*{-0.1cm}
    \caption{Feasibility study for $pp \rightarrow pp \phi(2170)$ using the CBM detector. 
    Left: Invariant-mass distributions for the signal and three background channels -- two dedicated modes and generic background events. 
    The expected signal detection efficiency is about 20\%, and a suppression of the generic background of order $10^{-8}$ appears feasible. 
    Right: For the decay $\phi(2170) \to \phi(1020) (\to K^+K^-) \pi^+\pi^-$ the acceptance is flat across the complete kinematic phase space, providing excellent conditions for a full partial-wave analysis. Take note of the following disclaimer~\cite{cbm_feasibility_note}.
    \label{fig:phi2170-Feasi}}
\end{figure}

\begin{itemize}
\item \textbf{Pentaquark states}. 
The $P_{c\bar{c}}$ and $P_{cs}$ states carry non-zero baryon number, making them ideal candidates to search for at this facility. 
The reaction $pp \to pp\, J/\psi$ can be used to search for the $P_c$ states via their discovery mode, namely their decay to $J/\psi\, p$. 
A sensitive probe of their internal structure comes from their coupling to open-charm channels. 
Therefore, it is critical to search for them in reactions such as $pp \to p\, \Lambda_c\, \bar{D}^{(*)}$ and $pp \to p\, \Sigma_c^{(*)}\, \bar{D}^{(*)}$. 
These reactions have already been discussed in some detail in previous chapters.

Moreover, SIS100 holds the potential to discover and establish pentaquark states predicted to exist near the $D^{(*)}\Sigma_c^*$ thresholds in both molecular approaches and the BOEFT framework.  
These states are absent in some diquark models. 
Here, $pp$ reactions at FAIR offer a high discovery potential and are far superior to, \textit{e.g}., photoproduction on nucleons, where the vector-meson–nucleon coupling of the pentaquarks is suppressed compared to their coupling to open-charm channels. 
In addition, in $\Lambda_b$ decays studied at LHCb, the $\bar{D}\Sigma_c^*$ structures appear suppressed, and for higher states the phase space is limited.

The proximity of the observed $P_c$ pentaquarks to the $\Sigma_c \bar{D}^{(*)}$ thresholds in $\Lambda_b$ decays suggests that their production is intimately connected to these open-charm channels. 
The production cross sections for open-charm final states are expected to be significantly larger than those for charmonium final states. 
This is because the charm and anticharm quarks must move with a small relative velocity to form charmonium, which is highly suppressed owing to limited phase space. 
For instance, the production cross sections of $J/\psi\, \pi\pi$ and $h_c\, \pi\pi$ are of the order of tens of picobarns in the resonance region and even smaller outside, while the cross sections of $\pi\, D^{(*)} \bar{D}^{(*)}$ remain of the order of hundreds of picobarns just above threshold.  A compilation of data is available, \textit{e.g}., in Ref.\,\cite{Nakamura:2023obk}). 

Thus, it is anticipated that the cross sections for $pp \to p\, \Lambda_c\, \bar{D}^{(*)}$ and $pp \to p\, \Sigma_c^{(*)}\, \bar{D}^{(*)}$ could be one to two orders of magnitude larger than that for $pp \to pp\, J/\psi$ shown in Fig.\,\ref{fig:pp_Jpsi_sim}.

Initial simulations have been performed assuming a proton beam energy of 29\,GeV. With a 100\,MHz $pp$ interaction rate and a 1\,nb cross section for $pp \to pp\, J/\psi$, the CBM detector can reconstruct roughly 1000 $J/\psi$ mesons per day in either dilepton decay channel. 
Projections for 100 days of running are shown in Fig.\,\ref{fig:pp_Jpsi_sim} and Fig.\,\ref{fig:dalitz_Jpsi_sim}, which illustrate the excellent ability to measure the $P_{c\bar{c}}$ states if they are produced in this reaction.

One can also search for flavour-exotic pentaquarks decaying to final states such as $p\, \bar{D}^{(*)}$. The strange-quark partner $P_{cs}$ states decaying to $\Lambda\, J/\psi$ can be hunted in similar reactions, such as $pp \to p\, K^+\, \Lambda\, J/\psi$.

\item \textbf{Light-quark exotic mesons}. 
Most exotic quantum number light-quark hybrid mesons are expected to decay into final states including either a $\pi^0$ or an $\eta$, making states like the $\pi_1(1600)$ and $\eta_1(1855)$ difficult to reconstruct. 
Other exotic hadron candidates that are narrower and decay to all charged final states, such as the $\phi(2170) \to \phi(1020)\, \pi^+ \pi^-$, are better suited for study. 

A first glimpse has been obtained from Fast-Simulation (FastSim) studies of $pp$ reactions at the highest nominal beam momentum of 30\,GeV/$c$ for $pp \to \phi(2170) pp$, with $\phi(2170) \to \phi(1020) (\to K^+K^-) \pi^+ \pi^-$, reconstructed with the CBM detector. 
Signal detection efficiencies of 20\%, background suppression of about $10^{-8}$, and a flat acceptance can be expected; see Fig.\,\ref{fig:phi2170-Feasi}. 
A mass resolution for the $\phi(2170)$ of about 6\,MeV/$c^2$, together with a flat acceptance, provide excellent conditions to address this exotic candidate in the $pp$ production mechanism. 

One complication is that the production mechanisms for these light-quark states at SIS100 energies are not well known. 
At higher energies, light meson production in $pp$ collisions proceeds through a combination of diffractive Pomeron and Reggeon exchanges. 
If similar processes contribute at SIS100, then the target proton will scatter at large angles and will either need to be treated as a missing particle, or a new recoil proton detector will need to be constructed. 
If meson production is instead dominated by high-momentum transfer processes, then all final-state particles are more likely to fall within the CBM experimental acceptance.

\item \textbf{Charmonium-like exotics}. 
The $X(3872)$ is notable for being an exotic state observed in numerous experiments and production reactions, with a distinctive decay to $\pi^+ \pi^- J/\psi$. 
This makes it an important benchmark for charmonium-like exotic searches. 
Currently, no predictions exist for its production cross section in $pp$ collisions at SIS100 energies, but it is natural to expect the cross section will be lower than that of $J/\psi$ production. 
A useful reference is the production of the $\psi(2S)$, which also decays prominently to $\pi^+ \pi^- J/\psi$, and is expected to have a production cross section that is a factor of $5$–$10$ smaller than that of the $J/\psi$ in $pp$ collisions \cite{Cassing:2000vx}. 
Identification of $X(3872)$ production at SIS100/CBM would serve as a crucial reference point for estimating the production cross sections of other charmonium-like exotics, such as the $Z_c$ states.

\end{itemize}

\newpage
\section{Probing matter: from baselines to medium modifications}
\label{sec.DenseMatter}

{\small {\bf Convenors:} \it J. Aichelin, E. Bratkovskaya, M. Lorenz} 

\noindent
This chapter addresses the importance of nucleon--nucleon ($pp$, $pn$), pion-nucleon ($\pi p(n)$), and proton-\textcolor{blah}{ and pion-}nucleus ($pA$\textcolor{blah}{, $\pi A$}) collisions in probing cold and dense nuclear matter, which are key areas of investigation for the HADES and CBM experiments at FAIR.

Compared to relativistic heavy-ion collisions (HICs), such reactions offer better control over initial conditions and are therefore indispensable for quantifying nuclear medium effects and testing microscopic reaction models. While nucleus-nucleus $AA$ collisions provide access to a wide range of baryon densities and temperatures, their interpretation hinges critically on benchmark data from smaller systems.

Many of the available datasets for $pp$, $pn$, $\pi p$ and $pA$ reactions at few-GeV energies stem from experiments performed decades ago with limited precision and statistics. Yet, modern theoretical developments, \textit{e.g}., on the production of strange hadrons or the in-medium modification of hadronic spectral functions, call for new, high-quality data to test these predictions.

In particular, reactions involving nuclear targets at low to intermediate energies enable the study of hadron properties in a well-defined environment near normal nuclear density and low temperature, avoiding the complexities of evolving many-body systems as in $AA$ collisions.

Modern accelerator facilities cover a wide span of CM energies, from a few~GeV to several~TeV; and this enables systematic studies of the QCD phase diagram. 
Regardless of beam energy, the hadronic phase is always a key component of the reaction evolution. Consequently, a precise understanding of hadronic interactions across systems and energies is essential for drawing reliable conclusions, especially concerning QCD matter properties such as chiral symmetry restoration or quark-gluon plasma formation.

At FAIR, the SIS18/SIS100 synchrotrons uniquely bridge the hadronic and partonic domains. The long reaction times and high baryon densities reached in $AA$ collisions at these energies make this regime particularly sensitive to key QCD phenomena, such as the search for the critical point or a first-order phase transition.

To fully exploit this physics potential, high-precision data on $pp$, $pn$, $\pi p$, and $pA$ reactions are needed as theoretical baselines and for constraining transport and statistical models. 
The large and complementary acceptance of the HADES and CBM detectors, together covering polar angles from $3^\circ$ to $85^\circ$ in the laboratory frame, enables (nearly) full event reconstruction and the extraction of multi-differential observables. This comprehensive angular coverage is essential for quantifying medium effects, hadron formation mechanisms, and transport properties of strongly interacting matter.

\textcolor{blah}{
This chapter reviews the present experimental and theoretical knowledge of elementary and proton-/pion-induced nuclear reactions relevant for FAIR and discusses selected future measurements with HADES and CBM, which can provide important new constraints.}
\subsection{$\pi p$, $pp$ and $pn$ reactions}
\label{sec:ppi-induced-reactions}

\noindent
It is worth beginning with a discussion of the presently open questions in the physics of $\pi p$, $pp$, and $pn$ reactions, focusing on their relevance to $pA$ ($\pi A$) and heavy-ion reactions. The theoretical modelling of hadron-hadron reactions is discussed in Chapter~\ref{sec.HadrHadrInter}.

\subsubsection{Transition from hadronic to partonic interactions}

\noindent
The GSI/FAIR energy range (SIS18 and SIS100) for $\pi p$, $pp$, and $pA$ collisions is presumably characterised by the transition from hadron- to parton-dominated reactions. In hadron-dominated interactions, the final-state particles are predominantly distributed according to the full three-dimensional phase space. In contrast, parton-dominated reactions are expected to exhibit a more longitudinal phase-space distribution, insofar as one would find from extrapolations of the Lund string fragmentation model \cite{NilssonAlmqvist:1986rx} to this intermediate energy regime. 
In the Lund model, the transverse momentum of the produced particles is limited by the transverse size of the string.

Measuring the excitation function of event topologies in $\pi p$ and $pp$ collisions may provide insights into the nature and onset of the underlying particle production mechanisms in this transition region. 
These studies should be complemented by measurements of multi-strange (anti-)baryons, such as $\Omega$ and $\bar{\Omega}$, whose production rates have been proposed as potential indicators for distinguishing between hadron- and parton-dominated dynamics \cite{Bleicher:2001nz}.

\subsubsection{Particle multiplicities in $pp(n)$ reactions as a function of $\sqrt{s}$}

\noindent
The hadron multiplicities as a function of CM energy, from production threshold up to the highest FAIR energies, are poorly known for many species.

\begin{figure}[t]
\begin{center}
\includegraphics[width=0.49\linewidth]{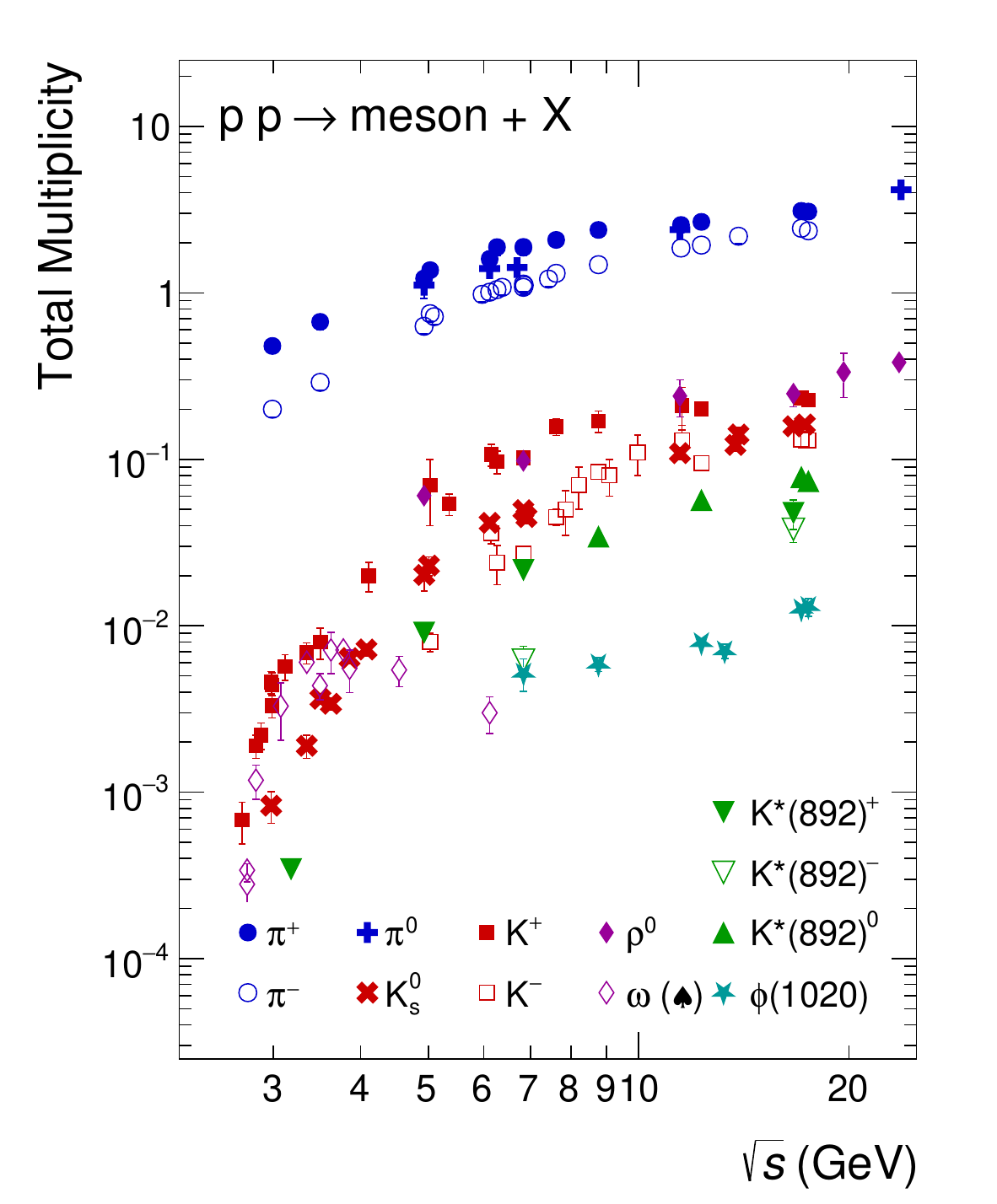}
\includegraphics[width=0.49\linewidth]{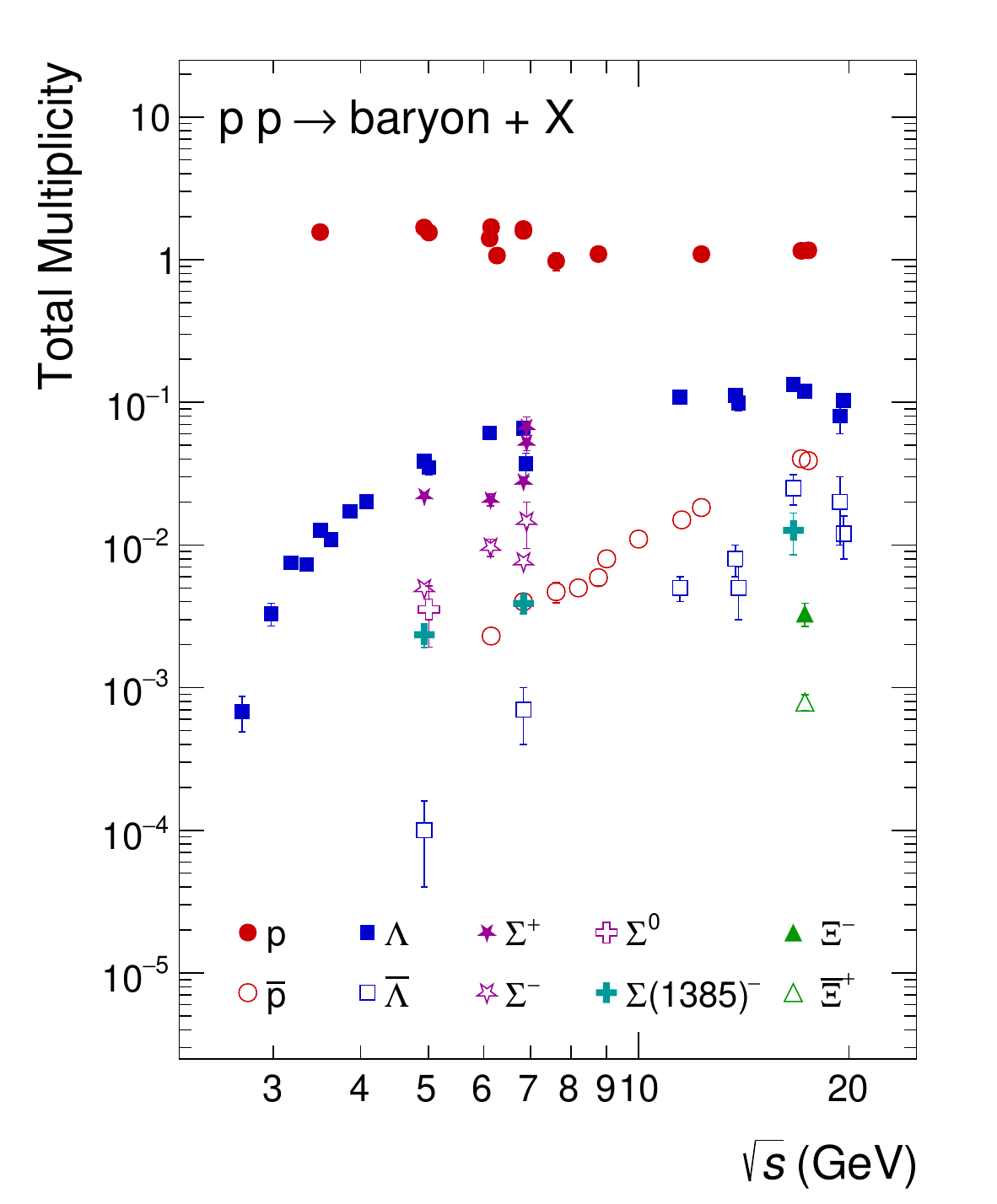}
\end{center}
\vspace*{-4ex}

\caption{
Compilation of measured, inclusive meson (left panel) and baryon (right panel) multiplicities in $pp$ collisions as a function of the CM energy. The data are taken from \cite{Antinucci:1972ib, French-Soviet:1976tin, NA61SHINE:2013tiv, NA49:2005qor, NA61SHINE:2017fne, Schopper:1988hwx, Gazdzicki:1996pk, NA61SHINE:2019xkb, NA49:2009wth, NA61SHINE:2021iay, NA49:2011bfu, Tef, HADES:2015ybo, NA61SHINE:2019gqe, NA49:2000jee, NA49:2009brx, NA61SHINE:2015haq, HADES:2016pau, NA61SHINE:2020dwg, NA61SHINE:2021uyb}. 
For the $\omega$ meson ($\spadesuit$), the exclusive reaction $pp \rightarrow \omega\, pp$ is shown. 
\label{fig:mult_pp_summary}}
\end{figure}

Figure~\ref{fig:mult_pp_summary} compiles available experimental data on inclusive meson (left) and baryon (right) yields in $pp$ collisions over a CM energy range of 2.5--25\,GeV. A significant lack of precise measurements is evident across this range, particularly for strange hadrons, including single- and multi-strange (anti-)baryons ($\Lambda, \bar\Lambda, \Sigma, \Xi, \bar\Xi, \Omega, \bar\Omega$), and resonances involving light and strange quarks ($\rho, \omega, \phi, K^*$).

This lack of precise reference data limits understanding of hadron production mechanisms in HICs. 
As demonstrated for kaons \cite{Hartnack:2011cn}, multiplicity differences between $pp$ and heavy-ion ($AA$) collisions are particularly insightful. Close to the energetic threshold, the excitation functions for $K^+$ and $K^-$ mesons in $pp$ differ substantially. 
In contrast, the yields in $AA$ collisions exhibit a similar energy dependence. 
The enhanced strangeness production in $AA$ is explained in transport models by multistep processes, such as $NN \to \Delta N$ followed by $\Delta N \to K^+\Lambda N$, which allow the required energy for kaon production to be accumulated more efficiently. 
These secondary reactions depend on the density reached in the collision and thus connect the $K^+$ yield to the nuclear equation of state (EoS) \cite{Hartnack:2001zs, Hartnack:2011cn}.

In addition, $K^-$ mesons in $AA$ collisions can be produced via secondary interactions such as $Y N \to N N K^-$ and $\pi Y \to \bar{K} N$, where $Y = \Lambda, \Sigma$ \cite{Barz:1985xc, Hartnack:2011cn}. 
These channels link the $K^-$ yield to the abundance of hyperons, and thus indirectly to the $K^+$ production. 
As the beam energy available at SIS18 is insufficient to study heavier strange baryons in detail, corresponding measurements at FAIR energies are essential. 
Based on these investigations, one expects similar or even larger differences between $pp$ and $AA$ multiplicities for these hadrons, which may help to identify the yet poorly understood production mechanisms in $p A$ and HICs.

\begin{figure}[h!]
\begin{center}
\includegraphics[width=0.4\linewidth]{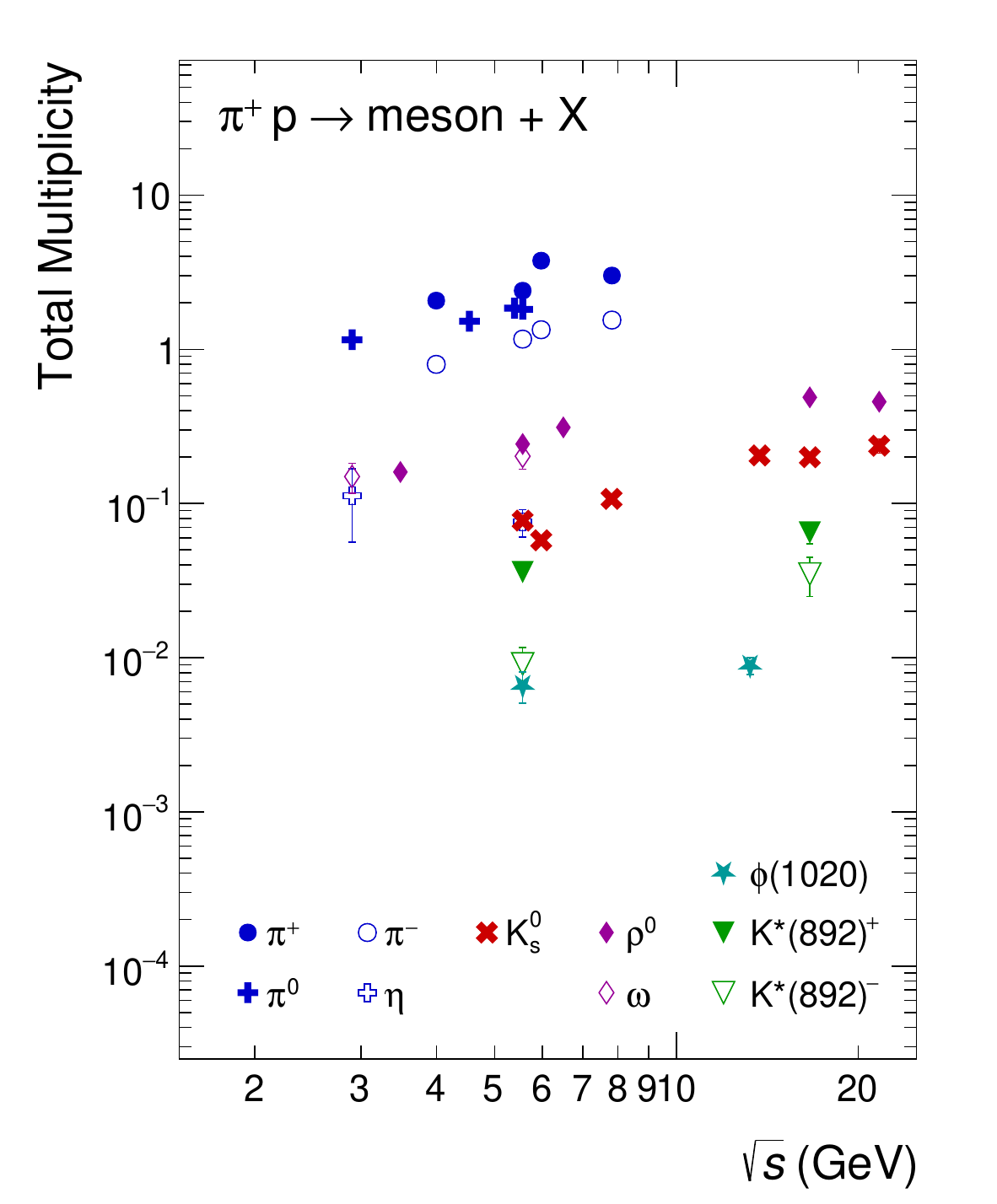}
\includegraphics[width=0.4\linewidth]{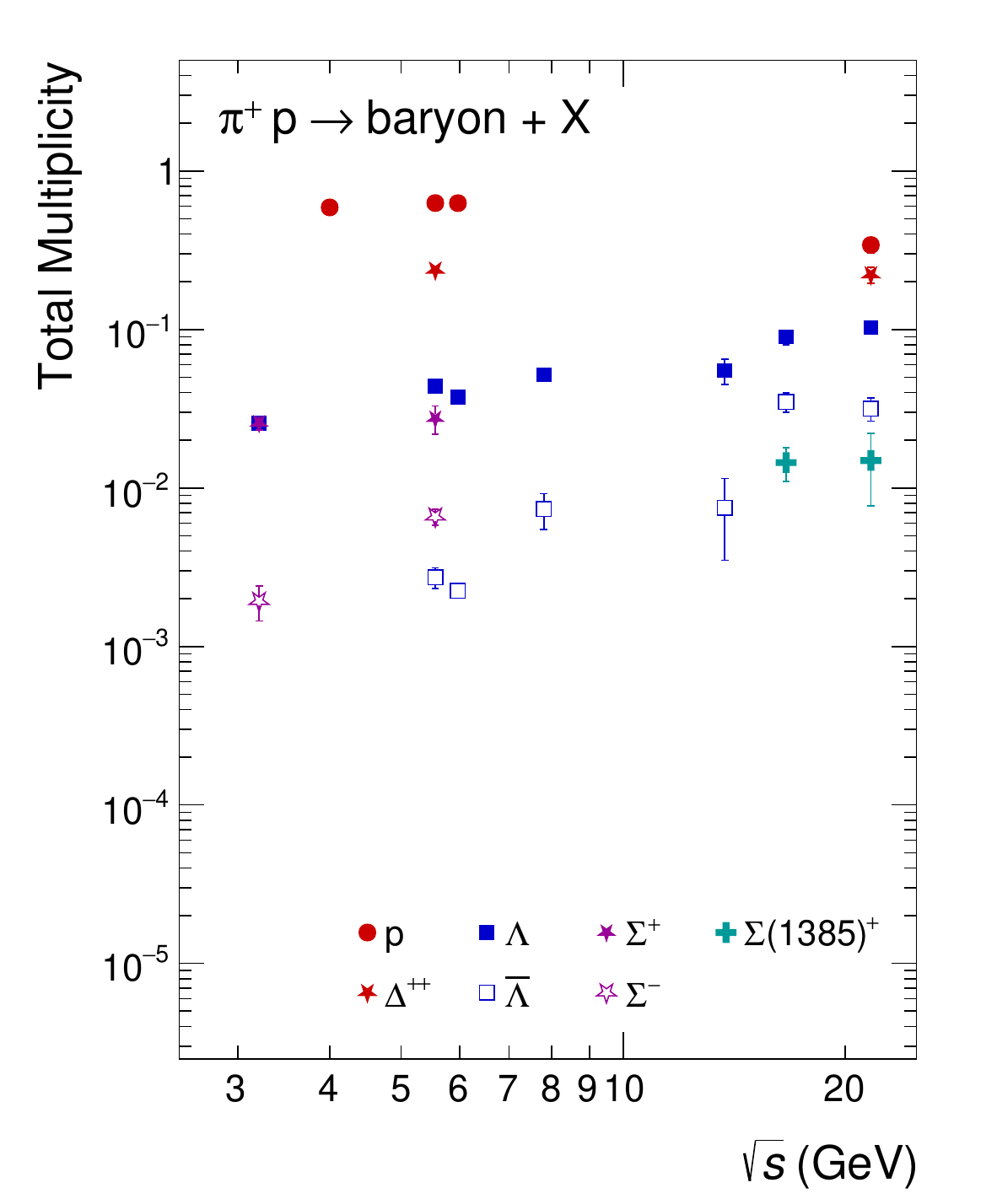}
\end{center}

\vspace*{-4ex}

\caption{Compilation of measured, inclusive meson (left panel) and baryon (right panel) multiplicities in $\pi^{+}p$ collisions as a function of the CM energy. The data are taken from~\cite{Schopper:1988hwx, Azhinenko:1979hk, EHSNA22:1990otw, EHSNA22:1990vem, EHSNA22:1989gyx}.
\label{fig:mult_pip_p_summary}}
%

\begin{center}
\includegraphics[width=0.4\linewidth]{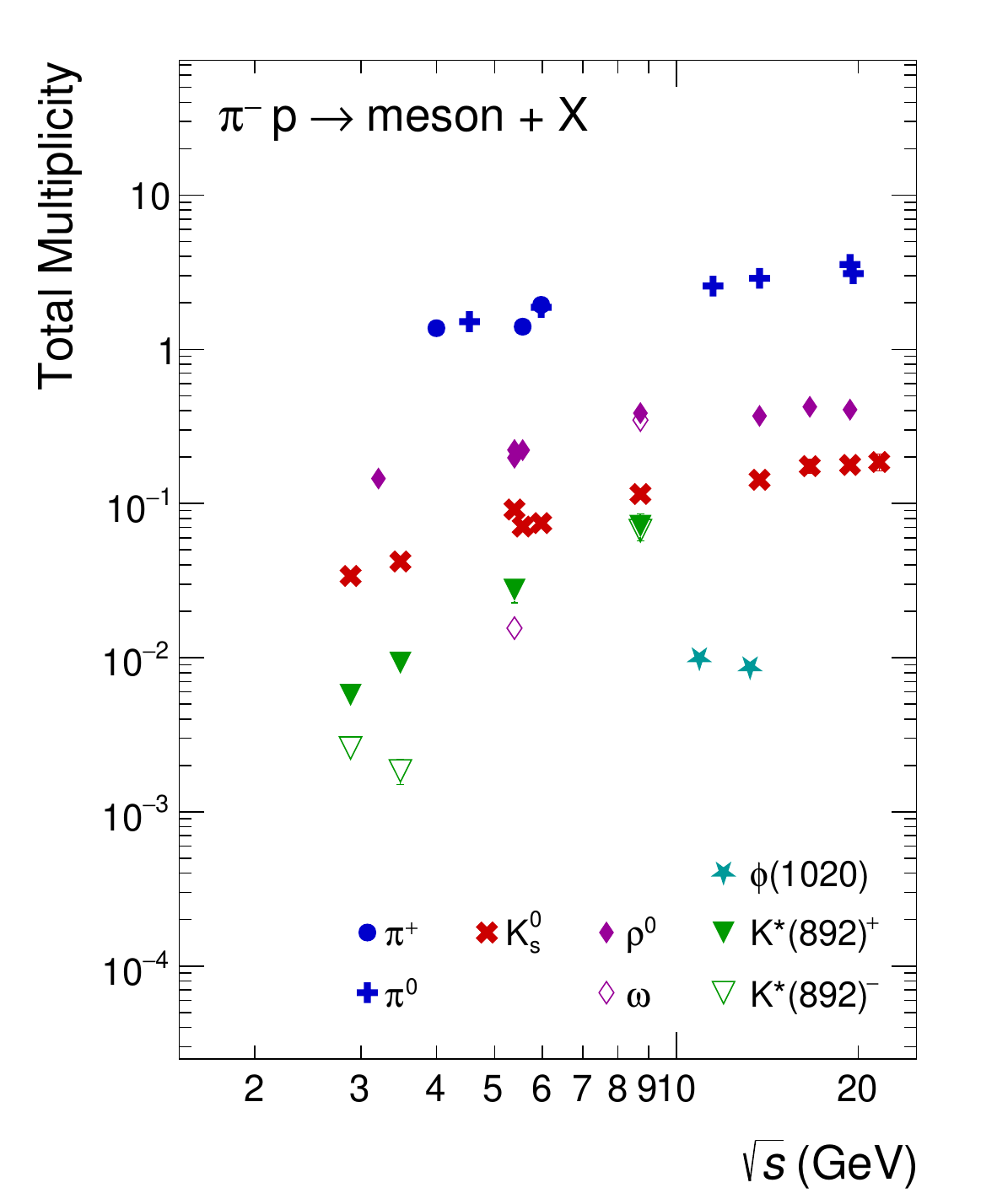}
\includegraphics[width=0.4\linewidth]{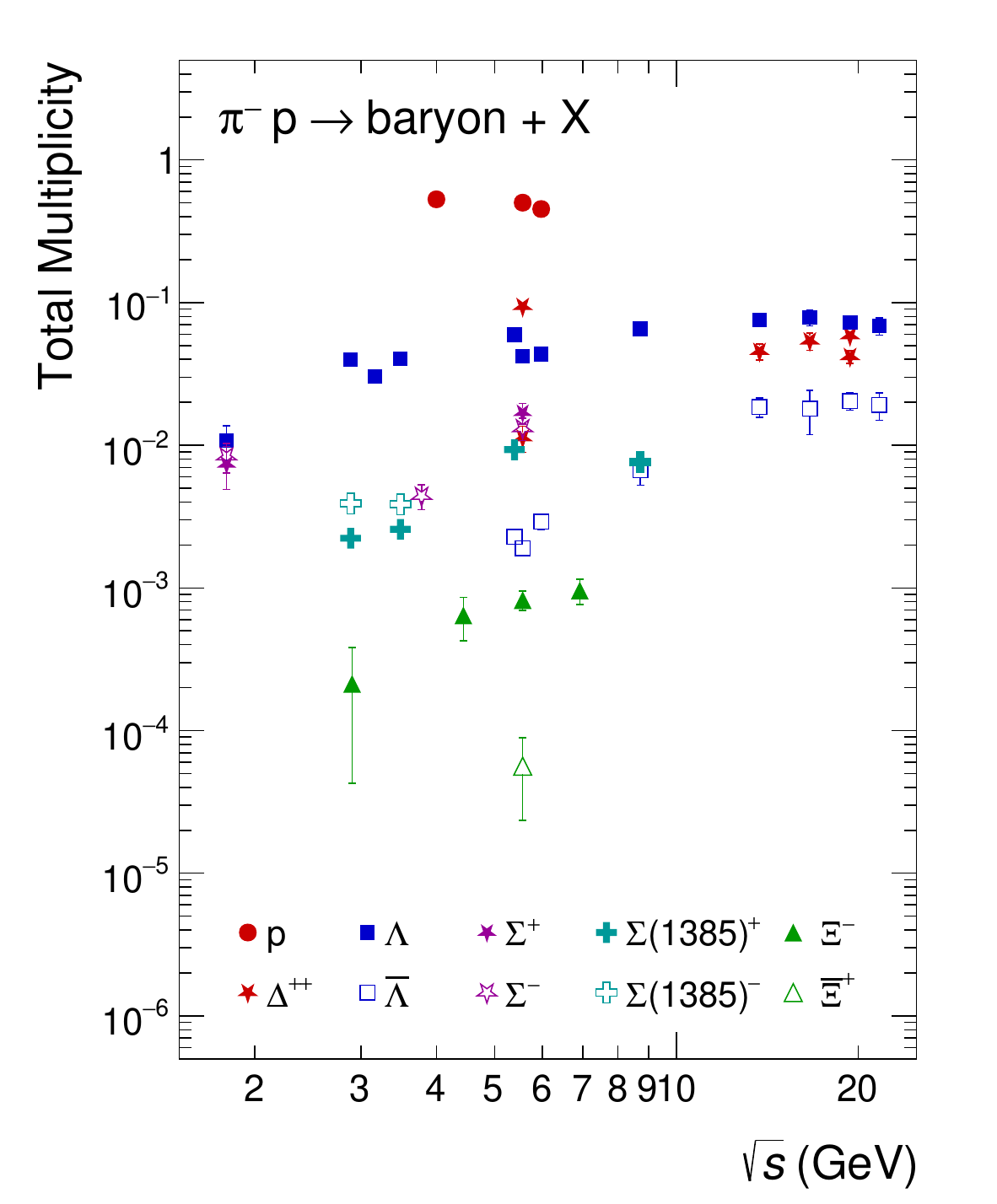}
\end{center}

\vspace*{-4ex}

\caption{Compilation of measured, inclusive meson (left panel) and baryon (right panel) multiplicities in $\pi^{-}p$ collisions as a function of the CM energy. The data are taken from \cite{Schopper:1988hwx}.
\label{fig:mult_pim_p_summary}}
\end{figure}

\subsubsection{Resonance studies in $\pi$-induced reactions}

The FAIR experiments HADES and CBM offer a unique opportunity to investigate pion-induced reactions, in particular $\pi p$ interactions, with high precision. 
These measurements significantly advance the understanding of hadron production mechanisms and serve as essential reference data for heavy-ion physics. 
The importance of meson-baryon reactions increases with collision energy, as they contribute substantially to the total hadron yield. 
Moreover, such reactions play a central role in subthreshold dynamics, notably in the production of antikaons via $\pi + Y \to \bar{K} + N$ processes \cite{Hartnack:2011cn}.

Figures~\ref{fig:mult_pip_p_summary} and \ref{fig:mult_pim_p_summary} show compilations of available world data on $\pi^+{+}p$ and $\pi^-{+}p$ collisions, respectively. 
At high energies, hadron production proceeds predominantly via string fragmentation. At lower energies, relevant for the FAIR programme, the dominant mechanism is the excitation and decay of baryon resonances.

Furthermore, $\pi p$ reactions exhibit larger cross sections than $pp$ in the same energy range and selectively excite baryon resonances in narrow mass windows. 
These features make them ideally suited to probe the structure and decay modes of resonances above the $\Delta(1232)$, a region where experimental data remain scarce. 
Such information is crucial for constraining hadronic models and for interpreting the complex interplay of resonance dynamics and medium effects in HICs at FAIR energies.

Compared to $pp$, the $\pi p$ system offers several key advantages. The absence of initial-state baryon number in the meson beam results in reduced final-state multiplicity and background, simplifying event reconstruction and analysis. 
In particular, $\pi p$ reactions in the energy range covered by the HADES pion beam are dominated by $s$-channel processes, making them well suited for PWA. This contrasts with $pp$ collisions, where complex multiparticle final states hinder resonance identification.

Pion-induced reactions thus allow for clean measurements of the partial decay widths of baryon resonances into specific final states. 
This is of particular relevance for testing theoretical predictions, \textit{e.g}., Ref.\,\cite{Steinheimer:2015sha}, which suggest that some resonances may decay into strange hadrons with small, yet nonnegligible, branching ratios. 
Such decays have not yet been observed \cite{ParticleDataGroup:2024cfk}, but their existence could provide a natural explanation for the unexpectedly high yields of $\phi$ mesons \cite{Steinheimer:2015sha} and strange baryons \cite{Graef:2014mra} near production threshold in HICs \cite{STAR:2021hyx}.

Furthermore, $\pi p$ reactions exhibit larger cross sections than $pp$ in the same energy range and selectively excite baryon resonances in narrow mass windows. 
These features make them ideally suited to probe the structure and decay modes of resonances above the $\Delta(1232)$, a region where experimental data remain scarce. Such information is crucial for constraining hadronic models and for interpreting the complex interplay of resonance dynamics and medium effects in HICs at FAIR energies.

\subsection{$pA$ ($\pi A$) reactions}
\label{sec:piA-reactions}

Proton- and pion-induced reactions on heavy nuclear targets ($p A$ and $\pi A$) provide essential access to the in-medium properties of hadrons in cold nuclear matter at near normal baryon density ($\rho_0$). 
These reactions enable the study of possible modifications of hadron masses, widths, and spectral functions owing to the surrounding nuclear environment, which are key measures for understanding the manifestation of QCD symmetries, such as partial restoration of chiral symmetry, in the medium.
Additionally, $p A$ and $\pi A$ collisions allow detailed investigations of short-range correlations between nucleons within the nuclear target, offering insights into the underlying nuclear structure and dynamics, such as multinucleon effects and high-momentum components of the nuclear wave function.

Equally important, these reactions serve as an intermediate benchmark between elementary nucleon-nucleon  and $\pi$-nucleon  reactions on the one hand, and $AA$ collisions creating hot and dense matter on the other. This step is crucial for disentangling genuine in-medium effects from those arising in the highly dynamical environment of HICs.

In all these areas, experimental data at FAIR energies are eagerly anticipated by the theory community, as they offer a much-needed testing ground for advanced many-body theoretical calculations developed in recent years. 
These measurements have the potential to either confirm or challenge current theoretical models describing hadron behaviour in dense environments, contributing substantially to a deeper understanding of QCD in the nonperturbative regime.

\subsubsection{Short range neutron-proton correlation}

\textcolor{blah}{Short-range correlations (SRC) are discussed here not primarily as a nuclear-structure observable, but because they modify the momentum distribution of the nucleons participating in hadron–nucleus reactions. In particular close to particle-production thresholds, these high-momentum components can substantially modify the available collision energy and therefore represent an important baseline for interpreting hadron production in nuclei.}
SRC in nuclei arise because of the repulsive core and tensor components of the nuclear force, which dominate at distances below approximately $1\,$fm. 
As a result, nucleons can form high-momentum pairs where their relative momenta significantly exceed the average momentum of nucleons in the nucleus. 
Proton–neutron ($pn$) pairs are the dominant type, reflecting the isospin-dependent nature of the nuclear force. 
SRC pairs exist for a very short time and within a small spatial region of the nucleus. 
They account for a significant fraction of the high-momentum distribution of nucleons and influence the dynamics and properties of nuclear matter, including binding energies and density distributions. 
SRC are essential for understanding the structure of nuclei and the behaviour of dense nuclear matter in extreme environments, such as neutron stars and HICs.

At GSI/FAIR, SRC are best studied via measurements of exclusive $A(p,2pN)$ reactions ($N = \text{neutron or proton}$), utilising a 4.5\,GeV/$c$ high-intensity proton beam in combination, \textit{e.g}., with the large-acceptance HADES spectrometer and parts of the NeuLAND detector to measure the recoil neutron and detect the two protons emitted in the reaction.

The scientific goals of the measurements of neutron-proton correlations are as follows.
\begin{enumerate}
    \item Investigate the Migdal-Luttinger jump \cite{Migdal1957} in the nucleonic momentum distribution, which describes the transition from the mean-field region to the high-momentum tail dominated by SRC. 
    The main advantage of HADES is high statistics, which will allow accurate mapping of the expected transition.
    
    \item Study factorisation of the reaction mechanisms at low energies to understand the limits of factorisation in hard breakup reactions, paving the way for a FAIR programme to study SRC in short-lived nuclei in inverse kinematics.
    
    \item Investigate the influence of SRC on hadron multiplicities close to threshold, where the effective CM energy changes when the impinging nucleon ($\pi$) interacts with a nucleon in a correlated pair \cite{Reichert:2025egt}.
\end{enumerate}

Estimates based on detailed simulations predict approximately twenty times more events than currently available JLab $(e,e^{\prime}pn)$ data. 
For more details on the proposed measurement, see \url{https://www.hen-lab.com/srchades}. 
SRC may also play an important role in subthreshold particle production \cite{Reichert:2025egt}.

\subsubsection{Study of the in-medium properties of vector mesons by dileptons} 
\label{subsub:inmediumpropvect}

Dileptons provide a particularly clean and penetrating probe of hot and dense nuclear matter as, once produced, they essentially do not interact with the surrounding medium.
The main dilepton sources are hadronic decays and bremsstrahlung, thermal QGP radiation (including $q+\bar{q} \to e^+e^-$, $q+\bar{q} \to g+e^+e^-$, $q+g \to q+e^+e^-$ processes), primary Drell-Yan production, and semileptonic decays from correlated charm and bottom pairs.
The study of the electromagnetic response of such a medium is closely tied to the investigation of in-medium modifications of vector meson properties. 
Since vector mesons can decay directly into a lepton-antilepton pair, the invariant mass spectra of dileptons offer a means of extracting information on medium-induced changes to specific properties of vector mesons, such as their mass and/or width.

Ultra-relativistic HIC experiments in the 1990s observed an enhancement in dilepton production at low invariant mass compared to conventional hadronic models \cite{Agakishiev:1995xb, Mazzoni:1994rb}. This was later explained by in-medium modifications of the $\rho$ meson, with two main scenarios proposed: a reduction in its mass, as suggested by Brown-Rho scaling \cite{Brown:1991kk} and the Hatsuda-Lee sum rule \cite{Hatsuda:1991ez}, or a broadening of its spectral function, as predicted by many-body hadronic models \cite{Rapp:1997fs, Friman:1997tc, Peters:1997va, Lutz:2001mi}.
Early experiments highlighted the need for in-medium effects but could not distinguish between mass reduction and spectral broadening. 
Higher-resolution NA60 data \cite{Arnaldi:2006jq} strongly favoured broadening (see, \textit{e.g}., Refs.\,\cite{Linnyk:2015rco, Bleicher:2022kcu} and references therein), a conclusion supported by CERES \cite{Adamova:2006nu} and later RHIC experiments by PHENIX \cite{PHENIX:2009gyd} and STAR \cite{STAR:2023wta}.
Theoretical calculations have successfully described dilepton data at intermediate invariant masses only by incorporating in-medium effects; see, again, Refs.\,\cite{Linnyk:2015rco, Bleicher:2022kcu} and references therein.

\begin{figure}[t]
\centering
\includegraphics[width=7.cm]{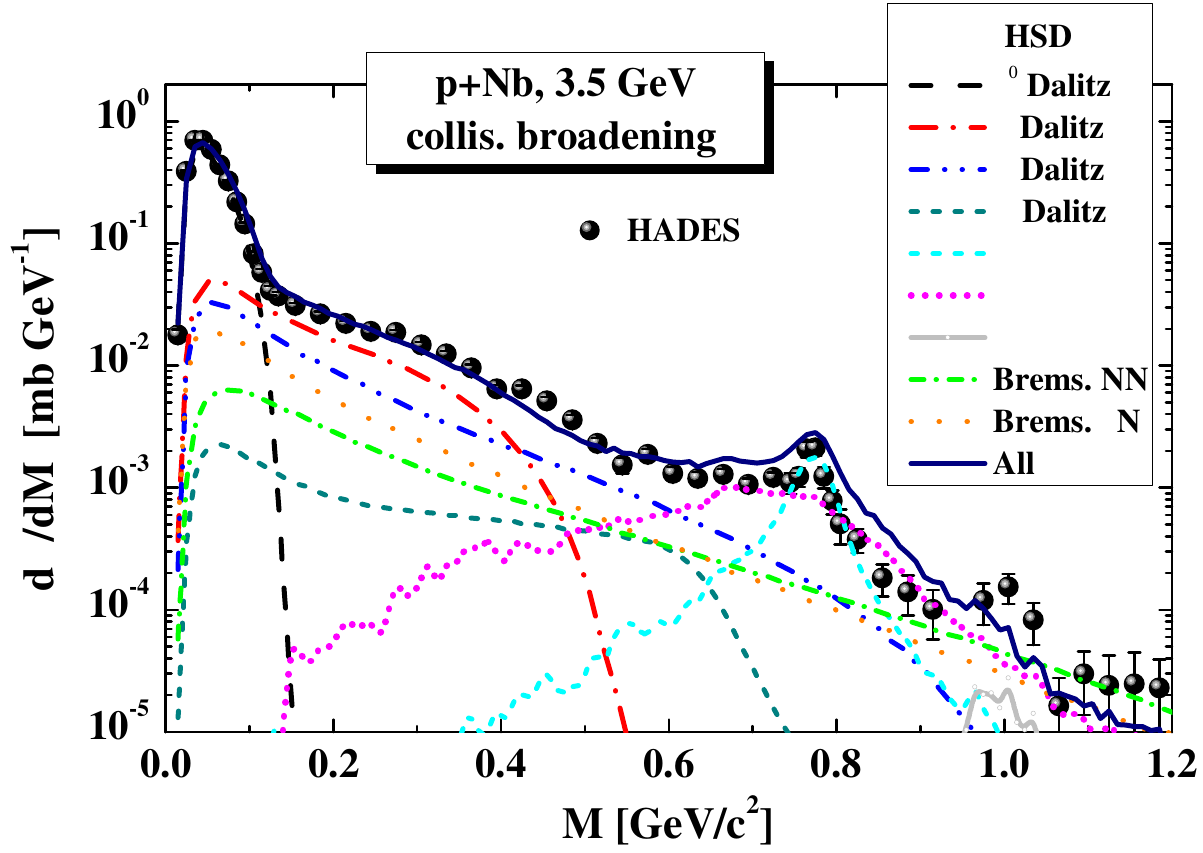}
\hspace*{5mm}
\includegraphics[width=7.cm]{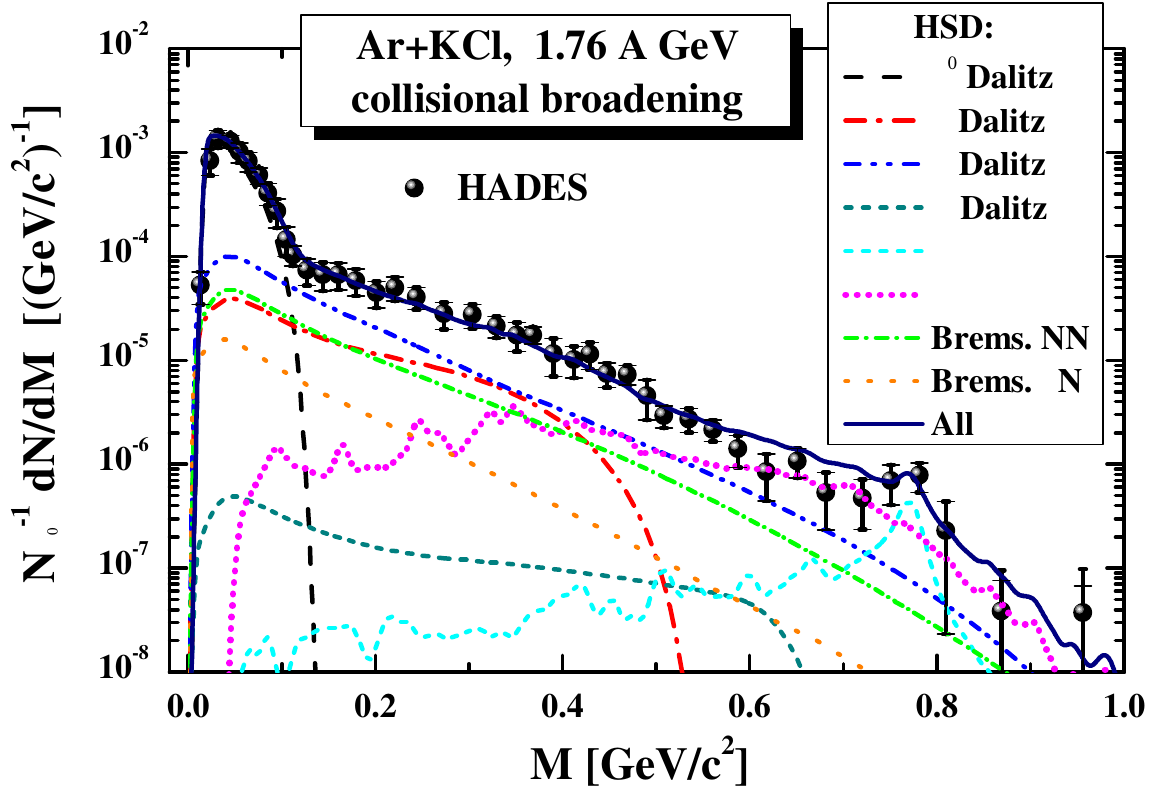}
\caption{ 
Left: Differential cross section $d\sigma/dM$ from PHSD for $e^+e^-$ production in the $p+\text{Nb}$ reaction at a bombarding energy of 3.5\,$A$GeV, in comparison with HADES data \cite{HADES:2012sui}.
Right: Mass-differential dilepton spectra, normalised to the $\pi^0$ multiplicity, from PHSD for Ar+KCl at 1.76\,$A$GeV, in comparison with HADES data \cite{HADES:2011nqx}. The PHSD calculations include the 'collisional broadening' scenario for vector mesons. 
(Figures are adapted from Ref.\,\cite{Bratkovskaya:2013vx}.)
\label{Fig_Dil}}
\end{figure} 

Evidence for in-medium modifications of vector meson properties was also observed in $pA$ and $AA$ reactions at low collision energies of a few~GeV during the 1990s by the DLS Collaboration \cite{Matis:1994tg,Wilson:1997sr}. 
Later, high-statistics measurements by the HADES Collaboration \cite{Agakichiev:2006tg, Pachmayer:2008yn} revealed dilepton spectra with a nearly exponential shape, consistent with microscopic transport model calculations that include in-medium spectral functions for vector mesons \cite{Schmidt:2008hm, Bratkovskaya:2013vx}. 
Figure~\ref{Fig_Dil} shows the parton hadron string dynamics model (PHSD) results from Ref.\,\cite{Bratkovskaya:2013vx} for dilepton spectra including collisional broadening of vector meson spectral functions (lower plots) versus free spectral functions (upper plots). (PHSD is a non-equilibrium microscopic transport approach for the description of HICs \cite{Cassing:2008sv, Cassing:2009vt}.)

The investigation of in-medium modifications of hadrons benefits uniquely from the combination of dilepton spectroscopy with pion-induced reactions. Owing to their lower incident momentum compared to proton beams, pion beams produce secondary particles with reduced recoil momenta. 
This leads to longer interaction times within the nuclear medium, thereby increasing sensitivity to potential modifications of hadronic properties in nuclear matter.

For penetrating probes, such as the decay of vector mesons reconstructed via their dielectron decays, the inside-to-outside emission fraction is significantly enhanced. 
Here, the HADES experiment has a unique advantage over others, such as CLAS and KEK E325 \cite{CLAS:2007dll, Naruki:2005kd}. 
It provides extensive coverage of dilepton pairs with low momentum relative to the nuclear medium, thereby enhancing sensitivity to the expected modifications of hadrons in nuclear matter \cite{HADES:2012sui, Post:2003hu}. 
This makes HADES ideally suited to investigate line-shape and line-strength modifications owing to in-medium effects on the light vector mesons $\rho$, $\omega$, and $\phi$.

FAIR energies provide excellent conditions for probing medium effects on vector mesons in $p(\pi)A$ reactions, in addition to HICs. Simultaneous measurements of $p(\pi)p$, $p(\pi)A$, and $AA$ reactions at the same bombarding energies will allow robust constraints to be placed on production mechanisms, from zero baryon density (elementary reactions), to $\rho_0$ (in $p(\pi)A$), up to high baryon densities (in $AA$ collisions).

\begin{figure*}[t]
    \centering
     \includegraphics[width=0.37\linewidth]{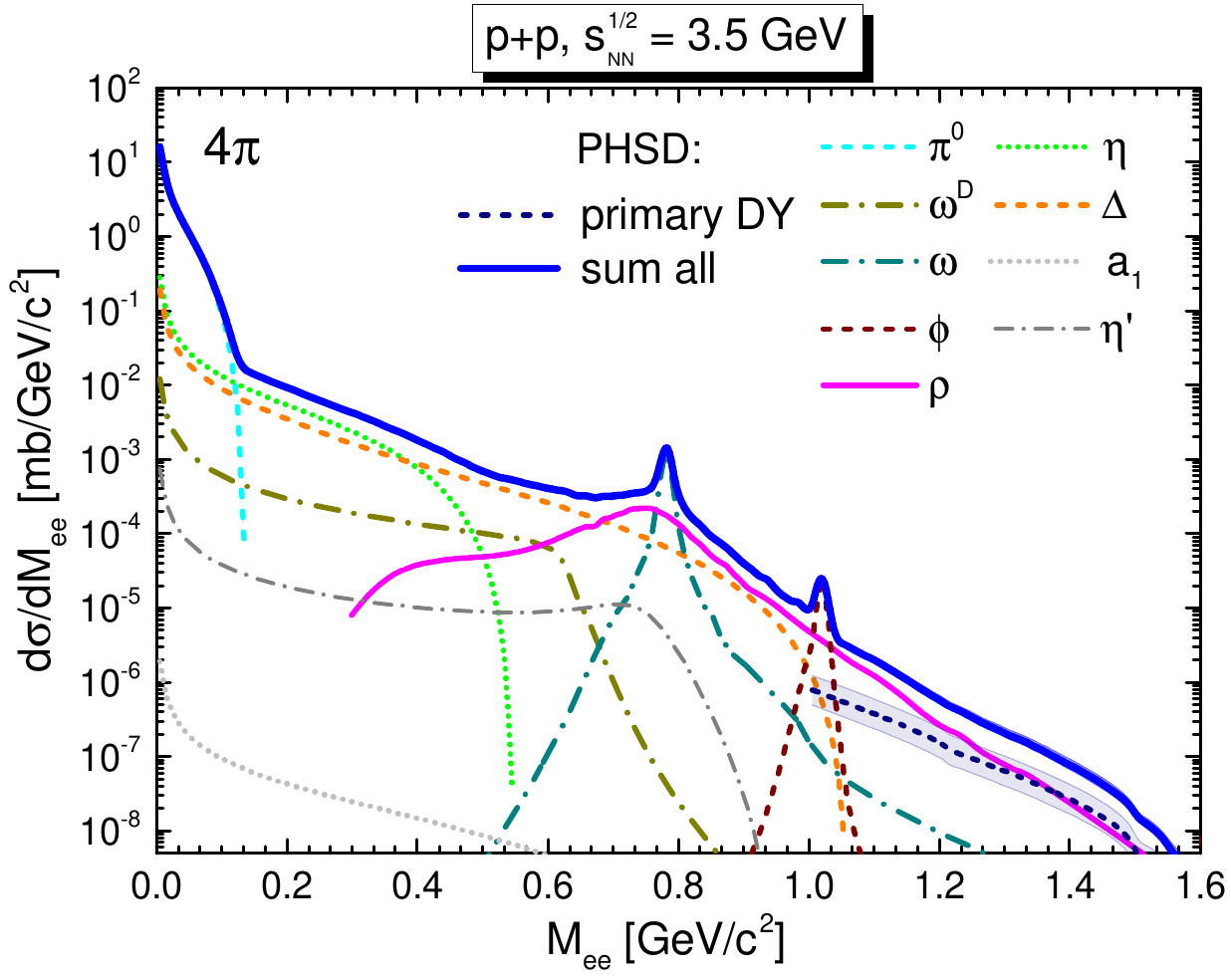}
     \includegraphics[width=0.37\linewidth]{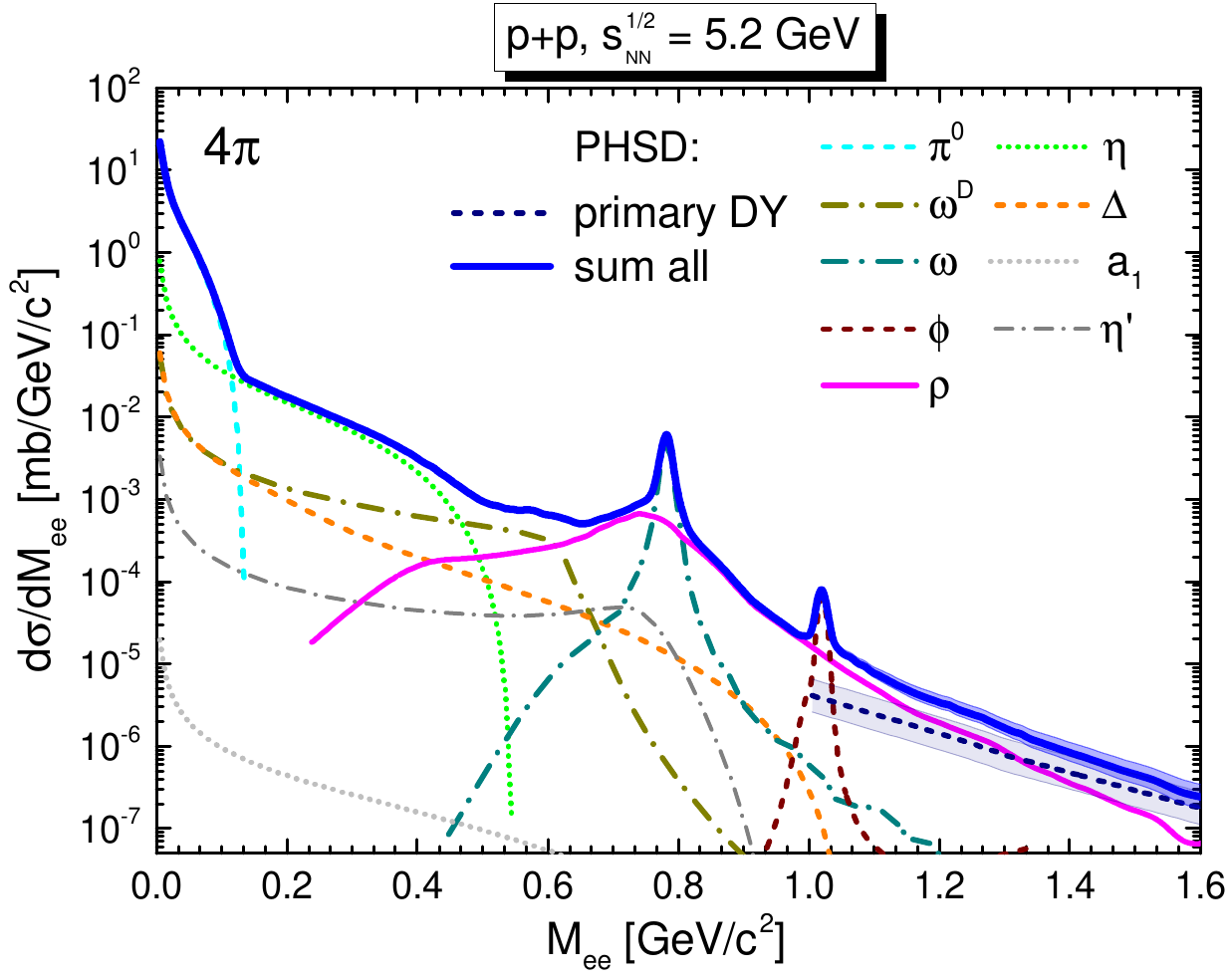}
     \includegraphics[width=0.37\linewidth]{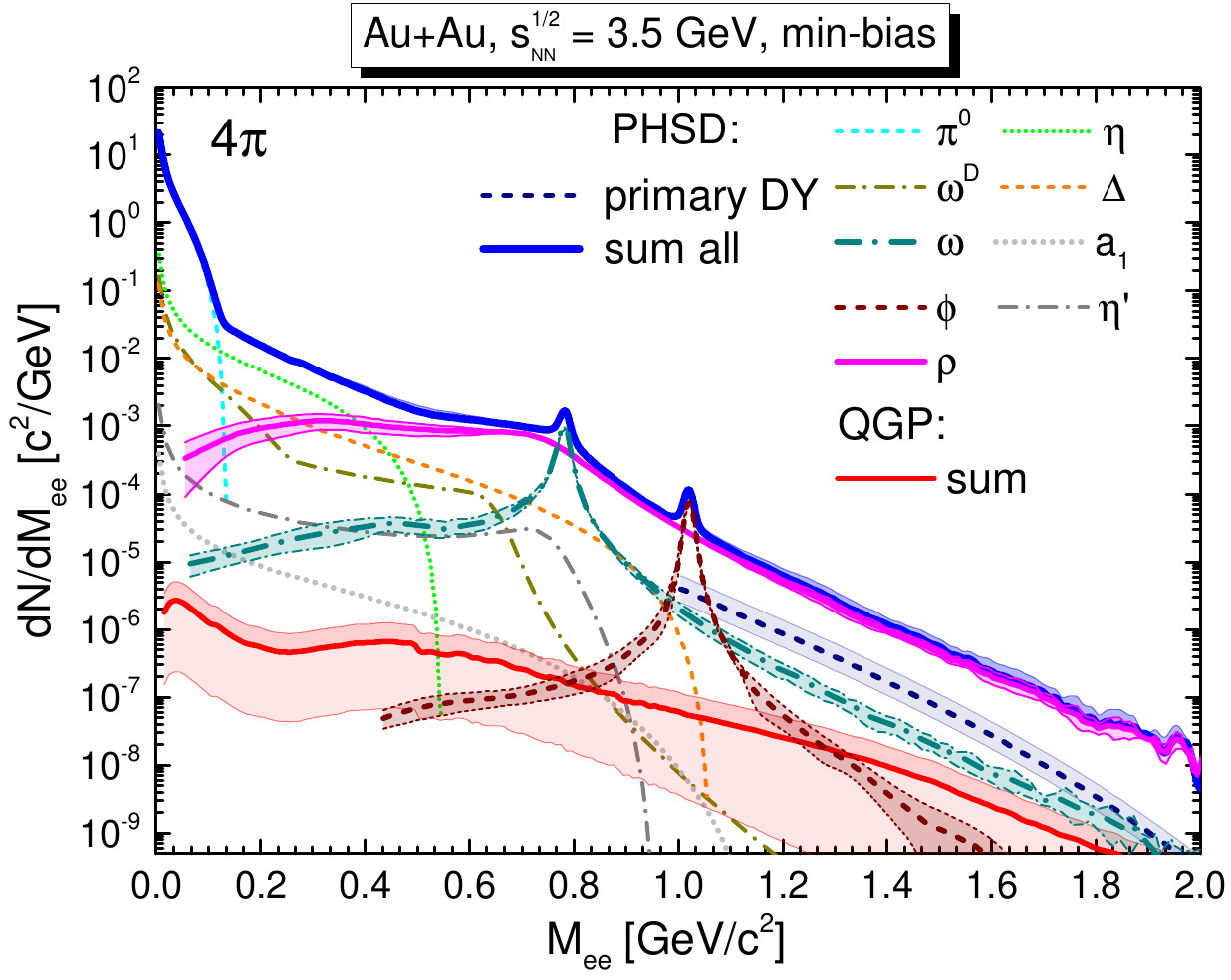}
     \includegraphics[width=0.37\linewidth]{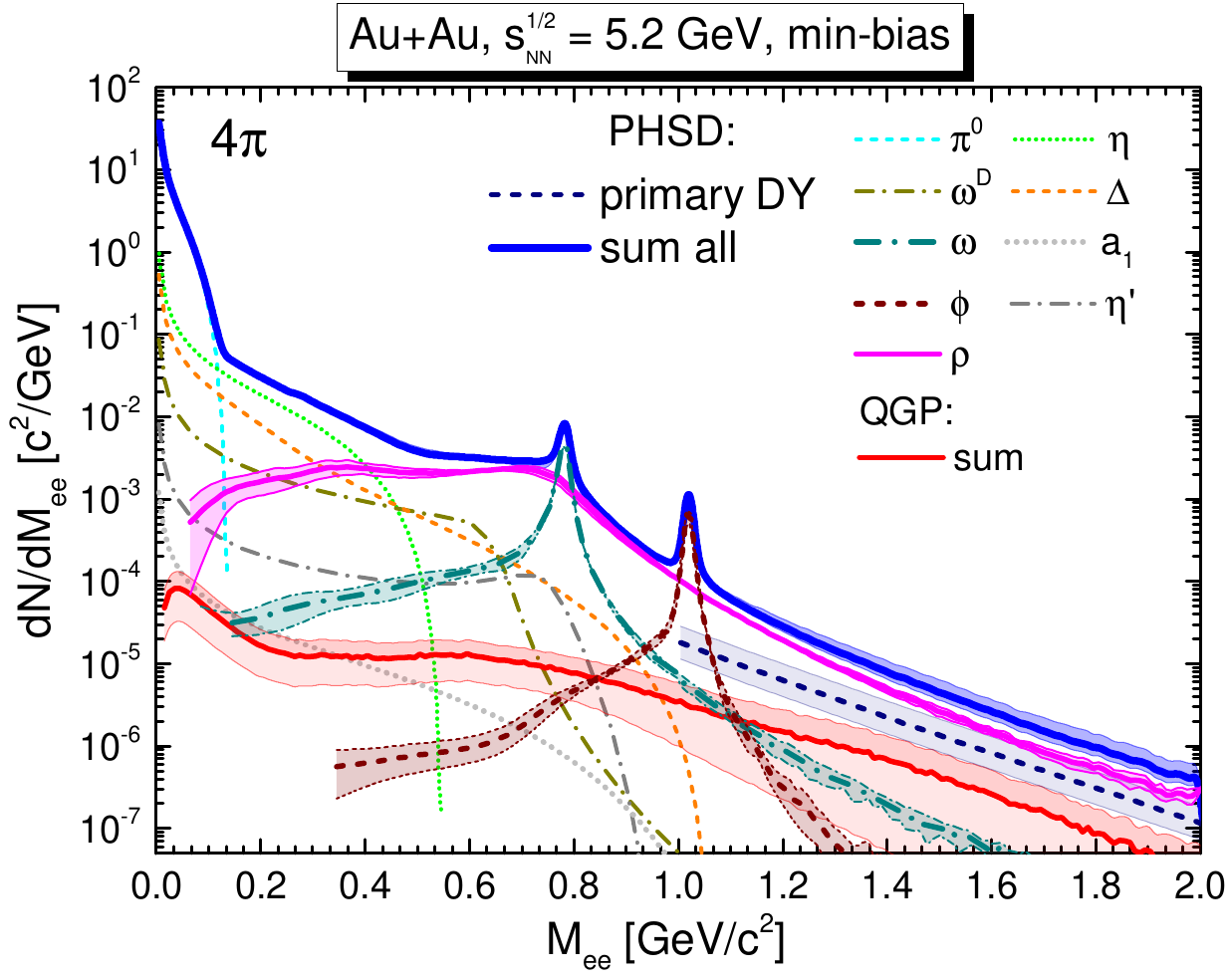}     
    \caption{Invariant mass spectra $d\sigma/dM_{ee}$ of dileptons from PHSD for $pp$ collisions at $\sqrt{s_{NN}} = 3.5\,$GeV (upper left) and 5.2\,GeV (upper right), and mass-differential dilepton spectra $dN/dM_{ee}$ for Au+Au collisions for minimum bias at $\sqrt{s_{NN}} = 3.5$\,GeV (lower left) and 5.2\,GeV (lower right). 
The total yield is shown by the blue lines, while the various contributions are specified in the legends. 
(Figures are adapted from Ref.\,\cite{Jorge:2025wwp}.)
    \label{dil_pp_SIS100}}
\end{figure*}

Fig.~\ref{dil_pp_SIS100} provides the PHSD predictions for the future CBM experiment at SIS100 energies at FAIR, showing the invariant mass spectra for $pp$ and Au+Au collisions at $\sqrt{s_{NN}} = 3.5$ and 5.2~GeV. One observes that the variety of hadronic channels, including vector meson production, is complemented by partonic contributions (QGP), which may become accessible after subtraction of the primary Drell--Yan contributions.


\subsubsection{Drell-Yan processes in $pp$ and $pA$ collisions}

To reliably extract information on the vector meson spectral function at intermediate and high masses, as well as on QGP emission from dilepton spectra in $pA$ and $AA$ collisions, the contribution of primary Drell-Yan (DY) processes ($q + \bar{q} \to \ell^+ \ell^-$, $\ell = e, \mu$), originating from binary nucleon-nucleon ($NN$) collisions, must be well understood and accurately controlled.

As shown in the recent PHSD study \cite{Jorge:2025wwp}, the DY contribution is one of the dominant sources of dileptons at $M > 1\,$GeV in $pp$, $pA$, as well as $AA$ collisions at FAIR energies. This is illustrated in Fig.\,\ref{dil_pp_SIS100} for $pp$ and Au+Au collisions at $\sqrt{s_{NN}} = 3.5, 5.2\,$GeV.

An accurate measurement of high-mass dilepton spectra in $pp$ collisions is mandatory, while a theoretical modelling of the DY process at such low energies remains an open question. 
As long as the invariant mass of the dilepton pair is sufficiently large, perturbative QCD and the factorisation formalism are applicable and the DY process can be estimated. 
However, for $\sqrt{s_{NN}}$ of a few GeV, DY dileptons are produced by large-$x$ partons, where the PDFs are poorly constrained and associated with large uncertainties \cite{Jorge:2025wwp}.

The FAIR experiment will provide a unique opportunity to study DY production at intermediate and low energies, \textit{i.e}., at the boundary between partonic and hadronic degrees of freedom. 
High-precision measurements of dilepton invariant mass spectra, transverse momentum ($p_T$) distributions, and, in particular, dilepton angular distributions in different mass bins will help disentangle the DY contribution from other sources.

Moreover, dimuons from DY processes in SIS100 $pA$ collisions have recently been identified as a useful probe for the unambiguous determination of the stopping power of cold nuclear matter for energetic partons; see Ref.\,\cite{Bhaduri:2017ptw} for details. 
In these reactions, the projectile quarks propagate through the cold nuclear matter of the target nucleus and may undergo energy loss owing to multiple scattering before the $q\bar{q}$ annihilation takes place. 
As final-state interactions are negligible for the DY process, the resulting dimuons are considered an ideal tool to probe initial-state quark energy loss in nuclear interactions. 
Modifications of the DY dimuon spectrum as a function of Bjorken-$x$ or $x_F$ in $pA$ collisions with varying target nucleus size are regarded as suitable observables for this purpose. 
However, strong shadowing effects introduce significant uncertainties in determining the magnitude of energy loss extracted from DY data.

The DY data measured by the Fermilab E866 (NuSea) Collaboration in $pA$ collisions can be simultaneously described with or without including energy loss effects, provided that nuclear modifications of parton distribution functions are taken into account. 
The situation has recently improved with the availability of 120\,GeV $pA$ DY data from the Fermilab E906 (SeaQuest) experiment. 
The lower beam energy and the selected dimuon mass range allow probing a kinematic region where shadowing effects play a minimal role.

The observed suppression of the DY dilepton production cross section with increasing Feynman-$x$ ($x_F$) can unambiguously be attributed to quark energy loss in the nuclear medium. However, the currently available data do not permit a determination of the path-length dependence of the average energy loss inside the nucleus. 
Various phenomenological models that assume linear or quadratic dependence of the energy loss on the nuclear path length can both describe the E906 data within uncertainties.

High-precision DY data to be collected at FAIR/SIS100 will improve this situation. 
Theoretical calculations \cite{Giri:2025bfq} predict distinguishable suppression patterns in the DY cross section for models with different path-length dependence of the beam quark energy loss. 
Such measurements would provide key insights into the underlying energy loss mechanisms in cold nuclear matter.

\textcolor{blah}{We note, that Drell-Yan measurements at SIS100 will be experimentally challenging because of the small production cross sections and the presence of sizeable competing dilepton sources in the relevant invariant-mass region. The achievable sensitivity will therefore depend strongly on the collected statistics, detector acceptance, and the efficiency of background and source separation. Consequently, the prospects discussed here should be regarded as a long-term opportunity whose quantitative reach has to be established by dedicated feasibility studies.
}

 \subsubsection{Anisotropy of dilepton emission}

Apart from the differential $e^+e^-$ spectra, the investigation of lepton angular distributions is also promising, since the virtual photon produced in hadronic interactions is polarized. 
The coupling of a virtual photon to hadrons induces a dynamical spin alignment of both the resonances and the virtual photons owing to conservation laws. 
Consequently, the angular distribution of the lepton pair with respect to the virtual photon momentum becomes anisotropic \cite{Hoyer:1986pp, Bratkovskaya:1995kh, Ipp:2007ng}.

Isotropisation of dimuon angular distributions up to the GeV mass range, associated with thermalisation, was experimentally observed for the first time by the NA60 collaboration \cite{NA60:2008iqj}. 
Lepton angular distributions at high invariant masses (above the $J/\psi$ region) have been studied both experimentally \cite{Saclay:1979yqw, Anderson:1979tt} and theoretically \cite{Stroynowski:1980kc}. 
At these scales, the dominant process is quark-antiquark annihilation via the DY mechanism. 
Earlier measurements showed small deviations from quark-parton model predictions, which were found to be consistent with detailed QCD calculations. 
These angular characteristics have proven effective in discriminating between theoretical models.

References~\cite{Bratkovskaya:1995dg, Bratkovskaya:1995my, Bratkovskaya:1996nf, Speranza:2018osi} proposed to use lepton pair angular distributions in $pp$, $pA$, and $AA$ collisions to distinguish between different sources in the low invariant mass region ($M < 1$~GeV), where many dilepton channels contribute. 
It has also been shown that, owing to the spin alignment of the virtual photon and of the colliding or decaying hadrons, the decay anisotropy of the lepton pair is sensitive to the specific hadronic production mechanism.

The first measurement of dilepton anisotropy at low energies was performed by the HADES collaboration \cite{HADES:2011nqx}, which observed a substantial transverse polarisation in Ar+KCl collisions at 1.76\,AGeV, originating from hadronic channels, in agreement with theoretical predictions.

The high-luminosity data expected at FAIR will allow for the measurement of dilepton angular distributions in different invariant-mass bins. Extraction of the anisotropy coefficients with high precision will provide information on the polarisation of the dilepton sources and aid in the differentiation of production channels and the validation of theoretical models.

\subsubsection{In-medium properties of strange hadrons} 
\label{subsub:inmediumpropstrbar}
\noindent {\bf Strange pseudoscalar mesons in nuclear matter} \\
\noindent The in-medium properties of strange pseudoscalar mesons, namely kaons ($K$) and antikaons ($\bar{K}$), have been extensively studied over the years using a variety of theoretical approaches, including Nambu-Jona-Lasinio (NJL) models \cite{Lutz:1994cf}, relativistic mean-field (RMF) theories \cite{Schaffner:1996kv}, and quark-meson coupling (QMC) models \cite{Tsushima:1997df}. 

The behaviour of kaons in dense matter can be reasonably well described using the low-density theorem, whereby the $KN$ interaction is approximated by free-space $KN$ scattering. This approximation is justified by the fact that the propagation of kaons in nuclear matter does not differ significantly from their propagation in vacuum.

However, the low-density theorem is not applicable to antikaons, as the $\bar{K}N$ interaction near threshold is strongly influenced by the presence of the $\Lambda(1405)$ resonance. In this context, unitarised coupled-channel approaches in dense matter have proven to be effective tools for describing the properties of $\bar{K}$ mesons in the medium. 
The idea is to address meson-baryon scattering in matter by solving an equivalent equation to the Bethe-Salpeter in a dense medium, introducing medium corrections in the intermediate coupled-channel meson-baryon propagators.

The $KN$ effective interaction has also been obtained using unitarised coupled-channel approaches, starting from an SU(3) chiral effective Lagrangian ($\chi$EFT) as the interaction kernel \cite{Kaiser:1995eg,Oset:1997it}, and incorporating medium corrections to the $KN$ amplitude, such as Pauli blocking, Fermi motion in the kaon dispersion relation, and a self-consistent treatment of the kaon propagator. 
These studies report small changes, about 10\% or less, in the kaon mass \cite{Waas:1996fy, Korpa:2004ae, Tolos:2005jg}, thereby confirming the validity of the low-density approximation.

For antikaons in matter, coupled-channel unitarised theories have been developed, starting either from ChEFT with strangeness \cite{Waas:1996fy, Lutz:1997wt, Ramos:1999ku, Tolos:2006ny, Tolos:2008di} or from meson-exchange models \cite{Tolos:2000fj, Tolos:2002ud}. 
The in-medium $\bar{K}N$ interaction includes effects such as Pauli blocking and the self-energy (or single-particle potential) of the mesons and baryons involved in the intermediate propagators. 
In particular, the $\bar{K}$ self-energy is obtained by solving a Dyson--Schwinger equation, typically at lowest order, coupled to the in-medium $\bar{K}N$ interaction. 
This is because the dressed $\bar{K}$ propagator, and thus the self-energy, enters the meson-baryon intermediate loops involving antikaons and nucleons; and the self-energy itself depends on the in-medium $\bar{K}N$ interaction, necessitating a self-consistent procedure. 

Once the $\bar{K}$ self-energy is determined, the $\bar{K}$ spectral function can be computed from the imaginary part of the dressed propagator. For further details, see Ref.\,\cite{Tolos:2020aln}.

\begin{figure}[t]
\begin{center}
\includegraphics[height=0.4\textwidth, width=0.45\textwidth]{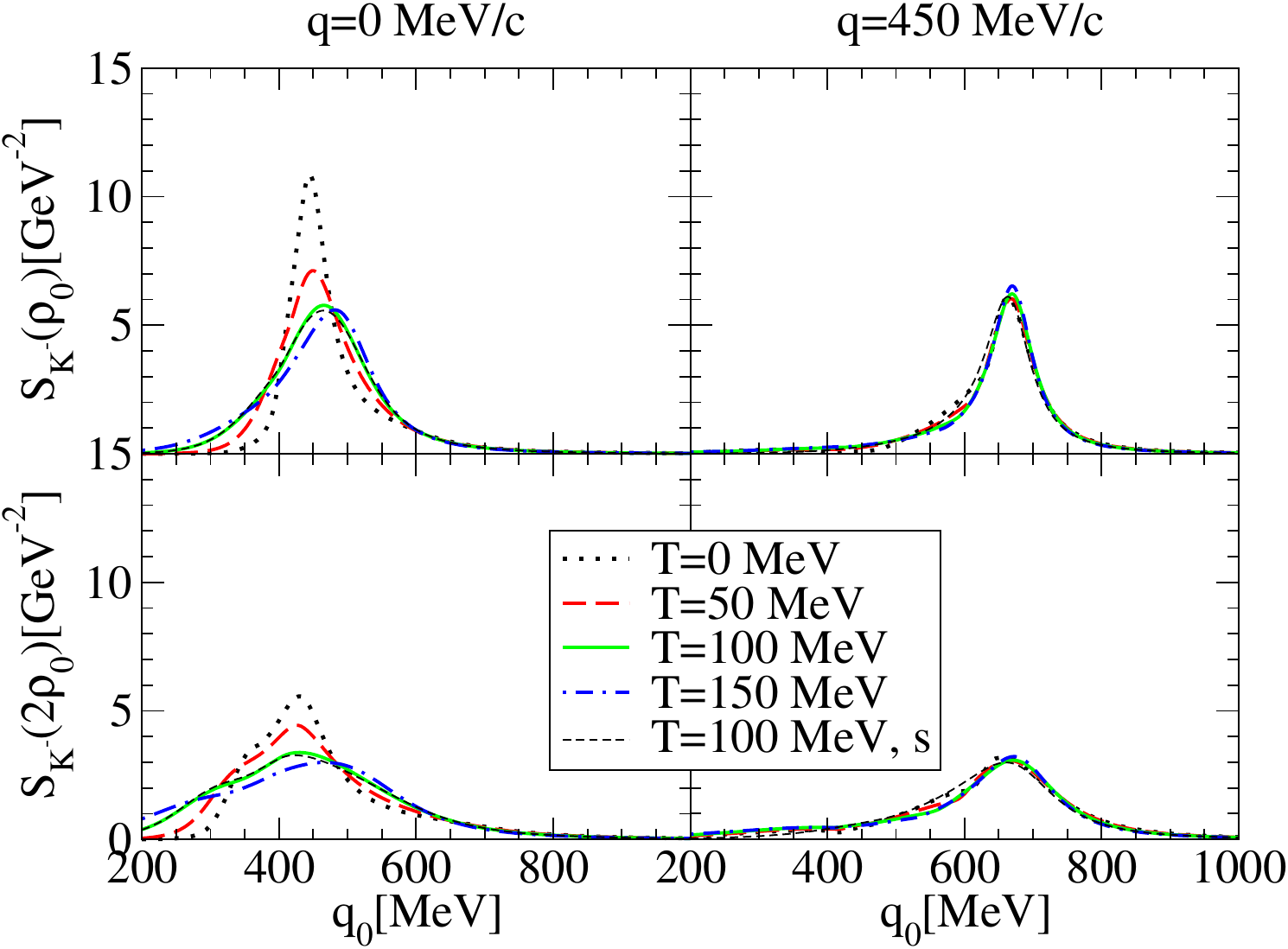}
\hfill
\includegraphics[height=0.4\textwidth, width=0.45\textwidth]{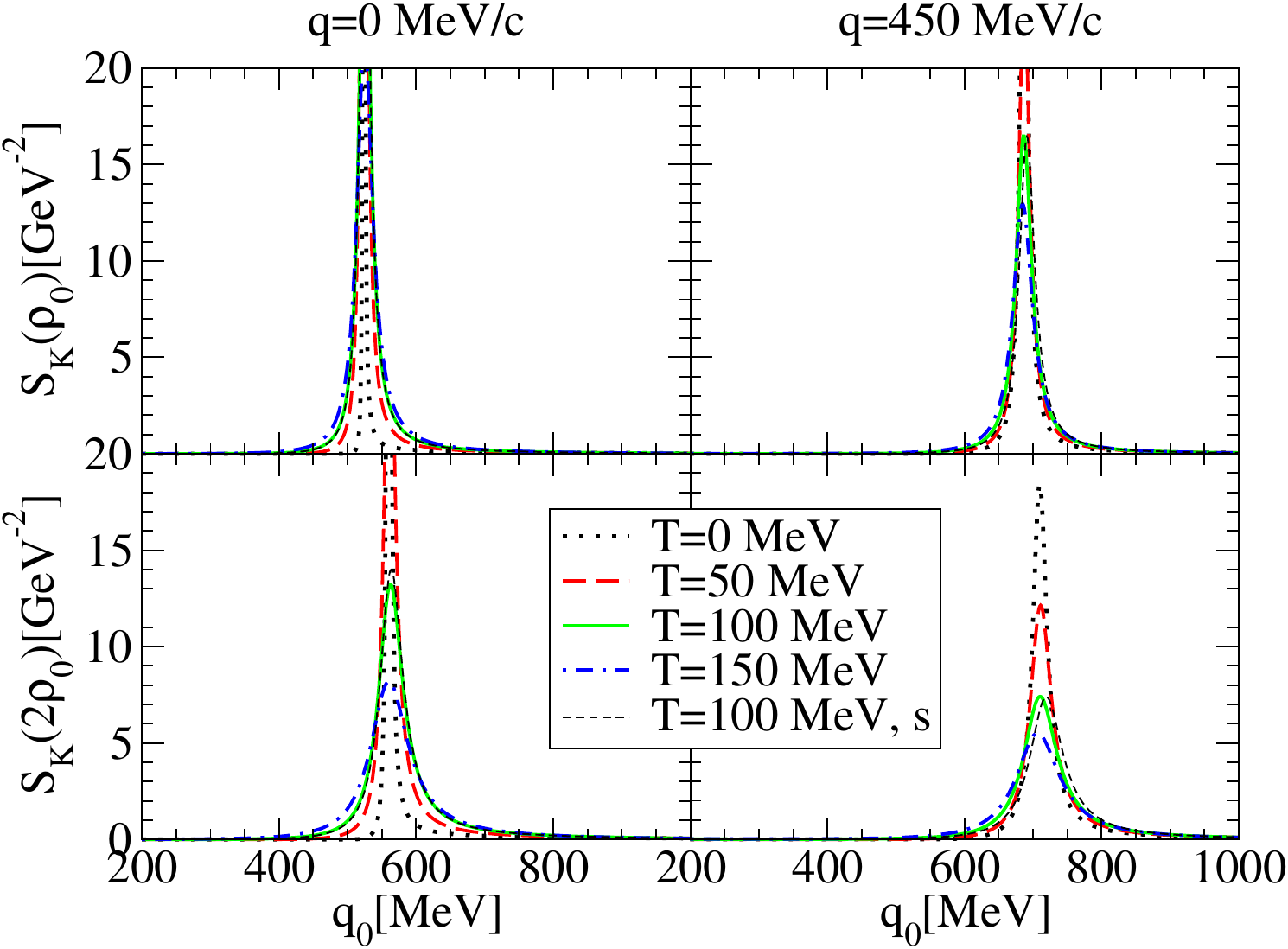}
\caption{ 
The $\bar{K}$ (left) and $K$ (right) meson spectral functions for $q = 0, 450\,$MeV/$c$ at $\rho_0$ and $2\rho_0$ are shown as functions of the meson energy, and for different temperatures, based on a self-consistent calculation including the dressing of baryons and pions. 
(Figures adapted from Ref.\,\cite{Tolos:2008di}.)
 \label{fig1}}
\end{center}
\end{figure}

In Fig.\,\ref{fig1}, the $\bar{K}$, $K$ spectral functions are shown, obtained from the $s$- and $p$-wave in-medium $KN$ and $\bar{K}N$
interactions within the chiral unitary approach of Refs.\,\cite{Tolos:2006ny, Tolos:2008di}. 
The $\bar{K}$ spectral function (Fig.\,\ref{fig1}\,-\,left) displays a broad peak resulting from the mixing between the quasi-particle peak and the $\Lambda(1405)N^{-1}$, as well as the $YN^{-1}$ excitations, where $Y = \Lambda,\Sigma$ and $N^{-1}$ denotes a nucleon hole. 
These $p$-wave $YN^{-1}$ subthreshold excitations appear mainly for $\bar{K}$ at finite momentum. 
Medium effects, such as temperature and density, soften the $p$-wave contributions to the spectral function at the quasiparticle energy. 
The $K$ spectral function (Fig.\,\ref{fig1}\,-\,right) is characterised by a narrow quasiparticle peak that diminishes with increasing temperature and density as the phase space for $KN$ states increases, while also shifting to higher energies at larger densities owing to the repulsive nature of the $KN$ interaction in matter.

\begin{figure}[t]
\centering
\includegraphics[width=0.60\textwidth,clip]{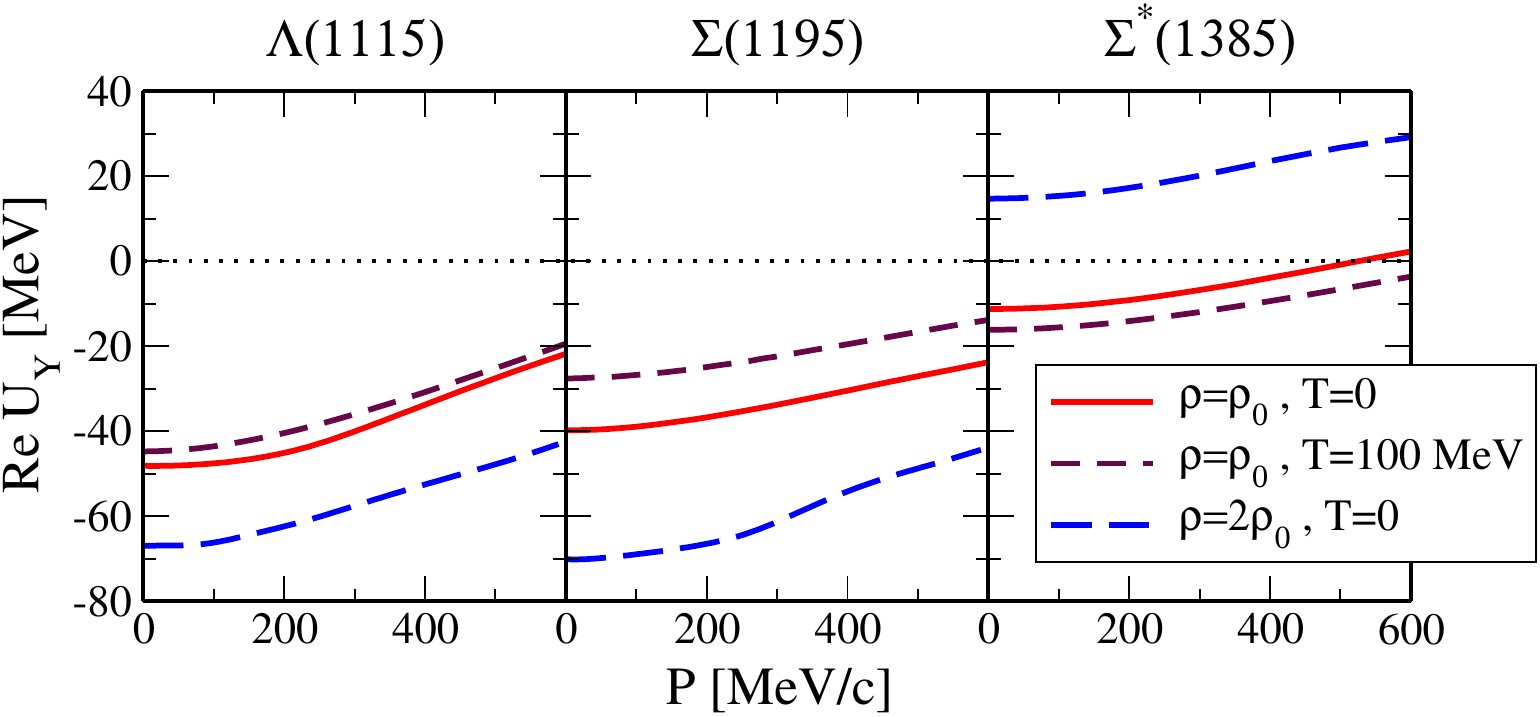} 
\\[2ex]

\includegraphics[width=0.60\textwidth,clip]{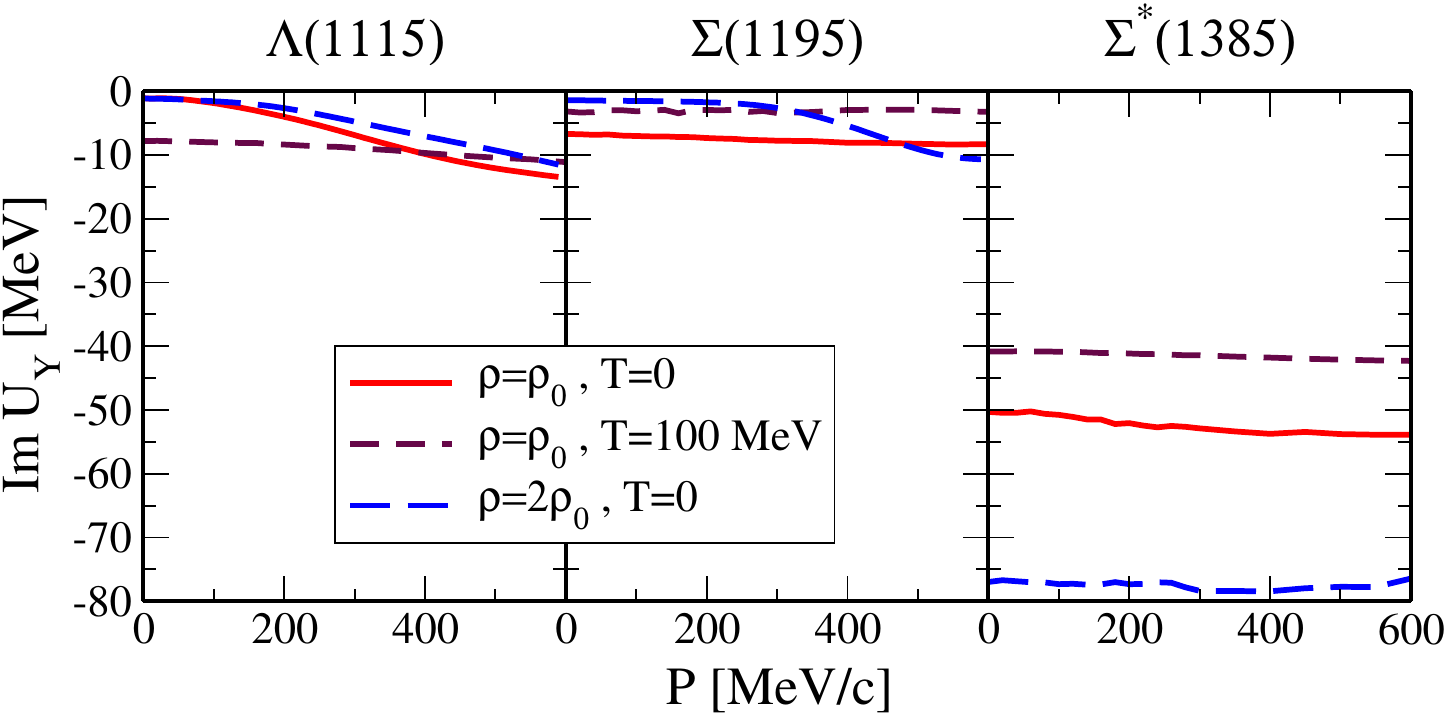} 
\caption{ Real (upper) and imaginary part (lower) of the hyperon potentials. (Images adapted from Ref.\,\cite{Cabrera:2014lca}.)
\label{fig:kmp-vs-kmn}}       
\end{figure}

Furthermore, it is possible to extract information on the behaviour of $\Lambda(1115)$, $\Lambda(1405)$, $\Sigma(1195)$ and $\Sigma^*(1385)$ in matter by analysing the in-medium amplitudes. 
Figure~\ref{fig:kmp-vs-kmn} shows the single-particle potentials of $\Lambda(1115)$, $\Sigma(1195)$ and $\Sigma^*(1385)$ hyperons at finite momentum, density and temperature. 
As for the $\Lambda$ and the $\Sigma$ potentials, these are roughly $-50$ and $-40$~MeV, respectively, at $\rho_0$, and zero temperature. 
Both hyperons develop a finite decay width with increasing density and temperature, as they can be absorbed in the nuclear medium or undergo quasielastic scattering processes. 
With regard to the $\Sigma^*$, it develops an attractive potential of about $-10\,$MeV at $\rho_0$ and zero temperature, which becomes repulsive at higher densities, while its decay width is enhanced at finite density owing to the opening of new absorption channels. 
The temperature effect in this case is moderate owing to the substantial phase space already available at zero temperature. 
Moreover, the $\Lambda$, $\Sigma$ and $\Sigma^*$ potentials show a smooth behaviour with momentum \cite{Cabrera:2014lca}.

In order to access the medium-modified properties of the antikaon, one can study the formation and propagation of antikaons in the dense medium created in HICs at intermediate beam kinetic energies (in the GeV regime). 
To do so, the use of transport models is essential, as these provide the link between the experimental observables and the underlying physical processes; see, \textit{e.g}., Ref.\,\cite{Hartnack:2011cn}.

In Refs.\,\cite{Cassing:1999es, Hartnack:2001zs}, the first transport calculations for antikaons were performed, neglecting the finite width of the antikaon spectral function. 
Later, antikaon production was studied using offshell dynamics with in-medium spectral functions in the hadron string dynamics (HSD) transport model \cite{Cassing:2003vz}, employing the J\"ulich meson-exchange model for the $\bar{K}N$ bare interaction \cite{Tolos:2000fj, Tolos:2002ud}. More recently, the in-medium effects on strangeness production in HICs at (sub)threshold energies have been analysed using the microscopic PHSD transport approach, incorporating in-medium antikaon properties from the ChEFT scheme of Refs.\,\cite{Tolos:2006ny, Cabrera:2014lca}. 
The rapidity distributions, $p_T$ spectra, and polar and azimuthal angular distributions, as well as the directed and elliptic flow of strange mesons in C$+$C, Ni$+$Ni, and Au$+$Au collisions, have been studied and compared with experimental data from the KaoS, FOPI and HADES Collaborations \cite{Song:2020clw}. 
This study concluded that modifications of strange meson properties in nuclear matter are necessary to explain the data consistently.

\begin{figure}[t]
\includegraphics[width=0.55\textwidth]{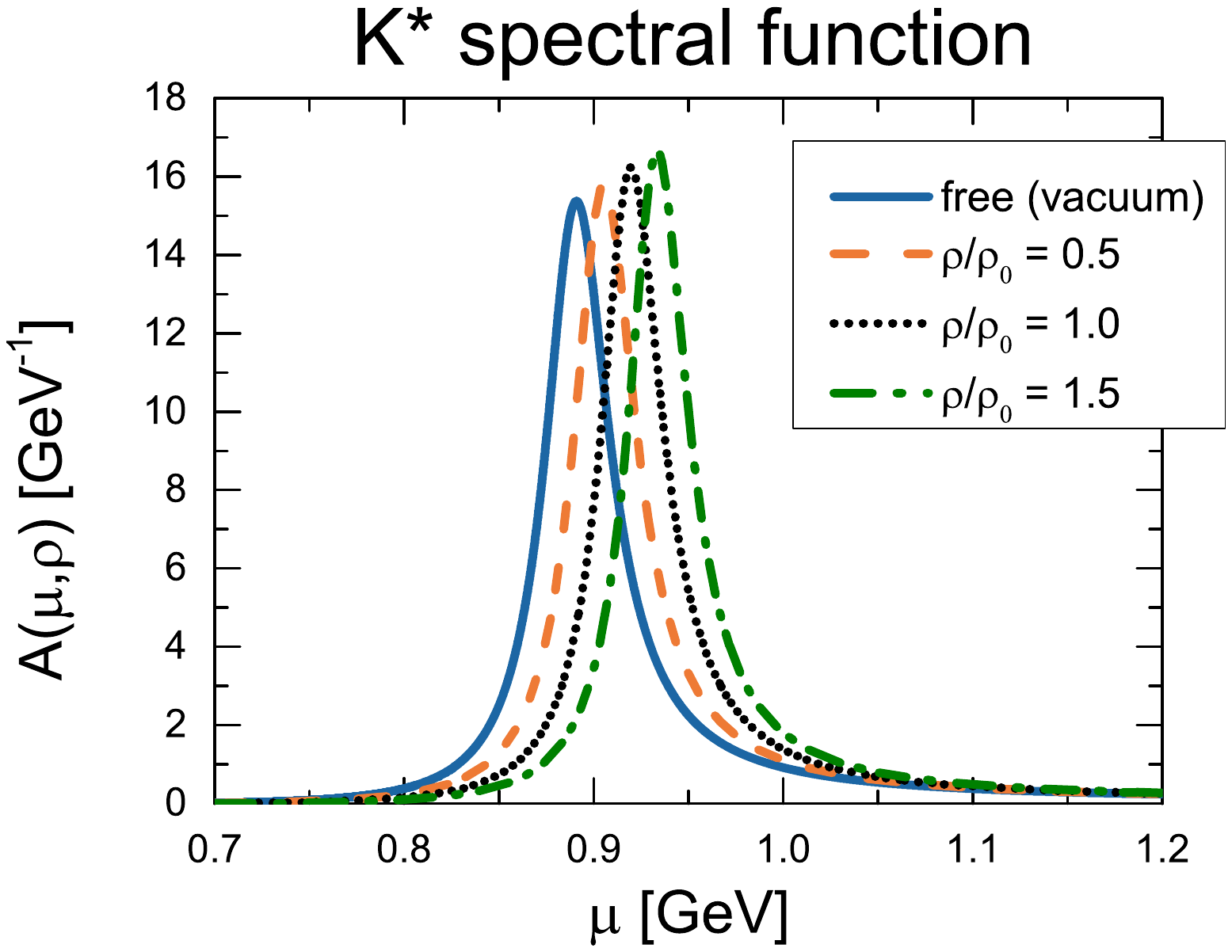}
\includegraphics[width=0.3\textwidth]{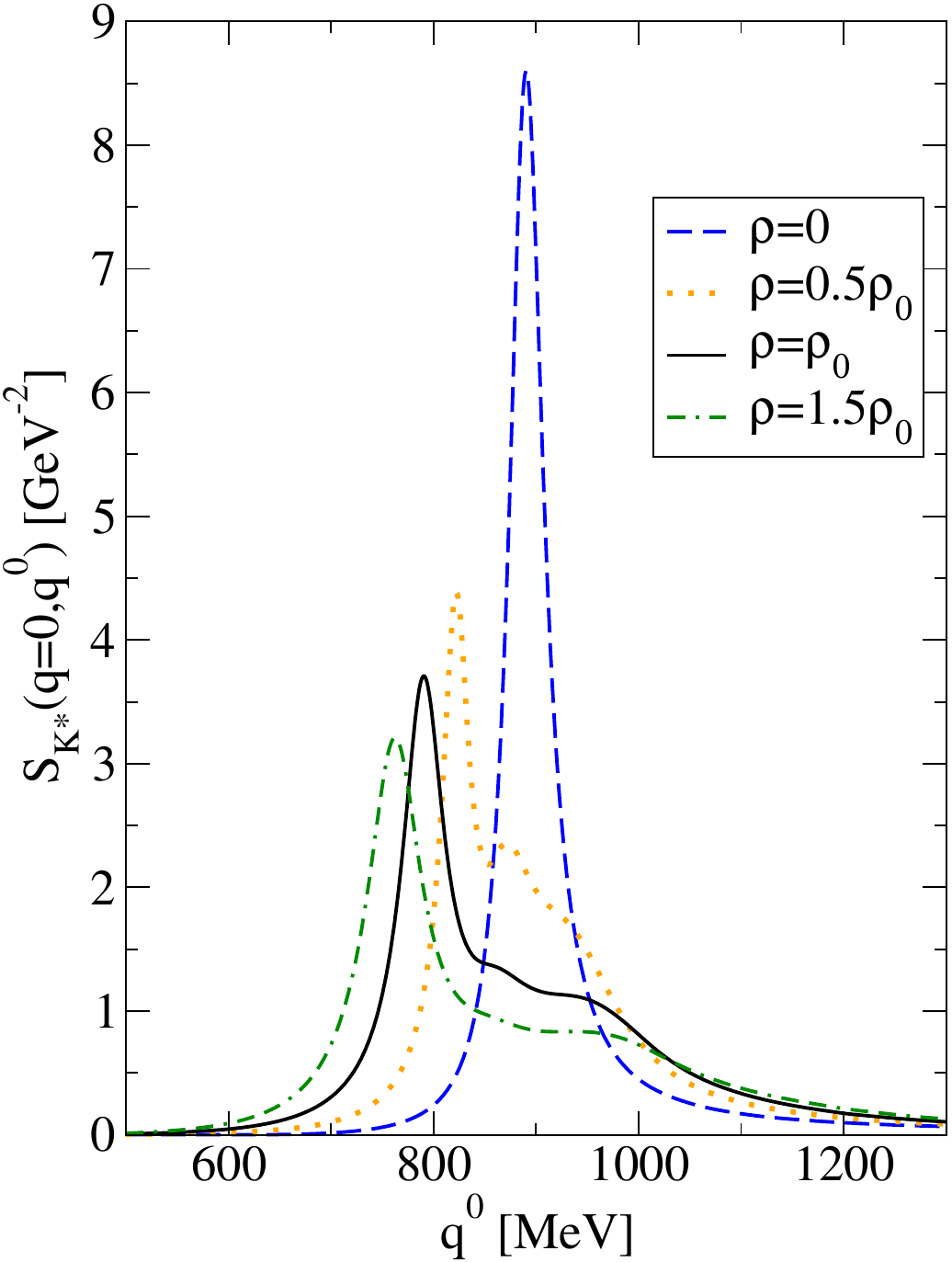}
\caption{Left.  $K^*$ spectral function at zero momentum for different densities. (Adapted from Ref.\,\cite{Ilner:2013ksa}.) 
Right. $\bar K^*$ spectral function at  zero momentum for different  densities, including $\rho=0$ (vacuum case). (Adapted from Ref.\,\cite{Tolos:2010fq}. \label{fig:vector_spectral}}
\end{figure}

\vspace*{2mm}\noindent
{\bf Strange vector mesons in nuclear matter}\\
\noindent With regard to strange vector mesons ($K^*$ and $\bar{K}^*$) in dense matter, Ref.\,\cite{Ilner:2013ksa} obtained the $K^*$ spectral function by adopting the relativistic Breit-Wigner prescription. 
The nuclear medium modifications arose from two sources. 
First, the modification of the dominant decay mode, $K^* \rightarrow K \pi$, induced by medium effects on the light pseudoscalar mesons. 
Second, the quasielastic interaction of the strange meson with nucleons, $K^* N \rightarrow K^* N$, and related absorptive channels with vector mesons. 
As in the $K$ case, the $K^*$ self-energy was evaluated considering the low-density theorem, owing to the lack of resonant states in the $s = 1$ channel \cite{Ilner:2013ksa}. 
The resulting $K^*$ spectral function is shown in the Fig.\,\ref{fig:vector_spectral}\,-\,left for zero momentum and different densities.

As for the $\bar{K}^*$ spectral function, the hidden gauge formalism was considered as the starting point to extract the bare interactions. Two sources for the in-medium modification of the $s$-wave $\bar{K}^*$ self-energy in nuclear matter were then taken into account~\cite{Tolos:2010fq}. On the one hand, a unitarised theory in coupled channels was employed due to the presence of resonant states in vector meson--baryon amplitudes in the strange $S = -1$ sector (states with an $s$-quark), such as $\Lambda^*(1800)$ and $\Sigma^*(1750)$, in order to compute the in-medium $\bar{K}^* N$ interaction coupled to different vector meson--baryon channels, i.e.\ $\bar{K}^* N \to \bar{K}^* N, \rho Y, \omega Y, \phi Y, \dots$. The importance of the $\bar{K}^* \to \bar{K} \pi$ channel in determining the properties of $\bar{K}^*$ necessitated the implementation of medium corrections in $\bar{K}$ and $\pi$, thus accounting for $\bar{K}^* N \to \bar{K} N, \pi Y, \bar{K} \pi N, \pi \pi Y \dots$, where $Y = \Lambda, \Sigma$.

The $\bar{K}^*$ meson spectral function is shown in Fig.\,\ref{fig:vector_spectral}\,-\,right for zero momentum and different densities. 
The long-dashed line refers to the calculation in free space, where only the $\bar{K} \pi$ decay channel contributes, while the other three lines correspond to calculations which incorporate $\bar{K}^* \rightarrow \bar{K} \pi$ in the medium, as well as the quasielastic $\bar{K}^* N \to \bar{K}^* N$ and other $\bar{K}^* N \to V B$ processes, with $V$ and $B$ denoting vector-mesons and baryons, respectively. 
The $\Lambda^*$-hole and $\Sigma^*$-hole excitations appear above the quasiparticle peak. The effect of increasing density leads to a broadening of the quasiparticle peak and a dilution of these resonant-hole states.

The modified properties of $K^*$ and $\bar{K}^*$ can be tested in different experimental scenarios. 
Whether or not the broadening could be tested through the transparency ratio in different nuclei, as done in Ref.\,\cite{Tolos:2010fq} for $\bar{K}^*$, HICs are also an ideal venue for studying the behaviour of strange vector mesons under extreme conditions. 
In particular, the dynamics of $K^*$ and $\bar{K}^*$ in HICs has been investigated using the PHSD transport approach, which implements in-medium effects on $K^*$ and $\bar{K}^*$ arising from their production from quark-gluon plasma as well as from the hadronic phase \cite{Ilner:2016xqr}. 
These calculations were performed for Au+Au at $\sqrt{s_{NN}} = 200\,$GeV (STAR/RHIC) and Pb+Pb at $\sqrt{s_{NN}} = 2.76\,$TeV (ALICE/LHC), together with Au+Au for $\sqrt{s_{NN}} = 5-60\,$GeV (CBM/FAIR, BMN/NICA, or BESII/RHIC). 
Several conclusions were drawn from this analysis. 
First, the main production channel is the resonant annihilation of $\pi + K$ ($\bar{K}$) in the final hadronic phase at LHC or RHIC energies. Second, at lower energies (CBM, BMN, BESII), the in-medium hadronic effects could be relevant owing to the expected longer reaction time and higher densities, whereas at LHC or RHIC the baryon densities are too low for in-medium effects to play a significant role. 
Finally, the rescattering and absorption of the $K^*$ and $\bar{K}^*$ decay channels make it difficult to extract the in-medium properties of strange vector mesons.

Knowledge of the properties of strange mesons in dense nuclear matter is important to understand the early universe and neutron stars.
\begin{figure}[t]
\begin{center}
\includegraphics[width=0.8\textwidth]{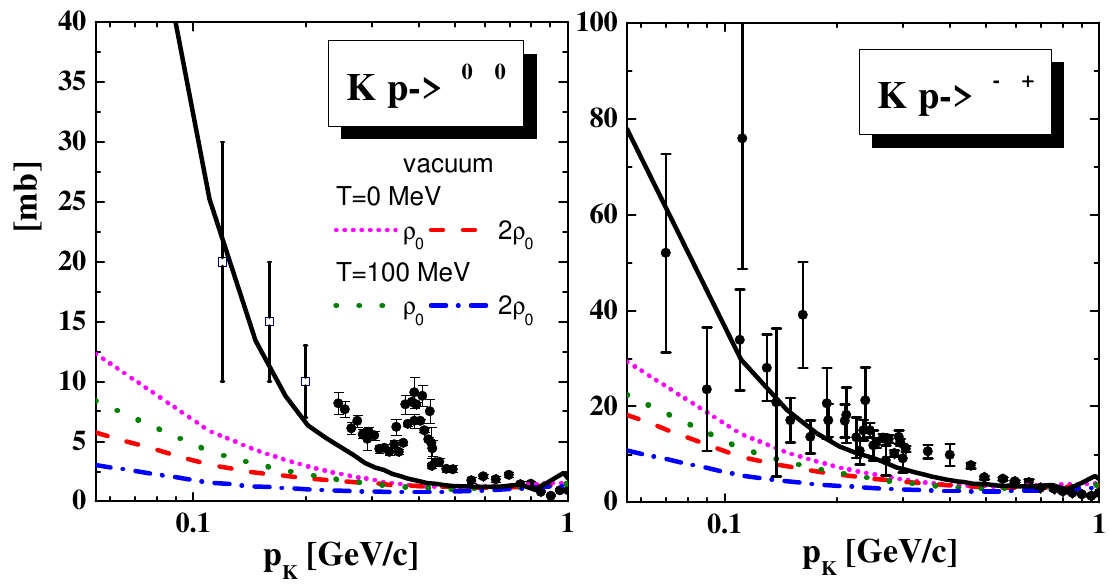}
\caption{$K^-p$ cross sections to the final states $\pi^0\Sigma^0$, $\pi^-\Sigma^+$, according to G-matrix calculations at $\rho=\rho_0, 2\rho_0$ and $T=0, 100\,$MeV in comparison with the data \cite{Schopper1988}. (Figures adapted from Ref.\,\cite{Song:2020clw}.)
\textcolor{blah}{The solid lines are a parametrization of the vacuum cross sections while colored lines show the in-medium cross sections at $T=0$ at $\rho_0$ (pink short dotted line) and $2\rho_0$ (red dashed line) as well as for $T=100$ MeV at $\rho_0$ (cyan dotted line) and $2\rho_0$ (blue dot-dashed line).}
 \label{fig1str}}
\end{center}
\end{figure}
The aforementioned in-medium modifications strongly alter the in-medium masses and in-medium cross sections compared to their vacuum values. 
This is demonstrated in Fig.\,\ref{fig1str}, which displays the $K^-p$ cross sections to the final states $\pi^0\Sigma^0$, $\pi^-\Sigma^+$ according to the aforementioned $G$-matrix calculations at $\rho = \rho_0, 2\rho_0$ and $T = 0, 100\,$MeV, in comparison with data \cite{Schopper1988}. 
One sees that even in cold nuclear matter, the cross section $K^-p \to \pi^0\Sigma^0$ is reduced (compared to the vacuum) by a large factor for small $K^-$ momenta. For the $K^-p \to \pi^-\Sigma^+$ reaction, the reduction is slightly more than a factor of two.

A few years ago, $\bar{K}$ production in a self-consistent coupled-channel unitarised $G$-matrix approach based on an SU(3) chiral effective Lagrangian \cite{Cabrera:2014lca} was implemented in the PHSD scheme \cite{Song:2020clw}. 
%
Since $K$, unlike $\bar{K}$, does not form a resonance with the nucleon, a linear repulsive potential is introduced for $K$ in medium, which effectively increases the $K$ mass in nuclear matter. 
The produced strange meson propagates according to the offshell dynamics based on the Kadanoff-Baym many-body theory, which ensures that the in-medium spectrum eventually returns to the vacuum spectrum \cite{Cassing:2003vz}.

\begin{figure}[t]
\begin{center}
\includegraphics[width=0.49\linewidth]{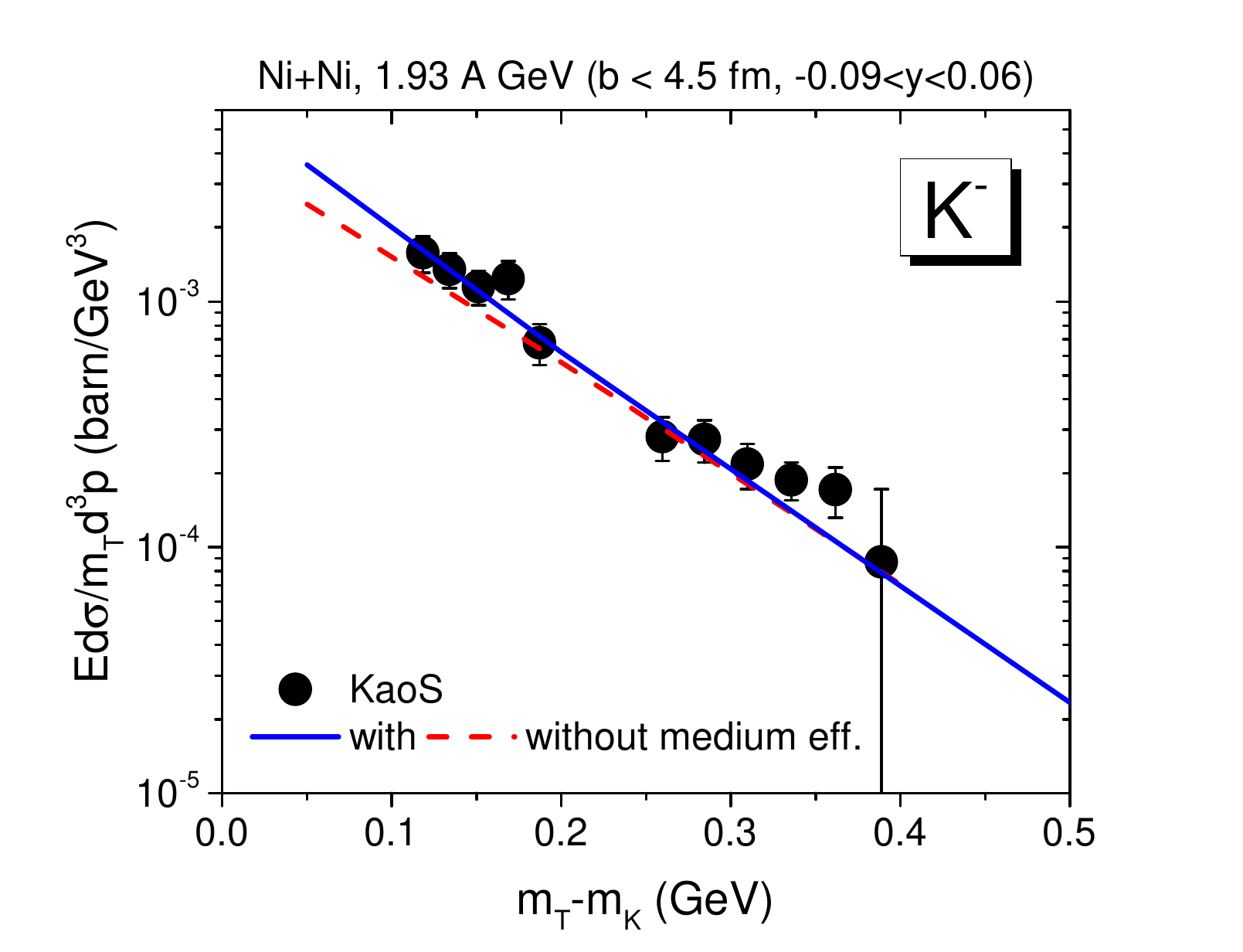}
\includegraphics[width=0.49\linewidth]{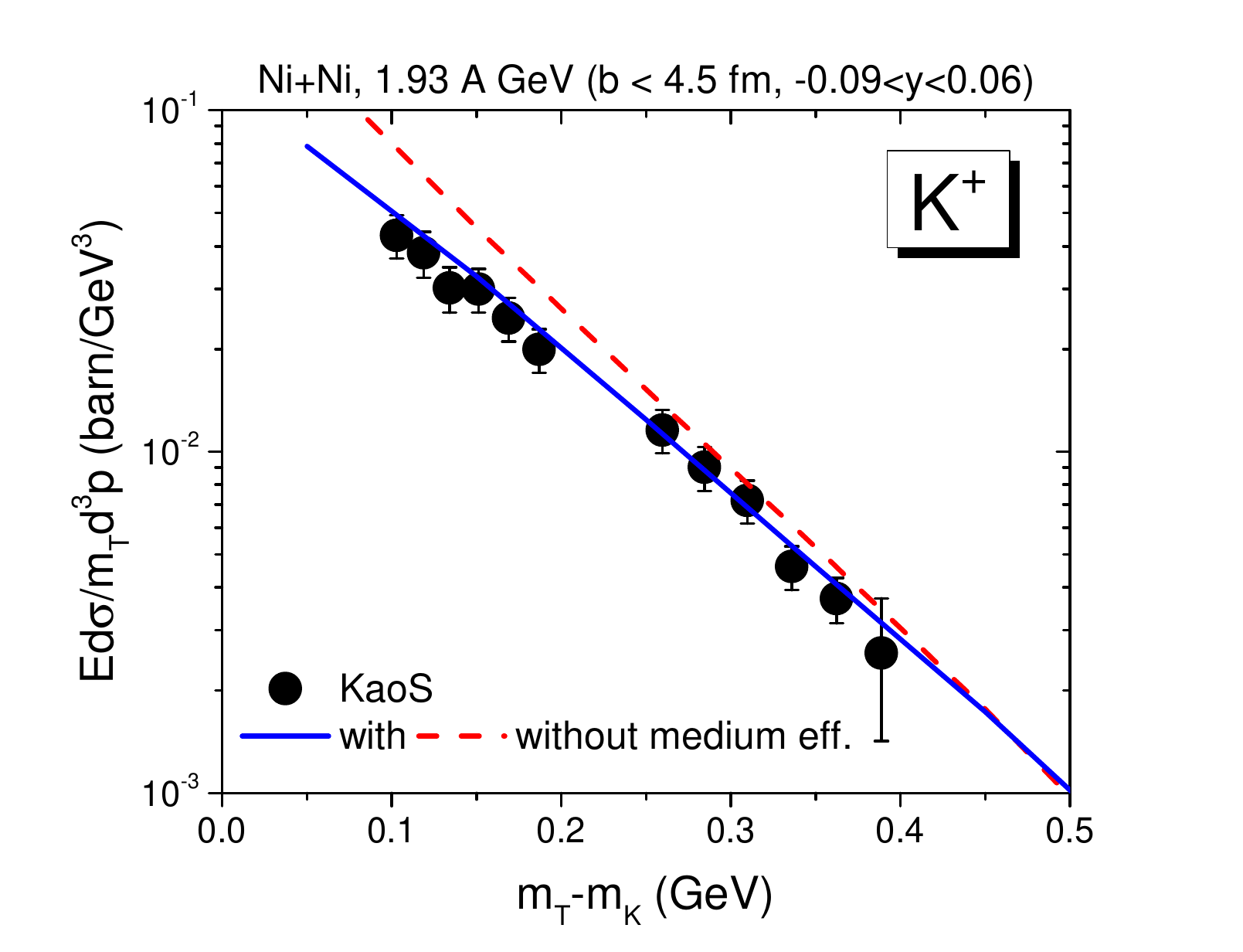}
\end{center}
\vspace*{-0.5cm}
\caption{$m_T$ spectra of $K^-$ (left) and $K^+$ (right) in Ni+Ni collisions at 1.93\,A,GeV with and without medium effects.
The PHSD results are compared with data from the KaoS Collaboration \cite{KaoS:2000eil}. (Figures adapted from Ref.\,\cite{Song:2020clw}.)
\label{pt-spectra}       }
\end{figure}

Figure~\ref{pt-spectra} shows the PHSD results for the $p_T$ spectra of $K^-$ and $K^+$ in Ni+Ni collisions at 1.93\,A\,GeV, with and without medium effects \cite{Song:2020clw}. 
One can see that in-medium modifications of cross sections and interaction potentials have strong effects on the final observables. 
They enhance the $K^-$ production and soften its $p_T$ spectrum. 
On the other hand, $K^+$ production is suppressed and its spectrum becomes harder due to the repulsive in-medium potential. 
The comparison with the data from the KaoS Collaboration \cite{KaoS:2000eil} supports the opposite behaviour of the modifications of $K^\pm$ in the medium.

\begin{figure}[t]
\begin{center}
\includegraphics[width=0.49\linewidth]{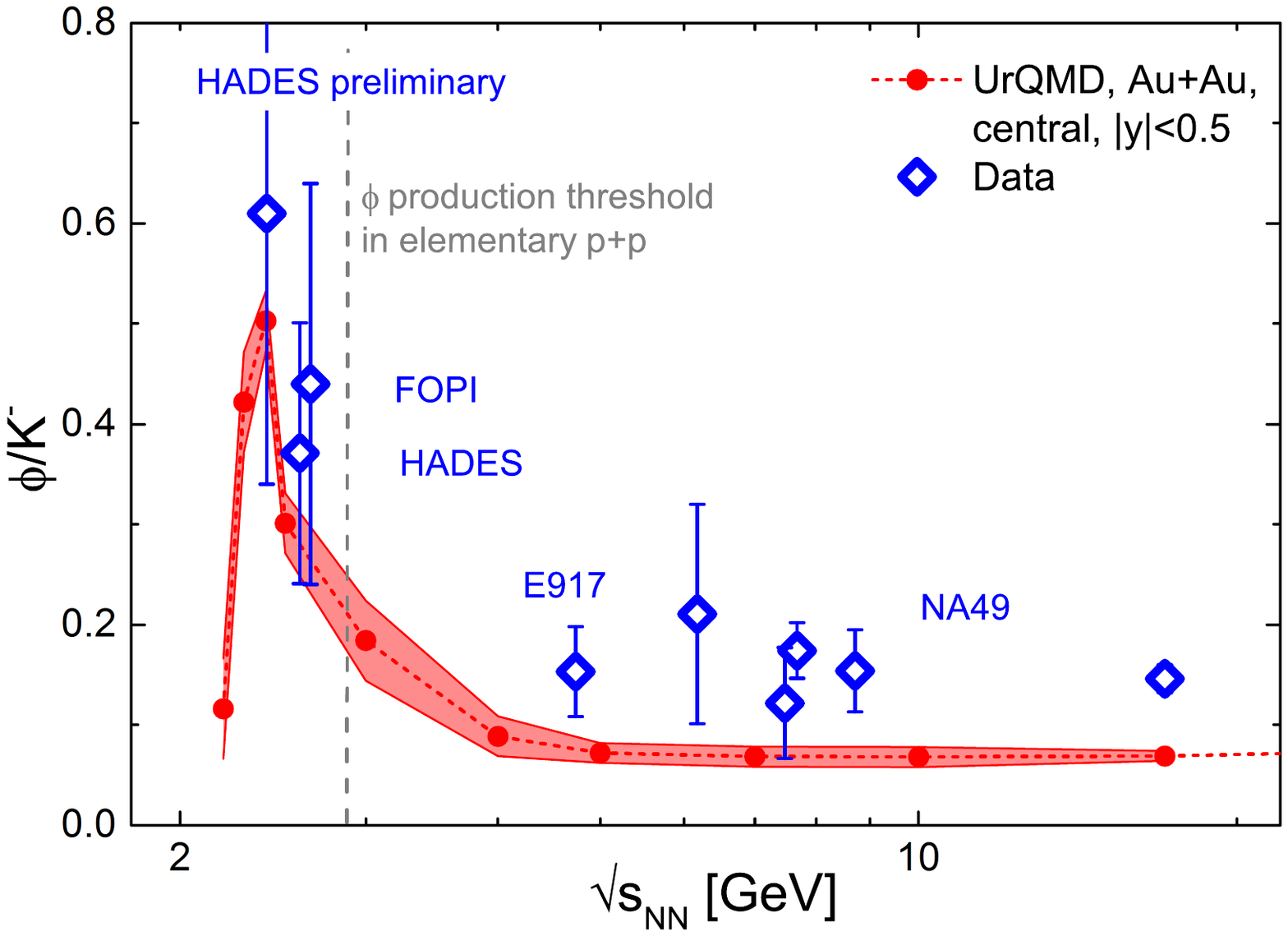}
\end{center}
\vspace*{-0.3cm}
\caption{Excitation function of the $\phi/K^-$ ratio for central ($b < 5$ fm) Au+Au collisions, calculated with a UrQMD model including the new $N^*$ decays (red line).
(Figure adapted from Ref.\,\cite{Steinheimer:2017dtk}.)}
\label{phi-km-UrQMD}       
\end{figure}

\vspace*{2mm}\noindent
{\bf Hidden strangeness: $\phi$ mesons}\\
\noindent The experimentally observed high $\phi$ multiplicity and the unexpected behaviour of the $\phi/K^-$ ratio (along with the $\phi/\Xi^-$ ratio) \cite{E917:2003gec, NA49:2002pzu, NA49:2007stj, NA49:2008goy, STAR:2004yym, STAR:2021hyx, FOPI:2016jgt}
have triggered theoretical efforts, as both exceed transport model predictions. 
A possible explanation was proposed in Ref.\,\cite{Steinheimer:2015sha}, assuming that high-mass nucleon resonances ($N^\ast$ with $M > 2$~GeV) decay to $\phi$ via $N^\ast \to N + \phi$. 
This scenario was realised in \cite{Steinheimer:2015sha, Steinheimer:2017dtk} by introducing decays of $N^*(1990)$, $N^*(2080)$, $N^*(2190)$, $N^*(2220)$, and $N^*(2250)$ to $\phi$ with a branching ratio of approximately $0.2$\% in a UrQMD model; see Fig.\,\ref{phi-km-UrQMD}. Formation of such states is enhanced in $AA$ collisions owing to Fermi motion and secondary meson-baryon interactions, while their role in near-threshold $pp$ data, \textit{e.g}., ANKE \cite{ANKE:2007dyc}, is negligible.
This idea was later adopted by the SMASH group \cite{Steinberg:2018jvv}, with a larger branching ratio of $0.5$\%. 
Decays to other vector mesons, such as the $\omega$, are discussed in Ref.\,\cite{Fabbietti:2015tpa}. 
However, decays of heavy $N^*$ resonances to $\phi$ mesons remain unobserved experimentally \cite{ParticleDataGroup:2020ssz}.

Alternatively, $\phi$-meson production in the nuclear medium was also realised in PHSD \cite{Song:2022jcj} by employing the same self-consistent, coupled-channel unitarised method, but introducing a width broadening to account for medium effects, which is related to the partial restoration of chiral symmetry.

\begin{figure}[t]
\begin{center}
\includegraphics[width=0.5\linewidth]{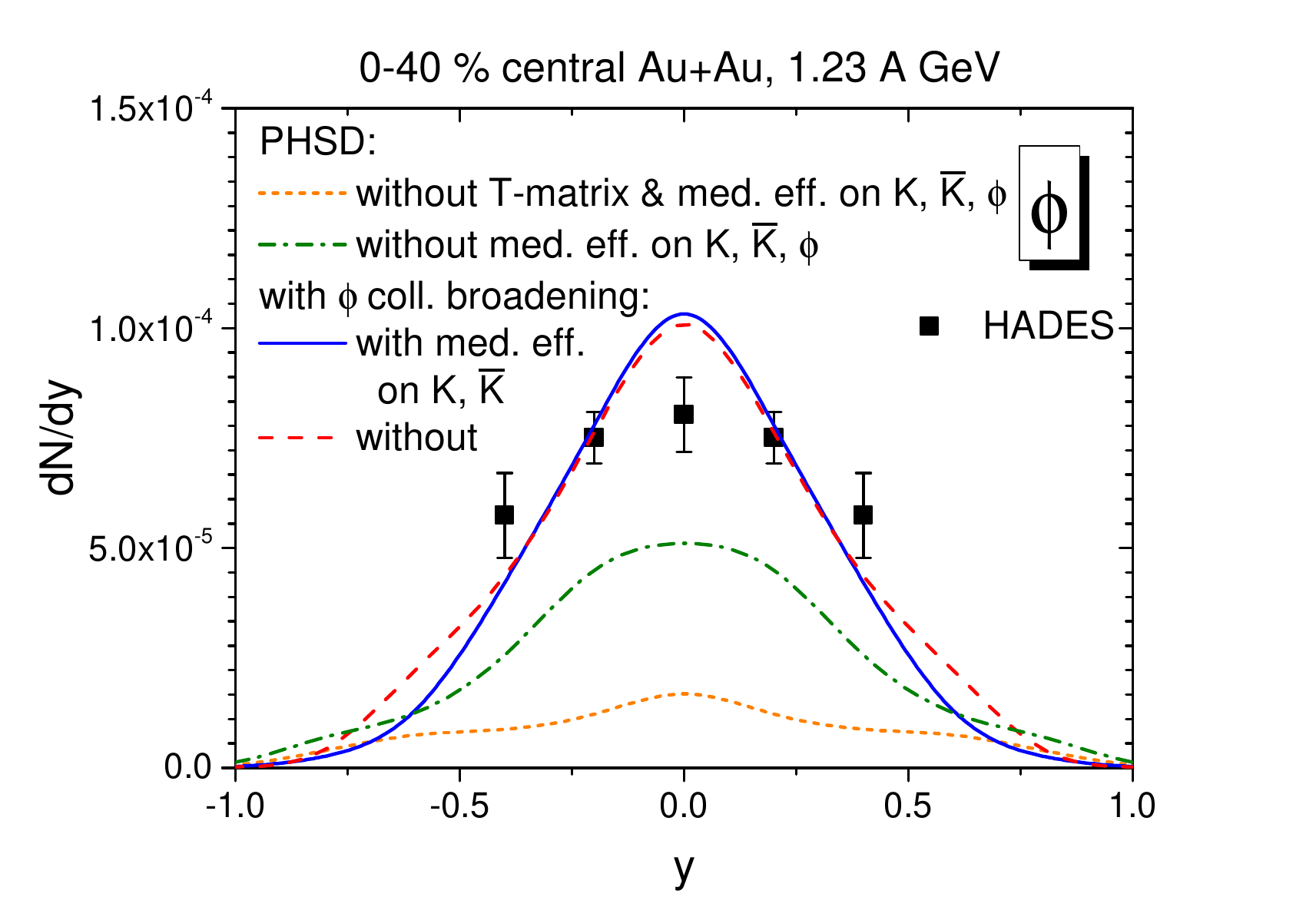}
\includegraphics[width=0.48\linewidth]{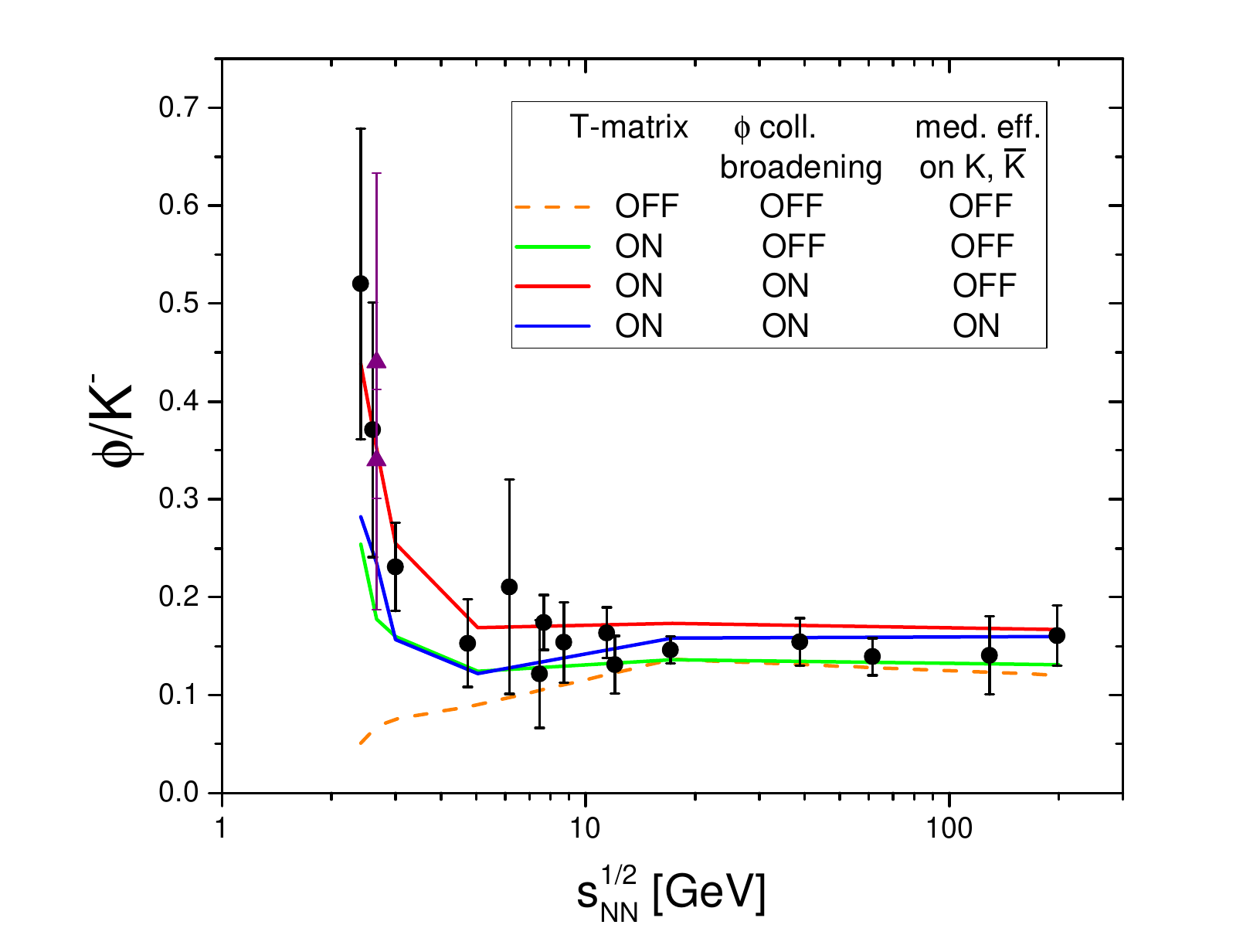}
\end{center}
\vspace*{-0.5cm}
\caption{Left.  PHSD results for rapidity distribution of $\phi$ mesons in $0-40$\% central Au+Au collisions at 1.23\,A\,GeV, compared with the data from the HADES Collaboration \cite{HADES:2017jgz}. 
Right. Ratio of $\phi$ meson to $K^-$ in HICs as a function of collision energy \cite{Song:2022jcj}. The PHSD simulations were carried out in four different scenarios. (Figures adapted from Ref.\,\cite{Song:2022jcj}.)}
\label{phi-km}       
\end{figure}

The self-consistent coupled-channel unitarised method introduces numerous production channels for $\phi$ mesons and, consequently, enhances $\phi$ production, as shown by the difference between the orange dotted line and the green dot-dashed line in the left panel of Fig.\,\ref{phi-km}. 
A further enhancement, corresponding to the solid blue line, is attributed to width broadening of the $\phi$ meson in medium:
\begin{eqnarray}
\Gamma^*(M,p,\rho)=\Gamma_0(M,p)+\Gamma_{\rm coll}(\rho)\approx \Gamma_0(M,p)+\alpha_{\rm coll}(\rho/\rho_0),
\end{eqnarray}
where the in-medium spectral width of the $\phi$ meson, $\Gamma^*$, is the sum of the vacuum width, $\Gamma_0$, and the collisional width, $\Gamma_{\rm coll}$, with $\alpha_{\rm coll}$ taken to be 25\,MeV \cite{Song:2020clw,Song:2022jcj}. 
It can be seen that both the additional production channels from the self-consistent coupled-channel unitarised method and the in-medium width broadening of the $\phi$ meson are necessary to reproduce the data from the HADES Collaboration \cite{HADES:2017jgz}. 
The dashed red line includes medium effects on $K$ and $\bar{K}$, which, however, have little influence on $\phi$ production.

The right panel of Fig.\,\ref{phi-km} shows the ratio of $\phi$ to $K^-$ mesons in HICs as a function of collision energy. 
The experimental data indicate that the ratio remains constant at high energies but increases rapidly as the collision energy decreases. 
This behaviour can be explained by both many production channels of $\phi$ mesons from the self-consistent coupled-channel unitarised method and the width broadening of $\phi$ meson in medium. 
The ratio, however, decreases when in-medium effects on $K^-$ are included, since these enhance $K^-$ production in HICs, as shown in Fig.\,\ref{pt-spectra}. 

Thus, multistep meson–baryon reactions involving newly produced mesons and baryonic resonances, along with in-medium modifications of $\phi$ meson properties, lead to an enhancement of $\phi$ production in HICs compared to $pp$ reactions, where such secondary processes and high-density conditions are absent.

The forthcoming FAIR data on $\bar{K}$, $K$, and $\phi$ production in $pp$, $\pi p$, and $pA$ reactions will provide essential constraints on strangeness production mechanisms and support the interpretation of the $\phi/K^-$ ratio observed in $AA$ collisions. 
In this context, experimental searches for possible decays of baryon resonances into strange hadrons are of particular interest and possess significant discovery potential.

\vspace*{2mm}\noindent
{\bf $K_1/K^*$ ratio in connection to chiral symmetry restoration }\\[2mm]
%
\noindent Chiral symmetry restoration may also be probed with strange chiral partners, such as the axial-vector meson $K_1$(1270), which has $J^P=1^+$, and the $K^*$(892), with $J^P=1^-$ \cite{Sung:2023oks}. 
In contrast to the well-known chiral partners, $a_1$ and $\rho$, both kaon resonances can, in principle, be measured experimentally in a much simpler way: the $K_1$(1270) decays with a branching ratio of 16\% to $K^\ast\pi$, with a short lifetime corresponding to a decay width of 90\,MeV. 
The $K^*$ subsequently decays with 100\% branching ratio to $K\pi$, with a decay width of 47\,MeV. 
Assuming chiral restoration in the dense medium, the $K_1/K^*$ ratio is predicted to differ significantly from the case without restoration \cite{Sung:2023oks}, which, crucially, requires data in the elementary proton–proton channel as a reference.

\vspace*{2mm}\noindent
{\bf Probing isospin effect in dense matter by the measurement of isospin partners}\\
\noindent In order to make use of data from HICs for the calculation of neutron stars, one important, but often overlooked, aspect is the isospin dependence of potentials. 
One proposal \cite{Yong:2022pyb} is to measure isospin effects by studying ratios of isospin partners in various collision systems and centralities, such as $n/p$, $\pi^-/\pi^+$, $K^0_S/K^+$, $\Xi^-/\Xi^0$, or $\Sigma^-/\Sigma^+$. 
In particular, a significant difference in the ratio between stiff and soft symmetry energies is expected for most of the pairs, with the $\Sigma^-/\Sigma^+$ ratio showing especially strong sensitivity according to Ref.\,\cite{Yong:2022pyb}.
Experimentally, $\Sigma^{\pm}$ are challenging to identify because they decay predominantly into $N\pi$, with one decay product being neutral. However, the CBM experiment has the potential to detect the kink in $n\pi^{\pm}$ decays owing to its excellent track reconstruction capabilities.

\subsubsection{Study of the in-medium properties of open and hidden charmed hadrons} 
\label{subsub:inmediumpropopphidd}


\noindent
{\bf Open charm in nuclear matter:} \\
\noindent The properties of open-charm mesons ($D$, $\bar D$, $D^*$, $\bar D^*$) in nuclear matter have been intensively studied over the past few decades. 
The various analyses range from phenomenological studies based on a quark–meson coupling model \cite{Guichon:1987jp, Sibirtsev:1999js}, Polyakov-loop extensions of Nambu–Jona-Lasinio models \cite{Blaschke:2011yv}, mean-field computations in nuclear matter \cite{Mishra:2003se, Kumar:2011ff, Pathak:2014vra}, approaches relying on $\pi$-exchange that incorporate heavy-quark symmetries \cite{Yasui:2012rw}, QCD sum-rule (QSR) models \cite{Gubler:2018ctz}, and self-consistent unitarised coupled-channel schemes \cite{Tolos:2013gta}.

Within unitarised coupled-channel schemes, one can determine the full spectral characteristics of open-charm mesons in nuclear matter from the behaviour of the transition amplitudes in dense environments. 
From the exploratory work on the $D$-meson spectral function in Refs.\,\cite{Tolos:2004yg, Tolos:2005ft}, the spectral features of $D$ mesons in nuclear matter have been determined using $t$-channel vector-meson exchange unitarised schemes \cite{Lutz:2005vx, Mizutani:2006vq, Tolos:2007vh,Jimenez-Tejero:2011dif}. 
More recently, the necessity of explicitly implementing heavy-quark spin symmetry (HQSS) constraints in the unitarised coupled-channel approach has become apparent. 
Schemes that explicitly implement this symmetry have been developed \cite{Garcia-Recio:2008rjt, Gamermann:2010zz, Romanets:2012hm,Garcia-Recio:2013gaa}, enabling the determination of the transition amplitudes and, consequently, the open-charm spectral functions in nuclear matter \cite{Tolos:2009nn, Garcia-Recio:2010fiq, Garcia-Recio:2011jcj}.

\begin{figure}[t]
\begin{center} 
\includegraphics[width=0.5\textwidth,height=7cm]{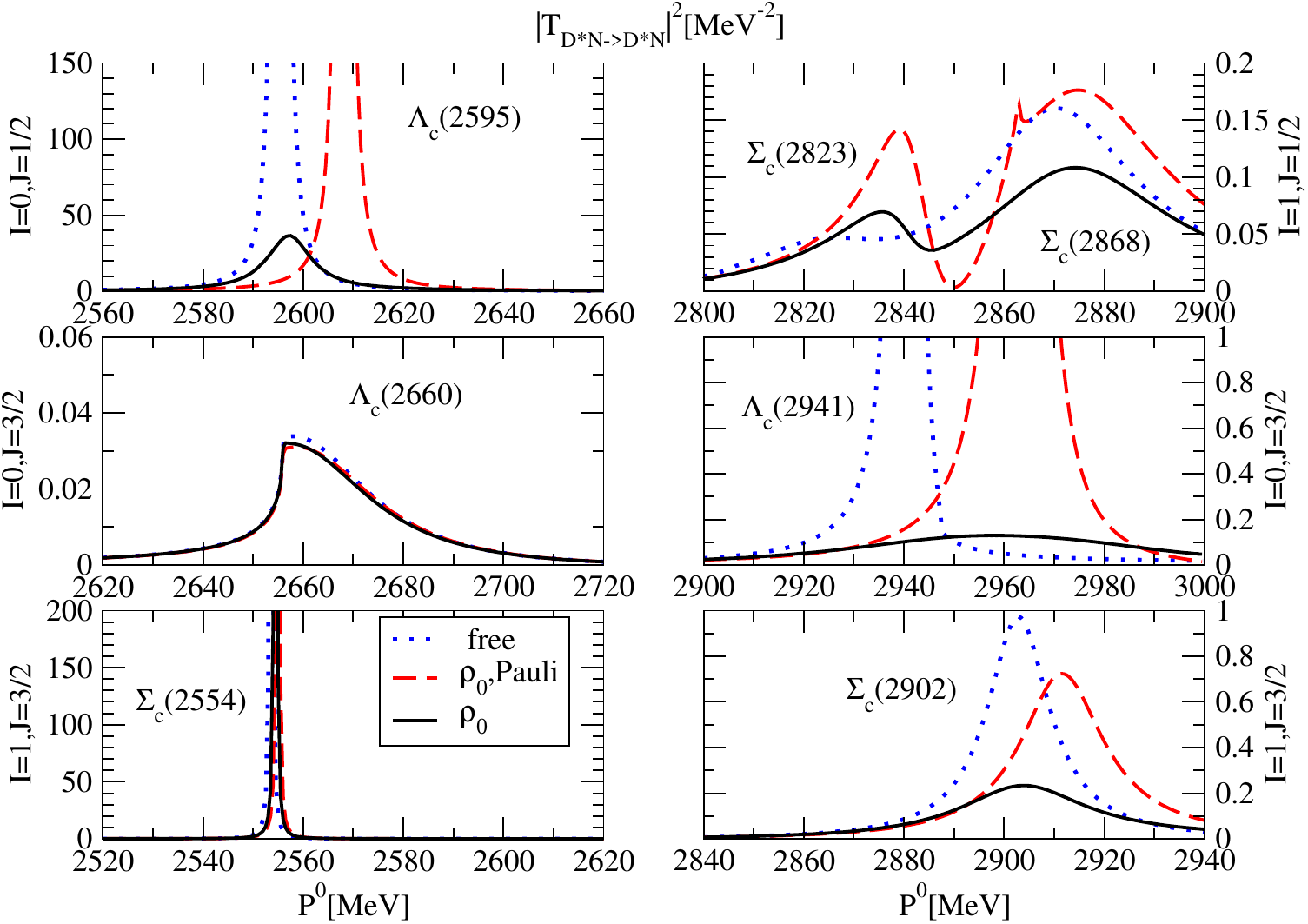}
\includegraphics[width=0.4\textwidth,height=6cm]{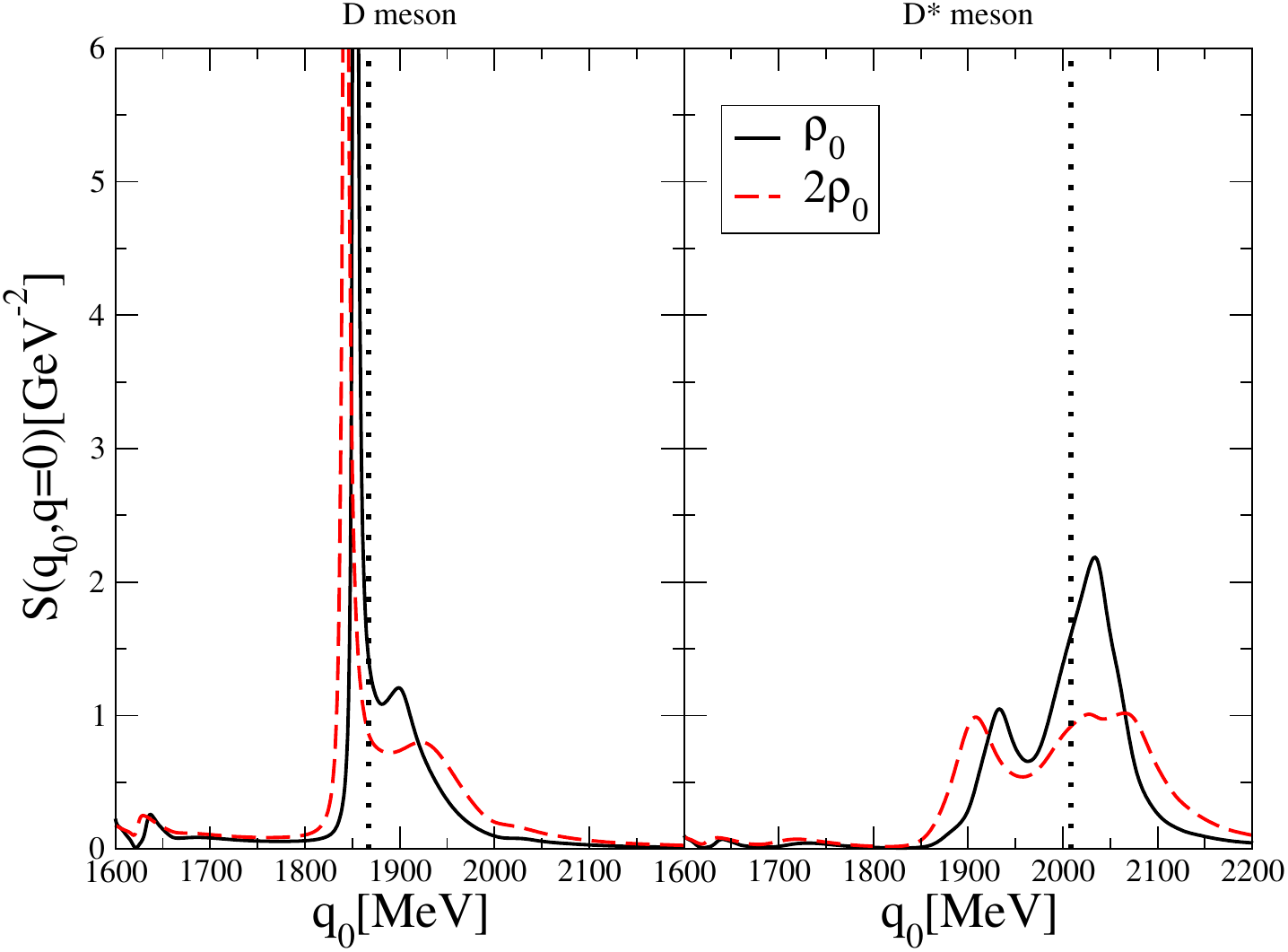}
\caption{
Left. Squared amplitude of the $D^* N \to D^* N$ transition for different partial waves as a function of CM energy, $P^0$, at fixed total momentum $|\vec{P}| = 0$. 
Several resonances are visible: (I = 0, J = 1/2) $\Lambda_c(2595)$; (I = 1, J = 1/2) $\Sigma_c(2823)$ and $\Sigma_c(2868)$; (I = 0, J = 3/2) $\Lambda_c(2660)$ and $\Lambda_c(2941)$; (I = 1, J = 3/2) $\Sigma_c(2554)$ and $\Sigma_c(2902)$.
Right. $D$ and $D^*$ meson spectral functions as a function of meson energy, $q_0$, for different densities at $\vec{q} = \vec{0}$. 
The positions of the $D$ and $D^*$ meson masses in vacuum are indicated by dotted vertical lines. 
(Figures adapted from Ref.\,\cite{Tolos:2009nn}.)
\label{fig:opencharm}}
\end{center}
\end{figure}

Figure~\ref{fig:opencharm}\,-\,left displays the squared amplitude of $D^*N \to D^*N$ in dense matter for different in-medium scenarios and partial waves as a function of CM energy, for zero total momentum. 
One sees how dense matter modifies the properties (masses and widths) of charmed baryonic states generated by the scattering of $DN$ and coupled channels, such as the $\Lambda_c(2595)$. 
Figure~\ref{fig:opencharm}\,-\,right displays the $D$ and $D^*$ spectral functions for two densities at zero three-momentum. 
One observes that the $D$ and $D^*$ quasiparticle peaks mix with dynamically generated state–hole excitations. 
In both cases, increasing density smears out the particle–hole states and broadens the spectral functions.


Medium modifications of $D$ and $D^*$ mesons in nuclear matter, as well as the density dependence of the $D^{(*)}N$ transition amplitudes, influence the dynamical evolution of heavy mesons in the nuclear medium. 
This effect can locally be characterised by the heavy-flavour spatial diffusion coefficient, $D_s$, which quantifies the spread of heavy particles in the medium owing to Brownian diffusion; see Ref.\,\cite{Das:2024vac} and references therein. 
While finite-temperature effects on heavy mesons have been analysed for the $D_s$ coefficient, \textit{e.g}., in Ref.\,\cite{Torres-Rincon:2021yga}, its dependence on baryon density remains largely unexplored, aside from the trivial density effect of collision centres \cite{Tolos:2013kva}, which leads to enhanced $D$-meson stopping in dense media. 

Potential modifications of transition amplitudes in $D^{(*)}N$ scattering could also be probed by femtoscopy measurements, such as those for $DN$ pairs in high-multiplicity $pp$ collisions at $\sqrt{s}=13\,$TeV measured by the ALICE collaboration \cite{ALICE:2022enj}. 
According to the Koonin--Pratt formula (see Sec.\,\ref{subsec.femto}), both the source function and the interacting-pair wave function determine the correlation function. 
In low-energy collisions, one expects a rather different scale of the source function compared to that in LHC collisions, combined with possible modifications of the $DN$ wave function owing to the density effects discussed in this section.

\vspace*{2mm}\noindent
{\bf Constraining the gluon condensate:} \\[2mm]
%
\noindent The gluon condensate $\langle \alpha_s G^2 \rangle$ is important for QSR phenomenology, influencing heavy-quark hadron masses via a representation of non-perturbative gluon dynamics. 
Its vacuum value is fitted via QCR analyses of the $J/\psi$, while an in-medium value can be inferred from charmed vector meson masses. 
At normal nuclear density, a 5\% reduction in the gluon condensate induces a downward mass shift in charmonium states owing to the second-order Stark effect \cite{Lee:2003}. 
Studying the $J/\psi$ and $\psi(3686)$ with a $5-10\,$GeV antiproton or $10-20\,$GeV pion/proton beam, using dileptonic decays at CBM, allows investigation of these effects. 
Transport models predict two dilepton yield contributions: one from the nuclear centre and another from vacuum decays, with negligible surface effects. 
The $J/\psi$ mass shift is too small to separate from the vacuum contribution, while the $\psi(3686)$ is expected to shift by over 100\,MeV, producing a second peak below the vacuum peak. 
This peak is experimentally detectable, as background contributions from DY and open-charm decays remain minimal near threshold. 

\vspace*{2mm}\noindent
{\bf $J/\psi$ interaction cross section with hadrons:}\\
Knowledge of the charmonium absorption/dissociation cross section on hadrons is important for the description of $J/\psi$ production in HICs. However, the inelastic cross section of the $J/\psi$ remains an open question owing to the lack of precise experimental data. 
Existing studies from high-energy proton–nucleus collisions primarily probe the evolution of preresonance coloured $c\bar{c}$ states rather than fully formed charmonium. 
At the FAIR SIS100 accelerator ($15-30\,$~GeV), however, $J/\psi$ production occurs near threshold, where the mesons predominantly form before interacting with nuclear matter \cite{Bhaduri:2017ptw}. 
Unlike at higher energies, these colour-neutral states will undergo direct dissociation in the medium.

\begin{figure}[t]
\centering
\includegraphics[width=0.42\textwidth]{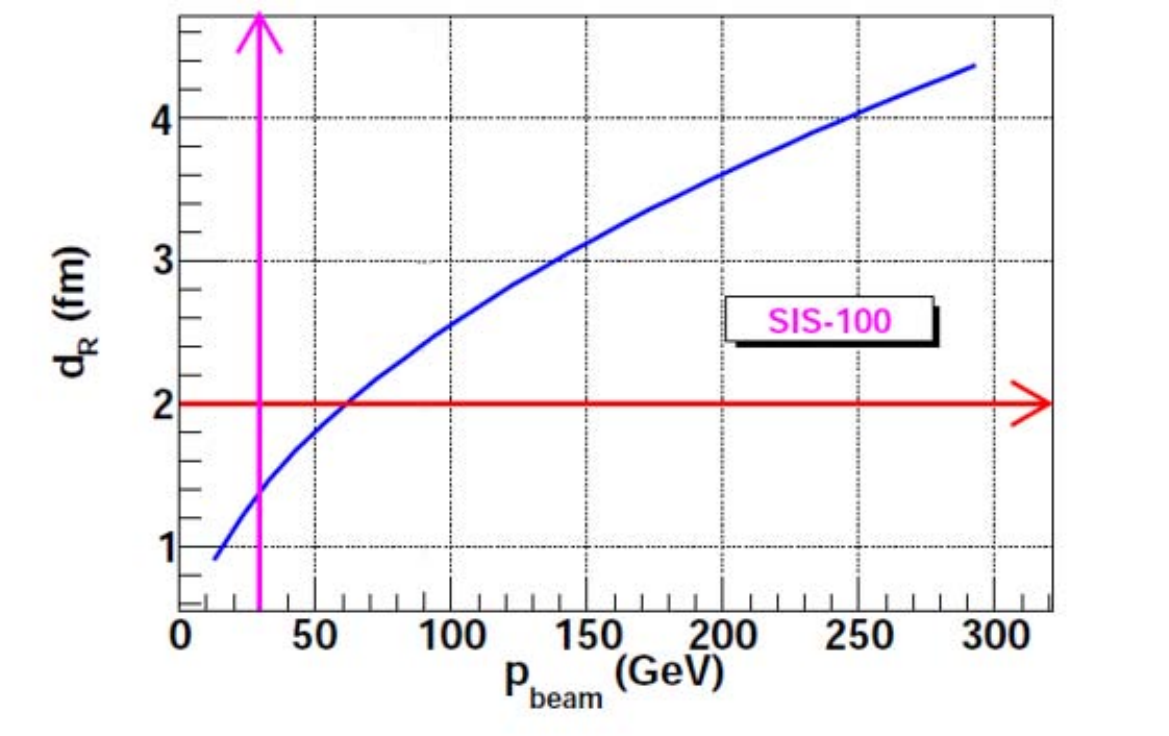}
\includegraphics[width=0.4\textwidth]{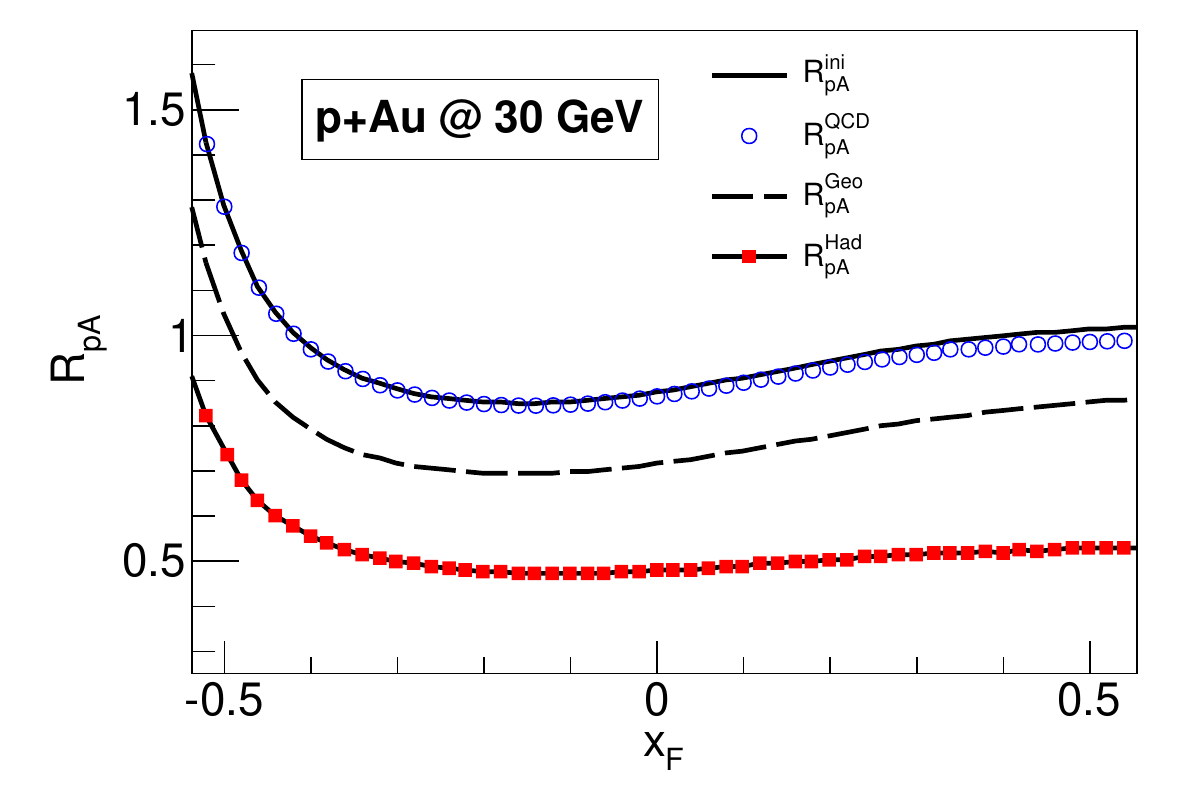}
\caption{(Left) J/$\psi$ resonance formation length ($d_{R}$) in the laboratory frame as a function of beam momentum in p+A collisions. (Right) $x_{F}$ dependence of $J/\psi$ transparency ratio ($R_{pA}$) in 30\,GeV p+A collisions for different models of inelastic collision (dissociation) in cold nuclear medium. (Images adapted from Ref.\,\cite{Bhaduri:2017ptw}.)
\label{fig:inel_jpsi}}
\end{figure}

Theoretical models predict widely varying dissociation cross sections: perturbative QCD (pQCD) suggests small values owing to limited hard gluon interactions, whereas non-perturbative models anticipate larger cross sections, peaking near threshold, driven mainly by the reaction $J/\psi + N \to \Lambda_c + \bar{D}$. 
These differences significantly impact the $J/\psi$ survival probability and the experimentally accessible transparency ratio ($R_{pA}$), which encapsulates cold nuclear matter (CNM) effects. 
Predictions for $R_{pA}$ at FAIR reveal distinct suppression patterns for different dissociation mechanisms, making the upcoming SIS100 $p+A$ experiments critical to resolving these theoretical uncertainties. \textcolor{blah}{The corresponding $J/\psi$ formation length and predictions for the transparency ratio $R_{pA}$ for different dissociation scenarios are shown in Fig.~\ref{fig:inel_jpsi}}.

\vspace*{2mm}\noindent
{\bf Production of charmed hadrons from secondary $\bar p +p$ annihilations:}\\
%
A novel way to investigate the production of exotic charmed hadrons, such as $\Lambda_c$, $\Sigma_c$, $\Xi_c$, $D$ and $D_s$, is to use secondary antibaryon–baryon ($\overline{B} + B$) annihilation processes in proton–nucleus ($p+A$) reactions at FAIR proton beam energies of $10-30$ GeV \cite{Reichert:2025iwz}. 
This approach is motivated by the current experimental capabilities of the CBM detector and aims to expand the accessible charm production channels in the FAIR energy regime.

Direct production of charmed hadrons in primary $pp$ or $pA$ interactions is severely limited at these energies owing to kinematic thresholds. However, the production of antibaryons in such reactions opens the possibility for secondary annihilation processes within the nuclear medium. 
These $\overline{B} + B$ interactions create localised, gluon-rich environments that are favourable for the formation of exotic multiquark and heavy-flavour states. 
Such mechanisms are supported by earlier experimental observations \cite{JETSET:1998akg}, including anomalously high yields of exotic mesons in $\bar{p}p$ collisions and indications of strong deviations from the expected OZI suppression.

To quantitatively assess the potential of this production channel, the UrQMD transport model has been employed to simulate $p+\mathrm{Au}$ collisions at FAIR energies. 
The model provides the invariant mass distributions of secondary $\overline{B} + B$ annihilations, revealing a non-negligible fraction of events occurring within the nuclear environment and showing $\sqrt{s_{\overline{B}B}}$ values exceeding charm production thresholds, particularly at $30$ GeV.

Charm production cross sections in $\bar{p}p$ annihilation are taken from previously developed theoretical models \cite{Haidenbauer:2014rva, Haidenbauer:2016pva}, including meson/baryon exchange and constituent quark approaches. 
By folding these cross sections with the simulated $\sqrt{s_{\overline{B}B}}$ spectra, one can estimate the yields of various charmed hadrons. 
The estimates are shown in Table~\ref{tab:tab_charm_had}. 
These predictions represent conservative lower limits because additional charm production mechanisms, \textit{e.g}., direct $pp$, $pn$ interactions or higher-order processes, are not included.
\textcolor{blah}{We note that the proton beam energies listed in the Table can be achievable  with SIS100. 
}

\begin{table}[t]
    \centering
    \renewcommand{\arraystretch}{1.5}
    \begin{tabular}{|l|c|c|c|c|}
         \hline
       Hadron  & \multicolumn{4}{c|}{Rate per 90 days $p$+Au} \\
         \hline
         \hline
         & \multicolumn{2}{c|}{$E_\mathrm{lab}=30$~GeV} & \multicolumn{2}{c|}{$E_\mathrm{lab}=20$~GeV} \\ \hline
          & M/B exch. & Quark model & M/B exch. & Quark model \\  
         \hline
    $ \Lambda_c^+ \overline{\Lambda_c^-}$    & $3.1\cdot10^4$ & $9.0\cdot10^3$  & $4.9 \cdot 10^3$ & $1.4 \cdot 10^3$ \\ \hline
    $  \Sigma_c^+ \overline{\Lambda_c^-}$    & $2.2\cdot10^3$ & $8.0\cdot10^1$  & $2.7 \cdot 10^2$ & $1.0 \cdot 10^1$ \\ \hline
    $ \Sigma_c^+ \overline{\Sigma_c^-}$      & $5.8\cdot10^2$ & $2.0\cdot10^0$  & $4.0 \cdot 10^1$ & $0.2 \cdot 10^{0}$ \\ \hline
    $  \Sigma_c^{++} \overline{\Sigma_c^{--}}$& $2.1\cdot10^3$& $2.0\cdot10^0$  & $1.4 \cdot 10^2$ & $0.2 \cdot 10^{0}$ \\ \hline
    $  \Sigma_c^0 \overline{\Sigma_c^0}$      & $8.0\cdot10^2$& $2.0\cdot10^0$  & $5.5 \cdot 10^1$ & $0.2 \cdot 10^{0}$ \\ \hline
    $  \Xi_c^+ \overline{\Xi_c^-}$            & $1.5\cdot10^3$& $1.2\cdot10^1$  & $1.5 \cdot 10^2$ & $1.2 \cdot 10^{0}$ \\ \hline
    $  \Xi_c^0 \overline{\Xi_c^0}$            & $6.5\cdot10^2$& $1.2\cdot10^1$  & $6.5 \cdot 10^1$ & $1.2 \cdot 10^{0}$ \\ \hline
    $  D^0 \overline{D^0}$                    & $3.1\cdot10^3$& $1.1\cdot10^3$  & $1.3 \cdot 10^3$ & $4.6 \cdot 10^2$ \\ \hline
    $  D^+ {D^-}$                             & $3.8\cdot10^3$& $8.8\cdot10^2$  & $1.5 \cdot 10^3$ & $3.6 \cdot 10^2$ \\ \hline
    $  D_s^+ {D_s^-}$                         & $1.2\cdot10^3$& $1.6\cdot10^3$  & $4.9 \cdot 10^2$ & $6.0 \cdot 10^2$ \\ \hline\hline
    $  \phi  \phi$                            & $3.9\cdot10^7$ & --   & $1.7 \cdot 10^7$ & -- \\ \hline
    \end{tabular}
    \caption{Charmed (hidden strange) hadron yields per 90 days at full luminosity in min. bias $p$+Au reactions at $E_\mathrm{lab}=30$ GeV (left columns) and $E_\mathrm{lab}=20$ GeV (right columns) \cite{Reichert:2025iwz}. 
    \label{tab:tab_charm_had}}
\end{table}

The results demonstrate that secondary ${\overline{B}B}$ annihilation provides a viable mechanism for accessing the charm sector at FAIR energies, even prior to the operation of PANDA. 
The proposed approach offers a unique opportunity to study exotic charm dynamics in a baryon-rich environment and may serve as a complementary probe for understanding QCD in the nonperturbative regime.

\subsubsection{Production of light nuclei and hypernuclei}
\label{subsec.hypernuclei_in_piA}

The formation of clusters, especially hyperclusters, in nuclear matter is interesting in itself and at the same time a crucial reference measurement for the interpretation of cluster formation in $AA$ collisions. 
This is especially true for pion-induced reactions.  
Hypernuclei production offers an additional way to investigate the hyperon-nucleon ($Y$-$N$) interaction, as elaborated in Chapter~\ref{sec.HadrHadrInter}. 

Pion-induced reactions with heavy nuclei, as exemplified by $\pi^-+\mathrm{C}$ and $\pi^-+\mathrm{W}$ collisions studied with the HADES experiment at GSI\cite{HADES:2017mzn}, offer a unique opportunity to investigate the production of light and heavy nuclei as well as hypernuclei in a controlled environment. 
\textcolor{blah}{The 2014 HADES measurement provided differential production cross sections for pions, protons, $\Lambda$ hyperons, and strange mesons in $\pi^-+\mathrm{C}$ and $\pi^-+\mathrm{W}$ reactions at $p_{\mathrm{lab}}=1.7$~GeV/$c$~\cite{hades24}. The comparison of the two target systems provides important constraints on hadron production and propagation in cold nuclear matter and constitutes an experimental baseline for calculations of cluster and hypernucleus formation in pion-induced reactions.}
Compared to proton-induced reactions, pion-induced reactions provide more free energy for particle production at the same incident momentum, thus enabling the creation of $K\Lambda$ pairs and even $KK\Xi$ triples through the decay of heavy nucleon resonances \cite{Graef:2014mra, Steinheimer:2015sha}.

The produced $\Lambda$ or $\Xi$ hyperons may further interact within the target nucleus, allowing for bound-state formation during the fragmentation of the system. 
The deceleration of $\Lambda$ or $\Xi$ within the nucleus evidently depends on the target size, making the target-size dependence of hypernucleus formation a sensitive probe of the spacetime evolution of strange hadrons and their in-medium interactions.

\begin{figure} [t]
    \centering
    \includegraphics[width=0.8\columnwidth]{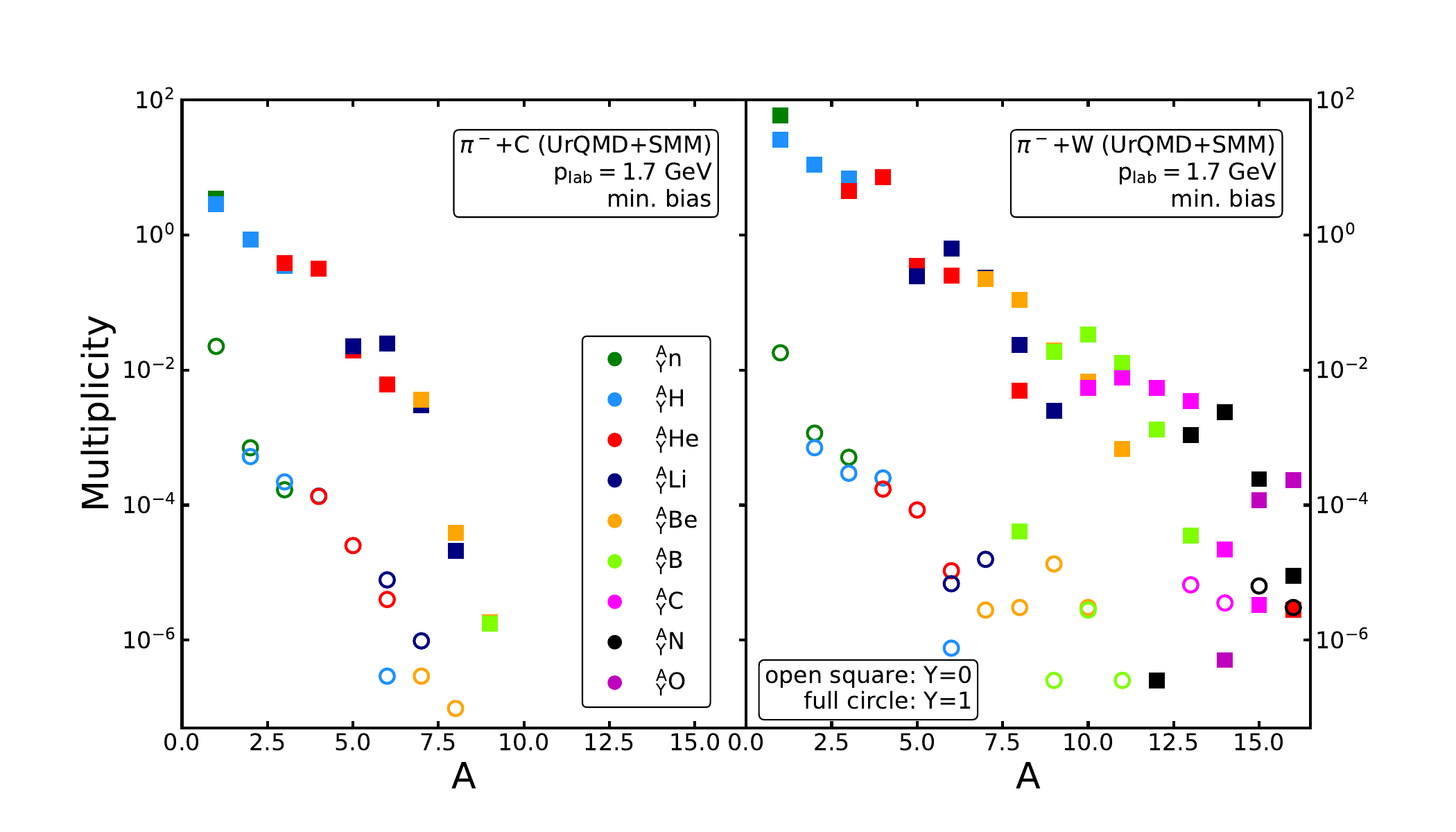}
    \vspace*{-0.2cm}
    \caption{Predictions for light nuclei (full squares) and hypernuclei (open circles) production in $4\pi$ for minimum bias pion-induced reactions, \textit{i.e}., $\pi^-+\mathrm{C}$ (left panel) and $\pi^-+\mathrm{W}$ (right panel) at an incident beam momentum of $p_\mathrm{lab}=1.7\,$GeV$/c$. 
    The results are from a SMM analysis of reactions simulated with UrQMD (v3.5). (Figure adapted from Ref.\,\cite{Kittiratpattana:2023atz}.)
    \label{fig:mult_nuclei_hypernuclei_pA_reactions}}
\end{figure}

The production of light nuclei and hypernuclei up to mass number $A=16$ has been explored in a complementary study \cite{Kittiratpattana:2023atz, Ergun:2024jwi} using the UrQMD \cite{Bass:1998ca, Bleicher:1999xi} transport model combined with the statistical multifragmentation model (SMM) \cite{Bondorf:1995ua}. Figure~\ref{fig:mult_nuclei_hypernuclei_pA_reactions} shows predictions for light nuclei (full squares) and hypernuclei (open circles) production in $4\pi$ acceptance for minimum-bias pion-induced reactions on carbon ($\pi^-+\mathrm{C}$, left panel) and tungsten ($\pi^-+\mathrm{W}$, right panel) at an incident beam momentum of $p_\mathrm{lab} = 1.7\,$GeV$/c$.
The results indicate substantial production of both light and heavier (hyper)nuclei.

At higher incident momenta, channels for $\Xi$ and double-$\Lambda$ production will open, enabling the study of double-strange hypernuclei in a clean environment. 
The expected luminosity at SIS100 will further allow for detailed investigations of hypertriton (${}^3_\Lambda$H), hyperhydrogen-4 (${}^4_\Lambda$H), hyperhelium-4 (${}^4_\Lambda$He), and heavier or double-strange hypernuclei, significantly advancing the study of hypernuclear binding energies, lifetimes, and two- and three-body hyperon–nucleon interactions.

Based on preliminary rate estimates from $\pi^- + \text{W}$ data taken in 2014 with the HADES detector \cite{HADES:2017mzn}, combined with significant improvements in pion beam properties since then, forthcoming measurement campaigns are expected to offer substantial enhancements over the 2014 experiment, resulting in an overall gain factor of 100. 
This notable increase in efficiency and data collection capability translates into an expected production of approximately 10{,}000 hypertritons. 
These yields represent an unprecedented amount -- nearly an order of magnitude increase -- in hypernuclei statistics at GSI, enabling statistically robust studies of hypernuclei properties, including lifetimes, branching ratios, and formation dynamics.

\subsubsection{Determination of momentum dependence of the optical potential}

In infinite nuclear matter, momentum and position are uncorrelated, allowing the equation of state (EoS) of cold nuclear matter to be directly calculated from the interaction potential. 
In contrast, for finite, expanding nuclear matter, such as that formed in HICs, the interaction depends explicitly on both the momentum and the local density (via the position) of the particles.

\begin{figure}[t]
\centering
    \includegraphics[height=0.27\textwidth]{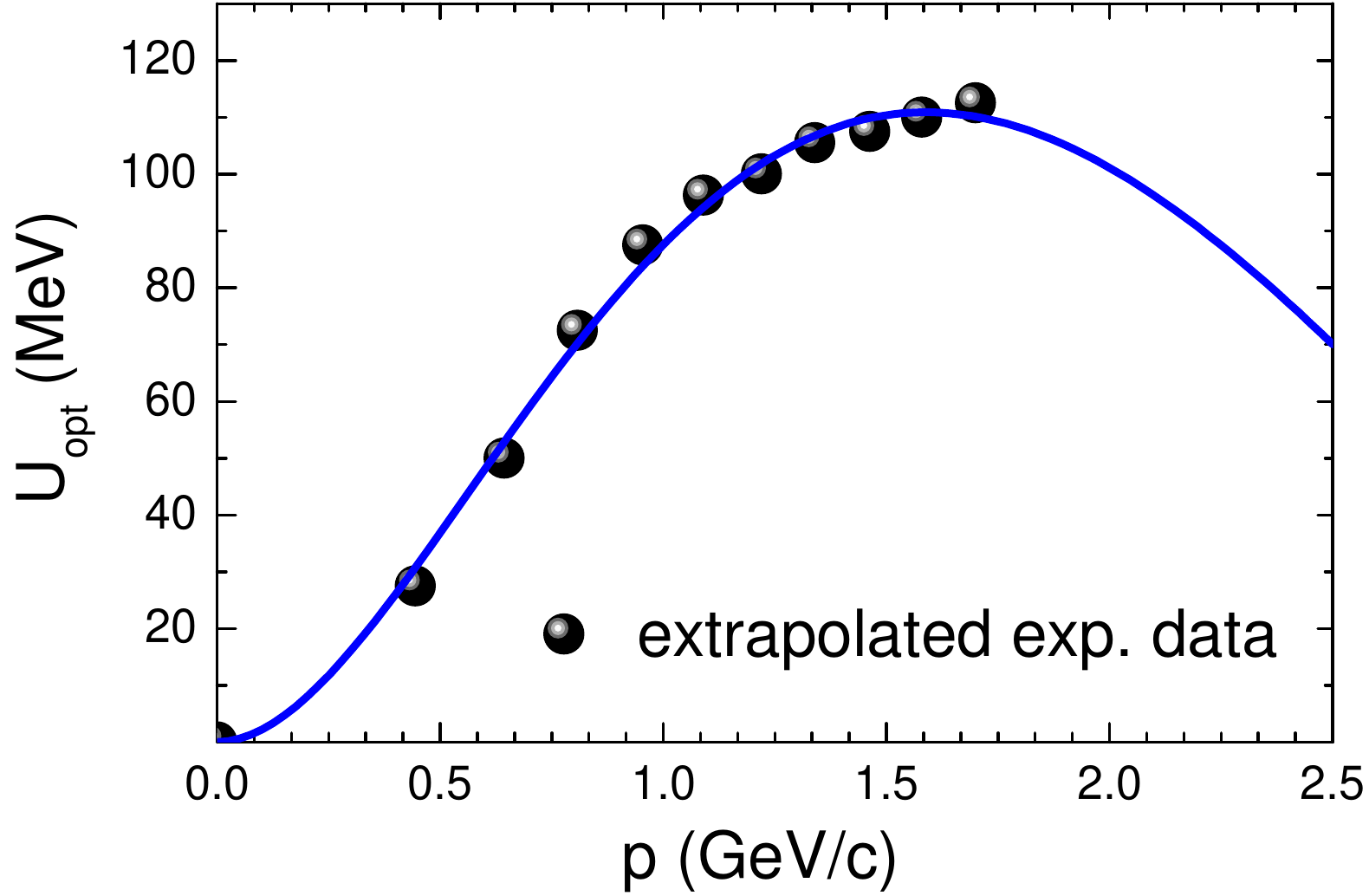} 
    \includegraphics[height=0.3\textwidth]{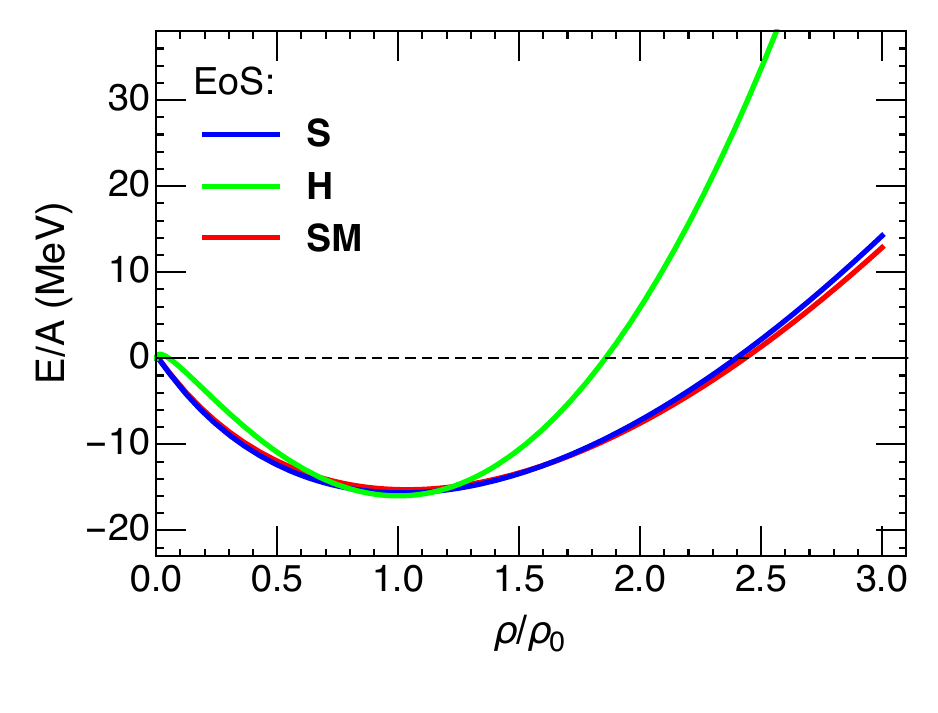}       
    \caption{Left. 
    Schr\"odinger equivalent optical potential $U_{opt}$ versus total momentum $p$ of the proton extracted from $pA$ collisions \cite{Clark:2006rj, Cooper:1993nx}. 
    Right. EoS for $T=0$ for the hard (green line), soft (blue line) and the soft momentum-dependent potential (red line).      
    (Figures adapted from Ref.\,\cite{Kireyeu:2024hjo}.)
    \label{Fig_Uopt}}
\end{figure}

The \textcolor{blah}{non-relativistic} two-body interaction potential is typically decomposed into a static component (Skyrme-type interaction) and a momentum-dependent component.
The hard, soft, and soft momentum-dependent EoS employed in the PHQMD transport model for studying flow coefficients \cite{Kireyeu:2024hjo} are illustrated in Fig.\,\ref{Fig_Uopt}\,-\,right. 
Notably, while the soft and soft momentum-dependent EoS are nearly identical in cold infinite matter, they lead to markedly different predictions for observables such as particle spectra and flow coefficients in HICs. For detailed discussions, see Refs.\,\cite{Tarasovicova:2024isp, Steinheimer:2024eha, Kireyeu:2024hjo}.

The momentum dependence of \textcolor{blah}{the non-relativistic two-body interaction, called optical potential, can be extracted from elastic $pA$ scattering data} \cite{Kireyeu:2024hjo}.
Currently elastic pA data only extend up to momenta of 1.7\,GeV$/c$ (see Fig.\,\ref{Fig_Uopt}\,-\,left), which is significantly below the beam momenta reached in HICs at top GSI and FAIR energies.  To reduce systematic uncertainties in determining the EoS from transport models, knowledge of the optical potential at higher momenta is required. This can be achieved through dedicated $pA$ collision measurements at FAIR.

\textcolor{blah}{Experimentally, the momentum dependence of the optical potential can be constrained by systematic measurements of proton-nucleus reactions over a broad range of beam momenta and target masses. In particular, elastic $pA$ scattering observables, including differential cross sections and angular distributions, provide sensitivity to the real and imaginary parts of the proton-nucleus optical potential. Extending such measurements at FAIR to substantially higher projectile momenta than presently available would allow the momentum dependence of the mean field to be constrained in the kinematic region directly relevant for transport calculations of heavy-ion collisions.}





\subsubsection{Influence of the electromagnetic fields on particle
dynamics in nuclear matter}

The magnetic field strength, $B$, generated in heavy-ion collisions has been estimated to reach up to $eB \sim 50\,m_\pi^2 \sim 10^{19}\,$Gauss, exceeding the field strength on the surface of magnetars by an order of magnitude \cite{Kharzeev:2007jp, Skokov:2009qp}. 
However, such intense fields exist only briefly at the early stages of the reaction \cite{Voronyuk:2011jd}. 
In symmetric systems, like Au+Au collisions, the electric and magnetic forces acting on charged particles largely compensate each other, resulting in minimal impact on observables such as rapidity distributions, particle spectra, and flow harmonics \cite{Toneev:2011aa}.

In asymmetric collisions, this compensation is reduced owing to the imbalance in proton numbers between the colliding nuclei \cite{Hirono:2012rt, Voronyuk:2014rna}. 
One can expect the impact of the electromagnetic (EM) field to increase in extremely asymmetric systems, such as proton-nucleus collisions. 
This issue was explored in Ref.\,\cite{Oliva:2019kin}, where the directed flow $v_1$ of light mesons—$\pi^+$, $\pi^-$, and $K^+$, $K^-$—in $p$+Au collisions at the top RHIC energy was studied within the transport framework using the PHSD approach.

\begin{figure}[t]
\centering
\includegraphics[width=0.4\textwidth]{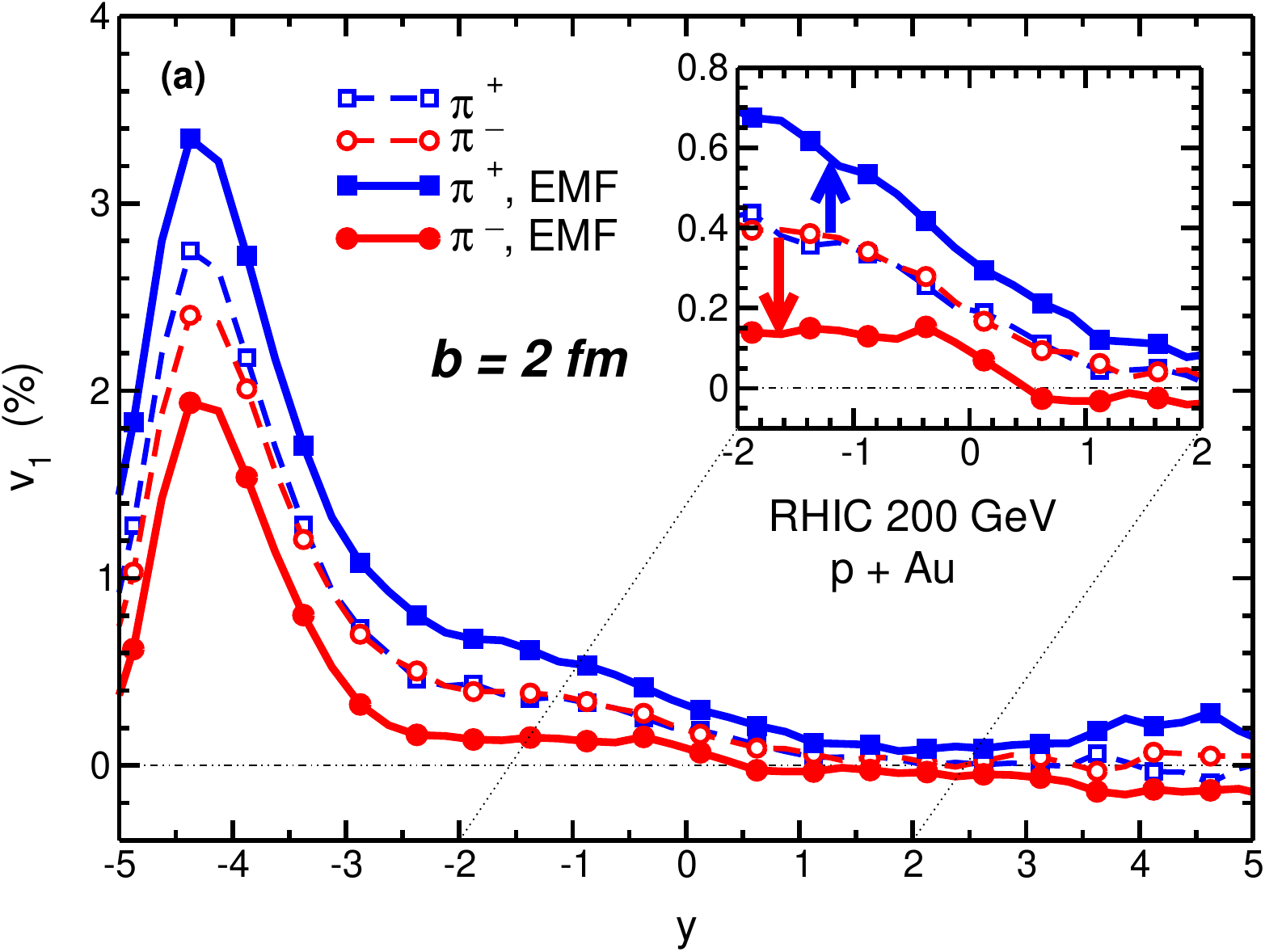}
\includegraphics[width=0.4\textwidth]{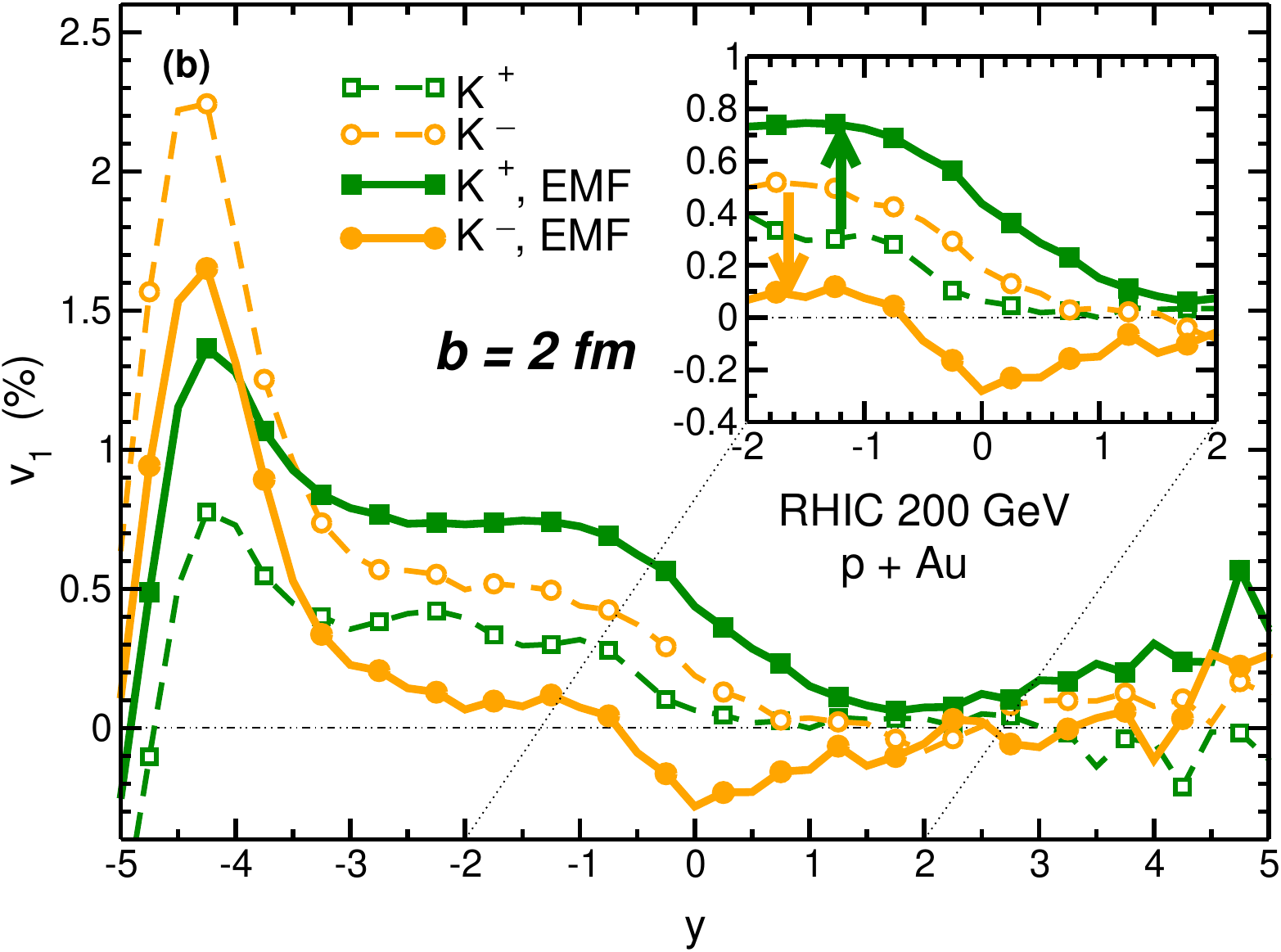}
\caption{Directed flow in percentage of pions (left) and kaons (right) as a function of rapidity for $b=2\,$fm of $p$+Au collisions at $\sqrt{s_{NN}}=200\,$GeV obtained with PHSD simulations with (solid curves) and without (dashed curves) electromagnetic fields. 
The inset panels are zooms of the rapidity window $\left|y\right|<2$, with arrows highlighting in which direction the presence of the electromagnetic fields affects the $v_1$ observable. 
(Figure adapted from Ref.\,\cite{Oliva:2019kin}.)
\label{fig:v1_y_pAu200GeV}}
\end{figure}
Figure~\ref{fig:v1_y_pAu200GeV} illustrates the splitting of the $v_1$ for pions (left) and kaons (right) under the influence of electromagnetic  fields at RHIC energies.
\textcolor{blah}{While quantitative calculations of electromagnetic-field effects in $p+A$ collisions at FAIR energies are presently not available, studies of asymmetric nucleus--nucleus collisions suggest that charge-dependent effects may become even more pronounced in the FAIR/NICA energy range~\cite{Toneev:2016qjx}.
}

The upcoming high-statistics CBM experiment with a proton beam provides an opportunity to study the influence of electromagnetic fields on the ratios of charged particles, like $\pi^+/\pi^-$ and $K^+/K^-$, as well as on collective observables, $v_n$.

\subsubsection{Probing isospin symmetry violation in kaon production}

\begin{figure} [t!]
    \centering
    \includegraphics[width=0.45\columnwidth]{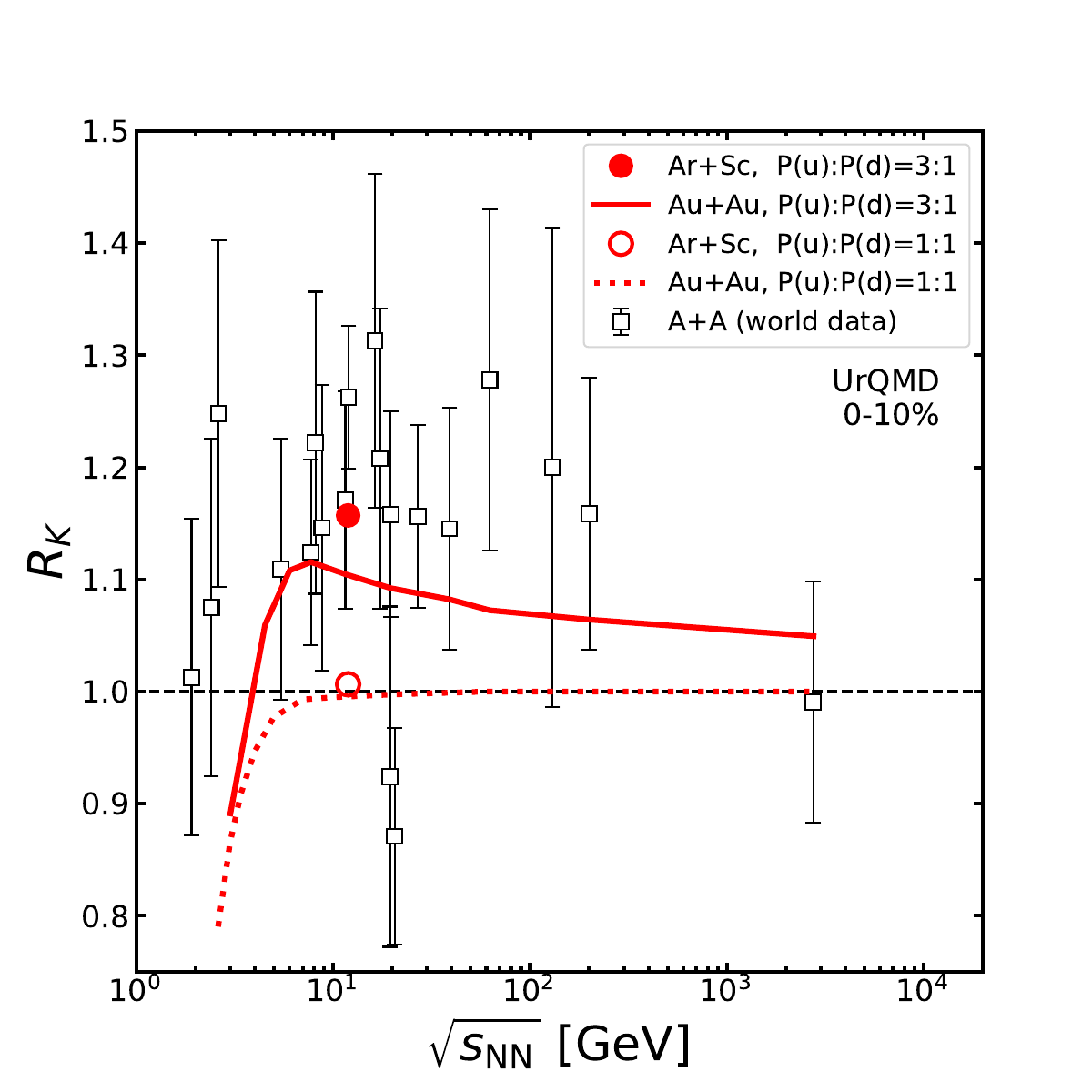}
    \caption{Comparison of simulations with data on the ratio $R_K=(K^+ + K^-)/(K^0 + \bar K^0)$. 
    Dotted line. Central Au+Au reactions with the standard parametrisation of the colour field fragmentation parameters. 
    Full lines. Results for the refitted parameters which allow for an asymmetry between up- and down-quark production in the colour field. Black squares. Experimental data in nucleus-nucleus collisions, taken from the compilation in Ref.\,\cite{NA61SHINE:2023epu}. 
    (Figure adapted from Ref.\,\cite{Reichert:2025znn}.) }
    \label{fig:excfct}
\end{figure}

Isospin symmetry is a near-exact feature of the strong interaction, reflecting the approximate equality of up and down quark masses. 
In the absence of electromagnetic and weak effects, the strong interaction treats up and down quarks nearly identically, making isospin a generally good symmetry of QCD. 
However, it should also be noted that SU(3)-flavour symmetry is significantly broken for the other components, namely U-spin ($d\leftrightarrow s$) and V-spin ($u\leftrightarrow s$). 
Recent measurements from the NA61/SHINE collaboration (and earlier data from other experiments) \cite{NA61SHINE:2023azp, Brylinski:2024uei} have revealed a statistically significant excess of charged kaons over neutral kaons in central HICs, challenging the conventional understanding of isospin conservation in QCD processes. 
The full energy excitation function is shown in Fig.\,\ref{fig:excfct}.

The ratio $R_K = (K^+ + K^-)/(K^0 + \overline{K}^0)$ is expected to be close to unity in collisions involving isospin-symmetric nuclei (and less than one for nuclei with more neutrons than protons). 
Deviations of more than 10\% from this expectation, particularly the 18.4\% excess observed at $\sqrt{s_{NN}} = 11.9\,$GeV in Ar+Sc collisions, suggest a nontrivial source of isospin symmetry violation. 
Traditional sources, such as quark mass differences, Coulomb effects, and weak decays, are insufficient to account for the observed discrepancy.

Several recent investigations \cite{Giacosa:2017pos, Brylinski:2023nrb, Giacosa:2024bup, Reichert:2025znn, Xing:2025eip} attribute the observed violation to an inherent asymmetry either in the sea-quark content of the nuclei or in the fragmentation of colour fields. 
In this picture, up-quark pair production is favoured over down-quark production during string breaking, leading to a bias in the formation of charged kaons. 
This asymmetry, though subtle, is proposed to be universal, affecting both elementary reactions, such as $e^+e^-$ annihilation, and complex nuclear collisions.

The hadron physics programme at CBM is well suited to explore this isospin-symmetry breaking in kaon production, as it enables investigation of the onset of this violation in the beam energy range from 5 to 30\,GeV per nucleon. 
This covers the critical transition region where the ratio $R_K$ is expected to turn from $R_K < 1$ to $R_K > 1$, potentially related to the transition from resonance to string-dominated dynamics. 
Especially promising are studies of different collision systems with varying initial isospin, \textit{e.g}., $pp$, $p$+C, and $p$+Au, as well as pion-induced reactions, which may allow a test of the (speculated) universality of the enhancement across different systems.

If CBM confirms a persistent $R_K > 1$ across various systems and energies, it would lend strong support to the hypothesis that quark pair production during string fragmentation is flavour asymmetric. 
This would imply that some standard assumptions in hadronisation models used in Pythia, UrQMD, and similar frameworks, may need to be reconsidered. 
In particular, the string breaking probabilities for $u\bar{u}$ versus $d\bar{d}$ pairs would need to be reparameterised to reflect experimental findings.

The hadron physics programme at CBM can therefore provide new insights to test the hypothesis of asymmetric up/down quark production in string fragmentation. These measurements will not only clarify the origin of the observed $R_K$ excess but may also reveal deeper, non-perturbative QCD mechanisms governing hadron formation, isospin, SU(3) flavour symmetry, and the microscopic dynamics of the strong interaction.


\subsection{The Nuclear Guide to $\nu$--A: Constraints for Long-Baseline Neutrinos}\label{sec.lblref}

The Long Baseline (LBL) neutrino oscillation experiments T2K (Tokai to Kamioka) and NOVA (NuMI Off-axis $\nu_{e}$ Appearance) are entering a new era of precision physics. The next generation experiments Hyper-K (Hyper Kamiokande) and DUNE (Deep Underground Neutrino Experiment) which are under construction will enable an increase of the neutrino rates by factors between 25 to 100 in statistics depending on the channel. This unprecedented statistics will give access to CP violation in the neutrino sector and to the neutrino mass hierarchy. In these experiments, neutrinos are produced by high-intensity accelerators (e.g., J-PARC, Fermilab), with neutrino rates expected to increase by roughly a factor of 20. The oscillation probability at a given distance from the source depends on neutrino energy, mixing angles, and mass splittings, and is measured by comparing rates, reconstructed energies, and flavors at near detectors (ND, close to the source) and far detectors (FD, hundreds of kilometers away).\par

With statistics no longer the main limitation, the field is transitioning from statistics- to systematics-dominated datasets. Accurate oscillation measurements hinge on precise neutrino energy reconstruction and on reducing total systematics from today’s $\sim 5$--$10\%$ to about $1$--$2\%$. The largest and most challenging uncertainties arise from modelling neutrino-nucleus interactions and final-state hadrons, primarily via hadronic models in GEANT4/transport frameworks. Longstanding tensions between models and $\nu$ cross-section measurements are most visible in exclusive final states with protons and pion production.\par

At T2K/Hyper-K, neutrinos in the 500--700~MeV range are often identified through CCQE-like topologies ($\nu_\mu + n \to \mu^- + p$ on nuclei), inferring $E_\nu$ from lepton kinematics. 
However, the precise reconstruction, based on the exclusive reconstruction of the muon and the proton (or the neutron for antineutrinos), is a major asset at the near detector to improve the precision on the neutrino flux and cross-section constraints. Moreover, single-pion production is a major background inducing a bias in the $E\nu$ reconstruction and becomes relevant in the high-energy tail (above $\sim 1$~GeV). For DUNE, with a harder flux, pion production channels are even more important; energy reconstruction must incorporate leptons \emph{and} hadrons ($\nu_\mu + N \to \mu^- + N' + n\pi$, $n \ge 1$).\par

Nuclear effects are critical, both at the interaction vertex and through re-scattering of produced protons and pions inside the nucleus (final-state interactions, FSI). Additional ingredients include Fermi motion, short-range correlations and secondary interactions in detector materials. Precise hadronic modelling must account for missing energy carried by undetected neutrons, re-absorbed pions and protons, or sub-threshold hadrons, which bias $E_\nu$ and feed energy-migration matrices.
Monte Carlo simulations attempt to correct for the mentioned above effects by tuning the cross sections from dedicated pion/proton-nucleus cross-section measurements. The access to the new pion-nucleus cross sections is crucial to improve the precision of the neutrino oscillation measurements in present and future LBL experiments. To tune the models and to test the dynamics besides the inclusive or total cross sections also differential spectra  
are important. This requires large statistics and versatile detection system, like the HADES spectrometer. \par

\paragraph{Separating the constraints: $\pi{+}A$ versus $p{+}A$}\par
For LBL flux and interaction predictions, two complementary hadron-production inputs are required.\par
(i) $\pi{+}A$: Pion--nucleus data constrain in-medium propagation and intra-nuclear transport of pions that dominate secondary and tertiary cascades and govern re-scattering/absorption systematics relevant to $\nu$ final states and energy migration. While the primary vertex differs between $\nu{+}A$ and $\pi{+}A$, the in-nucleus dissipation proceeds through similar processes (baryon-resonance excitation, rescattering, absorption). Hence $\pi{+}A$ directly improves neutrino--nucleus interaction modelling and the treatment of hadronic final states.\par
(ii) $p{+}A$: Proton--nucleus data provide baseline constraints on primary hadron production in target and horn materials (thin and replica targets), which are essential for beam-flux calibration. These measurements anchor $d^2\sigma/dp\,d\Omega$ for parents ($\pi$, $K$, $p$) across materials and phase space relevant to focusing optics. \par

\paragraph{State of $\pi{+}A$ data and proposed program} 
Existing pion-beam data are sparse and concentrated near the $\Delta(1232)$ resonance ($p \lesssim 500$~MeV/$c$). Critically, there is a lack of measurements in the energy--material combinations most relevant for LBL systematics, notably $\pi^{-}{+}\mathrm{O}$ and $\pi^{-}{+}\mathrm{C}$ (T2K/Hyper-K environment) and $\pi^{-}{+}\mathrm{Ar}$ (DUNE). This gap can be addressed with targeted measurements.\par

In a pioneering pion-beam run in 2014, HADES measured inclusive $p,\,\pi^{+},\,\pi^{-},\,d,\,t$ yields and semi-exclusive two- and three-particle coincidences (e.g., $2\pi^{-}$, $2\pi^{+}$, $2p$, $\pi^{+}\pi^{-}p$) on a carbon target at $p_\pi$=0.69\,GeV/$c$~\cite{hoj23,ram24}, with additional high-statistics datasets at 0.61, 0.66, 0.75, and 0.80~GeV/$c$. The analysis of data taken at 0.69~GeV/$c$ served as a benchmark for transport models (SMASH,  JAM2.1$/$RQMD.RMF, GiBUU) and the intranuclear cascade model INCL++, used across heavy-ion and neutrino communities. The model predictions exhibit an unexpectedly large spread, underscoring the need for targeted adjustments. \par

We propose, as a potential first phase of the programme, to complete the $\pi^{-}{+}\mathrm{C}$ set with $p_\pi$=0.50~GeV/$c$ and to extend to $\pi^{-}{+}\mathrm{Fe}$ and $\pi^{-}{+}\mathrm{Pb}$ at \SIlist{0.50;0.60;0.70}{GeV}/$c$. These energies are strategic for T2K/Hyper-K. At higher pion beam momenta $\sim1$ GeV/$c$, measurement with KCl target, already utilized in HADES, would be beneficial for DUNE, which will use liquid Ar-TPC for the neutrino detection. Since KCl ($^{39}$K$^{35}$Cl) has  similar to $^{40}$Ar atomic number, and it is easy to produce, hence measurement of $\pi$+KCl scattering is still valuable.  The resulting datasets will selectively probe QE-like (including 2p2h), multi-pion production, re-scattering and absorption/charge-exchange, thereby providing strong model discrimination and reducing the systematics that dominate LBL precision.\par 

In a second phase, the programme can be supplemented by rapid, high-statistics $p{+}A$ runs with CBM, leveraging the high-intensity proton beam at FAIR to strengthen flux calibration constraints.\par

Combined, these datasets enable joint ND--FD fits with reduced degeneracies between flux, cross sections, and detector response, unlocking the target precision at the percent level.\par

\subsection{Summary}
$p(\pi)+A$ reactions are a very promising research field. 
They allow for the study of many-body effects under defined conditions, including in-medium modifications of hadrons, subthreshold particle production, two-step reactions, and (hyper)nuclei production. 

Moreover, the production and in-medium behaviour of open and hidden charmed hadrons can be examined. 
Together with $p(\pi)+p$ collisions, they also provide input for phenomenological and transport approaches, necessary to interpret the results of HICs, and even offer possibilities to search for dark matter.

Beyond these QCD–nuclear physics goals, dedicated $p{+}A$ and $\pi{+}A$ reference measurements provide  in-nucleus transport constraints required for future LBL neutrino experiments. 
They enable flux calibration (via $p{+}A$ parents, $d^2\sigma/dp\,d\Omega$) and reduce interaction/FSI systematics (via $\pi{+}A$), thereby lowering dominant uncertainties in neutrino–energy reconstruction.

\newpage
\section{Connections and input to astro(particle) physics} 
\label{sec.AstroConnection}

{\small {\bf Convenors:} \it K.-H. Kampert, T. Saito, I. Vida\~na}



\noindent The GSI/FAIR facility offers increasing links to astrophysics and astroparticle physics. 
In this context, high-intensity proton and pion beams are of key interest. Proton beams enable precise measurements of nuclear reaction rates relevant to stellar and explosive nucleosynthesis, \textit{e.g}., $p$-, $rp$-, and $\nu p$-processes, helping to constrain models of element formation in stars, novae, and supernovae. 
Pion beams provide access to the nuclear EoS, relevant for neutron stars (NSs), core-collapse supernovae, and compact-star mergers. 
They also allow studies of in-medium hadron modifications, offering insights into QCD phase transitions and chiral symmetry restoration under extreme astrophysical conditions.
Furthermore, proton, nuclear, and pion beams support the modelling of cosmic ray interactions in galactic environments and dense astrophysical sources, and contribute to understanding extensive air showers; so, linking nuclear physics with high-energy astrophysics.
Last but not least, SIS100 can contribute to the experimental search for DM via the vector and axion portals, with dark photons and light mesons acting as mediators.

This section outlines the astrophysical motivation behind the experiments proposed in this document. The focus is on how accelerator-based measurements can improve our understanding of astrophysical phenomena, particularly through their connection with the nuclear EoS, cosmic ray physics, and dark matter searches.

 
\subsection{The nuclear equation of state: experimental and astrophysical constraints} 


The nuclear EoS is a key ingredient in modelling NSs, core-collapse supernovae, and compact-star mergers \cite{Oertel:2016bki, Rezzolla:2018jee}. {\color{blah} Its determination remains challenging owing to the wide range of densities, temperatures, and isospin asymmetries involved, the limited knowledge of hadron properties in the nuclear medium (see Secs.~\ref{subsub:inmediumpropvect}, \ref{subsub:inmediumpropstrbar}, \ref{subsub:inmediumpropopphidd}), and the complexity of the nuclear many-body problem \cite{mbp}. Many-body approaches are generally well constrained for finite nuclei and homogeneous nuclear matter around nuclear saturation density, $\rho_0 \simeq 0.16$ fm$^{-3}$, where they can be tested against a broad range of nuclear-structure and scattering data. At higher densities, larger isospin asymmetries, and finite temperatures, relevant for the interiors of neutron stars, core-collapse supernovae, and compact-star mergers, the predictive power of these approaches becomes more model dependent. In particular, microscopic calculations may involve expansions or approximations whose convergence and uncertainties become increasingly difficult to control as the density increases, while the relevant in-medium interactions and many-body correlations are less directly constrained by experiment \cite{mbp,Drischler:2021kxf}.} Astrophysical observations are therefore crucial for testing EoS models. This subsection reviews experimental and observational constraints, common theoretical approaches, and highlights open questions the proposed physics programme aims to address.

\subsubsection{Nuclear experimental constraints}

Around nuclear saturation density, $\rho_0$ and isospin asymmetry $\beta = (\rho_n - \rho_p)/(\rho_n + \rho_p) = 0$, the nuclear EoS can be characterised by a set of a few isoscalar ($E_0, K_0, Q_0$) and isovector ($S_0, L, K_{\mathrm{sym}}, Q_{\mathrm{sym}}$) parameters that can be constrained by laboratory experiments and astrophysical observations. 
They are related to the coefficients of the Taylor expansion of the energy per particle of nuclear matter in density and isospin asymmetry:
\begin{equation}
\frac{E}{A}(\rho, \beta) = E_0 + \frac{1}{2}K_0 x^2 + \frac{1}{6}Q_0 x^3 + \left(S_0 + Lx + \frac{1}{2}K_{\mathrm{sym}}x^2 + \frac{1}{6}Q_{\mathrm{sym}}x^3\right)\beta^2 + \mathcal{O}(4) \ , 
\qquad x = \frac{\rho - \rho_0}{3\rho_0} \ .
\label{ec:expansion}
\end{equation}

Measurements of density distributions \cite{DeVries:1987atn} and nuclear masses \cite{Audi:2002rp} give access to the values of $\rho_0 =$0.15-0.16\,fm$^{-3}$ and the binding energy of symmetric nuclear matter at saturation, $E_0 = -16 \pm 1\,$MeV, respectively. 
The nuclear incompressibility, $K_0$, is a key parameter characterising the stiffness of nuclear matter and can be extracted from the analysis of Isoscalar Giant Monopole Resonances (ISGMR) in heavy nuclei, as well as from HIC data. However, its extraction is complex and remains somewhat ambiguous. 
Analyses of ISGMR typically suggest values for $K_0$ in the range of approximately 210 to 250 MeV \cite{Blaizot:1980tw, Colo:2004mj, Piekarewicz:2003br}. 
In contrast, analyses of HICs often indicate a lower value, typically in the range of $180-220\,$MeV \cite{Fuchs:2000kp,Senger:2021cfo,Senger:2024fwc}, with some experiments pointing to values around $220 \pm 40\,$MeV \cite{Wang:2018hsw}, suggesting a softer EoS. 

Earlier, it was thought that ISGMR and HICs provided fundamentally different compressibility moduli. However, this distinction has largely been superseded. Early interpretations, such as those based on plastic ball data \cite{Molitoris:1986pp}, initially indicated the need for a hard EoS, but later studies found that both soft momentum-dependent and hard static EoS parametrisations can reproduce key flow observables in HICs \cite{Aichelin:1987ti, Hillmann:2019wlt, Kireyeu:2024hjo}. 
The importance of momentum-dependent interactions has since been reinforced by studies of the optical potential in proton-nucleus scattering experiments \cite{Cooper:1993nx, Cooper:2009zza}. 
Additionally, detailed analyses of kaon production, particularly the work initiated by Ref.\,\cite{Fuchs:2000kp}, further support a soft, momentum-dependent EoS as consistent with HIC data \cite{Hartnack:2005tr, Hartnack:2011cn}. 

Nowadays, there is broad consensus that HIC data collectively favour a soft, momentum-dependent EoS. 
However, this introduces a tension with neutron star physics, which typically requires a much stiffer EoS to account for observed $2\,M_\odot$ NSs \cite{Demorest:2010bx, Antoniadis:2013pzd, NANOGrav:2019jur}. 
A significant theoretical challenge in determining the EoS arises because most microscopic calculations rely on expansions that lose their validity at densities between $\rho_0$ and $2\rho_0$. 
For instance, in the case of the Bethe-Goldstone \cite{mbp,Day:1967zza} and Dirac-Brueckner-Hartree-Fock \cite{TerHaar:1986xpv, TerHaar:1987ce, Brockmann:1990cn} theories, the expansion parameter $k_F r_c$ (where $k_F$ is the Fermi momentum and $r_c$ is the range of the nucleon-nucleon interaction) approaches unity in this density regime, signalling the breakdown of the expansion’s reliability. 
Similar limitations affect ChEFTs. 

As a result, extracting the EoS from experimental data remains essential. Transport model simulations with varying EoS parametrisations are commonly used, though this method still faces considerable uncertainties. 
The extracted EoS often depends on model-specific inputs, such as nucleon-nucleon cros section parametrisations, interaction ranges, and treatments of Pauli blocking. 
Although current approaches can constrain $K_0$ within about 20\% uncertainty, achieving greater precision will require more systematic studies and closer collaboration between research groups to reconcile differences in model implementations and assumptions.

To improve the reliability of transport simulations, it is crucial to anchor these models with precise and comprehensive reference data from elementary reactions. 
In this context, systematic studies of hadronic interactions using proton and pion beams
are essential, as they provide the microscopic input needed to constrain transport models. 
This, in turn, enables a more accurate extraction of nuclear matter properties from HICs. 
While HICs can significantly enhance our understanding of the nuclear EoS, a persistent challenge is that the high densities achieved in such reactions are always accompanied by high temperatures. 
Such conditions differ markedly from the cold, dense environment of NSs. 
{\color{blah}However, HICs can provide complementary information on the nuclear EoS in regions of the density-temperature plane that partially overlap with those encountered in compact-star mergers. In particular, both systems can reach densities above nuclear saturation density and temperatures of several tens of MeV, with merger simulations reaching temperatures of order $\sim 80-100$ MeV locally and transiently \cite{Sekiguchi2011,Kastaun2016,Perego2019}. HICs therefore provide an experimentally accessible laboratory in which the properties of hot and dense nuclear matter can be studied \cite{BraunMunzinger2016,Sorensen2024}. The correspondence should not, however, be interpreted as implying that the thermodynamic conditions are identical: neutron-star mergers involve much more neutron-rich matter, different chemical composition and weak-interaction equilibrium, and a very different dynamical evolution. HIC measurements can instead provide complementary constraints on the nuclear interactions and finite-temperature EoS relevant for interpreting merger dynamics and, ultimately, gravitational-wave and electromagnetic observations \cite{Most2023}.}
Further research is also necessary to understand the compressibility of deformed nuclei, and the trend in compressibility among neutron-rich systems, as these factors are critical for modelling nuclear matter in extreme environments. 
In addition, the skewness parameter $Q_0$, one of the less constrained isoscalar parameters, remains highly uncertain, with current estimates ranging from about $-500$ to $300\,$MeV, highlighting the need for continued experimental and theoretical efforts to narrow this range.

Experimental information on the isovector parameters of the nuclear EoS can be obtained from several sources, such as the analysis of Isovector Giant \cite{Garg:2006vc} and Pygmy \cite{Klimkiewicz:2007zz, Carbone:2010az} Dipole Resonances, isospin diffusion measurements \cite{Chen:2004si}, isobaric analogue states \cite{Danielewicz:2008cm}, isoscaling \cite{Shetty:2007zg}, measurements of the neutron skin thickness in heavy nuclei \cite{Brown:2000pd, Typel:2001lcw, Horowitz:1999fk,Roca-Maza:2011qcr, Brown:2007zzc, Centelles:2008vu, Warda:2009tc}, or meson production in HICs \cite{Li:2004cq,Fuchs:2005zg}. 
Astrophysical observations can also be used to constrain these parameters. 
It has been shown, for instance, that the slope parameter, $L$, of the symmetry energy is correlated with the radius \cite{Hu:2020ujf} and the tidal deformability, $\lambda=\frac{2}{3}k_2R^5$, of a $1.4 M_\odot$ NS.
(Here, with $k_2$ is the tidal Love number and $R$ the radius of the star \cite{Zhao:2018nyf, Tsang:2019vxn}.)
Moreover, that precise and independent measurements of the radius and the tidal deformability from multiple observables of NSs can potentially pin down the correlation between $K_{sym}$ and $L$, and thus the high-density behaviour of the nuclear symmetry energy \cite{Li:2020ass}. 

However, while $S_0$ is more or less well established ($\approx 30$ MeV), the values of $L$, and in particular those of $K_{sym}$ and $Q_{sym}$, are still uncertain and poorly constrained. 
For example, combining different data, Ref.\,\cite{Lattimer:2012xj} gives $29.0 < S_0/{\rm MeV} < 32.7$ and $40.5 < L/{\rm MeV} < 61.9$, while Ref.\,\cite{Danielewicz:2013upa} suggests $30.2 < S_0/{\rm MeV} < 33.7$ and $35 < L/{\rm MeV} < 70$. 
The range of variation of $S_0$ and $L$ given by these two works, while overlapping, shows noticeable differences, particularly in $L$, highlighting the significant uncertainty still existing in the isovector part of the nuclear EoS, and underscoring the need for more precise data and refined theoretical models to better constrain these key parameters.

\subsubsection{Astrophysical observations}

The main astrophysical constraints on the nuclear EoS arise from observations of NSs. After more than fifty years of observations, a vast amount of data on various NS observables has been gathered. This data provides valuable insights into the internal structure of NSs, and, consequently, the nuclear EoS, the sole input required to solve the structural equations of NSs. 
The principal observables are listed here.
\begin{itemize}
\item {\it Masses.} 
They can be inferred with high precision for pulsars in binary systems by measuring post-Keplerian parameters, \textit{e.g}., the relativistic Shapiro delay \cite{Shapiro:1964uw}, an increase in light travel time through the curved space-time near a massive body. 
When measured over long periods of time, precise masses can be obtained for both the millisecond pulsar and its companion.
Examples are those of the millisecond pulsars 
PSR J1614–2230 ($1.928 \pm 0.017 M_\odot$) \cite{Demorest:2010bx}, 
PSR J0348+0432 ($2.01 \pm 0.04 M_\odot$) \cite{Antoniadis:2013pzd}, 
and PSR J0740+6620 ($2.14^{+0.10}_{-0.09} M_\odot$) \cite{NANOGrav:2019jur}. Given these measurements, any reliable nuclear EoS should predict maximum NS masses greater than $2M_\odot$. 

\item {\it Radii.} 
They can be determined either from the thermal emission of low-mass X-ray binaries or by tracking the X-ray emission from ``hot spots'' on the NS surface as the star rotates. 
Recent analysis of the NICER mission observations \cite{Riley:2019yda, Miller:2019cac}, combined with ChEFT models, estimated the mass of 
PSR J0030+0451 to be approximately $1.4 M_\odot$ and its radius near 13\,km \cite{Luo:2024lbz}. 
The equatorial radius of PSR J0740+6620 was found to be $13.7^{+2.6}_{-1.5}\,$km with a 68\% confidence interval \cite{Miller:2021qha}. This value was updated in 2024 with the refined radius of this millisecond pulsar measured as $12.92^{+2.09}_{-1.13}\,$km \cite{Dittmann:2024mbo}.

\item {\it Gravitational Waves.} 
The GW170817 event \cite{LIGOScientific:2017vwq}, detected on 17$^{\text{th}}$ August 2017, was a groundbreaking observation in astrophysics.
It has enabled further constraints on the EoS through direct measurements of the NS mass and tidal deformability; see, \textit{e.g}., \cite{LIGOScientific:2019eut}.
In particular, the value of the tidal deformability inferred from this event has imposed constraints on NS radii: they mustlie between about 12 and 13\,km, in agreement with the mass-radius measurements by NICER, and compatible with a moderately soft nuclear EoS. 
Additionally, the electromagnetic counterparts of the merger event have been used for EoS constraints employing different approaches. 
For instance, they indicate that very soft EoS models are likely incompatible with the brightness of the kilonova \cite{Bauswein:2017vtn}.
\end{itemize}

Other NS observables that constrain the nuclear EoS include rotational periods, moments of inertia, surface temperatures, gravitational redshifts, quasiperiodic oscillations, magnetic fields, pulsar glitches, and timing noise. (See, \textit{e.g}., Ref.\,\cite{Vidana:2018lqp}.)

\subsubsection{Theoretical approaches to the nuclear EoS}

Theoretically, the nuclear EoS has been determined by many authors using both phenomenological and microscopic many-body approaches. 
Phenomenological approaches, either non-relativistic or relativistic, are based on effective interactions that are frequently built to reproduce the properties of nuclei. 
Skyrme interactions \cite{Skyrme:1959zz} and relativistic mean-field (RMF) models \cite{Serot:1984ey} are among those most widely used. 
Many such interactions are built to describe nuclear systems close to the isospin-symmetric case and, therefore, predictions at high isospin asymmetries should be taken with care. 
Most Skyrme forces are, by construction, well behaved close to $\rho_0$ and moderate values of the isospin asymmetry. 
However, only certain combinations of the parameters of these forces are well determined experimentally. 
As a consequence, there is a large proliferation of different Skyrme interactions that produce a similar EoS for symmetric nuclear matter but predict a very different EoS for pure neutron matter. 
Relativistic mean-field (RMF) models are based on effective Lagrangian densities where the interaction between baryons is described in terms of meson exchange. 
The couplings of nucleons with mesons are usually fixed by fitting masses and radii of nuclei, and the properties of nuclear bulk matter, whereas those of other baryons, like hyperons, are fixed by symmetry relations and hypernuclear observables.

Microscopic approaches, on the other hand, are based on realistic two- and three-body forces that describe scattering data in free space and the properties of the deuteron. 
These realistic interactions have mainly been constructed within the framework of a meson-exchange theory \cite{Nagels:1973ku, Machleidt:1987hj, Nagels:1977ze, Holzenkamp:1989tq, Maessen:1989sx, Stoks:1999bz, Haidenbauer:2005zh, Rijken:2006en, Rijken:2006ep, Rijken:2010zzb}.
 {\color{blah}In the last two decades, chiral effective field theory (ChEFT) has emerged as a complementary and systematic framework for constructing nuclear interactions based on the symmetries and low-energy degrees of freedom of QCD \cite{Weinberg:1990rz, Weinberg:1991um, Bedaque:2002mn, Entem:2003ft, Epelbaum:2004fk, Epelbaum:2005pn, Polinder:2006zh, Haidenbauer:2013oca, Epelbaum:2014efa, Haidenbauer:2023qhf, Petschauer:2020urh, Kalantar-Nayestanaki_2012}. It provides two- and three-nucleon interactions that can be employed in many-body calculations of homogeneous nuclear matter, including symmetric nuclear matter and pure neutron matter \cite{Drischler:2016kuj,Alp:2025eqs}.
In particular, ChEFT calculations of neutron matter provide important constraints on the density dependence of the nuclear symmetry energy and its uncertainty at densities around and moderately above saturation 
\cite{Kruger:2013iza,Drischler:2020hue}. Recent calculations have further improved the theoretical control of the neutron-matter EoS and its uncertainties 
\cite{Drischler:2020hue,Tews:2024qnb}. The predictive range of ChEFT is nevertheless limited as the density increases, since its expansion parameter grows with the characteristic momentum of the system. Consequently, ChEFT constraints are particularly valuable at densities up to approximately $2\rho_0$, while higher-density neutron-star matter requires additional theoretical input and/or extrapolation \cite{Drischler:2020hue,Drischler:2021yqb}.

To obtain the nuclear EoS from these microscopic interactions, one must solve the complicated many-body problem \cite{mbp}, whose main difficulty lies in the treatment of the repulsive core that dominates the short-range part of the interaction.}
Different microscopic many-body approaches have been extensively used to study the nuclear EoS. 
These include: the Brueckner--Bethe--Goldstone \cite{mbp,Day:1967zza} and the Dirac--Brueckner--Hartree--Fock \cite{TerHaar:1986xpv, TerHaar:1987ce, Brockmann:1990cn} theories, the variational method \cite{Akmal:1998cf}, the correlated basis function formalism \cite{Fabrocini:1993eaz}, the self-consistent Green function technique \cite{Rios:2008fz, Carbone:2013eqa, Rios:2020oad}, the $V_{\mbox{low}\,k}$ approach \cite{Bogner:2003wn}, Quantum Monte Carlo techniques \cite{Wiringa:2000gb, Carlson:2003wm, Gandolfi:2009fj, Gezerlis:2013ipa}, renormalisation group methods \cite{Furnstahl:2013oba}, or Hartree--Fock calculations with chiral interactions \cite{Tews:2012fj, Kruger:2013kua, Keller:2020qhx}. 
All these methods have been employed to describe the EoS of homogeneous nuclear matter, {\color{blah}with differences between approaches providing an important measure of the theoretical uncertainty.}

Nonuniform matter, present in the crust of NSs and in supernova cores, has been described using either single-nucleus approximation (SNA) or nuclear statistical equilibrium (NSE) models \cite{Oertel:2016bki}. 
In SNA models, the composition of matter is assumed to be made of one representative heavy nucleus (the one energetically favoured), plus light nuclei (typically $\alpha$ particles) or unbound nucleons. 
In NSE models, on the other hand, matter composition is assumed to be a statistical ensemble of different nuclear species and nucleons in thermodynamic equilibrium.

Improving current theoretical approaches to the nuclear EoS is essential for an accurate description of nuclear matter under extreme conditions. 
Ongoing and future experimental advances offer promising pathways to enhance the accuracy, consistency, and predictive power of these approaches.


\begin{figure}[t]
\begin{center}
\includegraphics[width=0.7\linewidth]{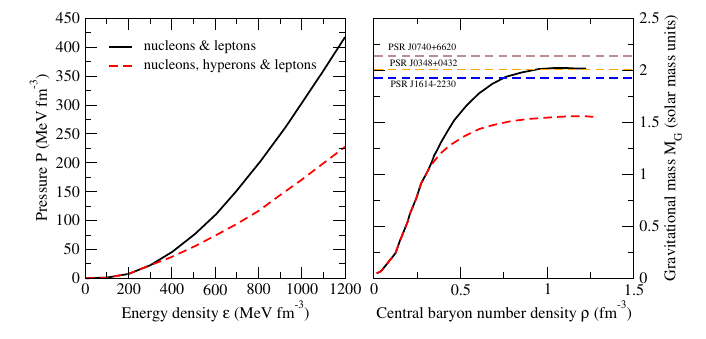}
\end{center}
\vspace*{-0.5cm}
\caption{
{\color{blah} Illustration of the possible effect of the presence of hyperons on the nuclear EoS (left) and on the mass of a neutron star (right). The curves are schematic and are intended only to illustrate the possible softening of the EoS associated with the appearance of hyperonic degrees of freedom and its potential impact on the maximum neutron-star mass. They do not correspond to any specific calculation based on microscopic interactions or on a quantitative EoS model. The magnitude of the effect depends on the poorly constrained YN, YY, and hyperonic three-body interactions. The horizontal lines show the observed masses of the millisecond pulsars PSR J1614-2230 ($1.928 \pm 0.017 M_\odot$) \cite{Demorest:2010bx}, PSR J0348+0432 ($2.01 \pm 0.04 M_\odot $) \cite{Antoniadis:2013pzd} 
and PSR J0740+6620 ($2.14^{+0.10}_{-0.09} M_\odot$) \cite{NANOGrav:2019jur}.}
}
\label{fig:MASS_RADIUS}
\end{figure}

\subsubsection{Open questions and future perspectives}
\label{subsubsect.astro}


{\color{blah}Some important questions regarding the nuclear EoS remain unanswered. One issue is particularly significant in the context of NSs: the so-called \textit{hyperon puzzle} \cite{Chatterjee:2015pua}. At high densities, such as those expected in the core of NSs, the nucleon chemical potential can become large enough to make the conversion of nucleons into hyperons energetically favourable. The appearance of hyperons can then relieve the Fermi pressure exerted by the baryons and soften the EoS, as illustrated schematically in the left panel of Fig.\ \ref{fig:MASS_RADIUS}.
We clarify that this figure is intended only as a schematic illustration of the possible qualitative effect of hyperons; it does not show the results of any specific calculation based on microscopic interactions or on a quantitative EoS model. The magnitude of this effect, in fact, depends sensitively on the poorly constrained hyperon-nucleon (YN) and hyperon-hyperon (YY) interactions, as well as on possible three-body hyperonic forces. In particular, the conventional formulation of the \textit{hyperon puzzle} relies on specific assumptions or estimates for these hyperon couplings.

For a range of such commonly adopted interaction models (see e.g., Ref.\ \cite{Chatterjee:2015pua} and references therein), the additional softening associated with the appearance of hyperons leads to a substantial reduction of the maximum mass of the corresponding NS, as illustrated also schematically in the right panel of Fig.\ \ref{fig:MASS_RADIUS}. In these models, the resulting maximum mass can be incompatible with the recent observations of two-solar-mass millisecond pulsars \cite{Demorest:2010bx, Antoniadis:2013pzd, NANOGrav:2019jur}. This tension is what is commonly referred to as the hyperon puzzle. It should nevertheless be stressed that whether hyperons necessarily lead to such a conflict with observations remains an open question, given the current uncertainties in the YN and YY interactions and, in particular, in their two- and three-body couplings. More accurate constraints on these interactions are therefore essential to determine whether the hyperon puzzle represents a genuine problem for the description of NS matter.
}

Compared to two- and three-nucleon interactions, the hyperon-nucleon (YN) and hyperon-hyperon (YY) interactions are still poorly constrained (see Sec.\,\ref{subsec.BaryonBaryonInt}), mainly owing to the scarcity of YN scattering data and the complete lack of YY data. 
This is because of experimental difficulties associated with the short lifetimes of hyperons and the low beam intensities presently available. 
Three-baryon forces involving hyperons (YNN, YYN, and YYY) \cite{Petschauer:2015elq, Haidenbauer:2016vfq, Kohno:2018gby, Logoteta:2019utx, Gerstung:2020ktv} are even less well constrained experimentally, although they could play a fundamental role in solving the hyperon puzzle.

New insights into two-body hyperon-nucleon (YN) interactions, beyond those from scattering experiments and hypernuclear studies, have recently come from femtoscopy measurements. 
This method, described in Sec.\,\ref{subsec.femto}, uses momentum-space correlations of YN pairs \cite{ALICE:2020mfd, ALICE:2021njx}, YY pairs \cite{ALICE:2019eol}, and YNN triplets \cite{ALICE:2022boj} produced in proton-proton collisions at the LHC. 
These data also help to constrain hyperon-hyperon (YY) interactions and three-body forces. 
For background on scattering and hypernuclei studies, see Secs.\,\ref{subsec.scatfromprod}, \ref{subsec.hypernuclei}, \ref{subsec.hypernuclei_in_piA} \cite{Botta:2012xi, Gal:2016boi}.

The comparison of the measured two- and three-body correlation functions with theoretical predictions permits one to test, improve, and constrain the existing two- and three-body hyperon interactions.

Notably, recent progress in LQCD (see Sec.\,\ref{sec:toolslatqcd}), particularly by the HAL QCD Collaboration \cite{HALQCD:2019wsz}, has provided a first-principles determinations of baryon-baryon interactions involving hyperons. 
These calculations have started to deliver meaningful constraints on the $\Lambda\Lambda$ and $\Xi N$ interactions, which were previously inaccessible, and have shown remarkable agreement with femtoscopic data, offering an increasingly robust bridge between QCD and the phenomenology of hypernuclear matter. 
This progress significantly enhances the reliability of theoretical models in the strangeness sector, even though further refinements are still needed to reach higher precision.

Very recently, the EoS of hypernuclear matter and the structure of NSs have been studied \cite{Vidana:2024ngv} employing a chiral YN interaction \cite{Haidenbauer:2019boi} tuned to femtoscopic $\Lambda p$ data \cite{ALICE:2021njx, Mihaylov:2023pyl}, and $\Lambda\Lambda$ and $\Xi N$ interactions determined from LQCD \cite{HALQCD:2019wsz} that reproduce the femtoscopic $\Lambda\Lambda$ and $\Xi^- p$ data \cite{ALICE:2020mfd}. 

Further complementary information on the YN and YY interactions can also be obtained by using the Dalitz plot analysis method \cite{Beringer:2012}. 
This approach has successfully been employed in reactions such as $pp \rightarrow pK^+\Lambda$ carried out at COSY-TOF \cite{COSYTOF:2012inb}, where it has revealed significant features, such as the $\Sigma N$ cusp, providing insights into YN interactions. 
The application of this technique to exclusive reactions, such as $pp \rightarrow \Sigma^+\Sigma^+ K_S K_S p$, offers a powerful method for extracting low-energy scattering parameters of YY interactions.

More broadly, open questions concerning the nuclear EoS include the following.
(\textit{i}) The connection between nuclear physics experiments and the properties of neutron-rich matter expected to be present in the crust and cores of NSs.
(\textit{ii}) A better understanding of the systematic uncertainties in the analysis of experimental data and astrophysical observations.
(\textit{iii}) 
{\color{blah}
The combination of complementary constraints on the nuclear EoS obtained at very different characteristic scales and under different physical conditions. These include microscopic information from nucleon scattering and nuclear masses and radii, constraints on the isovector sector from neutron-rich nuclei and collective excitations, information on dense matter from HICs, and macroscopic constraints from neutron-star masses, radii, tidal deformabilities, gravitational-wave observations, and electromagnetic counterparts of compact-star mergers. Combining these complementary data sets is essential because they probe different density, temperature, and isospin regimes and therefore constrain different aspects of the nuclear interaction and the EoS.
}
(\textit{iv}) The robustness and reliability of the different theoretical models and approaches to the nuclear many-body problem as the conditions of density, temperature and isospin asymmetry become more and more extreme.
(\textit{v}) The actual composition and state of matter of NS interiors. 

The understanding of the underlying interactions between the matter constituents of NSs, from which the nuclear EoS arises, hinges on a detailed comparison between theoretical models and experimental data. 
Especially, the use of proton and pion beams at GSI/FAIR would allow for controlled studies of nuclear reactions and matter under extreme isospin and density conditions, providing essential constraints on the in-medium nuclear interactions that are critical for accurately modelling NSs interiors. 

Future experimental studies of very neutron-rich nuclei near the limits of existence of rare isotopes, such as those that will be carried out at the SIS100 at FAIR, along with HICs, will provide crucial information to better understand the nuclear EoS and reduce systematic uncertainties in the interpretation of both experimental data and astrophysical observations.
Forthcoming measurements at GSI/FAIR using high-intensity 30\,GeV proton beams and heavier projectiles will probe hadronic matter under extreme conditions and contribute key input for transport model simulations. 
Moreover, elementary hadron physics studies at FAIR, such as those exploring in-medium modifications of hadrons and the onset of chiral symmetry restoration, will play a vital role in constraining the strong interaction at high densities and temperatures, thereby enriching existing theoretical models of dense matter and their applications in astrophysics. 
Observationally:
\begin{enumerate}[label=(\textit{\roman*}),itemsep=0.2em, parsep=0pt, topsep=0.3em]
    \item new X-ray telescopes, such as IXPE, XRISM, or Athena, will enable further advances and improve the modelling of X-ray sources;
    
    \item the improvement of the sensitivity of the LIGO, Virgo, and KAGRA \cite{IGWN_Observing_Plans_2025} as well as LISA \cite{Amaro-Seoane2023LISA_Astrophysics} detectors in the next decade will increase the prospects for dense-matter science with gravitational waves;
    
    \item the third generation of gravitational-wave detectors (Einstein Telescope, Cosmic Explorer) in the 2030s will enable the detection of a post-merger gravitational-wave signal with the potential to discriminate purely nucleonic matter from matter with hyperonic degrees of freedom \cite{Blacker2024};
    
    \item more observations of electromagnetic counterparts of compact object mergers will succeed and provide EoS constraints; and
    
    \item future observations of young NSs, \textit{e.g}., Cas~A or the potential remnant of SN~1987A, will constrain theoretical cooling models and possibly set limits on the energy gaps of superfluid matter.
\end{enumerate}

\subsection{High-energy nuclear fragmentation reactions and their relevance to cosmic ray astrophysics}


\subsubsection{Understanding and simulating galactic cosmic rays}

The latest generation of space-based and balloon-borne cosmic-ray (CR) experiments, most notably the Alpha Magnetic Spectrometer (AMS-02) \cite{Battiston:2008zza}, operated on the International Space Station, has brought GeV to TeV range CR physics into an era of precision.

CRs up to about 100\,PeV ($10^{17}$\,eV) are thought to be of galactic origin and are commonly referred to as galactic cosmic rays (GCRs). 
However, their sources have not yet been identified, nor is their propagation through the galactic magnetic field (GMF) and interstellar medium (ISM) well understood. 
In fact, the highly complex GMF is not well known either, although significant progress has recently been made in measuring and modelling the GMF \cite{Beck:2000dc, Unger:2019xct}.
Typical field strengths in the galactic plane are found to be on the order of 3\,\textmu G. 
The Larmor radius $r_{\rm L}$  of charged particles, measured in parsec (pc), can be parameterised as
\begin{equation}
r_{\rm L} \simeq 1.08~\textrm{pc} \times \left( \frac{E}{\textrm{PeV}} \right) \times \left( \frac{1}{Z} \right) \times \left( \frac{\text{\textmu G}}{B} \right),
\end{equation}
with $E$ and $Z$ being the energy and atomic number of the CR. Even for protons at PeV energies, $r_{\rm L}$ is much smaller than the scale height of the galactic disk, which is on the order of a few hundred parsec. 

Numerical models that simulate the propagation of GCRs, such as \textsc{DRAGON} \cite{Evoli:2017vim} or \textsc{GALPROP} \cite{Vladimirov:2010aq}, incorporate various physical processes that affect the transport, interaction, and energy loss of GCRs as they travel through the ISM. 
Since the Galactic dimensions are orders of magnitude larger than the typical Larmor radii, CRs propagate primarily by diffusion, which can be modelled by solving diffusion equations. The diffusion coefficients, to be determined by comparison with observational data, are critical parameters that influence how cosmic rays spread through the galactic volume.

Unlike in man-made particle accelerators, the deflection of charged particles in astrophysical magnetic fields actually modifies the magnetic field that bends the particles, thereby amplifying it through cosmic dynamo effects and pushing it out into the halo of the galaxy.

On their journey to Earth, CRs and electrons are not only deflected by the GMF, but also lose energy through several mechanisms. 
Protons and nuclei undergo nuclear interactions with interstellar gas, producing secondary particles. 
Interactions with photon fields give rise to photo-nuclear interactions, including photo-disintegration of nuclei or photo-pion production at high CM energies. 
High-energy electrons lose energy through Bremsstrahlung and inverse Compton scattering with low-energy photons, mostly CR microwave background (CMB) radiation, producing photons typically on the keV--MeV and TeV scales, respectively. In the following, only interactions involving protons and nuclei will be discussed.

The complex interplay of nuclear and particle physics interaction processes enables one to infer key quantities by comparing simulations with data. These quantities include the residence time of CRs in our galaxy, which provides information about the luminosity of galactic CR sources required to maintain a constant CR energy density despite the steady escape of CRs from the Milky Way. Other quantities include the age of GCRs, which provides information about the mean density of the interstellar medium (ISM) through which GCRs propagate. This quantity is related to the size of the galactic halo. 
Extracting this information from CR data requires precise modelling of cosmogenic particle production, especially in $n$-, $p$-, and He-induced reactions across a range of nuclei and with beam energies ranging from approximately 100\,MeV/n to 50\,GeV/n. 
This puts GSI/FAIR in an ideal position to provide such data.

Another motivation for studying GCRs, besides learning about their sources, acceleration mechanisms, and the galactic medium through which they propagate, is the search for indirect signatures of dark matter (DM) or dark photons. 
This possibility has recently attracted much interest. 
Relics from the early Universe could annihilate or decay in the halo of our galaxy, leaving a feeble footprint in charged GCRs and/or in photons or positrons. 
For example, Galactic cosmic-ray antiprotons are expected to be produced mainly by CR interactions in the ISM. 
However, a mild excess could also arise from DM particles annihilating with a thermal cross section into quarks \cite{Cuoco:2016eej, Cui:2016ppb}.

In view of these different processes contributing to antiprotons and antinuclei, it has been recognised that any claim about possible signatures of antimatter and dark matter in cosmic-ray data requires a precise understanding of CR propagation in our Galaxy, which in turn requires a precise understanding of nuclear interactions in the GeV to TeV regime.

\subsubsection{The need for more precise cross sections measurements}
\label{sec:precise-crosssections}

The main constraints on propagation scenarios are provided by the two main groups of CR species: the so-called \textit{primary} species (in particular H, He, C, O, Ne, Mg, Si, and Fe), which are mostly produced directly by CR sources; and the \textit{secondary} species (including Li, Be, and B, as well as elements just below the Mg, Si, and Fe groups), which are strongly enriched by spallation processes of primary species with the ISM, \textit{e.g}, C + p$_{\rm ISM} \to$ B + $X$.
The boron-to-carbon (B/C) ratio observed on Earth is a classic example: Li, Be, and B were produced in only minute amounts during Big Bang nucleosynthesis (BBN) and are not enriched by stellar burning, so the observed fluxes in CRs result effectively from spallation reactions of CNO elements with the ISM. Thus, the larger the observed B/C ratio, the more the primary CR species have undergone spallation interactions on their way to Earth, \textit{i.e}., the more matter they have passed through.

Measurements of the $^{10}$Be/$^9$Be isotopic ratio complement these observations by encoding information about the residence time of CRs within the galaxy, \textit{viz}.\ how long it took them to pass through the matter. 
This is because $^9$Be is stable, whereas $^{10}$Be undergoes $\beta$-decay with $t_{1/2} \simeq 1.39 \times 10^6$\,y, which is comparable to the residence time of GCRs. Thus, the smaller the $^{10}$Be/$^9$Be ratio is, the longer it took for the GCRs to reach Earth.

The combination of the two quantities, namely, the B/C and the $^{10}$Be/$^9$Be ratios, allows estimation of the mean density of matter traversed, which in turn, knowing the ISM density in the galactic plane, indicates that GCRs spend most of their time in the extended galactic halo.

\begin{figure}[t]
\begin{center}
\includegraphics[width=0.4\linewidth]{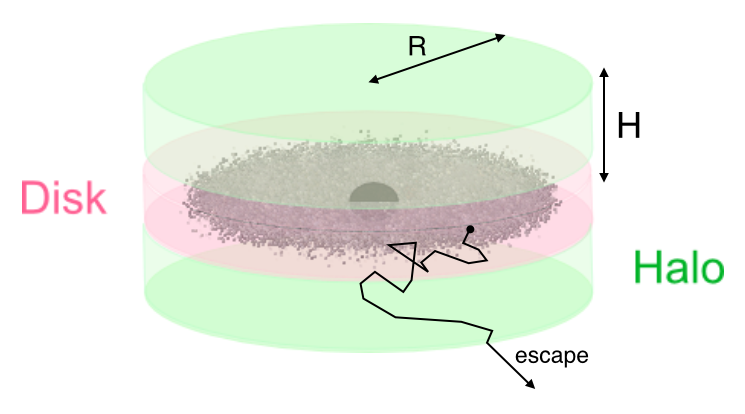}
\vspace*{-5mm}\includegraphics[width=0.59\linewidth]{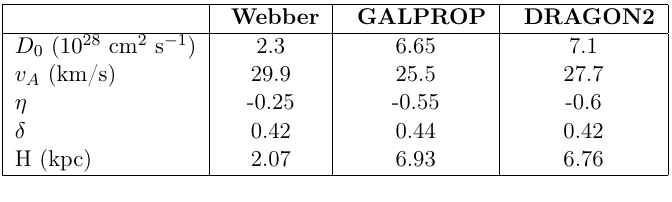}
\end{center}
\caption{Left. Sketch of a simple 2D leaky box model used for the Galaxy structure with halo size $H$. 
Right. Diffusion parameters, $D_0$, and Halo size, $H$, inferred from CR propagation models based on different cross section parameterisations. 
(Figure and table adapted from Ref.\,\cite{DeLaTorreLuque:2021yfq}.)
\label{fig:leaky-box}}
\end{figure}

To illustrate the impact of cross-section uncertainties on such astrophysical interpretations, Ref.\,\cite{DeLaTorreLuque:2021yfq} used precise measurements from AMS-02 and different cross-section models to estimate the thickness of the Galactic halo ($H$), and found results varying between 2.07\,kpc and 6.93\,kpc. This variation is largely due to uncertainties in the nuclear fragmentation cross-section, as shown in Fig.\,\ref{fig:leaky-box}. 
A similarly strong dependence on nuclear cross sections was found for the diffusion parameter, $D_0$, which characterises CR propagation in the Milky Way.

Reference~\cite{Maurin:2022gfm} showed that CR data contributes only 5-10\,\% to the uncertainty in $\Delta H / H$, while nuclear fragmentation uncertainties contribute about 40\,\%. 
Historically, nuclear fragmentation cross sections were measured in the 1980s at the Bevalac \cite{Webber:1990kc}, but precise data on the production of $^{10}$Be and $^9$Be in $pA$ interactions are lacking \cite{Evoli:2019wwu}, especially for C, N, and O targets, and at energies of about 1-50\,GeV matching AMS-02 measurements.
These gaps lead to significant uncertainty in CR propagation models, with an accuracy deficit exceeding 50\,\%, contributing about 40\,\% to the uncertainty in $H$. 
The recent R3B experiment at GSI has resumed these measurements and revealed discrepancies of up to an order of magnitude compared to Bevalac data, which are likely due to measurement inaccuracies. 
If the R3B data are accurate, they would significantly affect the calculation of the magnetic halo thickness $H$, emphasising the need for improved precision in nuclear fragmentation cross sections.

\begin{figure}[t]
\begin{center}
\includegraphics[width=0.99\linewidth]{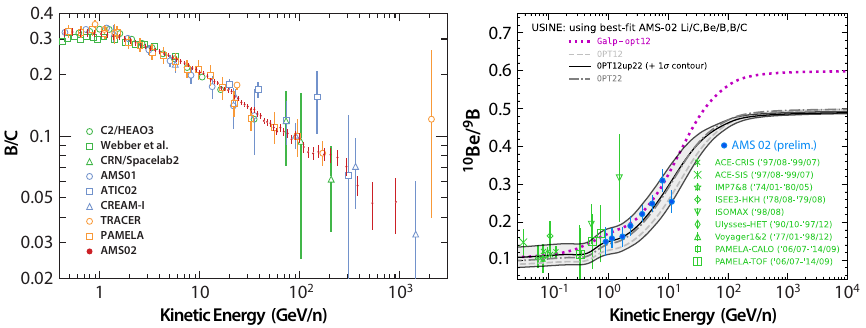} 
\end{center}\vspace{-5mm}
\caption{Boron-to-carbon ratio (left) and $^{10}$Be/$^9$Be ratio (right) as a function of kinetic energy per nucleon, measured by balloon and satellite born experiments since 1980. 
The different lines in the right image represent propagation using different cross section models and the gray band illustrates the uncertainties imposed by nuclear cross sections. (Left figure adapted from Ref.\,\cite{AMS:2016brs}) and right figure from Refs.\,\cite{Maurin:2022gfm, Lipari:2022imm}).
\label{fig:CR-isotopes-ratios}}
\end{figure}

Figure~\ref{fig:CR-isotopes-ratios} shows an exemplary compilation of data on the B/C and $^{10}$B/$^{9}$B ratios from AMS-02 and other experiments. Compared to early balloon-borne measurements, which were very limited in exposure time, AMS-02, which has been operating on the International Space Station for over a decade, has achieved remarkable statistical and systematic precision up to kinetic energies of about 10\,GeV per nucleon.

Unfortunately, the precision and energy range of current cross-section measurements do not match that of the CR data obtained from space-based experiments. This limits the (astro)physics that can be derived from such data. Most of the cross-section data were measured a long time ago, and the uncertainties for important reactions are in the range of 10-20\,\% \cite{Genolini:2018ekk, Genolini:2023kcj}, much larger than those found in current high-precision CR data sets.
Moreover, for some astrophysically important reaction channels, cross section measurements are limited to low energies or are missing altogether, meaning that semiempirical parameterisations must be used to approximate the cross sections; see, \textit{e.g}., Refs.\,\cite{Webber:1990ka,Summerer:1999ae}.

Current models for CR propagation assume equal cross sections for $^9$Be and $^{10}$Be, but experimental data suggest otherwise. 
Measurements of nuclear charge radii in Ref.\,\cite{Nortershauser:2008vp} reveal a 12\% difference between the two isotopes, which could have significant implications for CR propagation models. This once again demonstrates that accurate measurements of the nuclear fragmentation cross sections for $^9$Be and $^{10}$Be on a proton target, at the level of a few percent, are critical to improving these models.

Additionally, recent studies suggest that low-energy $^{10}$Be may be produced by neutrinos from low-mass core-collapse supernovae (CCSN), with meteorite observations showing a $^{10}$Be/$^9$Be ratio higher than previously expected \cite{Banerjee:2016tzd}. 
A reliable nuclear fragmentation model with 5\% accuracy will enable more precise calculations of CR propagation. 
This, in turn, will enable isolation of the contribution of CCSN neutrinos and gain insight into their role in the formation of the solar system.

As AMS-02 and similar missions continue to provide more accurate high-energy CR data, improving the precision of nuclear fragmentation cross section measurements will become critical. 
Achieving 5\% accuracy will enable the development of precise models of the galactic magnetic field and enhance CR propagation calculations. 
These models can then be compared with gamma-ray observations. 
Any excess intensity could indicate DM annihilation or other phenomena. 
Thus, refining nuclear fragmentation data is essential for both galactic studies and the DM search.

In fact, in the context of multi-messenger astrophysics, it is worth noting that the present uncertainties in the cross sections also affect the calculated astrophysical gamma-ray and neutrino fluxes \cite{Dorner:2025egk} by up to a factor of two, again limiting the quantitative conclusions that can be drawn from data-model comparisons. 
Similarly, the $\bar{\rm p}$, $\bar{\rm d}$, and $\overline{\rm He}$ production and interaction cross sections are of key importance in assessing whether their observed fluxes in CR can be explained by ordinary CR interactions, which represent the main background, or require an additional DM contribution.
Moreover, the aforementioned uncertainty in the galactic halo thickness, $H$, which is largely induced by the cross section uncertainties, directly affects the expected flux of cosmic particles from DM annihilation, since it determines how much of the injected DM flux is retained within the diffusive halo volume; see, \textit{e.g}., Ref.\,\cite{Genolini:2021doh}.
  
A precise understanding of space radiation, radiation protection, and the production of secondary particles is also crucial for human spaceflight and the exploration of the Moon, Mars, and beyond, where a minimum amount of shielding material must be used to ensure maximum biological protection.

\subsubsection{UHECR propagation and the modelling of extensive air showers}

The flux of CRs at energies greater than one~PeV is low, approximately 100 particles per square metre per year. 
This means that CRs beyond the PeV energy scale can only be studied by detecting the extensive air showers (EASs) they produce. 
Simulating the properties of EASs, such as the longitudinal profile of the electromagnetic and muonic components, or the footprint of secondary particles that reach the ground, requires modelling of high-energy nuclear interactions at CM energies that exceed those of man-made accelerators, like the LHC.

Enormous progress has been made in recent years, both in measuring the properties of EASs and in modelling the proton, pion, kaon, and nuclear interactions that drive the evolution of an EAS. The event generators used for this purpose are tuned to accelerator data and are often also employed in relativistic HICs, providing another vital link between the two communities.

As with the AMS-02 data discussed above, further progress in understanding ultra-high-energy cosmic rays (UHECRs) is largely limited by our understanding of nuclear collisions and the subsequent interactions of $p$, $n$, $\pi$, $K$, \ldots\ with nuclei in the atmosphere \cite{Kampert:2012mx}.

Measuring different properties of EASs simultaneously has led to the conclusion that all the presently available hadronic interaction models underestimate the number of muons observed at ground level by about $20 -40$\%. This is known as the \textit{Muon Puzzle}~\cite{Albrecht:2021cxw}.
Efforts have begun to understand and solve this problem, both by improving the physics of event generators \cite{Botti:2024upz} and by tuning event generators to accelerator and EAS data simultaneously \cite{Albrecht-tuning}.

Experiments at the CERN LHC and SPS, particularly at very forward rapidities and with good particle identification capabilities, have started to play an important role in the tuning of interaction models applied to the first very-high-energy interactions.

However, it is important to realise that the properties of EASs are not only determined by the first high-energy interaction, but, above all, by the large number of secondary particles, most importantly protons, neutrons, and pions, interacting at several tens of GeV with nuclei in the atmosphere. 
Pions are of particular interest: they are the most abundant particles produced in EASs, but $\pi+A$ data are very scarce above 1\,GeV. 
Two-dimensional cross sections of secondary particles emerging from such interactions need to be measured with a precision at least at the level of p+A interactions.

There are also a variety of other measurement needs, the most important being double-differential cross sections for neutron production, especially for pion-, proton-, and nucleus-induced interactions above 1\,GeV/nucleon, where there are essentially no data. 
For energies as low as 1\,MeV, neutrons start to diffuse in the atmosphere, causing delayed spurious signals in particle detectors, which must be understood for precise EAS measurements.

Thus, in summary, there is a great need for proton- and pion-induced interactions on light nuclei in the energy range uniquely covered by the FAIR facility; see also Ref.\,\cite{Maurin:2025gsz}.

Pion-beam data are essential as well, to constrain hadronic models that govern neutrino-nucleus interactions and thus neutrino‐energy reconstruction for next-generation LBL experiments (T2K/Hyper-K, DUNE), where nuclear effects and missing energy from FSI dominate the systematics. A concise justification and the implications for flux, cross-section modelling, and reconstruction are discussed in Sec.~\ref{sec.lblref}.


A precise understanding of the production of charmed particles in EASs is of particular interest for a number of reasons: they are of key importance for describing the prompt flux of high-energy muons and neutrinos, and thereby the onset of astrophysical neutrinos \cite{Albrecht:2021cxw}; despite their low production cross section, they affect the longitudinal evolution of EASs at high energies \cite{Guiot:2017wek}; and they influence the muon charge ratio measured at GeV energies at ground level \cite{Ismail:2023vpf}.
The latter is directly dependent on the energy and mass composition of the primary cosmic flux. 
As discussed, \textit{e.g}., in Ref.\,\cite{Reichert:2025iwz}, measurements at FAIR could help to improve the charm production cross sections, particularly at threshold energies.

\subsubsection{Experimental opportunities}
\label{sec.astro-opportunities}

{\color{blah}As discussed above, there is an urgent need for accurate measurements of nuclear and particle production and interaction cross sections in the GeV energy domain, in order to fully exploit the physics potential of the latest generation of space- and balloon-borne experiments and to improve the modelling of particle induced air showers.
GSI has a long history of conducting this type of measurements. We will begin with a review of some key measurements, before moving on to discuss opportunities and plans for future GSI/FAIR experiments.}

Between 1996 and 2011, several experiments were performed at GSI based on the inverse kinematics technique and using liquid hydrogen targets; see, \textit{e.g}., Ref.\,\cite{Villagrasa-Canton:2006exk}.
The cross sections were obtained with a stated uncertainty of 4\%, although this may be underestimated, as the transmission of the fragment separator (FRS) spectrometer was rather crudely evaluated.
Another set of experiments was carried out using the large-acceptance ALADIN spectrometer; see, \textit{e.g}., Ref.\,\cite{LeGentil:2005wx}. 
Given the uncertainties imposed by secondary particle production cross sections on important astrophysical parameters, there is a clear need to reduce these uncertainties to the level of a very few percent.

At higher energies, pilot runs for carbon fragmentation measurements were recently performed by NA61/SHINE at CERN. 
However, precise data on a variety of nuclei and ranges of energies are still lacking. 
Currently, no facility can accurately measure the forward-moving fragments from nuclear fragmentation reactions at approximately 10\,GeV per nucleon, including the CBM experiment at FAIR. 
Measuring the cross section for specific isotopes, such as $^{10}$Be, requires a high-energy beam production and separation facility.

The FRS at GSI is currently the only facility capable of nuclear fragmentation measurements up to 2\,GeV per nucleon, including $^{10}$Be beams. 
However, extrapolation to higher energies lacks precision, so additional data on different hadrons produced alongside nuclear fragments are needed to develop a more accurate model up to 10\,GeV per nucleon, necessary to support calculations of cosmic-ray propagation.

In the future, the new recoil separator, the Super-FRS \cite{Geissel:2003lcy}, will take over the role of the current FRS and become an excellent tool for measuring cross sections with high accuracy, using an even greater variety of secondary beams at higher intensities. 
Currently, there is no liquid-hydrogen target at the entrance of the Super-FRS, but early experiments can be conducted using CH$_2$ and C targets in the meantime.

A second direction is to set up a measurement in the future high-energy cave (HEC), where the new superconducting large-acceptance magnet GLAD (GSI Large Acceptance Dipole) will be installed as the backbone of the R3B setup. This will enable exclusive experiments, similar to the SPALADIN (Spallation at ALADIN) measurements mentioned above, to be performed.

Another interesting possibility will be offered by be the CBM fixed-target experiment, which will be supplied with SIS100 beams up to their maximum energy. 
While CBM's focus is on identifying light hadrons and baryons, significant modifications and additions to its setup would be required for high-quality isotope identification in the forward region. 
The WASA-FRS experiment, which was set up at GSI in 2022, could however fill this gap. It enables the simultaneous measurement of beam fragments and hadrons from nuclear fragmentation reactions \cite{Saito:2021gao, Saito:2023fnx}.

In the HypHI experiment, beam fragments at 2\,GeV per nucleon were measured with the FRS, while protons and pions were detected with the WASA detector. 
A similar setup can used for fragmentation reactions, although the WASA magnet will be upgraded to provide a stronger magnetic field. 
The aim of the upgraded WASA-FRS setup is to accurately measure nuclear fragmentation reactions up to 2\,GeV per nucleon, including hadron production, in order to construct a high-energy nuclear fragmentation model.

{\color{blah} At SIS18, the WASA-FRS programme will provide detailed measurements of nuclear fragmentation reactions up to about 2 GeV per nucleon. In addition to identifying nuclear fragments and determining fragmentation cross sections, the WASA setup enables coincident measurements of charged particles, including mesons. This makes it possible to go beyond a purely phenomenological description of inclusive fragmentation yields and to develop reaction models that are more directly constrained by the underlying reaction dynamics.

The energy reach of SIS18, however, does not cover the full range relevant for cosmic-ray propagation, which extends to approximately 10 GeV per nucleon. Measurements at SIS100 with CBM can therefore provide an essential test of whether the reaction models constrained by the detailed WASA-FRS data remain valid when extrapolated to much higher energies. Although the acceptance for such measurements with CBM will be predominantly in the forward region and will not provide the same degree of exclusivity as the WASA-FRS setup, suitable forward observables can be used to validate the energy dependence of the fragmentation model in the cosmic-ray-relevant regime. The resulting model can then be used for cosmic-ray propagation calculations, improving our understanding of galactic structure and contributing to dark-matter searches.}
%


{\color{blah}Secondary pion beams have been made available for measurements of hadron production at the CERN PS and SPS (see \textit{e.g}.\ \cite{NA61SHINE:2022tiz}), J-PARC (see \textit{e.g.}\ \cite{Ohnishi:2019cif}), and have also successfully  provided to HADES at SIS18 with momenta of 0.4 to 2.8\,GeV/$c$ \cite{PionBeamHADES:2002, PionBeamHADES:2017}. 
Such beams are of enormous interest and are proposed for systematic experimental studies at high intensity in the near future. The HIHR beamline at J-PARC will provide pion beams of up to 2\,GeV/$c$ with intensities of up to $10^8 \pi$ per spill offering a rich experimental program (see \textit{e.g}.\ \cite{Takahashi:2025wqw}). HADES will continue its pion-beam programme in the near future, after which it is planned to be reinstalled in the CBM cave for measurements at SIS100. Higher pion-beam energies may become accessible if a production target is constructed. Moreover, the usage of secondary hadron beams with Super-FRS, such as pions and antiprotons, are considered as well; see Chapter~\ref{sec.OverviewFacilities}.
This could offer a rich and unique scientific programme
in its own right, as well as enabling $\pi+A$ measurements with SIS100. Such measurements are of great interest also to event generators employed in simulations of EASs; see above.}

In the meantime, while waiting for more precise data from accelerator experiments, a virtue has been made of necessity in very specific cases. Attempts have been made to constrain unknown or poorly known cross sections using observed isotopic ratios of CRs;  see, \textit{e.g}., Ref.\,\cite{Zhao:2024qbj}). 
The need for precise cross-sections to advance CR physics and other applications was also discussed in an ongoing series of CERN workshops \cite{Maurin:2025gsz}.

\subsection{DM constraints from GSI/FAIR physics}
\label{sec.astro_dm}

{\color{blah}Astrophysical and cosmological observations suggest the existence of matter (DM) that acts gravitationally but has not yet been observed. 
Key evidence for DM comes from galaxy rotation curves, gravitational lensing, CMB measurements, and large-scale structure formation.  All of these phenomena indicate the presence of a significant unseen mass. Systems such as the Bullet Cluster demonstrate that this mass is collisionless and distinct from ordinary matter. 
Further constraints from dwarf galaxy dynamics and cosmic shear confirm that DM is cold or warm, non-luminous, and essential to cosmic structure formation. Particle and nuclear physics experiments play a crucial role in ongoing searches DM and in theoretical frameworks, as they may provide evidence for particles beyond the standard model of particle physics. This makes them complementary to astrophysical searches, as astrophysicists could optimize their instruments to search for such particles in the Universe. In this subsection, we present key areas in which experiments at FAIR could address this issue.

\subsubsection{DM and dark photons}

Dark Matter is a proposed form of matter thought to account for approximately 85\% of the matter in the Universe. Since DM cannot be incorporated into the Standard Model, new interactions between DM particles and ordinary SM particles must be introduced via unknown dark-sector forces.
DM mediators are proposed to interact with Standard Model (SM) particles through four ``portals'': vector, Higgs, neutrino, and axion; see Refs.\,\cite{Alexander:2016aln, Battaglieri:2017aum, Agrawal:2021dbo}. }



The vector portal introduces a gauge symmetry mixing between U(1) and U(1)$^\prime$ \cite{Holdom:1985ag}, described by a Lagrangian involving the hypercharge field-strength tensor of the SM photon field and the DM vector boson field: ${\cal L} \sim (\epsilon^2/2)\, F_{\mu\nu}{F^{\mu\nu}}^\prime$. 
Mediators in this framework, known as $U$-bosons, dark photons, hidden photons, or $A^\prime$, have an unknown mass, $M_U$. 
The kinetic mixing parameter, $\epsilon^2$, defines the interaction strength between SM and DM  articles \cite{Fayet:1980ad, Fayet:2004bw, Boehm:2003hm, Pospelov:2007mp, Batell:2009di, Batell:2009yf}, enabling $U$-bosons to decay into lepton pairs, such as $e^+e^-$ or $\mu^+\mu^-$.

$U$-bosons can be produced via Dalitz decays of SM particles, such as pseudoscalar mesons ($\pi^0$, $\eta$), baryonic resonances, like the $\Delta$, and other decay channels, providing opportunities for detection in dilepton experiments. 
This has generated significant experimental and theoretical interest \cite{Alexander:2016aln, Battaglieri:2017aum, Beacham:2019nyx, Billard:2021uyg}.

Experimental constraints on dark photons are derived from beam dump experiments, fixed-target setups, collider searches, and rare meson decay studies, supplemented by cosmological and astrophysical bounds from the cosmic microwave background and stellar cooling \cite{Fabbrichesi:2020wbt}.
Search efforts have systematically tightened exclusion limits on the kinetic mixing parameter, $\epsilon^2$, reaching sensitivities near $10^{-6}$ for masses between 20\,MeV and a few GeV.

For instance, the HADES Collaboration conducted a dark photon search using dilepton experiments at the SIS18 accelerator with proton and heavy-ion beams \cite{Agakishiev:2013fwl}. 
Based on measurements of $e^+e^-$ pairs from $pp$ and $p$+Nb collisions at 3.5\,GeV, as well as Ar+KCl collisions at 1.76\,A\,GeV, HADES provided an upper limit on $\epsilon^2$ in the mass range of $M_U = 0.02-0.55\,$GeV.

Subsequent measurements via fixed-target experiments, including A1 \cite{Merkel:2014avp}, NA48/2 \cite{NA482:2015wmo}, and APEX \cite{APEX:2011dww}, have superseded the HADES result.
The NA48/2 experiment investigated a large sample of $\pi^0$ Dalitz decays obtained from in-flight weak decays of kaons; 
the BaBar collider experiment used its accumulated $e^+e^-$ data sets to survey a wide mass range up to $M_U = 8$~GeV; 
and the MAMI-A1 experiment \cite{Merkel:2014avp} investigated electron scattering off a $^{181}$Ta target at energies between 180 and 855\,MeV to search for a dark photon signal.
In the mass range discussed here, $M_U = 20-500\,$MeV, the limit on $\epsilon^2$ has thus been pushed down to approximately $10^{-6}$.
Furthermore, a recent measurement of excess electronic recoil events by the XENON1T Collaboration might also be interpreted in favour of dark matter sources, with dark photons among the possible candidates \cite{XENON:2020rca}.

\begin{figure*}\phantom{a}
\hspace*{-1cm}
\includegraphics[width=.54\linewidth]{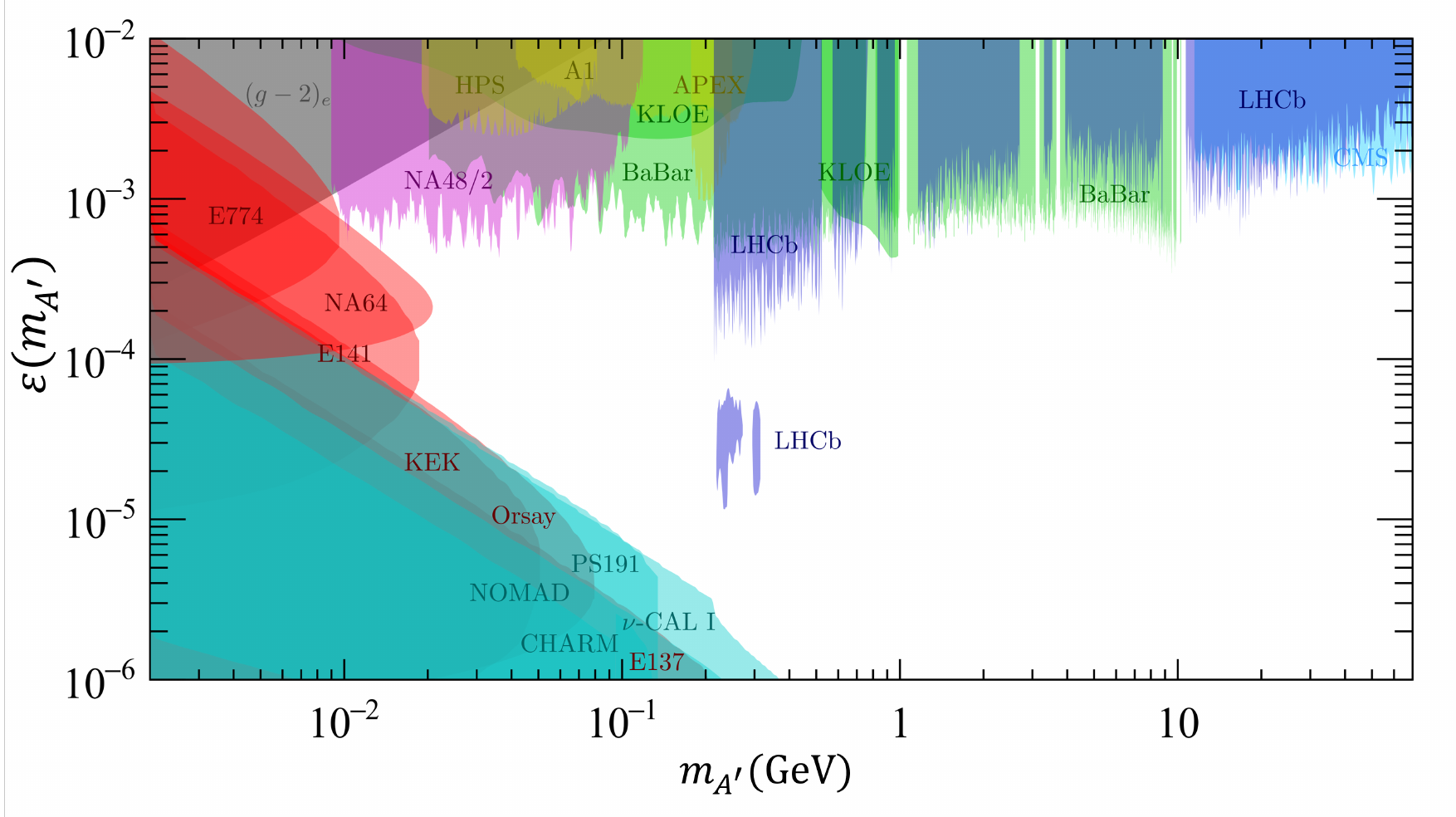}
\includegraphics[width=.45\linewidth]{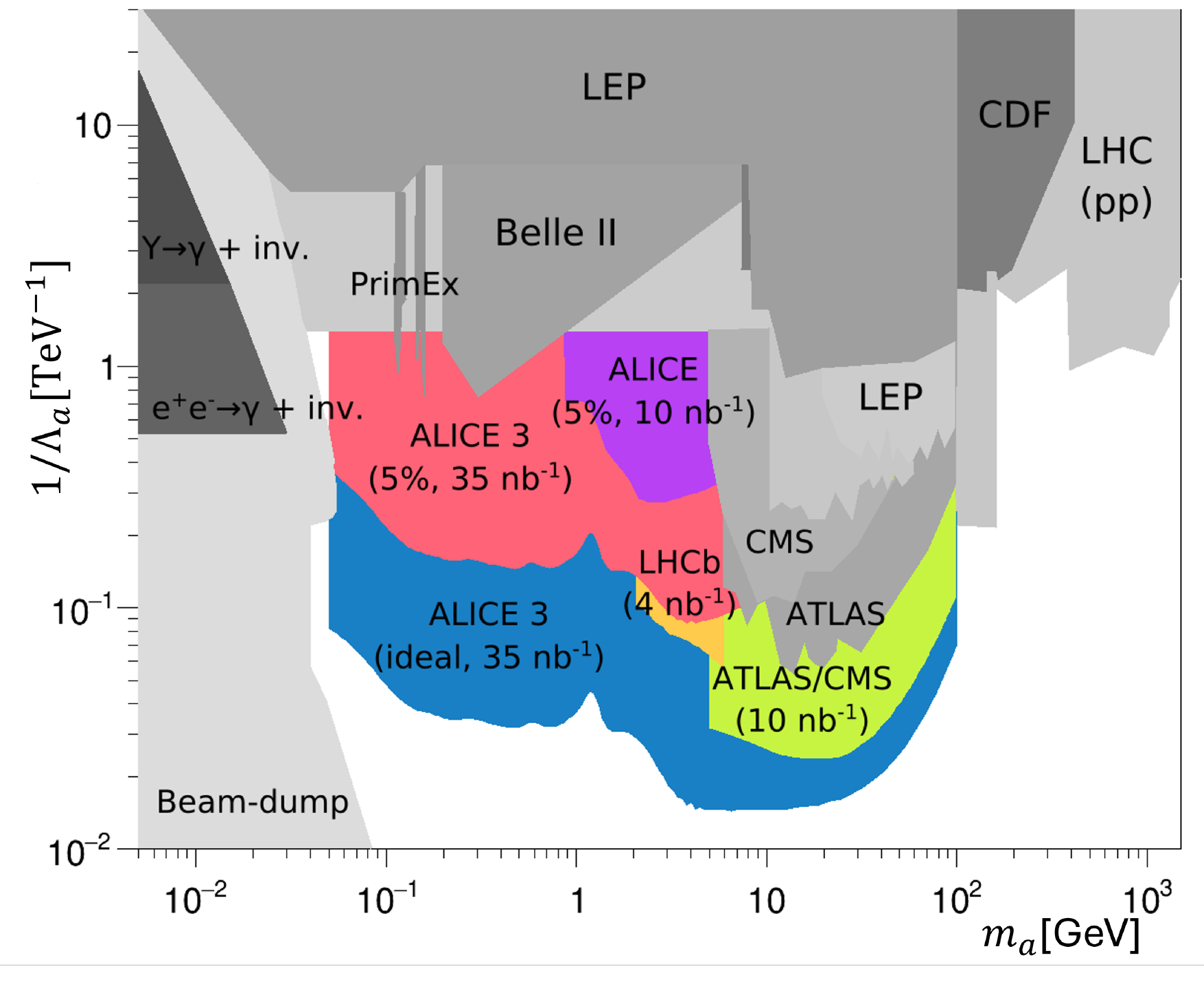}
\caption{
Left. Compilation of experimental upper limits for the kinetic mixing parameter, $\epsilon^2$, versus the mass of dark photons set by worldwide experiments. 
For orientation, the yellow band indicates $\epsilon^2=10^{-6}$.
(Figure adopted from Ref.\,\cite{LHCb:2019vmc}.)
Right. Bounds in the ($m_a$, $1/{\rm \Lambda}_a$) plane from existing (grey) and future (coloured, with heavy-ions at the LHC) ALP searches. 
(Figure adapted from Ref.\,\cite{dEnterria:2022sut}.
\label{Fig_epsil2}}
\end{figure*}

Moreover, the experimental search for dark photons, as well as other candidates for dark matter, is a central focus of collider experiments at the LHC \cite{dEnterria:2022sut, Batell:2022dpx}, including LHCb, as well as at BaBar \cite{BaBar:2009lbr, BaBar:2014zli}, which explore higher mass ranges with exceptional precision. 
Additional constraints from experiments such as KLOE \cite{KLOE-2:2014qxg} and CMS \cite{CMS:2023hwl} further refine the parameter space, imposing stringent limits on the existence of dark photons.
The left panel of Fig.~\ref{Fig_epsil2} shows a compilation of the experimental upper limits on $\epsilon^2$ as a function of the dark photon mass, as set by a wide range of international experiments \cite{Battaglieri:2017aum, Ilten:2018crw,LHCb:2019vmc}.

{\color{blah}
\subsubsection{Axions and Axion Like Particles}
\label{sect.dm.axions}

Originally, the existence of an axion particle was proposed by Peccei and Quinn (PQ) \cite{Peccei:1977hh}, to solve the strong CP problem of QCD. It arises from the spontaneous breaking of a global $\mathrm{U(1)_{PQ}}$ symmetry featuring a common breaking mechanism for the electroweak and PQ symmetries. 
The theory was later developed by Wilczek \cite{PhysRevLett.40.279}, who incorporated constraints from the low-energy QCD regime and postulated the existence of a QCD axion.

Axion-like particles (ALPs) emerge as pseudo-Nambu–Goldstone scalar bosons of a new spontaneously broken global symmetry in various beyond-the-Standard-Model (BSM) scenarios, such as supersymmetry, Higgs extensions, and composite dynamics models \cite{Bauer:2017ris, dEnterria:2021ljz}. }

Light pseudoscalars have also been proposed as DM candidates or dark-sector mediators. 
In many models, ALPs couple to photons via the effective Lagrangian:
\begin{equation}
\mathcal{L} = -\frac{1}{4}g_{a\gamma} a F^{\mu\nu} \tilde{F}_{\mu\nu},
\end{equation}
where $a$ is the ALP field, $F^{\mu\nu}$ is the photon field-strength tensor, and $g_{a\gamma} = 1/\Lambda_a$ is the ALP–photon coupling constant, linked to the high-energy scale $\Lambda_a$. 
The production and decay rates of ALPs are governed by the axion mass, $m_a$, and $g_{a\gamma}$. 
The mass region $m_a$ in the $50\,\mathrm{MeV}$ to $5\,\mathrm{GeV}$ range is particularly intriguing, as ALPs in this regime could help explain the anomalous magnetic moment of the muon \cite{Marciano:2016yhf, Bauer:2017ris}.  
Today, however, the muon $g-2$ may no longer be an issue \cite{Cho:2025}.

The most stringent limits on ALPs in the $m_a$ range from $5-100\,$GeV were set by light-by-light scattering measurements in Pb+Pb ultra-peripheral collisions (UPCs) at $\sqrt{s_{NN}} = 5.02\,$TeV by CMS \cite{CMS:2018erd} and ATLAS \cite{ATLAS:2020hii}. 
Figure~\ref{Fig_epsil2}\,-\,right shows the global compilation of bounds in the ($m_a$, $1/\Lambda_a$) plane from existing (grey) and future (coloured, based on heavy-ion collisions at the LHC) ALP searches.

Recently, a new set of calculations has restored the QCD axion with mass $m_a = \mathcal{O}(1-100)\,$MeV and decay constant $f_a = \mathcal{O}(1-10)\,$GeV \cite{Alves:2017avw}, along with a more general class of ALPs incorporating additional PQ-breaking contributions to their masses \cite{Alves:2020xhf, Alves:2024dpa}.
These calculations were motivated by a claim from the ATOMKI collaboration \cite{Krasznahorkay:2015iga}, which reported the emission of $e^{+}e^{-}$ pairs in several transitions of the $^8$Be isotope. 
In these processes, a ``bump-like'' excess in the invariant mass and opening angle distributions of the $e^+ e^-$ pairs was observed and attributed to a new particle with a mass around 17\,MeV (denoted $X_{17}$). 
One of the first proposed explanations involved a protophobic gauge boson \cite{Feng:2016ysn}. 
Subsequently, an alternative explanation suggested a piophobic particle with suppressed mixing with the neutral pion \cite{Alves:2017avw}.

This discovery has triggered searches in other physical systems, where the hadronic couplings of the new piophobic ALP could lead to more stringent constraints. 
In this context, one of the most promising processes involves rare and very rare $\eta$ and $\eta^{\prime}$ meson decays, which are suppressed within the SM. 
In particular, possible axio-hadronic decays include the three-body final states $\eta \to \pi^{+}\pi^{-}a$ and $\eta \to \pi^{0}\pi^{0}a$.
The leading-order potential term in the chiral Lagrangian contributes directly to the amplitudes of these processes, potentially resulting in significantly enhanced branching ratios. 
Moreover, assuming that the ALP is short-lived and predominantly decays into $e^+e^-$ ($BR \approx 1$), this channel provides a means to evade bounds from beam-dump and fixed-target experiments. 
Estimates based on ChPT predict branching ratios spanning two orders of magnitude: $BR(\eta \to \pi\pi a) \approx 10^{-6}$–$10^{-8}$.

Experimentally, to probe for the possible existence of $a$, one can study the $\eta \rightarrow \pi^{+}\pi^{-}e^{+}e^{-}$ final state. 
This decay has already been observed by several experiments and investigated in the context of possible CP violation in flavour-conserving reactions \cite{KLOE:2008arm, Adlarson:2015zta}. 
Most recently, this decay was measured by the BESIII experiment, which found no evidence of CP violation within the collected statistical sample \cite{BESIII:2025cky}. 
Within the same BESIII study, limits were also placed on the possible existence of an ALP in the process $\eta \to \pi^{+}\pi^{-}a$ for the mass range $5 – 200\,$MeV/$c^2$, with branching ratio limits established at the level of $BR = 1.5 \times 10^{-5}-6.1 \times 10^{-7}$.

To further investigate these decays, the HADES detector has collected a high-statistics data sample, yielding an integrated luminosity of approximately 6.0\,pb$^{-1}$ in proton–proton collisions at a kinetic energy of $T = 4.5\,$GeV \cite{HADES:2020pcx}, providing comparable upper limits. 
These constraints could be significantly improved by future measurements of $\eta/\eta^{\prime}$ decays at SIS100, benefiting from higher luminosities and production cross sections.


\subsubsection{Input from FAIR/SIS100}

{\color{blah}The future FAIR will contribute to the search for DM candidates from both hadronic and heavy-ion perspectives.  
The elementary beam programme at FAIR/SIS100 will improve the sensitivity to dark photons and other light dark-sector states thanks to its substantially increased interaction rates and larger production cross sections for light mesons. This will enhance searches for DM candidates, including ALPs and dark photons.
For example, HADES can examine the electron-positron invariant mass in very large samples of $\pi^0$ Dalitz decays ($\pi^0 \rightarrow \gamma A^{\prime}$, $A^{\prime} \rightarrow {\rm e^+e^-}$), in the range $20 \le M_{\rm ee} \le 100$\,MeV/$c^2$. Furthermore, high-statistics measurements of $\eta$ and $\eta^\prime$ decays in proton-induced reactions can extend the search for narrow $e^+e^-$ resonances and improve the existing limits on the kinetic-mixing parameter in the low-mass region. Such measurements would complement the existing HADES results and other fixed-target searches.

NS cooling can provide astrophysical constraints on hypothetical particles such as axions. 
If axions couple to nucleons, they can be produced efficiently in NS cores via nucleon-nucleon bremsstrahlung, as well as Cooper pair breaking and formation processes \cite{Iwamoto:1984ir, Sedrakian:2015krq, Sedrakian:2018kdm}. 
Conversely, if axions couple to photons, they can be generated through the Primakoff effect, in which thermal photons convert into axions in the presence of electric fields from charged particles \cite{Raffelt:1985nj, Raffelt1996}.
In both cases, axions can escape the stellar interior, providing an additional channel for energy loss and thereby enhancing the NS cooling rate. 
Observations of young NSs, such as the one in Cassiopeia~A, match standard cooling predictions and thus constrain the strength of axion-nucleon and axion-photon couplings; see Fig.\,\ref{Fig_epsil2}\,-\,right. 
By comparing theoretical cooling models that include axion emission with the observed surface temperatures of NSs, one can exclude axion models that would cause excessive cooling. This provides valuable constraints on axion properties that complement DM studies.

As the above examples show, even though GSI/FAIR experiments will not detect axions directly, they will provide crucial data for testing astrophysical scenarios involving them.  This will advance our understanding of dense nuclear matter relevant to NS interiors, thereby refining models of axion production in astrophysical environments.

Finally, as discussed in Sec.\,\ref{sec:precise-crosssections}, precise production cross sections of anti-protons, anti-deuterons, anti-helium and even heavier anti-nuclei will aid the search for DM in space-based experiments such as AMS-02. This is because classes of DM candidates could annihilate into antimatter \cite{Delahaye:2007fr, Lavalle:2008zb, Evoli:2011id, Huang:2016tfo} so that their signal would be superimposed on that of anti-particles originating from ordinary galactic cosmic ray propagation. As outlined above, CBM and HADES would help improve both the modeling of the galactic cosmic ray propagation and the production of anti-nucleons and anti-nuclei in galactic cosmic ray interactions.
}
\newpage
\section{Theoretical tools and techniques}
\label{sec.Tools}
{\small  {\bf Convenors:} \it V. Cred\'e, A.~Pilloni, A. Szczepaniak}

\noindent 
This chapter summarises analysis techniques used for directly examining strong interaction data and various theoretical methods in nonperturbative QCD. 
PWA is a vital technique for unravelling the complex dynamics of hadronic interactions and connecting them with QCD, especially in the nonperturbative regime, where the strong coupling prevents the use of standard perturbation theory. 
PWA decomposes scattering amplitudes into components with definite angular momentum, allowing the identification of resonance structures and the quantum numbers of intermediate states \cite{Hoehler:1983}. 
It plays a key role in interpreting experimental data from high-energy physics facilities, where multiparticle final states are common.

PWA based on the analytic properties required by $S$-matrix theory enables the extraction of resonance parameters. 
These parameters are crucial for spectroscopy and ultimately assist in understanding nonperturbative QCD dynamics, and they can be compared with theoretical predictions based on QCD.

LQCD provides a first-principles approach by simulating QCD on a discrete spacetime grid, enabling the calculation of hadron masses, decay constants, and form factors. 
Nonperturbative continuum techniques, such as the DSE/BSE approach, further improve our ability to expose and understand the structure and dynamics of hadrons. 
EFTs, such as chiral perturbation theory, offer analytical insights by systematically expanding a set of quantities in some small parameter and exploiting known symmetries.  
Together, these methods create a comprehensive framework for linking QCD theory with experimental observations.

\subsection{Partial-wave analysis}
\label{subsec.FSI-PWA}

At FAIR, the combination of high-intensity pion and proton beams with cutting-edge detectors, such as HADES, PANDA, and CBM, facilitate extensive PWA studies. 
The capability to investigate pion-nucleon scattering at SIS18 and proton-induced reactions up to 30\,GeV at SIS100 allows for the precise extraction of resonance parameters, broadening the scope of hadron spectroscopy beyond previous efforts at CERN and BNL \cite{Beringer:2012}.
Furthermore, HADES offers a unique environment in which to study in-medium hadron modifications, which are essential for understanding QCD matter under extreme conditions \cite{HADES:2024}.

\subsubsection{Reaction mechanisms for partial-wave analysis}

Different reaction mechanisms and final states offer complementary insights into the structure and properties of hadronic resonances. 
The choice of reaction strongly influences the types of resonances studied and the precision with which their parameters can be extracted.

\paragraph{Pion-nucleon scattering and in-medium effects}
    Pion-nucleon scattering remains a fundamental aspect of baryon resonance research owing to its direct interaction with the strong force. 
    FAIR's SIS18 accelerator enables precise studies of $\pi N \to MB$ (meson--baryon) final states \cite{HADES:2024}, which are essential for resolving nucleon resonances that exhibit complex decay patterns \cite{Crede:2013kia}. 
    Pion-induced reactions also facilitate investigations into in-medium effects, including modifications of vector meson spectral functions ($\rho$, $\omega$, $\phi$), which are crucial for understanding QCD matter at high baryon densities \cite{HADES:2024}. 
    These studies have significant implications for both heavy-ion physics and astrophysical environments, such as neutron star interiors.

\paragraph{Proton-induced reactions}
Proton-induced reactions enable access to higher-energy resonance states and the production of exotic hadrons. FAIR’s SIS100 accelerator, with proton beam energies reaching up to 30\,GeV, supports the investigation of highly excited nucleon states \cite{Guo:2017jvc}. 
These states often decay into multimeson final states, such as $\pi\pi N$ and $\pi\eta N$, requiring coupled-channel analyses for precise parameter extraction \cite{Kamano:2016}.

\paragraph{Photoproduction experiments}
Photon-induced reactions serve as a clean method for investigating electromagnetic transition form factors and nucleon resonances. 
Although photoproduction experiments at facilities such as JLab (CLAS, GlueX), ELSA, and MAMI provide high-precision data, the limited electromagnetic coupling of some resonances restricts their applicability \cite{Ronchen:2018}. FAIR enhances these efforts by offering direct hadronic probes.

\paragraph{Proton-antiproton collisions}
The upcoming PANDA experiment at FAIR is designed to study exotic hadrons through proton-antiproton annihilation. 
This environment is particularly well suited for investigating hybrid mesons and glueballs \cite{Anisovich:2017}.

\subsubsection{Angular coordinate systems}
Understanding angular distributions provides valuable insight into interaction dynamics and resonance decays. 
Two commonly used reference frames for such analyses are the helicity frame and the Gottfried-Jackson frame. 
In this section, a brief overview and definitions of these frames is provided.

The helicity frame is defined in the rest frame of the decaying particle, where the motion of its decay products is examined relative to its momentum direction. 
Consequently, the $\hat{z}$-axis is aligned with the momentum of the parent particle, and the helicity angle indicates the orientation of a decay product relative to this axis. 
The helicity $\hat{z}_H$-axis is defined as
\begin{equation}
    \hat{z}_{H} = \vec{p}_{\,\text{parent}}\,,
\end{equation}
where $\vec{p}_{\,\text{parent}}$ denotes the momentum of the parent particle in the CM frame.


\begin{figure}[t]
\begin{center}
\includegraphics[width=.45\textwidth]{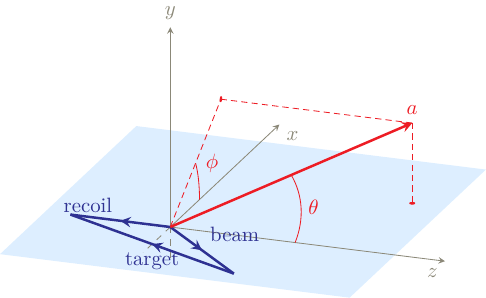}
\includegraphics[width=.45\textwidth]{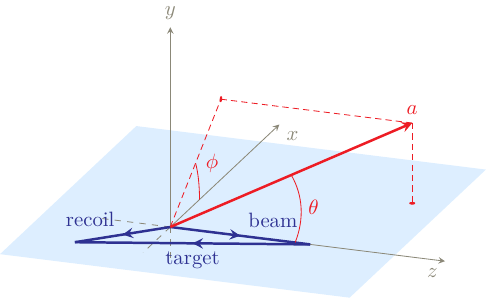}
\caption{(Left) Definition of the helicity frame. The reaction plane $xz$, 
shown in blue, containing the momenta of the incident beam, the nucleon target, and the recoiling nucleon, is shown in blue. 
The $z_H$-axis is aligned with the direction of the produced resonance in the CM frame. 
The angles $\theta$ and $\phi$ are the polar and azimuthal angles of particle~$a$ in the helicity frame. 
(Right) Definition of the Gottfried-Jackson frame. 
The reaction plane is the same as in the helicity frame. 
The $z_{GJ}$-axis is aligned with the direction of the incident beam.
\label{fig:angularframes}}
\end{center}
\end{figure}

In contrast, the $z$-axis, $\hat{z}_{GJ}$, is aligned along the beam momentum in the Gottfried-Jackson frame,
\begin{equation}
    \hat{z}_{GJ} = \vec{p}_{\,\text{beam}}\,,
\end{equation}
where $\vec{p}_{\,\text{beam}}$ represents the beam momentum in the CM frame, which is typically aligned with the direction in the laboratory frame. Alternatively, the $\hat{z}_{GJ}$-axis may be chosen along the target momentum to examine target fragmentation processes.

In both frames, the $\hat{y}$-axis is defined by the cross product between the beam and the produced resonance:
\begin{equation}
    \hat{y} = \frac{\vec{p}_{\text{beam}} \times \vec{p}_{\text{parent}}} { |\vec{p}_{\text{beam}} \times \vec{p}_{\text{parent}}| }\,,
\end{equation}
and the $\hat{x}$-axis is defined such that it completes a right-handed coordinate system:
\begin{equation}
    \hat{x} = \hat{y} \times \hat{z}\,.
\end{equation}

\subsubsection{Partial-wave analysis frameworks}
Generally, the scattering amplitude can be written as an expansion using Wigner $D$-functions. 
Specifically, for $2 \to 2$ scattering of spinless particles, it takes the form \cite{Workman:2012}
\begin{equation}
    T(s, \cos{\theta}) = \frac{1}{4\pi} \sum_{\ell=0}^{\infty} (2\ell + 1)\, P_\ell(\cos{\theta})\, T_\ell(s)\,, \label{eq:pwa}
\end{equation}
where $s$ and $\theta$ denote the total energy squared and the scattering angle in the CM frame, respectively. 
Here, $\ell$ is the orbital angular momentum, $T_{\ell}(s)$ are the partial-wave amplitudes, and $P_{\ell}(\cos\theta)$ are the Legendre polynomials.
(The normalisation of partial-wave amplitudes is conventional; different conventions exist in the literature. 
The partial-wave amplitudes are related to the phase shifts $\delta_{\ell}(s)$ and inelasticity $\eta(s)$ via
\begin{equation}
    T_{\ell}(s) = \frac{\eta(s)\, e^{2i\delta_{\ell}(s)} - 1}{2i\, \rho(s)}\,,
\end{equation}
where $\rho(s)$ is the two-body phase space factor. 
Resonances correspond to poles of the partial waves $T_\ell(s)$ in the complex energy plane \cite{JPAC:2021rxu, Mai:2022eur, Doring:2025sgb}. The mass and half-width of a resonance are given by the real and imaginary parts of the pole position, respectively.

Several PWA frameworks have been developed to systematically study hadronic interactions. 
These frameworks rely on different theoretical assumptions and mathematical methods, which influence the accuracy and reliability of the extracted resonance parameters~\cite{JPAC:2021rxu, Mai:2022eur, Doring:2025sgb}. 
Although the angular dependence is well understood for any $\ell$, the energy dependence cannot be derived from first principles. 
The literature presents a wide range of parametrisations, depending on the specific features of the problem under consideration.

\paragraph{Breit-Wigner}.
A pole associated with a narrow-width resonance is located close to the real axis. 
Near an isolated resonance far from thresholds, the corresponding partial wave can be described using the traditional Breit-Wigner form, which in the elastic case reads
\begin{equation}
    T_\ell(s) = \frac{g^2\, p^{2\ell}(s)}{m^2 - s - i g^2\, p^{2\ell}(s)\, \rho(s)}\,,
\end{equation}
where $p(s)$ denotes the scattering momentum, $m$ is the Breit-Wigner mass, and $g$ is a real coupling constant. 
This partial wave saturates unitarity, with
$\operatorname{Im} T_\ell(s) = \rho(s)\, \left|T_\ell(s)\right|^2$\,.
The Breit-Wigner model is commonly used as a reference, even when the underlying assumptions are not fully satisfied. 
In practice, partial waves are often constructed as a sum of Breit-Wigner functions with complex couplings, which violates unitarity. 
This approach is not ideal for broad, overlapping resonances.
In such cases, alternative methods are generally preferred.

\paragraph{Coupled channels}.
The coupled-channel formalism evaluates multiple reaction channels simultaneously by promoting the partial wave to a matrix $T^\ell_{ij}(s)$ that describes the process $i \to j$. 
Time-reversal symmetry implies $T^\ell_{ji}(s)=T^\ell_{ij}(s)$.

\paragraph{$K$-matrix formalism}.
The $K$-matrix method guarantees unitarity of the scattering amplitude by defining
\begin{equation}
    T^\ell(s) = K(s) \left[1 - i\, \rho(s)\, K(s)\right]^{-1},
\end{equation}
where \( K(s) \) encodes the scattering dynamics and is typically parametrised as a sum of ``bare poles'' and a background polynomial:
\begin{equation}
    K(s) = \sum_{n} \frac{g_n g_n^T}{m_n^2 - s} + C(s)\,.
\end{equation}


\paragraph{Dispersion relations and Chew-Mandelstam phase space}.
Dispersion relations enforce analyticity constraints on scattering amplitudes by connecting their real and imaginary parts through integral equations. Although applying these relations rigorously to scattering processes is practically feasible only in the simplest cases \cite{Ananthanarayan:2000ht,Garcia-Martin:2011iqs, Hoferichter:2015hva, Pelaez:2020gnd}, improved analytic properties can be obtained by replacing the phase space in $K$-matrix parametrisations with the so-called Chew-Mandelstam phase space (also known as the ``loop function''),
\begin{equation}
    i\rho(s) \to G(s) = \frac{s}{\pi} \int_{s_\text{th}}^\infty \frac{\rho(s')}{s'(s' - s)}\, \mathrm{d}s'\,.
\end{equation}

\paragraph{Regge theory}.
Regge theory examines the analytic properties of partial waves as functions of angular momentum. 
It states that partial waves are analytic functions in the complex angular momentum plane, with singularities, poles, and cuts constrained by unitarity. This is analogous to the singularities that arise in the complex energy plane. In particular, Regge poles have the same interpretation as poles in the energy plane: both represent resonances. 
However, a Regge pole corresponds to an entire family of resonances and therefore encodes more information about the underlying dynamics than a pole in the energy plane, which describes a single resonance at fixed angular momentum.
Moreover, the analytic structure of the partial wave in the angular momentum plane enables, in principle, the resummation of the partial wave series to reconstruct the full scattering amplitude. 
This approach is particularly useful in high-energy, low-momentum-transfer reactions, where the amplitude can be systematically expanded in terms of contributions from Regge poles and cuts. 
These contributions are ordered by the real part of the pole or branch point and can be formulated within an EFT framework.




\subsubsection{PWA approaches for two-body final states: $\pi N\to MB$}
\label{sec:PWAs}

The scattering amplitude of spinless particles, which depends on the CM energy $\sqrt{s}$ and the scattering angle $\theta$, is expanded in a series of Legendre polynomials, as given by Eq.\,\eqref{eq:pwa}.

However, the relevant reaction processes for accessing nucleon and $\Delta$ resonances involve particles with spin.
In such cases, the coupling of the orbital angular momentum $L$ and the total spin $S$ of the system to the total angular momentum $J$ must be taken into account, according to the rule $|L - S|\leq J \leq L + S$.

The scattering process is conveniently expressed in terms of helicity eigenstates $|\lambda\rangle$, and, since the helicity operator and the angular momentum operator commute, these states can be expanded in terms of the eigenstates of $J$ using the Wigner $D$-matrices, which provide a matrix representation of the finite rotation operator \cite{Jacob:1959at}:
\begin{equation}
    \langle \lambda' |T(s, \cos{\theta})| \lambda \rangle = \frac{1}{4\pi} \sum_J (2J+1)\, D^{J}_{\lambda' \lambda}(\Omega)\, \langle \lambda' |T^{J}(s)| \lambda \rangle \,.
\end{equation}
The dependence on the scattering angle $\theta$ is entirely contained in the Wigner $D$-matrices, which depend on the solid angle $\Omega$ of the scattering process and can be expressed in terms of the reduced rotation matrices $d^{J}_{\lambda' \lambda}(\theta)$.

Instead of the helicity basis, partial-wave amplitudes are often expressed in the $JLS$ basis, where the states are characterised by their total angular momentum $J$, orbital angular momentum $L$, and spin $S$, rather than by their helicity. 
Traditional PWAs, such as the Karlsruhe–Helsinki \cite{Hohler:1984ux} and Carnegie Mellon–Berkeley \cite{Cutkosky:1979fy} analyses, focused almost exclusively on elastic $\pi N$ scattering owing to the limited availability of inelastic experimental data at the time. 
In these analyses, resonance states were identified according to their appearance in specific $\pi N$ partial waves, using the spectroscopic notation $L_{2I\, 2J}$, where $I$ denotes the isospin.
With the inclusion of increasingly many inelastic reaction channels -- including photon- and electron-induced processes -- a reaction-independent naming scheme was introduced. 
This scheme relies on intrinsic properties of the resonance states, such as their spin and parity $J^P$.

Over the years, not only has the naming scheme changed, but also the notion of what defines a resonance state. 
Historically, a description in terms of Breit-Wigner parameters, namely the ``mass'' and ``width'' extracted from a Breit-Wigner parameterisation of the amplitude $T^J(s)$, was generally accepted, despite significant channel and model dependences.
In contemporary analyses, it is common to define resonances as poles of the $S$-matrix in the complex energy plane on the unphysical Riemann sheet. 
To access their properties, specific requirements must be imposed on the analytic structure of the analysis framework.

One of the most influential recent PWA efforts is the GWU/SAID analysis \cite{Workman:2012hx}, which employs a Chew-Mandelstam $K$-matrix parameterisation of the partial-wave amplitude. 
Several independent studies of pion-, kaon-, and photon-induced processes are available; see, \textit{e.g}., Ref.\,\cite{Briscoe:2023gmb}. 
The partial-wave amplitudes for elastic and charge-exchange $\pi N$ scattering serve as experimental input for many further phenomenological analyses.

Another well-known PWA programme is MAID \cite{Tiator:2018heh}, a unitary isobar model focused on the analysis of photo- and electroproduction of pions, $\eta$ mesons, and kaons. 
By combining fixed-$t$ dispersion relations and Laurent-Pietarinen techniques, the Mainz-Tuzla-Zagreb Collaboration also determined pole positions based on the MAID Breit-Wigner amplitudes \cite{Osmanovic:2021rck}.

In recent decades, photoproduction reactions \cite{Crede:2013kia, Ireland:2019uwn, Thiel:2022xtb} have become the primary source of information for the light baryon resonance spectrum, owing partly to their superior quality and quantity of data, although they have not entirely supplanted pion-induced data. 
Modern analysis frameworks now treat photon- and pion-induced reactions on an equal footing within coupled-channel fits.
Pioneering work in this area was carried out by the Gie\ss{}en group \cite{Penner:2002ma, Penner:2002md}, employing a multichannel $K$-matrix approach that incorporates a microscopic background. 
Building on a phenomenological background parameterisation, the Bonn-Gatchina group \cite{Anisovich:2011fc} utilises the $K$-matrix (N/D) framework to determine resonance parameters from coupled-channel fits to an extensive experimental database of pion- and photon-induced reactions with two- and three-body final states.
Assuming quasi-two-particle unitarity, the latter approach enables the study of sequential decays of resonances, such as in the reactions $\pi^- p \to \pi^+ \pi^- n$ and $\pi^0 \pi^- p$, recently measured by the HADES Collaboration \cite{HADES:2020kce}.

So-called dynamical coupled-channel models \cite{Doring:2025sgb} are based on effective Lagrangians and solve a scattering equation that includes off-shell intermediate states and the real dispersive parts of the amplitude. 
Examples include the ANL/Osaka approach (formerly EBAC) \cite{Kamano:2013iva} and the J\"ulich--Bonn model \cite{Ronchen:2012eg}. 
The latter has recently been extended to electroproduction reactions in the semi-phenomenological Jülich--Bonn--Washington framework \cite{Mai:2023cbp}, which was employed to extract baryon transition form factors at the pole for several resonances \cite{Wang:2024byt}.
Electroproduction reactions have also been investigated by the Yerevan/JLab group \cite{CLAS:2009ces}, using a unitary isobar model in combination with dispersion-relation techniques. 
For a recent analysis of two-pion electroproduction, see Ref.\,\cite{Mokeev:2023zhq}.

Despite advances in understanding the light baryon resonance spectrum enabled by high-precision electromagnetic probes, all analysis frameworks continue to rely critically on input from pion-induced reactions to parameterise the hadronic FSI. 
Hadronic data that match the electromagnetic data in both quality and quantity are therefore indispensable for a comprehensive determination of resonance properties, particularly in strengthening the statistical significance of resonance signals.

\subsection{Moment expansion of observables}
PWAs form a bridge between measurements of scattering processes and the QCD spectrum. 
Partial waves expressed in the $JLS$ basis possess well-defined quantum numbers, such that a given resonance contributes to a single partial wave. Moreover, at a fixed energy, only a finite number of partial waves contribute to the scattering amplitude.

However, the extraction of partial waves from observables suffers from ambiguities \cite{Barrelet:1971pw, JointPhysicsAnalysisCenter:2023gku}. 
In general, the different mathematical solutions are distinguished by imposing continuity across consecutive energy bins. 
Alternatively, one can expand the observables directly in a complete basis of angular functions. 
In this case, the coefficients of the expansion are referred to as \emph{moments}. 
By construction, the moments are unambiguous, as they are defined as integrals of the observables weighted by angular functions.
Moments are quadratic forms of partial waves, and their interpretation in terms of resonances is therefore less straightforward than that of the partial waves themselves. 
However, they can be advantageous in enhancing the small contributions of a resonance when it interferes with a dominant signal.

Moment expansions are typically employed in reactions with three-particle final states, where standard techniques from two-body production are more difficult to implement. 
To appreciate this point, consider the photoproduction of two pions on a nucleon target. 
The reaction $\gamma p \to \pi^+\pi^- p$ was measured by the CLAS Collaboration \cite{CLAS:2008ycy, CLAS:2009ngd}. 
The moments published by CLAS were described in terms of a reaction model in Ref.\,\cite{JointPhysicsAnalysisCenter:2024qld}. 
The unpolarised intensity is expressed as a product of scattering amplitudes:
\begin{equation}
    \label{eq:int2}
    I^0(\Omega)  = \frac{\kappa}{2}
    \sum_{\lambda, \lambda_1, \lambda_2} A_{\lambda; \lambda_1 \lambda_2} (\Omega)\, A^*_{\lambda; \lambda_1 \lambda_2} (\Omega),
\end{equation}
where $\kappa$ is a phase-space factor (defined in Ref.\,\cite{Mathieu:2019fts}), $\Omega$ denotes the angles of one of the pions, and $\lambda$, $\lambda_1$, and $\lambda_2$ are the helicities of the photon, target proton, and recoil proton, respectively. 
The angles and helicities are defined in the rest frame of the two-pion system.

\begin{figure}[t]
    \centering
    \includegraphics[width=0.9\linewidth]{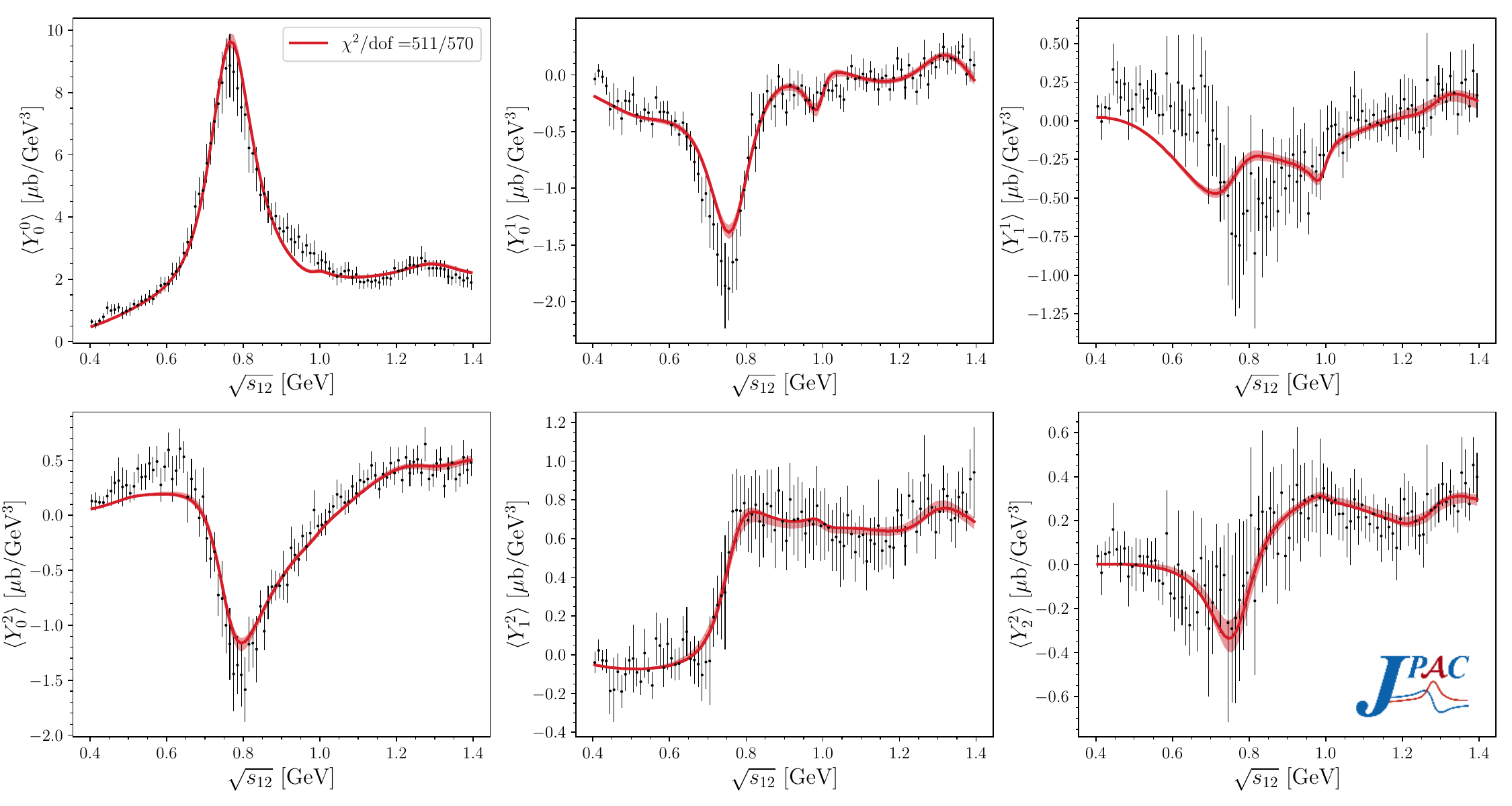}
    \caption{Low-order moments of the \(\pi^+\pi^-\) angular distribution measured by CLAS at \(t = -0.65\,\text{GeV}^2\). 
    Reprinted figure with permission from Ref.\,\cite{JointPhysicsAnalysisCenter:2024qld}. 
    Copyright (2024) by the American Physical Society.}
    \label{Fig:Moments}
\end{figure}

The production helicity amplitudes can be expanded in partial waves to include the relevant contributions. 
In the present example, the two-pion system spans from threshold up to 1.4\,GeV. 
The dominant resonances within this energy range are the scalar \(f_0(500)\) and \(f_0(980)\), the vector \(\rho(770)\), and the tensor \(f_2(1270)\) states. 
Each of these states is modelled via Regge exchange, and the scattering amplitude is expressed as
\begin{align}
    A_{\lambda; \lambda_1\lambda_2} (\Omega)  =  \sum_{\ell m} T^\ell_{\lambda m; \lambda_1\lambda_2} Y^{m}_\ell(\Omega),
    \label{ampl_pwe}
\end{align}
where the sum is restricted to \(\ell \leq 2\). 
In addition, Ref.\,\cite{JointPhysicsAnalysisCenter:2024qld} incorporates a Deck-like mechanism, in which a pion scatters off the target to produce nucleon excitations.

The unpolarised moments are defined as
\begin{align}
    \langle Y^L_M\rangle & =  \int  I^0(\Omega) \, d^L_{M0}(\theta) \cos (M\phi)\, \diff \Omega \,,
\end{align}
where \(\Omega = (\theta,\phi)\). 
By inserting the partial wave expansion into the moment definition, one obtains explicit relations between the moments and the partial waves. 
The first few moments are shown in Fig.\,\ref{Fig:Moments} for a fixed value of the momentum transfer between the nucleons, \(t = -0.65\,\text{GeV}^2\).

Of particular interest is the moment \(\langle Y^1_0\rangle\), which contains interference terms between the \(S\) and \(P\) waves. Schematically, the first moments read
\begin{subequations}
\begin{align}
    \langle Y^0_0\rangle & \propto |S|^2 + 3|P|^2 + \ldots \\
    \langle Y^1_0\rangle & \propto SP^* + S^*P + \ldots
\end{align}
\end{subequations}
Exact expressions, along with the definitions of the polarised moments, are provided in Refs.\,\cite{Mathieu:2019fts, Mathieu:2019gxo}.
In Fig.\,\ref{Fig:Moments}, one observes a dominant \(\rho(770)\) peak in \(\langle Y^0_0\rangle\). The much smaller \(f_0(980)\) signal is nearly invisible in the differential cross section (\(\langle Y^0_0\rangle\)), but becomes clearly visible through its interference with the long tail of the \(\rho(770)\) in \(\langle Y^1_0\rangle\).

\subsection{Spin alignment for multibody processes}
The intermediate energy range at FAIR provides a setting where the dominant reaction mechanism for exclusive hadron production is $t$-channel exchange that excites resonances in both top and bottom interaction vertices. 
This regime is similar to those at GlueX and CLAS with hadronic beams, and to COMPASS/AMBER, but the energy is not high enough to be purely diffractive. 
As a result, both top and bottom resonance production can occur simultaneously, leading to overlapping processes in the final-state phase space.

This overlap offers an opportunity to study correlations between the physics of the top and bottom vertices. It also poses a challenge in constructing the total reaction amplitude, since one must combine the two processes and account for their interference. 
A key complication arises from the spin of the involved particles: the spin state is an intrinsic part of the reaction's coherence and must be carefully included. 
For a proton beam, the $pp$ scattering reaction is symmetric under permutation of the top and bottom vertices, so symmetrizing the amplitude is an integral part of the construction.

\begin{figure}[t]
    \centering
    \includegraphics[width=0.7\linewidth]{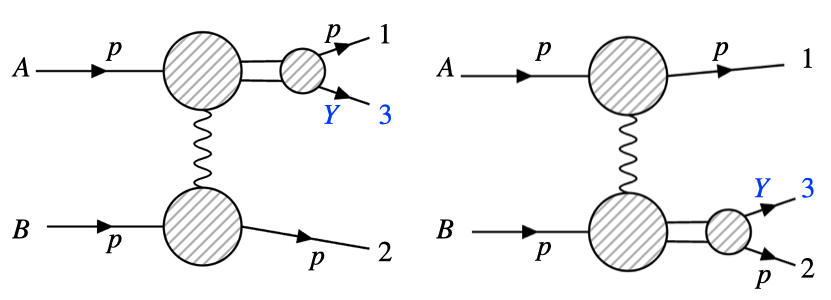}
    \caption{%
        The $t$-channel exchange mechanisms for exclusive meson production in $pp$ collisions.
        Left: forward resonance production (top vertex).
        Right: target resonance excitation (bottom vertex).
    \label{fig:production_diagrams}}
\end{figure}

Figure~\ref{fig:production_diagrams} illustrates possible diagrams for resonance production in $pp$ collisions with a single meson in the final state. The process may proceed through an excited baryon state in the top vertex; a similar process can occur in the bottom vertex; hence, the processes will be referred to as top and bottom production, respectively. 
In what follows, we discuss an application of general helicity-formalism considerations for exclusive multi-particle states, following Ref.\,\cite{Habermann:2024sxs}.

The helicity formalism is a powerful tool for making the angular dependence
of the reaction amplitude explicit. 
In principle, a fully covariant approach  would follow naturally from a known Lagrangian or Hamiltonian, with well-defined vertex functions. 
However, at the hadronic scale, no fundamental Lagrangian specifies how mesons and baryons couple to each other. 
Consequently, building interaction vertices must rely on general principles and models.
In such a situation, the helicity formalism provides a ``minimalistic'' but general framework:
it encodes universal scattering theory constraints on spin and angular momentum without needing explicit details of the underlying QCD dynamics. 
The missing parts can then be modelled, guided by symmetry principles and experimental data.

Although no fundamental Lagrangian exists for these hadronic-level interactions, it is still possible to construct a covariant, phenomenological description using general rank-$j$ tensors for spin-$j$ particles \cite{Filippini:1995yc, Rarita:1941mf}.
In such approaches, spin alignment does not arise as a separate concern, since all spin indices are contracted in a fully Lorentz-invariant manner.
However, the complexity of tensor structures grows quickly for higher-spin particles and multi-body decays, and separating spin-orbit couplings or isolating partial-wave components can become unwieldy.
Moreover, additional energy-dependent factors often appear in covariant decay amplitudes, potentially leading to mismatches with canonical treatments at higher energies while vanishing in the non-relativistic limit \cite{JPAC:2017vtd, JPAC:2018dfc}.
Despite these drawbacks, covariant frameworks are still advocated by several groups \cite{Filippini:1995yc, Chung:2007nn, Chen:2017gtx} and have been implemented in analysis packages \cite{Anisovich:2006bc, d_argent_2024_11212410, Bellis:2007zz}.

The helicity formalism produces simple, direct formulae by expressing reaction matrix elements via reduced amplitudes, explicit helicity couplings, and well-defined angular functions. 
Because spin projections are kept explicit, care must be taken to maintain consistent quantisation axes throughout the amplitude construction. 
This consistency is critical for correctly describing interference effects, especially in overlapping resonance regimes at intermediate energies.

For single-meson production in proton-proton scattering, any amplitude describing the process carries five helicity indices:
$\lambda_A$ and $\lambda_B$ are the helicities of the incoming protons, $\lambda_1$ and $\lambda_2$ are those of the outgoing protons, and $\lambda_3$ is the helicity of the produced meson.
One usually chooses the most convenient rest frame for partial-wave expansions and the definition of helicity states. 
However, this differs between the two production mechanisms shown in Fig.\,\ref{fig:production_diagrams}. 
For the top process, one typically adopts a Gottfried-Jackson frame (Fig.\,\ref{fig:angularframes}\,-\,right) aligned with the produced resonance at the top vertex. 
In the bottom process, an analogous Gottfried-Jackson frame is defined with respect to the resonance at the bottom vertex. 
Consequently, the spin quantisation axes for the top and bottom processes differ, leading to additional subtleties when their amplitudes are summed.

\newcommand{\orig}{\ensuremath{\mathfrak{o}}}
\newcommand{\targ}{\ensuremath{\mathfrak{c}}}

The helicity state for a particle with spin $j$ in a given frame is unambiguously defined by a set of transformations from its rest frame:
\begin{equation} \label{eq:helicity_state}
    \left|p_\orig,\lambda_\orig\right\rangle = R(\Omega_\orig) B_{z}(\gamma_\orig) \left|0,\lambda_\orig\right\rangle_{0}\,,
\end{equation}
where $\left|0,\lambda\right\rangle_{0}$ is a canonical spin state in the rest frame, $B_{z}(\gamma_\orig)$ is the boost along the chosen $z$-axis, and $R(\Omega_\orig) = R_z(\phi_{\orig}) R_y(\theta_{\orig})R_z(0)$ is a rotation in the ZYZ convention, needed to align a vector along the $z$ axis with a particle momentum $p_\orig$. 

A generic Lorentz transformation $\Lambda$ acting on a helicity state induces a Wigner rotation, $R_{\targ(\orig)}$:
\begin{equation}
    \Lambda \left|p_\orig,\lambda_\orig\right\rangle
    =
    \sum_{\lambda_\targ} \left|p_\targ,\lambda_\targ\right\rangle
    D^j_{\lambda_\targ,\lambda_\orig}(R_{\targ(\orig)}),
\end{equation}
where $p_\targ=\Lambda p_\orig$, the basis state in the new frame, and $\left|p_\targ,\lambda_\targ\right\rangle$ is defined according to Eq.\,\eqref{eq:helicity_state}, with transformation parameters $\gamma_\targ$ and $\Omega_\targ$.
The $D^j_{\lambda_\targ,\lambda_\orig}$ are Wigner $D$-functions. The Wigner rotation of the $Y$ state with respect to the $X$ state, is defined as,
\begin{equation}
    R_{\targ(\orig)} = \left[R(\Omega_\targ) \, B_{z}(\gamma_\targ)\right]^{-1} \Lambda \left[R(\Omega_\orig) \, B_{z}(\gamma_\orig)\right]\,,
\end{equation}
with the resulting rotation decomposed into three rotations in the ZYZ convention,
\begin{equation}
    R_{\targ(\orig)} = R_z(\phi_{\targ(\orig)})R_y(\theta_{\targ(\orig)}) R_z(\psi_{\targ(\orig)})\,.
\end{equation}
Importantly, for baryons, the azimuthal angle $\psi_{\targ(\orig)}$ must be determined in an extended range, $[-\pi, 3\pi]$ using the $\mathrm{SL}(2,\mathbb{C})$ representation of the Lorentz group \cite{Habermann:2024sxs}.


When summing amplitudes from the two mechanisms, one must account for the frame-dependent definitions of helicity states. 
Denote the top amplitude by $A^{\text{top}}$ and the bottom amplitude by $A^{\text{bot}}$. 
The total amplitude is then given by
\begin{align}
    A_{\lambda_A,\lambda_B,\lambda_1,\lambda_2,\lambda_3} & =
    A^{\text{top}}_{\lambda_A,\lambda_B,\lambda_1,\lambda_2,\lambda_3} +
    \sum_{\lambda'_1,\lambda'_2,\lambda'_3}
    A^{\text{bot}}_{\lambda_A,\lambda_B,\lambda'_1,\lambda'_2,\lambda'_3} \nonumber \\
    & \quad \times
    D^{1/2*}_{\lambda'_1,\lambda_1}(R^{1}_{\text{bot}(\text{top})})\,
    D^{1/2*}_{\lambda'_2,\lambda_2}(R^{2}_{\text{bot}(\text{top})})\,
    D^{j*}_{\lambda'_3,\lambda_3}(R^{3}_{\text{bot}(\text{top})})\,,
\end{align}
where $R^{i}_{\text{bot}(\text{top})}$ are the Wigner rotations that transform the helicity states of the bottom mechanism into the reference frame of the top mechanism, for each final-state particle $i$.
These rotations ensure that both production mechanisms are expressed in a consistent helicity basis, allowing for correct interference. 
The Wigner rotations $R^{i}_{\text{bot}(\text{top})}$ can be computed numerically from the polar and azimuthal angles, \textit{e.g}., using the \texttt{DecayAngle} package \cite{habermann_2024_13741241}.

To parametrise a reaction with more than two particles in the final state, one can consider a cascade of subsequent two-body decays, with each intermediate state carrying a well-defined spin. 
In principle, each topology in such a cascade can be expanded in partial waves, requiring an infinite number of terms to describe a realistic process. 
In practice, the amplitude is constructed by combining several cascade topologies, each truncated to a finite partial-wave sum.

For a particle of four-momentum $p$, its helicity angles are defined via the spherical coordinates of its three-momentum:
\begin{align}
    \cos\theta & = \frac{p_z}{|\vec{p}\,|}\,, \qquad
    \phi        = \arctan(p_y, p_x)\,, \qquad \text{and} \qquad
    \gamma      = \frac{E}{m}\,,
\end{align}
where $E$ and $m$ are the energy and mass of the particle. 
The polar angle $\theta$ spans $[0,\pi]$, and the azimuthal angle $\phi$ spans $[-\pi,\pi]$.
In the standard helicity convention, one reaches the rest frame of the particle $p$ via the transformation
\begin{equation}\label{eq:undo.BRR}
    B^{-1}(\gamma)\, R_y^{-1}(\theta)\, R_z^{-1}(\phi)\,,
\end{equation}
where $B^{-1}(\gamma)$ is the inverse boost, $R_y^{-1}(\theta)$ is a rotation about the $y$-axis, and $R_z^{-1}(\phi)$ is a rotation about the $z$-axis. 
Two alternative conventions, the \textit{minus\_$\phi$} and \textit{canonical} conventions, are discussed in Ref.\,\cite{Habermann:2024sxs}.

When combining two particles with spin in the helicity formalism, the order in which the particles are considered also matters. 
Two key points arise.
\begin{enumerate}
    \item The transformation to the second particle's rest frame may differ from the standard sequence in Eq.\,\eqref{eq:undo.BRR}.
    \item An additional phase for the second particle must be included when applying parity constraints or comparing with an $\ell s$ basis.
\end{enumerate}
Both effects originate from the definition of the two-particle state:
\begin{equation}
    \left|0;\lambda_1,\lambda_2 \right\rangle_0 =
    \left[ R B_z(\gamma_1)\,|0,\lambda_1\rangle_0 \right] \otimes 
    \left[ R R_z(\pi)\, B_z(\gamma_2)\,|0,\lambda_2\rangle_0 \right] 
    \times (-1)^{j_2 - \lambda_2}\,,
\end{equation}
where $\left| 0,\lambda_i \right\rangle_0$ are single-particle canonical spin states. 
This construction ensures that under rotations, the combined system behaves like a canonical state with total spin projection $\lambda_1 - \lambda_2$.

\subsection{Regge approaches in $pp$ collisions}
One useful feature of collider reactions is the ability to kinematically separate one or both of the initial-state protons in quasi-elastic production at energies above the baryon resonance region. 
This separation arises owing to the dominance of Reggeised hadron exchanges in collisions at higher energies and small momentum transfers, which lead to a rapidity gap between the beam and target protons.
Given this kinematic separation, the dynamics of particle subsystems in the final state may be studied with limited influence from FSIs involving spectator particles -- an advantage not available in other production mechanisms, such as multi-body decays or $e^+e^-$ collisions. 
As such, high-energy production reactions have served as cornerstones for spectroscopy programmes at pion- and photon-beam facilities, such as those of the COMPASS \cite{Ketzer:2019wmd} and GlueX \cite{GlueX:2021} experiments.


Two reaction topologies, corresponding to different kinematic regimes, are of particular interest for hadron spectroscopy in $pp$ collisions.

\begin{figure}[t]
    \centering
    \includegraphics[width=0.7\linewidth]{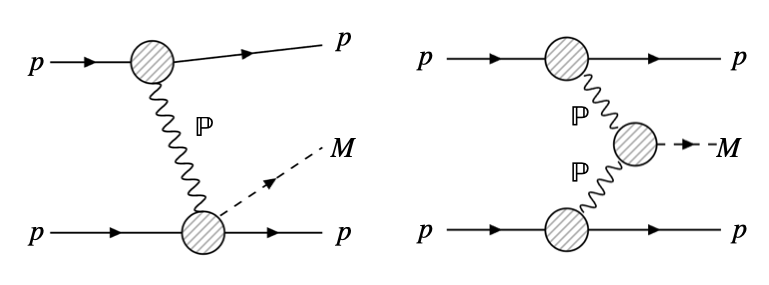}
    \caption{Diagrammatic representation of meson production through single and (left) and the double (right) Pomeron exchanges.
    \label{Fig:Regge_Pomeron_exchange}}
\end{figure}

The first is the so-called peripheral or ``single Regge'' production region, whereby a multiparticle system with baryon quantum numbers, $\mathcal{B}$, is produced in $pp \to \mathcal{B}p$ with large total invariant mass, $s$, and small momentum transfer, $t$. 
In the limit $\sqrt{s} \gg m_\mathcal{B} \geq m_p$, the scattering amplitude for the overall process is expected to factorise, schematically as:
\begin{equation}
    \label{eq:singe_regge}
    \langle \mathcal{B} p |T(s,t)|p p \rangle \propto \beta_{\mathcal{B}p}(t) \, \beta_{pp}(t) \left(\frac{s}{s_0}\right)^{\alpha(t)}\,.
\end{equation}
This is illustrated diagrammatically in the left panel of Fig.\,\ref{Fig:Regge_Pomeron_exchange}.
Here, the dependence on the total CM energy enters only as a power law in terms of a so-called ``Regge trajectory'', $\alpha(t)$. 
The trajectory is an analytic function of $t$, as required by the structure of the scattering amplitude in the complex angular momentum plane, and corresponds to the exchange of not just a single particle but an infinite tower of possible orbital excitations in the $t$-channel \cite{Gribov:1971zn, Collins:1977jy}. Typical trajectories for such $pp$ collisions include those of vector mesons and the Pomeron.

The functions $\beta_{\mathcal{B}p}(t)$ and $\beta_{pp}(t)$ are typically referred to as the ``Regge residues'' and contain the remaining dynamical dependences of the scattering amplitude. 
The factorisation of these residues into two pieces implies that one of the protons is kinematically separated and acts as a spectator to the creation of the system $\mathcal{B}$ via beam fragmentation. 
Thus, any nontrivial dependence on $m_\mathcal{B}$ and other kinematic variables related to its subsequent decays is entirely contained in $\beta_{\mathcal{B}p}(t)$, which does not depend on the total invariant mass of the collision. 
As such, studies of the dynamics of $\mathcal{B}$ and its subsequent decays are simplified by decoupling its production, which can be effectively modelled.

Because both beam and target are protons, one may also consider the complementary peripheral region of small Mandelstam $u$, \textit{i.e}., large $t$. 
In this region, we have a very similar factorisation, whereby the beam proton is a spectator while $\mathcal{B}$ is produced in the backward direction via target fragmentation, in which case the amplitude is described by a form analogous to Eq.\,\eqref{eq:singe_regge} with $t \leftrightarrow u$.

This factorisation is a general feature of scattering amplitudes at high energies and a fundamental prediction of Regge theory. 
Formally, the dominance of a few Regge poles holds as $s \to \infty$ and $t \to 0$, but Regge-motivated exchange models have been shown to describe scattering amplitudes well across a wide array of reactions at finite kinematics \cite{Nys:2018vck}.
The proton beams at SIS100 provide ample kinematic range to study the light baryon spectrum in this manner and even allow access to heavier sectors relevant to hidden-charm pentaquark candidates. 
Specifically, the light baryon spectrum can be accessed through the production of $pp\pi^0$, $pp\pi^+ \pi^-$, $pp\eta$, or even open-strangeness final states such as $p\Lambda K^+$ and $p\Sigma^+ K^0$.

In addition to light baryon production, SIS100 allows for the possibility of searching for hidden-charm pentaquark candidates, \textit{i.e}., the so-called $P_{c\bar{c}}$ states, in produced $ppJ/\psi$ final states. 
Although the available phase space in this case is likely not large enough to strictly satisfy the conditions of the Regge limit, $P_{c\bar{c}}$ states can be produced via gluon exchange, which is effectively modelled, at least phenomenologically, as a Pomeron.

The second reaction topology relevant for spectroscopy efforts in $pp$ collisions involves the production of mesons in more central kinematics via so-called ``double Regge'' processes \cite{Shimada:1978sx, Brower:1974yv}; see the right diagram in Fig.\,\ref{Fig:Regge_Pomeron_exchange}. 
In particular, a system with meson quantum numbers, $\mathcal{X}$, is produced in $pp \to \mathcal{X}pp$, where the invariant masses of both $\mathcal{X}p$ systems individually are large. 
These kinematics feature rapidity gaps allowing both protons to be kinematically separated. 
Because the reaction of interest is now a $2 \to 3$ process, the factorisation is more complex but schematically has the form:
\begin{equation}
    \label{eq:double_regge}
    \langle \mathcal{X}pp | T(s,t,s_1,t_1)| pp \rangle \propto \beta_{pp}(t_1) \beta_{pp}(t) \left(\frac{s_1}{s_0}\right)^{\alpha(t_1)} \left(\frac{s}{s_0}\right)^{\alpha(t_1) - \alpha(t)} V_\mathcal{X}(t_1, t, \eta) ~,
\end{equation}
where $s$ and $t$ remain the invariant mass and momentum transfer of the whole reaction, while $s_1$ and $t_1$ are the invariant mass and momentum transfer of the $\mathcal{X}$ and beam-proton system. 
This is represented in the right diagram of Fig.\,\ref{Fig:Regge_Pomeron_exchange}. 
Analogous to Eq.\,\eqref{eq:singe_regge}, we have two power-law factors and residues emerging from the exchange of Reggeons at both top and bottom vertices.

Here, just as above, the entirety of the dependence on the produced meson $\mathcal{X}$ and its decay products is contained in the ``double Regge'' vertex $V_\mathcal{X}$, which depends on the ratio of invariant masses $\eta = s_1 s_2 / s$. 
Amplitudes of this form, and subsequent models for the vertex functions, have recently been of interest in applications to double meson production at COMPASS \cite{Bibrzycki:2021rwh, Ketzer:2019wmd} and GlueX \cite{Gleason:2020mtb}. 
The central exclusive particle production programme at the LHC \cite{Albrow:2010yb} has also investigated double Regge production reactions, specifically double Pomeron exchange, as seen in Fig.\,\ref{Fig:Regge_Pomeron_exchange}. 
At much lower energies, production at FAIR may be sensitive to subleading Reggeon exchanges and can be used to produce the spin-exotic $\pi_1(1600)$ in $pp \to \eta^{(\prime)}\pi pp$.

\subsection{Lattice QCD}
\label{sec:toolslatqcd}
As mentioned previously, hadron interactions are governed by a quantum field theory known as quantum chromodynamics, whose Lagrangian density is given in Eq.\,\eqref{QCDdefine}. 
Although this theory appears very similar to quantum electrodynamics (QED), the standard perturbative method of calculating physical quantities in QED using Feynman diagrams fails badly when attempting to describe low-energy QCD processes, such as hadron formation. 
An efficacious approach to calculating low-energy quantities in QCD is known as lattice QCD \cite{Wilson:1974sk}. In this approach, QCD is formulated on a hypercubic space-time lattice so that all requisite path integrals can be evaluated in a non-perturbative manner using Markov-chain Monte Carlo methods.
The general steps of this approach -- following, \textit{e.g}., an early review \cite{Gupta:1997nd} -- include:
\begin{enumerate}
    \item Discretisation of space-time and Wick rotation (Euclidean metric). Thus, quarks cannot move freely but are constrained to discrete space-time points in discrete (Euclidean) times $\tau$, separated by lattice spacing $a$. 
    (Hereafter, following common LQCD practice, $t$ will be used to denote Euclidean time $\tau$.)
    Owing to computational limitations, the number of possible space-time points is necessarily restricted to a finite hypercubic volume of size $L_x \times L_y \times L_y \times L_t$. 
    Currently, typical calculations relevant for hadron interactions rely on volumes with site lengths of 48, 64, etc., in lattice units, corresponding to a few femtometres in physical units. 
    The finiteness of the lattice volume requires the definition of boundary conditions on the sides of the hypercubic lattice, including, \textit{e.g}., periodic, antiperiodic, or twisted \cite{Sachrajda:2004mi}.

    \item Formulation of the gauge and fermion degrees of freedom and choice of LQCD action. Important considerations include discretisation effects, implementation of QCD symmetries, and numerical cost.

    \item Definition of the measure of the path integral and its sampling through numerical simulation. This is typically done using the Hybrid Monte Carlo algorithm.

    \item Construction of operators corresponding to the hadron system under study and calculation of correlation (Schwinger) functions. These correlation functions generally decay exponentially (with respect to Euclidean time) to discrete, real-valued energy eigenvalues, which encode the dynamics of the hadron system in question.
\end{enumerate}

In a more practical implementation, predictions for physical observables are extracted from any quantum field theory through the vacuum expectation values of its quantum field operators acting at different space-time points, known as the $n$-point correlation (Schwinger) functions. 
In LQCD, the stationary-state energies are obtained from Hermitian matrices of temporal correlations
    $C_{ij}(t) = \langle 0\vert\, O_i(t\!+\!t_0)\, \overline{O}_j(t_0)\ \vert 0\rangle$,
where $t_0$ is the time at which each source operator acts and $t + t_0$ is the time at which each sink operator acts. 
The judiciously designed operators $\overline{O}_j$ create the states of interest, and the correlators are obtained from path integrals over the quark $\psi, \overline{\psi}$ and gluon $U$ fields of the form
\begin{equation}
    C_{ij}(t) = \frac{ \int {\cal D}(\overline{\psi}, \psi, U)\ 
        O_i(t+t_0)\ \overline{O}_j(t_0)\ \exp\left(-S[\overline{\psi}, \psi, U]\right)}{
        \int {\cal D}(\overline{\psi}, \psi, U)\ 
        \exp\left(-S[\overline{\psi}, \psi, U]\right)},
\end{equation}
where $O_j(t) = O_j[\overline{\psi}(t), \psi(t), U(t)]$.
The LQCD action is formulated in imaginary time:
\begin{equation}
    S[\overline{\psi}, \psi, U] = \overline{\psi}\, K[U]\, \psi + S_G[U],
\end{equation}
with $K[U]$ being the fermion Dirac matrix and $S_G[U]$ the gluon action. The integrals over the Grassmann-valued quark fields can be carried out exactly, yielding integrals over the gluon fields:
\begin{equation}
    C_{ij}(t) = \frac{ \int {\cal D}U\ \det K[U]\ 
        K^{-1}[U] \cdots K^{-1}[U]\ \exp\left(-S_G[U]\right)}{
        \int {\cal D}U\ \det K[U]\ 
        \exp\left(-S_G[U]\right)}.
\end{equation}
These integrals are then estimated using the Monte Carlo method. The Metropolis algorithm~\cite{Metropolis:1953am}, together with highly sophisticated update proposals~\cite{Duane:1987de, Clark:2006fx, Clark:2006wp}, is normally used. Such calculations must be performed on high-performance computers. 
To obtain physical observables reliably, computations using several lattice spacings must be performed, ensuring sufficiently large volumes to permit extrapolations to the continuum limit.

In finite volume, each correlator matrix element has a spectral representation given by
\[
    C_{ij}(t) = \sum_n Z_i^{(n)} Z_j^{(n)\ast} e^{-E_n t},
\]
where $E_n$ are the stationary-state energies and the overlap factors are $Z_i^{(n)} = \langle 0 \vert O_i(0) \vert n \rangle$, which yield information about the nature of each eigenstate. 
This expression neglects small temporal wrap-around effects. 
It is not practical to fit an entire correlation matrix to the above form. 
It can be shown that the eigenvalues of the matrix
    $C(t_0)^{-1/2}\ C(t)\ C(t_0)^{-1/2}$,
with $t_0$ fixed at some small value, tend to the eigenenergies of the lowest $N$ states with which the $N$ operators overlap as $t$ becomes large \cite{Luscher:1990ck}. 
This fact is used to rotate to a suitably diagonalised matrix in order to carry out separate fits to the individual diagonal elements to obtain the energy estimates. 
The overlap factors can then be determined.

Studying hadron–hadron scattering and transition form factors in LQCD requires extracting the energies of finite-volume multi-hadron states, which are excited states. 
Correlation matrices of significant size are needed, and operators with very good overlap onto the states of interest are crucial. 
To reliably extract the energies of multi-hadron states, multi-hadron operators composed of constituent hadron operators with well-defined relative momenta are required. 
The computation of temporal correlation functions involving such operators necessitates quark propagators from all sites on one time slice to all sites on another time slice, especially for a large number of starting times. 
Until recently, such computations were not feasible, but novel techniques, such as those presented in Refs.\,\cite{HadronSpectrum:2009krc, Morningstar:2011ka}, have allowed such calculations to be successfully performed.

The idea that the finite-volume energies obtained in LQCD can be related to the infinite-volume scattering $S$-matrix was first discussed in Refs.\,\cite{Luscher:1990ux, Luscher:1991cf}. 
These calculations were later revisited in Refs.\,\cite{Rummukainen:1995vs, Kim:2005gf} in the case of a single channel of identical spinless particles, and subsequent studies have generalised their results to treat multi-channels with different particle masses and nonzero spins \cite{Luu:2011ep, Fu:2011xz, Leskovec:2012gb, Hansen:2012tf, Gockeler:2012yj, Briceno:2014oea, Morningstar:2017spu}. 

One finds that the total lab-frame energy $E$ in an $L^3$ volume is related to the scattering $K$-matrix through a quantisation condition of the form
\begin{equation}
    \det \left[F(E,\bm P,L)^{-1} + {\cal K}(E_{\rm cm}) \right] = 0,
    \label{eq:lqcdquant1}
\end{equation}
where the CM energy is $E_{\rm cm} = \sqrt{E^2 - \bm{P}^2}$, with $\bm{P}$ being the total three-momentum of the system. 
The elements of the matrix $F$ are known but complicated functions of the energy, momentum, and lattice volume; and ${\cal K}$ is related to the $K$-matrix with threshold factors removed. 
The quantisation condition in Eq.\,\eqref{eq:lqcdquant1} is a single relation between an energy $E$ determined in finite volume and the entire scattering matrix. 

To proceed, the $K$-matrix is suitably parametrised. 
The goal then is to find best-fit values of these parameters such that the discrete spectrum produced by solving the above quantisation condition matches the spectrum obtained in LQCD. 
In this way, LQCD can determine resonance masses and widths, even though the finite-volume stationary states of LQCD are all discrete.
Of course, the $K$-matrix can also be easily connected to a broader class of scattering amplitudes such as, \textit{e.g}., the UChPT or dynamical coupled-channel approaches, and similar; for details, see Refs.\,\cite{Doring:2025sgb, Mai:2022eur, Mai:2025wjb}.

Some recent examples of such computations can be found in Refs.\,\cite{Guo:2018zss, Rodas:2023gma, Wilson:2019wfr, Mai:2019pqr, Woss:2019hse, BaryonScatteringBaSc:2023zvt, Fischer:2020yvw, BaryonScatteringBaSc:2023ori, Yan:2024gwp}, although such computations are becoming commonplace. 
This procedure has also been generalised to three-particle systems \cite{Hansen:2014eka, Hansen:2019nir, Hammer:2017uqm, Hammer:2017kms, Mai:2017bge, Feng:2024wyg}, and some recent studies in LQCD using these formalisms can be found in Refs.\,\cite{Mai:2018djl, Mai:2019fba, Draper:2023boj, Blanton:2021llb, Yan:2024gwp, Mai:2021nul, Brett:2021wyd, Alexandru:2020xqf,Yan:2025mdm}.

The techniques in LQCD for determining form factors of stable particles are well known; see, \textit{e.g}., Ref.~\cite{Alexandrou:2020okk}.
They involve computations of so-called three-point correlation functions, correlators involving a source operator at time $t_0$, the current operator at time $t_0+t_I$, and a sink operator at later time $t_0+t_{\rm sep}$:
\begin{eqnarray}
    \label{eq:3pt_spec}
    C_\Gamma(t_{\rm sep},\bm{p}^\prime;t_I,\bm{p}) &=&
    \langle 0| N(t_0+t_{\rm sep},\bm{p}^\prime)\, j_\Gamma(t_0+t_I,\bm{x}_c)
    \, N^\dagger(t_0,\bm{p}) | 0\rangle, \\
    &=& \sum_{m,n} z_m(\bm{p}^\prime) z^\dagger_n(\bm{p})
    e^{-E_m(t_{\rm sep}-t_I)} e^{-E_n t_I}
    \Gamma_{mn}(q^2), \\
    \Gamma_{mn}(q^2) &=& \langle m(\bm{p}^\prime)| j_\Gamma |n(\bm{p})\rangle,
\end{eqnarray}
where momentum conservation constrains the momentum of the initial state $\bm{p}=\bm{p^\prime-q}$, $N^\dagger(t_0,\bm{p})$ creates a hadron at time $t_0$ having definite three-momentum $\bm{p}$, and $\bm{x}_c$ is an arbitrarily chosen spatial site. 
To minimise contamination from higher-lying states, a variationally optimised hadron operator, expressed as a linear superposition of smeared, judiciously chosen three-quark or quark-antiquark operators, should be employed at the source and sink. 
This is usually done by evaluating a matrix of such three-point correlators and optimising the coefficients in the final analysis. 
To extract the form factors, two-point functions are also needed, suitable ratios of the three- and two-point functions must be evaluated, and renormalisation of the current operator must be taken into account. 

Generalised parton distribution (GPD) functions have also been computed in various lattice QCD studies. 
Some very recent computations of form factors, GPDs, and other matrix elements can be found in Refs.\,\cite{Alexandrou:2023qbg, Tsuji:2023llh, Djukanovic:2023beb, HadStruc:2024rix, Rodekamp:2023wpe}. 

The generalisation of the above elastic form factor method to transitions is also known, and a recent exploratory calculation \cite{Briceno:2016kkp} computed the $P$-wave $\pi\pi\rightarrow\pi\gamma^\ast$ transition amplitude, albeit on a very small lattice and for a very heavy pion mass $m_\pi\sim 400\,$MeV.

LQCD is a powerful method for carrying out calculations of low-energy observables in QCD, such as resonance energies and widths, axial and electromagnetic form factors, GPD, and transition amplitudes. 
Unfortunately, the method does require exorbitant amounts of computing resources. 
For this reason, many studies are performed using quark masses that are unphysically large in order to reduce the
computing costs. 
Multigrid inverters and the advent of supercomputing systems with graphics processing unit (GPU) acceleration have greatly helped to improve LQCD computations in recent years.

  \begin{figure}[t]
 	\centering
 	\includegraphics[width=1\textwidth]{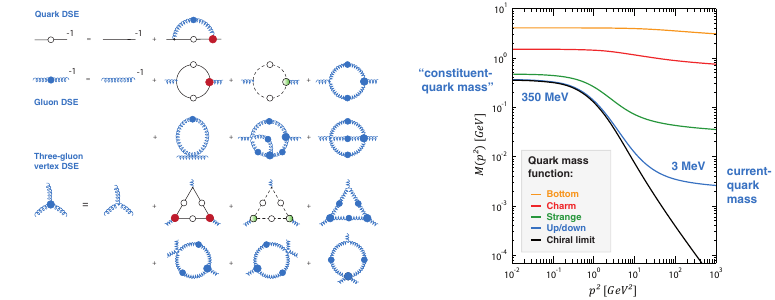}  
 	\caption{Left: Dyson-Schwinger equations for the quark and gluon propagator and three-gluon vertex. 
      The solid and dashed lines represent quarks and ghosts,
 	  the springs are gluons. Two-loop diagrams in the three-gluon  DSE are omitted.
      Right: Quark mass function for different quark flavors obtained from the quark DSE. Reprinted from Ref.~\cite{Eichmann:2016yit}, Copyright (2016), with permission from Elsevier.}
 	\label{fig:dses}
 	\vspace{-0mm}
 \end{figure} 

\subsection{DSE/BSE}\label{sec:functional-methods}
Continuum Schwinger function methods provide a nonperturbative approach to QCD that is based upon calculation of the $n$-point correlation functions defined by derivatives of the QCD generating functional.
(These are the same quantities computed using LQCD.)
Taking the quantum expectation values of the classical equations of motion
yields the Dyson–Schwinger equations (DSEs), which relate different $n$-point functions to each other \cite{Roberts:1994dr, Alkofer:2000wg,Roberts:2000aa,  Roberts:2000aa, Fischer:2006ub}. 
Similar systems of equations can be derived using the functional renormalisation group \cite{Dupuis:2020fhh}.
Therefore, in principle, functional methods are \textit{ab initio} approaches like LQCD:
while on the lattice one calculates the correlation functions directly from the path integral,
the strategy with the continuum methods is to calculate them \textit{from each other} using equations derived from the path integral.

The left plot in Fig.\,\ref{fig:dses} shows the DSEs for the quark and gluon propagator and the three-gluon vertex.
Each loop represents a four-momentum integration, and because the quantities on the left appear again on the right inside the integrals, these are integral equations which can be solved recursively. 
In this process, the classical propagators and vertices from the Lagrangian get \textit{dressed} and acquire structure.
Since the ingredients satisfy their own DSEs involving higher $n$-point functions, this yields an infinite tower of equations which, for practical calculations, must be truncated.  
The truncations employed so far cover a broad landscape, ranging from QCD-based models to \textit{ab initio} calculations without any model input; see, \textit{e.g}., Refs.\,\cite{Bashir:2012fs, Eichmann:2016yit, Huber:2018ned, Qin:2020jig} and references therein and thereto.

A key feature of QCD is dynamical mass generation.
As elucidated in Sec.\,\ref{sec.intro}, the sum of the light quark current masses, produced by Higgs boson couplings into QCD, only amounts to 1\% of the proton mass, so the rest must somehow emerge from QCD dynamics. 
This can be seen in the quark DSE, whose solutions show that the Higgs-produced quark current masses, which are inputs to QCD, are dynamically transmogrified into large constituent-quark masses at low momenta; see Fig.\,\ref{fig:dses}\,-\,right panel.
This nonperturbative effect is an expression of the dynamical breaking of chiral symmetry, which slowly becomes less important as the quark current mass increases, \textit{i.e}., for heavy quark flavours. 
The resulting quark mass function at low momenta carries a signature imprint of the quark chiral condensate \cite{Lane:1974he, Politzer:1976tv, Langfeld:2003ye} and sets the scale for the pion decay constant as well as the masses of the nucleon and other hadrons. 
The same effect is also seen in gauge-fixed LQCD computations; see, \textit{e.g}., Refs.\,\cite{Bowman:2005vx, Oliveira:2018lln, Kamleh:2023gho}. 

Another robust prediction from both continuum and lattice Schwinger function methods is the disappearance of the massless pole in the gluon propagator, 
\textit{i.e}., the gluon develops a mass gap; see \textit{e.g}., Refs.\,\cite{Braun:2007bx, Aguilar:2008xm, Fischer:2008uz, Cyrol:2017ewj, Huber:2018ned, Falcao:2020vyr, Eichmann:2021zuv, Binosi:2022djx, Ferreira:2023fva} and references therein.

 \begin{figure}[t]
	\centering
	\includegraphics[width=0.65\textwidth]{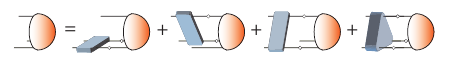}
	\caption{Covariant Faddeev equation for a baryon.}
	\label{fig:faddeev}
	\vspace{-0mm}
\end{figure} 
 
Hadrons appear as poles in QCD's correlation functions.
The hadron spectrum can be extracted from Bethe-Salpeter equations (BSEs), which hold at a given hadron pole and can be viewed as the quantum field-theory analogues of the Lippmann-Schwinger equation in quantum mechanics.
The BSE for a three-quark system is the covariant Faddeev equation, sketched in Fig.\,\ref{fig:faddeev}, whose kernel consists of all two- and three-body irreducible interactions.
The Faddeev amplitudes carry a rich structure, with
64 (128) independent Lorentz--Dirac tensors for $J=1/2$ ($J=3/2$) baryons~\cite{Eichmann:2009qa, Sanchis-Alepuz:2011egq},
whose dressing functions depend on five momentum variables.
In general, BSEs apply equally to bound states and resonances, but to extract a resonance mass 
one must analytically continue the eigenvalues of the equation to unphysical Riemann sheets, which is technically challenging \cite{Santowsky:2020pwd}.

So far many spectroscopy studies have used the rainbow-ladder truncation,
where the BSE kernel is an effective gluon exchange between the quarks \cite{Maris:1997tm, Maris:1999nt, Qin:2011dd}.
This preserves chiral symmetry and its dynamical breaking pattern:
the quark DSE dynamically generates quark masses, as in Fig.\,\ref{fig:dses},
and the pseudoscalar mesons calculated from their DSE-coupled BSEs are the massless Goldstone bosons in the chiral limit \cite{Maris:1997hd}.
The approach has also been fruitful for baryon spectroscopy, where diquark correlations are important for the binding of baryons;
one can reduce the three-quark Faddeev equation to a quark-diquark equation, which yields very similar spectra \cite{Eichmann:2016yit, Barabanov:2020jvn}.
In particular, with the mass-scale set by the pion decay constant, the predicted nucleon mass is $940\,$MeV in both approaches without any tuning.

The DSE/BSE approach has also been applied to various form factor calculations,
\textit{e.g}., 
nucleon electromagnetic and axial form factors \cite{Eichmann:2011vu, Eichmann:2011pv, Chen:2021guo, Yao:2024uej}, 
various baryon transition form factors \cite{Eichmann:2011aa, Segovia:2014aza, Segovia:2015hra, Sanchis-Alepuz:2017mir, Chen:2023zhh}, 
and gravitational form factors \cite{Yao:2024ixu}.
As with the spectrum, the three-quark and quark-diquark calculations
show good overall agreement \cite{Eichmann:2016yit, Barabanov:2020jvn, Cheng:2025yij}.
A typical common feature is missing strength at low $Q^2$, which in some cases can be attributed to missing meson-baryon effects that are not yet captured by the truncations.
Especially important in the timelike region ($Q^2 < 0$) are (vector-)meson poles, which are automatically generated from the quark-gluon interactions \cite{Maris:1999bh}. 
Recent studies also include the $\rho\to\pi\pi$ resonance mechanism, which turns the $\rho$ meson into a resonance \cite{Williams:2018adr, Miramontes:2021xgn}.

Apart from meson and baryon spectroscopy and form factors, the DSE/BSE approach has been applied to a range of problems including four- and five-body systems \cite{Eichmann:2015cra, Santowsky:2020pwd, Hoffer:2024alv, Eichmann:2025gyz}, glueballs \cite{Huber:2020ngt, Huber:2021yfy, Souza:2019ylx},
parton distributions \cite{Roberts:2021nhw, Ding:2022ows, Xu:2025cyj, Yu:2025fer}, 
the muon $g-2$ problem \cite{Aliberti:2025beg}, 
and the QCD phase diagram \cite{Fischer:2018sdj}.
In parallel, various systematic improvements are under way, such as
going beyond rainbow-ladder by implementing higher $n$-point functions \cite{Chang:2011ei, Williams:2015cvx, Eichmann:2016yit, Huber:2020ngt, Qin:2020jig}
or studying the multihadron components of resonances.
Finally, hadron spectroscopy and structure calculations using the functional renormalisation group are also beginning to take shape; see, \textit{e.g}., Refs.\,\cite{Fu:2025hcm, Zhang:2025ofc}.

\subsection{Effective field theories}
Quantum EFTs are a powerful tool for addressing multiscale systems. 
They are based on scale separations and factorisations, and exhibit a power counting that allows one to attach well-defined errors to derived physical prediction.
Owing to the complexities of QCD, many different EFTs have been developed 
to simplify the dynamics depending on the system or the process under consideration.
An EFT is typically organised as an expansion in a ratio of energy scales -- small over large -- introducing a suitable expansion parameter.
While low-energy degrees of freedom are treated fully dynamically, high-energy (short-distance) degrees of freedom are integrated out and absorbed into strength parameters of local operators.
Depending on the EFT, these are fixed either from the fundamental theory via a systematic procedure called matching, or must be determined from experiment or other theory, often LQCD results.
At a given order in the expansion, the number of such operators is finite, making the framework systematically improvable to the desired level of accuracy -- at least in principle.

In QCD, quarks may be divided into two fundamental sets: heavy quarks (charm, bottom, top), whose masses, $m_Q$, are much larger than the QCD mass scale,
$\Lambda_{\rm QCD}$, and light quarks (up, down, strange), whose masses
$m_q$ are smaller. 
The cases $m_Q \gg \Lambda_{\rm QCD}$ and $m_q \ll \Lambda_{\rm QCD}$, enable an EFT treatment of hadrons that exploit symetries of the fundamental interaction.
Since these symmetries are sometimes ``hidden'' via strong interaction dynamics, the EFT is typically simpler than a full QCD treatment, at least at low orders in the effective expansion; hence, able to deliver predictions currently inaccessible to \textit{ab initio} QCD studies.
The following discussion focuses on chiral EFTs and EFTs for heavy quarks.
%
Numerous other EFTs have been developed to study particular hadronic processes or in-medium interactions believed to be described by QCD.

\subsubsection{Chiral EFTs for two-hadron systems}
It is well known that QCD in the light sector has an approximate chiral symmetry, which would be exact in the case of massless quarks. 
As this symmetry is dynamically broken, eight Goldstone bosons are expected. 
In the low-energy regime, one can define a ChPT \cite{Gasser:1983yg, Gasser:1984gg} by introducing explicit hadron degrees of freedom. 
While this procedure formally leads to infinitely many terms and spoils renormalisability, a systematic power counting, exploiting the mass gap given by the small Goldstone boson masses and momenta, allows for a finite number of terms at each order and a practicable renormalisation at a given order \cite{Weinberg:1978kz}.

Low-energy constants (LECs) effectively encode the details of the QCD dynamics, including the presence of heavier quarks. 
Since the EFT is defined at the hadron level, such constants are non-perturbative quantities. 
They can be calculated with continuum or lattice Schwinger function methods, or extracted via comparison with experiments. 
These constants are universal, \textit{i.e}., once known, any other process involving these LECs can be predicted. 
For a summary of our current knowledge, see Ref.\,\cite{FlavourLatticeAveragingGroup:2019iem}. 

One may summarise the steps involved as
\begin{align}
    \mathcal{L}_{\rm QCD}
    \overset{\rm EFT}{\longrightarrow}
    \mathcal{L}_{\rm ChPT}
    \overset{\rm LECs}{\longleftrightarrow}
    \text{Observables}
\label{eq:ch5:maxim:UCHPT}
\end{align}
The procedure described above relies, roughly speaking, on an expansion in the small masses of the (pseudo) Goldstone bosons and small momenta, $p$. 
This means that the convergence of the series can break down when one of these quantities becomes large. 
Moreover, in doubly-heavy systems like nucleon–nucleon scattering or the scattering of two states containing heavy quarks, there is a kinematic enhancement of certain topologies of the type $(\pi \mu / p)$,
where $\mu = m_1 m_2 / (m_1 + m_2)$ denotes the reduced mass of the two-hadron system. 
In both cases, unitarisation or resummation of certain types of EFT diagrams, defining the “interaction kernel”, is called for. 
Although some model dependence is introduced through this scheme, the connection to ChPT (and QCD) persists through the interaction kernel. 

This so-called UChPT method can be summarised through 
\begin{align}
	\mathcal{L}_{\rm QCD}
	\overset{\rm EFT}{\longrightarrow}
	\mathcal{L}_{\rm CHPT}
	\overset{V_{\rm CHPT}}{\longrightarrow} 
	T=V_{\rm CHPT}+\int V_{\rm CHPT}\,G\,T\,
	\longrightarrow 
	\text{Observables}
	\label{eq:ch5:maxim:UCHPT2}
\end{align}
Here, in the first step (left to right), the QCD degrees of freedom (quarks and gluons) are replaced by effective fields (hadrons) in a model-independent way. 
In the second step, a unitary meson–baryon scattering matrix is introduced through an integral equation -- usually a Bethe–Salpeter or Lippmann–Schwinger equation -- using the chiral potential $V_{\rm ChPT}$ of a fixed order. 
As such, it sums up an infinite subset (bubble chain) of all possible Feynman diagrams necessary to calculate the full meson-baryon 4-point function in the EFT. 
This procedure ensures two main strengths of the approach, namely, that at a fixed chiral order, the amplitude matches ChPT exactly, while simultaneously satisfying the unitarity condition of $S$-matrix theory. 
These two features allow one to study, \textit{e.g}., not only the excited baryon spectrum -- connecting theory ($\rm QCD \to \rm ChPT$) with phenomenology/experiments -- but also unphysical pion mass finite-volume spectra such as those often obtained using LQCD. 
Examples of the latter connection can be found, \textit{e.g}., in Refs.\,\cite{Doring:2013glu, Guo:2023wes}.

Additionally, for resonances in two-hadron systems that contain a light Goldstone boson, unitarisation can be employed and still lead to a systematically improvable formalism for two-hadron systems. 
This is achieved by the observation that the chiral expansion can be employed as subtraction terms for a dispersive treatment of the inverse scattering amplitude \cite{Dobado:1996ps, Oller:2000fj}.
The equations that emerge are analogous
to Eq.\,\eqref{eq:ch5:maxim:UCHPT2}. 
This kind of formalism has been applied successfully to $\pi\pi$, $\pi K$ and $\bar KK$ scattering to study the light scalar mesons \cite{Oller:1997ti}, as well as their quark mass \cite{Hanhart:2008mx} and $N_c$ dependence \cite{Pelaez:2003dy}, $\pi \Sigma$ and $\bar KN$ scattering \cite{Mai:2009ce}, and many other processes.

The simplicity and success of these types of models do not come without a price, namely that terms of higher order than that of $V_{\rm ChPT}$ are not fully accounted for.
This introduces a certain model dependence. 
Fortunately, there are various options to gain control over the latter. Necessarily, these require either input from new experiments or independent theoretical approaches at the level of observable quantities (phase shifts, line shapes, etc.). 
A systematic study of the model dependence and its possible reduction is discussed in Refs.\,\cite{Mai:2014xna}. 
For a review of the accomplishments of ChPT and its unitarisations in improving our understanding of hadron interactions, see, e.g., Ref.\,\cite{Meissner:2024ona}.

\subsubsection{Nonrelativistic effective field theories}
\label{NREFT}
Heavy quarks play a key role in investigations of strong interactions and in the  search for physics beyond the Standard Model at $B$ factories.
For hadrons made of one heavy quark, such as heavy-light mesons and baryons, the appropriate nonrelativistic EFT is called heavy quark effective theory (HQET)
\cite{Isgur:1991wq, Eichten:1989zv, Isgur:1990yhj, Isgur:1989vq}, reviewed, \textit{e.g}., in Refs.\,\cite{Neubert:1993mb, Manohar:2000dt}.
Heavy-light hadrons are systems characterised by just two relevant energy scales, $m_Q$ and $\Lambda_{\rm QCD}$. 
HQET follows from QCD by integrating out modes associated with the heavy quark mass and exploiting the hierarchy $m_Q \gg \Lambda_{\rm QCD}$.
In the context of HQET, one deals with heavy-light hadrons made of either a charm or a bottom quark (the top quark does not live long enough to form a bound state before decaying weakly into a $b$ quark).
HQET was the first nonrelativistic EFT of QCD and has enabled a wide range of physical applications to the spectrum and decays of heavy-light mesons and baryons.

Systems composed of more than one heavy quark, such as quarkonia (\textit{e.g}., charmonium and bottomonium), quarkonium-like states, or doubly-heavy baryons, are characterised by more energy scales. 
The typical distance between the heavy quarks is of the order of $1/(m_Q v)$, with $v \ll 1$ being the relative velocity of the heavy quarks.
This implies that the typical momentum transfer is of order $m_Q v$ (which corresponds to the inverse of the usual radius of the state), and the typical binding and kinetic energy are of order $m_Q v^2$. 
The inverse of $m_Q v^2$ sets the time scale of the bound state. 

Taking into account this hierarchy of dynamically generated scales, $m_Q \gg m_Q v \gg m_Q v^2$, one can construct nonrelativistic EFTs: specifically, 
nonrelativistic QCD (NRQCD) \cite{Caswell:1985ui, Lepage:1992tx, Bodwin:1994jh}
at the scale $m_Q v$, obtained by integrating out QCD modes associated with the scale $m_Q$, and potential NRQCD (pNRQCD) \cite{Brambilla:1999xf, Brambilla:2004jw, Pineda:1997bj} at the scale $m_Q v^2$, obtained by integrating out modes associated with the energy scale $m_Q v$. 

NRQCD and pNRQCD have enabled precise and systematic studies of heavy quarkonium spectrum, decay, production, propagation in medium, and extraction of Standard Model parameters; see, \textit{e.g}.,
Refs.\,\cite{Brambilla:2010cs, QuarkoniumWorkingGroup:2004kpm, Brambilla:2014jmp}.

The pNRQCD scheme implements the Schrödinger equation for the quarkonium field as the zeroth-order problem, with the potential derived by direct considerations of QCD, and is particularly useful for phenomenological applications. 
The only parameters appearing in this EFT are those of QCD, namely, the strong coupling constant and the quark masses. 
When the soft scale $m_Q v$ is perturbative, singlet and colour-octet quarkonia must be considered; when the soft scale is nonperturbative, $m_Q \sim \Lambda_{\rm QCD}$, only colour-singlet quarkonia are dynamical, and the calculation of the potential requires non-perturbative methods, typically LQCD.
In this latter case, one refers to the Born–Oppenheimer EFT (BOEFT) \cite{Berwein:2015vca, Berwein:2024ztx, Brambilla:2017uyf, Brambilla:2018pyn, Soto:2020xpm}.
BOEFT can describe, within a unified framework, quarkonium and all the XYZ exotic states discovered in recent decades that are composed of two 
heavy quarks (or antiquarks): hybrids, tetraquarks, pentaquarks, and doubly-heavy baryons.
 
When focusing on the energy region around hadron–hadron thresholds, one can construct an NREFT with hadrons as the effective degrees of freedom, where their velocities in the CM frame serve as small dimensionless expansion parameters. 
In particular, for $S$-wave interactions, the leading-order interaction Lagrangian contains only constant contact terms, while higher-order terms involve derivatives of the hadron fields. 
The coefficients in the effective Lagrangian also act as counterterms to cancel ultraviolet divergences in loop diagrams.
They are usually determined by fitting to experimental data or LQCD results.
Such NREFTs (either pionless or pionful) have extensively been developed to study the $NN$ interaction; see, \textit{e.g}., Refs.\,\cite{Epelbaum:2008ga, Hammer:2016xye, Meissner:2022cbi}. 

Similar NREFTs have been used to study the spectrum and/or line shapes of hadronic molecules, including the hidden-charm $XYZ$ states \cite{AlFiky:2005jd, Fleming:2007rp, Braaten:2010mg,Baru:2011rs, Nieves:2012tt, Guo:2013sya, Dai:2019hrf, Habashi:2020qgw}, the hidden-charm $P_c$ states \cite{Du:2019pij, Du:2021fmf}, and the double-charm tetraquark $T_{cc}$ \cite{Du:2021zzh, Albaladejo:2021vln}; see also Ref.\,\cite{Guo:2017jvc}.
Heavy quark spin symmetry plays an important role in predicting new hadronic molecules in this context \cite{Guo:2009id}, while moving from double-charm to double-bottom systems induces scale dependence \cite{Baru:2018qkb}.
See also Section~\ref{subsubsec.cusp} for a description of threshold cusps in the NREFT framework.
\newpage
\section{Experimental facilities at GSI/FAIR}
\label{sec.OverviewFacilities}
{\small {\bf Convenors:} \it J.~Ritman, C.~Sturm} 


\noindent This chapter provides a concise, primarily technical overview of the accelerator complex and the experiments at GSI/FAIR that are relevant to the hadron physics topics discussed in the preceding chapters.

\subsection{The GSI/FAIR accelerator complex}
\label{sec.FairGsi}
\begin{figure}[h!]
\begin{center}
\includegraphics[width=0.98\linewidth]{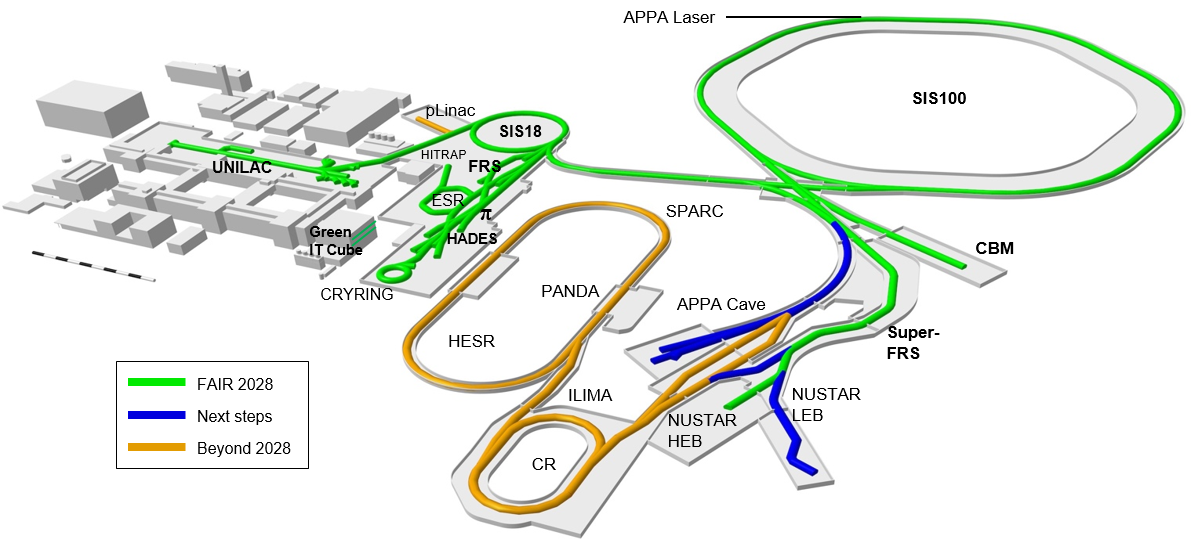}
\end{center}
\caption{Schematic overview of GSI/FAIR.}
\label{Ch9-fig:FAIR}
\end{figure}

A schematic overview of the GSI/FAIR accelerator complex is presented in Fig.\,\ref{Ch9-fig:FAIR}. 
The GSI accelerator facility currently in operation comprises the 120\,m-long universal linear accelerator 
(UNILAC) \cite{GSI:UNILAC}, which accelerates all ion species, from protons to uranium, up to 11.4\,MeV/u, 
while intermediate ion energies of 3.6, 4.8, 5.9, and 8.6\,MeV/u are achieved by operating selected Alvarez cavities 
without acceleration. 
The UNILAC beam is injected into the normal-conducting synchrotron SIS18 \cite{GSI:SIS18}, 
with a maximum rigidity of 18.6\,T$\cdot$m 
and a circumference of 216\,m, which accelerates ions up to 2\,GeV/u (carbon) or 1\,GeV/u (uranium), 
and protons up to 4.5\,GeV kinetic energy. 
Fast, knock-out, or resonant extraction modes from SIS18 are available, 
providing beam spills of up to 20\,s duration. The spill profile and microstructure can be further shaped and smoothed 
by a spill feedback loop system driven by experiment counter signals.
Depending on the charge state, the intensities achieved at SIS18 for uranium ions are 4.5\,$\times$\,10$^{10}$ per cycle 
for U$^{28+}$ and 3\,$\times$\,10$^{9}$ per cycle for U$^{73+}$. 
The maximum SIS18 intensity for protons is 2\,$\times$\,10$^{11}$ per cycle. 

The primary SIS18 beam is transported either directly to the experimental areas and/or to one of the following: 
the FRS (18\,T$\cdot$m maximum magnetic rigidity), the experimental storage ring (ESR, 10\,T$\cdot$m maximum rigidity, 
with electron cooling), or CRYRING, a low-energy super-conducting storage ring 
(0.054 – 1.44\,T$\cdot$m rigidity). This allows beam delivery to multiple target stations spill-by-spill, 
enabling high-level parallel operation. 

In addition, the GSI/FAIR accelerator complex includes four target stations for the generation of secondary beams: 
pions and rare isotope beams (RIBs) at SIS18, and antiprotons and RIBs at SIS100. 
The RIB target stations are located within the FRS (served by SIS18) and the Super-FRS (SFRS), which is served by both SIS18 and SIS100 beams.

\begin{figure}[t]
\begin{center}
\includegraphics[width=0.95\linewidth]{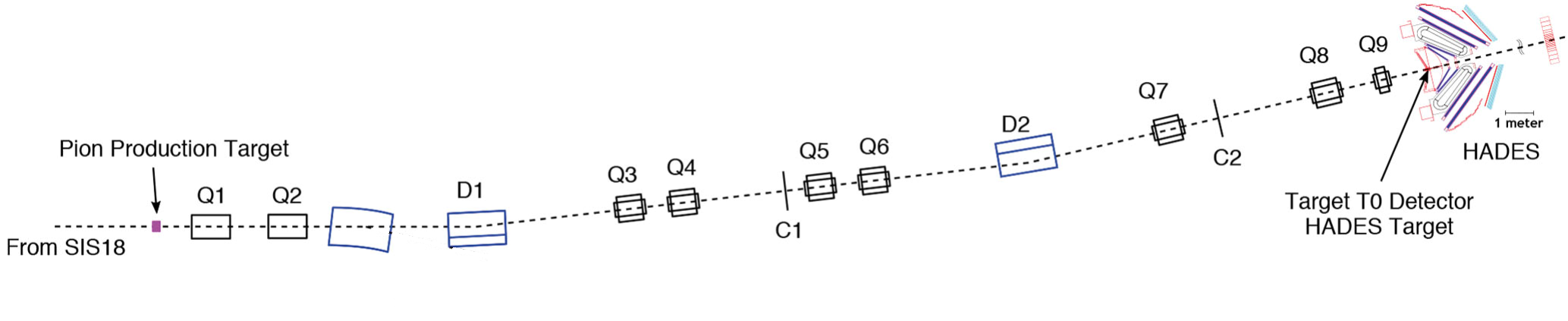}
\end{center}
\caption{Schematic overview of the 33.5\,m long pion beam facility \cite{PionBeamHADES:2002, PionBeamHADES:2017}.
\label{Ch9-fig:GSI-piFacility}}
\end{figure}

\paragraph{The SIS18 pion beam facility.}
A dedicated target station generates secondary pion beams 
with momenta of 0.4 to 2.8\,GeV/$c$ \cite{PionBeamHADES:2002, PionBeamHADES:2017}. 
The secondary beam facility is depicted in Fig.\,\ref{Ch9-fig:GSI-piFacility}. 

The high-acceptance dielectron spectrometer (HADES) experiment is connected via a 33.5\,m-long beam line comprising two dipole magnets (D1, D2), three quadrupole doublets (Q1Q2, Q3Q4, Q5Q6), and a quadrupole triplet (Q7Q8Q9), located inside the HADES cave to focus the beam onto the HADES target. 
The quadrupole settings are adjusted according to second-order beam optics calculations to optimise the acceptance and focus of the beam line. 
Additionally, the dipole magnet pair D1 and D2 are slightly tilted vertically to raise the beam by 0.5\,m above the standard beam height, aligning it with the central axis of the HADES experiment. 
The rigidity $B\rho$ of the beam line defines the central momentum within a momentum acceptance window of approximately 8\%. 
Position-sensitive, double-sided silicon strip detectors (10\,$\times$\,10\,cm$^2$), C1 and C2, are positioned near the intermediate focal planes to enable reconstruction of the momentum of individual beam particles with a resolution of about $\Delta p / p \sim 0.3\%$ ($\sigma$), at particle rates up to $10^7$\,s$^{-1}$. 
The (secondary) pion beam, with momenta up to 1.5\,GeV/$c$, is generated by bombarding an intense (primary) $^{14}$N beam with 2.0\,GeV/u kinetic energy onto a 10\,cm-long beryllium target. 
Higher yields for pion momenta above 1.5\,GeV/$c$ are achieved with a primary proton beam at 3.5\,GeV, enhancing the pion yield by up to an order of magnitude at 2.8\,GeV/$c$ momentum. 
Consequently, pion rates of approximately $10^6$\,s$^{-1}$ are available for the maximum primary beam intensities provided by SIS18.

\paragraph{The superconducting synchrotron SIS100 at FAIR.}
The UNILAC and the synchrotron SIS18 serve as the injector line for the fast-ramping (4\,T/s), superconducting synchrotron SIS100 -- the driver accelerator for FAIR -- with a circumference of 1100\,m and 100\,T$\cdot$m maximum rigidity. 
The SIS100 is a novel type of synchrotron \cite{SIS100PSpiller}, explicitly designed for the acceleration of intermediate charge-state heavy ions. 
As the FAIR accelerators serve a broad spectrum of experiments, SIS100 provides high flexibility, enabling operation in fast-extraction mode with a cycle length of about 1\,s, as well as cycles with long injection plateaus for stacking, or long extraction plateaus for slow extraction with spill lengths of up to 100\,s. 

The use of super-conducting magnets in SIS100 is driven primarily by the requirement for a cryogenic ultra-high vacuum (UHV) system, which functions as a super-pump and stabilises the beam dynamics by reducing the effects of residual gas pressure. 
Like the existing GSI accelerator facilities, SIS100 can accelerate all ion species from protons to uranium. 
Through the use of various stripper stations, the injector line provides a wide range of charge states for all ion species.
The maximum kinetic energy and intensity for protons is 29\,GeV with 2.5\,$\times$\,10$^{13}$ protons per cycle, while for uranium ions, depending on the charge state, it ranges from 2.7\,GeV/u for U$^{28+}$ with 
5\,$\times$\,10$^{11}$ ions per cycle, to 11\,GeV/u for U$^{92+}$ with 4\,$\times$\,10$^{10}$ ions per cycle. 
In addition, SIS100 will be equipped with a unique laser cooling system. 
Even ion beams with an initially large momentum spread of $\Delta p/p = 10^{-3}$ can be captured by the laser light and cooled down to $\Delta p/p = 10^{-7}$ within a few seconds.

\begin{figure}[t!]
\begin{center}
\includegraphics[width=0.40\linewidth]{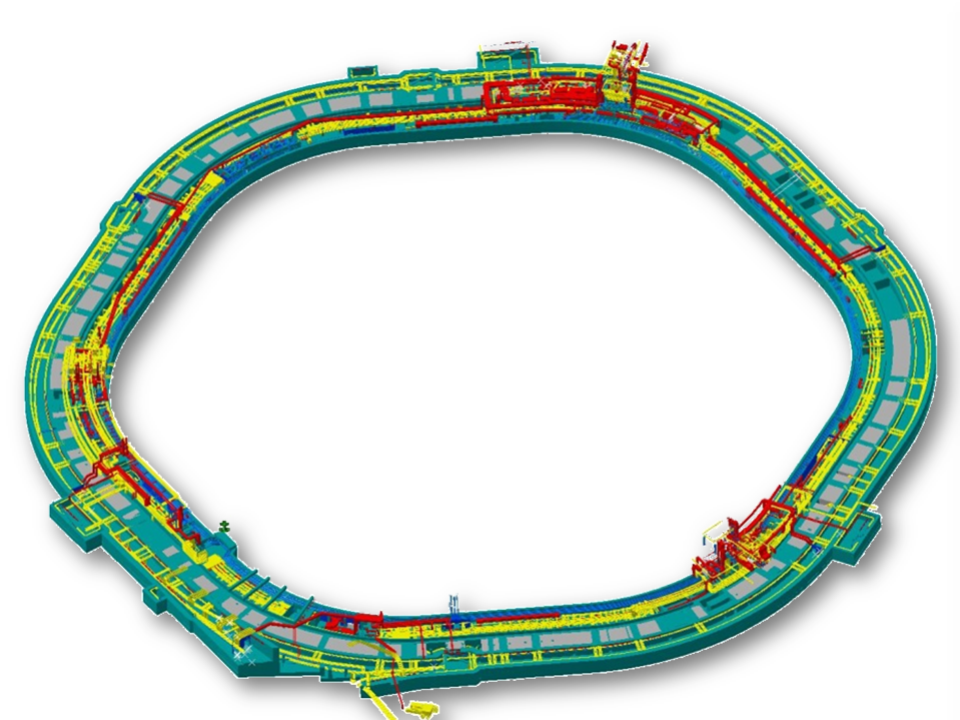}
\hspace{1cm}
\includegraphics[width=0.38\linewidth]{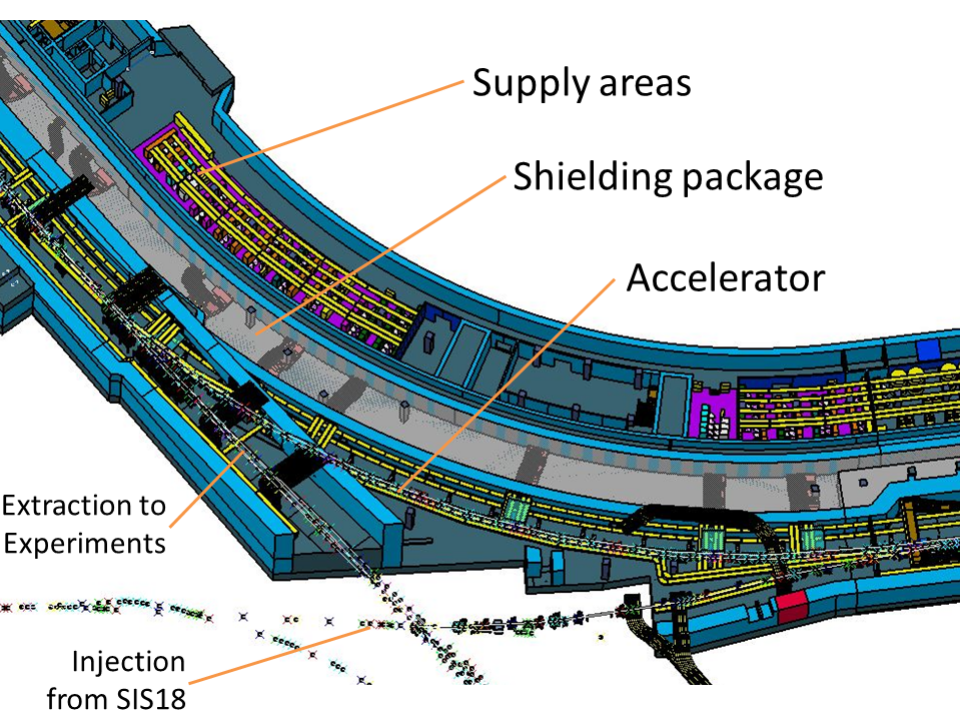}
\end{center}
\caption{Detailed view of the SIS100 tunnel with 1100 m circumference and located about 17\,m underground.  
\label{Ch9-fig:SIS100Tunnel}}
\end{figure}

The SIS100 is installed in a tunnel approximately 17\,m underground. As shown in Fig.\,\ref{Ch9-fig:SIS100Tunnel}, the tunnel design incorporates an inner supply ring, separated from the accelerator tube by substantial shielding, which protects the electronics at high beam intensities and allows extended access for maintenance. 
Furthermore, both the FAIR design and the shell construction enable an upgrade option to install an additional superconducting stretcher ring within the SIS100 tunnel, increasing the energy of primary beams and allowing parallel operation modes. 
This significantly enhances the available beam time for experiments.

\begin{figure}[t]
\begin{center}
\includegraphics[width=0.8\linewidth,angle=0]{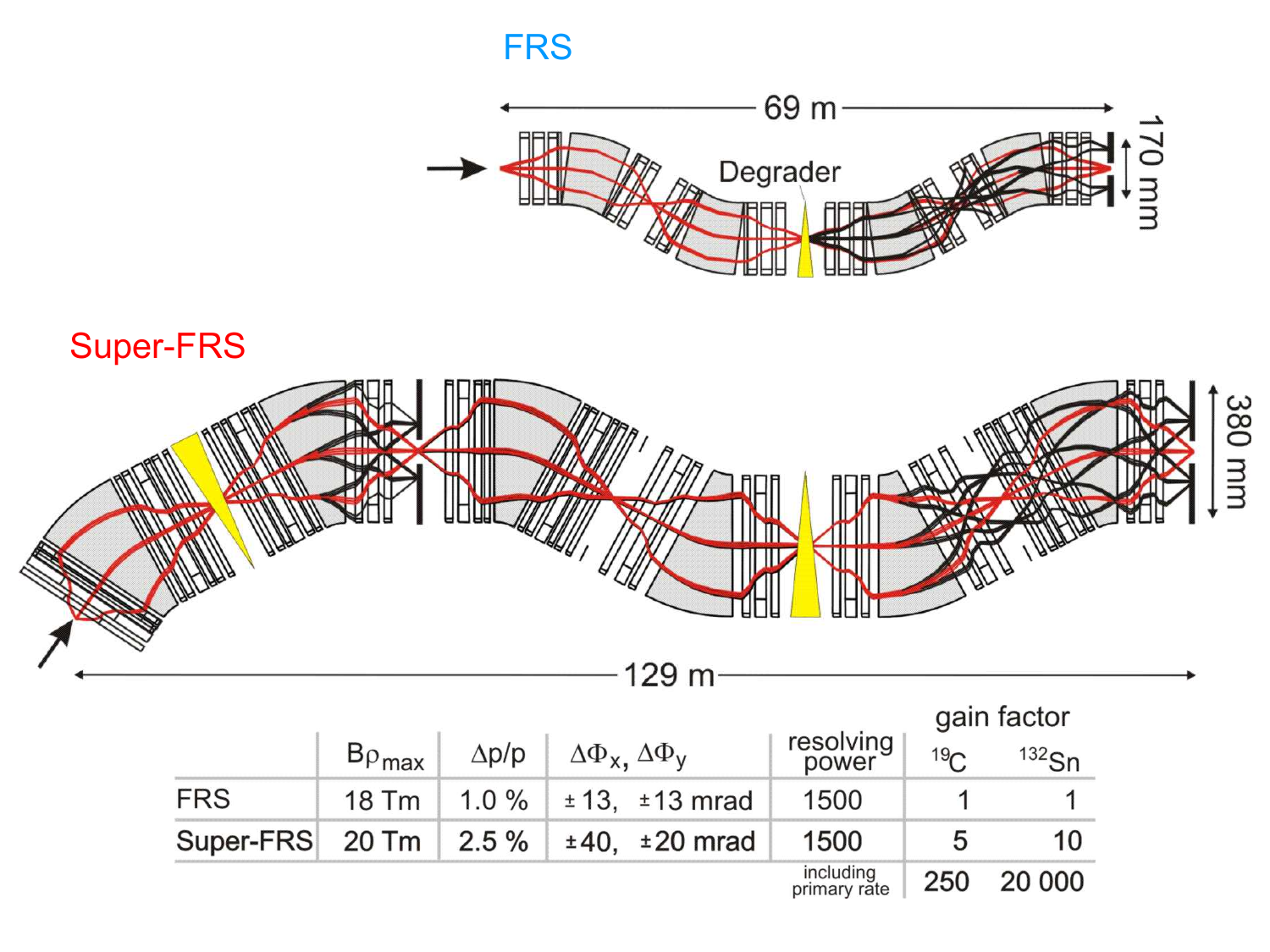}
\end{center}
\vspace*{-1cm}
\caption{Comparison of FRS and Super-FRS at GSI/FAIR.}
\label{fig:ch9-FRS-SFRS}
\end{figure}

\paragraph{Fragment separators.}
Fragment separators are employed at various radioactive ion beam facilities worldwide for the production, in-flight separation, and identification of exotic nuclei. 
They are high-resolution momentum spectrometers capable of spatially separating particle beams with different momenta, even when the phase space is large, as is characteristic of secondary beams. 
For the FRS \cite{Geissel:1991zn} at GSI and the superconducting fragment separator (Super-FRS) \cite{Geissel:2003lcy} at FAIR, this capability even applies to relativistic, multi-GeV beams. 
Both the FRS and Super-FRS, shown in Fig.\,\ref{fig:ch9-FRS-SFRS}, offer selected but globally unique opportunities for hadron physics experiments. 

The salient features of the FRS and Super-FRS include access to ions 
of all chemical elements (from hydrogen to uranium), exotic nuclei, as well as production of $\pi^\pm$ and $\overline{p}$, with a momentum resolving power of $p/\Delta p = 1500$. 
Furthermore, both separators provide dispersion matching, momentum spectroscopy with a resolution of $\delta p/p \approx 10^{-4}$, as well as missing-mass spectroscopy and selectivity for dedicated reaction channels that enhance signal-to-background ratios. 
Compared with the FRS, the Super-FRS introduces several new features: 
super-conducting magnets with a higher magnetic rigidity of 20\,T$\cdot$m, increased aperture and acceptance (both longitudinal and transverse), 
preseparation of secondary beams to improve beam purity, 
and enhanced ion-optical flexibility owing to multi-stage operation. Additionally, the SIS100 delivers primary beams with higher intensity and energy.

\paragraph{The antiproton production scheme.}
The antiproton production chain will commence at the planned proton linac (pLinac), which will accelerate an intense proton beam to 70\,MeV. 
This beam will be further boosted to 4\,GeV by SIS18 and subsequently to 29\,GeV by SIS100. 
At SIS100, beam bunches containing up to 2.5\,$\times$\,10$^{13}$ protons will be compressed to 50\,ns. 
Antiprotons will then be produced via inelastic collisions of the 29\,GeV protons with nucleons in a metallic antiproton production target. 
A pulsed magnetic horn will focus the resulting antiprotons towards the $\overline{\text{p}}$ separator, which selects antiprotons with a momentum of approximately 3\,GeV and transports them to the collector ring (CR).

Rapid cooling in the CR enables efficient collection of antiproton bunches, which are then injected into the high energy storage ring (HESR) for accumulation, stochastic cooling, and post-acceleration. 
The HESR is a slow-ramping synchrotron and storage ring with a maximum rigidity of 50\,T$\cdot$m and a circumference of 575\,m. 
The beam kinetic energy can be varied between 3\,GeV and 14.5\,GeV. 
A single interaction region in the HESR will host an internal target, surrounded by a large detector system, namely, an antiproton annihilation experiment at HESR in form of a multi-purpose detector, such as a modernised version based on the design proposed in Ref.\,\cite{Barucca2021PANDA}.
Two operational modes are foreseen for the HESR: a high-luminosity mode, with peak luminosities of up to 2$\times$10$^{32}$\,cm$^{-2}$\,s$^{-1}$, and a high-resolution mode, with a relative momentum spread of approximately 5$\times$10$^{-5}$. 
These capabilities will allow measurement of hadronic state masses and widths with an accuracy of $50–100\,$keV.

\subsection{FAIR computing facility}
\label{sec:greencube}

Scientific computing for FAIR research will use a centrally orchestrated infrastructure built on the existing Green-IT Cube \cite{GSI_GreenITCube, Kollegger2015_GreenCube_FAIR_Tier0}, which will serve as Tier-0 and handle the vast majority of the computing workload. 
The Green-IT Cube is an operational computing centre located on the GSI campus and is internationally recognised for its eco-friendly design \cite{BlueAngel_GreenITCube}. 
Conceptually~\cite{messchendorp2025conceptualdesignreportfair}, it optimises the sharing of computing resources across the various research pillars and between online and offline processing, thereby minimising costs and power consumption while respecting and incorporating the needs of the different research groups. 
Accordingly, no dedicated online clusters for individual experiments are envisaged; instead, resources will be shared within a central system. Additional large-scale computing infrastructures, particularly those governed by FAIR's international partners, will be part of a federated computing network, following the F.A.I.R. (Findable, Accessible, Interoperable, Reusable) principles.

The compute facility will provide about 2.2\,MHEPSpec06 of capacity to accommodate both online and offline processing in the phase when SIS100 becomes available.
(One physical core of an Intel E5-2680v4@2.4\,GHz processor corresponds to $\sim$22\,HEPSpec06 \cite{gridpp-hepspec06}.)
Here, online data processing alone already amounts to 1.1~MHEPSpec06. 
For offline activities, primarily the production of derived data, higher-level analyses, Monte Carlo studies, and theoretical calculations, a continuously available capacity of about 1.9\,MHEPSpec06 will be foreseen. Looking ahead, an increase in capacity by roughly a factor of two is anticipated. 
These substantial compute requirements go hand in hand with a significant increase in data volumes compared with current needs at GSI. 
This is driven by the high data rates of the large-scale experiments, producing up to $\sim$30\,PB/year of raw data. In addition, the associated Monte Carlo and other simulated datasets required to determine efficiencies, control systematic uncertainties, and interpret the acquired data are of a comparable order of magnitude ($\sim$15\,PB/year).

FAIR computing explores the usage of new methods to reduce costs and/or enhance the physics output from harvested data. 
A promising avenue is the application of machine-learning (ML) and artificial-intelligence (AI) techniques. 
In particular, the use of ML algorithms to accelerate simulations; the ambition to develop smart experiment and accelerator control to optimise beam time and streaming-readout processes; potential incorporation of techniques to improve tracking and particle identification into the computing models; and the use of AI in (open-science) data management schemes. 
On the hardware side, significant performance gains are foreseen by deploying data-processing algorithms on alternative architectures, such as accelerator cards (GPUs) and data-acquisition-specific field-programmable gate arrays (FPGAs). 
Exploratory activities in this direction are ongoing at GSI within several collaborations, including CBM in the context of the miniCBM demonstrator test facility. 
Explorations within open-science-driven and digitally innovative projects and programmes at European and national levels are continuing, in order to further enhance the physics reach.

\subsection{The HADES experiment at SIS18} 
\label{sec.hades}

HADES is a versatile magnetic spectrometer designed to study dielectron production in pion-, proton-, deuteron-, and heavy-ion–induced reactions. Its main features include a ring-imaging gas Cherenkov detector for electron–hadron discrimination, a tracking system composed of six super-conducting coils producing a toroidal magnetic field, drift chambers, and a time-of-flight array for further electron–hadron discrimination and a forward spectator detector for event characterisation. 

The HADES physics programme focuses on investigating hadron properties in nuclei as well as in hot and dense QCD matter. The detector system provides 85\% azimuthal coverage over a polar angle range from 18$^{\circ}$ to 85$^{\circ}$, a single-electron efficiency of 70\%, and a vector meson mass resolution of $\Delta M / M < \SI{2.5}{\percent}$. 
Identification of charged pions, kaons, and protons is achieved by combining time-of-flight and energy-loss measurements over a broad momentum range ($0.1 < p < 1$~GeV/$c$) \cite{GSI:HADES-EPJA2009}.

\begin{figure}[t]
\begin{center}
\includegraphics[width=0.80\linewidth,angle=0]{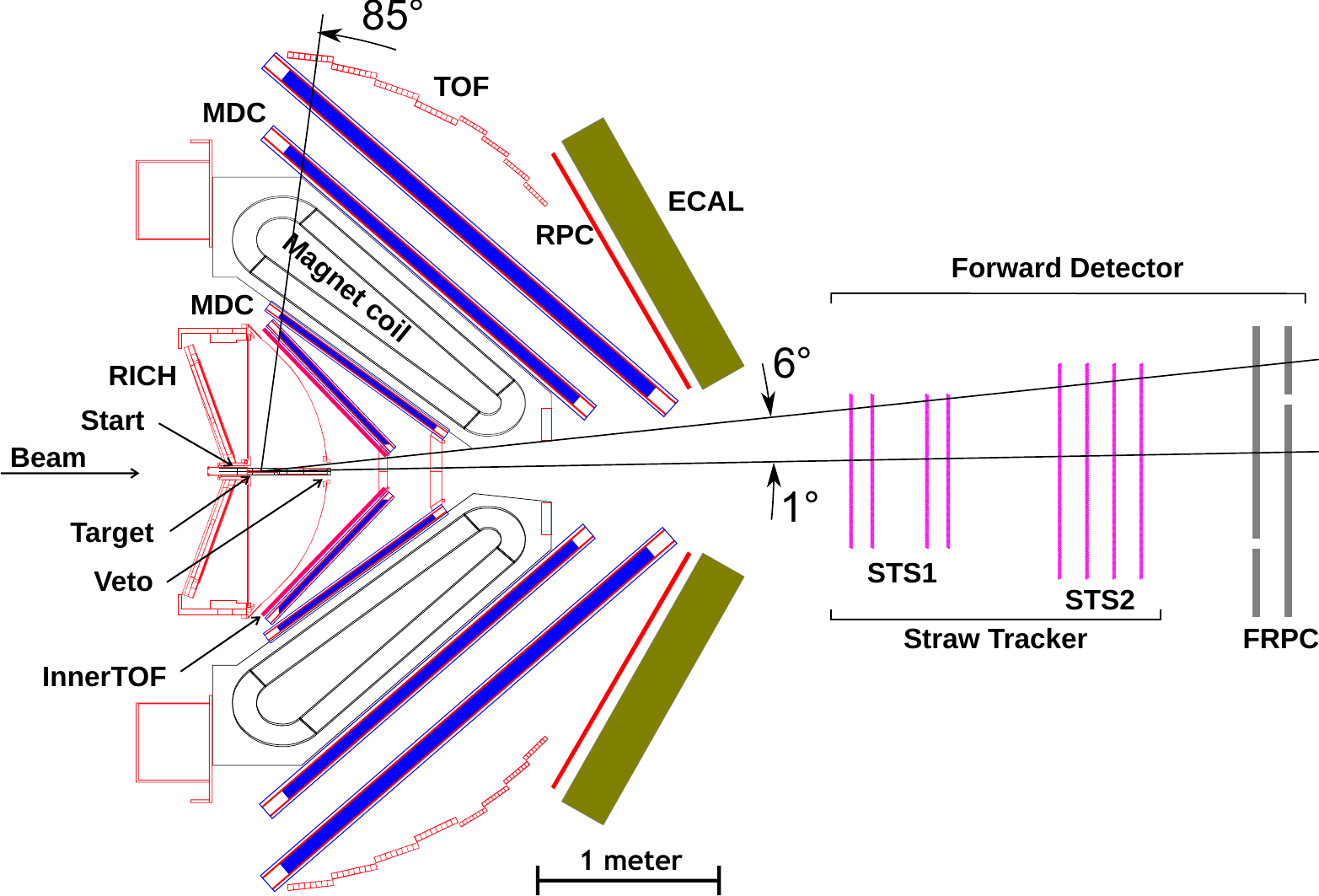}
\end{center}
\vspace*{-0.5cm}
\caption{Schematic cross-sectional view of the HADES spectrometer in the elementary beam configuration.}
\label{fig:HADES-FD}
\end{figure}

The HADES spectrometer has recently undergone significant upgrades, including an increase in the photoefficiency of the ring imaging Cherenkov (RICH) detector ($\gtrsim$ 20 photoelectrons per electron ring), an enhanced data acquisition (DAQ) rate (currently 50\,kHz trigger rate), the implementation of an electromagnetic calorimeter, and the addition of a forward detector (FD) based on PANDA straw tube technology instrumenting HADES particle detection in the forward region. 
A schematic overview of the upgraded HADES setup is shown in Fig.\,\ref{fig:HADES-FD}. 
The FD extends the acceptance to cover the forward region with polar angles from $1^{\circ}$ to approximately $6^{\circ}$. 
This phase-space region is particularly important, {\it e.g.}, for the exclusive reconstruction of hyperon channels, in which the daughter baryon has a high probability of being emitted at low polar angles.

The FD consists of two straw tracking stations (STS1, STS2), each comprising four double-layers of 1\,cm diameter straw tubes located at 3.3\,m and 4.6\,m downstream of the target, respectively. 
These are based on the design developed for the PANDA experiment \cite{PANDA:STT-TDR,HADES:STS1,HADES:STS2}. 
Additionally, a forward resistive plate chamber (FRPC) for time-of-flight (ToF) measurements is placed 5.7\,m downstream of the target. 
An inner time-of-flight detector (iTOF), a $3\times6$ segmented 5\,mm thick plastic scintillator, has also been installed between the RICH and the first MDC chamber. 
The iTOF provides event multiplicity information for triggering and an additional ToF measurement for particle identification. Further details can be found in Ref.\,\cite{HADES:2022vus}.

The liquid hydrogen (LH$_2$) target is a 5\,cm long, 2.5\,cm diameter cylindrical vessel operated at atmospheric pressure and 20\,K. It is enclosed by a 100\,$\mu$m Mylar foil, with an additional 100\,$\mu$m foil located 6\,mm downstream of the target. A proton beam has a nuclear interaction probability of 0.7\% within the target material.

The maximum proton beam intensity is limited to $7.5 \times 10^7$\,p/s owing to the constraints of the HADES start detector. 
Combined with the LH$_2$ target, this results in a maximum average luminosity of $1.5 \times 10^{31}$\,cm$^{-2}$s$^{-1}$. 
An ongoing upgrade of the DAQ system will support operation rates up to 200\,kHz. 
Moreover, replacing the LH$_2$ target with a polyethylene (PE) target of equal length could yield up to a factor of seven increase in luminosity at the same beam current.

The large acceptance and high resolution of HADES enables precision measurements of both inclusive and exclusive final states, as discussed in previous sections. 
Current analyses from recent proton–proton data at $T=4.5$\,GeV reveal strong signals for a broad range of final states. 
In the meson sector, this includes $\pi^0$, $\eta \to (\gamma\gamma,\ \gamma e^+e^-,\ \pi^+\pi^-\pi^0,\ \pi^+\pi^-e^+e^-)$, $\rho$, $\omega$, $\phi$, as well as $K$, $K^0_S$, and $K^*$. 
Numerous $N^\ast$ and $\Delta^\ast$ baryon resonances are observed up to $M \lesssim 2$\,GeV. 
Additionally, hyperon states are identified, including $\Lambda$, $\Sigma^0 \to (\Lambda \gamma,\ \Lambda e^+e^-)$, $\Sigma^\pm$, $\Sigma^{\pm,0}(1385)$, $\Lambda(1405) \to (\Sigma^0\pi^0,\ p K^-)$, and $\Lambda(1520) \to (p K^-,\ \Lambda\pi^+\pi^-)$.

\subsection{(S)FRS in combination with the WASA central detector} 
\label{sec.frs-wasa}
\begin{figure}[t]
\begin{center}
\includegraphics[width=1.0\linewidth,angle=0]{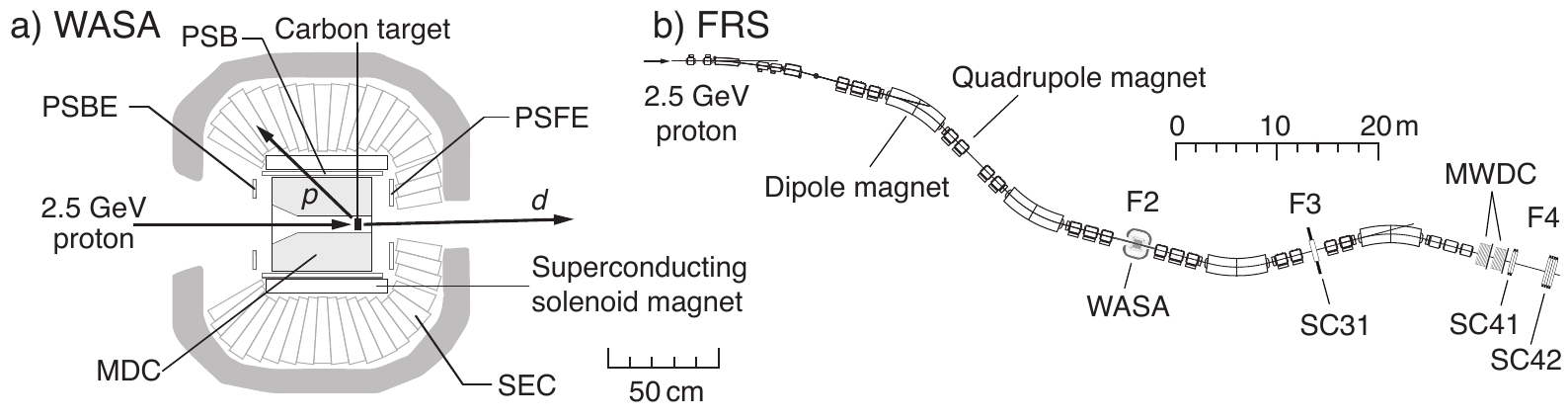}
\end{center}
\vspace*{-0.7cm}
\caption{The WASA detector setup 
(a) at the fragment separator 
(b) was used for the experiment searching for 
$\eta^\prime$-mesic nuclei. 
A 2.5\,GeV proton beam was directed onto a $^{12}$C target for the $(p, d)$ reaction. 
The emitted deuteron was momentum-analysed in the F2–F4 section. 
A set of two MWDCs was installed at F4, 
a plastic scintillation counter (SC31) at F3, 
and two counters (SC41, SC42) at F4. 
WASA consists of a super-conducting solenoid magnet, a cylindrical tracking detector (MDC), plastic scintillation counters (PSB, PSBE, and PSFE), and an EM calorimeter (SEC).
\label{fig:ch9-WASA-FRS}}
\end{figure}

%
A dedicated programme at the FRS addresses specific topics at the borderline of nuclear and hadron physics. 
This includes sub threshold antiproton  production \cite{Gillitzer:1992jj, Schroeter:1993ps},
the discovery and study of deeply bound pionic 
states \cite{Geissel:2002ur, Suzuki:2002ae},
the search for $\eta^\prime$-mesic nuclei \cite{n-PRiMESuper-FRS:2016vbn, sekiya2025excitationspectra12rmcpd},
the study of nucleon resonances in asymmetric nuclear matter \cite{Super-FRS:2020grv}, 
and the production and study of hypernuclei \cite{Saito:2021gao}. 
In a recent series of FRS experiments, the WASA central detector was 
integrated into the central focal plane of the FRS, see Fig.\,\ref{fig:ch9-WASA-FRS}, 
further enhancing the scientific opportunities.

This combination opens up possibilities for new experiments involving high-resolution spectroscopy at forward angles, \textit{i.e.}, at 0$^\circ$, and measurements of light decay particles with nearly full solid-angle acceptance in coincidence. 
The first series of WASA–FRS experiments was successfully carried out in 2022, aiming to produce and study light hypernuclei and to search for $\eta^\prime$-mesic nuclei.

The WASA–FRS (and later, Super-WASA@Super-FRS) combines a segmented 4$\pi$ open-geometry array and calorimeter with a high-resolution magnetic separator-spectrometer for reaction studies of relativistic particles: the first stage of the FRS (or Super-FRS) can serve as an analyser (\textit{e.g}., for fragment selection), while the second stage acts as a spectrometer for high-resolution momentum spectroscopy ($\delta p/p \approx 10^{-4}$). 
In the case of hypernuclei studies, this allows for missing-mass spectroscopy with a resolution on the order of 1\,MeV; alternatively, the second stage can serve as a selective trigger.

For instance, the search for $\eta^\prime$-mesons bound to nuclei was performed via missing-mass spectroscopy employing the $^{12}$C($p$,\,$d$) reaction near the $\eta^\prime$ production threshold. 
Here, the outgoing deuteron was selected and identified with the FRS, and only those events were recorded in which light particles (such as protons emitted in the decay of the $\eta^\prime$-mesic nuclei) were simultaneously detected in coincidence with the WASA detector. 
This significantly reduces background and enhances experimental sensitivity.

The hypernuclei studies used 2\,GeV/u beams of $^6$Li and $^{12}$C on a $^{12}$C target, with the kinematics chosen such that the hypernuclei could be identified via their decay into backward-emitted pions (detected with WASA) and forward particles such as $d$, $^3$He, and $^4$He, which were momentum-analysed with the FRS.

It is also noteworthy that the combination of a $\sim 4\pi$ detector and Super-FRS will enable the use of secondary hadron beams, such as pions and antiprotons, to tag decay particles from rare events, like the formation of $\eta^\prime$-mesic nuclei, and to perform missing-mass spectroscopy of nuclear reactions. 
According to the Sanford–Wang empirical formula, the production rate of a pion beam will be about $10^8$/s using a 30\,GeV proton beam with an intensity of $10^{12}$/s and a 40~g/cm$^2$ Be (equivalent) target. 
For $\pi^+$ beams, new optical settings of the Super-FRS may need to be developed based on the B$\rho$-$\Delta E$-B$\rho$ mode to reduce proton contamination.

Currently, considerations are underway for forthcoming experiments using a new, more compact setup, and several Letters of Intent have been submitted to the latest G-PAC.

\subsection{The CBM experiment at SIS100}
\label{sec.cbm}
\begin{figure}[t]
\begin{center}
\includegraphics[width=0.98\linewidth]{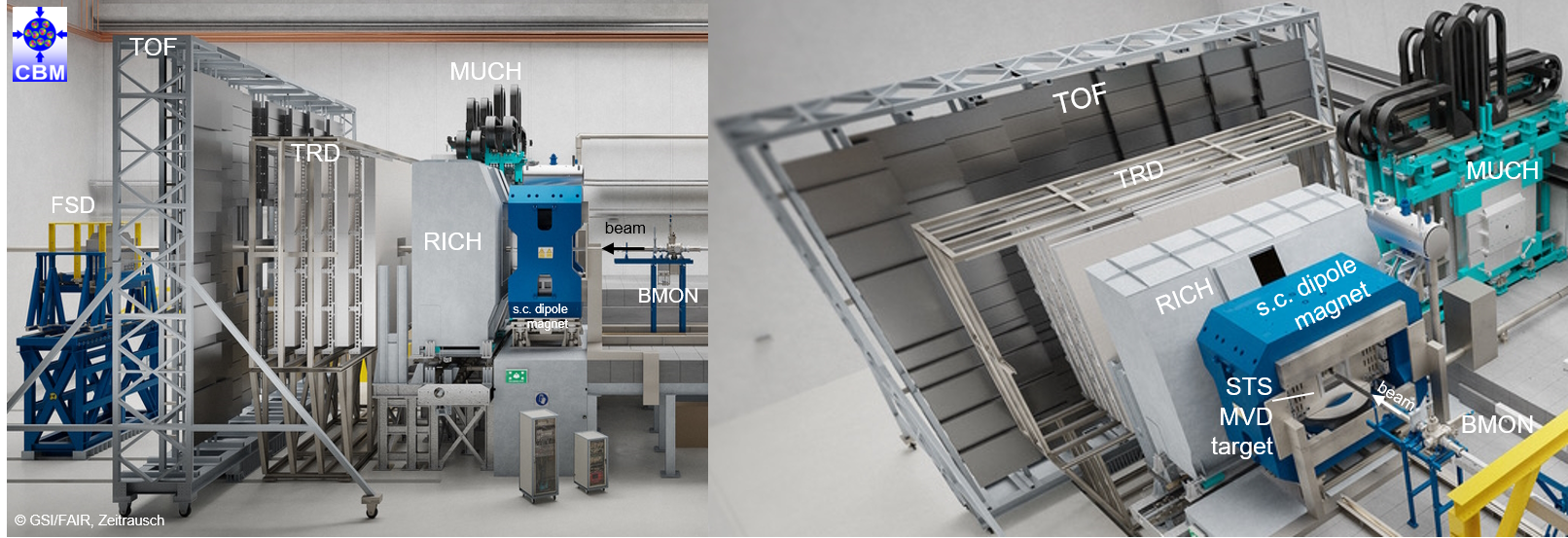}
\end{center}
\vspace*{-0.5cm}
\caption{The CBM setup inside the CBM cave at SIS100; the beam enters from the right. 
Shown is the configuration dedicated to electron identification, with a RICH detector located directly behind 
the superconducting dipole magnet, whereas the MuCH, containing massive hadron absorbers, is shifted to the parking position.
\label{fig:CBM}}
\end{figure}

Figure~\ref{fig:CBM} shows the CBM setup at SIS100. It is designed as a high-rate, multi-purpose 
fixed-target experiment capable of handling interaction rates of up to 10\,MHz 
\cite{CBM:PhyProg2017,CBM:PhyProg2026,CBM:Website}. 
It supports two detector configurations tailored to different physics cases: a setup with a RICH 
and transition radiation detector (TRD) for electron identification in hadron and dielectron measurements 
(depicted in Fig.\,\ref{fig:CBM}), and a configuration with a muon chamber (MuCh) system for muon detection. 
Both share a common tracking and are selected based on the final states of interest.

The identification of rare probes, such as charm hadrons and thermal dileptons, requires efficient background suppression 
and high interaction rate capability. To enhance the detection of multistrange hyperons and multi-strange 
(and charm) hypernuclei, an online event selection is required. 
CBM employs a triggerless data acquisition scheme based on self-triggered readout electronics and data transport \cite{CBM:TDR-DAQ-2022}, 
fast reconstruction algorithms, and a high-throughput computing system. Radiation-hard detectors and real-time data processing 
are essential for handling the extreme interaction rates.

The SIS100 beam delivered to the CBM cave reaches average intensities of up to $10^{12}$ protons or $10^{9}$ Au ions per second. 
Key quality parameters include: 
(\textit{i}) a beam halo fraction below $10^{-5}$ for the integrated intensity beyond \SI{5}{\milli\metre} from the beam axis; 
(\textit{ii}) spill structure fluctuations below \SI{50}{\percent}, down to the nanosecond scale; 
and (\textit{iii}) a transverse emittance (4$\sigma$) of $1 \times 0.6$\,mm\,mrad, as well as 
a momentum spread (2$\sigma$) of $5 \times 10^{-4}$.




%

\subsubsection{Setup}
Charged-particle tracking is centred around a \textbf{superconducting dipole magnet} that provides a field integral 
of \SI{1}{\tesla\meter}, enabling a momentum resolution of $\Delta p/p < \SI{2}{\percent}$. 
The magnet features an H-type design with a warm iron yoke and cylindrical superconducting Nb-Ti coils, 
which operate at \SI{686}{\ampere} and generate a maximum field of \SI{3.25}{\tesla}. 
The magnet gap accommodates the full tracking system, providing a vertical acceptance of \SI{\pm25}{\degree} 
and a horizontal acceptance of \SI{\pm30}{\degree}.

The first detector downstream of the target is the micro-vertex detector (\textbf{MVD}), which offers high spatial resolution 
and minimal material budget for reconstructing displaced decay vertices. It consists of four planar stations equipped with thin, 
large-area monolithic active pixel sensors (MAPS). The detector geometry can be configured for either vertexing or tracking, 
with station positions ranging from 5 to 20\,\si{\centi\meter} downstream of the target in vacuum. 
In the vertexing configuration, a resolution of \SIrange{50}{100}{\micro\meter} along the beam axis is expected.

Downstream of the MVD, the silicon tracking system (\textbf{STS}) serves as the central tracking detector. 
It consists of eight layers of double-sided silicon micro-strip sensors mounted on lightweight supports and read out 
via self-triggering electronics located at the detector periphery. 
Together with the MVD, the STS reconstructs charged-particle trajectories within the magnetic field over a distance 
of about \SI{1}{\meter}.

Electron identification is provided by a RICH detector located directly behind the dipole magnet, 
approximately \SI{1.6}{\meter} downstream of the target. 
The RICH uses a \SI{1.7}{\meter} long gaseous radiator, focusing mirrors, and multianode photomultiplier tubes (MAPMTs) 
for efficient detection of Cherenkov photons with high granularity and sensitivity down to the near-UV region. 
Additional electron identification and tracking at high momenta is provided by the TRD, which targets $e^\pm$ 
with $p > \SI{1.0}{\giga\electronvolt\per\textit{c}}$ ($\gamma \gtrsim 1000$). 
It consists of four layers grouped into a single station, located \SIrange{4.1}{6.2}{\meter} downstream of the target. 
The TRD covers an active area of about \SI{114}{\square\meter} and uses rectangular pad readout with a resolution of 
approximately \SI{300}{\micro\meter} across and \SIrange{3}{30}{\milli\meter} along the pad. Every second layer 
is rotated by \ang{90}, and a 2D readout design is foreseen in the innermost region to enhance low-$p_T$ tracking 
and $e/\pi$ separation in combination with the RICH.

For muon identification, the MuCH combines hadron absorbers with gaseous tracking detectors. Positioned downstream of the STS, 
it identifies muons by filtering hadrons in four iron absorbers interleaved with 12 tracking layers arranged in triplets. 
An optional fifth absorber can further suppress background from hadronic decays. 
MuCH is optimised for the detection of muons from vector meson and charmonium decays, such as $J/\psi \rightarrow \mu^+ \mu^-$.

Hadron identification is achieved via a large-area TOF wall composed of multi-gap resistive plate chambers (MRPCs). 
The TOF system covers approximately \SI{100}{\square\meter} and is flexibly positioned \SIrange{6}{10}{\meter} downstream 
of the target to optimise particle separation over the full SIS100 energy range. 
It provides a time resolution of about \SI{80}{\pico\second}, enabling efficient $\pi/K$ and $K/p$ separation.

Additional timing information, as well as beam diagnostics, is provided by the beam monitor (BMON) system, positioned upstream 
of the target. It consists of two diamond-based detector stations: the Time-0 station delivers the reaction start time 
with a precision of about \SI{50}{\pico\second} at moderate rates, while the HALO station (diamond or low-gain-avalanche-diodes based) 
monitors beam conditions such as the beam halo profile.

For event characterisation in nucleus-nucleus collisions, the forward spectator detector (FSD) detects forward-going, 
noninteracting projectile nucleons to determine collision centrality and reaction plane orientation. An extension of the FSD 
to include neutron detection, {\it e.g.}, via a neutron calorimeter, is foreseen as a possible option to enhance its capabilities.

\subsubsection{Optimisations for proton beam measurements}
\label{subsec.SetupModifications}

In the initial phase of operation, only minimal modifications of the CBM setup will be performed in order to maximise synergies with the heavy-ion collision experiments at CBM and bring the base version of CBM into full operation as soon as possible. 
Nevertheless, certain minimal modifications will be required at an early stage to enable measurements of exclusive final states from proton-proton reactions. These modifications will affect the following items, listed in order from most upstream to most downstream:

The much higher beam current (measured in particles per second) for the proton beam compared to the heavy-ion beam will severely compromise the information from the currently foreseen start detector. 
As a consequence, the start detector will need to be retracted, and the event time will be reconstructed via an overall event hypothesis. 
For this procedure, the measured stop time, path length, and momentum for each track are used to calculate an assumed event start time for a given mass hypothesis. 
The choice of tested mass hypotheses will be restricted by the sign of the curvature of the track in the magnetic field and information from the particle identification detectors, {\it e.g.}, RICH, MuCH and/or TRD. 
The joint assignment of mass hypotheses to tracks in an event that gives the most consistent determination of the assumed start time is then selected. 
This method has successfully been employed at, {\it e.g.}, the HADES experiment.

Implementation of a liquid hydrogen (LH$_2$) target at CBM will require significant modifications to the region surrounding the primary vertex. Thus, for initial measurements, it is foreseen to include a CH$_2$ target in the base CBM target chamber. 
The LH$_2$ target will then be implemented after the smooth running of the base version of CBM has been established.

The current design of the CBM beam pipe is based on the concept that the dipole magnet is always operated at maximum field for all beam momenta. This concept results in the need for a significantly higher material budget at small polar angles to enable the beam pipe to be moved to account for the different deflection of the primary, unscattered beam at different beam rigidity. 
Since this phase-space region is very important for the measurement of exclusive events, it is being considered to replace the central beam pipe for proton-beam experiments. 
In this case, the dipole field will be scaled with the beam momentum to ensure nearly constant deflection of the primary beam, and then the bellows that enable the beam pipe to swivel will be replaced with a fixed, low-mass construction to connect the target region with the downstream beam pipe. This change will also be of significant operational advantage when performing excitation function studies that entail frequent modification of the beam energy.

CBM will include the newly designed FSD, which is an array of 5\,cm thick scintillating detectors located about 10\,m downstream of the target. 
This system will be decisive in reconstructing fast forward-emitted particles, such as those emitted in elastic events as well as inelastic exclusive events with low momentum transfer. 
Moreover, this system will measure the quasifree spectator in deuteron-induced reactions, which provide, {\it e.g.}, access to neutron-proton reactions to investigate isospin effects in the production of many of the hadronic systems described above. 
Furthermore, this system contributes significantly to the overall acceptance of CBM to measure hyperon decays, in which the daughter baryon is generally emitted at low polar angles. 
Directly downstream of the FSD, it will be supplemented by a neutron calorimeter (NCAL) consisting of about 100 elements of 45\,cm thick plastic detectors. This detector, previously used by the COSY-TOF collaboration, has about a 40\% neutron detection efficiency and will enable investigations of exclusive events including a forward-emitted neutron, as well as supplement the event characterisation of heavy-ion reactions.

\newpage

\section{International context}
\label{sec.IntContext}
{\small {\bf Convenors:} \it J.~Ritman, C.~Sturm} 

\noindent The physics programme outlined in this document is embedded in a global effort to deepen our understanding of hadron structure and interactions, the emergence of mass, spin, and the physics of dense gluonic systems.
Numerous international facilities contribute complementary or partially overlapping capabilities to explore the topics discussed above, 
and new facilities are either under construction or discussion; see, \textit{e.g}., Refs.\,\cite{AbdulKhalek:2021gbh, Anderle:2021wcy, Arrington:2021biu, Accardi:2023chb}.
To place the GSI/FAIR programme in a broader perspective, this section provides a concise overview of selected major initiatives in the field.

\subsection{Comparative overview of accelerator facilities}
\label{subsec.Fac}
There are numerous facilities worldwide that operate accelerator infrastructure \cite{List:accelerators}. 
However, only a subset of them is directly relevant to the hadron physics topics addressed in this document. Therefore, the comparison of selected performance parameters provided in Table~\ref{tab:ch9-Overview-Facilities} is restricted to those facilities that are either in operation or soon will be. 

\begin{table}[htb]
\centering
\begin{scriptsize}
\begin{tabular}{ | l |c|c|c|c|c| }
\hline
\bf {Facility} & \bf{Beam Type} & \bf{$P_{beam}$ / $\sqrt{s}$} & \bf{Intensity / Luminosity} & \bf{Operational} \\
\hline
BEPC-BESIII         &   $e^+e^-$     & 2–5 GeV              & $10^{33}$/cm$^2$/s          & yes   \\
CERN-LHC            &   $pp$         & 0.9–14 TeV           & $2\times10^{33}$/cm$^2$/s   & yes   \\
CERN-SPS North Area &   $\mu^\pm$    & 20–380 GeV           & $1\times10^8$/s             & yes   \\
                    &   $\pi^\pm$    & 10–380 GeV           & $5\times10^8$/s           & yes   \\
                    &   $K^\pm$      & 10–380 GeV           & $2.5\times10^7$/s           & yes   \\
                    &   $p$ primary          & 400-450 GeV           &  up to $10^{13}$/s       & yes   \\
                    &   $p$ secondary          & 10–380 GeV           & $5\times10^{8}$/s       & yes   \\
CERN-PS East Area   &   $\mu^\pm$    & 0.1–15 GeV           & $1\times10^4$/s             & yes   \\
                    &   $\pi^\pm$    & 0.1–15 GeV           & $1\times10^6$/s           & yes   \\
                    &   $K^\pm$      & 0.1–15 GeV           & $2\times10^4$/s           & yes   \\
                    &   $p$ primary          & 2-24 GeV           &  up to $10^{11}$/s       & yes   \\
                    &   $p$ secondary          & 0.1–15 GeV           & $1\times10^{6}$/s       & yes   \\                    
ELSA                &   $\gamma$     & 0.2–3.1 GeV/$c$        & $10^7$–$10^8$/s             & yes   \\
                    &   $e$          & 0.8–3.2 GeV/$c$        & 0.1 nA                      & yes   \\
FAIR-SIS100         &   $p$          & 5–30 GeV/$c$           & $10^{13}$/s                 & 2028  \\
GSI-SIS18           &   $\pi^-$      & 0.5–2 GeV/$c$          & $10^6$/s                    & yes   \\
                    &   $p$          & 0.6–6 GeV/$c$          & $10^{11}$/s                 & yes   \\
HIAF                &   $p$          & 11 GeV/$c$             & $6\times10^{12}$/s          & 2025  \\
JLab GlueX          &   $\gamma$     & 7–12 GeV/$c$           & $10^8$/s                    & yes   \\
JLab Halls A/B/C    &   $e$          & 11 GeV/$c$             & $10^{32}$–$10^{35}$/cm$^2$/s& yes   \\
JLab KLF            &   $K^0_L$      & 0–10 GeV/$c$           & $10^4$/s                    & 2028  \\
J-PARC K1.8         & $K^\pm/\pi^\pm$& $<$2.0 GeV/$c$         & $\sim10^6$ $K^-$/spill      & yes   \\
J-PARC K1.8BR       & $K^\pm/\pi^\pm$& $<$1.1 GeV/$c$         & $\sim5\times10^5$ $K^-$/spill & yes \\
J-PARC high-p       & $p$           & 31 GeV/$c$             & $10^{10}$/spill             & yes   \\
J-PARC $\pi 20$     & $\pi^\pm/K^\pm/\overline{p}/\mu^\pm$ & $<$31 GeV/c & $10^5$–$10^7$/spill & partly \\
J-PARC HIHR         & $\pi^\pm$     & $<$2 GeV/$c$           & $2\times10^8$/spill         & 2034  \\
J-PARC K10          & $\pi^\pm, K^\pm,\overline{p}$ & $<$10 GeV/$c$ & $2\times10^6$ $K^-$/spill & 2039  \\
MESA               & $e^-$         &  5-155 MeV/$c$ & 150 $\mu$A (EB), 10 mA (ERL) & 2026 \\
NICA                & $pp$          & $\leq$27 GeV              & $10^{32}$/cm$^2$/s          & 2026  \\
SuperKEKB Belle~II  & $e^+e^-$     & 7 GeV / 4 GeV         & $5\times10^{34}$/cm$^2$/s   & yes   \\
EIC                 & $e^-p/e^-\mathrm{ion} $  & $20-140$ GeV      & $10^{33}-10^{34}$/cm$^2$/s   & 2035   \\
\hline
\end{tabular}
\caption{Overview of beam types, energies and intensities/luminosities at selected accelerator facilities relevant for hadron physics, excluding possible heavy-ion beams.}
\label{tab:ch9-Overview-Facilities}
\end{scriptsize}
\end{table}


\subsubsection*{BEPC}
\label{subsec.BEPC}

The BEPCII electronpositron collider in Beijing operates in the 
$\sqrt{s} = 2.0$ to $5\,\text{GeV}$ energy range and serves the BESIII
detector, which investigates hadron spectroscopy, charm physics, and
light meson interactions in the tau-charm sector. 
With a peak luminosity of up to $1 \times 10^{33}\,\text{cm}^{-2}\,\text{s}^{-1}$, BESIII enables precision studies of charmonium and open-charm states, electromagnetic form factors, and light hadron decays \cite{Ablikim:2019hff}.
Among the large data samples already available are $10^{10}$ $J/\psi$ and $2.7\times 10^9$ $\psi(2S)$ events, and $20\;\text{fb}^{-1}$ collected on the $\psi(3770)$ resonance.

The facility has provided important data on scalar mesons, hyperons, the so-called XYZ states, and rare decays, offering valuable input to nonperturbative QCD studies. 
Future operational plans include the replacement of the inner part of the drift chamber and upgrades to the injecting LINAC, RF system, and other accelerator systems, enlarging the range in $\sqrt{s}$ up to $5.6\;\text{GeV}$ and extending the scientific reach of the experiment through 2030.

\subsubsection*{CERN}
\label{subsec.CERN}
The CERN accelerator complex provides diverse opportunities for hadron physics in both fixed-target and collider modes. The super proton synchrotron (SPS) delivers high-intensity secondary beams of pions, kaons, protons, and muons with momenta up to 400\,GeV/$c$, enabling precision studies of hadron structure, spectroscopy, and interactions. Key experiments include COMPASS, NA61/SHINE, and AMBER, which address fundamental questions related to confinement, the meson and baryon spectrum, and the structure of the nucleon, pion, and kaon in terms of parton distributions \cite{NA61:2014xbx, COMPASS:2022cda, AMBER:2023afc}.

The LHC, operating at CM energies up to $\sqrt{s}=14$~TeV for $pp$ collisions and $\sqrt{s_{\mathrm{NN}}}=5.02\,$TeV for Pb+Pb collisions, complements these efforts at much higher energies. 
ALICE, ATLAS, CMS, and LHCb all contribute to hadronic physics, exploring topics ranging from baryon production and parton energy loss in the QCD medium to the hadronisation of heavy quarks and exotic hadron searches \cite{ALICE:2022wpn, LHCb:2021bjt}.

Ongoing developments include the AMBER experiment as a successor to COMPASS and the implementation of fixed-target programmes at LHCb and ALICE using internal gas targets. 
These setups offer access to the backward rapidity region and enable hadron structure and spin studies at high momentum transfers \cite{Citron:2018lsq, LHCb:2021bjt}.

\subsubsection*{EIC}
\label{subsec.EIC}

The Electron-Ion Collider (EIC) is a new experimental facility under construction at Brookhaven National Lab in the USA~\cite{AbdulKhalek:2021gbh}.  
The EIC is a versatile, high-energy machine, colliding electron beams of energy $5-18$\,GeV with beams of either protons of energy $100-275$\,GeV or nuclear beams, ranging from deuterium to uranium, of energy up to 100\,GeV/$u$.  
These collisions cover a CM energy range of $20-140$\,GeV, with electron and proton beam polarisations of over 70\%.   
The expected luminosity is $10^{33}-10^{34}\,\mathrm{cm}^{-2}\,s^{-1}$, which is several orders of magnitude more than that reached at HERA.

The ePIC Collaboration was formed to build the first detector, which includes a solenoidal spectrometer with excellent tracking, calorimetry, and particle ID, as well as tagging of the far-forward scattered electrons and hadrons, with the ability to reconstruct all the particles produced in these interactions.  
Concepts for second detector at a different interaction point are being developed.  
The main physics programme is expected to begin around 2035.

The primary focus of EIC physics is understanding the quark-gluon structure of hadrons and atomic nuclei.
Key science questions to be addressed at EIC include 
how are the nucleon mass and spin generated by QCD and apportioned among its partonic constituents, 
what are the properties of dense systems of gluons, 
what is the internal structure of light pseudoscalar mesons and how does it differ from that of the nucleon, 
and how do quarks and gluons interact in the nuclear medium at these high energies.
With the excellent detector that is planned, the potential for a second, and expected increases in CM energy and luminosity, additional topics in hadron physics may also addressed.

\subsubsection*{ELSA}
\label{subsec.ELSA}
The future experiment \textsc{INSIGHT}@ELSA will enhance the present capabilities to study baryon resonances at the electron accelerator ELSA \cite{Hillert:2017nzr}. 
It will feature a unique combination of almost complete angular coverage for high-resolution photon measurements,
charged-particle detection, 
and the ability to perform measurements using a polarised beam and a polarised target. 
The measurement of single and double polarisation observables is an indispensable prerequisite for performing an unambiguous PWA to extract the resonances from the data. 
Non-strange ($N^\ast$, $\Delta^\ast$) and, in particular, strange baryon resonances ($\Lambda^\ast$, $\Sigma^\ast$) will be studied in photoproduction experiments off the proton and neutron, addressing a large variety of final states.



\subsubsection*{HIAF}
\label{subsec.HIAF}
The high intensity heavy-ion accelerator facility (HIAF) is a new scientific infrastructure under construction in Huizhou, China.
It is slated to go into operation by 2026 \cite{HIAF}.
HIAF aims to advance nuclear physics and astrophysics by producing and studying exotic nuclei, particularly short-lived, neutron-rich, and proton-rich isotopes, that are not typically found on Earth.  
The facility's design integrates a superconducting linac and a high-energy synchrotron with a maximum magnetic rigidity of 34\,T$\cdot$m, enabling the acceleration of ions up to approximately 4.2\,GeV/u, depending on the ion species.
HIAF is expected to achieve a maximum pulsed heavy-ion beam intensity of up to $10^{12}$ particles per pulse.
Several experimental setups will be installed in the high-energy beam experimental area.

\subsubsection*{JLab}
\label{subsec.JLAB}
The Continuous Electron Beam Facility (CEBAF) \cite{Leemann:2001, Adderley:2024}, located at the Thomas Jefferson National Accelerator Facility (Jefferson Lab), delivers electron and secondary photon beams of the world's highest intensities and spin polarisations, with energies up to 12\,GeV, to its four experimental halls.
The CEBAF large acceptance spectrometer at 12\,GeV (CLAS12) is located in Hall~B and aims to further the global understanding of hadron structure and QCD \cite{C12Overview, Achenbach:2025kfx}. 
The gluonic excitations experiment (GlueX) in Hall~D \cite{GlueX:2021} has been designed to study meson photoproduction reactions. 
Halls~A and C utilise a wide variety of detectors to perform measurements, generally at higher precision but smaller acceptance than Halls~B and D.

A recent upgrade project intends to develop, over the coming years, a high-duty-cycle, high-intensity, and high-polarisation positron beam to serve a unique experimental programme \cite{PWG:2022}. 
Furthermore, a proposal has been formulated to increase the CEBAF electron beam energy from the present 12\,GeV to above 20\,GeV~\cite{Accardi:2023chb}.

The planned K-long facility is located in Hall~D and generates a tertiary beam of $K_L$ mesons. 
This beam is aimed at a target of liquid hydrogen or deuterium inside the existing GlueX detector. 
The new $K_L$ beamline will use 12\,GeV electrons with a total current of 5\,$\mu$A directed onto the compact photon source (CPS), with the resulting photon beam aimed at a beryllium kaon production target (KPT). 
Charged particles are magnetically swept away, leaving a beam consisting mostly of $K_L$ mesons and neutrons. 
The beam of neutral kaons will cover a broad energy range up to approximately 10\,GeV.

\subsubsection*{J-PARC}
\label{subsec.JParc}

The J-PARC Hadron Experimental Facility delivers high-intensity beams of pions, kaons, and muons produced by a 30\,GeV proton beam. 
Its physics programme spans strangeness nuclear physics, hadron physics, and flavour physics, with a strong focus on understanding the formation and interactions of hadrons \cite{Ohnishi:2019cif}. 
The current hadron experimental facility operates four beam lines: a charged kaon beam line (K1.8/K1.8BR), a neutral kaon beam line (KL), a primary 30\,GeV proton beam line (high-p), and a primary 8\,GeV proton beam line for a muon-electron conversion experiment (COMET).

The K1.8 beam line has a maximum momentum of 2\,GeV/$c$ and is dedicated to hypernuclear studies, particularly the spectroscopy of $S = -2$ hypernuclei through $(K^-, K^+)$ reactions. 
The high-p beam line delivers intensities up to $10^{10}$ per pulse (2\,s pulse every 4.2\,s). Its primary focus is on vector meson in-medium studies, particularly $\phi$-meson spectral changes in nuclear matter through $p+A$ reactions. 
Secondary beam production up to 20\,GeV/$c$ is planned at high-p ($\pi$20 beam line), facilitating charmed baryon spectroscopy. 
Negative beam operation is expected to become available after Japanese Fiscal Year (JFY) 2025--26. An intensity upgrade for $10^7\,\pi$/spill is planned to be completed in JFY~2034.

The KL beam line utilises neutral kaon beams for precision measurements of the rare kaon decay $K_L \to \pi^0 \nu \bar{\nu}$, to search for beyond-SM physics. 
The COMET beam line delivers primary 8\,GeV protons to generate intense muon beams for the study of charged lepton flavour violation, {\it i.e.}, the $\mu \to e$ conversion process, aiming for unprecedented sensitivity.

A major extension is planned to double the area and add new beam lines, including the high-intensity high-resolution $\pi$ beam line (HIHR) for ultra-high-resolution $\Lambda$ hypernuclei spectroscopy, and the K10 beam line for multistrangeness baryon studies \cite{Aoki:2021cqa}. 
The current plan estimates a 6--7 year construction period, including a 3-year no-beam phase, with a tentative completion around JFY~2034, depending on budget availability.

\subsubsection*{MESA}

The Mainz energy-recovering superconducting accelerator (MESA), at Johannes Gutenberg University Mainz, is a next-generation electron linac that uses energy-recovery (ERL) technology to deliver extremely intense, low-energy beams while recycling much of the beam’s energy, dramatically improving efficiency. 
Designed to run in both external-beam and ERL modes, it will pioneer multiturn energy recovery in a superconducting environment and serve as a test bed for future large facilities. 
Scientific highlights center on precision SM tests and searches for new physics: the P2 experiment will measure the weak mixing angle via parity-violating electron scattering, while MAGIX (an internal-target spectrometer) and DarkMESA (a beam-dump setup) probe hadron structure, nuclear astrophysics, and light dark-sector particles. 
MESA is a flagship facility of the PRISMA+ Cluster of Excellence in Mainz, enabling a new generation of high-luminosity, low-energy experiments.

\subsubsection*{NICA}
\label{subsec.NICA}
A complementary programme to study the structure and properties of hadronic matter is planned for implementation at the Nuclotron-based ion collider facility (NICA), the construction of which continues at JINR-Dubna \cite{Kekelidze:2016hhw, Kekelidze:2017ual, Syresin:2021kiy}. 
NICA will provide a variety of heavy-ion beams up to Au$^{79+}$ with a kinetic energy up to 4.5\,GeV/u, as well as polarised proton and deuteron beams possessing a high degree of longitudinal or transverse polarisation and with total energy up to 13.5\,GeV.

The final goal of the SPD experiment \cite{SPD:2024gkq} is to provide access to the gluon TMDs in the proton and deuteron via the measurement of specific single and double spin asymmetries in the production of charmonia, open charm, and high-$p_T$ prompt photons \cite{Arbuzov:2020cqg}.



\subsubsection*{SuperKEKB}
\label{subsec.Belle2}
SuperKEKB, located at the high energy accelerator research organisation (KEK) in Tsukuba, Japan, is the world’s highest-luminosity electron-positron collider, operating with 7\,GeV electrons and 4\,GeV positrons.
Its target luminosity of $6 \times 10^{35}\,\text{cm}^{-2}\,\text{s}^{-1}$, about 30-times higher than that achieved at its predecessor, KEKB, enables precision measurements of weak interaction parameters, study of exotic hadrons, and searches for beyond-SM physics. 
SuperKEKB is mainly operated at around the $\Upsilon(4S)$ energy, which decays into a $B\bar{B}$ pair, hence it is known as a B-factory.

SuperKEKB and Belle II provide a unique platform for hadron physics: the background is low, heavy quarks are produced as abundantly as light quarks, and direct production of $J^{PC}=1^{--}$ mesons with hidden flavor, as well as production of other spin states via two-photon processes, is possible. In addition, $B$ meson decays preferably yield charmed hadrons, both open and hidden, leading to the discovery of $X(3872)$ in $B\rightarrow X(3872)K\rightarrow J/\psi \pi \pi$ at Belle. 
Belle II \cite{Belle-II:2018jsg} is planned to operate until around 2043, with the goal of collecting 50~ab$^{-1}$ of data, about 50-times more than Belle.



\begin{table}[t]
\begin{scriptsize}
\centering
\begin{tabular}{ | l ||c|c|c|c|c|c|c|c|c|c |c|c|c|c|c|c|c|c|c|c|c|  } 
\hline
{ \bf Topic \hfill \begin{sideways}{\hspace{5 mm} Facility}    \end{sideways} } & 
\begin{sideways}{BESIII}        \end{sideways} &
\begin{sideways}{LHCb}          \end{sideways} & 
\begin{sideways}{ALICE}         \end{sideways} & 
\begin{sideways}{AMBER}         \end{sideways} & 
\begin{sideways}{INSIGHT}       \end{sideways} & 
\begin{sideways}{HADES ($\pi$)} \end{sideways} & 
\begin{sideways}{CBM}           \end{sideways} & 
\begin{sideways}{WASA@SFRS}     \end{sideways} & 
\begin{sideways}{HIAF}          \end{sideways} & 
\begin{sideways}{Halls A/B/C}   \end{sideways} & 
\begin{sideways}{GlueX}         \end{sideways} & 
\begin{sideways}{KLF}           \end{sideways} & 
\begin{sideways}{J-PARC K1.8}   \end{sideways} & 
\begin{sideways}{J-PARC K1.8BR \hspace{2 mm}} \end{sideways} & 
\begin{sideways}{J-PARC high-p/$\pi20$ \hspace{1 mm}} \end{sideways} & 
\begin{sideways}{J-PARC HIHR}\end{sideways} & 
\begin{sideways}{J-PARC K10}    \end{sideways} & 
\begin{sideways}{BelleII}       \end{sideways} & 
\begin{sideways}{NICA}          \end{sideways} &
\begin{sideways}{MESA}          \end{sideways} &
\begin{sideways}{EIC}          \end{sideways} 
\\
\hline \hline
$N^*$, $\Delta^*$ spectroscopy           &\+&  &\+&  &\+&\+&\+&  &  &\+&\+&\+&\+&\+&\+&  &  &  &  &  &  \\ \hline
$|S|=1$ spectroscopy                     &\+&  &\+&  &\+&\+&\+&  &  &  &\+&\+&\+&\+&\+&  &\+&  &  &  &  \\ \hline
$|S|=2,3$ spectroscopy                   &\+&  &\+&  &  &  &\+&  &  &  &  &\+&\+&  &\+&  &\+&  &  &  &  \\ \hline
$C\ne 0$ spectroscopy                    &\+&\+&\+&\+&  &  &\+&  &  &  &  &  &  &  &\+&  &  &\+&  &  &  \\ \hline
Meson-Meson interactions                 &  &  &\+&\+&\+&\+&\+&  &  &  &  &  &  &  &  &  &  &  &  &  &  \\ \hline
Meson-Baryon interactions                &  &  &\+&\+&\+&\+&\+&  &  &\+&\+&\+&  &\+&\+&  &\+&  & &   &  \\ \hline
Baryon-Baryon interactions               &  &  &\+&  &  &\+&\+&  &\+&  &  &  &\+&\+&\+&\+&\+&  & \+ &\+&  \\ \hline
Dibaryons                                &  &  &\+&  &  &  &\+&  &  &  &  &  &\+&  &\+&  &\+&  & \+ &  &  \\ \hline
Hypernuclei                              &  &  &\+&  &  &\+&\+&\+&\+&  &  &\+&\+&\+&  &\+&\+&  & \+ &  &  \\ \hline
Mesic atoms                              &  &  &  &  &  &  &  &\+&  &  &  &  &  &\+&  &  &  &  &  &  &  \\ \hline
Baryon Transition Form Factors           &\+&  &  &\+&  &\+&\+&  &  &\+&\+&  &  &  &\+&  &  &\+&  &  &  \\ \hline
Semileptonic Baryonic transitions        &\+&  &  &  &  &  &\+&  &  &  &  &  &  &  &  &  &  &\+&  &  &  \\ \hline
Intrinsic Charm of Nucleon               &  &  &  &\+&  &  &\+&  &  &\+&  &  &  &  &\+&  &  &\+&   & &\+  \\ \hline
Exotic hadrons with charm                &\+&  &\+&\+&  &  &\+&  &  &  &  &  &  &  &\+&  &  &\+&  &  &\+  \\ \hline
Charmed hadron--Nucleon FSI              &  &\+&\+&  &  &  &\+&  &  &  &  &  &  &  &  &  &  &  &  &  &  \\ \hline
Light--quark exotics                     &\+&\+&  &\+&\+&  &\+&  &  &\+&\+&\+&\+&  &  &  &  &\+&  &  &  \\ \hline
Particle multiplicities in $pp$ reactions&  &  &  &  &  &\+&\+&  &  &  &  &  &  &  &  &  &  &  &   & &  \\ \hline
Resonance production with $\pi$ beam     &  &  &  &\+&  &\+&  &  &  &  &  &  &\+&\+&\+&  &  &  &  &  &  \\ \hline
In-medium hadron properties              &  &  &  &  &  &\+&\+&  &\+&  &  &  &  &  &\+&  &\+&  &\+ & &\+  \\ \hline
In-medium strange hadron properties      &  &  &  &  &  &\+&\+&  &  &  &  &  &\+&\+&\+&\+&\+&  & \+ &  &\+  \\ \hline
Short Range Correlations                 &  &  &  &  &  &\+&\+&  &\+&\+&  &  &\+&  &  &  &  &  &\+ & &\+  \\ \hline
Structure functions via $\pi$/p + A reactions      &  &  &  &\+&  &  &\+&  &  &  &  &  &  &  &\+&  &  &  & \+ &  & \+ \\ \hline
Isospin effects in dense matter          &  &  &  &  &  &\+&\+&  &\+&  &  &  &\+&\+&  &\+&  &  &\+ & &  \\ \hline
Dark matter searches                     &\+&\+&\+&\+&  &  &\+&  &  &  &\+&  &  &  &  &  &  &\+&  & \+ &   \\ \hline
\end{tabular}
\caption{Comparison of the physics topics for which the corresponding approved or proposed experiments expects to make significant contributions.}
\label{tab:ch9-expt-Phys-Matrix}
\end{scriptsize}
\end{table}

\subsection{Comparative overview of physics programmes}
\label{sec.competition}

All experimental setups at these accelerator facilities are designed to explore a broad spectrum of hadronic processes. 
However, each experiment has been optimised for different observables, resulting in varying resolutions, detection efficiencies, acceptances, and 
statistical sensitivities for the study of a given process. These differences make the programmes complementary.  
Where there is overlap, it guarantees constructive competition; and no facility is redundant because each has its particular strengths.

Table~\ref{tab:ch9-expt-Phys-Matrix} highlights some of the key areas where individual experiments are expected to provide significant advances.

\newpage

\section{Conclusions and outlook}
This document has been compiled through an international collaborative effort involving experimentalists and theorists working in, or connected to, the interdisciplinary fields of hadron, nuclear, and (nuclear) astro- and astroparticle physics. It serves to define a comprehensive hadron physics programme for the next decade at GSI/FAIR in Darmstadt, Germany.

Referring to the roadmap of the hadron physics programme at GSI and FAIR (Fig.~\ref{fig:roadmap}), the short-, medium-, and long-term perspectives can be structured into three strongly overlapping phases, primarily guided by the available accelerators. The primary and secondary beams -- offered across different energy ranges -- together with the availability of detector setups, determine the key physics questions that can feasibly be addressed. The comprehensive programme presented here places particular emphasis on measurements that utilise the SIS100 accelerator with proton beams, which is expected to become operational in the near future as part of FAIR Phase \textit{First Science +}, which is expected to begin in 2028.

The initiative and campaign remain open to further international collaborators, who are warmly encouraged to engage and contribute.  This invitation addresses both the international experimental and theoretical communities and extends to those working in areas such as hardware, software, and computational developments.

This newly established network within the global hadron physics community will continue to support the experimental programme well into the next decade and beyond.  It is prepared to exploit the opportunities that will arise with the future availability of antiprotons at GSI/FAIR.
\newpage
\section*{Acknowledgments}
\addcontentsline{toc}{section}{Acknowledgments}
The efforts of the authors of this document were supported (in part or entirely) by various funding agencies worldwide, as identified below.
%
F.N., J.G.M., J.R. \-- ExtreMe Matter Institute EMMI at GSI Helmholtzzentrum für Schwerionenforschung, Helmholtz Research Academy Hesse for FAIR (HFHF), and the Ministry of Culture and Science of the State of Northrhine Westphalia (Netzwerke 2021, NRW-FAIR) for the generous financial support for the QCD at FAIR workshops.
A.B.\ -- European Union ERC Synergy Grant HeavyMetal no.\ 101071865;
and Deutsche Forschungsgemeinschaft (DFG, German Research Foundation) through Project-ID 279384907 -- SFB 1245 (subproject B07).
M.A.\ -- Spanish Ministerio de Ciencia e Innovaci\'on (MICINN) under contracts PID2023-147458NB-C21 and CEX2023-001292-S; 
Generalitat Valenciana under contracts PROMETEO/2020/023 and  CIPROM/2023/59;
Ram\'on y Cajal program by MICINN grant no.\,RYC2022-038524-I;
and Atracci\'on de Talento programme by CSIC PIE 20245AT019.
V.C. and S.D.\ -- U.S. Department of Energy, Office of Science, Office of Nuclear Physics, under Contract No.\ DE-FG02-92ER40735.
B.D -- Bundesministerium f\"{u}r Bildung und Forschung through ErUM-FSP T01, F\"{o}rderkennzeichen 05P21RFCA1 and 05P21RFFC3.
G.E.\ -- Austrian Science Fund (FWF) under grant no.\ 10.55776/PAT2089624.
C.S.F.\ -- DFG under grant number FI 970/11-2.
F.K.G.\ -- National Natural Science Foundation of China (NSFC) under grant nos.~12125507, 12447101;
and Chinese Academy of Sciences (CAS) under grant no.\ YSBR-101.
C.H.\ -- CAS President's International Fellowship Initiative (PIFI) under Grant No. 2025PD0087; and MKW NRW under funding code NW21-024-A (also D.R.)
N.H.\ -- Helmholtz-Institute Mainz, Section SPECF.
P.H.\ -- United Kingdom’s Science and Technology Facilities Council grant no.\ ST/Y000315/1.
K.H.K.\ -- ``Netzwerke 2021'', an initative of the MKW NRW;
and Bundesministerium f\"{u}r Forschung, Technologie und Raumfahrt (BMFTR) Verbundforschung under grant no.\ 05P24PX1.
M.M.\ -- Heisenberg Programme by the Deutsche Forschungsgemeinschaft (DFG, German Research Foundation) – 532635001.
V.M.\ -- Serra H\'unter Professor; and Spanish national grant nos.\ PID2023-147112NB-C21, CNS2022-136085.
A.P.\ -- GeV-AI project (CUP I57G21000110007), in the context of the ICSC Spoke 2 Open Calls, project funded by European Union -- NextGenerationEU -- and National Recovery and Resilience Plan (NRRP) -- Mission 4 Component 2.
J.R.P.\ -- Spanish Grant PID2022-136510NB-C31 funded by MCIN/AEI/ 10.13039/501100011033;
and European Union Horizon 2020 research and innovation program under grant agreement no.\ 824093 (STRONG2020).
C.R.\ -- PID2022-140162NB-I00 and CNS2022-135768 funded by the Spanish MICIU/AEI/10.13039/501100011033/;
European Union NextGenerationEU/PRTR; 
and grants 2019-T1/TIC-13194 and 2023-5A/TIC-28925 (Atracci\'on de Talento Investigador of the Community of Madrid).
C.D.R.\ -- NSFC grant no.\ 12135007.
D.R.\ -- DFG through CRC 1639 NuMeriQS - 511713970.
%
T.R.\- Main-Campus-Doctus fellowship provided by the Stiftung Polytechnische Gesellschaft (SPTG) Frankfurt am Main; Samson AG; The Branco Weiss Fellowship - Society in Science, administered by the ETH Z\"urich. 
L.T.\ -- CEX2020-001058-M (Unidad de Excelencia ``Mar\'{\i}a de Maeztu'') and PID2022-139427NB-I00 financed by the Spanish MCIN/AEI/10.13039/501100011033/FEDER,UE;
Generalitat de Catalunya under contract 2021 SGR 171;
Generalitat Valenciana under contract CIPROM/2023/59;
and CRC-TR 211 ``Strong-interaction matter under extreme conditions''- project Nr. 315477589 - TRR 211. 
A.S.\ --  U.S. Department of Energy under Grant No.~DE-AC05-06OR23177, under which Jefferson Science Associates, LLC, manages and operates Jefferson Lab, and No.~DE-FG02-87ER40365. Furthermore, this work contributes to the aims of the U.S.\ Department of Energy \mbox{ExoHad} Topical Collaboration, contract \mbox{DE-SC0023598}. 
H.Z.\ -- National Science Centre of Poland grant nos. 2021/41/B/ST2/02409, 2020/38/E/ST2/00019;
and IDUB-POB projects granted by WUT (Excellence Initiative: Research University (ID-UB)).
G.W.\ -- Hungarian OTKA fund K138277.
S.L.\ -- Swedish Research Council (Vetenskapsr\aa det) (grant number 2019-04303).
J.M.T-R. \-- Project No. PID2023-147112NB-C21, financed by the Spanish MCIN/ AEI/10.13039/501100011033/; and
from Contract 2021 SGR 171 by the Generalitat de Catalunya. 
M.A. \--  Spanish Ministerio de Ciencia e Innovaci\'on (MICINN) under contracts PID2023-147458NB-C21 and CEX2023-001292-S; by Generalitat Valenciana under contracts PROMETEO/2020/023 and  CIPROM/2023/59, Ramón y Cajal program by MICINN Grant No.\,RYC2022-038524-I, and Atracción de Talento program by CSIC PIE 20245AT019.
N.B. \-- DFG cluster of excellence ORIGINS funded by the Deutsche Forschungsgemeinschaft under Germany’s Excellence Strategy-EXC-2094- 390783311; Advanced ERC grant ERC-2023-ADG-Project EFT- XYZ.
L.A. \-- National Key Research and Development Program of China under grant No. 2024YFA1610501.
A.K. \-- Polish National Science Centre through the grant 2024/53/B/ST2/00975.
K.B.M. \-- U.S. Department of Energy, Office of Science, Office under Award Number(s) DE-SC0015903.
L.F. \-- U.S. Department of Energy, Office of Science, Office under Award Numbers DE-SC0023197 and DE-SC0022529.

\newpage
\addcontentsline{toc}{section}{References}
\bibliographystyle{elsarticle-num} 
 \bibliography{references.bib}
%
%
%

\newpage
\section*{Glossary of abbreviations}
\addcontentsline{toc}{section}{Glossary of abbreviations}


\setlength{\LTpre}{4pt}   
\setlength{\LTpost}{6pt}  

\begin{longtable}{@{}ll@{}}
\toprule
\textbf{Abbreviation} & \textbf{Full form} \\
\midrule
\endfirsthead

\multicolumn{2}{l}{\small\textit{(continued from previous page)}}\\
\toprule
\textbf{Abbreviation} & \textbf{Full form} \\
\midrule
\endhead

\midrule
\multicolumn{2}{r}{\small\textit{(continued on next page)}}\\
\endfoot


\bottomrule
\endlastfoot
AA      & Nucleus–nucleus (A\,+\,A) collision \\
AD      & Antiproton Decelerator \\
AI      & Artificial Intelligence \\
ALADIN  & A Large Acceptance DIpole magNet \\
ALICE   & A Large Ion Collider Experiment \\
ALICE-PUBLIC & ALICE Public Data Releases \\
ALP     & Axion‑Like Particle \\
AMBER   & Apparatus for Meson And Baryon Experimental Research \\
AMS     & Alpha Magnetic Spectrometer \\
ANKE    & Apparatus for Studies of Nucleon and Kaon Ejectiles \\
ANL     & Argonne National Laboratory \\
APEX    & A Prime EXperiment \\
ATLAS   & A Toroidal LHC ApparatuS \\
ATOMKI  & Institute for Nuclear Research (Atommagkutató Intézet) \\
BABAR   & B\,$\bar{\text{B}}$ (BaBar) Experiment \\
BATH    & University of Bath \\
BBN     & Big Bang Nucleosynthesis \\
B/C     & Boron to Carbon ratio \\
BE      & Bose-Einstein \\
BEPC    & Beijing Electron–Positron Collider \\
BEPCII  & Beijing Electron–Positron Collider II \\
BES     & Beijing Spectrometer \\
BESII   & Beijing Spectrometer II \\
BESIII  & Beijing Spectrometer III \\
BGO     & Bismuth Germanate (scintillator) \\
BGOOD  & Bismuth Germanate – Open Dipole detector \\
BIC     & Binary Intra‑nuclear Cascade model \\
BM@N/BMN & Baryonic Matter at Nuclotron \\
BMON    & Beam MONitor \\
BNL     & Brookhaven National Laboratory \\
BOEFT   & Born–Oppenheimer Effective Field Theory \\
BR      & Branching Ratio \\
BSE     & Bethe-Salpeter equation \\
BSM     & Beyond the Standard Model \\
CATS    & Correlation Analysis Tool using the Schr\"odinger equation \\
CB      & Crystal Barrel \\
CBM     & Compressed Baryonic Matter experiment \\
CCSN    & Core-Collapse SuperNovae \\
CDF     & Collider Detector at Fermilab \\
CEBAF   & Continuous Electron Beam Accelerator Facility \\
CECA    & Correlation Emission function with Cluster Algorithm \\
CERES  & CErenkov Ring Electron Spectrometer \\
CERN    & European Organization for Nuclear Research \\
CF      & Correlation Function \\
ChEFT/$\chi$EFT   & Chiral Effective Field Theory \\
ChPT    & Chiral Perturbation Theory \\
CLAS    & CEBAF Large Acceptance Spectrometer \\
CLEO    & Cornell Electron Storage Ring Detector \\
CM      & Center-of-Mass \\
CMD     & Cryogenic Magnetic Detector \\
CMB     & Cosmic Microwave Background \\
CMS     & Compact Muon Solenoid \\
CNM     & Cold Nuclear Matter \\
COMET  & Coherent Muon to Electron Transition experiment (J‑PARC) \\
COMPASS & Common Muon and Proton Apparatus for Structure and Spectroscopy \\
COSY    & Cooler Synchrotron \\
COSY-TOF& Cooler Synchrotron Time‑Of‑Flight detector \\
CP(V)   & Charge Parity (Violation) \\
CR      & Cosmic Ray \\
CRYRING & Low‑energy heavy‑ion storage ring \\
CSB     & Charge-Symmetry Breaking \\
CSM     & Continuum Schwinger function Method \\
DAQ     & Data Acquisition \\
DAFNE   & Double Annular $\phi$‑Factory for Nice Experiments \\
DM      & Darm Matter \\
DRAGON  & Detector of Recoils And Gammas Of Nuclear reactions \\
DSE     & Dyson-Schwinger equation \\
DUNE  & Deep Underground Neutrino Experiment \\
DVCS    & Deeply Virtual Compton Scattering \\
DY      & Drell-Yan \\
EAS     & Extensive Air Shower \\
EFT     & Effective Field Theory \\
EIC     & Electron–Ion Collider \\
EMC     & European Muon Collaboration \\
EMFF    & Electromagnetic Form Factor \\
ELSA    & Electron Stretcher Accelerator \\
EOS     & Equation of State \\
EPECUR  & Experiment for the Pentaquark Search in the Elastic Scattering \\
EPICS   & Experimental Physics and Industrial Control System \\
ERBL    & Efremov–Radyushkin–Brodsky–Lepage \\
ESR     & Experiment and Storage Ring \\
ET      & Einstein Telescope \\
ETFF    & Electromagnetic Transition Form Factor \\
ETMC    & Extended Twisted Mass Collaboration \\
FAIR    & Facility for Antiproton and Ion Research \\
F.A.I.R.& Findable, Accessible, Interoperable, and Reusable \\
FASER   & ForwArd Search ExpeRiment \\
FD      & Forward Detector \\
FOPI    & FOur PI (4$\pi$) acceptance heavy‑ion detector \\
FPI     & Pion Electromagnetic Form Factor ($F_\pi$) \\
FPGA    & Field‑Programmable Gate Array \\
FRPC    & Forward Resistive Plate Chamber \\
FRS     & FRagment Separator \\
FS(+)   & First Science (Plus) \\
FSD     & Forward Spectator Detector \\
FSI     & Final-State Interaction \\
GALPROP & GALactic cosmic‑ray PROPagation code \\
GCR     & Galactic Cosmic Ray \\
GEM     & Gas Electron Multiplier \\
GFF     & Generalised Form Factors \\
GlueX   & Gluonic Excitations Experiment  \\
GMF     & Galactic Magnetic Field \\
GPD     & Generalised Parton Distribution \\
GPU     & Graphics Processing Unit \\
GRAAL   & GRenoble Anneau Accelerateur Laser experiment \\
GSI     & Gesellschaft für Schwerionenforschung \\
CSM     & Continuum Schwinger function Method \\
GW      & Gravitational Wave \\
GWU     & George Washington University \\
HADES   & High‑Acceptance Di‑Electron Spectrometer \\
HAL    & Hadron‑to‑Atomic‑nuclei from Lattice QCD collaboration \\
HESR    & High‑Energy Storage Ring \\
HIAF    & High Intensity heavy‑ion Accelerator Facility \\
HIC     & Heavy-Ion Collision \\
HLT     & High‑Level Trigger \\
HL-LHC  & High‑Luminosity LHC upgrade \\
HMS     & High Momentum Spectrometer \\
HQET    & Heavy Quark Effective Theory \\
HQSS    & Heavy‑Quark Spin Symmetry \\
HSD     & Hadron-String-Dynamics \\
Hyper-K & Hyper Kamiokande \\ 
HypHI   & Hypernuclear Spectroscopy with Heavy‑Ion beams experiment \\
IAEA    & International Atomic Energy Agency \\
IC      & Intrinsic Charm \\
IDE     & Integrated Development Environment \\
ILC     & International Linear Collider \\
INSIGHT & Investigations of the strong interaction in the light flavor sector \\
IRB     & InfraRed Background \\
ISGMR  & Isoscalar Giant Monopole Resonance \\
ISM     & InterStellar Medium \\
iTOF    & innter TOF \\
JPAC    & Joint Physics Analysis Center \\
J‑PARC  & Japan Proton Accelerator Research Complex \\
JLab    & Thomas Jefferson National Accelerator Facility \\
KaoS    & Kaon Spectrometer \\
KEK     & High Energy Accelerator Research Organization (Japan) \\
KLF     & $K_L$ Facility \\
KLOE    & K Long Experiment \\
LBL     & Long BaseLine \\
LEC     & Low-Energy Constant \\
LHC     & Large Hadron Collider \\
LHCb    & Large Hadron Collider beauty experiment \\
LH$_2$  & Liquid Hydrogen \\
LIGO    & Laser Interferometer Gravitational‑Wave Observatory \\
LINAC   & LInear ACcelerator (linear accelerator) \\
LL      & Lednick\'y-Lyuboshitz \\
LO      & Leading Order \\
LQCD    & Lattice Quantum Chromodynamics \\
MAID    & Mainz Unitary Isobar Model \\
MAMI    & Mainz Microtron \\
MC      & Monte Carlo \\
MDC     & Multiwire Drift Chamber \\
MESA    & Mainz Energy-Recovering Superconducting Accelerator \\
MPD     & Multi‑Purpose Detector experiment (NICA collider, JINR Dubna) \\
MRPC    & Multi‑gap Resistive Plate Chamber \\
MSVc    & Modular Start Version completion \\
MUCH    & MUon Chamber system \\
MVD     & Micro-Vertex Detector \\
MWPC    & Multi‑Wire Proportional Chamber \\
NCAL    & Neutron CALorimeter \\
NCSM    & No‑Core Shell Model \\
N/D     & Numerator–Denominator method \\
NICA    & Nuclotron‑based Ion Collider fAcility \\
NICER  & Neutron Star Interior Composition Explorer \\
NJL     & Nambu--Jona-Lasinio \\
NLEFT   & Nuclear Lattice Effective Field Theory \\
NLO     & Next‑to‑Leading Order \\
NN      & Nucleon Nucleon \\
NNF     & Nucleon-Nucleon Force \\
NNLO    & Next‑to‑Next‑to‑Leading Order \\
NOVA    & NuMI Off-axis $\nu_{e}$ Appearance \\
NREFT   & Non-Relativistic Effective Field Theory \\
NRQCD   & Non‑relativistic Quantum Chromodynamics \\
NS      & Neutron Star \\
NSE     & Nuclear Statistical Equilibrium \\
OZI     & Okubo–Zweig–Iizuka rule \\
PANDA   & antiProton ANnihilation at DArmstadt experiment \\ 
PDG     & Particle Data Group \\
PDF     & Parton Distribution Function \\
PE      & PolyEthylene \\
PHSD  & Parton–Hadron–String Dynamics (model)\\
pNRQCD  & potential Non‑relativistic Quantum Chromodynamics \\
$\pi A$ & pion-nucleus scattering \\
$\pi N$ & pion-nucleon scattering \\
$\pi p$ & pion-proton scattering \\
$pA$    & proton-nucleus scattering \\
$pN$    & proton-nucleon scattering \\
$pp$    & proton-proton scattering \\
PQ      & Peccei-Quinn \\
PSR     & Pulsating Source of Radio (Pulsar) \\
PWA     & Partial‑Wave Analysis \\
PYTHIA  & General‑purpose Monte Carlo event generator for high‑energy collisions \\
QCD     & Quantum Chromodynamics \\
QED     & Quantum Electrodynamics \\
QGP     & Quark–Gluon Plasma \\
QMC     & Quark-Meson Coupling \\
QSR     & QCD Sum-Rule \\
RF      & Radio Frequency \\
RHIC    & Relativistic Heavy‑Ion Collider \\
RIB     & Rare Isotope Beam \\
RICH    & Ring‑Imaging Cherenkov (Čerenkov) detector \\
RMF     & Relativistic Mean‑Field model \\
RSM     & Resonance Source Model \\
R3B (R$^3$B) & Reactions with Relativistic Radioactive Beams \\
SAID    & Scattering Analysis Interactive Dial‑in database \\
SFRS    & Super-FRS \\
SHINE   &  SPS Heavy Ion and Neutrino Experiment \\
SIS18   & Schwerionen‐Synchrotron 18 Tm \\
SIS100  & Superconducting Heavy‑Ion Synchrotron 100 Tm \\
SM      & Standard Model \\
SMOG    & System for Measuring Overlap with Gas \\
SNA     & Single-Nucleus Approximation \\
SOLID   & Solenoidal Large Intensity Device \\
SPALADIN & Spallation at ALADIN \\
SPD     & Spin Physics Detector \\
SPS     & Super Proton Synchrotron \\
SQL     & Structured Query Language \\
SRC     & Short Range Correlations \\
SSA     & Single‑Spin Asymmetry \\
STAR    & Solenoidal Tracker At RHIC \\
STS     & Straw Tracking Station/Silicon Tracking System \\
SQTM    & Single Quark Transition Model \\
SU      & Special Unitary \\
TCP     & Transmission Control Protocol \\
TFF     & Transition Form Factors \\
TOF     & Time‑Of‑Flight \\
TPC     & Time Projection Chamber \\
TRD     & Transition Radiation Detector \\
T2K     & Tokai to Kamioka \\
UCHPT   & Unitarized Chiral Perturbation Theory \\
UHECR   & Ultra-High Energy Cosmic Ray \\
UNILAC  & Universal Linear Accelerator \\
UPC     & Ultra‑Peripheral Collision \\
UrQMD  & Ultra‑relativistic Quantum Molecular Dynamics \\
VEPP    & e$^{+}$e$^{-}$ Collider (Novosibirsk) \\
VMD     & Vector Meson Dominance \\
WASA    & Wide‑Angle Shower Apparatus \\
WIMP    & Weakly Interacting Massive Particle \\
XENON   & XENON Dark‑Matter Experiment \\
XRISM   & X‑Ray Imaging and Spectroscopy Mission \\
YN      & Hyperon–Nucleon interaction \\
YY      & Hyperon–Hyperon interaction \\
ZDC     & Zero‑Degree Calorimeter \\
3BF     & Three Body Force \\
\end{longtable}

\end{document}